\documentclass[11pt,oneside]{uwthesis}

\usepackage{graphicx}
\graphicspath{{1B_floquet_ising_errors/}{Ising_scattering_momentum/}{Magic_in_the_Schwinger_Model/}{Phi4_Scattering/}{Scattering_Ising/}{Device_422/}}

\usepackage{multirow}
\usepackage{siunitx}
\usepackage{latexsym}
\usepackage{mathtools}

\usepackage{amsmath,amssymb,amsbsy,amsfonts}
\usepackage{array}
\usepackage{bm}
\usepackage{bbold}

\usepackage{graphics}
\usepackage[dvipsnames]{xcolor}
\usepackage{tikz}
\usetikzlibrary{quantikz2}

\usepackage{mathrsfs}
\usepackage{xcolor}
\usepackage{cancel}
\usepackage[normalem]{ulem}

\usepackage[unicode, pdfprintscaling=None, colorlinks]{hyperref}
\hypersetup{
    colorlinks,
    allcolors=blue!50!black,
}
\usepackage{xurl}
\usepackage{inconsolata}
\usepackage{makecell}
\usepackage{tabu}
\usepackage{tabularx}
\newcolumntype{Y}{>{\centering\arraybackslash}X}
\usepackage{diagbox}

\usepackage[capitalise]{cleveref}

\usepackage{relsize}
\usepackage{microtype}

\usepackage{xfrac}
\usepackage{orcidlink}
\usepackage{ulem} 
\usepackage{amssymb}
\usepackage{subcaption}
\usepackage{caption}
\usepackage{stackengine}

\usepackage{bbm}
\usepackage{rotating}

\usepackage{adjustbox}
\usepackage{placeins}
\usepackage{float}

\usepackage[title]{appendix}

\usepackage{dsfont}
\usepackage{extarrows}
\usepackage{physics}
\usepackage{textgreek}
\usepackage{soul}
\usepackage{listings}

\usepackage[utf8]{inputenc}
\usepackage[T1]{fontenc}
\usepackage{xspace}
\usepackage{nccmath}
\usepackage{lipsum}
\usepackage{algorithm2e}

\usetikzlibrary{decorations.markings, arrows, positioning, calc}
\newcommand{\specnorm}[1]{\left|\left|#1\right|\right|}
\makeatletter
\newcommand{\Biggg}[1]{\bBigg@{3.5}{#1}}
\newcommand{\Bigggg}[1]{\bBigg@{4}{#1}}
\makeatother

\DeclareMathOperator*{\argmin}{argmin}

\usepackage{cite}
\begin{document}

%
\prelimpages
\Title{Fundamental Physics at the Frontier of Noisy Quantum Computation}
\Author{Nikita Alekseyevich Zemlevskiy}
\Year{2026}
\Program{Physics}

\Chair{Martin J. Savage}{Professor}{Department of Physics}
\Signature{Lukasz Fidkowski}
\Signature{Isabel Garcia Garcia}

\copyrightpage

{\Degreetext{A dissertation\\
  submitted in partial fulfillment of the\\ requirements for the degree of}
 \def\thefootnote{\fnsymbol{footnote}}
 \let\footnoterule\relax
 \titlepage
 }
\setcounter{footnote}{0}

\setcounter{page}{-1}
\abstract{
\noindent
Quantum computing offers a new, orthogonal direction for investigating fundamental physics, extending beyond classical numerical methods and conventional observables. 
Realizing this potential requires directly confronting the noise limiting currently available quantum computers. 
Progress rests on advancing algorithms, interpreting their results, and managing their errors together.
This thesis presents several advancements in the use of quantum simulation and quantum information to probe fundamental physics.

The first is in the use of quantum computers to simulate collisions in quantum field theories. 
Central to these simulations are new wavepacket preparation and time evolution techniques that address the circuit depth bottleneck previously associated with these steps. 
Together with new error mitigation strategies well-suited to the quantum simulations, these developments allow for simulations with some of the largest effective circuit volumes to date.
These methods enable the first quantum simulation providing numerical evidence for inelastic particle production, a key process in fundamental physics.

Entanglement and magic (nonstabilizerness) generated in the dynamics of fundamental physical processes are simultaneously responsible for the classical and quantum hardness of Hamiltonian simulation. 
On quantum computers, entangling gates dominate the error budget of currently available devices, and non-Clifford operations that generate magic will consume the most resources in future fault-tolerant simulations.
Motivated by this duality, the second advancement centers on the role quantum-information-theoretic quantities play in physical processes.
Beyond mere correlations with the physics of the process, entanglement and magic are shown to probe the interactions present in scattering and hadronization dynamics.

A precision study requires a complete quantification of algorithmic and hardware uncertainties, an outstanding goal as quantum simulations mature.
The third advancement in this thesis addresses error management. 
A framework minimizing the effect of algorithmic errors in analog quantum simulations is presented. 
Device errors are confronted directly by performing an error-detected quantum computation.
In a step toward fault tolerance, encoded quantum simulations are shown to improve estimation of local observables relative to unencoded runs.

Together, the developments in this thesis mark practical progress toward fault-tolerant quantum simulations of fundamental physics capable of scientific discovery.
}

\tableofcontents
\listoffigures
\listoftables

\acknowledgments{
\noindent
I would like to first thank my parents, Aleksandra and Aleksey, for their support during my PhD and throughout my life. 
Your countless hours teaching me after school to stay ahead of the curriculum, and the projects we did around the house, instilled in me a sense that nothing is impossible if you roll up your sleeves and work hard.
Together with my brother, Georgiy, I've had the privilege of an upbringing that values hard-earned achievement and a tight-knit family with an open-door mindset.
I always look forward to our daily calls and hope to one day replicate these qualities in a family of my own.
I am also thankful to my grandparents Tamara and Grigoriy for their curiosity in my work and for listening to my endless explanations of quantum computing.
I appreciate the friends I made in high school that have remained in my life: Aleksandra Pasciak, Jessica Chen, Sara Hollenberg, Peter Lebedev, Andrew Kitain, Justo Karell, Kieran Duncan, Jesse Deluca and Zach Alper. 
My time in Stamford and visits since then would not be the same without you.
Special thanks to Aleks, Jess, and Sara for keeping me up to date on Stamford happenings; I am always excited to discuss the latest developments with you. 

I am fortunate to have many close friends from my undergraduate time at Duke University that have kept in touch. 
My time with Jimmy Shackford in Ken Brown's lab trapping ions made me realize my intellectual curiosity for quantum computing.
I have made countless memories with Jimmy since then, and deeply value our friendship through the years. 
My friends Elliott Bolzan, Ksenia Sokolova, and Maria ter Weele have supported me greatly during my PhD and I always appreciate our conversations, intellectual or otherwise.
From my freshman year on, I made friendships that have made my life rich and full: Hunter McNamara, Madeline Halpert, Amber Strange, Julia Villegas, Thara Veeramachaneni, Georgina Del Vecho, and many others. 
I am grateful that the club swim team has stuck together throughout these years. 
Jesse Yue, Maddy Bolger, Jaya Pokuri, Yingying Zhang, Hailey Prevett, Michelle Wei, Sarah Putney, Justin Kim, Lexx Pino, Grace Smith, Dennis Ling, I always look forward to our meetups and hope they will continue in the future. 

I am grateful to my undergraduate advisor Ken Brown for showing me the ropes of quantum computing and sparking my research curiosity. 
Reconnecting with you and former group members is my favorite part of March Meeting, and your continued support has made me feel at home in the wider quantum information community.
I would also like to thank my undergraduate professors Roxanne Springer and Thomas Barthel for teaching me how to learn and for first drawing me to quantum mechanics.
I always look forward to my visits back to Duke to chat about physics and life. 

My friends Jesse and Thara living in Seattle are part of the reason I chose UW for my PhD.
Since then, I am lucky to have made many friends in the area that have enriched my time as a graduate student.
When I look back on graduate school, I'll always think of my housemates Murali Saravanan, William Marshall, and Henry Froland. 
While living with your coworkers isn't always easy, I am glad we made the best of our four years together and will always think fondly of our cookouts and shared commiserations about graduate school. 
Roland Farrell instilled in me a love for mountains, and I look forward to the adventures we will have together in the future.
In the same spirit I thank Murali for accompanying me on climbs and humoring my teaching techniques.

Having friends from different parts of your life become friends with each other is one of my favorite feelings, and I am fortunate to have one large unified friend group in Seattle. 
I am glad to have shared countless trips, gatherings, and general fun with: Katie Mason, Zhiyao Li, Woody Ye, Emily Meng, Jiahui Liao, Yihui Liao, Michael Costello, Newton Kwan, Justin Wang, Vinith Sharma, Raphael Kim, Sarah Zhou, and many others.
I would like to especially thank Katie and Murali for managing to put up with me over the years.
My cohort gave me a sense of community when I started my PhD in a new city during the COVID pandemic, and I'm grateful for the highs and lows I've shared over the past six years with Kent Wilson, Jeremy Hartse, Ivan Chernyshev, Ellis Thompson, Sam Borden, Adina Ripin, and CJ Nave. 
 
I would like to thank the members of IQuS for fostering a collaborative research environment; I greatly enjoy a full office, our group meetings, and hallway chats.
I can't think of a better introduction to quantum simulation than my experience in this group; our workshops have greatly sharpened my academic interests.
I have particularly benefited from collaborations with Roland and Marc Illa, who showed me what it means to fully drive a project to completion and leave no stone unturned. 
Special thanks to my officemate and housemate Henry for making me appreciate the power of meticulous analytic calculations in our work together.
I also thank Katie Hennessey for supporting IQuS and making life easier as a PhD student.

I would like to thank my advisor Martin Savage for guiding me academically and professionally throughout my PhD.
Your persistence and willingness to hear everyone's ideas are qualities I wish to someday replicate in myself. 
The research insights and candid feedback you have given me over the years have helped me grow from a student solving problems with known answers into a researcher pursuing problems no one has attempted before.
Thank you for inspiring the tenacity to dig deeper and slight arrogance to attempt seemingly impossible research directions. 

The simulations on quantum devices in this thesis would not have been possible without the support of Sieglinde Pfaendler and the IBM Quantum Credits program. 
Finally, the work in this thesis has been supported in part, by U.S. Department of Energy, Office of Science, Office of Nuclear Physics, InQubator for Quantum Simulation (IQuS) under Award Number DOE (NP) Award DE-SC0020970 via the program on Quantum Horizons: QIS Research and Innovation for Nuclear Science.
It was also supported, in part, by the Department of Physics and the College of Arts and Sciences at the University of Washington.
}

\dedication{\begin{center}To my parents, Aleksandra and Aleksey, and my brother Georgiy\end{center}}

\textpages
\chapter{Introduction}
\label{chap:intro}
\noindent
Quantum computing seeks to harness the complexity of nature at its most fundamental scale to redefine the limits of computation.
In many ways, the field today occupies the same position classical computing did in its nascent stages, a developing technology whose ultimate reach could not have been anticipated from its earliest demonstrations.
Powered by the steady progress of Moore's law~\cite{moore1965}, classical computing has since produced some of the defining achievements of the modern era: it took humans to space~\cite{gerovitch,tomayko1988}, built the modern-day internet~\cite{bernerslee1990,leiner2009}, decoded the human genome~\cite{ihgsc2001}, learned to predict the weather~\cite{charney1950}, defeated the reigning world chess champion~\cite{campbell2002}, and, only a few years ago, solved the long-standing protein folding problem~\cite{jumper2021}.
Much like classical computing in its infancy, the greatest achievements of quantum computing are yet to be discovered.
Building on the most well-known quantum algorithm with an exponential advantage over any classical method, Shor's factoring algorithm~\cite{Shor:1994jg}, the field has since expanded well beyond cryptography.
Perhaps the most promising direction returns to Feynman's original proposal of simulating quantum systems themselves~\cite{Feynman1982,Feynman1986}, a task whose complexity overwhelms classical computers.
Quantum simulations are expected to help design less energy-intensive fertilizers~\cite{Reiher:2016apy}, accelerate the discovery of new drugs and materials~\cite{cao2019}, enable the search for room-temperature superconductors~\cite{McArdle:2018tza}, and simulate the strong interactions of quarks and gluons~\cite{Bauer:2022hpo}. 
Motivated by the applications still waiting to be found, this thesis pursues quantum simulation as a tool for fundamental physics.

Present-day hardware, however, is still far from the machines these applications will ultimately require. 
A useful benchmark for the capability of near-term devices is the notion of quantum utility~\cite{Kim:2023bwr}: a quantum computation that exceeds brute-force classical techniques and is only accessible to approximate classical methods.
With the availability of devices with increasingly large numbers of qubits on superconducting~\cite{Kim:2023bwr,Gao:2024fik,GoogleQuantumAIandCollaborators:2024efv} and neutral-atom platforms~\cite{Bluvstein:2023zmt,Manetsch:2024xrn}, demonstrations surpassing brute-force classical methods have become increasingly common.
A closely related but more demanding goal is practical quantum advantage: a computation whose result is out of reach for any classical method~\cite{Daley:2022eja}. 
Demonstrations of quantum advantage have so far been limited to contrived computations~\cite{Arute:2019zxq,Gao:2024fik}.
Claims of quantum utility and advantage are often challenged with rapidly advancing classical techniques~\cite{Tindall:2023cqi,Pan:2021krm}, representative of the evolving frontier of quantum capabilities.
Quantum simulation is widely regarded as one of the most promising directions on the path to scientifically useful quantum advantage~\cite{Daley:2022eja,Bauer:2023qgm}, since the systems of greatest scientific interest are often precisely those whose physics hinders an efficient classical representation~\cite{preskill2018simulating}.

Present-day devices operate in what is known as the noisy intermediate-scale quantum (NISQ) regime: quantum processors with tens to hundreds of qubits, capable of executing circuits of modest depth before the accumulation of hardware error overwhelms the signal of interest~\cite{Preskill:2018jim}. 
In this regime, the fault-tolerance (FT) assumed by many quantum algorithms with proven exponential speedups~\cite{10.1098/rspa.1998.0164}, including Shor's algorithm itself~\cite{Cain:2026rmb}, remains out of reach~\cite{Preskill:2018jim,Bharti_2022}. 
In contrast, quantum simulation has progressed through a combination of steadily improving hardware, in qubit count, gate fidelity, coherence time, and algorithms explicitly designed around the constraints of near-term devices. 

The value of NISQ simulations, however, is not measured only by the observables they produce. 
Much has been learned about physical systems from mapping them onto noisy devices and making sense of the imperfect results. 
Encoding a physical theory onto hardware and extracting a reliable signal from noisy data surfaces structure conventional analysis can leave implicit: which features of a system are robust to noise, how information is organized within it, and how that structure shapes what a computation can recover.
This effort is equally beneficial for quantum information science.
Confronting real hardware noise is what has driven the field to quantify the buildup of complexity in realistic circuits, and to develop the error management techniques that recover a signal from noisy data.
These are the same resources and methods that will set the cost of computation well into the FT era. 
Several results in this thesis began in exactly this way, as questions about what a simulation would cost or whether its output could be trusted, and matured into conclusions in physics and quantum information in their own right. 
In this sense near-term devices serve not only as computational instruments but as a lens that brings hidden features of familiar systems into view.

Two broad strategies exist for implementing a quantum simulation of dynamics. 
Digital quantum computing decomposes the desired unitary into a sequence of discrete logical gates, in direct analogy to a classical circuit, and is the strategy pursued throughout most of this thesis.
Analog quantum computing instead realizes a target Hamiltonian by continuously modulating the physical parameters of an experimental apparatus, and is possible whenever the physics of interest can be mapped onto a simulator's native interactions exactly, or with controllable error~\cite{RevModPhys.86.153}.
Hybrid digital-analog quantum simulation protocols have also been demonstrated~\cite{Arrazola_Pedernales_Lamata_Solano_2016, Lamata_2018,Andersen:2024aob}, which extend the reach of analog-only methods.
Analog platforms are typically more restrictive than fully digital ones, but their comparatively simple control requirements have enabled system sizes large enough to exceed classical simulability~\cite{https://doi.org/10.48550/arxiv.2204.13644,scholl_schuler_williams_eberharter_barredo_schymik_lienhard_henry_lang_lahaye_et_al._2021,ebadi_wang_levine_keesling_semeghini_omran_bluvstein_samajdar_pichler_ho_et_al._2021,Daley:2022eja}.

Noise intrinsic to present-day hardware affects analog and digital simulations alike, and noise management techniques are crucial in any quantum computation.
At one end of these techniques lies error mitigation, a near-term technique suited to today's noisy devices.
Error mitigation accepts that noise will corrupt a computation, and instead corrects for its effect in post-processing, at the cost of an overhead in the number of circuit repetitions required for a reliable estimate~\cite{Temme:2016vkz,Cai:2022rnq,Liao:2023eug,Kandala:2018kwe,Kim:2021gvc}.
Quantum error correction (QEC), at the opposite end, encodes logical information redundantly across many physical qubits so that errors can be actively detected and corrected during the computation itself.
The overhead required for full FT has so far placed this approach out of reach for practical simulations~\cite{Steane:1996ghp,Steane:1996va,Gottesman:1997zz,Gottesman:1997qd,Aliferis:2005ftz,GoogleQuantumAIandCollaborators:2024efv}.
Between these lies quantum error detection (QED), a lower-overhead simplification of QEC that still encodes logical information, but discards corrupted results rather than actively correcting them~\cite{Linke:2017bvn,Roffe:2018oim}.
This intermediate approach has recently enabled several demonstrations of an improvement over an unencoded computation, a milestone known as beyond-break-even performance~\cite{Gottesman:2016gef,Dasu:2026dwm,Perlin:2026mph,Rodriguez:2024bhh,Reichardt:2024xfs,Bluvstein:2025ped,Chen:2021num,GoogleQuantumAIandCollaborators:2024efv,Hetenyi:2024zvf,Caune:2024doa}.
These results mark a regime in which error management techniques on present-day hardware are beginning to deliver measurable benefit, bringing scientifically useful quantum simulation within closer reach.

Efforts in fundamental physics span three principal frontiers. 
One pushes toward smaller distances and higher energies, where the fundamental constituents of matter and their interactions are probed directly. 
A second pushes toward larger scales, concerning the evolution of stars, galaxies, and the universe as a whole. 
In contrast to the first two frontiers, the third pushes toward greater complexity, where many interacting degrees of freedom become so correlated that a system's information resides in these correlations instead of its constituents~\cite{Preskill:2012tg}.
Quantum computing, and quantum information science more broadly, is native to the third frontier.
Its resources scale with the same exponentially growing Hilbert space that renders generic many-body quantum systems intractable to classical methods, giving it direct access to real-time dynamics that classical methods generically struggle to capture~\cite{Bharti_2022,Banuls:2019bmf}.
However, the impact of quantum information on fundamental physics is not restricted to the complexity frontier.
In the high-energy frontier, quantum simulations promise first-principles access to scattering and hadronization ordinarily reached only in collider experiments~\cite{Bauer:2023qgm,Bauer:2025nzf}, as well as to nuclear structure and reactions~\cite{Bauer:2022hpo,Savage:2023qop,Klco:2021lap}.
In the large-scale frontier, quantum simulations offer a potential route to dynamics such as false-vacuum decay, collective neutrino oscillations in core-collapse supernovae, and the nonequilibrium evolution of the early universe, that is otherwise accessible only through indirect observation~\cite{Bauer:2022hpo,Bauer:2023qgm}. 

First-principles calculation is the pinnacle of theoretical scientific investigation: a prediction derived directly from a theory's fundamental degrees of freedom, without relying on phenomenological models.
Lattice field theory provides first-principles predictions for static, equilibrium quantities, from the masses of the light hadrons~\cite{BMW:2008jgk} to the neutron-proton mass splitting~\cite{Beane:2006fk} and the order of the finite-temperature transition of quantum chromodynamics (QCD)~\cite{Aoki:2006we}.
These successes, however, are largely confined to quantities that do not require following a system's dynamics in real time.

This restriction reflects a general strategy underlying most classical approaches to strongly interacting physics. 
A description is built around a simple reference point, with its accuracy degrading as the true physics moves away from that reference.
Effective field theory constructs a systematically improvable description by discarding the effects of physics at energies far above a chosen scale~\cite{PhysRevB.4.3174,Weinberg:1978kz}, while perturbation theory constructs an analogous expansion around a weakly coupled reference theory.
Both strategies organize the space of physical states around an axis defined by a tractable limit, with hardness increasing away from it.
Lattice Monte Carlo methods, the workhorse of ab initio equilibrium QCD, are organized around a closely related axis: theories whose path-integral weight can be interpreted as a probability distribution.
When this weight becomes complex, such as for real-time evolution or at finite baryon density, importance sampling breaks down and the cost of a reliable estimate grows exponentially with the spacetime volume simulated, an obstruction known as the sign problem~\cite{Troyer_2005,PhysRevD.86.105012}.

Another such axis concerns entanglement. 
Matrix product states (MPS) represent a quantum state efficiently whenever the entanglement across any spatial bipartition remains modest. 
This has been shown to hold for the ground states of gapped, one-dimensional systems~\cite{Hastings:2007iok,PhysRevLett.93.227205}, and conjectured to be true for a broad class of gapped systems~\cite{Eisert:2008ur}.
Matrix product states are the basis of one of the most successful classical tools for simulating strongly correlated matter~\cite{White:1992zz,Verstraete:2008qpa}, and have been applied to a wide range of systems. 
Real-time dynamics again proves to be an exception: the typical rapid growth of entanglement under time evolution eventually outpaces what a MPS can efficiently represent~\cite{Calabrese:2005in}.

Yet another axis along which the Hilbert space may be organized is nonstabilizerness, or magic. 
This axis organizes states by their distance from stabilizer states, the reference set that the Gottesman-Knill theorem guarantees can be simulated efficiently on a classical computer regardless of how entangled they become~\cite{Aaronson:2004xuh}.
Magic is the resource required to escape this efficiently simulable regime~\cite{Bravyi:2004isx,Emerson:2013zse,Chitambar:2018rnj,Howard:2017maw,Heinrich:2019aei}. 
In this sense, magic is a measure of hardness distinct from entanglement: a state may possess one in abundance while lacking the other entirely.

Entanglement and magic also pose challenges on quantum computers, and noise is what separates the regime in which each dominates.
Entangling gates account for most of the error budget of every present-day platform, making entanglement the limiting resource on hardware available today.
On the other hand, a fully FT quantum computer protects Clifford operations natively and non-Clifford gates are costly. 
This shifts the dominant cost from entanglement to magic as FT is approached~\cite{Bravyi:2004isx}.
These axes, and the classical methods organized around them, recur throughout this thesis.

\begin{figure}
    \centering
    \includegraphics[width=0.5\linewidth]{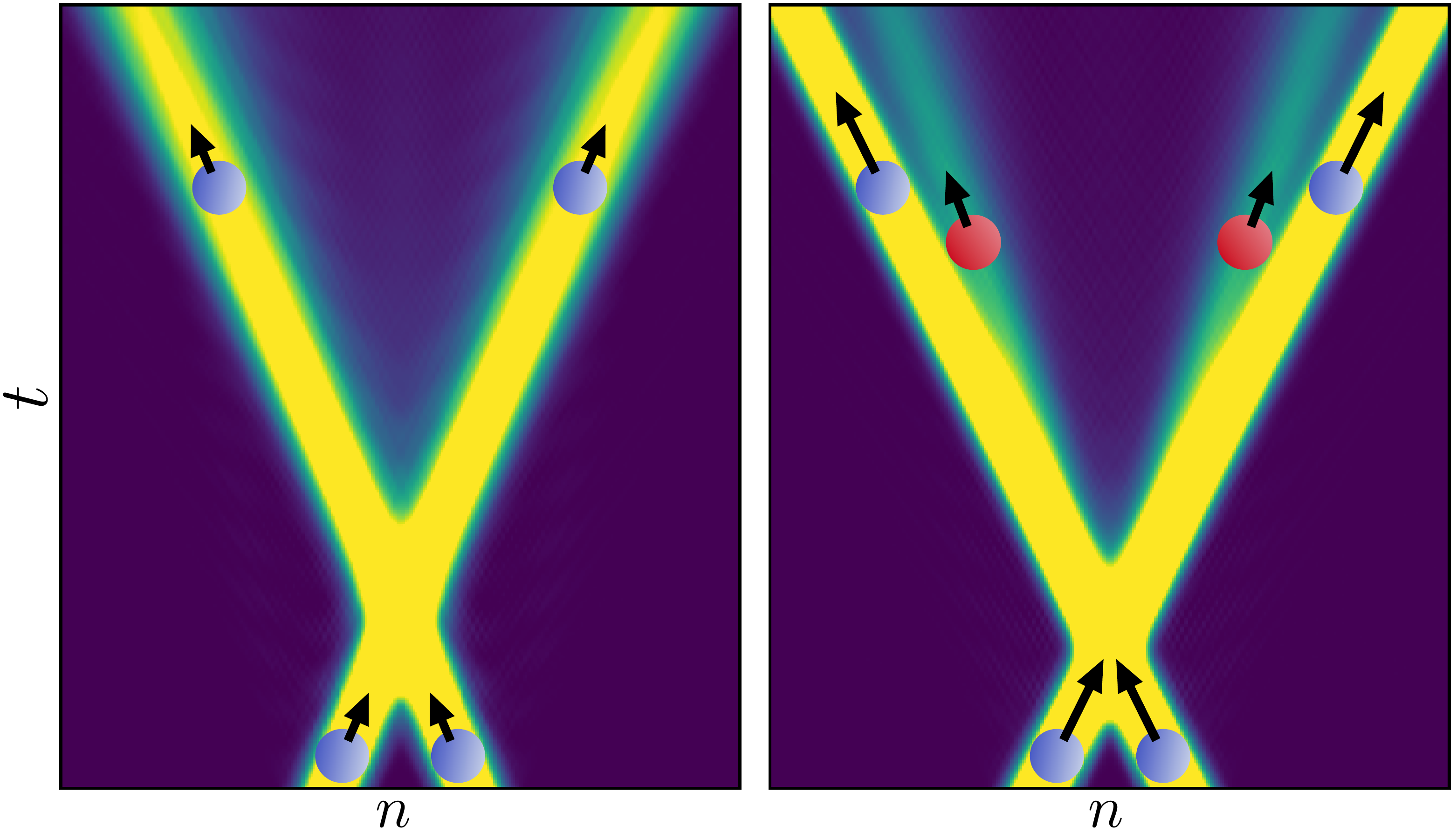}
    \caption{\textit{Low-energy and high-energy collisions.} 
    The heatmap shows the energy density in an example of elastic and inelastic collisions as a function of position $n$ and time $t$. 
    Propagating particles are identified as beams of energy with a constant velocity.
    Left: the elastic scattering of two light particles (blue). Right: an inelastic process that produces a heavy particle (red) which exits the collision slower due to momentum conservation.
    This figure is adapted from Ref.~\cite{Farrell:2025nkx}.}
\label{fig:elastic_vs_inelastic}
\end{figure}

The first part of this thesis explores the use of quantum computers to simulate scattering in quantum field theories (QFTs), a process whose dynamics generically lie far from these classically tractable axes.
High-energy particle collisions can convert energy into matter through the inelastic production of new particles.
The study of these collisions is relevant to the behavior of matter in extremely dense and hot environments and is probed experimentally at particle colliders.
Figure~\ref{fig:elastic_vs_inelastic} shows two examples of particle collisions.
Elastic scattering is shown on the left panel, and an inelastic collision producing a new particle is shown on the right.
Predicting the outputs of high-energy, high-multiplicity collisions is a central aim in fundamental physics.

Scattering makes it possible to probe the dynamics and interactions of a QFT in a controlled, idealized version of real particle collisions.
Particles initially in well-separated asymptotic states are brought into contact. 
The resulting transition amplitudes encode the full nonperturbative content of the theory, from the masses and couplings of its stable excitations at low energies to the inelastic production of new particles at high energies.
Classical lattice techniques have been remarkably successful at extracting the former class of observables directly from a theory's path integral, including particle masses, matrix elements, and scattering phase shifts inferred through finite-volume methods~\cite{Luscher:1985dn,Luscher:1986pf,Beane:2008dv,Beane:2010em,Luu:2011ep,Briceno:2013hya,Hansen:2014wea,Kaplan:1998tg,Kaplan:1998we,Beane:2006mx}.
These methods are intrinsically restricted to processes for which such finite-volume relations exist, and even where applicable they yield only asymptotic, infinite-time information rather than the dynamics of the collision itself~\cite{Beane:2010em}. 
Extending classical lattice methods to genuine real-time dynamics is obstructed by the same sign problem that limits efficient Monte Carlo simulations of finite-density fermionic systems. 
Further, real-time collision dynamics is exactly the regime where MPS simulations struggle because of the large amount of entanglement that is found to be generated in high-energy inelastic scattering~\cite{Milsted:2020jmf,Vanhecke:2021noi,Van_Damme_2021,Rigobello:2021fxw,Belyansky:2023rgh, Papaefstathiou:2024zsu}.
Quantum computers, whose resource requirements do not carry the exponential cost associated with classical techniques, offer a route to first-principles calculations of collision dynamics directly.
This motivates a broad effort to develop the algorithms and techniques necessary to realize this promise for problems in fundamental physics~\cite{Bauer:2023qgm,DiMeglio:2023nsa}.

A quantum simulation of a collision proceeds through several stages: after a theory's vacuum is prepared, wavepackets representing well-separated, asymptotically free particles are initialized with definite momenta, and the resulting state is time-evolved through the collision and beyond. 
The seminal work of Jordan, Lee, and Preskill (JLP)~\cite{Jordan:2011ci,Jordan:2012xnu,Jordan_2018} proposed creating wavepackets in the free theory where creation and annihilation operators are known and adiabatically turning on interactions.
While various alternatives and approximate methods were developed following the JLP proposal, they incurred a prohibitive cost in the form of exponential classical overhead~\cite{Farrell:2024fit,Zemlevskiy:2024vxt}, or polynomial circuit depth scaling with the size of the wavepacket~\cite{Davoudi:2024wyv,Chai:2023qpq,Jordan:2011ci,Hite:2025pvb,Turco:2023rmx,Turco:2025jot}. 
While polynomial resource scaling is formally efficient, it precludes the use of these methods for large-scale state preparation in near-term simulations.
Building on this foundation, quantum simulations of scattering have since been demonstrated across a range of theories and platforms~\cite{Dreher:2022scr,Turco:2023rmx,Kreshchuk:2023btr,Chai:2023qpq,Briceno:2023xcm,Sharma:2023bqu,Turro:2023dhg,Wang:2024scd,Wu:2024rod,Bennewitz:2024ixi,Turro:2024ksf,Yusf:2024igb}. 
Because both this preparation step and the subsequent time evolution must be implemented with circuit depths compatible with the coherence times of present-day hardware, the wavepacket sizes, propagation times, and lattice volumes accessible on real devices have so far been limited. 
Overcoming this circuit-depth bottleneck, for both state preparation and time evolution, is the problem addressed by the first part of this thesis.

Chapter~\ref{chap:phi4_scattering} addresses this problem in the context of scalar field theory in one spatial dimension.
This is the theory of coupled harmonic oscillators with a quartic potential that governs interactions between particles. 
Scalar and pseudoscalar particles play key roles in many areas, including nuclear physics~\cite{Weinberg:1978kz,Holland:2019zju}, condensed matter physics~\cite{Shankar:1993pf,Schafer:2006yf}, high energy physics~\cite{Higgs:1964pj,Higgs:1964ia,Caprini:2005zr}, as well as cosmology~\cite{PhysRevD.23.347} and physics beyond the Standard Model~\cite{Peccei:1977hh,Peccei:2006as,Rinaldi:2021jbg}.
Scalar fields have also been used as a venue to study phase transitions and spontaneous symmetry breaking~\cite{Rychkov:2014eea,Rychkov:2015vap,Thompson:2023kxz}. 
Moreover, scattering in scalar field theory has been shown to be BQP-complete~\cite{Jordan_2018}, meaning all ``quantum-efficient'' problems may be solved by mapping them to quantum simulations of scattering with various initial conditions. 

Variational algorithms have recently emerged as a promising avenue for such computations~\cite{McClean:2015vup,Cerezo:2020jpv,Biamonte:2021ntr}. 
These hybrid quantum-classical techniques employ parameterized circuits whose gates depend on a set of tunable parameters. 
The circuit is executed on the quantum computer to evaluate a cost function, and a classical optimizer uses the result to propose the next set of parameters, iterating this loop until the cost is minimized.
Variational algorithms have been developed for a variety of tasks, famously for ground-state preparation using the Variational Quantum Eigensolver (VQE)~\cite{Peruzzo:2013bzg}, but also for combinatorial optimization~\cite{Farhi:2014ych}, quantum machine learning~\cite{Biamonte:2016ugo}, linear algebra problems~\cite{Bravo-Prieto:2019kld}, quantum simulation~\cite{Yuan:2018jdl}, and other applications. 
The power of these algorithms stems from their ability to significantly compress circuits that would otherwise be too deep for current hardware. 
A central obstacle to training such circuits is the barren plateau phenomenon, in which the gradient of the cost function vanishes exponentially with system size for generic ansatze~\cite{Larocca:2024plh,Cerezo:2023nqf,Anschuetz:2022wvo}. 
Fortunately, this obstruction is substantially reduced by incorporating the physical structure of the problem being simulated directly into the ansatz~\cite{Barthel:2023ine}.

Guided by the symmetries and correlation structure of the states involved, Chapter~\ref{chap:phi4_scattering} describes the development of scalable variational circuits that compress all parts of a quantum simulation.
The vacuum of the theory is prepared using symmetry-preserving energy minimization circuits determined on classical computers through the variational algorithm ADAPT-VQE~\cite{Grimsley:2018wnd}.
Fixed-size wavepacket initialization and time evolution circuits for a scattering simulation are compressed using hardware-efficient brickwall circuits with depths independent of system size.
While this method still carries exponential classical overhead with the size of the initial wavepacket, it represents a significant reduction in the depth required over previous techniques.
Together with a new error mitigation strategy based on the classically known time evolution of the vacuum, these techniques are used to simulate the scattering of two wavepackets on 120 qubits of IBM's superconducting quantum computer {\tt ibm\_fez}.
Circuits with up to 4924 two-qubit gates and a two-qubit gate depth of 103 are run, significantly pushing the capabilities of the quantum computer.
The resulting observables are found to be in qualitative agreement with classical MPS simulations, and the effect of interactions on the collision is clearly resolved.
These simulations constitute the first quantum simulation of wavepacket scattering in an interacting quantum field theory carried out at this scale.

Chapter~\ref{chap:ising_scattering} targets wavepacket preparation directly, removing the exponential classical overhead that remained in the previous approach of Chapter~\ref{chap:phi4_scattering}. 
A new algorithm is introduced which leverages mid-circuit measurements and classical feedforward (MCM-FF) to prepare wavepackets in depth independent of their size. 
While any quantum operation can in principle be implemented using unitary gates alone, MCM+FF can substantially reduce the resources required for certain state-preparation tasks, and is required for QEC implementations. 
Qubits are measured partway through the circuit, and the outcomes are used to apply corrective gates to the remaining qubits, projecting the system onto the target state. 
The similarity of momentum eigenstates to W states~\cite{Dur:2000zz}, which have been studied at length in applications to quantum communication, sensing and optimization~\cite{Agrawal2006,Wang2007,Liu2011,Li_2007,Joo_2003,Wang_2020,9259949,Catalano:2024bdh}, allows for a great reduction in the resources required for wavepacket preparation.
The first step of the algorithm prepares a state establishing the spatial profile and momentum content of the target wavepacket, but containing high-energy and multi-particle contributions. 
Constant-depth MCM-FF circuits for W state preparation~\cite{Piroli:2024ckr,Buhrman:2023rft,Piroli:2021fjn,Smith:2022nbd,Cruz_2019} are adapted to prepare this wavepacket ``skeleton''. 
A single-particle wavepacket is then formed by projecting the skeleton state onto a single-particle eigenstate. 
This is done using ADAPT-VQE (the same approach used for vacuum preparation in Chapter~\ref{chap:phi4_scattering}). 
The resulting circuit depth is independent of both the wavepacket size and the spatial dimension of the lattice, a superexponential improvement over previous approaches.

This algorithm is broadly applicable across lattice models and Chapter~\ref{chap:ising_scattering} uses it to prepare wavepackets in one- and two-dimensional Ising field theory, scalar field theory, and the Schwinger model.
Chapter~\ref{chap:ising_scattering} then applies this algorithm on quantum hardware to simulate inelastic particle production in one-dimensional Ising field theory.
The Ising field theory is a striking illustration of how rich, nonperturbative physics can emerge from a simple nearest-neighbor spin interaction, making it a natural and versatile testbed for real-time quantum simulation. 
Despite its simplicity, it supports a nontrivial spectrum and phase structure, and recent quantum simulations have used it to probe phenomena including many-body localization~\cite{Shtanko:2023tjn}, discrete time crystals~\cite{Shinjo:2024vci}, Majorana edge modes~\cite{Mi:2022egw}, thermalization~\cite{Jaschke:2019jka}, confinement and string breaking~\cite{De:2024smi,Surace:2020ycc,Kormos:2016osj}, false-vacuum decay~\cite{Milsted:2020jmf,Pavesic:2025nwm,Borla:2026fdb}, and scattering~\cite{Jha:2024jan}.
Wavepacket preparation circuits determined through our two-step algorithm initialize a scattering simulation on 104 qubits of IBM's {\tt ibm\_marrakesh}.
Trotterization, the standard method of decomposing the time evolution unitary into digital gates~\cite{Lloyd1073,Suzuki:1991jtk}, is used to access dynamics well beyond the collision with the application of 45 Trotter steps and 5589 two-qubit gates.
The energy density of the post-collision state near the outgoing particles is skewed if a slower-moving particle is produced in an inelastic collision (see the right panel of Fig.~\ref{fig:elastic_vs_inelastic} for an example).
A heavy particle produced in the collision is identified from the resulting skewness, providing evidence for inelastic particle production in a quantum simulation for the first time.

At the time they were performed, the simulations in Chapters~\ref{chap:phi4_scattering} and~\ref{chap:ising_scattering} were among the largest carried out on quantum hardware by effective circuit volume. 
Although both of these one-dimensional simulations were classically simulable with MPS methods, the entanglement generated in the wake of inelastic collisions was a challenge for MPS.
Pushing quantum computations of scattering past the capabilities of approximate classical methods would likely require a simulation of inelastic collisions in two dimensions.
Further, the dense circuits used to simulate the post-collision states caused the reverse light cone of the final observables (i.e., which qubits are causally connected to the observable) to span almost the entire device. 
As a result, noise from every qubit in the device contributed to the measured signal.
This regime, in which circuit depth and qubit count are both pushed toward the limits of the hardware, is especially challenging. 
Noise accumulates precisely where entanglement must be built, and error mitigation is essential to recovering a physically meaningful signal from the resulting data. 
The techniques developed in Chapters~\ref{chap:phi4_scattering} and~\ref{chap:ising_scattering} are representative of a broader theme spanning the frontier of near-term quantum computation: the same physical regimes that challenge classical simulation are often the ones in which quantum hardware is most susceptible to noise.


The second part of this thesis turns from manipulating quantum information in simulations to studying its properties in fundamental physics processes. 
Measures of quantum complexity such as entanglement and magic are sensitive probes capable of revealing structure that is invisible to conventional observables (for a recent review, see Ref.~\cite{Robin:2026lqp}).
The first of these, entanglement, is captured by the Schmidt decomposition.
A pure state $|\psi\rangle$ defined on a bipartition of a system into subsystems $A$ and $B$ can be decomposed into Schmidt vectors $|\psi\rangle = \sum_i \lambda_i |i\rangle_A |i\rangle_B$.
The Schmidt coefficients $\lambda_i$ quantify the entanglement between the two subsystems. 
When a single coefficient dominates, the state is well approximated by a product state, while a broad distribution of coefficients indicates significant quantum correlations across the bipartition. 
Beyond quantifying the degree of entanglement, individual Schmidt vectors have been shown to carry direct physical meaning in a range of contexts, including string breaking and hadronization~\cite{Grieninger:2026bdq,Florio:2024aix}, the characterization of topological features~\cite{Li:2008kda,Fidkowski:2010nhf}, and the identification of quasiparticle excitations~\cite{Zauner-Stauber:2018gqr,Cocchiarella:2025mtv}.

A scattering event in a quantum field theory is a coherent superposition of every process consistent with its symmetries and kinematics.
Resolving this superposition into its individual channels (i.e., decomposing the inclusive final state into exclusive channels corresponding to each allowed process) is essential to identifying the underlying interaction mechanisms and extracting channel-specific observables. 
Collider facilities such as RHIC and the LHC are built on the principle that channel-specific observables carry far more information than inclusive measurements~\cite{ParticleDataGroup:2024cfk}.
For instance, the relative abundances of color-singlet configurations after a heavy-ion collision can reveal the mechanism of hadronization~\cite{Rafelski:1982pu,ALICE:2016fzo}.
The same principle applies across energy scales, from nuclear decay channels relevant to nucleosynthesis pathways~\cite{Burbidge:1957vc} to quantum chemical reactions where several processes occur in superposition~\cite{Kassal:2010xwg,Manthe2016Smatrix}.
As discussed above, classical simulations of collisions in one-dimensional systems with moderate growth of entanglement are possible using Matrix Product States (MPS), which faithfully represent the post-scattering state with a tractable bond dimension.
While such simulations provide access to the inclusive post-scattering wavefunction, extracting the contributions of individual channels from this many-body state is nontrivial. 

Resolving individual channels from inclusive measurement of energy density alone is impossible, and another ingredient is needed.
Chapter~\ref{chap:exclusive_channels} develops a method for this extraction in MPS simulations, building on the evidence for inelastic particle production established in Chapter~\ref{chap:ising_scattering}. 
Existing approaches to channel resolution in simulations of scattering require prior knowledge of the asymptotic particle content of the theory~\cite{Jha:2024jan,Vanderstraeten:2013xda,Haegeman:2013xcv}. 
Instead, an experimentally inspired method is introduced that isolates individual scattering channels directly from the entanglement structure of the late-time wavefunction.
The simplest example of this method is the case of channels of different outgoing particle species with distinct momenta. 
In this case, particles in different channels occupy distinct spatial regions at late times, and a Schmidt decomposition at the appropriate bipartition separates them naturally, with each Schmidt vector corresponding to a specific channel.
This way, the wavefunctions and associated abundances corresponding to exclusive channels are extracted.
Matrix product states maintain spatial locality of degrees of freedom in a wavefunction, so this method is naturally suited to collision events where the outgoing channels are spatially separated.
However, this channel isolation method can in principle be applied to any quantum number that distinguishes the channels, provided a Schmidt decomposition could be practically taken.
Extending this to more complicated scattering events, such as higher-order processes and collisions in higher dimensions is discussed in Chapter~\ref{chap:exclusive_channels}.
Channel isolation in quantum simulations of scattering could be done in an analogous manner, with ``particle detectors'' playing the role of Schmidt decompositions and classifying results into channels shot by shot.

The method to resolve channels is applied to MPS simulations of the same one-dimensional Ising field theory scattering process studied in Chapter~\ref{chap:ising_scattering}. 
Schmidt decompositions performed at strategically chosen spatial bipartitions separate the distinct outgoing particles, deterministically identifying the elastic and inelastic channels. 
This method independently confirms the production of the heavy particle in that process, this time through the entanglement structure of the post-collision state rather than the skewness of its energy density.
Branching ratios for the elastic and inelastic channels are found to be in agreement with previous results~\cite{Jha:2024jan}.

Beyond entanglement, magic offers a complementary lens through which the complexity of a state may be investigated.
A variety of magic monotones have been developed to quantify magic, including measures based on the flatness of the Schmidt spectrum $\{\lambda_i\}$~\cite{Tirrito:2023fnw,Leone:2021rzd} and the stabilizer Rényi entropy~\cite{Leone:2021rzd,Haug:2023hcs,Tarabunga:2023hau,Haug:2024ptu,Hamaguchi:2023zpb}, several of which can be computed directly from the MPS representation of a wavefunction.
Unlike entanglement entropy, which is inherently defined with respect to a bipartition, magic in its basic form requires no such division of the system.
A bipartition can nonetheless be introduced to resolve how magic is distributed spatially. 
Nonlocal magic isolates the portion of a state's magic that cannot be attributed to its individual subsystems alone, providing a probe of extended quantum correlations distinct from those captured by entanglement. 

\begin{figure}
    \centering
    \includegraphics[width=0.5\linewidth]{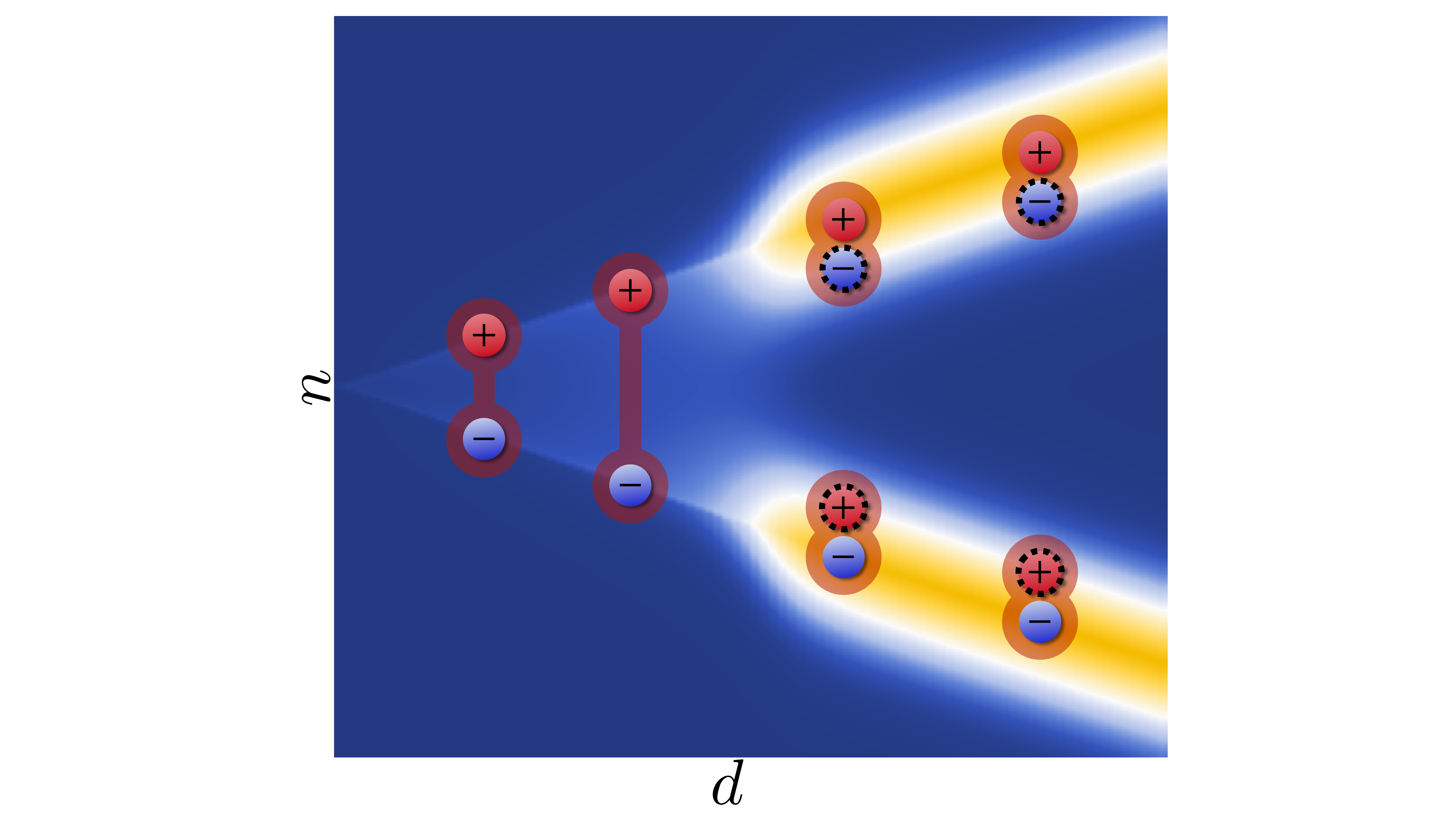}
    \caption{{\it The formation of hadrons in the string breaking process.}
    The energy density, indicated by the heatmap color, shows the linearly increasing potential with the separation $d$ of external charges (red and blue circles). A flux tube connecting the external charges develops (red tubes).
    At a certain separation, a pair of hadronic bound states is energetically favorable. 
    Dynamical charges (circles with dashed border) are extracted from the vacuum to screen the external charges. 
    This figure is adapted from Ref.~\cite{Grieninger:2026bdq}.
    }
    \label{fig:string_breaking_schematic}
\end{figure}

The formation of flux tubes, or chromoelectric strings, between color charges is a primary emergent feature of QCD~\cite{Gross:1973id,Politzer:1973fx}.
At modest separations between static quark-antiquark pairs, the nonlinearity of the gluon self-interactions~\cite{PhysRev.96.191}, combined with quantum fluctuations, confine the flux tubes into one-dimensional ``strings''.
As the separation increases, the energy stored in the flux tube grows approximately linearly until a two-hadron state becomes energetically favored.
Such final states correspond to dynamical quarks from the vacuum rearranging themselves into baryon number $\pm 1/3$ configurations around the static charges.
This process is known as string breaking and is schematically shown in Fig.~\ref{fig:string_breaking_schematic}.
In nature, string breaking occurs in high-energy collisions that produce high-multiplicity final states of strongly interacting particles.
While the dynamics of particle creation through string breaking, fragmentation and hadronization is well modeled~\cite{Field:1977fa,Andersson:1983ia,andersson1998lund,Sjostrand:2019zhc,GEANT4:2002zbu}, 
high-precision predictions in environments far from present-day experiments are challenging.
Such predictions will be crucial at the future Electron-Ion Collider for determining nucleon partonic distributions, measuring gluon helicity contributions, and understanding diffractive dijet production~\cite{Abir:2023fpo,AbdulKhalek:2021gbh}, 
and for the discovery of new fundamental physics at the LHC.

Significant theoretical and numerical results exist for string breaking processes 
(for recent overviews, see Refs.~\cite{Halimeh:2025vvp,Kharzeev:2026jkq}).
Classical lattice QCD simulations examine the electric potential between static charges as a function of their separation~\cite{PhysRevD.59.031501,CP-PACS:1998hwq,SESAM1998209,Bali:2005fu,Pennanen:2000yk,PhysRevD.63.111501,Bulava:2019iut}.
Hamiltonian and tensor network methods complement these techniques, providing access to real-time dynamics and quantum observables~\cite{Buyens:2015tea,Grieninger:2025rdi,Florio:2025hoc, Florio:2023dke, Florio:2024aix, Janik:2025bbz,Barata:2025hgx,Artiaco:2025qqq,Verdel:2019chj,Verdel:2023mmp,Mallick:2024slg} and allowing studies in two dimensions~\cite{Cochran:2024rwe,Gonzalez-Cuadra:2024xul, Borla:2025gfs,Cataldi:2025cyo,Xu:2025abo,DiMarcantonio:2025cmf,Ciavarella:2024fzw,Ciavarella:2024cyt}.
Recent advances in experimental quantum technologies have also enabled string breaking simulations on quantum hardware, e.g., Refs.~\cite{Crippa:2024hso,Liu:2024lut,De:2024smi,Surace:2024bht,Ciavarella:2024lsp,Alexandrou:2025vaj,Luo:2025qlg}.

Building on this progress, Chapter~\ref{chap:schwinger_magic} explores the evolution of quantum complexity in a strongly interacting system undergoing string breaking using MPS.
The Schwinger model~\cite{Schwinger:1962tp} is the one-dimensional theory of quantum electrodynamics which is a confining gauge theory sharing many features with QCD.
These include the presence of a chiral condensate, charge screening, and few-body hadronic bound states.
This has made the Schwinger model a well-established testbed for quantum simulations of gauge theories~\cite{Martinez:2016yna,PhysRevA.90.042305,Muschik:2016tws,Klco:2018kyo,Farrell:2023fgd,Farrell:2024fit,Nguyen:2021hyk}. 
Rather than simulating dynamical screening and vacuum rearrangement directly, string breaking is studied in a minimal setting.
External charges are inserted at various separations $d$, and the ground state at each $d$ encodes the rearrangement of degrees of freedom inherent to string breaking.
A suite of entanglement- and magic-based complexity measures is computed using MPS computations across the string breaking transition.
Quantum complexity of the state is observed to peak at the point of string breaking, and is found to act as an ``X-ray'', revealing hadronic structure hidden to conventional observables.
Further, computations of nonlocal magic and mutual information reveal the presence of previously unknown nonlocal quantum correlations along the string. 
Experimental efforts are underway to measure magic and entanglement in hadronization following high-energy quark collisions~\cite{White:2024nuc,Yazgan:2025pah}.
These correlations, if detected in experiment, may affect fragmentation and hadronization modeling. 

Since entanglement and magic are the very resources that limit classical simulability, characterizing their behavior in Chapters~\ref{chap:exclusive_channels} and~\ref{chap:schwinger_magic} indicates which physical processes will ultimately require the quantum simulations developed at the frontier of near-term hardware to study directly.
Perhaps more importantly, this work, originally motivated by understanding the resources a quantum simulation would consume, recasts entanglement and magic as probes of intrinsically quantum correlations in fundamental physics.


The third part of this thesis turns to a question left implicit in the preceding two: how precisely can these simulations be trusted?
A precision quantum simulation requires a complete quantification of the uncertainties affecting its result.
Errors enter quantum simulations at several stages. 
First, a continuum theory must be discretized onto a lattice, which introduces errors familiar to lattice field theory calculations~\cite{Davoudi:2012ya,Klco:2018zqz}.
Mapping the discretized problem onto the degrees of freedom of the device often introduces approximations, such as the truncation of a continuous field to a finite register of qubits~\cite{Klco:2018zqz}. 
These digitization errors are typically under more robust theoretical control than other sources of error, since their scaling can be bounded analytically or systematically improved by increasing the number of qubits used.
However, they sometimes require model- and coupling-dependent treatment, as is the case with formulations of gauge theory simulations~\cite{Kogut:1974ag,Bauer:2021gek,Ciavarella:2021nmj,Haase:2020kaj,Grabowska:2024emw}.
Approximate state preparation and time evolution implementations necessarily carry errors. 
The errors stemming from faulty physical-qubit operations and measurements on near-term devices are similarly unavoidable.
Finally, statistical uncertainty is introduced when the circuit is evaluated using a finite number of shots.
Achieving control over all sources of error is an outstanding goal as quantum simulations mature. 
Chapter~\ref{chap:spiral} takes one concrete step in this direction in the context of time evolution errors in analog quantum simulations.

Many of the platforms capable of analog simulations are described by the Ising model, including trapped ions~\cite{RevModPhys.93.025001, PhysRevB.95.024431, PRXQuantum.2.020328}, neutral atoms~\cite{Henriet_Beguin_Signoles_Lahaye_Browaeys_Reymond_Jurczak_2020, PRXQuantum.3.020303, Browaeys_Lahaye_2020, ebadi_wang_levine_keesling_semeghini_omran_bluvstein_samajdar_pichler_ho_et_al._2021}, nuclear spins~\cite{MADI1997300, PhysRevB.75.094415}, and superconducting qubits~\cite{PhysRevLett.112.200501, PhysRevX.5.021027, johnson_amin_gildert_lanting_hamze_dickson_harris_berkley_johansson_bunyk_et_al._2011}.
Chapter~\ref{chap:spiral} focuses on emulating dynamics in the Heisenberg model using Ising interactions.
The Heisenberg model, similar to the Ising model, is a widely studied spin model primarily relevant to condensed matter systems~\cite{gong_zhu_sheng_2014, ma_dakic_naylor_zeilinger_walther_2011}.
It has been shown to be a universal model for quantum computing~\cite{cubitt_montanaro_piddock_2018}, and its simulation can be used to study processes relevant for high-energy particle physics~\cite{Nachman_2021, Bauer:2022hpo, Caspar:2022llo, 2a}, quantum gravity~\cite{maldacena2023simple}, and QCD~\cite{Florio:2023dke,PhysRevD.105.083020}.
Realizing Heisenberg evolution on a native Ising-type device requires engineering the target Hamiltonian from the available one, a strategy pursued across a range of platforms~\cite{Martin:2022wyl, tyler2023higherorder, zhou2023robust, Zhou:2023xnx, Geier:2021uxg, PRXQuantum.3.020303, choi_zhou_knowles_landig_choi_lukin_2020, richerme_gong_lee_senko_smith_foss-feig_michalakis_gorshkov_monroe_2014, jurcevic_lanyon_hauke_hempel_zoller_blatt_roos_2014,PhysRevA.95.013602, PhysRevA.97.023611, 1a}.

Chapter~\ref{chap:spiral} addresses the errors that accompany this engineering. 
Two sources of algorithmic error are in tension on a real device: Trotter error, arising from the decomposition of the target evolution into a finite product of non-commuting unitaries, and idle error, stemming from the inability of an analog system to switch off its native interactions while local pulses are applied. 
On an idealized device these two errors could be treated independently, since reducing the Trotter step size always reduces Trotter error, and reducing pulse width always reduces idle error. 
On a real device, however, the two compete, since taking many short Trotter steps to suppress the former incurs a large cost from the latter if idle errors are significant, and pulse widths cannot be made arbitrarily small. 
A general framework for quantifying and optimizing this tradeoff is developed and applied to several practical methods for engineering Heisenberg-type evolution on an Ising-native device.
These include Trotter-like decompositions using field pulses and Floquet-engineered constant-field approaches. 
In each case, the choice of time evolution parameters that minimizes the combined algorithmic error given the constraints of currently available hardware is explained.
The resulting optimum depends on the device error rate and coherence time, showing the interdependence of algorithmic errors and noise.

Beyond algorithmic errors, a reliable quantum simulation must also contend with device errors.
Suppressing these has so far relied largely on error mitigation, which either incurs an exponential shot overhead to produce formally unbiased results, or trades that guarantee for cheaper, heuristic methods with inherent bias.
Demonstrating a net benefit of FT schemes has been an outstanding goal in quantum computing, and the present hardware and algorithms are beginning to make this a reality.
Leveraging high gate fidelities and flexible connectivity, recent demonstrations on trapped ions~\cite{Dasu:2026dwm,Perlin:2026mph} and neutral atoms~\cite{Rodriguez:2024bhh,Reichardt:2024xfs,Bluvstein:2025ped} have achieved ``beyond break-even'' performance, with the encoded error rate falling below the unencoded rate~\cite{Gottesman:2016gef}. 
Solid-state platforms have seen similar success, albeit in more restricted settings~\cite{Urbanek:2020cza,Chen:2021num,GoogleQuantumAIandCollaborators:2024efv,Hetenyi:2024zvf,Caune:2024doa,Gupta:2023zei,Vigneau:2025avm,Abraham:2026slx}.
These recent demonstrations of FT implementations suggest that, even on presently available hardware, quantum simulations can benefit from FT implementations beyond what error mitigation alone can achieve. 
Realizing this benefit requires carefully balancing the error-rate improvements from FT against its associated overhead.
Although topological codes such as the surface code~\cite{Kitaev:1997wr,Dennis:2001nw} and the heavy-hex code~\cite{Chamberland:2019zev} are naturally suited to limited-connectivity devices~\cite{Benito:2024mll}, the measurement and decoding overhead required for their logical operations places them out of reach for present-day simulations.

A motivation for the work presented in Chapter~\ref{chap:dev_422} is to use FT to advance quantum simulations of lattice gauge theories~\cite{Pato:2026wow,Rajput:2021trn,Spagnoli:2026qni,Turco:2026cte,Spagnoli:2024mib,Yao:2025cxs,Carena:2024dzu,Froland:2026tfx} describing the fundamental forces of Nature~\cite{Klco:2021lap,Bauer:2022hpo,Bauer:2023qgm,Davoudi:2022bnl,Beck:2023xhh,DiMeglio:2023nsa}. 
The mappings and dynamics of such simulations are organized by symmetries that group the degrees of freedom into repeating units.
As such, it is natural to study FT schemes whose code blocks coincide with these units.
In the Schwinger model (quantum electrodynamics in 1+1D), for example, a spatial site maps to two qubits, e.g. Refs.~\cite{Martinez:2016yna,Klco:2018kyo,Farrell:2023fgd,Farrell:2024fit}, suggesting the use of code blocks that implement two logical qubits.
While the full power of QEC may still be beyond reach, immediate improvements can be achieved through the adoption of lightweight, low-overhead quantum error detection (QED) encodings~\cite{Linke:2017bvn,Vuillot:2018tqy,Harper:2019upm,Takita:2017blo,Corcoles:2014imo,Yamamoto:2023xan,Wang:2023qcn}. 
The recently studied Iceberg codes are a family of $[[n+2,n,2]]$ codes that use two extra physical qubits to encode $n$ logical qubits into a single block~\cite{Gottesman:1997zz,Linke:2017bvn,Roffe:2018oim,Chao:2017owu,Self:2022lsx,Cao:2024mql,Chertkov:2025qzc,Vezvaee:2025yol,Dasu:2026dwm}.

Chapter~\ref{chap:dev_422} builds on this approach, encoding a quantum simulation across 21 blocks of the $[[4,2,2]]$ code, using up to 136 active physical qubits to represent 42 logical qubits on IBM's heavy-hexagonal quantum computer \texttt{ibm\_boston}.
Rather than pursuing a fully FT logical gate set, FT syndrome extraction is paired with non-FT logical rotations, retaining the benefit of error detection while avoiding the overhead a fully FT gate set would require.
For a physical error rate $p$, this partially FT approach yields a logical error rate $p_L=Ap+Bp^2$.
The non-FT rotations contribute a term linear in $p$ and the FT components contribute a term suppressed quadratically, and error suppression is possible provided the non-FT component is engineered so that $A\ll B \ll 1$.
Further, the encoding's flexible logical connectivity enables simulations on a two-dimensional lattice with less overhead than a direct, unencoded implementation of the same geometry would require.
A central obstacle to any error-detecting scheme is the exponential loss of shots surviving postselection as the number of syndrome extraction rounds increases. This obstacle is addressed with a selective-filtering technique which we call observable-ranked postselection (ORP). 
It correlates syndrome flips with the value of the logical observable of interest, recovering reliable results without the prohibitive shot loss of full postselection.
Retaining enough shots to control statistical uncertainty, while still removing the detection events most likely to signal an uncorrected logical error, is precisely what makes it possible to extract a reliable signal from a device operating just below the code's pseudothreshold.

These techniques are demonstrated using quench dynamics simulating false-vacuum decay in the one-dimensional and two-dimensional Ising model. 
False-vacuum decay describes how a metastable vacuum tunnels to a stable lower-energy one, with fluctuation-nucleated bubbles of true vacuum growing once they exceed a critical size and driving the conversion of the whole system~\cite{Coleman:1977py,Callan:1977pt}. 
Rather than modeling this tunneling and nucleation process directly, this chapter realizes an analogous scenario on the lattice by preparing a finite spin system in a polarized product state and studying its unitary relaxation dynamics, capturing the growth of an already-formed bubble rather than its formation~\cite{Lagnese:2021grb,Lagnese:2023xjg}.
A system prepared in a product state away from its true vacuum exhibits distinct melting and localized regimes of quench dynamics consistent with previous studies of false-vacuum decay in this model~\cite{Milsted:2020jmf,Balducci:2022zym,Balducci:2022kvd,Pavesic:2024ryc}. 
At the device's operating error rate, error detection  is found to improve the estimation of local observables relative to unencoded runs by 2-6\% in one dimension, and by over 200\% in two dimensions at the latest times studied.
These results demonstrate beyond-break-even performance achieved without any error mitigation, isolating the improvement due to QED alone from any gain that mitigation might separately provide.

The results presented in Chapters~\ref{chap:spiral} and~\ref{chap:dev_422} show that careful management of errors can go a long way in quantum simulations. 
Balancing different sources of error throughout the entirety of a simulation will be a persistent feature of the crossover from the NISQ era to full FT.
Understanding which simulations will benefit the most as this crossover regime develops will require tracking the shift from entanglement to magic as the dominant costly resource.
As this balance shifts, its impact will go beyond improvements to any single simulation's accuracy, making whole classes of simulations newly accessible.
Throughout this crossover, the effort to extract physics from imperfect hardware will remain a source of insight, valuable both for what it reveals about physical systems and the simulations it enables.
The contributions in this thesis represent small steps toward the goal of scientific discovery in fundamental physics with quantum computers.

\part{Quantum simulations of scattering}
\chapter{Scalable quantum simulations of scattering in scalar field theory on 120 qubits}
\label{chap:phi4_scattering}

\noindent
{\it This chapter is associated with Ref.~\cite{Zemlevskiy:2024vxt}: ``Scalable quantum simulations of scattering in scalar field theory on 120 qubits'' by Nikita A. Zemlevskiy.}

\section{Introduction}
\noindent
State preparation and time evolution for quantum simulations of scattering often require circuits whose depth exceeds what is feasible on present-day hardware. 
This chapter addresses this challenge in the context of scalar field theory in one spatial dimension.
Following the seminal work of Jordan, Lee, and Preskill (JLP)~\cite{Jordan:2011ci,Jordan:2012xnu,Jordan_2018}, there have been numerous efforts targeting aspects of quantum simulations of scattering in scalar field theory~\cite{Klco:2018zqz,Yeter-Aydeniz:2018mix,Klco:2019yrb,Klco:2019xro,Barata:2020jtq,Klco:2020aud,Kurkcuoglu:2021dnw, Macridin:2021uwn, Liu:2021otn,Illa:2022jqb,Li:2022ped,Ozzello:2023dzn,Hardy:2024ric}. 
Variational compression offers a route to executing these simulations with circuit depths compatible with near-term devices.
Variational algorithms on the basis of brickwall~\cite{Haghshenas:2021nsg,Leone:2022aux,Filippov:2022exc,Okada:2022fiy,Tepaske:2022uad,Barthel:2023ine,Miao:2023her,Tepaske:2023mfc,Robertson:2023jlp,Zhang:2024kuf,Wright:2024yxm} circuits\footnote{``Brickwall'', ``brickwork'', and ``hardware-efficient'' are commonly used to refer to dense variational ansatze using native single- and two-qubit gates. 
``Brickwall'' is used to describe this class of circuits in this work.} and other ansatze are used to simulate the scattering of two particle wavepackets on IBM's quantum computers. 
While the application of variational algorithms to quantum simulations has been proposed and executed on devices in the past~\cite{Ibrahim:2022liy,Causer:2023wpp,Wang:2024pap,Zhuang:2024mdh,Sachdeva:2024kob}, this work serves as the first example of a quantum simulation using physics-informed brickwall circuits at scale for state preparation and time evolution. 
Furthermore, it is the first quantum simulation of wavepacket scattering in an interacting quantum field theory. 
These proof-of-concept simulations are a demonstration of the techniques that will be foundational for simulations of high-energy, high-inelasticity scattering events in more complicated theories, where rich physical phenomena are expected. 
The simulations proceed in several steps: 
\begin{enumerate}
    \item Vacuum preparation
    \item Particle creation
    \item Time evolution
    \item Measurement
\end{enumerate}
By incorporating the symmetries and physical structure of the scalar field theory into the circuit design, the simulation is carried out in a scalable way, unlocking the full capabilities of state-of-the-art devices. 
Throughout this work, ``scalable'' will be used to indicate scalability for states whose structure does not change with increasing system size. 
Recent work has determined upper bounds of $O(10^{12})$ fault-tolerant operations required for simulations of scattering in one-dimensional scalar field theory~\cite{Hardy:2024ric}. 
The variational techniques to minimize circuit depth developed in this work enable approximate simulations of scattering processes with feasible gate counts for devices available today. 
While previous quantum simulations have taken a hybrid approach, (e.g., state preparation was done variationally, but time evolution was implemented using an algorithm with rigorous performance guarantees), this work establishes the application of physics-informed variational algorithms to all parts of the simulation. 
The variational methods developed in this work are expected to have broad relevance to simulating physical models of interest.

This chapter is organized as follows. 
An overview of lattice scalar field theory in one dimension and its mapping to qubits is given in Sec.~\ref{phi4:sec:lattice_scalar_field_theory}.
The algorithm for implementing scalable variational circuits (SVCs) for a class of quantum simulation problems is outlined in Sec.~\ref{phi4:sec:scalable_variational_circuits}. 
This algorithm may be seen as a general framework capable of compressing physical simulation circuits at scale. 
Section~\ref{phi4:sec:circuits} describes how knowledge of the physics of the theory is used to determine SVCs both for state preparation and time evolution. 
The errors stemming from these approximate circuit compression methods are quantified. 
In Sec.~\ref{phi4:sec:results}, SVCs are used to simulate scattering in scalar field theory using 120 qubits of IBM's superconducting quantum computer {\tt ibm\_fez}. 
Both the free and interacting cases are studied.
A new error mitigation strategy on the basis of simulations of vacuum evolution is introduced. 
With this technique, the results are found to be in qualitative agreement with MPS circuit simulations, and the effect of the interaction strength on the nature of the collisions is clearly seen. 
Improvements and comments on the extensibility of the developed methods are presented in Sec.~\ref{phi4:sec:discussion}. 

\section{Lattice scalar field theory}
\label{phi4:sec:lattice_scalar_field_theory}
The Hamiltonian of a real scalar field $\phi$ in one dimension with periodic boundary conditions (PBCs) defined on a lattice of $L$ spatial sites is written as
\begin{align}
    H \ &= \  \sum_{j=0}^{L-1} \frac{1}{2} \Pi_j^2 + \frac{1}{2}m^2 \phi_j^2 + \frac{1}{2} (\phi_{j+1} - \phi_j)^2 + \frac{\lambda}{4!} \phi_j^4 \label{phi4:eq:lattice_h} \\
    &\equiv \ H_\Pi + H_\phi + H_\text{kin} + H_\text{int}\ . \nonumber
\end{align}
Here $m$ is the bare mass, $\lambda$ is the coupling controlling the strength of $\phi^4$ interactions, $\Pi_j$ is the conjugate momentum operator obeying the canonical commutation relation $[\phi_i, \Pi_j] = i\delta_{ij}$, and the nearest-neighbor finite-difference operator $\nabla \phi = (\phi_{j+1} - \phi_j)$ is used in the lattice representation of the kinetic term. The lattice spacing is set to $a=1$ in this work. 

With PBCs, the eigenstates of $H$ are labeled by their spatial momenta $k=2\pi n/L$, where $n$ is an integer such that $k \in (-\pi,\pi]$. The vector of fields $\vec{\phi} = (\phi_0, \phi_1, \dots, \phi_{L-1})$ is related to the vector of spatial momentum eigenstates $\vec{\chi}$ by a spatial Fourier transform: $\vec{\chi} = V \vec{\phi}$, where $V = \frac{1}{\sqrt{L}}e^{i \vec{k}\cdot\vec{j}}$ is the Fourier transform matrix. In the free theory $(\lambda=0)$, the energy of a single-particle eigenstate with spatial momentum $k$ is 
\begin{equation}
    E_k^2 \ = \ m^2 + 4\sin^2\left(\frac{k}{2}\right)\ .
    \label{phi4:eq:dispersion}
\end{equation}

The spectrum of $H$ in the free theory with PBCs can be constructed by introducing creation and annihilation operators
\begin{equation}
    a_k \ = \ \sum_j e^{-ikj} \left(\sqrt{\frac{E_k}{2}} \phi_j + i \sqrt{\frac{1}{2E_k}} \Pi_j\right)\ .\label{phi4:eq:ap}
\end{equation}
This can be inverted to find expressions for $\phi$ and $\Pi$ in terms of $a_k$ and $a_k^\dagger$. Written in terms of the creation and annihilation operators, $H$ takes the following form:
\begin{equation}
    H \ = \ \frac{1}{L}\sum_k E_k \left(a_k^\dagger a_k + \frac{1}{2}\right)\ .
\end{equation}

The vacuum of the free theory can be written down analytically. In the $a_k$ Fock basis, the ground state of the system can be written as a tensor product of the $L$ harmonic oscillators $\vec{\chi}$, each in its ground state (denoted by the subscript 0):
\begin{equation}
    |\psi_\text{vac}\rangle \ = \ |\psi_{\chi_0}\rangle_0 \otimes |\psi_{\chi_1}\rangle_0 \otimes \dots \otimes |\psi_{\chi_{L-1}}\rangle_0\ .
    \label{phi4:eq:vacuum_ket}
\end{equation}
The eigenfunctions of the harmonic oscillator can be used to construct wavefunctions in the free scalar field theory. The momentum-space representation of the $n^\text{th}$ harmonic oscillator eigenfunction is given by 
\begin{align}
    \langle \chi | n\rangle \ = \ \frac{1}{\sqrt{2^n n!}} \left( \frac{E_n}{\pi} \right)^\frac{1}{4} e^{-\frac{1}{2}E_n \chi^2}\text{H}_n\!\left(\sqrt{E_n}\chi\right)\ ,
    \label{phi4:eq:sho_eigenfunctions}
\end{align}
where $\text{H}_n$ is the $n^\text{th}$ Hermite polynomial. Since $\text{H}_0=1$, the ground-state wavefunction of the scalar field theory in this basis can be written as a product of Gaussians: 
\begin{equation}
    \langle \vec{\chi} | \psi_\text{vac} \rangle \ = \ \frac{(\text{det} E)^{1/4}}{\pi^{L/4}} e^{-\frac{1}{2}\vec{\chi}^T E\vec{\chi}}\ ,
    \label{phi4:eq:vacuum_wavefunction}
\end{equation} 
where $E$ is the matrix consisting of the energies in Eq.~\eqref{phi4:eq:dispersion} on the diagonal. The position-space wavefunction can be found using the Fourier transform $\vec{\chi} = V\vec{\phi}$.

Similar to the vacuum, wavepackets in the free theory can be specified analytically. Using the vacuum state of Eq.~\eqref{phi4:eq:vacuum_ket}, the operator $a_k^\dagger$ defined in Eq.~\eqref{phi4:eq:ap} creates a single-particle excitation with momentum $k$:
\begin{equation}
    |k\rangle \ = \ a_k^\dagger|\psi_\text{vac}\rangle = |\psi_{\chi_0}\rangle_0 \otimes \dots\otimes |\psi_{\chi_k}\rangle_1 \otimes\dots\otimes |\psi_{\chi_{L-1}}\rangle_0\ .
\end{equation}
A general single-particle state is a superposition of these excitations:
\begin{equation}
    |\psi_\text{wp}(p)\rangle \ = \ \mathcal{N}\sum_k g(k,p) |k\rangle\ ,
    \label{phi4:eq:single_particle_superposition}
\end{equation}
where $g(k,p)$ is an envelope in $k$-space that specifies the form of the state, and $\mathcal{N}$ is a normalization constant. A localized wavepacket state can be created by choosing a Gaussian profile $g(k,p;\sigma_p)=\frac{1}{\sqrt{2\pi}\sigma_p}e^{-\frac{1}{2}(k-p)^2/\sigma_p^2}$. The wavefunction of a Gaussian wavepacket with momentum $p$ and spread $\sigma_p$ is found using Eq.~\eqref{phi4:eq:sho_eigenfunctions} to be
\begin{equation}
    \langle \vec{\chi}|\psi_\text{wp}(p;\sigma_p)\rangle \ = \ \mathcal{N} e^{-\frac{1}{2}\vec{\chi}^T E \vec{\chi}}\sum_k g(k,p;\sigma_p)\,\text{H}_1\!\left(\sqrt{E_k}\chi_k\right)\ .
    \label{phi4:eq:wp_wavefunction}
\end{equation}
This specifies a wavepacket centered at $j=0$; it can be moved elsewhere using a spatial translation operator. As before, the position-space wavefunction is found by the Fourier transform $\vec{\chi} = V\vec{\phi}$. Multiparticle states can be constructed in a similar fashion. 

Time-evolved observables in the free theory can be computed analytically by expanding the state in the appropriate basis of multiparticle excitations. For example, the time evolution of a single-wavepacket state can be written as $|\psi_\text{wp}\rangle(t) = \mathcal{N}\sum_k e^{-i t E_k} g(k,p) a_k^\dagger|\psi_\text{vac}\rangle$. By rewriting $\phi^2_j$ in terms of $a_k$ and $a_k^\dagger$, $\langle \phi^2_j\rangle(t)$ can be obtained using Wick contractions. In the interacting $(\lambda \neq 0)$ theory, $E_k,a_k^\dagger,$ and $|\psi_\text{vac}\rangle$ are all functions of $\lambda$. In cases where $\frac{\lambda}{4!}$ is small, observables can be obtained by treating the $\phi^4$ term as a perturbation.

In the context of quantum simulation, the interacting theory can be studied provided that the required states can be prepared. JLP introduced a procedure to create interacting particle wavepackets from their free counterparts~\cite{Jordan:2011ci,Jordan:2012xnu}. By the adiabatic theorem, if the $\phi^4$ interactions are turned on in a manner that is slow enough for states to adjust, then eigenstates in the free theory will flow into the corresponding eigenstates in the interacting theory. However, because wavepackets are superpositions of eigenstates, they will propagate and broaden during this adiabatic evolution. To counter this, the adiabatic evolution turning on $\lambda$ can be broken up into several steps, and the unwanted evolution can be ``undone'' at each step by evolving backwards. Let $s$ parameterize the interactions in the Hamiltonian
\begin{align}
    H(s) \ = \ H_\phi + H_\Pi + H_\text{kin} + s H_\text{int}\ ,\quad s \in [0,1]\ .
\end{align}
The adiabatic turn-on is then implemented in $N$ steps by interleaving forward time evolution with increasing $s$ at each step with reverse time evolution at fixed $s$:
\begin{equation}
\begin{split}
    U_\text{ad} \ = \ \prod_{n=0}^{N-1}\Bigg[
    &\, T\left\{\exp\left(-i\frac{t_\text{ad}}{2}\int_{\frac{(2n+1)}{2N}}^{\frac{(n+1)}{N}}H(s)\,ds\right)\right\} \exp\left(it_\text{ad}H\!\left(\frac{2n+1}{2N}\right)\right)\\
    &\times T\left\{\exp\left(-i\frac{t_\text{ad}}{2}\int_{\frac{n}{N}}^{\frac{(2n+1)}{2N}}H(s)\,ds\right)\right\}\Bigg]\ ,
\end{split}
\label{phi4:eq:u_adiabatic}
\end{equation}
where $T$ is the time-ordering operator. With a proper choice of $N$ and $t_\text{ad}$ depending on the parameters $m$ and $\lambda$~\cite{ Jordan:2011ci,Jordan:2012xnu,Li:2024lrl}, it can be shown that unwanted evolution can be sufficiently suppressed.  

\subsection{Qubit representation}
\label{phi4:sec:qubit_representation}
At each lattice site $j$, the field takes continuous values $\phi_j \in (-\infty, \infty)$. Since the Hilbert space at each $j$ is infinite, the field must be digitized to be represented on a quantum computer. The JLP digitization is used to map this system onto qubits~\cite{Jordan:2011ci,Jordan:2012xnu}. Given a register of $n_q$ qubits, a maximum magnitude $\phi_\text{max}$ is chosen,\footnote{There exists an optimal choice of $\phi_\text{max}$ for a given $n_q$ as a result of the Nyquist-Shannon Sampling Theorem~\cite{shannon_collected_1993,10.5555/3179430.3179434,Macridin:2018oli,Macridin:2018gdw}. The optimal $\phi_\text{max}$ also depends on $m$ and $\lambda$, since these parameters control the support of the eigenstates of the theory (see App.~\ref{phi4:sec:digitization_effects}). The optimal values of $\phi_\text{max}$ are determined numerically for small $n_q$ in Ref.~\cite{Klco:2018zqz}, and analytically in Refs.~\cite{Bauer:2021gek,Kane:2022ejm}.} and the field at each point in space is evenly discretized in $2^{n_q}$ steps of size $\delta_\phi$ in the range $[-\phi_\text{max}, \phi_\text{max}]$. At each site $j$, the digitized field can take on the values
\begin{equation}
    \phi_j \ = \ -\phi_\text{max} + \ell \delta_\phi, \quad \delta_\phi \ = \ \frac{2 \phi_\text{max}}{2^{n_q} - 1}\ , \quad \ell \in [0, 2^{n_q}-1]\ .
    \label{phi4:eq:phi_digitized}
\end{equation}
Twisted boundary conditions in field space~\cite{Klco:2018zqz} are used to preserve the symmetry in the digitizations of $\phi$- and $\Pi$-space. The allowed conjugate momenta $k_\phi$\footnote{The conjugate momenta $k_\phi$, which are used to construct the $\Pi_j$ operator, should not be confused with the spatial momenta $k$, which label eigenstates of the Hamiltonian.} are distributed symmetrically around 0:
\begin{equation}
    k_{\phi,j} \ = \ -\frac{\pi}{\delta_\phi} + \left(\ell + \frac{1}{2}\right)\frac{2\pi}{2^{n_q} \delta_\phi}\ , \quad \ell \in [0, 2^{n_q}-1]\ .
    \label{phi4:eq:pi_digitized}
\end{equation} 
The allowed state of the field at each site is represented by qubits using a binary encoding of the $2^{n_q}$ allowed values from Eqs.~\eqref{phi4:eq:phi_digitized} and~\eqref{phi4:eq:pi_digitized}:
\begin{equation}
    |\phi_j = -\phi_\text{max} + \ell \delta_\phi\rangle \ = \ |\ell\rangle\ ,
\end{equation} 
(e.g., $|\phi_j = -\phi_\text{max} + \delta_\phi\rangle = |1\rangle = |01\rangle$ for $n_q=2$).

This digitization is particularly convenient because the form of the operators in the Hamiltonian is simple when expressed in terms of their action on qubits:
\begin{equation}
    \phi_j \ = \ -\frac{\phi_\text{max}}{2^{n_q}-1}\sum_{\ell=0}^{n_q-1} 2^\ell Z_{n_q j + \ell}\ ,
    \label{phi4:eq:phi_qubits}
\end{equation}
and in $\Pi$-space:
\begin{equation}
    \Pi_j \ = \ -\frac{\pi}{2^{n_q}\delta_\phi} \sum_{\ell=0}^{n_q-1} 2^\ell Z_{n_q j + \ell}\ .
    \label{phi4:eq:pi_qubits}
\end{equation}

\begin{figure*}
\centering
{\includegraphics[width=0.8\linewidth]{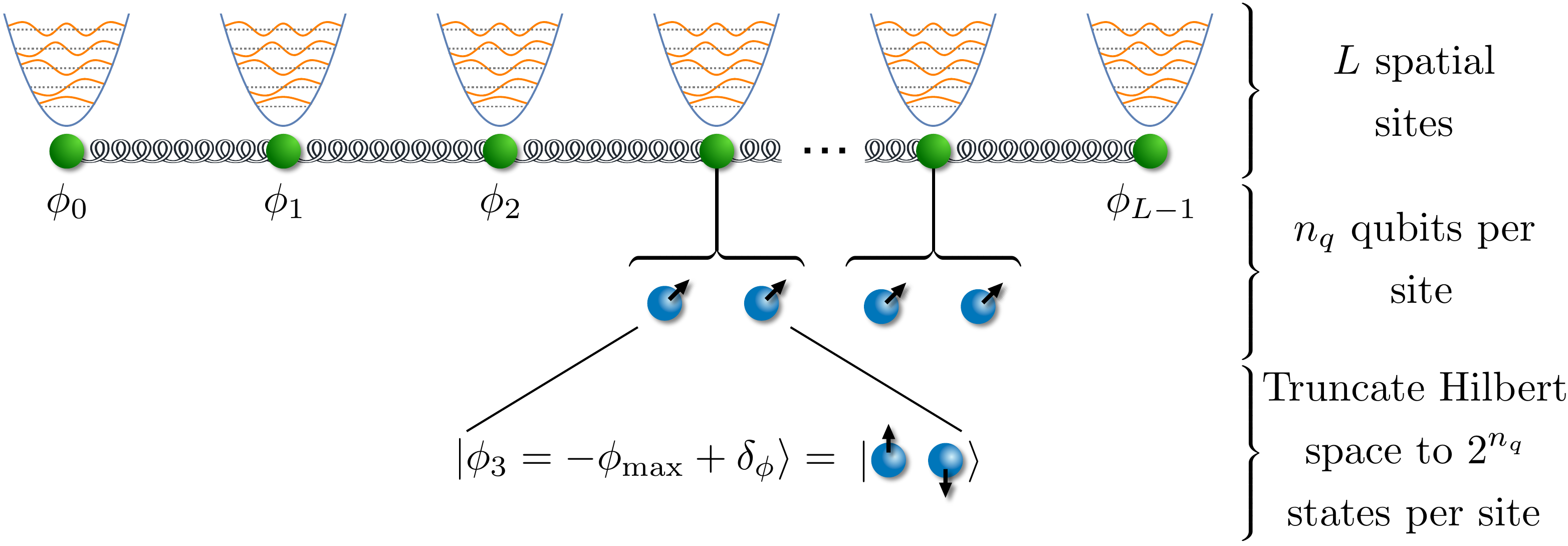}}
\caption{{\it Mapping the one-dimensional lattice scalar field theory to qubits.} For a system of $L$ spatial sites, $n_q$ qubits are used to represent the state of the field at each spatial site. The infinite-dimensional bosonic Hilbert space of $\phi$ at each site is truncated to $2^{n_q}$ values. A maximum magnitude of the field $\phi_\text{max}$ is chosen, and the values of the field in increments of $\delta_\phi$ are assigned to the possible qubit states.}
\label{phi4:fig:map_to_qubits}
\end{figure*} 

The representation of the lattice scalar field by qubits is shown in Fig.~\ref{phi4:fig:map_to_qubits}. This construction efficiently recovers the low-energy physics in a way that is systematically extendable to include higher-energy contributions. In Ref.~\cite{Klco:2018zqz} it was shown that, aside from device noise and errors stemming from approximations to the time evolution operator, the error in local observables decays double-exponentially with the number of qubits used in the site-wise digitization.\footnote{That is, for a local observable, the percent error $\epsilon$ in the low-energy eigenvalues of the digitized observable compared to their undigitized counterparts scales as $\epsilon \sim 2^{-2^{n_q}}$. In fact, it was found in Ref.~\cite{Klco:2018zqz} that $n_q=4$ provides sufficient precision in local observables, so that other sources of error dominate. Furthermore, it was also found that the $\phi$ basis is advantageous over other representations, such as the basis of harmonic oscillator eigenstates, and strategies associated with improved actions that are often used in lattice QCD. This advantage stems from the simplicity and efficiency of the implementations of operators in the theory (Eqs.~\eqref{phi4:eq:phi_qubits} and~\eqref{phi4:eq:pi_qubits}).}

The JLP prescription gives an efficient implementation of the operators using quantum gates~\cite{Jordan:2011ci,Jordan:2012xnu}. Operators in the Hamiltonian of Eq.~\eqref{phi4:eq:lattice_h} can be broken into two non-commuting groups: those that contain $\phi$, and those that contain $\Pi$. The operators made up of powers and tensor products of $\phi$ are diagonal in the $\phi$ basis, and are straightforward to implement. Aside from factors of the identity, $\phi^2_j$ and $(\phi_{j+1}-\phi_j)^2$ contain only $ZZ$ operators and can be implemented using $O(n_q^2)$ two-qubit gates. Similarly, $\phi^4_j$ can be implemented with $O(n_q^4)$ two-qubit gates, as it has additional $ZZZZ$ terms for $n_q \geq 4$. $\Pi^2_j$ is translated into gates by using a site-wise quantum Fourier transform (QFT) to switch into the conjugate momentum basis where $\Pi$ is diagonal. Using this, $\Pi^2_j$ can be implemented with $O(n_q^2)$ two-qubit gates as well.\footnote{Note that for a device with nearest-neighbor (NN) two-qubit gate connectivity, as is the case for most superconducting quantum computers, there is an extra two-qubit gate overhead to convert circuits from arbitrary connectivity to NN. To reduce the two-qubit gate depth, efficient methods for the decomposition of specific circuit elements to NN connectivity have been developed~\cite{10.5555/2011827.2011828,Park:2023goh,Farrell:2024fit}.}

Conventionally, the time evolution operator is implemented in quantum simulations using first-order Trotterization: 
\begin{align}
    e^{-itH} \ = \ \left( U_\text{trot}^{(1)}\left(\frac{t}{n}\right)\right)^n + O\left(\left(\frac{t}{n}\right)^2\right)\ , \quad \quad U_\text{trot}^{(1)}(t) \ = \ e^{-it(H_\phi + H_\text{kin} + H_\text{int})}e^{-itH_\Pi}\ ,
\end{align}
where the terms in the exponential are organized into commuting groups. Higher-order product formulas can be recursively constructed by symmetric combinations of $U_\text{trot}$ via the formula~\cite{Suzuki:1991jtk}
\begin{align}
    U_\text{trot}^{(2n)}(t) = U_\text{trot}^{(2n-2)}(p_nt)^2 U_\text{trot}^{(2n-2)}\left((1-4p_n)t\right) U_\text{trot}^{(2n-2)}(p_nt)^2\ , \quad \quad p_n = \frac{1}{4-4^{1/(2n-1)}} \ .
\end{align}
These formulas have more favorable error scaling with the step size $t/n$, at the cost of a larger gate depth. In this work, $U_\text{trot}^{(2)}$ is used as a point of comparison against variational methods. After two-qubit gate cancellations and reordering terms, one second-order Trotter step can be implemented with a two-qubit gate depth of 20 for $n_q=2$.

\section{Scalable variational circuits}
\label{phi4:sec:scalable_variational_circuits}
The preparation of an arbitrary state on a quantum computer requires a number of gates that grows exponentially with system size~\cite{Knill:1995kz}. Fortunately, states in physical theories have properties, such as symmetries, that can be utilized to simplify the state preparation task. This was studied in the context of using the locality of correlations to systematically truncate exact scalar field theory state preparation circuits in Refs.~\cite{Klco:2019yrb,Klco:2020aud}. As was recently demonstrated for state preparation in Refs.~\cite{Farrell:2023fgd,Farrell:2024fit}, the combination of these ideas with variational methods proves to be a powerful way to create compressed scalable circuits, while not sacrificing accuracy.

There always exists a unitary that takes one particular state $|\psi_\text{ansatz}\rangle$ to another target state $|\psi_\text{targ}\rangle$ in the Hilbert space. Implementing this unitary is the goal of state preparation (e.g., $|\psi_\text{ansatz}\rangle = |\psi_\text{vac}\rangle,\,|\psi_\text{targ}\rangle = |\psi_\text{wp}\rangle$) and time evolution (e.g., $|\psi_\text{ansatz}\rangle =|\psi_\text{wp}\rangle,\,|\psi_\text{targ}\rangle = e^{-itH}|\psi_\text{wp}\rangle$). While many methods exist to approximately construct these unitaries, they often produce circuits that are too deep to implement on currently available hardware. Furthermore, these methods are often general, even though only a single or several matrix elements are of interest in a simulation. In this sense, the problem may be reduced from finding an implementation of an operator that properly transforms all states in the Hilbert space, to that of finding an operator that implements $|\psi_\text{ansatz}\rangle \rightarrow |\psi_\text{targ}\rangle$. To take advantage of this simplification and compress unitaries given by conventional methods, variational circuits can be used to approximate the action of these unitaries in a systematically improvable way.

Suppose a quantum circuit $U(\vec{\theta})$ is parameterized by $\vec{\theta}$. The task of the variational algorithm is to choose the parameters $\vec{\theta}_\text{opt}$ that implement $|\psi_\text{ansatz}\rangle \rightarrow |\psi_\text{targ}\rangle$ as closely as possible. To measure the quality of the prepared state $|\psi_\text{ansatz}(\vec{\theta})\rangle = U(\vec{\theta})|\psi_\text{ansatz}\rangle$ against $|\psi_\text{targ}\rangle$, the local infidelity is used: 
\begin{align}
    I_d\left(\rho_\text{targ}, \rho_\text{ansatz}(\vec{\theta})\right) \ &= \ 1-\left(\text{tr} \sqrt{\sqrt{\rho_\text{targ}} \rho_\text{ansatz}(\vec{\theta}) \sqrt{\rho_\text{targ}}}\right)^2\ .
    \label{phi4:eq:local_infidelity}
\end{align}
Here, $\rho_\text{targ}$ and $\rho_\text{ansatz}$ are the density matrices of the target and prepared states, respectively, and the subscript $d$ indicates that reduced states spanning $d$ spatial sites are used.\footnote{For pure states $\rho_\text{targ}$ and $\rho_\text{ansatz}$ (i.e., $d=L$), this expression reduces to $I_L=1-|\langle \psi_\text{targ}|\psi_\text{ansatz}\rangle|^2$.} Since the global fidelity $|\langle\psi_\text{targ}|\psi_\text{ansatz}\rangle|^2$ will become arbitrarily small with increasing system size, the reduced state infidelity is used instead to measure the local quality of $|\psi_\text{ansatz}\rangle$. This quantity is particularly relevant for translationally invariant states, or translationally invariant states with local perturbations, since these have a repeating local structure. 

The mass gap and locality of interactions appearing in physical systems with Hamiltonians such as Eq.~\eqref{phi4:eq:lattice_h} enable scaling up quantum simulations of these systems. Correlations in the ground states of such systems decay exponentially with separation beyond the correlation length (i.e., for $|i-j|>\xi,\, \langle\phi_i\phi_j\rangle \sim e^{-|i-j|/\xi}$)~\cite{Hastings:2005pr}. Here, the correlation length $\xi$ is defined to be the inverse of the gap (equivalently, the inverse of the mass of the lightest particle), $\xi=1/m_\text{particle}$.\footnote{In the free theory, $m_\text{particle}\rightarrow m$ as $\delta_\phi \rightarrow 0$, while in the interacting theory, $m_\text{particle} = m_\text{particle}(m,\lambda)$. $m_\text{particle}\neq m$ in this work as a result of the digitization used $(n_q=2)$. This digitization also introduces interactions (see App.~\ref{phi4:sec:digitization_effects}).} Because of this, the finite-volume ground-state wavefunction for $L\gg\xi$ is exponentially close to its infinite-volume form. This, in turn, means that local observables measured in the ground state agree with their infinite-volume values up to corrections that are exponentially small in the system size, $\langle A \rangle_L = \langle A \rangle_\infty + \mathcal{O}(e^{-L/\xi})$. 

The exponential convergence of ground states with system size implies that the structure of the circuits that create these states also converges exponentially~\cite{Farrell:2024fit}. In turn, this means that the circuits that act on these ground states also exhibit the same exponential convergence. Importantly, this applies not only to ground states, but generally to states whose structure does not change as the system size increases (provided the correlations decay as above). In particular, fixed-width wavepackets of single particles meet these criteria due to their locality. This property enables the determination of circuits that can be scaled to arbitrary system sizes. If the target state $|\psi_\text{targ}\rangle$ is known for a small system size, the equivalent state can be created in a larger system with the following algorithm to create SVCs:
\begin{enumerate}
    \item Create a parameterized circuit ansatz that implements the unitary $U(\vec{\theta})$. The ansatz must have a structure that can be scaled up to arbitrary system sizes. \label{phi4:enum:ansatz}
    \item Optimize the parameters in $U(\vec{\theta})$:\label{phi4:enum:optimization}
    \begin{enumerate}
        \item Prepare the target state $|\psi_\text{targ}\rangle$ to a desired precision using a known method.\footnote{The classical methods to determine $|\psi_\text{targ}\rangle$ for ground-state preparation, particle creation, and time evolution are given in Sec.~\ref{phi4:sec:lattice_scalar_field_theory}. Their quantum counterparts are discussed in Sec.~\ref{phi4:sec:circuits}.}
        \item Initialize the initial state $|\psi_\text{ansatz}\rangle$.
        \item Optimize the parameters $\vec{\theta}$ by minimizing the local infidelity: $\displaystyle \min_{\vec{\theta}}\, I_d\left(\rho_\text{targ}, \rho_\text{ansatz}(\vec{\theta})\right)$, with $|\psi_\text{ansatz}(\vec{\theta})\rangle = U(\vec{\theta})|\psi_\text{ansatz}\rangle$.\label{phi4:enum:infidelity}
    \end{enumerate}
    \item Repeat steps~\ref{phi4:enum:ansatz}, ~\ref{phi4:enum:optimization} for a set of increasing system sizes $L$. Fit an exponential to the parameters $\vec{\theta}$ and extrapolate to $L$ of interest. 
\end{enumerate}
\begin{figure*}
\centering
{\includegraphics[width=0.9\linewidth]{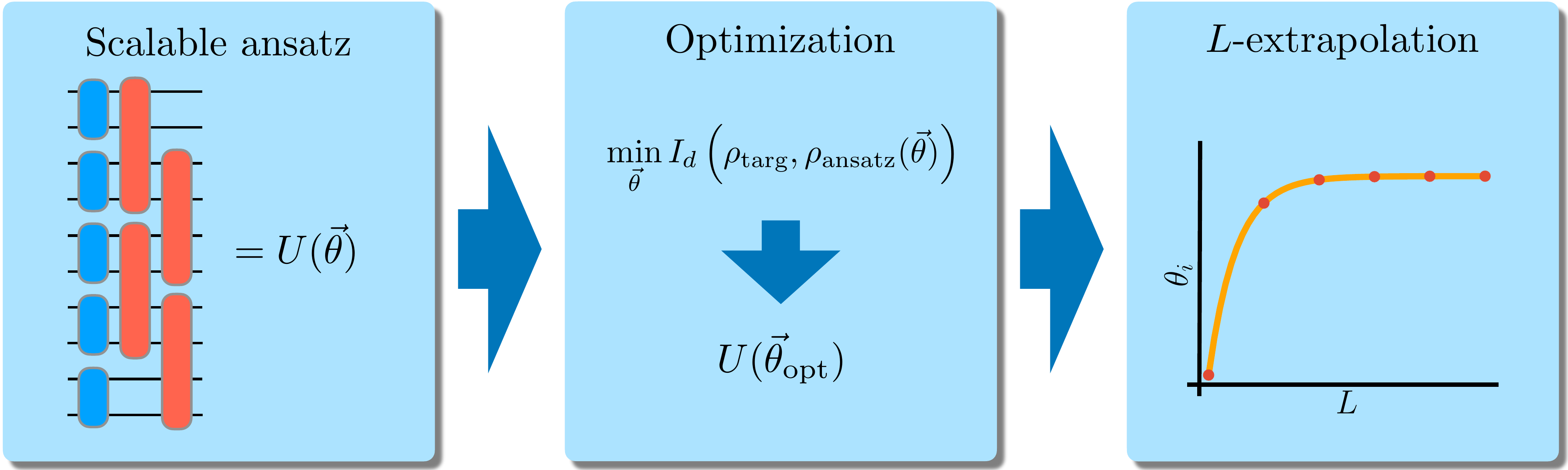}}
\caption{{\it Scalable variational circuits.} A parameterized ansatz with a scalable structure is chosen by considering symmetries of the system and properties specific to the target unitary. The infidelity between the target state and the produced state is minimized to optimize the variational circuit. This process is repeated for a series of increasing system sizes. The parameters are then extrapolated to an arbitrary system size of interest. }
\label{phi4:fig:scalable_variational_circuits_procedure}
\end{figure*} 
This algorithm is shown schematically in Fig.~\ref{phi4:fig:scalable_variational_circuits_procedure}. Depending on the convergence properties of the system being studied, this algorithm can be carried out using simulations on a classical computer, or on a small register of a quantum computer.\footnote{If running on a quantum computer, methods such as those described in Ref.~\cite{Huang:2024qmc} may be used to estimate the infidelity.}

Because this algorithm creates a variational unitary by optimizing a single matrix element $\langle\psi_\text{targ}|U(\vec{\theta})|\psi_\text{ansatz}\rangle$, there is no guarantee that the implemented $U$ will behave as expected for different starting states. However, since only initial states $|\psi_\text{ansatz}\rangle$ with very specific properties are of interest in quantum simulations, this algorithm is ideal for the simulation task. If an approximation to the unitary is needed that is valid for more general states, step~\ref{phi4:enum:infidelity} of the algorithm can be modified by minimizing the average of the infidelity over several states instead: 
\begin{equation}
    \min_{\vec{\theta}} \frac{1}{N}\sum_{i=0}^{N-1} I_d\left(\rho_{\text{targ},i}, \rho_{\text{ansatz},i}(\vec{\theta})\right)\ .
\end{equation}
The set of training states $\{|\psi_{\text{targ},i}\rangle,\,|\psi_{\text{ansatz},i}\rangle\}$ can be chosen to cover all the states of interest, or some representative subset of them~\cite{Mansuroglu:2021azm,Tepaske:2022uad,Guo:2024tnb}. This process takes advantage of the simplification of the problem of approximating a general unitary in the case of quantum simulation. Creating variational operators by optimizing matrix elements makes this algorithm much more tractable both for classical and quantum computing, than if the whole operator were optimized.

The power of this method lies within the choice of the ansatz circuit. The ansatz must be carefully chosen to have enough degrees of freedom and expressivity to have a chance at replicating the action of its target unitary. Greedy algorithms such as ADAPT-VQE have recently emerged as efficient and powerful ways to build the ansatz circuit~\cite{Grimsley:2018wnd,Feniou:2023gvo}. This method, in the form it is used in this work, is summarized below. 
\begin{enumerate}
    \item Begin with an initial state $|\psi_\text{ansatz}\rangle$, a target state $|\psi_\text{targ}\rangle$, and a pool of operators $\{O_i\}$ made up of commutators of terms in the Hamiltonian. These operators respect some or all of the symmetries of the system.
    \item Choose the next operator to append to the ansatz:
    \begin{enumerate}
        \item For each operator in the pool, optimize the local infidelity as a function of all parameters $\vec{\theta}$ in the ansatz, $\displaystyle \min_{\vec{\theta}}\, I_d\left(\rho_\text{targ}, \rho_\text{prep}(\vec{\theta})\right)$.\footnote{ADAPT-VQE was originally developed for general problems where $|\psi_\text{targ}\rangle$ is not known a-priori, so the optimization was implemented by minimizing the energy to find the ground state.}
        \item Append the operator that decreases the infidelity the most to the ansatz, and update all the parameters 
        \begin{align}
         |\psi_\text{ansatz}\rangle \ \rightarrow \ e^{-i\theta_iO_i}|\psi_\text{ansatz}(\theta_0,\theta_1,\dots,\theta_{i-1})\rangle \ = \ |\psi_\text{ansatz}(\theta_0,\theta_1,\dots,\theta_{i})\rangle \ .   
        \end{align}
    \end{enumerate}
    \item Repeat step 2 until the required convergence criterion is reached, such as a combination of circuit depth and infidelity. Use the parameters determined at the previous step as an initial guess for the next step.
\end{enumerate}
This algorithm corresponds to steps 1-2 of the SVC framework as one way to determine the ansatz and optimize the variational parameters at a fixed system size. When used with SVC, this algorithm is known as SC-ADAPT-VQE~\cite{Farrell:2023fgd,Farrell:2024fit}. Incorporating symmetries directly into the ansatz makes it more optimizable, and guarantees that the resulting state obeys the symmetries of the system (provided the input state also has the symmetries). This method and its numerous variants have been shown to perform extremely well for ground-state preparation. In particular, it has been used for ground-state preparation and particle creation in the context of translationally invariant systems~\cite{Farrell:2023fgd,Feniou:2023gvo,Farrell:2024fit,VanDyke:2022ffj,Gustafson:2024bww}. Impressively, it was also shown that although greedy methods may not find the best sequence of operators, almost all other sequences perform substantially worse, which removes the need to optimize over all possible sequences. 

It is important to consider the tradeoff between ease of optimization and circuit depth when selecting a variational ansatz. The simplest shallow and expressive ansatz is the brickwall ansatz, where each layer consists of parameterized single-qubit rotations followed by entangling gates. Compared to ADAPT-VQE, this approach trades the gate depth on the quantum device for the classical computational complexity involved in optimizing a large set of parameters for a set of circuits of increasing size. Optimizing a circuit with many parameters is a computationally intensive task that is susceptible to the well-known problem of barren plateaus~\cite{Cerezo:2023nqf,Larocca:2024plh}. Properties of the target state or target operator can be used to simplify the ansatz structure and reduce the number of parameters that need to be optimized. To ease the classical optimization task, new brickwall layers can be added and optimized iteratively using the same algorithm as described above.   

As such, finding the right balance between circuit depth and structure is imperative to create a trainable variational operator that mimics the target unitary with a shallow circuit. Using the physics of the system and the operator being approximated to inform the design of the ansatz proves to be a powerful way to increase the performance of this variational method~\cite{Sim:2019yyv,Seki:2020nnj,Funcke:2020vkw,Anschuetz:2022wvo,Gibbs:2024ggs}.

\section{Quantum simulation of scalar field theory using scalable variational circuits}
\label{phi4:sec:circuits}
\begin{figure*}
    \centering
    \includegraphics[width=0.9\linewidth]{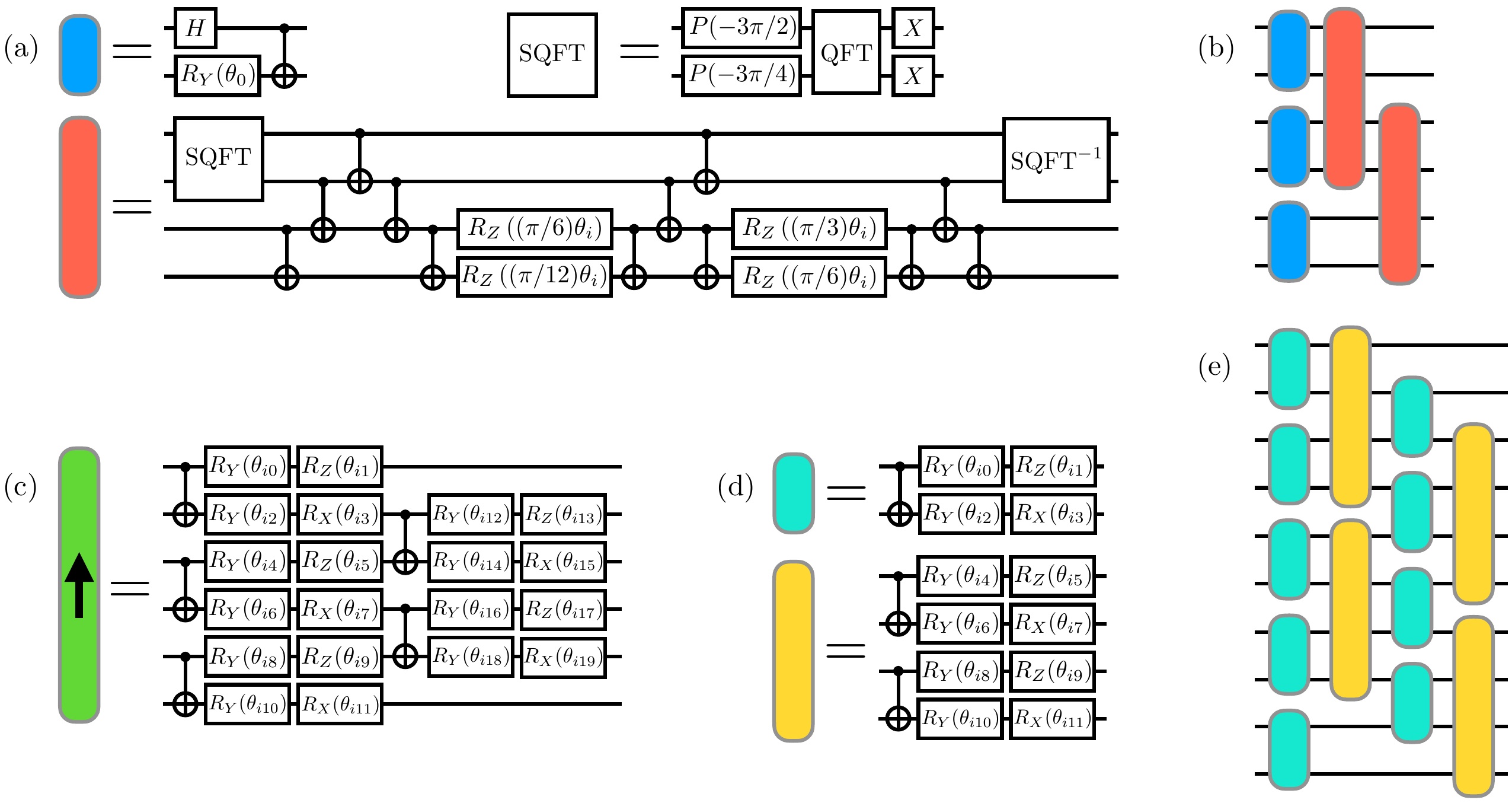}
    \caption{{\it Circuit elements used for state preparation and time evolution in simulations of scattering.} (a): The circuit elements used to prepare the vacuum. The input state to SC-ADAPT-VQE is a real, symmetric tensor product state over the spatial sites prepared by a single-site variational operator (blue). The circuit implementing $e^{-i\theta O_1}$ (red) is used to couple neighboring sites. The symmetric version of the QFT defined in Ref.~\cite{Klco:2018zqz}, SQFT, is used to account for the symmetric digitization of $\phi$ and $\Pi$ around 0. (b): The translationally invariant circuit to prepare the vacuum. One layer of SC-ADAPT-VQE is Trotterized into separate terms coupling even-odd and odd-even spatial sites. This circuit has depth 25 and 3 variational parameters (one for the input state, and one for each application of $O_1$). (c): One layer of the brickwall circuit that is used to create particle wavepackets with negative momentum (indicated by the up arrow) on top of the vacuum. To create the corresponding positive-momentum particles, the circuit is flipped and site-wise SWAP gates are added. Each layer has 20 parameters and is depth 2. The building block of this circuit is the variational gate introduced in Ref.~\cite{Madden:2021dax}. (d): The circuit elements that are used to implement time evolution. Both single- (cyan) and two-site (yellow) operators are used, mimicking the terms present in the Hamiltonian of Eq.~\eqref{phi4:eq:lattice_h}. The building block of these circuits is the same as (c). (e): Two layers of the translationally invariant circuit used to implement time evolution. Each layer has depth 2 and 12 parameters. Single- and two-site operators are applied in an iterative fashion.}
    \label{phi4:fig:circuit_elements}
\end{figure*}

The scalable variational circuits procedure can be carried out on a classical computer or a small quantum computer. Throughout this work, classical computational resources are leveraged to determine parameters that are extrapolated to system sizes of interest to be run on quantum computers. This work uses $n_q=2$ qubits to represent the state of $\phi$ at each site, for the theory with bare mass $m=1/2$. The method of Ref.~\cite{Klco:2018zqz} is used to choose the cutoff $\phi_\text{max}=1.5$ to minimize field digitization effects, which are studied in App.~\ref{phi4:sec:digitization_effects}. The dynamics of free $(\lambda=0)$ and interacting $(\lambda=2)$ wavepackets is compared. For the parameters chosen, the correlation lengths $\xi_\lambda$ of the free and interacting theory are $\xi_0=2.4503$ and $\xi_2=1.5817$, respectively. These are determined, as in the previous section, by computing the inverse gap for small but increasing system sizes $L$, fitting an exponential, and extrapolating to the infinite-volume limit. The circuit elements that are used to compose the variational circuits throughout this section are shown in Fig.~\ref{phi4:fig:circuit_elements}. Details of the optimization strategies and choices are given in App.~\ref{phi4:sec:optimization}. The resulting parameters $\vec{\theta}_\text{opt}$ for all components of the variational circuit are given in App.~\ref{phi4:sec:variational_params}.

\subsection{Vacuum preparation}
\label{phi4:sec:circuits_vac_prep}
As a first step toward simulating a scattering process, the SVC framework is used to prepare the ground state of the scalar field theory. Together with a variational input state, SC-ADAPT-VQE is used to prepare the vacuum of the theory both in the free and interacting cases. Compared to other methods for preparing ground states, such as Quantum Imaginary Time Evolution~\cite{Motta:2019yya} or subspace methods~\cite{Yoshioka:2024lle}, SC-ADAPT-VQE gives shallow circuits with few variational parameters, making optimization easy and enabling implementation on modern-day devices.

In the free theory, $|\psi_\text{vac}\rangle$ defined in Eq.~\eqref{phi4:eq:vacuum_wavefunction} can be used as $|\psi_\text{targ}\rangle$. Approximations to the vacuum of the interacting theory can be found analytically using standard perturbative methods for small $\lambda$ as is discussed in Sec.~\ref{phi4:sec:lattice_scalar_field_theory}. Fortunately, for the chosen system parameters $m$ and $\lambda$, the variational parameters of the circuits initializing these states show convergence for small-enough system sizes where exact diagonalization (ED) is viable. Taking advantage of this, the interacting ground state determined from ED is used as $|\psi_\text{targ}\rangle$. 

It is important to choose a good initial input for the SC-ADAPT-VQE algorithm. There has been recent work using MPS or other ansatze as input to variational algorithms to improve their performance~\cite{Ravi:2022cwu,Khan:2023uhz,Fomichev:2023mtt}. Time reversal symmetry (implying the reality of the ground-state wavefunction), $\phi \rightarrow -\phi$ symmetry, and translational invariance can be used to choose a judicious input state to the greedy algorithm. The circuit to initialize the input state, shown in blue in Fig.~\ref{phi4:fig:circuit_elements}a, uses one variational parameter to create an arbitrary real, symmetric state on each spatial site. This product state can be viewed as the vacuum $| \widetilde{\psi}_\text{vac}\rangle$ of the theory with the correlation matrix $K=V^\dagger E V$ truncated to only contain elements on its diagonal: 
\begin{equation}
    \langle{\vec{\phi}} | \widetilde{\psi}_\text{vac} \rangle \ = \ \mathcal{N} e^{-\frac{1}{2}\vec{\phi}^T \text{diag}(K)\vec{\phi}}\ . \label{phi4:eq:adapt_vqe_input_state}
\end{equation}
By choosing this state, or a state like this, the variational algorithm only has to build out the correlations between spatial sites.

The Hamiltonian of the theory is real and time reversal invariant, so it must have real eigenstates. It is found that an operator pool built out of commutators of terms in the Hamiltonian, similar to the pools used in Refs.~\cite{Farrell:2023fgd,Farrell:2024fit}, is effective. A pool with this structure ensures the reality of the prepared vacuum by applying only real unitaries, and maintains the symmetries of the system. The lowest-order operator in $\phi$ and $\Pi$ that acts over more than one site and is not included in the Hamiltonian is derived from the commutator of terms in $H_\text{kin}$ and $H_\Pi$, $O_1 = -\frac{i}{2}\sum_j[\phi_j \phi_{j+1},\Pi^2_j]$. The canonical commutation relation $[\phi_i, \Pi_j] = i\delta_{ij}$ only holds in the limit $\delta_\phi \rightarrow 0$ (i.e., only for the undigitized operators $\phi_j$, $\Pi_j$). Nevertheless, this relation can still be used at nonzero $\delta_\phi$ to simplify the operators in the SC-ADAPT-VQE pool because the requisite symmetries are maintained. This makes operators such as $O_1$ easier to implement using quantum gates. Using this, $O_1$ can be rewritten as 
\begin{align}
    O_1 \ = \ \sum_j\Pi_j \phi_{j+1}\ , \label{phi4:eq:pi_phi}
\end{align}
which can readily be implemented on a quantum device using the techniques described in Sec.~\ref{phi4:sec:qubit_representation}. To build out longer correlations, the separation between the sites that $O_1$ acts on can be increased: 
\begin{equation}
    O_d \ = \ \sum_j\Pi_j \phi_{j+d}\ . \label{phi4:eq:pi_phi_d}
\end{equation}
This operator can similarly be obtained from the commutator of $\Pi_j^2$ with a longer-range version of $H_\text{kin}$. The pool used for SC-ADAPT-VQE consists of the operators $\{O_d\}$  for a range of increasing $d$. Note that the operators $O_d$ are imaginary in the $\phi$-basis. This is because $\phi_j$ is real and $\Pi_j$ is imaginary since it is the conjugate momentum operator. As a result, $e^{-i\theta O_d}$ is a real unitary.

The circuit implementing $e^{-i\theta O_1}$ on a pair of sites is given in red in Fig.~\ref{phi4:fig:circuit_elements}a. Because $O_1$ acts on all pairs of neighboring sites, it must be Trotterized into an operator acting on even-odd sites, and an operator acting on odd-even sites. This is shown in Fig.~\ref{phi4:fig:circuit_elements}b, where there are two layers of the $O_1$ circuit. This Trotterization breaks translational invariance, so one variational parameter is used for each Trotter layer to partially mitigate this effect. 

\begin{figure*}
    \centering
    \includegraphics[width=\linewidth]{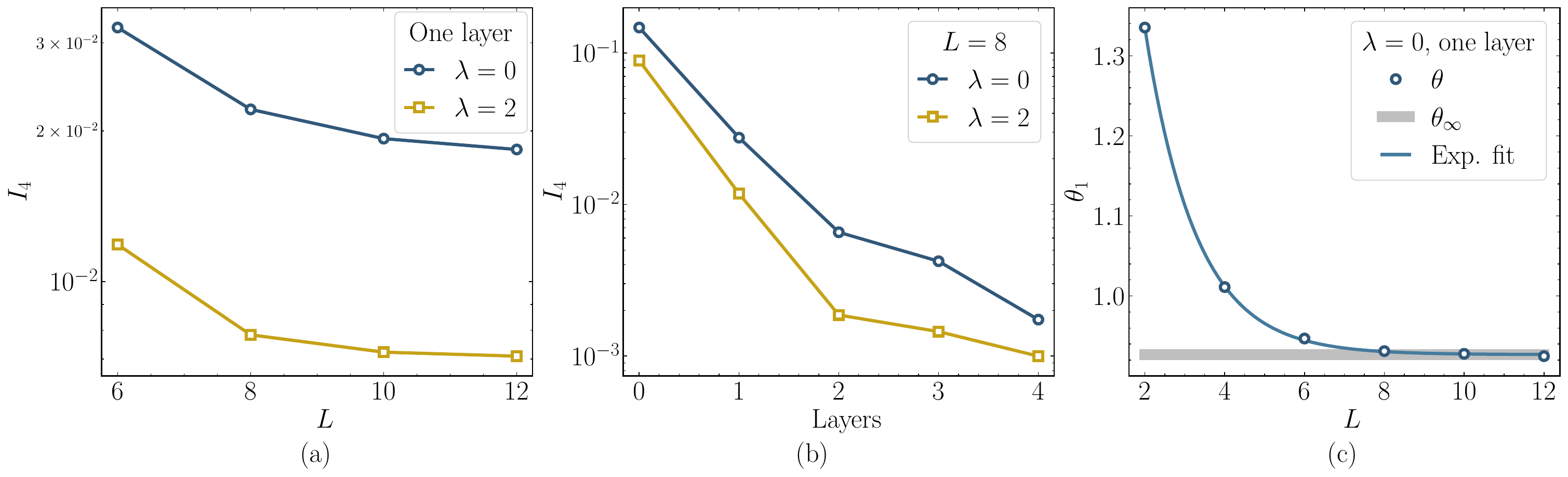}
    \caption{{\it Convergence of wavefunction and state preparation parameters.} 
    Convergence is shown corresponding to the circuit in Fig.~\ref{phi4:fig:circuit_elements}b for both the free and interacting theory. (a): $I_4$ taken over a region of four contiguous lattice sites (eight qubits) as a function of system size. One layer of SC-ADAPT-VQE is used. (b): $I_4$ as a function of number of SC-ADAPT-VQE layers used in the ansatz for a $L=8$ system. Each layer uses a longer-range operator $O_d$ given in Eq.~\ref{phi4:eq:pi_phi_d} to create correlations that span more lattice sites. (c): An example of the exponential convergence of one of the angles parameterizing the $\lambda=0$ vacuum preparation circuit with one layer of SC-ADAPT-VQE, as a function of system size. The remainder of the angles are given numerically in App.~\ref{phi4:sec:variational_params}.
    }
    \label{phi4:fig:vac_prep_figs}
\end{figure*}

Figure~\ref{phi4:fig:vac_prep_figs} shows the convergence of the variationally prepared vacuum. The local infidelity over four sites, $I_4$, is used to measure the quality of the prepared state, spanning roughly 2-3$\times$ the correlation length of the infinite-volume vacuum. With just one SC-ADAPT-VQE layer, $I_4$ can be seen to be non-increasing in Fig.~\ref{phi4:fig:vac_prep_figs}a, meaning that the local quality of the prepared wavefunction is not degrading with system size for both $\lambda=0$ and $\lambda=2$. The infidelity is systematically improved by applying more layers of $O_d$ with successively larger $d$. The result of longer-range operators creating correlations that span more lattice sites is seen in the plots of $I_4$ in Fig.~\ref{phi4:fig:vac_prep_figs}b. In both of these figures, it can be seen that the $\lambda=2$ wavefunction converges faster as a result of the smaller correlation length in the interacting theory. Figure~\ref{phi4:fig:vac_prep_figs}c is a representative example of the convergence of the parameters with increasing system size. These plots indicate that the $L$-extrapolation is applicable and the parameters to initialize the ground state for an arbitrary system size can be determined using modest $L$, where classical simulation is possible.\footnote{If the SVC method is run using classical computation, the correlation length of ground states that can be prepared with the presented method is limited by the maximum number of qubits that can be simulated classically.}

The circuit shown in Fig.~\ref{phi4:fig:circuit_elements}b, consisting of the circuit preparing the symmetric input state and one layer of SC-ADAPT-VQE, is used to initialize the vacuum. It has a two-qubit gate depth of 25, making it shallow enough to save circuit depth for wavepacket preparation and time evolution. The extrapolated parameters $\vec{\theta}_\text{opt}$ that implement the vacuum preparation are given in Table~\ref{phi4:tab:gs_prep_angles}, and numerical $I_4$ values are given in Table~\ref{phi4:tab:error_budget}.

\subsection{Wavepacket creation}
\label{phi4:sec:circuits_wp_prep}
The creation of particles on top of the prepared vacuum requires initializing local excited states. JLP proposed a method of creating wavepackets in the free theory by applying unitary evolution under $a_k^\dagger$ and $a_k$ defined in Eq.~\eqref{phi4:eq:ap}, and then adiabatically turning on the interaction as described in Sec.~\ref{phi4:sec:lattice_scalar_field_theory}.\footnote{Whereas JLP specify the form of their wavepackets in position space, the method used in this work specifies the target profile in $k$-space. See App.~\ref{phi4:sec:wp_creation_details} for a comparison.} Instead of this, SVC is used to compress the wavepacket preparation circuit by taking advantage of classical computation. This step of the algorithm prepares a wavepacket whose size remains fixed as the total system size increases. While this method does not allow for the scalable preparation of wavepackets of increasing size (as will be necessary for simulations approaching the continuum), it utilizes the scalability in total system size, including vacuum regions.\footnote{In this case, the classical computing capability limits the maximum size of the wavepacket.} This is useful because a finer spacing in $k$-space increases the spatial momentum resolution of the wavepacket, building a more accurate representation of wavepackets in the continuum. Furthermore, wavepackets may only scatter between different lattice momenta in a lattice quantum field theory. Increasing the system size increases the total number of lattice momenta and thus more faithfully reproduces the target continuum process.

\begin{figure*}
\centering
    \includegraphics[width=\linewidth]{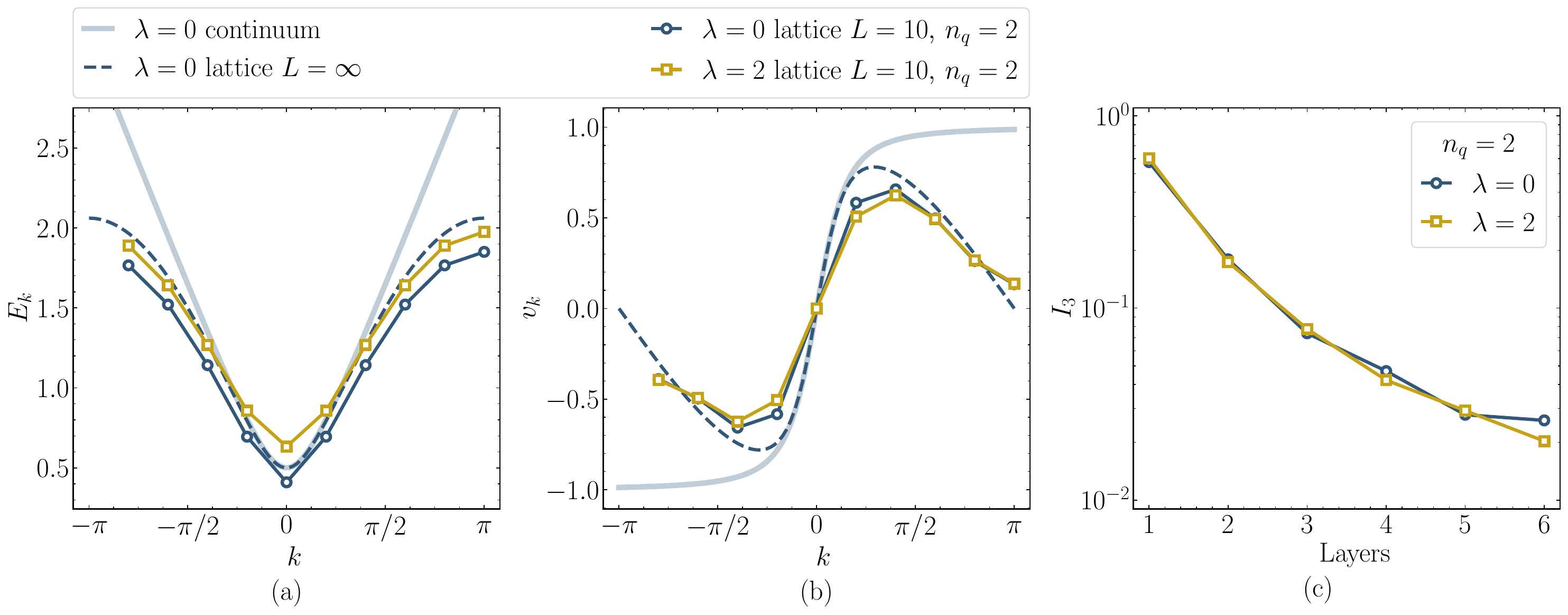}
    \caption{{\it Group velocity, dispersion relation, and wavepacket convergence.} (a): The dispersion relation $E_k$ and (b): the group velocity $v_k$ for $m=1/2$ and $\lambda=0,2$. Exact computation is possible in the continuum (faint solid lines) and in the free lattice theory (dashed lines). The values of $E_k$ and $v_k$ are computed numerically by using ED on the Hamiltonian of Eq.~\eqref{phi4:eq:lattice_h} and projecting onto each momentum sector (solid lines with circle and square markers). The numerical computations are done on a lattice of $L=10$ spatial sites with $n_q=2$ qubits per site. Discrete gradients are used for the numerical computations of $v_k$. (c): The convergence of the prepared particle wavepackets occupying three spatial sites with momentum $k=-\pi/3$, as a function of number of layers of the circuit from Fig.~\ref{phi4:fig:circuit_elements}c. The local infidelity spanning the width of the wavepacket, $I_3$, is used to measure the quality of the prepared wavefunction for $\lambda=0,2$.}
    \label{phi4:fig:e_k_v_k_wp_prep_convergence}
\end{figure*}

To reduce the impact of lattice artifacts, the momenta of the wavepackets must be carefully chosen. The infinite-volume lattice dispersion relation (Eq.~\eqref{phi4:eq:dispersion}) mimics that of the continuum, $E_{k,\text{cont.}} = \sqrt{m^2+k^2}$, for small $k$. These are shown by the faint solid and dashed lines respectively in Fig.~\ref{phi4:fig:e_k_v_k_wp_prep_convergence}a. The group velocity of a particle wavepacket with momentum $k$ is found in the infinite-volume lattice theory to be
\begin{align}
    v_k \ &= \ \frac{\partial E_k}{\partial k} \ = \ \frac{\sin{k}}{\sqrt{m^2 + 4 \sin^2{\frac{k}{2}}}}\ , \label{phi4:eq:group_velocity}
\end{align}
and in the continuum to be 
\begin{align}
    v_{k,\text{cont.}} \ &= \ \frac{\partial E_{k,\text{cont.}}}{\partial k} \ = \ \frac{k}{\sqrt{m^2 + k^2}}\ . \label{phi4:eq:group_velocity_cont}
\end{align}
These are plotted as the dashed and faint solid lines respectively in Fig.~\ref{phi4:fig:e_k_v_k_wp_prep_convergence}b. $E_k$ and $v_k$ can be computed numerically in the digitized theory for any value of $\lambda$ by using ED and projecting onto each momentum sector. These numerically calculated values are shown in Figs.~\ref{phi4:fig:e_k_v_k_wp_prep_convergence}a and b by the solid lines with circle and square markers for $L=10$.

To simulate a high-energy collision, the highest momenta should be selected where the (digitized) lattice dispersion relation is still close to that of the continuum. With $L=10$ spatial sites, the resolution is poor and the digitized $v_k$ only match the continuum values at small $k$. The agreement can be improved with finer resolution by increasing the number of lattice sites. Finite-volume effects are reduced by using many lattice sites outside the interaction region. Based on this and on considerations of circuit depth and spread in $k$-space, wavepackets with $k=\pm\pi/3$ are chosen, with spread $\sigma_k=\pi/3$. The choice of $\sigma_k$ determines the support of the wavepacket in both position and momentum space. To limit the quantum resources required, the wavepacket width is truncated so that the circuits that create these states only act where the wavepackets have nonvanishing support. In this work, the spatial extent of the wavepackets is fixed to three lattice sites, and the circuits that prepare them act over $3n_q=6$ qubits.\footnote{Similar to the field digitization, choosing the extent of the wavepackets in position and momentum space according to the Nyquist-Shannon Sampling Theorem will ensure a double-exponential convergence in precision for future calculations.} A final choice in the setup of the simulation is the initial location of the wavepackets. The wavepackets should be well-separated at $t=0$ so they can be approximately treated as asymptotic ``in'' states. While this is strictly true only for infinite separation, the wavepackets are local, and their extent is fixed by the width of the circuit preparing them. Taking advantage of the locality of the $\phi^4$ interaction, the wavepackets are initialized to be separated by one spatial site. This construction may need to be modified for theories where interactions have less locality.

Leveraging the analytic knowledge of the free theory, the wavefunction of the wavepacket $|\psi_\text{wp}\rangle$ defined in Eq.~\eqref{phi4:eq:wp_wavefunction} is used as $|\psi_\text{targ}\rangle$ for the variational optimization. The adiabatic turn-on of interactions given by Eq.~\eqref{phi4:eq:u_adiabatic} is simulated classically, producing interacting wavepacket wavefunctions that are then used for $|\psi_\text{targ}\rangle$ in SVC for the interacting theory. See App.~\ref{phi4:sec:wp_creation_details} for details on the classical determination of states used as $|\psi_\text{targ}\rangle$ for this step of state preparation. With this method, interacting wavepackets are prepared directly on top of the interacting vacuum. This way, in addition to reducing the circuit depth, SVC also removes the necessity for implementing the adiabatic evolution to turn on interactions. 

Because the region where the desired wavepacket has support does not grow with increasing system size $L$, the number of qubits over which the variational circuit has to act is fixed by the spatial width of the wavepacket. Scaling these circuits is simple and straightforward since the number of parameters and structure do not change with system size. This is another example of using physical properties of the state, in this case locality of the wavepacket, to guide the design of the variational circuit. One layer of the brickwall ansatz used for this step of the state preparation is shown in Fig.~\ref{phi4:fig:circuit_elements}c. The building blocks of this circuit are the variational gates defined in Ref.~\cite{Madden:2021dax}. This block is particularly useful because any product of gates from the universal gate set $\{\text{CNOT},R_X,R_Y,R_Z\}$ can be written in terms of products of the block with different parameters. Each layer of the brickwall ansatz has a two-qubit gate depth of 2 and uses 20 variational parameters.\footnote{The number of parameters can be reduced by incorporating the physical structure of the wavepackets (e.g., the spatial symmetry about the peak) into the circuits.}

Since spatially separated particles are required for the ``in'' states of the scattering simulation, the wavepacket creation circuits are executed for both wavepackets in parallel. The same set of parameters is used to initialize both the left and right wavepackets to maintain symmetry. The circuit preparing the left $(k=+\pi/3)$ wavepacket is created in such a way that the resulting wavepacket is the mirror image of the right $(k=-\pi/3)$ wavepacket. This is indicated by the up (down) arrows for $k=-\pi/3$ $(k=+\pi/3)$ respectively, seen in the full circuit shown in Fig.~\ref{phi4:fig:full_circuit}. Site-wise SWAP gates are used to flip the qubit mapping in Eq.~\ref{phi4:eq:phi_digitized}, increasing the two-qubit gate depth of this part of the circuit by six.

Figure~\ref{phi4:fig:e_k_v_k_wp_prep_convergence}c shows the convergence of the prepared wavepacket state as a function of the number of layers of the brickwall ansatz of Fig.~\ref{phi4:fig:circuit_elements}c. It is seen that the local infidelity spanning the wavepacket width, $I_3$, is systematically improved by adding more layers. Four layers of the circuit shown in Fig.~\ref{phi4:fig:circuit_elements}c are used to prepare the free and interacting wavepackets.\footnote{Note that with the parameters set to zero, inserting CNOTs between layers of this circuit will collapse the circuit to the identity. This is used in vacuum simulations to maintain the structure of the circuit.} These brickwall circuits have two-qubit gate depths of 14 (including extra site-wise SWAP gates), and 80 parameters in total. While it is possible to run an extrapolation of the wavepacket parameters, this is not necessary. The angles parameterizing the vacuum creation circuits are observed to change on the scale of $10^{-4}-10^{-3}$ for $L=10-12$ (see Table~\ref{phi4:tab:gs_prep_angles}), which is at the precision tolerance of IBM devices. The extrapolated angles for the vacuum preparation are used to prepare the equivalent vacuum state on a smaller system size for purposes of training the wavepacket circuit. The vacuum preparation angles extrapolated to $L=60$ are used in an $L=12$ system to optimize the wavepacket preparation parameters. The resulting parameters $\vec{\theta}_\text{opt}$ are given in Table~\ref{phi4:tab:wp_prep_angles}, and numerical $I_3$ values are given in Table~\ref{phi4:tab:error_budget}.

\subsection{Time evolution}
\label{phi4:sec:circuits_time_evolution}
Scalable variational circuits can also be used to produce compressed time evolution circuits in place of non-variational methods such as Trotterization, QDrift~\cite{Campbell:2019fez}, or Linear Combination of Unitaries~\cite{Childs:2012gwh}. There has been much work in recent years in this direction. The structure of the ansatze that have been used varies widely, from circuits with parameterized Trotter terms~\cite{Mansuroglu:2021azm,Tepaske:2022uad,Kotil:2022vmv,Tepaske:2023mfc}, to brickwall ansatze~\cite{Mizuta:2022rrz,Keever:2022hfy,Miyakoshi:2023zzc,Causer:2023wpp,Kanasugi:2024ivt,Gacon:2024bsc,Gibbs:2024emw}. The related fast-forwarding techniques~\cite{Commeau:2020aab,Gu:2021hyo} involve using variational circuits to approximately diagonalize a Hamiltonian to implement evolution in a number of gates that is constant with $t$ but scales with $L$ because of the diagonalization step.

\begin{figure*}
\centering
    \includegraphics[width=0.65\linewidth]{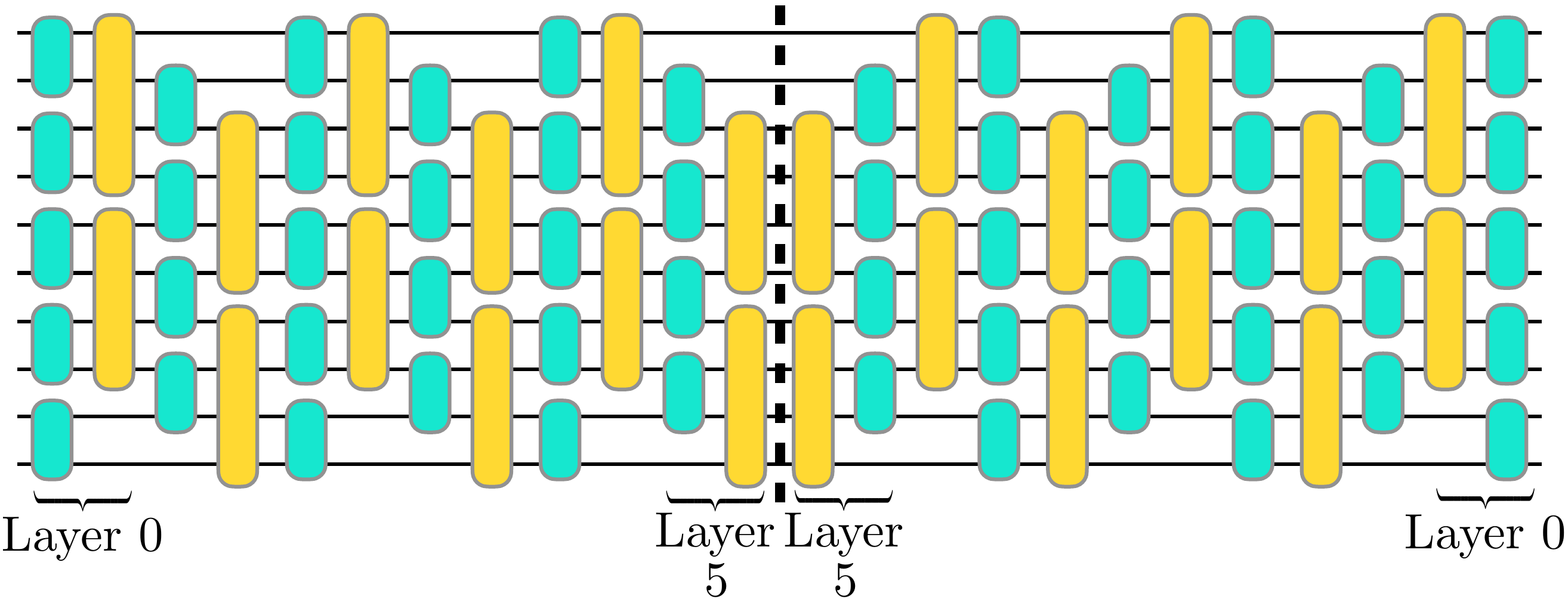}
    \caption{{\it Variational time evolution circuit.} One step consists of six forward layers followed by six reverse layers of the circuits of Fig.~\ref{phi4:fig:circuit_elements}d. Single- and two-body terms are used in an alternating brickwall fashion, mimicking the terms present in the system Hamiltonian.}
    \label{phi4:fig:time_evolution_circ_step}
\end{figure*}

In this work, a translationally invariant brickwall ansatz is used to compress the time evolution operator. Note that while there is no guarantee that a shallow brickwall circuit can approximate the time evolution operator well, time evolution under any Hamiltonian with local interactions is subject to Lieb-Robinson bounds~\cite{Lieb:1972wy}. Moreover, there is a maximum value for the group velocity of excitations from the lattice dispersion relation (Eq.~\eqref{phi4:eq:group_velocity})~\cite{Farrell:2024mgu} so it is reasonable to expect that low-depth brickwall circuits can reliably compress evolution under the Hamiltonian of Eq.~\eqref{phi4:eq:lattice_h} for sufficiently early times. The ansatz is made up of layers of the circuit shown in Fig.~\ref{phi4:fig:circuit_elements}e, arranged in a symmetric way, seen in Fig.~\ref{phi4:fig:time_evolution_circ_step}. Each variational step is made up of six forward layers and six reverse layers, with the reverse layers having the same parameters as the forward layers. This structure is chosen to mimic the structure of a second-order Trotter step. Each layer uses 12 parameters and has a two-qubit gate depth of two. The time evolution of the two-wavepacket state $e^{-itH}|\psi_\text{2wp}\rangle$ with 100 second-order Trotter steps on an $L=12$ system is used as $|\psi_\text{targ}\rangle$ to determine the variational circuits.

Similar to the wavepacket preparation step, the $L$-extrapolation is not necessary to implement time evolution for the times simulated in this work. However, in this case this simplification is only valid for early times $(t\sim L)$, because the time evolution circuits act on the whole lattice. Since the vacuum is not prepared exactly, and the time evolution is not implemented exactly, the vacuum regions will evolve in time~\cite{Hayata:2024fnh}. This evolution will be captured by the variational optimization procedure, up to finite-size effects that scale as $O(t/L)$.\footnote{These effects are estimated and the vacuum evolution is used for error mitigation. See App.~\ref{phi4:sec:error_mitigation} for details.} To get around this limitation, the optimization can be run on larger systems for later times, or an $L$-extrapolation can be implemented. Propagating particles in a scattering simulation show more drastic effects from finite-size effects, such as wrapping around the lattice due to PBCs and re-scattering. Furthermore, larger wavepackets (needed for high-energy simulations or simulations approaching continuum) require more time to overlap, scatter, and fully separate. In these situations, it would be necessary to run the optimization for later simulation times on a larger lattice. While this is a limitation of the method in general, the scattering simulations are set up to require minimal time before scattering is observed (see Sec.~\ref{phi4:sec:circuits_wp_prep} for details). This places a limit both on the time required in the simulation, as well as the size of the interaction region that needs to be used for training the variational circuits. 

One promising way to use variational circuits for time evolution without running into these finite-size effects is to compress a single time evolution step, and use that step repeatedly in place of a Trotter step. Similar methods have been investigated in Refs.~\cite{Tepaske:2023mfc,Causer:2023wpp}. While the resulting circuits will be deeper than those produced by the current method, they will be more scalable with $L$. Another interesting direction is using MPS simulations of the time evolution operator for training~\cite{Miyakoshi:2023zzc,Causer:2023wpp,Gibbs:2024emw}. This would remove the limitations placed on this method by finite-size effects. In this setup, the time evolution of one or many states may be used for training, as described in Sec.~\ref{phi4:sec:scalable_variational_circuits}. Methods that target time evolution operators in specific subspaces, such as restricted to the low-energy subspace, are currently being developed~\cite{Kanasugi:2024ivt,Li:2024lrl}. The combination of subspace methods with variational time evolution has the potential to simplify the optimization task while also improving the quality of the produced circuits.

\begin{figure*}
\centering
    \includegraphics[width=\linewidth]{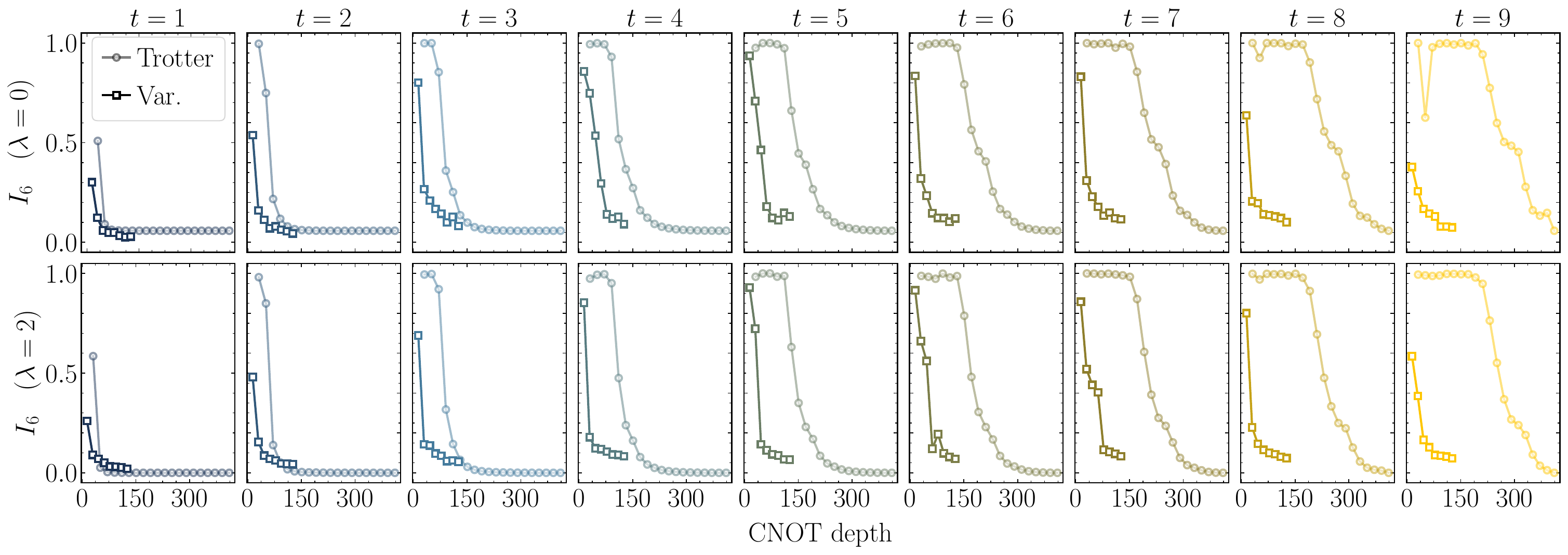}
    \caption{{\it Convergence of the time-evolved single-wavepacket state.} An $L=6$ system is shown for $\lambda=0$ (top row) and $\lambda=2$ (bottom row) with $t=1-9$. The second-order Trotter time evolution (circles) is compared to the variational steps (squares) as a function of two-qubit gate depth. Two variational steps consisting of an increasing number of layers shown in Fig.~\ref{phi4:fig:circuit_elements}d are used. The infidelity of the whole $L=6$ system with the exact time-evolved state is used to measure the quality of the wavepacket state as it propagates through the lattice.}
    \label{phi4:fig:trot_vs_variational_fidelity}
\end{figure*}

Figure~\ref{phi4:fig:trot_vs_variational_fidelity} shows the convergence of the time evolution variational circuits, compared to second-order Trotter circuits as a function of circuit depth. The plots show the infidelity as a function of two-qubit gate depth for the time evolution of a single wavepacket on an $L=6$ system. The Trotter time evolution circuits are constructed by adding more Trotter steps as described in Sec.~\ref{phi4:sec:qubit_representation}. Variational circuits fix the number of variational steps to two while increasing the depth of each step by adding layers of the circuit elements in Fig.~\ref{phi4:fig:circuit_elements}d following the structure of Fig.~\ref{phi4:fig:time_evolution_circ_step}. The results for $t=1-9$ show that the brickwall circuits are able to consistently compress the Trotter circuits. This can especially be seen at late times, where considerably deeper Trotter circuits are required to accurately implement the time evolution operator, whereas the variational circuits show a modest growth in depth. Sharp dips in the curves of both the variational and Trotter infidelities are results of finite-size effects, as well as of numerical optimization (for the variational lines). A difference in the infidelity between $\lambda=0$ (top row) and $\lambda=2$ (bottom row) can be seen, which is attributed to using fixed-depth circuits to describe states with different correlation lengths. These plots show that by increasing the number of layers (and so, the number of parameters), the variational ansatz is able to represent the time-evolved state with increasing accuracy. 

For the runs on the device, the number of variational steps is increased with the simulation time: $t=\{1,2,3\}$ use one variational step, $t=\{4,5,6\}$ use two, and $t=\{7,8,9\}$ use three. The number of layers in each step is kept fixed to six forward and six reverse layers, as in Fig.~\ref{phi4:fig:time_evolution_circ_step}. The steps are optimized separately for each simulation time to find better quality solutions for the variational circuit. They are applied in a Trotter-like fashion, so that the number of parameters is the same for all time steps. The time evolution portion of the variational circuit uses 72 parameters in total. The parameters $\vec{\theta}_\text{opt}$ that implement the time evolution are given in Tables~\ref{phi4:tab:time_evolution_angles_1}-~\ref{phi4:tab:time_evolution_angles_9} and the local infidelities $I_{10}$ for each time step are shown in Table~\ref{phi4:tab:error_budget}.

\subsection{Summary of variational methods and error budget}
\label{phi4:sec:circuits_error_budget}
The choices in the design of the variational circuits and methods of determining parameters for a system of $L=60$ lattice sites are summarized below.
\begin{itemize}
    \item The vacuum $|\psi_\text{vac}\rangle$ is created using an input state preparation circuit and one step of SC-ADAPT-VQE, shown in Figs.~\ref{phi4:fig:circuit_elements}a and b. The circuits are trained on $|\psi_\text{targ}\rangle$ determined from ED with $L=2-12$, and parameters are extrapolated to $L=60$. This step of the state preparation has a two-qubit gate depth of 25 and three variational parameters.
    \item The two-wavepacket state $|\psi_\text{2wp}\rangle$ is initialized using two layered brickwall circuits in parallel (one layer is shown in Fig.~\ref{phi4:fig:circuit_elements}c). A single set of parameters is used for wavepackets with opposite momenta by spatially flipping the circuit and introducing site-wise SWAP gates. These circuits are determined using the $L=60$ vacuum parameters on an $L=12$ system, and the angles are not extrapolated. The prepared wavepackets span three lattice sites and have $k=\pm\pi/3$ and $\sigma_k=\pi/3$. The continuum expression for the $\lambda=0$ wavepacket (Eq.~\eqref{phi4:eq:wp_wavefunction}) and the resulting state after the adiabatic turn-on of $\lambda$ (Eq.~\eqref{phi4:eq:u_adiabatic}) are used as $|\psi_\text{targ}\rangle$ for the free and interacting theories, respectively. This step of the state preparation has a two-qubit gate depth of 14 and 80 parameters.
    \item Time evolution for $t=1-9$ is implemented using the translationally invariant brickwall circuit shown in Fig.~\ref{phi4:fig:time_evolution_circ_step}. It has a second-order Trotter step-like structure, and the number of steps is increased with time. The circuits are optimized using time evolution of the two-wavepacket state $|\psi_\text{2wp}\rangle$ as $|\psi_\text{targ}\rangle$, determined using 100 second-order Trotter steps. They are trained on a lattice of size $L=12$ with the vacuum preparation parameters for $L=60$ and the wavepacket preparation parameters determined in the previous step. The parameters are not extrapolated. This part of the circuit has a two-qubit gate depth of $22\lceil\frac{t}{3}\rceil$ and 72 variational parameters.
\end{itemize}

The systematic errors introduced by the variational approximations used in this work can be quantified by running exact classical simulations using {\tt qiskit}~\cite{Javadi-Abhari:2024kbf} (see App.~\ref{phi4:sec:digitization_effects} for details on the digitization effects). The local infidelity $I_d$ (Eq.~\eqref{phi4:eq:local_infidelity}), which is observed to be independent of system size as a result of the SVC framework, is used to quantify the local quality of the prepared state compared to the state of interest. The width of the region, $d$, over which the reduced state infidelity is computed, is chosen to be representative of the given step of the algorithm. For vacuum preparation, a subsystem spanning lattice sites covering several correlation lengths is used. Because of the translational invariance of the vacuum, this region can be located anywhere on the lattice, and $I_d$ is observed to be uniform throughout.\footnote{Translational invariance is seen over pairs of sites. This is a result of the vacuum preparation scheme using only operators spanning two neighboring sites. See Sec.~\ref{phi4:sec:circuits_vac_prep} for a discussion.} For  wavepacket preparation, the wavepacket width is used. A subsystem spanning the area where the interaction takes place from $t=1-9$ is used for all steps of time evolution. The values of $I_d$ at each step of the simulation algorithm are shown in Table~\ref{phi4:tab:error_budget}.

\begin{table}
\centering
\begin{tabularx}{\linewidth}{|c|c||Y||Y||Y|Y|} \hline
\multicolumn{2}{|c||}{} &  \multicolumn{2}{c||}{$I_d$} & \# of parameters & Parameter values \\ \hline
\multicolumn{2}{|c||}{Simulation step} & $\lambda=0$ & $\lambda=2$ & &\\ \hline\hline
\multicolumn{2}{|c||}{Vacuum preparation $(d=4)$} & 0.0181 & 0.0071 & 3 & Table~\ref{phi4:tab:gs_prep_angles}\\ \hline\hline
\multicolumn{2}{|c||}{Wavepacket preparation $(d=3)$} & 0.0555 & 0.0550 & 80 & Table~\ref{phi4:tab:wp_prep_angles}\\ \hline\hline
\multirow{10}{*}{\makecell{Time\\evolution\\$(d=10)$}} & $t=1$ & 0.1680 & 0.1275 & 72 & Table~\ref{phi4:tab:time_evolution_angles_1}\\ \cline{2-6}
& $t=2$ & 0.3307 & 0.2024 & 72 & Table~\ref{phi4:tab:time_evolution_angles_2}\\ \cline{2-6}
& $t=3$ & 0.5412 & 0.3157 & 72 & Table~\ref{phi4:tab:time_evolution_angles_3}\\ \cline{2-6}
& $t=4$ & 0.5692 & 0.3674 & 72 & Table~\ref{phi4:tab:time_evolution_angles_4}\\ \cline{2-6}
& $t=5$ & 0.6930 & 0.3675 & 72 & Table~\ref{phi4:tab:time_evolution_angles_5}\\ \cline{2-6}
& $t=6$ & 0.6447 & 0.4214 & 72 & Table~\ref{phi4:tab:time_evolution_angles_6}\\ \cline{2-6}
& $t=7$ & 0.6106 & 0.4032 & 72 & Table~\ref{phi4:tab:time_evolution_angles_7}\\ \cline{2-6}
& $t=8$ & 0.6087 & 0.3937 & 72 & Table~\ref{phi4:tab:time_evolution_angles_8}\\ \cline{2-6}
& $t=9$ & 0.5932 & 0.4207 & 72 & Table~\ref{phi4:tab:time_evolution_angles_9}\\ \cline{2-6}
\hline
\end{tabularx}
\renewcommand{\arraystretch}{1}
\caption{{\it Values of the local infidelity $I_d$ for all steps of the simulation algorithm for both the free and interacting theories.} The infidelity is computed using the target states $|\psi_\text{targ}\rangle$ specified at the beginning of this section. For vacuum preparation, a range of 4 lattice sites is used, which is roughly 2-$3\times$ the correlation length of the state. The local infidelity is uniform throughout the lattice because of translational invariance. For wavepacket preparation, the reduced state of the wavepacket (occupying three spatial sites) is used. For time evolution, the reduced state on 10 spatial sites is used, which contains the region where the interaction takes place. The number of variational parameters used for each step is shown in the third column. The table where the corresponding parameters are specified is given in the rightmost column.}
\label{phi4:tab:error_budget}
\end{table}

It can be seen that the infidelities for $\lambda=0$ are consistently larger than those for $\lambda=2$. This is a result of the interacting theory having a smaller correlation length than the free theory. Because circuits with equal numbers of parameters and equal two-qubit gate depths are used to initialize and time-evolve states for both values of $\lambda$, this difference in quality is expected. This is an indication that the circuit depth needs to be increased for states with larger $\xi$ in order to spread correlations over a greater number of lattice sites.

It should be noted that $I_d$, as well as other measures on the wavefunction, is a conservative estimate of the quality of observables calculated from these simulations. In fact, states with a modest overlap with the target state are seen to reproduce observables in the target state quite well. Appendix~\ref{phi4:sec:exact_vs_variational} shows the similarity of observables computed using exact and variational methods.

\section{Results}
\label{phi4:sec:results}
\begin{figure*}[!ht]
\centering
\includegraphics[width=0.5\linewidth,scale=0.75]{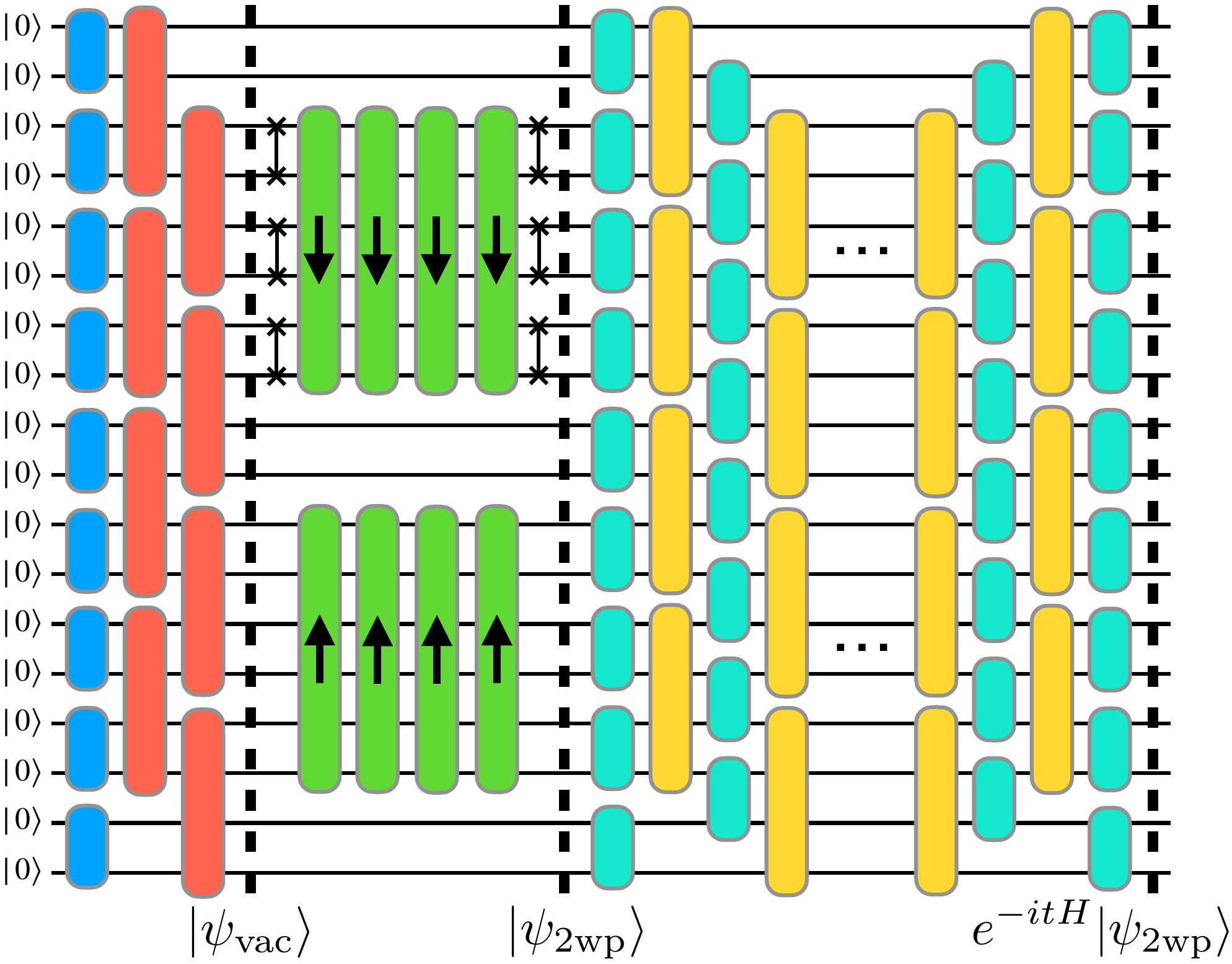}
\caption{{\it Physics-informed SVC used to simulate wavepacket scattering in scalar field theory.} The individual circuit elements are defined in Fig.~\ref{phi4:fig:circuit_elements}. First, the vacuum of the theory is created using SC-ADAPT-VQE. Wavepackets are initialized on top of this vacuum using a local brickwall ansatz. Then, a translationally invariant brickwall ansatz is used to implement the time evolution. }
\label{phi4:fig:full_circuit}
\end{figure*} 

\begin{table}
\centering
\begin{tabularx}{\linewidth}{|c||Y|Y|Y|Y|Y|Y|} \hline
$t$ & \# variational steps (per $t$) & \# two-qubit gates & Two-qubit gate depth & \# PTs/layouts (per circuit) & \# TREX twirls (per circuit) & \# shots (per circuit) \\\hline\hline
1-3 & 1 & 2284 & 59 & 80 & 2 & 8000\\\hline
4-6 & 2 & 3604 & 81 & 80 & 2 & 8000\\\hline
7-9 & 3 & 4924 & 103 & 80 & 2 & 8000\\\hline
\end{tabularx}
\renewcommand{\arraystretch}{1}
\caption{{\it Quantum simulation details.} The Quantum simulations are performed on 120 qubits of IBM's {\tt ibm\_fez} superconducting quantum computer. For a given simulation time $t$ (first column), the number of variational steps is given by $\lceil \frac{t}{3} \rceil$ (second column). The total number of two-qubit gates and corresponding two-qubit gate depths are given in columns three and four respectively, accounting for gate cancellations from different parts of the circuits and possible optimizations. Each circuit is executed using a number of Pauli-twirled circuits, with each one being mapped onto randomly chosen rotation of the selected qubit layout; this number is given in column five. Measurement mitigation is applied using two TREX twirls for each Pauli-twirled circuit (sixth column). The number of shots per circuit is given in the rightmost column.}
\label{phi4:tab:device_run_params}
\end{table}

Using the SVC framework, the circuit shown in Fig.~\ref{phi4:fig:full_circuit} is scaled up to $L=60$ (120 qubits) and executed on {\tt ibm\_fez} to simulate the scattering of two wavepackets in the $\phi^4$ theory. The theory is digitized onto $n_q=2$ qubits with a cutoff of $\phi_\text{max}=1.5$ to represent the state of the field at each spatial site with a bare mass of $m=1/2$. Error mitigation is vital for the successful extraction of observables from noisy quantum simulations~\cite{Kim:2023bwr}. An overview of the error mitigation methods used in this work is given here, and the implementation details and choices can be found in App.~\ref{phi4:sec:error_mitigation}. Once the bare circuits are created, several error mitigation layers are added. Dynamical decoupling (DD)~\cite{Viola:1998jx,Ezzell:2022uat} is used to mitigate idle errors and crosstalk between qubits. Pauli twirling (PT)~\cite{Wallman:2015uzh} is added to all two-qubit gates (CZ for {\tt ibm\_fez}) to convert coherent errors to incoherent stochastic noise, after which a depolarizing noise model is assumed for each observable of interest. Similar to PT for two-qubit gates, Twirled Readout Error eXtinction (TREX)~\cite{Berg:2020ibi} is used to mitigate measurement errors. Observables are then estimated using Operator Decoherence Renormalization (ODR)~\cite{Farrell:2023fgd,Farrell:2024fit,Urbanek:2021oej,ARahman:2022tkr}, which involves running a ``mitigation'' circuit for every Pauli- and TREX-twirled ``physics'' circuit. Circuits used for mitigation must be classically simulable, but must have similar error profiles to the physics circuits. In this work, circuits implementing the time evolution of the vacuum state $e^{-itH}|\psi_\text{vac}\rangle$ are used as mitigation circuits. Although the vacuum preparation and time evolution are implemented approximately and $e^{-itH}|\psi_\text{vac}\rangle \neq |\psi_\text{vac}\rangle$, the vacuum time evolution can be determined classically via an exponential extrapolation. See App.~\ref{phi4:sec:error_mitigation} for details on the classical calculation of this time evolution. Compared to methods that have been used in the past, it is found that the vacuum evolution circuits more accurately reflect the noise in the wavepacket circuits. This is especially noticeable for the case of brickwall circuits, where the gates are densely packed and noise spreads faster than in Trotter time evolution circuits. Using the mitigation circuits, the depolarization parameters are computed using the ratio of the measured outcome $\langle O_j\rangle_\text{meas}$ to the expected noiseless outcome $\langle O_j\rangle_\text{true}$ for each local observable $O_j$
\begin{align}
    p_j \ = \ \frac{\langle O_j\rangle_\text{meas}}{\langle O_j\rangle_\text{true}} \ = \ \frac{\langle \psi_\text{vac}|e^{itH} O_j e^{-itH}|\psi_\text{vac}\rangle_\text{meas}}{\langle \psi_\text{vac}|e^{itH} O_j e^{-itH}|\psi_\text{vac}\rangle_\text{true}}.
    \label{phi4:eq:odr_p_j}
\end{align}
The $p_j$ are then used to adjust for the noise in the physics circuits using the same equation in reverse. They are also used to filter out outlying measurements caused by local device imperfections. In addition to these methods, layout randomization is used to drive the noise closer to depolarizing noise. 

\begin{figure*}
\centering
    \begin{subfigure}{\linewidth}
        \includegraphics[width=\linewidth]{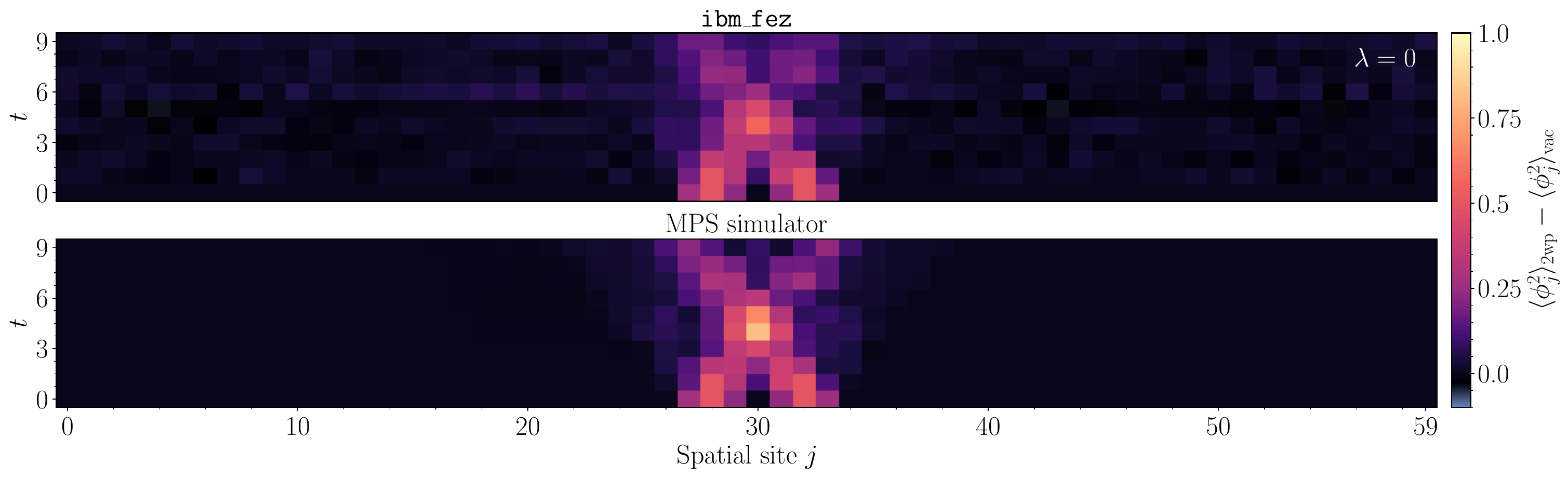}
        \caption{}
        \label{phi4:fig:scattering_heatmap_l_0}
    \end{subfigure}
    \begin{subfigure}{\linewidth}
        \includegraphics[width=\linewidth]{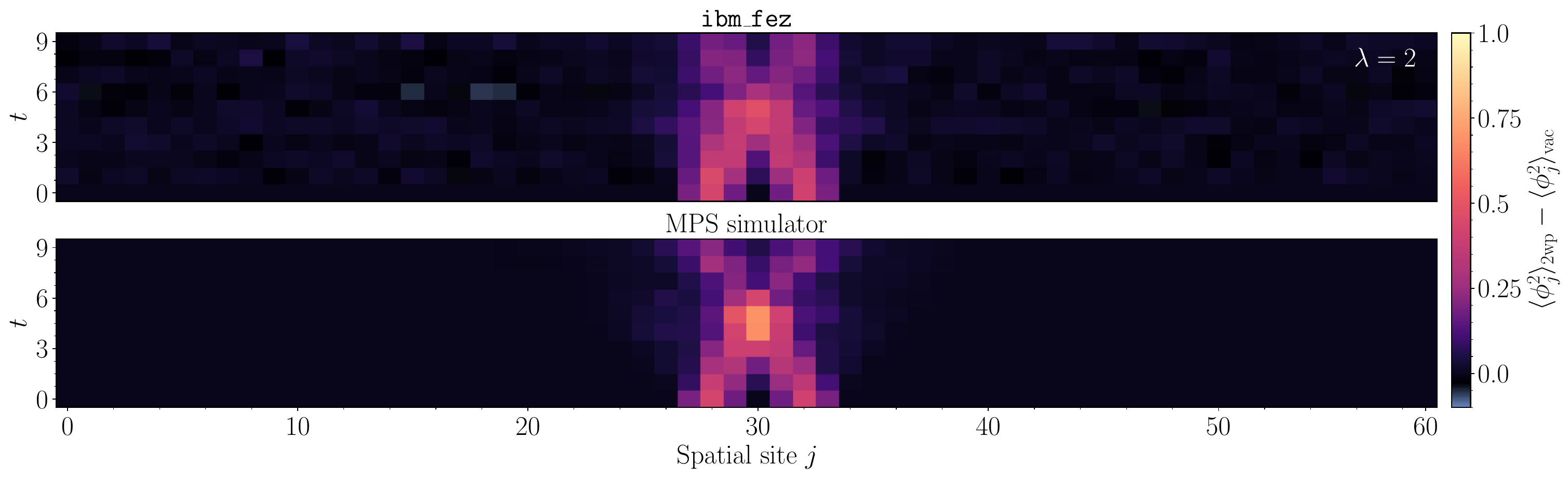}
        \caption{}
        \label{phi4:fig:scattering_heatmap_l_2}
    \end{subfigure}
\caption{{\it Simulations of scattering in scalar field theory.} The time evolution of the vacuum-subtracted $\langle\phi^2_j\rangle$ is shown for a $L=60$ (120 qubits) lattice for (a) the free $(\lambda=0)$ and (b) interacting $(\lambda=2)$ theories. Two wavepackets are initialized with opposite momenta $(k=\pm\pi/3)$ at $t=0$ and centered at sites $j=28,\,32$. The particles travel toward each other, collide, and propagate as time goes on. Error-mitigated results from IBM's quantum computer {\tt ibm\_fez} (top panels) are compared to noiseless MPS circuit simulations (bottom panels). The vacuum of the theory is built using variational circuits determined by SC-ADAPT-VQE, while the wavepacket preparation and time evolution is implemented by variational brickwall circuits. In the device data, the vacuum evolution is evaluated via a classically determined exponential extrapolation. A broken-down view for each time slice is shown in Fig.~\ref{phi4:fig:results_by_time}, and a zoomed-in comparison of the scattering region between the free and interacting theories is given in Fig.~\ref{phi4:fig:scattering_heatmap_small}. Details on the error mitigation strategies are given in the main text and in App.~\ref{phi4:sec:error_mitigation}.}
\label{phi4:fig:scattering_heatmap}
\end{figure*} 

\begin{figure*}
\centering
\includegraphics[width=\linewidth]{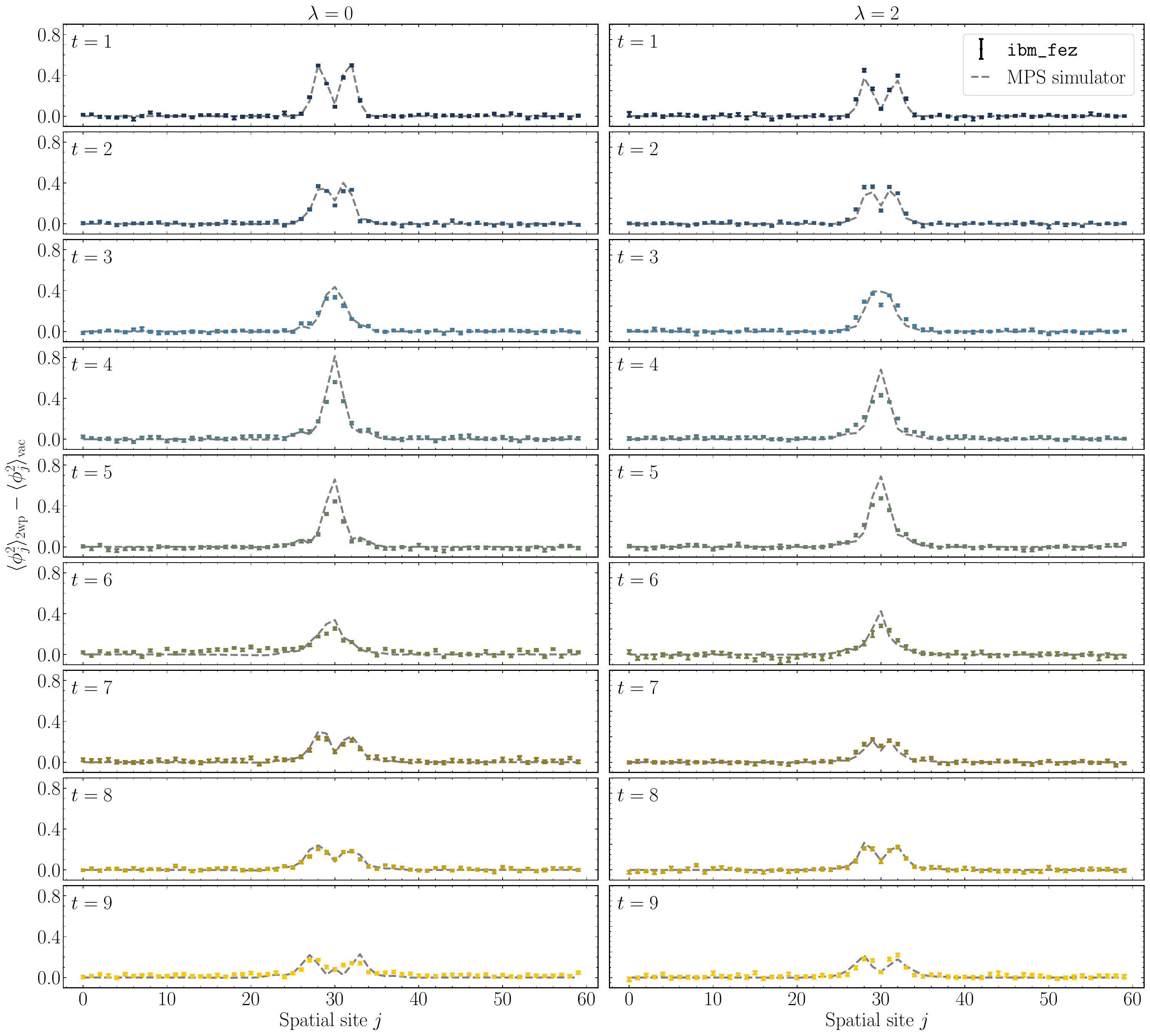}
\caption{{\it Simulations of scattering in scalar field theory by $t$.} The vacuum-subtracted $\langle\phi^2_j\rangle$ is shown for the scattering of two wavepackets, as in Fig.~\ref{phi4:fig:scattering_heatmap}, by time step.  Noiseless MPS circuit simulations (grey dashed lines) are compared to data from IBM's quantum computer {\tt ibm\_fez} after error mitigation (points) for $\lambda=0$ (left column) and $\lambda=2$ (right column). In the device data, the vacuum evolution is evaluated via a classically determined exponential extrapolation. The error bars represent one standard deviation of the bootstrap-resampled data.}
\label{phi4:fig:results_by_time}
\end{figure*} 

The two-qubit gate depths and counts, as well as other parameters of the runs on {\tt ibm\_fez} for times $t=1-9$, are shown in Table~\ref{phi4:tab:device_run_params}. Circuits with a maximum two-qubit gate depth of 103 using 4924 two-qubit gates (for $t=7,8,9$) are used. For time $t$, $\lceil \frac{t}{3} \rceil$ variational steps are used (the variational step is defined in Sec.~\ref{phi4:sec:circuits_time_evolution} and shown in Fig.~\ref{phi4:fig:time_evolution_circ_step}). For each time $t$, four circuits are executed:  $e^{-itH}|\psi_\text{2wp}\rangle$ and $e^{-itH}|\psi_\text{vac}\rangle$ for $\lambda=0$ and $\lambda=2$. 80 Pauli-twirled circuits are run for each physics and mitigation circuit. For each Pauli-twirled circuit, two TREX twirled circuits are run for measurement mitigation. Each circuit executed on the device is evaluated using 8000 shots.

In these simulations, fluctuations of $\langle \phi^2_j \rangle$ above the vacuum are measured on the quantum device.\footnote{Note that it is possible to efficiently measure $\langle H_j\rangle$ using the same setup. Besides $\Pi^2_j$, all terms in the Hamiltonian of Eq.~\eqref{phi4:eq:lattice_h} are in the $\phi$ basis, so their expectation values can be computed using measurements in the $\phi$ basis. To measure in the $\Pi$ basis, a second set of circuits can be run with the $k_\phi$ local Fourier transform appended. The results can then be added together to compute $\langle H_j\rangle$. From this, it is straightforward to estimate quantities that are accessible to analytic computations, such as cross sections computed using perturbative quantum field theory.} The results from the device and from MPS circuit simulations for $\langle \phi^2_j\rangle_\text{2wp}-\langle \phi^2_j\rangle_\text{vac}$ for $\lambda=0$ and $\lambda=2$ are shown in Figs.~\ref{phi4:fig:scattering_heatmap_l_0} and~\ref{phi4:fig:scattering_heatmap_l_2} respectively. A breakdown of the results by time step is shown in Fig.~\ref{phi4:fig:results_by_time}, and numerical values for these runs are given in Tables~\ref{phi4:tab:results_t_1}-~\ref{phi4:tab:results_t_9}. The uncertainties in the results from the quantum computer are estimated using bootstrap resampling. The expected results are determined using the {\tt qiskit}~\cite{Javadi-Abhari:2024kbf} MPS circuit simulator. This classical method works by converting the circuit to a tensor network acting on a MPS and applying the circuit to the MPS gate by gate while enforcing the cutoff and maximum bond dimension constraints at each step. See Ref.~\cite{qiskit_mps} for more information. In this work, a maximum bond dimension of 100 is used. For the circuits considered, it is seen that the results of these simulations are converged to $10^{-2}-10^{-3}$. 

The scattering process is seen by examining $\langle \phi^2_j\rangle_\text{2wp}-\langle \phi^2_j\rangle_\text{vac}$ in Fig.~\ref{phi4:fig:scattering_heatmap}. The propagation of particles is clearly identified as disturbances in the vacuum-subtracted $\langle\phi^2_j\rangle$ in the interaction region in the center (spatial sites 24-36). These are indicated as bright spots on the plots. The wavepackets, centered to the left and right of the center of the lattice at $t=0$, are initialized with opposite momenta. They travel toward each other and collide in the center. After the collision takes place, a light cone is seen to develop, indicating that the particles have propagated past each other and are traveling away from the point of the collision.

\begin{figure*}[t]
\centering
\includegraphics[width=\linewidth]{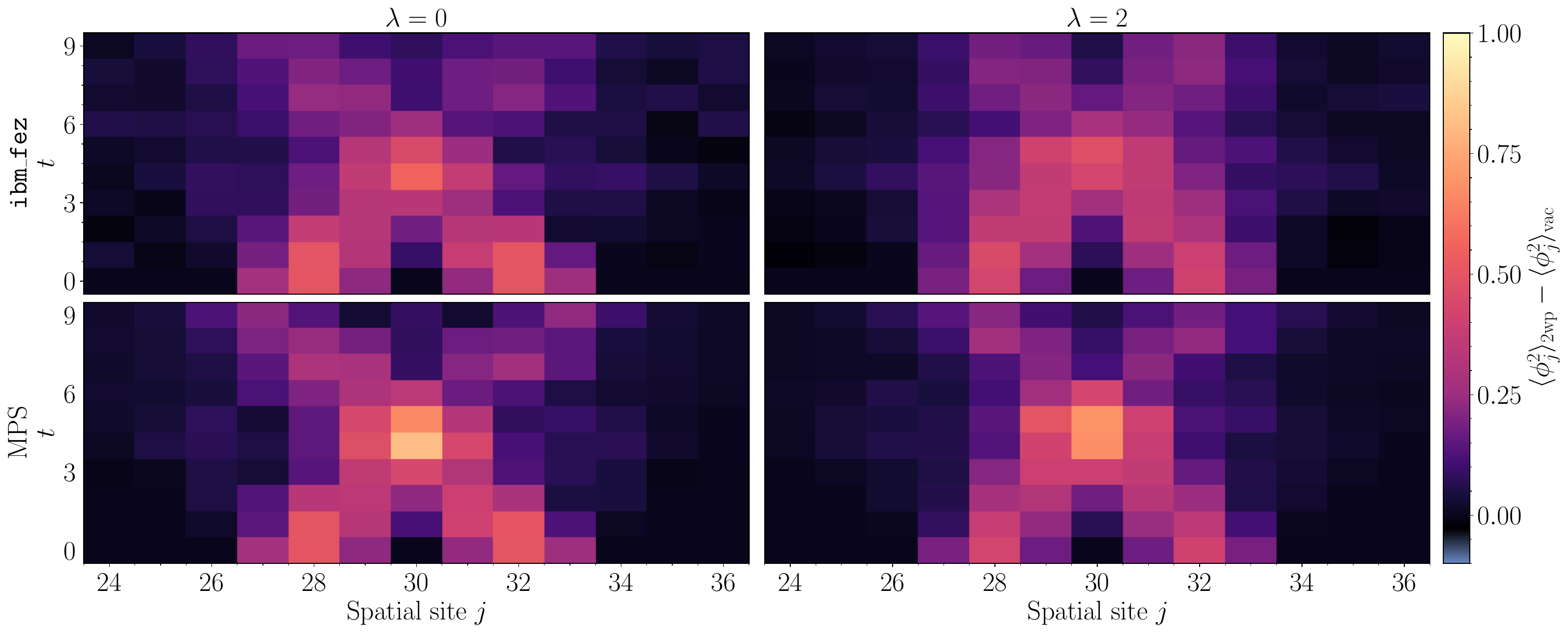}
\caption{{\it The interaction region in simulations of scattering.} A zoomed-in view of the interaction region of Fig.~\ref{phi4:fig:scattering_heatmap} is shown highlighting the effect of interactions. The vacuum-subtracted $\langle \phi_j^2\rangle$ in the free theory (left panels) is compared to the interacting theory (right panels). The top panels show error-mitigated results from IBM's quantum computer {\tt ibm\_fez}, while the bottom panels show noiseless MPS circuit simulations.}
\label{phi4:fig:scattering_heatmap_small}
\end{figure*}

The effect of interactions is clearly seen when comparing results in the interaction region between the free and interacting theories as in Fig.~\ref{phi4:fig:scattering_heatmap_small}. In the free theory, the collision is seen to peak around $t=4$, after which the particles travel away from each other. When interactions are included in the theory, the peak of the collision is observed later (around $t=5$), and a time delay due to the interaction is visible. The difference between $\lambda=0$ and $\lambda=2$ can also be seen at late times $(t=9)$, where the free particles can be seen having traveled one more spatial site than the interacting particles. A time delay is expected because the $\phi^4$ potential is repulsive, particles with overlapping wavefunctions incur an energy penalty. In this sense, the time delay can be viewed as the interacting particles encountering a potential step and thus losing kinetic energy when their wavepackets overlap. As a result, the particles spend a longer time in the central region where they are able to interact. It is expected that this effect will be even more prominent at greater interaction strengths. In the case of the digitized simulations done in this work with $n_q=2$, the interactions and time delay are caused by digitization effects. See App.~\ref{phi4:sec:digitization_effects} for a detailed examination of these interactions.

The results in Figs.~\ref{phi4:fig:scattering_heatmap}, ~\ref{phi4:fig:results_by_time}, and~\ref{phi4:fig:scattering_heatmap_small} show that by using a combination of variational circuit methods (SC-ADAPT-VQE and brickwall), together with error mitigation techniques, quantum computers are able to simulate scattering processes of interacting particles. The scalability of each step of this simulation algorithm enables the simulation of collision events on arbitrarily large system sizes in a systematically improvable manner. It should be noted that the observation of $\langle \phi^2_j\rangle_\text{2wp}-\langle \phi^2_j\rangle_\text{vac}=0$ in the vacuum regions is a nontrivial result, as it requires the wavepacket simulations from the quantum computer to agree closely with extrapolations from classical computations. In Figs.~\ref{phi4:fig:scattering_heatmap} and~\ref{phi4:fig:results_by_time}, the data from the quantum computer is seen to qualitatively agree with the MPS circuit simulations. Increasing levels of noise (as well as uncertainties) can be seen in the results for later times, which is expected as a result of deeper circuits. There exist points where the error-mitigated results still disagree from the MPS data by several standard deviations (see Fig.~\ref{phi4:fig:results_by_time}). This is particularly noticeable in the interaction region during the collision of the particles, but occurs in vacuum regions as well. This highlights a weakness of the chosen error mitigation method: because the state chosen for error mitigation is $e^{-itH}|\psi_\text{vac}\rangle$, its evolution does not capture potential changes to wavepackets. Since the noise is state-dependent, these mitigation circuits cannot mimic the noise in the circuits simulating scattering perfectly. This reflects the simplicity of the assumed error model, and is an area where much improvement is expected in the future. Although errors persist even after the application of error mitigation, it is clear that quantum computers today are capable of probing scattering events, potentially shedding light on inelastic collisions in more complicated theories in the future.

\section{Summary and outlook}
\label{phi4:sec:discussion}
Quantum simulations hold the potential of reliably uncovering the dynamics of complex inelastic collision events, such as those that take place in extremely dense and hot environments. As a first step toward such computations, this work simulates elastic scattering in one-dimensional lattice scalar field theory. Scalable variational quantum algorithms are developed for both state preparation and time evolution, paving the path for future simulations at larger scales. The simulation begins by using a translationally invariant SC-ADAPT-VQE ansatz to prepare the vacuum of the theory. Wavepackets representing particles are then created on top of the vacuum using a brickwall circuit, and a translationally invariant brickwall ansatz is used to implement time evolution of the system. The scalability of all circuits in these simulations is a result of the physical properties of states in the system, namely their translational invariance, locality, and the presence of a mass gap. The circuits are determined classically on modest system sizes, and extrapolated to arbitrarily large system sizes to be run on quantum devices. These methods are used to simulate the collision of two wavepackets using 120 qubits of IBM's 156 qubit superconducting quantum computer {\tt ibm\_fez}. The use of error mitigation and suppression techniques is crucial for the extraction of results that qualitatively agree with classical MPS simulations. Signatures of scattering and effects of the interaction strength are clearly identified in the vacuum-subtracted expectation value of the field determined from data from the quantum computer.

Overall, the utility of variational quantum algorithms in systematically compressing circuits is clear; the specialization of these methods to quantum simulations enables the use of physics to provide further performance improvements. These newly developed scalable variational methods, together with existing techniques and new modifications to error mitigation strategies, enable the first simulation of wavepacket scattering in an interacting quantum field theory on a quantum computer. The resulting gate counts and resource requirements are many orders of magnitude below upper bounds for fault-tolerant algorithms. Furthermore, these results are the first demonstration of the extraction of qualitatively correct observables from quantum devices using variational brickwall circuits at scale. This work establishes variational algorithms as tools that are useful for the compression of all parts of a quantum simulation. In particular, it is demonstrated that the problem of barren plateaus that plagues many variational quantum computations is reduced through knowledge of the system being simulated (e.g., its symmetries and mass gap). Efficient variational circuits are determined by using physics to remove redundancies in the circuit parameterization. The development of variational circuit compression methods that are both scalable and amenable to error mitigation is useful for minimizing noise in the quantum device through reduced two-qubit gate depth. These methods and results constitute an important stepping stone toward the quantum simulation of inelastic scattering processes at high energies, where known classical methods are expected to fail. 

Improvements to the optimization step of the variational algorithm, as well as more efficient ansatze, are expected in the near future. This work uses the most simple iterative gradient-based optimization method. Recent work~\cite{Stokes:2019pmg,Harrow:2021tta} has shown that it is possible to incorporate more sophisticated optimization protocols to improve the quality of the approximate quantum state. In parallel, ansatze for state preparation that take advantage of mid-circuit measurements~\cite{Chen:2023tfg,Baumer:2023vrf,Malz:2023xve,Baumer:2024jng,Piroli:2024ckr} can be immediately used in simulations today; the combination of these methods with variational quantum algorithms is a promising direction. These modifications would allow for greater precision in variational calculations, while maintaining the same circuit depth, or even reducing it.

Highly inelastic scattering events are believed to generate a large amount of entanglement, and as a result are prime candidates for demonstrations of quantum utility. Toward the simulation of inelastic collisions, several improvements to the current algorithm will be necessary. Initial states with energy higher than the two-particle threshold will need to be created, requiring more tightly peaked wavepackets in $k$-space. In turn, wavepackets that occupy a larger number of qubits will be required. The methods presented will need to be extended to the scalable creation of wavepackets of increasing sizes and their time evolution. The presence of a large number of lattice sites occupied by the vacuum will continue to provide the resolution in lattice momentum that will be necessary to observe inelastic effects. More qubits per site (i.e., increasing $n_q$) will be needed to achieve higher precision in these calculations, as well as to simulate genuine interactions. Similarly, larger system sizes and correlation lengths will be needed to approach continuum physics. All of these changes will require circuits with greater two-qubit gate depths, and improvements to the variational methods will be necessary. As the capabilities of state of the art quantum computers to realize large-scale system sizes improve, more sophisticated techniques for particle detection will be required. Further in the future, the extension of these methods to include more complex interactions, such as those described by non-Abelian gauge theories, will be required for the simulation of realistic particle collisions. 

\clearpage

\begin{subappendices}

\section{Digitization effects}
\label{phi4:sec:digitization_effects}
The Nyquist-Shannon Sampling Theorem~\cite{shannon_collected_1993,10.5555/3179430.3179434,Macridin:2018oli,Macridin:2018gdw} guarantees that with a proper choice of $\phi_\text{max}$, the digitized field theory represents the continuous lattice field theory with errors that are exponentially suppressed with an increasing number of wavefunction sample points. In particular, the eigenfunctions of the continuous theory can be reconstructed from the digitized eigenfunctions up to errors that scale as $\epsilon \sim 2^{-2^{n_q}}$. The single-site wavefunction $\psi(\phi)$ can be approximately reconstructed from $n$ samples at points $\phi_n$ using the Whittaker–Shannon interpolation formula~\cite{shannon_collected_1993}: 
\begin{align}
    \psi(\phi) &= \sum_{n=0}^{2^{n_q}-1} \psi(\phi_n)\, \text{sinc} \left(\frac{\phi+\phi_\text{max}-n\delta_\phi \phi_\text{max}}{\delta_\phi \phi_\text{max}}\right).\label{phi4:eq:shannon_interpolation}
\end{align}
Its multidimensional extensions can be used to reconstruct multi-site wavefunctions. 

A poorly chosen $\phi_\text{max}$ can introduce unwanted interactions due to the digitization and break the double-exponential convergence guaranteed by the Nyquist-Shannon Sampling Theorem. The optimal choice of $\phi_\text{max}$ for a given theory depends on $n_q$, as well as on the parameters $m$ and $\lambda$, since these control the strength of the potential. As a result, states with higher $m$ and $\lambda$ are more localized in $\phi$-space and less localized in $\Pi$-space. Previous approaches determined the optimal $\phi_\text{max}$ by minimizing the error in the commutator compared to the continuum value $||[\phi_i,\Pi_j]-i\delta_{ij}||$~\cite{10.5555/3179430.3179434,Macridin:2018oli,Macridin:2018gdw,Bauer:2021gek,Kane:2022ejm}, or minimizing the error in the eigenenergies by numerically solving the continuum theory for a small number of lattice sites~\cite{Klco:2018zqz}. However, single-site considerations do not capture additions to the potential from neighboring sites, and numerical solutions become intractable for large systems due to the exponential resources required in the discretization of the system. In this work, $\phi_\text{max}$ is approximately chosen by including the single-site contributions from $H_\text{kin}$, so that the only term that is omitted from the Hamiltonian of Eq.~\eqref{phi4:eq:lattice_h} is $-\sum_{j=0}^{L-1}\phi_{j+1}\phi_j$. Similar to the previous methods, this is an approximation to the optimal $\phi_\text{max}$ for large system sizes, but it is closer to the true value.\footnote{In principle, the optimal $\phi_\text{max}$ can be determined following an approach similar to that of Sec.~\ref{phi4:sec:scalable_variational_circuits}, by finding $\phi_\text{max}$ as a function of the system size $L$ for small $L$, and extrapolating to $L$ of choice.} Following these considerations, the optimal $\phi_\text{max}$ is determined by using Eq.~\eqref{phi4:eq:shannon_interpolation} to maximize the overlap between the low-energy eigenstates in the digitized theory and the continuum theory solved numerically. With this method, the best choice for $\phi_\text{max}$ is 1.5 for $n_q=2$, $m=1/2$, $\lambda=0$, and 1.43 for $n_q=2$, $m=1/2$, $\lambda=2$. However, for $\lambda=2$ the difference between $\phi_\text{max}=1.5$ and the ideal value is minimal in terms of the effect on the representation of the continuum, and as a result $\phi_\text{max}=1.5$ is used for both the free and interacting theories throughout this work. In this case, the errors due to the suboptimal choice of $\phi_\text{max}$ are small compared to digitization errors stemming from a small $n_q$. 

\begin{figure*}
\centering
\begin{subfigure}{\linewidth}
    \includegraphics[width=\linewidth]{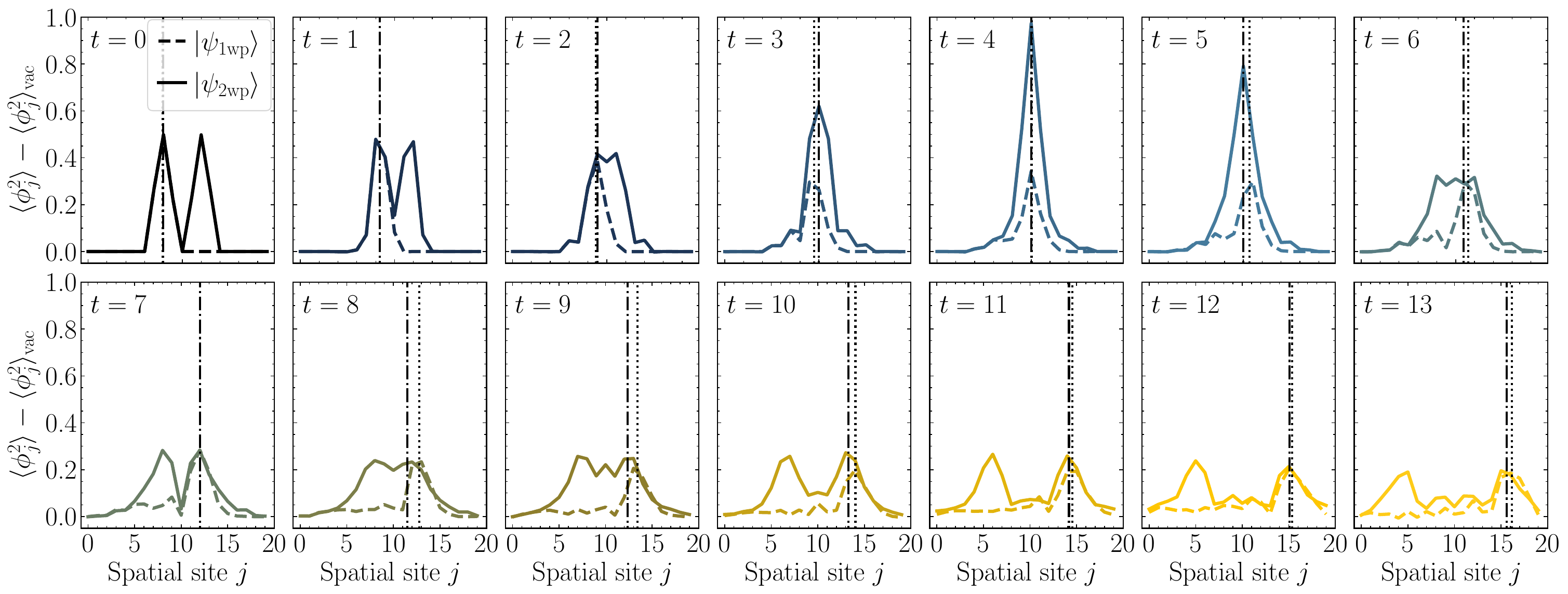}
    \caption{}
    \label{phi4:fig:single_vs_two_wp_free}
\end{subfigure}
\begin{subfigure}{\linewidth}
    \includegraphics[width=\linewidth]{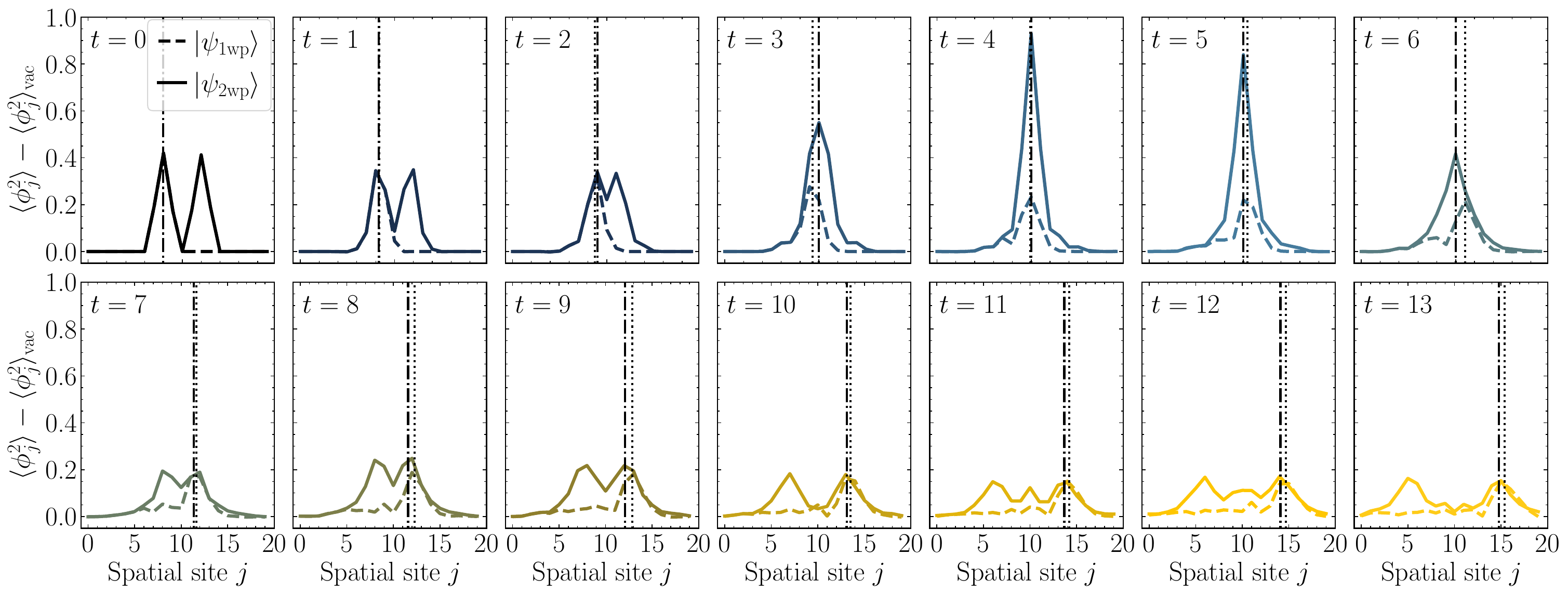}
    \caption{}
    \label{phi4:fig:single_vs_two_wp_int}
\end{subfigure}
\caption{{\it Comparison of the time evolution of the single- and two-wavepacket states} Time evolution of  $|\psi_\text{1wp}\rangle$ and $|\psi_\text{2wp}\rangle$, for (a) the free $(\lambda=0)$ and (b) the interacting $(\lambda=2)$ theories is shown. The center of the left wavepacket is tracked for the single-wavepacket state with the dotted vertical line, and for the two-wavepacket state with the  dash-dotted vertical line. The centers are determined by fitting a Gaussian around the wavepacket region, and by taking the maximum value when the wavepackets are interacting. A system of $L=20$ spatial sites with $n_q=2,m=1/2$ is used. The simulations are done using a MPS circuit simulator with size $1/10$ Trotter steps and a maximum bond dimension of 100.}
\label{phi4:fig:single_vs_two_wp}
\end{figure*}

Two qubits per lattice site is the coarsest nontrivial digitization. As a result, the digitization effects are significant and must be considered. With $n_q=2$ and $\delta_\phi$ chosen as in Eq.~\eqref{phi4:eq:phi_digitized}, the $\phi^2_j$ and $\phi^4_j$ operators may be written as 
\begin{align}
    \phi^2_j = \frac{\phi_\text{max}^2}{9}\left(5 + 4Z_{2j}Z_{2j+1}\right),\\
    \phi^4_j = \frac{\phi_\text{max}^4}{81}\left(41 + 40Z_{2j}Z_{2j+1}\right).
\end{align}
From these expressions, it is seen that the non-identity portions of $H_\phi$ and $H_\text{int}$ in the Hamiltonian of Eq.~\eqref{phi4:eq:lattice_h} are proportional to each other. This implies that simulating a theory with $\lambda \neq 0$ is equivalent to simulating one with a larger mass, 
\begin{align}
    m' = \sqrt{m^2 + \frac{\lambda}{4!}\frac{20}{9}\phi_\text{max}^2}.
\end{align}
As a result of this, the interactions present in the simulations in this work are due to digitization effects. In principle, these interactions may be mapped to an expansion of the form $\sum_n c_{2n}\phi^{2n}$, but determining the coefficients $c_{2n}$ is difficult in practice. These interactions are examined in Fig.~\ref{phi4:fig:single_vs_two_wp}, where the time evolution of a single wavepacket is compared to that of a two-wavepacket state for both values of the coupling. By considering the single-wavepacket evolution, which is truly noninteracting, it can be seen that there is a time delay due to the interaction of two wavepackets for both couplings, with the time delay for the $\lambda=2$ theory being larger.

\begin{figure*}
\centering
    \includegraphics[width=\linewidth]{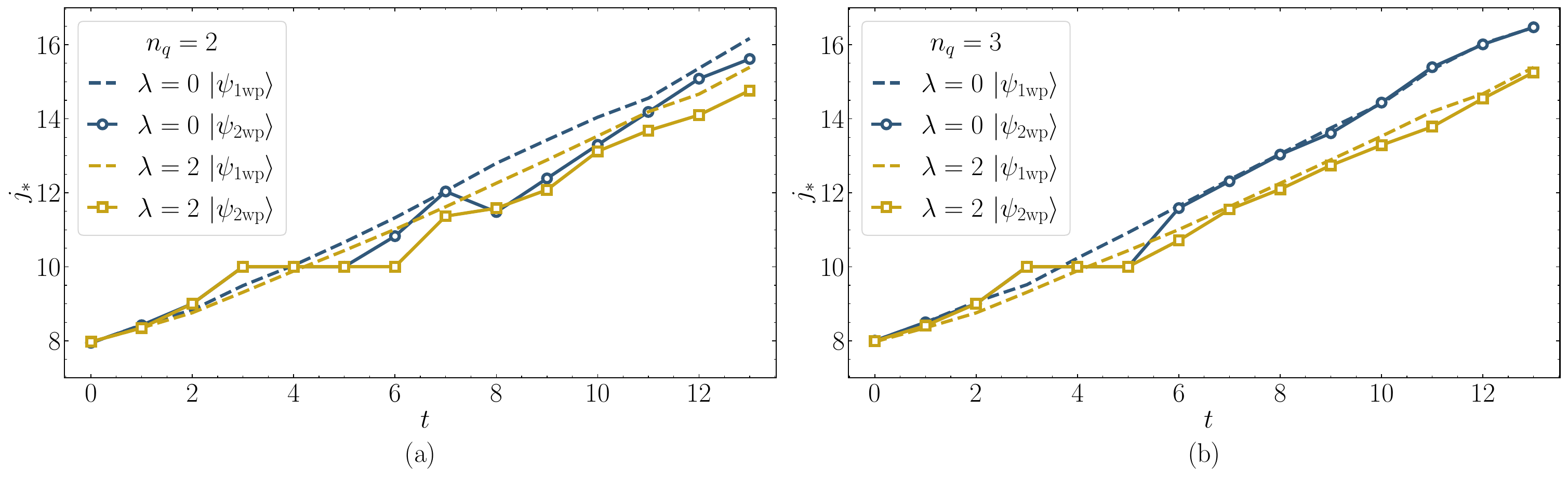}
    \caption{{\it The effect of digitization on wavepacket propagation.} The center $j_*$ of the left wavepacket propagating in time with (a) $n_q=2$ and (b) $n_q=3$ digitization is shown. The free and interacting theories are shown in blue and yellow, respectively. The single-wavepacket states $|\psi_\text{1wp}\rangle$ are marked with dashed lines, while the two-wavepacket states $|\psi_\text{2wp}\rangle$ are marked with solid lines. A system of $L=20$ spatial sites with $m=1/2$ is used. For $n_q=2$, $\phi_\text{max}=1.5$ is used for both $\lambda=0$ and $\lambda=2$ as in the simulations run on quantum computers in this work, while for $n_q=3$, the optimal $\phi_\text{max}$ values of 2.52 and 2.43 are used for the free and interacting theories, respectively. The simulations are done using a MPS circuit simulator with size $t=1/10$ Trotter steps and maximum bond dimension of 100.}
    \label{phi4:fig:center_tracking}
\end{figure*}

The centers $j_*$ of the wavepackets for all four cases are plotted in Fig.~\ref{phi4:fig:center_tracking}a. The dashed lines show the single-wavepacket time evolution. The slight difference in group velocity $v_k$, as expected from Fig.~\ref{phi4:fig:e_k_v_k_wp_prep_convergence}, is seen in the slopes of the $|\psi_\text{1wp}\rangle$ lines. By comparing the single-wavepacket and the two-wavepacket lines, the relative time delay between the $\lambda=0$ and $\lambda=2$ theories as a result of differences in $v_k$ is eliminated. The collision takes place between times $t=3-6$ indicated by the plateau, and the time delay is seen in the $\lambda=2$ plateau extending later in time than the $\lambda=0$ one. After an initial settling time, the ``asymptotic'' time delay is also seen in the difference in the center points at a given time. The difference of $j_*$ between the $|\psi_\text{1wp}\rangle$ and $|\psi_\text{2wp}\rangle$ is larger for $\lambda=2$, indicating that the time delay due to interactions is greater in this theory.

To complete the discussion of digitization effects and interactions caused by them, simulations of the same process with $n_q=3$ are shown in Fig.~\ref{phi4:fig:center_tracking}b. These simulations are expected to more accurately capture the continuum physics because of the added precision from doubling the number of points where the continuum wavefunctions are sampled. In addition, the non-identity parts of $H_\phi$ and $H_\text{int}$ are not proportional with $n_q=3$, making the $\phi^4$ operator a genuine interaction term not present in the free Hamiltonian. Using the methods described above, the optimal $\phi_\text{max}$ is determined to be 2.52 for $\lambda=0$, and 2.43 for $\lambda=2$. All other parameters of the simulations are the same as for $n_q=2$. In the $\lambda=0$ simulations for $n_q=3$ there is no time delay, and a very small time delay for $\lambda=2$, which shows the strength of the digitization effects for $n_q=2$.

\section{Details on wavepacket creation}
\label{phi4:sec:wp_creation_details}

Section~\ref{phi4:sec:circuits_wp_prep} describes wavepacket preparation by learning an approximate shallow-depth variational circuit. This circuit is optimized to maximize the overlap with $|\psi_\text{targ}\rangle$ being the wavepacket produced by the non-digitized analytical wavefunction (Eq.~\eqref{phi4:eq:wp_wavefunction}) for $\lambda=0$, and after adiabatically turning on the interactions (Eq.~\eqref{phi4:eq:u_adiabatic}) for $\lambda=2$. The wavepackets represented by $|\psi_\text{targ}\rangle$ above are compared to those determined from ED by block-diagonalizing the Hamiltonian into momentum sectors in Table~\ref{phi4:tab:adiabatic_vs_ed}. The local infidelity over the three sites spanning the wavepacket, $I_3$, is seen to decrease with increasing $n_q$ indicating that the digitized wavefunction approximates the continuous-field case with increasing quality.\footnote{The wavepackets produced from ED can also be used directly as $|\psi_\text{targ}\rangle$, eliminating the need for the continuous-field approximation (Eq.~\eqref{phi4:eq:wp_wavefunction}) and adiabatic turn-on (Eq.~\eqref{phi4:eq:u_adiabatic}). This approach would encounter the same classical limitations in terms of scalability and is conceptually simpler. However, the direct ED method was not used in determining the parameters for circuits run on quantum hardware in this work.}

\begin{table}
\centering
\begin{tabularx}{\linewidth}{|c||Y|Y|} \hline
 & $I_3\,\, (n_q=2)$ & $I_3\,\, (n_q=3)$ \\\hline\hline
$\lambda=0$ & 0.03922 & 0.01471 \\\hline
$\lambda=2$ & 0.03973 & 0.01537 \\\hline
\end{tabularx}
\renewcommand{\arraystretch}{1}
\caption{{\it Infidelity of prepared wavepackets.} Wavepackets produced from the continuous-field approximation (Eq.~\eqref{phi4:eq:wp_wavefunction}) with adiabatic turn-on using Eq.~\eqref{phi4:eq:u_adiabatic} for $\lambda=2$ are compared to those determined from ED, as a function of the number of qubits used in the digitization $n_q$. The free and interacting theories are shown in the center and bottom rows, respectively. The wavepackets are created on an $L=6$ system for both methods. The local infidelity over the three sites spanning the wavepacket, $I_3$ is used. For $n_q=\{2,3\}$, the values of $\phi_\text{max}=\{1.5,3.1\}$ are used as determined in Ref.~\cite{Klco:2018zqz}.}
\label{phi4:tab:adiabatic_vs_ed}
\end{table}

The adiabatic turn-on is implemented as follows. The time evolution is implemented using Eq.~\eqref{phi4:eq:u_adiabatic} for a total time of $t_\text{ad}=100$, broken into $N_\text{ad}=10=O(\sqrt{\lambda t_\text{ad}})$ steps (as determined in Refs.~\cite{Jordan:2011ci,Jordan:2012xnu}). Each adiabatic step in Eq.~\eqref{phi4:eq:u_adiabatic} is Trotterized into 1000 ``forward'' and 1000 ``backward'' steps. The forward steps linearly increase the value of $\lambda$ once every 10 steps, while the backward steps evolve with fixed $\lambda$. As explained in Sec.~\ref{phi4:sec:lattice_scalar_field_theory}, the backward evolution steps serve to undo the unwanted time evolution to individual eigenstates composing the wavepacket.

The profile of $|\psi_\text{targ}\rangle$, and thus of the produced wavepackets is specified to be a Gaussian in $k$-space, as in Eqs.~\eqref{phi4:eq:single_particle_superposition} and~\eqref{phi4:eq:wp_wavefunction}. This is slightly different from the approach JLP take in Refs.~\cite{Jordan:2011ci,Jordan:2012xnu}, where the form of the state is specified in position space. By specifying a Gaussian in $k$-space, the method used in this work also specifies a Gaussian in position space, where its size is then truncated to optimize device resources. The two approaches are equivalent, since JLP also truncate the spatial extent of the wavepackets.

\section{Comparison of exact and variational time evolution}
\label{phi4:sec:exact_vs_variational}
\begin{figure*}
\centering
\begin{subfigure}{\linewidth}
    \includegraphics[width=\linewidth]{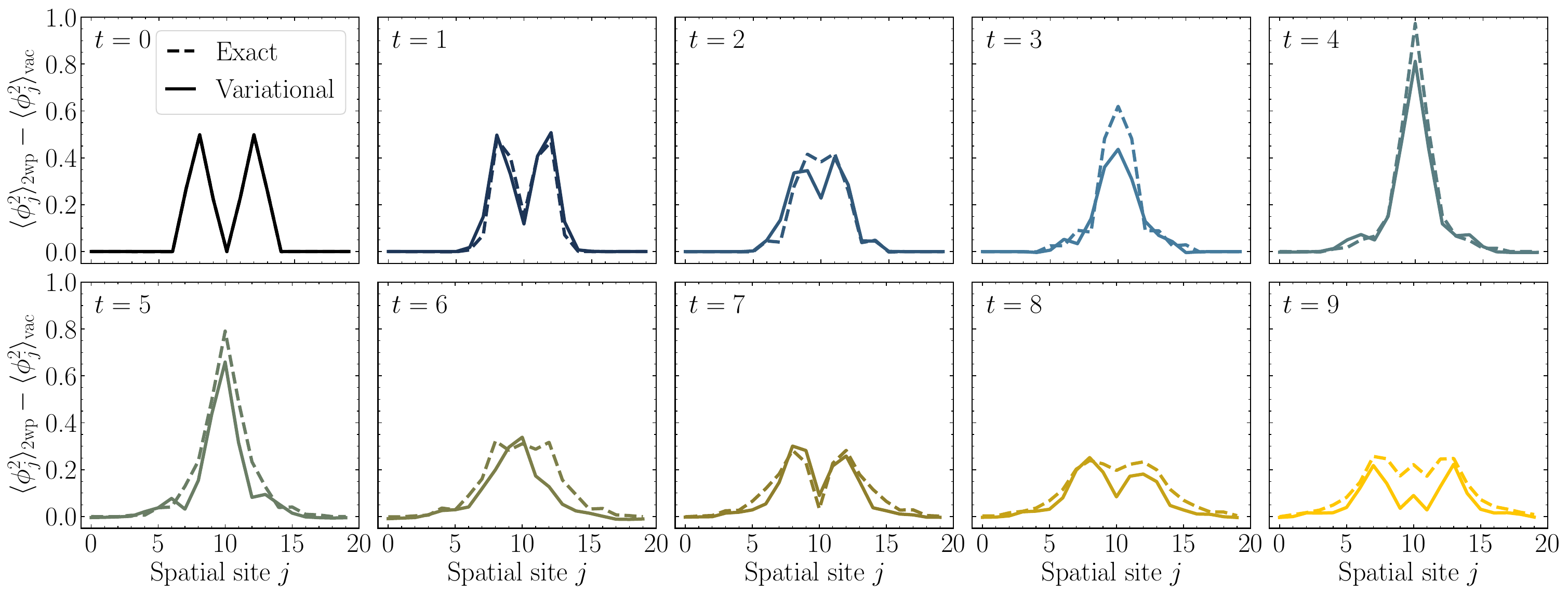}
    \caption{}
    \label{phi4:fig:exact_vs_variational_l_0}
\end{subfigure}
\begin{subfigure}{\linewidth}
    \includegraphics[width=\linewidth]{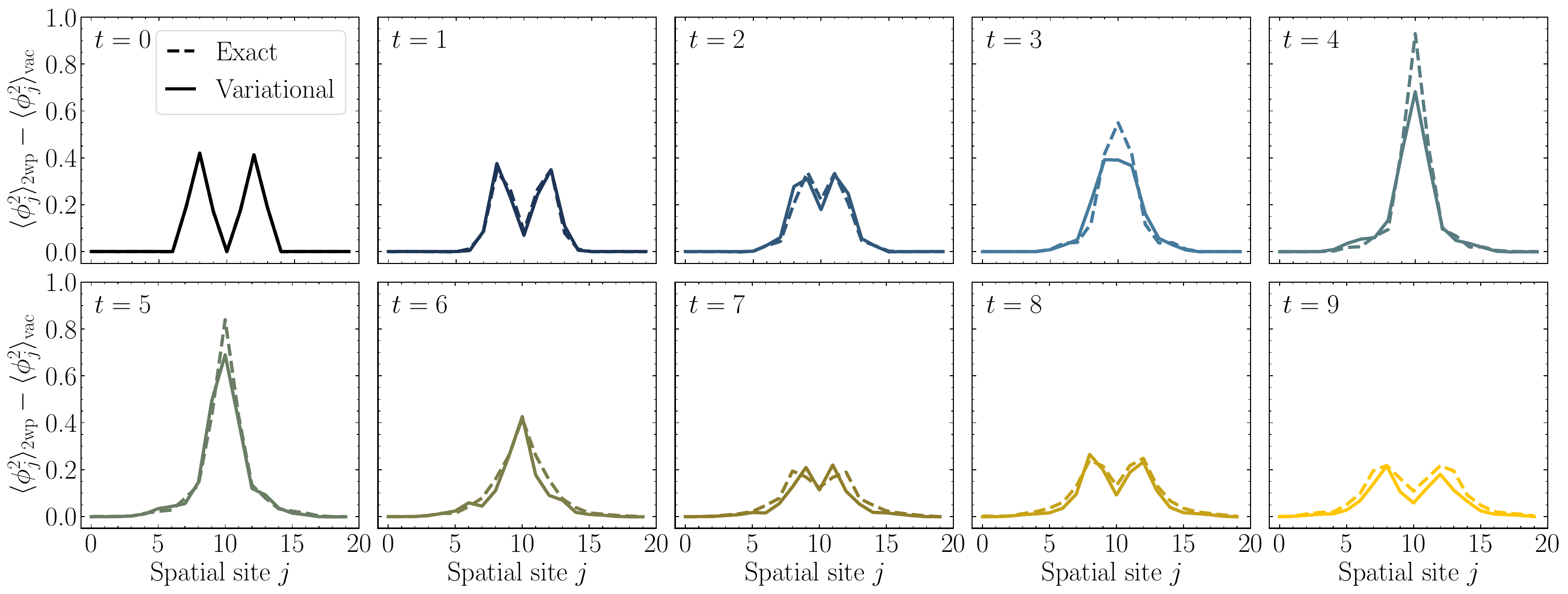}
    \caption{}
    \label{phi4:fig:exact_vs_variational_l_2}
\end{subfigure}
\caption{{\it Trotter vs. variational time evolution.} The vacuum-subtracted $\langle \phi^2_j \rangle$ for different times $t=0-9$ computed using a MPS circuit simulator for Trotterized time evolution (dashed lines) is compared to its approximation using SVC (solid lines) for (a) the free $(\lambda=0)$ theory and (b) the interacting $(\lambda=2)$ theory. An $L=20$ (40 qubits) system is used with maximum bond dimension 100. The exact results use Trotterization with size $1/10$ Trotter steps.}
\label{phi4:fig:exact_vs_variational}
\end{figure*} 
In this appendix, noiseless MPS circuit simulations of an $L=20$ system are used to compare exact time evolution in the digitized theory with $n_q=2$ to time evolution approximated by SVC. Exact calculations are done using matrix exponentiation applied to the $t=0$ $|\psi_\text{2wp}\rangle$ state prepared via SVC methods. Figure~\ref{phi4:fig:exact_vs_variational} compares $\lambda=0$ and $\lambda=2$. The SVC variational results can be seen to qualitatively track the exact time evolution, but degrade with time. The quality of the $\lambda=0$ variationally determined observables is seen to be consistently worse than those for $\lambda=2$, which is again a consequence of the $\lambda=0$ theory having a longer correlation length. Although the local infidelity spanning the interaction region $I_{10}$ is high (see Table~\ref{phi4:tab:error_budget}), the observables calculated from the SVC states still follow the exact values, suggesting that fidelity is a rough bound on the quality of prepared variational states. Since quantum devices compute observables, even states that have high infidelity may be used to approximately simulate the desired process.

In the last two $\lambda=0$ plots for $t=8,9$, the SVC observables are seen to deviate significantly from the exactly computed observables. This is an indication that the circuit used to simulate these time steps is not deep enough and more variational layers should be used. The number of steps is incremented at $t=4$ and $t=7$, and it can be seen in Fig.~\ref{phi4:fig:exact_vs_variational} that the variational approximations get closer to the true values at these times, than for the preceding time steps.

\section{Optimization}
\label{phi4:sec:optimization}
Optimization of the vacuum and wavepacket preparation circuits proceeds in a straightforward way. The SC-ADAPT-VQE ansatz has only three parameters, and its symmetry-preserving structure results in fast convergence to the true vacuum. The locality of the wavepacket state enables the general brickwall circuit used in state preparation to function well. Because of its expressivity, it is able to initialize the wavepacket with high accuracy using a small number of layers. Both of these optimization steps are implemented in an iterative manner, where additional layers are added as needed until desired criteria are met, such as a combination of maximum circuit depth and desired value of $I_d$.

The time evolution circuit optimization is done in a sequential manner over the simulation times. Because of the Trotter step-like nature of the ansatz, results from optimizations of previous steps are used as initial guesses for future optimizations. For example, since $t=4$ uses two variational steps, and $t=2$ uses one step, the angles implementing the $t=2$ evolution are used as initial guesses for the $t=4$ optimization. Table~\ref{phi4:tab:initial_guess_map} specifies how the initial guesses utilize parameters that have been optimized for past simulation times. The Broyden–Fletcher–Goldfarb–Shanno (BFGS) gradient-based optimization method~\cite{Fletcher_1986} is used, with a tolerance of $10^{-5}$.

\begin{table}
\centering
\begin{tabularx}{\linewidth}{|c||Y|Y||c||Y|Y||c||Y|Y|} \hline
$t$ & \# variational steps & Initial guess & $t$ & \# variational steps & Initial guess & $t$ & \# variational steps & Initial guess\\\hline\hline
1 & 1 & -     & 4 & 2 & $t=2$ & 7 & 3 & $t=5$\\\hline
2 & 1 & $t=1$ & 5 & 2 & $t=4$ & 8 & 3 & $t=7$\\\hline
3 & 1 & $t=2$ & 6 & 2 & $t=5$ & 9 & 3 & $t=8$\\\hline
\end{tabularx}
\renewcommand{\arraystretch}{1}
\caption{The map specifying the parameters from completed variational optimizations used as initial guesses for optimizations of circuits implementing later simulation times.}
\label{phi4:tab:initial_guess_map}
\end{table}

Gradient-based circuit optimization methods such as BFGS encounter challenges when the number of parameters is increased due to the problem of barren plateaus. Barren plateaus occur as a result of circuit overparameterization, when gradients of the cost function with respect to the parameters become vanishingly small, preventing gradient-based local optimization algorithms from effectively making progress.\footnote{Interestingly, it was recently shown that the absence of barren plateaus indicates that there exists an efficient way to simulate the problem classically~\cite{Cerezo:2023nqf}.} The barren plateau problem has been carefully studied (for a review, see Ref.~\cite{Larocca:2024plh}), as well as methods to overcome it in special cases~\cite{Ostaszewski:2019vnn,Nakanishi:2020wok,Wierichs:2021nwf,Barison2021efficientquantum,Nadori:2024twv}. In the context of the optimizations run to approximate the time evolution operator using the ansatz circuit of Fig.~\ref{phi4:fig:time_evolution_circ_step}, examination of the optimization landscape for small systems shows many flat regions. Despite this, it can be seen (e.g., in Fig.~\ref{phi4:fig:trot_vs_variational_fidelity}), that by using knowledge of the physics of the system to design variational circuits, these circuits are able to capture the time evolution well.

Initial guesses are a crucial ingredient to tame barren plateaus when using gradient-based optimization methods. So-called ``warm starts'' to standard optimization methods have been investigated in the context of parameterized circuit optimization~\cite{Puig:2024rtm,Wang:2024pap}. In this light, it may be beneficial to use gradient-free methods, such as those taking advantage of the periodic dependence of the cost function on the parameters, to produce initial guesses for gradient-based optimizers. The barren plateau problem can be further reduced by introducing more structure into the circuit, such as through physical considerations as described in Sec.~\ref{phi4:sec:scalable_variational_circuits}. This can be viewed as a way to remove redundancy in the circuit.

The ansatz for time evolution for a specific simulation time is also built up in a greedy way, iteratively adding layers and using the parameters found for the previous layer as initial guesses. This way, with each layer of the variational ansatz, the algorithm creates a more precise approximation to the true time evolution. An important property of the circuit in Fig.~\ref{phi4:fig:time_evolution_circ_step} is that it implements the identity operator when all parameters are set to 0. Because of this, the layer-wise iterative optimization method ensures that additional layers increase the quality of the solution, since the optimizer can set all parameters to 0 if no other choice increases the overlap with the target state. This method helps to address the problem of barren plateaus by providing optimized initial guesses for future layers of the circuit. This is found to perform much better than optimizing all parameters from scratch, where most initial guesses land in a part of parameter space with vanishing gradients. This method is used as a heuristic when selecting the number of layers per variational step and the number of steps for each simulation time.

In addition to using variational parameters optimized for earlier simulation times as initial guesses, parameters optimized for Hamiltonians with different $\lambda$ may be used. It can be seen in Table~\ref{phi4:tab:error_budget} that the $\lambda=0$ optimizations consistently perform worse than the $\lambda=2$ ones due to the larger correlation length in the free theory. Parameters implementing the time evolution in theories with smaller correlation lengths may be used as initial guesses for those with larger correlation lengths. For instance, the $\lambda=2$ parameters may help drive the $\lambda=0$ optimization to a better region of parameter space and produce a higher quality result. Overall, it is expected that there will be improvements to the optimization techniques, both based on the structure of the circuits, and on the hyperparameters and initial guesses.

\section{Error mitigation}
\label{phi4:sec:error_mitigation}
A suite of error mitigation techniques is used in this work, enabling extraction of meaningful results from noisy data from quantum devices. The central ingredient of the error mitigation strategy used in this work is ODR, which is a local version of decoherence renormalization~\cite{Urbanek:2021oej,Farrell:2022wyt,Farrell:2022vyh}. In this method, known noise-free results of circuits that are easily simulated classically are compared to results of the same circuits from the quantum device. This ratio is then used in a spatially local (operator-wise) fashion to measure the noise that has impacted each local expectation value. ODR assumes a Pauli noise model. The standard method of PT is used to convert the coherent noise in each circuit to Pauli noise, $\rho \rightarrow \sum_i p_i P_i \rho P_i$, where $\rho$ is a reduced state used to measure a local observable. The expectation value of an observable $O$ under this noise channel is given by $\langle O \rangle = \sum_i p_i\text{Tr}(P_i\rho P_i O)$. For Pauli observables, $P_i O P_i = \pm O$, and this channel rescales the measured values $\langle O\rangle_\text{meas}$ compared to their noise-free values $\langle O\rangle_\text{true}$, $\langle O\rangle_\text{meas} = p \langle O\rangle_\text{true}$. The error mitigation scheme works by estimating the parameter $p_j$ for this channel for each local observable $O_j$. 

In this work, ODR is run at the level of $\langle ZZ\rangle$ measurements, from which $\langle \phi^2_j \rangle$ is computed using Eq.~\eqref{phi4:eq:phi_qubits}. In the specific case of $n_q=2$ and $\phi_\text{max}=1.5$, $\langle \phi^2_j\rangle$ is given by
\begin{align}
    \langle \phi^2_j\rangle \ = \ 1.25 + \langle Z_{2j}Z_{2j+1}\rangle.
\end{align}
ODR estimates expectation values in the physics circuit $\overline{\langle ZZ \rangle}_\text{phys}$ twirl by twirl, using knowledge of the noiseless expectation values of the mitigation circuit $\langle ZZ\rangle_\text{mit,true}$ and its measurements from the device $\langle ZZ\rangle_\text{mit,meas}$ using Eq.~\eqref{phi4:eq:odr_p_j}:
\begin{align}
    \overline{\langle Z_{2j}Z_{2j+1}\rangle}_\text{phys} \ = \ \langle Z_{2j}Z_{2j+1}\rangle_\text{phys,meas}\frac{1}{p_j} \ = \ \langle Z_{2j}Z_{2j+1}\rangle_\text{phys,meas} \frac{\langle Z_{2j}Z_{2j+1}\rangle_\text{mit,true}}{\langle Z_{2j}Z_{2j+1}\rangle_\text{mit,meas}}.
\end{align}
ODR hinges on selecting mitigation circuits whose noise is as similar to the noise in the physics circuits as possible. The brickwall circuits used in this work are much more dense than those used in previous works where ODR was successfully applied. This results in errors spreading much faster throughout the qubits. As a result, it is found in this work that previous choices of mitigation circuits, such as applying the time evolution operator $U(-t/2)U(t/2) = \mathbbm{1}$ (self-mitigation~\cite{ARahman:2022tkr}) or $U(0) = \mathbbm{1}$ were not able to reflect the noise present in the physics circuit well. Similarly, using various ``Cliffordized'' versions of the physics circuits as mitigation circuits did not capture the noise accurately for use in ODR.\footnote{Proposals to combine Zero Noise Extrapolation~\cite{Temme:2016vkz} with ODR have seen fruitful results~\cite{Ciavarella:2024fzw}.} This is particularly evident at late times, where the circuits are deeper and the mitigation circuits used previously deviate more from the physics circuit being implemented.

\begin{figure*}
\centering
\includegraphics[width=\linewidth]{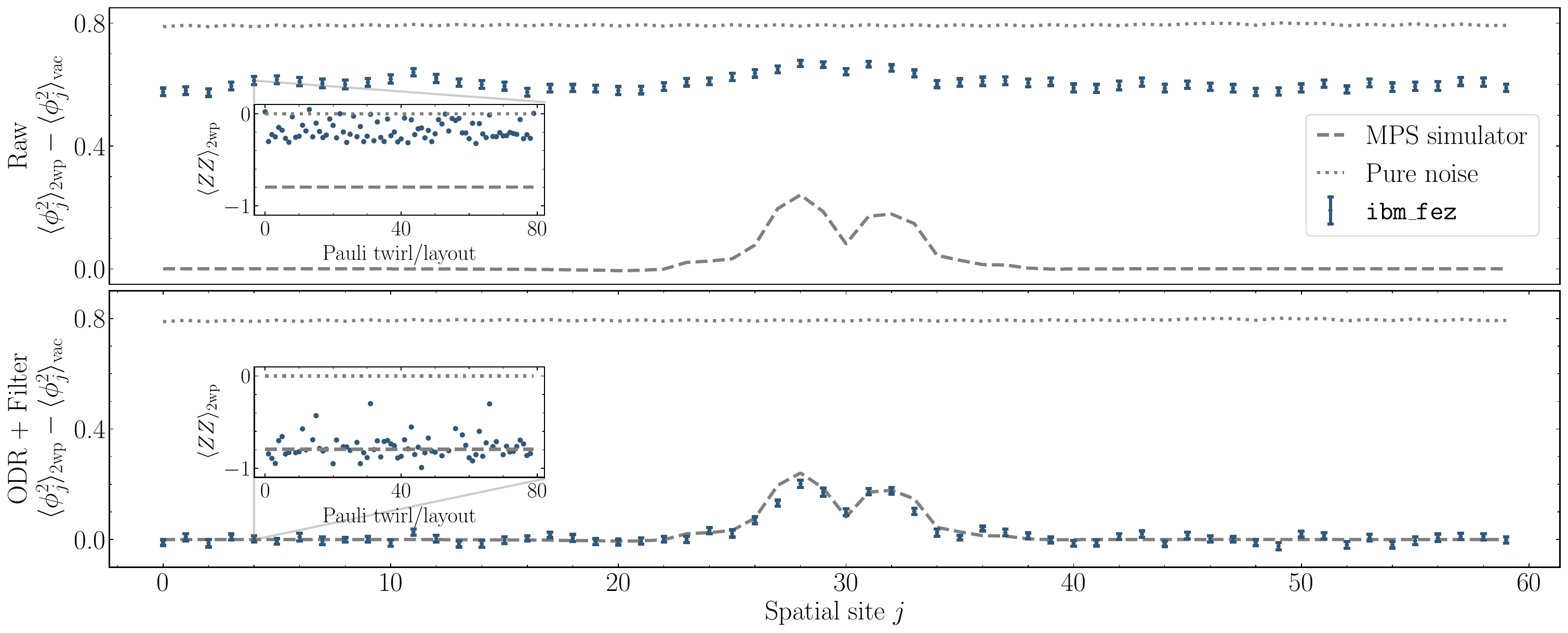}
\caption{{\it The effect of ODR and filtering on the data from {\tt ibm\_fez} for $\lambda=0,\,t=8$.} The points show data collected using 80 PTs, two TREX twirls, and 8000 shots for each circuit. The expectation value $\langle\phi^2_j\rangle_\text{2wp}-\langle\phi^2_j\rangle_\text{vac}$ is shown before (top panel) and after (bottom panel) ODR and filtering. The grey dashed line represents classical noiseless MPS circuit simulations, and the grey dotted line shows the results if the state had fully decohered to the completely mixed state. The error bars represent one standard deviation of the bootstrap-resampled data. ODR and filtering act on the $\langle ZZ\rangle_\text{2wp}$ expectation value twirl by twirl (insets).}
\label{phi4:fig:t_8_l_0_mitigation}
\end{figure*}

Instead, the circuits simulating the vacuum time evolution are used as mitigation circuits in this work. If the vacuum were prepared exactly, and if the time evolution operator were implemented exactly, time evolution would have no effect as the vacuum is an eigenstate of the Hamiltonian, $e^{-itH}|\psi_\text{vac}\rangle = |\psi_\text{vac}\rangle$. Since both the vacuum state preparation and the time evolution are implemented approximately using variational circuits, the prepared $|\psi_\text{vac}\rangle$ has some time evolution. However, this vacuum evolution can be computed classically, owing to the translation-invariant nature of $|\psi_\text{vac}\rangle$. Up to finite-size effects, the time evolution of translationally invariant states does not change with system size. The vacuum time evolution for a given system size can be estimated following the same procedure as for the convergence of the variational parameters described in the SVC algorithm. Due to the fact that correlations in the vacuum decay exponentially in this theory, finite-size effects have an exponentially small contribution. As a result, the infinite-volume vacuum expectation values $\langle \psi_\text{vac}|ZZ|\psi_\text{vac}\rangle_\infty$ (as well as those for any $L$) can be estimated by extrapolating from an exponential fit to the vacuum expectation values at a series of $L$. This provides a way to compute the $\langle ZZ\rangle_\text{mit,true}$. The values of these extrapolations for all simulation times in both the free and interacting theories, as well as the values at $L=60$ determined by MPS, are provided in Table~\ref{phi4:tab:extrapolated_vacuum_evolution}. For the parameters and simulation times considered, this method is seen to give much better estimates of the noise, as a result of the same time evolution being applied in both the physics and mitigation circuits. The effect of ODR can be seen in Fig.~\ref{phi4:fig:t_8_l_0_mitigation}. The unmitigated data shown in the upper panel is much closer to the completely decohered result (grey dotted line), than the true result from MPS circuit simulations (grey dashed line). After the error mitigation scheme is applied (bottom panel) to measure the noise and adjust the results, measurements of $\langle\phi^2_j\rangle_\text{2wp}-\langle\phi^2_j\rangle_\text{vac}$ are much closer to the expected results from MPS. The action of ODR on $\langle ZZ\rangle_\text{2wp}$ measurements, from which $\langle \phi^2_j\rangle_\text{2wp}$ is calculated, is shown in the insets of Fig.~\ref{phi4:fig:t_8_l_0_mitigation} for lattice site $j=4$. 

\begin{table}
\centering
\begin{tabularx}{\linewidth}{|c||Y|Y|Y|Y||Y|Y|Y|Y|}
\hline
&  \multicolumn{4}{c||}{$\lambda=0$}  &  \multicolumn{4}{c|}{$\lambda=2$} \\\hline
\multirow{2}{*}{\makecell{$t$}} & \multicolumn{2}{c|}{Even sites} & \multicolumn{2}{c||}{Odd sites} & \multicolumn{2}{c|}{Even sites} & \multicolumn{2}{c|}{Odd sites} \\
\cline{2-9}
 & MPS & Extrap. & MPS & Extrap. & MPS & Extrap. & MPS & Extrap. \\
\hline\hline
 1 & 0.3949 & 0.3944 & 0.3882 & 0.3877 & 0.3580  & 0.3576 & 0.3469 & 0.3465 \\\hline
 2 & 0.4677 & 0.4670  & 0.4678 & 0.4671 & 0.3692 & 0.3688 & 0.3621 & 0.3617 \\\hline
 3 & 0.4970  & 0.4955 & 0.5073 & 0.5058 & 0.3477 & 0.3471 & 0.3551 & 0.3546 \\\hline
 4 & 0.4678 & 0.4627 & 0.4768 & 0.4717 & 0.3959 & 0.3941 & 0.3767 & 0.3748 \\\hline
 5 & 0.5339 & 0.5251 & 0.5621 & 0.5528 & 0.3620  & 0.3617 & 0.3672 & 0.3671 \\\hline
 6 & 0.5275 & 0.5331 & 0.5404 & 0.5201 & 0.4027 & 0.4017 & 0.3995 & 0.3985 \\\hline
 7 & 0.4349 & 0.4292 & 0.4227 & 0.4170  & 0.3693 & 0.3686 & 0.4010  & 0.4004 \\\hline
 8 & 0.4620  & 0.4596 & 0.4617 & 0.4551 & 0.3910  & 0.3902 & 0.3768 & 0.3757 \\\hline
 9 & 0.4780  & 0.4606 & 0.4987 & 0.4844 & 0.4143 & 0.4136 & 0.3946 & 0.3939 \\\hline
\end{tabularx}
\renewcommand{\arraystretch}{1}
\caption{{\it Vacuum expectation values for error mitigation.} The vacuum expectation values $\langle\psi_\text{vac}|e^{itH}\phi^2_je^{-itH}|\psi_\text{vac}\rangle$ that are used as ``true'' values in the error mitigation scheme for the $L=60$ (120 qubits) simulations determined by MPS are compared to those determined via $L$-extrapolation from classical statevector simulations to a $L=60$ system. A maximum system size of $L=16$ (32 qubits) was used to compute the extrapolation. The even and odd sites have slightly different values because of the Trotterized state preparation, see Sec.~\ref{phi4:sec:circuits_vac_prep} for details.}
\label{phi4:tab:extrapolated_vacuum_evolution}
\end{table}

This method, similar to self-mitigation and other methods used earlier, is limited to early simulation times. In the case of the previous methods, the limiting factor was the growth of the difference in the noise between the physics and mitigation circuits with increasing circuit depth. Mitigation using the extrapolated vacuum evolution is limited by the quality of extrapolation. Finite-size effects grow as $O(t/L)$, so larger system sizes must be used to get reliable extrapolated values at later times. In previous choices of mitigation circuit, the physics circuit was modified to implement the identity and applied to the same initial state. In this work it is found that applying the same exact unitary mitigates errors in dense circuits (like the brickwall circuits of Fig.~\ref{phi4:fig:full_circuit}) much better, even though the starting state is different. This method is expected to work especially well in settings where perturbations on top of the vacuum are small, so that the state-dependent noise is similar in the physics and mitigation circuits. This is the case for wavepackets in scalar field theory (as seen in Fig.~\ref{phi4:fig:results_by_time} for example).

Note that it is possible to directly calculate an approximation to the vacuum evolution using MPS. In this work, results from classical MPS circuit simulations are used as ground truth to compare to results from {\tt ibm\_fez}. For the system parameters chosen and the times considered, it can be seen that the extrapolated vacuum evolution is quite close to the MPS predictions (see Table~\ref{phi4:tab:extrapolated_vacuum_evolution}), with the differences only noticeable at late times. It is expected that highly inelastic collisions will generate large quantities of entanglement and magic, thus being genuinely out of reach for simulations using classical methods such as MPS or Clifford+T. In this setting, the MPS values for the vacuum evolution may be used directly for mitigation. This is a reasonable expectation, as the ground states of one-dimensional gapped systems, such as the scalar field theory considered in this work, are known to obey area law entanglement~\cite{Hastings:2007iok}, so their short-time evolution may be simulated with a MPS with a relatively small bond dimension.\footnote{The approximate variational methods may yield states that are more entangled than the true, non-variational states. This may be averted by using techniques similar to Refs.~\cite{Miyakoshi:2023zzc,Causer:2023wpp,Gibbs:2024emw} to train the variational circuits with MPS in the first place. Furthermore, MPS approximations to the $|\psi_\text{2wp}\rangle$ evolution circuit may be used as mitigation circuits in the future as well, provided they can be constructed in a way that preserves the structure of the noise.}

The ratio $p_j$ calculated from the results of the mitigation circuit is used to remove observations that have been particularly affected by noise. This procedure, referred to as filtering~\cite{Farrell:2024fit}, discards measurements of a particular local observable if the ratio of the measured expectation value to the true expectation value is below a certain threshold; in this work the threshold is set to $p_j \geq 0.01$ for all runs. Once the ratio is applied to adjust the measurements of the physics circuit, a second round of filtering is done. This filter removes values of $\overline{\langle ZZ \rangle}_\text{phys} \geq 1$. Since the maximum possible value of $\langle ZZ\rangle=1$, any values above 1 are the result of the more decoherence being present for that observable in the mitigation circuits; these observations are discarded as well. These two filters work by removing outliers and extremely noisy data points, causing the size of the bootstrap samples for each local observable to be different. 

A loop of qubits connected by two-qubit gates is necessary because of the PBCs used in this work. As a result of the scalability of the simulation methods outlined in previous sections, it is possible to run circuits on the largest possible loop on {\tt ibm\_fez} (120 qubits). A specific loop is chosen based on a combination of the smallest errors and longest coherence time throughout the device. It is found in this work that the randomization of mapping from circuit qubits to device qubits (layout randomization) helps to tailor the noise closer to depolarizing noise, in addition to averaging out effects from noise due to localized imperfections on the device, such as specific noisy qubits, gates, and readout channels. In this work, a loop is fixed for all runs and the starting point of the loop is chosen at random for each circuit. For each run, the circuits are executed in an order that alternates between the mitigation (vacuum) and physics (wavepacket) circuits, for the same PT and layout. Since the noise on the device is expected to fluctuate with time, this is intended to more accurately measure the noise by running the mitigation circuits directly before the corresponding physics circuits. The same set of twirls and layouts is used for the physics and mitigation circuits.

To mitigate measurement errors, TREX is used. This method involves applying Pauli gates on all qubits at random prior to measurement, flipping the value of the classical result if the Pauli applied is $X$ or $Y$, and doing nothing in the case of $Z$ or $I$. Two TREX twirls are used for each Pauli-twirled circuit in this work. 

The IBM Heron processors have seen a 3-5$\times$ increase in performance over their Eagle predecessors, and have significantly decreased qubit crosstalk~\cite{Gambetta_2023}. Despite this, it is still seen in this work that crosstalk remains a non-negligible source of error. DD is used to minimize this effect and remove erroneous evolution during qubit idle times. To mitigate these effects, the crosstalk-robust CRXY4 DD sequence~\cite{PhysRevLett.131.210802} is used. This sequence consists of staggered XY4 sequences applied to neighboring qubits to cancel out collective evolution caused by crosstalk.\footnote{Recent work found that asymmetric implementations of $Y$ gates can have an adverse impact on the effectiveness of DD sequences~\cite{Vezvaee:2024ywq}. As a result, all $Y$ gates used in this work, including the DD stage, are converted to the native gate set using the symmetric implementation $Y = R_Z(\pi/2)XR_Z(-\pi/2)$.}

In particular, the effects of crosstalk are noticeable in regions where qubits are idle for numerous circuit layers directly adjacent to ones that have gates acting on them. These types of errors are most prevalent in the wavepacket creation step of the simulation algorithm, where the wavepacket creation circuits are acting in limited regions of the quantum device (green blocks of Fig.~\ref{phi4:fig:full_circuit}). To mitigate these effects, barriers are inserted to isolate the times where both qubits are idle or single-qubit gates are being run on the wavepacket qubits, from the times where two-qubit gates are acting in the wavepacket region. The barriers instruct the circuit transpilation to add DD separately for these two cases because the noise is different during each process. 

\section{Variational parameters}
\label{phi4:sec:variational_params}
In this appendix, the variational parameters determined using the SVC framework presented in Sec.~\ref{phi4:sec:scalable_variational_circuits} are given for vacuum preparation, wavepacket excitation, and time evolution. Exponential extrapolation is run for the vacuum preparation parameters. The $L$-extrapolated vacuum parameters are used at $L=12$ to optimize the wavepacket preparation and time evolution circuits. 

\begin{table}[h]
\centering
\begin{tabularx}{\linewidth}{|c||Y|Y|Y||Y|Y|Y|}
\hline
&  \multicolumn{3}{c||}{$\lambda=0$}  &  \multicolumn{3}{c|}{$\lambda=2$} \\\hline
$L$ & $\theta_0$ & $\theta_1$ & $\theta_2$ & $\theta_0$ & $\theta_1$ & $\theta_2$ \\ 
\hline \hline
2 & 2.5517 & 1.3354 & 0.5659 & 2.6685 & 1.0626 & 0.4370 \\\hline
4 & 2.6346 & 1.0111 & 0.9131 & 2.7164 & 0.7923 & 0.7438 \\\hline
6 & 2.6462 & 0.9469 & 0.8703 & 2.7206 & 0.7547 & 0.7141 \\\hline
8 & 2.6490 & 0.9312 & 0.8580 & 2.7213 & 0.7491 & 0.7088 \\\hline
10 & 2.6497 & 0.9278 & 0.8536 & 2.7215 & 0.7478 & 0.7087 \\\hline
12 & 2.6499 & 0.9248 & 0.8536 & 2.7214 & 0.7466 & 0.7088 \\\hline\hline
$\infty$ &  2.6495 & 0.9273 & 0.8534 & 2.7214 & 0.7476 & 0.7085 \\\hline
\end{tabularx}
\renewcommand{\arraystretch}{1}
\caption{The variational parameters to prepare the vacuum as a function of system size, for both the free and interacting theory. They correspond to the angles $\theta_0, \theta_1, \theta_2$ defined in the circuit elements of Fig.~\ref{phi4:fig:circuit_elements}a. These parameters are determined using classical simulations and optimization. The infinite-volume values shown in the last row are calculated by an exponential extrapolation.}
\label{phi4:tab:gs_prep_angles}
\end{table}

\begin{table}
\centering
\begin{tabularx}{\linewidth}{|c||Y|Y|Y|Y||Y|Y|Y|Y|}
\hline
&  \multicolumn{4}{c||}{$\lambda=0$}  &  \multicolumn{4}{c|}{$\lambda=2$} \\\hline
$j$ & $i=0$ & $i=1$ & $i=2$ & $i=3$ & $i=0$ & $i=1$ & $i=2$ & $i=3$\\
\hline\hline
  0 & -0.1531 &  0.8023 &  0.5352 & -0.6361 &  3.0113 & -0.2089 &  1.8433 &  1.2508 \\\hline
  1 & -2.6753 &  1.7051 & -0.0746 &  1.5072 & -0.1898 &  1.9292 &  0.0173 & -1.3358 \\\hline
  2 &  0.9219 &  0.4939 &  0.0057 &  0.3665 &  2.0861 &  1.3400   &  0.1340  &  1.8122 \\\hline
  3 &  0.2726 & -0.3087 &  0.5203 & -0.2217 &  2.9159 &  1.3803 & -0.9289 &  0.8222 \\\hline
  4 &  1.6515 &  1.5371 &  1.3394 &  1.4023 &  1.4625 & -1.0540  &  1.5245 &  1.5592 \\\hline
  5 & -0.3017 & -0.2190  & -0.7053 & -0.0079 &  1.0102 &  1.5805 &  1.4121 & -0.0310  \\
\hline
\end{tabularx}
\caption{The parameters $\theta_{ij}$ used to initialize Gaussian wavepackets of momenta $k=\pm\pi/3$ and spread in $k$-space $\sigma_k=\pi/3$ in the free and interacting theories. The $\theta_{ij}$ correspond to the angles defined in the brickwall layer shown in Fig.~\ref{phi4:fig:circuit_elements}c: $i$ labels the layer, and $j$ labels the parameter within layer $i$. These parameters are determined using classical simulations and optimization.}
\label{phi4:tab:wp_prep_angles}
\end{table}

\begin{table}
\centering
\setlength{\tabcolsep}{3pt} 
\scriptsize
\begin{tabularx}{\linewidth}{|c||Y|Y|Y|Y|Y|Y||Y|Y|Y|Y|Y|Y|}
\hline
&  \multicolumn{6}{c||}{$\lambda=0$}  &  \multicolumn{6}{c|}{$\lambda=2$} \\\hline
$j$ & $i=0$ & $i=1$ & $i=2$ & $i=3$ & $i=4$ & $i=5$ & $i=0$ & $i=1$ & $i=2$ & $i=3$ & $i=4$ & $i=5$ \\
\hline\hline
  0 &  0.0843 & -0.0609 &  0.6955 &  0.2961 & -0.0480  & -0.0191 &  0.1727 &  0.0329 & -1.1632 &  0.0605 &  0.6813 &  0.1975 \\\hline
  1 & -0.2860  & -3.4066 & -2.9043 &  0.0167 &  2.2841 & -0.0292 &  0.8868 & -2.6254 &  1.6950  & -0.1890  & -3.8622 &  0.1428 \\\hline
  2 &  1.5618 &  0.1327 &  1.4740  & -0.1091 & -1.4579 & -1.3838 & -0.8353 &  0.1261 & -1.1718 & -0.1111 &  1.0946 & -0.7012 \\\hline
  3 &  0.2173 &  1.3044 & -1.1338 & -1.4324 & -1.3960  & -1.6492 & -0.2904 & -2.0721 &  1.6594 &  0.5606 &  1.0016 & -0.4645 \\\hline
  4 &  0.0121 & -1.1901 &  1.9423 & -1.4593 &  0.1350  & -1.1644 & -0.6027 &  3.2926 &  0.2786 & -0.8616 & -1.5139 &  0.5588 \\\hline
  5 & -1.5731 &  2.2241 &  1.3656 &  0.1695 & -0.6027 & -1.1536 &  1.4649 & -1.5636 &  0.7272 &  1.4701 &  2.0278 &  1.0915 \\\hline
  6 & -0.1173 &  0.7041 & -1.0110  &  1.5841 & -0.6495 & -0.0152 &  0.0666 & -1.5045 &  0.8148 &  0.6901 & -0.1011 &  0.0370  \\\hline
  7 &  1.8137 &  1.2280  &  1.5580  & -0.1221 &  0.3547 & -0.0793 &  1.7586 & -1.2382 &  0.9962 & -0.5063 &  0.1100   &  0.0887 \\\hline
  8 &  0.0636 & -1.0966 &  1.9644 & -1.4114 &  3.3265 &  1.0599 & -0.6402 &  2.2965 & -0.0883 & -1.1814 & -2.0575 &  0.8626 \\\hline
  9 & -1.6840  &  2.0991 &  1.5450  &  0.3095 &  0.5833 & -1.1253 &  1.3265 & -2.0403 &  0.9642 &  1.9077 &  1.3393 &  1.1265 \\\hline
 10 & -0.0975 &  0.7022 & -1.0872 &  1.6051 & -0.6503 & -0.0324 & -0.1070  & -1.1265 &  1.2493 & -0.1006 & -0.5110  & -0.0454 \\\hline
 11 &  1.8584 &  1.1814 &  1.5940  & -0.0322 &  0.3009 & -0.0683 &  1.6193 & -1.1086 &  0.9842 & -0.5982 & -0.0874 &  0.0748 \\\hline
\end{tabularx}
\caption{The parameters $\theta_{ij}$ used to implement time evolution for $t=1$ in the free and interacting theories. The $\theta_{ij}$ correspond to the angles defined in the brickwall layer shown in Fig.~\ref{phi4:fig:circuit_elements}d: $i$ labels the layer, and $j$ labels the parameter within layer $i$. These parameters are determined using classical simulations and optimization.}
\label{phi4:tab:time_evolution_angles_1}
\end{table}

\begin{table}
\centering
\setlength{\tabcolsep}{3pt} 
\scriptsize
\begin{tabularx}{\linewidth}{|c||Y|Y|Y|Y|Y|Y||Y|Y|Y|Y|Y|Y|}
\hline
&  \multicolumn{6}{c||}{$\lambda=0$}  &  \multicolumn{6}{c|}{$\lambda=2$} \\\hline
$j$ & $i=0$ & $i=1$ & $i=2$ & $i=3$ & $i=4$ & $i=5$ & $i=0$ & $i=1$ & $i=2$ & $i=3$ & $i=4$ & $i=5$ \\
\hline\hline
  0 & -0.0127 & -0.0303 &  0.8574 &  0.2545 & -0.1172 &  0.0181 &  0.0398 &  0.0648 & -0.8169 &  0.3516 &  0.7947 &  0.1075 \\\hline
  1 & -0.7944 & -4.0568 & -3.6008 &  0.0228 &  3.4077 &  0.7981 &  1.1739 & -1.9846 &  1.7857 &  0.4659 & -4.8039 & -0.2786 \\\hline
  2 &  1.5237 &  0.3437 &  1.6710  & -0.1623 & -1.6623 & -1.6914 & -1.5506 &  0.1435 & -0.4705 & -0.0216 &  1.8417 & -1.0534 \\\hline
  3 &  0.0042 &  1.5494 & -1.0016 & -1.4038 & -1.9429 & -1.6069 & -0.0645 & -0.9712 & -1.0871 &  1.1405 &  1.3022 & -0.4437 \\\hline
  4 & -0.2237 & -0.8308 &  1.5671 & -1.4545 & -0.1409 & -0.5384 & -0.4929 &  2.1439 & -0.2443 & -1.7539 & -1.5463 & -0.2068 \\\hline
  5 & -1.5910  &  2.5192 &  1.4323 &  0.2501 & -0.8664 & -0.5929 &  1.2547 & -1.7316 & -0.1444 &  2.1785 &  1.9204 &  1.2651 \\\hline
  6 & -0.0115 &  0.6926 & -0.7056 &  2.0844 & -0.5686 &  0.0572 &  0.2956 & -1.9191 &  0.9858 & -0.5653 &  0.1543 &  0.0046 \\\hline
  7 &  2.6181 &  1.3091 &  1.4943 &  0.0006 & -0.3561 & -0.0597 &  0.4242 &  1.1723 &  1.2054 & -0.0798 &  0.1706 &  0.0134 \\\hline
  8 & -0.2804 & -0.9134 &  1.7664 & -1.2446 &  3.1131 &  0.5317 & -0.6196 &  2.0314 & -0.3388 & -1.9148 & -1.7189 & -0.1011 \\\hline
  9 & -1.5676 &  2.3869 &  1.5213 &  0.3573 &  0.8461 & -0.5576 &  1.1632 & -1.6832 &  0.0979 &  1.9924 &  2.1347 &  1.2702 \\\hline
 10 & -0.0285 &  0.6670  & -0.6825 &  2.0468 & -0.5351 &  0.0372 &  0.1732 & -2.0512 &  0.9512 & -0.7355 &  0.0777 &  0.0141 \\\hline
 11 &  2.5855 &  1.3686 &  1.4927 &  0.0083 & -0.3019 & -0.0333 &  0.3224 &  1.0980  &  1.3002 & -0.0688 &  0.3608 &  0.0228 \\\hline
\end{tabularx}
\caption{The parameters $\theta_{ij}$ used to implement time evolution for $t=2$ in the free and interacting theories. The $\theta_{ij}$ correspond to the angles defined in the brickwall layer shown in Fig.~\ref{phi4:fig:circuit_elements}d: $i$ labels the layer, and $j$ labels the parameter within layer $i$. These parameters are determined using classical simulations and optimization.}
\label{phi4:tab:time_evolution_angles_2}
\end{table}

\begin{table}
\centering
\setlength{\tabcolsep}{3pt} 
\scriptsize
\begin{tabularx}{\linewidth}{|c||Y|Y|Y|Y|Y|Y||Y|Y|Y|Y|Y|Y|}
\hline
&  \multicolumn{6}{c||}{$\lambda=0$}  &  \multicolumn{6}{c|}{$\lambda=2$} \\\hline
$j$ & $i=0$ & $i=1$ & $i=2$ & $i=3$ & $i=4$ & $i=5$ & $i=0$ & $i=1$ & $i=2$ & $i=3$ & $i=4$ & $i=5$ \\
\hline\hline
  0 &  0.0021 & -0.0586 &  1.5776 &  0.0003 &  0.5290  &  0.4828 &  0.1081 &  0.0477 & -1.1605 &  0.5728 &  1.1499 &  0.2074 \\\hline
  1 & -1.5784 & -4.6994 & -3.1345 &  0.0540  &  3.4056 &  0.0368 &  0.1861 & -2.7631 &  2.0690  &  0.1181 & -3.8915 &  2.7622 \\\hline
  2 &  1.4004 &  0.6131 &  0.7863 & -0.1804 & -1.6037 & -0.9163 & -1.5665 &  0.1879 & -0.5834 &  0.0316 &  1.8911 & -1.1214 \\\hline
  3 &  0.1212 &  1.6162 & -1.5701 & -1.7525 & -1.5181 & -0.8125 & -0.0984 & -1.5692 & -0.2535 &  3.0750  &  1.6547 & -0.5287 \\\hline
  4 & -0.3352 & -1.5919 &  1.2529 & -0.4186 &  1.6014 & -1.5573 & -0.1697 &  1.5929 & -0.2217 & -1.5676 & -1.8083 & -0.4096 \\\hline
  5 & -1.5505 &  3.1467 &  1.4259 &  0.3900   & -1.4477 &  0.2438 &  1.6468 & -1.5093 & -0.3895 &  2.8569 &  1.8389 &  1.1812 \\\hline
  6 & -0.0018 &  1.1808 & -0.0031 &  1.5923 & -2.6329 & -0.9154 & -0.0441 & -1.8221 &  0.9875 &  1.0029 & -0.2364 &  0.0088 \\\hline
  7 &  3.1182 &  1.5498 &  1.5514 &  0.3567 & -0.0363 & -0.0068 &  1.0255 &  0.1220  &  1.6884 & -0.0175 & -1.8634 & -0.0213 \\\hline
  8 & -0.4566 & -1.4424 &  1.4231 & -0.5025 &  4.7225 &  1.5587 & -0.1532 &  1.6630  & -0.1710  & -1.5943 & -1.7456 & -0.4261 \\\hline
  9 & -1.5682 &  3.1396 &  1.4702 &  0.0032 &  1.4704 & -0.2056 &  1.6425 & -1.5363 & -0.3713 &  2.9048 &  1.9027 &  1.1827 \\\hline
 10 &  0.0028 &  0.9779 &  0.0008 &  1.5751 & -0.6327 &  0.9388 & -0.0570  & -1.7169 &  0.9733 &  0.9879 & -0.1879 & -0.0017 \\\hline
 11 &  3.1055 &  1.5655 &  1.6030  &  0.2539 & -0.0241 & -0.0084 &  1.0308 &  0.1463 &  1.6700   & -0.0398 & -1.8732 & -0.0112 \\\hline
\end{tabularx}
\caption{The parameters $\theta_{ij}$ used to implement time evolution for $t=3$ in the free and interacting theories. The $\theta_{ij}$ correspond to the angles defined in the brickwall layer shown in Fig.~\ref{phi4:fig:circuit_elements}d: $i$ labels the layer, and $j$ labels the parameter within layer $i$. These parameters are determined using classical simulations and optimization.}
\label{phi4:tab:time_evolution_angles_3}
\end{table}

\begin{table}
\centering
\setlength{\tabcolsep}{3pt} 
\scriptsize
\begin{tabularx}{\linewidth}{|c||Y|Y|Y|Y|Y|Y||Y|Y|Y|Y|Y|Y|}
\hline
&  \multicolumn{6}{c||}{$\lambda=0$}  &  \multicolumn{6}{c|}{$\lambda=2$} \\\hline
$j$ & $i=0$ & $i=1$ & $i=2$ & $i=3$ & $i=4$ & $i=5$ & $i=0$ & $i=1$ & $i=2$ & $i=3$ & $i=4$ & $i=5$ \\
\hline\hline
  0 & -0.1073 &  0.0847 &  0.4879 &  0.3093 & -0.2566 & -0.0338 & -0.0025 & -0.0237 & -0.7062 &  0.2692 &  0.5673 & -0.2870  \\\hline
  1 & -0.5637 & -5.1000    & -2.6777 & -0.0863 &  3.3271 &  1.2604 &  1.0773 &  0.7424 &  2.4129 &  0.3060  & -4.2993 &  0.5121 \\\hline
  2 &  1.5109 &  0.4379 &  1.0296 & -0.1679 & -1.6439 & -0.8366 & -1.6468 &  0.3683 & -0.4062 &  0.0612 &  1.7473 & -1.9365 \\\hline
  3 &  0.0692 &  1.5616 & -1.0566 & -2.0557 & -2.3746 & -1.5466 & -0.0042 & -1.4611 &  0.0559 &  0.7209 &  0.9852 & -0.7805 \\\hline
  4 & -0.4893 & -1.0224 &  2.3117 & -1.6204 &  0.2698 & -0.5523 & -0.4652 &  2.3685 & -0.7806 & -1.5357 & -1.5061 &  0.0209 \\\hline
  5 & -1.6233 &  3.2011 &  1.8399 & -0.7357 & -0.8480  & -0.2008 &  1.5544 & -1.6221 &  0.9499 &  2.2305 &  2.2893 &  1.1904 \\\hline
  6 & -0.0555 &  1.5751 & -1.0267 &  1.3182 & -1.8387 &  0.1214 & -0.0479 & -1.2671 &  0.6778 & -0.0874 & -0.0155 & -0.0129 \\\hline
  7 &  3.6119 &  1.2176 &  1.5505 &  0.2732 & -0.9677 & -0.0461 & -2.3391 &  0.0142 &  1.3246 & -0.0462 & -0.7364 &  0.0004 \\\hline
  8 & -0.5405 & -0.9474 &  2.3937 & -1.5040  &  3.4139 &  0.5829 & -0.4848 &  2.2950  & -0.7543 & -1.5413 & -1.4418 &  0.0648 \\\hline
  9 & -1.4556 &  3.2751 &  1.6913 & -0.9322 &  0.7875 & -0.3782 &  1.4431 & -1.6935 &  0.9909 &  2.1455 &  2.0475 &  1.1984 \\\hline
 10 & -0.0087 &  1.5019 & -1.0804 &  1.3444 & -1.9556 &  0.0493 &  0.0243 & -1.2489 &  0.7149 &  0.0038 & -0.0335 & -0.0282 \\\hline
 11 &  3.4239 &  1.1735 &  1.5976 &  0.2317 & -1.0280  & -0.0450  & -2.4800   & -0.0288 &  1.4062 & -0.0953 & -0.7767 &  0.0393 \\\hline
\end{tabularx}
\caption{The parameters $\theta_{ij}$ used to implement time evolution for $t=4$ in the free and interacting theories. The $\theta_{ij}$ correspond to the angles defined in the brickwall layer shown in Fig.~\ref{phi4:fig:circuit_elements}d: $i$ labels the layer, and $j$ labels the parameter within layer $i$. These parameters are determined using classical simulations and optimization.}
\label{phi4:tab:time_evolution_angles_4}
\end{table}

\begin{table}
\centering
\setlength{\tabcolsep}{3pt} 
\scriptsize
\begin{tabularx}{\linewidth}{|c||Y|Y|Y|Y|Y|Y||Y|Y|Y|Y|Y|Y|}
\hline
&  \multicolumn{6}{c||}{$\lambda=0$}  &  \multicolumn{6}{c|}{$\lambda=2$} \\\hline
$j$ & $i=0$ & $i=1$ & $i=2$ & $i=3$ & $i=4$ & $i=5$ & $i=0$ & $i=1$ & $i=2$ & $i=3$ & $i=4$ & $i=5$ \\
\hline\hline
  0 & -0.1326 &  0.0372 &  0.5814 &  0.3058 & -0.2071 & -0.1162 &  0.0000      & -0.0105 & -1.3237 &  0.6746 &  0.6885 &  1.3835 \\\hline
  1 & -0.7267 & -5.7599 & -3.2273 & -0.0727 &  3.6763 &  1.1732 & -0.0002 &  1.0968 &  1.5905 &  1.5804 & -5.7955 &  1.5715 \\\hline
  2 &  1.5629 &  0.4892 &  1.0416 & -0.1845 & -1.7186 & -1.1372 & -1.6402 &  0.3005 & -0.4687 & -0.0184 &  2.7857 & -1.6367 \\\hline
  3 & -0.0209 &  1.5764 & -1.1759 & -2.1123 & -2.2658 & -1.4966 &  0.0821 & -1.5130  &  0.0071 &  0.2494 &  1.5453 & -1.5708 \\\hline
  4 & -0.5286 & -1.0685 &  2.2987 & -1.7233 &  0.1708 & -0.5050  & -0.0255 &  1.5761 &  0.0117 & -3.1069 & -1.1159 &  0.2522 \\\hline
  5 & -1.6923 &  3.3742 &  2.0409 & -0.7240  & -0.6956 & -0.1600   &  1.4987 & -1.0560  & -0.2535 &  3.5926 &  1.5549 &  0.2101 \\\hline
  6 & -0.1063 &  1.4380  & -0.8177 &  1.2692 & -1.6112 &  0.0894 &  0.0001 & -0.9552 &  0.4211 & -1.2032 &  0.2553 &  0.4862 \\\hline
  7 &  4.3371 &  1.3833 &  1.5239 &  0.3667 & -0.7124 &  0.0020  & -2.6630  & -0.0029 &  1.5707 & -0.0319 & -1.5476 &  0.1618 \\\hline
  8 & -0.6487 & -0.9463 &  2.4319 & -1.6415 &  3.3190  &  0.5682 &  0.0047 &  1.6030  &  0.0016 & -3.1070  & -0.9698 &  0.2296 \\\hline
  9 & -1.3430  &  3.6764 &  1.7624 & -1.0195 &  0.8142 & -0.4189 &  1.3910  & -1.0438 & -0.2541 &  3.6141 &  1.5336 &  0.2097 \\\hline
 10 & -0.0081 &  1.1308 & -1.0171 &  1.4978 & -1.9940  &  0.0378 & -0.0011 & -0.9338 &  0.4016 & -1.1540  &  0.2346 &  0.6202 \\\hline
 11 &  3.9799 &  1.1996 &  1.6125 &  0.2875 & -1.0267 & -0.0509 & -2.6706 & -0.0246 &  1.5695 & -0.0200   & -1.5884 &  0.1631 \\\hline
\end{tabularx}
\caption{The parameters $\theta_{ij}$ used to implement time evolution for $t=5$ in the free and interacting theories. The $\theta_{ij}$ correspond to the angles defined in the brickwall layer shown in Fig.~\ref{phi4:fig:circuit_elements}d: $i$ labels the layer, and $j$ labels the parameter within layer $i$. These parameters are determined using classical simulations and optimization.}
\label{phi4:tab:time_evolution_angles_5}
\end{table}

\begin{table}
\centering
\setlength{\tabcolsep}{3pt} 
\scriptsize
\begin{tabularx}{\linewidth}{|c||Y|Y|Y|Y|Y|Y||Y|Y|Y|Y|Y|Y|}
\hline
&  \multicolumn{6}{c||}{$\lambda=0$}  &  \multicolumn{6}{c|}{$\lambda=2$} \\\hline
$j$ & $i=0$ & $i=1$ & $i=2$ & $i=3$ & $i=4$ & $i=5$ & $i=0$ & $i=1$ & $i=2$ & $i=3$ & $i=4$ & $i=5$ \\
\hline\hline
  0 &  0.0411 & -0.1769 &  0.5015 &  0.8762 &  0.3464 & -0.0139 &  0.0030  & -0.0092 & -1.4508 &  0.6506 &  0.8001 &  1.1527 \\\hline
  1 & -0.0942 & -4.5314 & -4.1438 & -0.0091 &  2.9710  &  2.2259 & -0.0019 &  1.1037 &  1.8273 &  1.5725 & -4.2827 &  1.5692 \\\hline
  2 &  1.5761 &  0.5747 &  0.6118 & -0.5092 & -1.5826 & -1.5449 & -1.5997 &  0.2543 & -0.3165 & -0.0120  &  2.8019 & -1.7114 \\\hline
  3 & -0.0601 &  1.5681 & -1.4277 & -1.6061 & -1.5133 & -1.8533 &  0.0943 & -1.5557 &  0.0137 &  0.2642 &  1.6118 & -1.5714 \\\hline
  4 & -0.3745 & -0.7460  &  3.0021 & -2.1116 & -0.0530  & -1.3761 & -0.0057 &  1.5674 & -0.0024 & -3.1086 & -0.9774 &  0.1754 \\\hline
  5 & -1.5873 &  4.5537 &  1.6295 & -1.0644 & -0.4700   & -0.8731 &  1.5394 & -0.9324 & -0.2696 &  3.3930  &  1.5974 &  0.0565 \\\hline
  6 &  0.0100   &  1.0453 &  0.2529 &  1.4969 & -0.9874 &  0.0076 & -0.0109 & -0.6916 &  0.6663 & -1.5556 &  0.1951 &  0.6161 \\\hline
  7 &  3.0271 &  0.2209 &  1.6037 &  0.1586 & -0.9758 &  0.1598 & -2.6591 & -0.0200   &  1.5562 &  0.0239 & -1.5693 &  0.1799 \\\hline
  8 & -0.3817 & -0.7077 &  3.0607 & -2.039  &  3.1999 &  1.3527 &  0.0047 &  1.5856 &  0.0037 & -3.1055 & -1.0376 &  0.2258 \\\hline
  9 & -1.5157 &  4.5203 &  1.6534 & -1.0739 &  0.4792 & -0.8984 &  1.4434 & -0.9179 & -0.2720  &  3.3773 &  1.5878 &  0.0779 \\\hline
 10 &  0.0023 &  1.0685 &  0.2252 &  1.4997 & -0.9731 &  0.0103 & -0.0042 & -0.7239 &  0.5917 & -1.5104 &  0.2549 &  0.5399 \\\hline
 11 &  3.0216 &  0.1465 &  1.5569 &  0.2173 & -0.9865 &  0.1577 & -2.6715 & -0.0153 &  1.5556 &  0.0135 & -1.6080  &  0.1801 \\\hline
\end{tabularx}
\caption{The parameters $\theta_{ij}$ used to implement time evolution for $t=6$ in the free and interacting theories. The $\theta_{ij}$ correspond to the angles defined in the brickwall layer shown in Fig.~\ref{phi4:fig:circuit_elements}d: $i$ labels the layer, and $j$ labels the parameter within layer $i$. These parameters are determined using classical simulations and optimization.}
\label{phi4:tab:time_evolution_angles_6}
\end{table}

\begin{table}
\centering
\setlength{\tabcolsep}{3pt} 
\scriptsize
\begin{tabularx}{\linewidth}{|c||Y|Y|Y|Y|Y|Y||Y|Y|Y|Y|Y|Y|}
\hline
&  \multicolumn{6}{c||}{$\lambda=0$}  &  \multicolumn{6}{c|}{$\lambda=2$} \\\hline
$j$ & $i=0$ & $i=1$ & $i=2$ & $i=3$ & $i=4$ & $i=5$ & $i=0$ & $i=1$ & $i=2$ & $i=3$ & $i=4$ & $i=5$ \\
\hline\hline
  0 & -0.0038 &  0.0528 & -0.0417 &  0.2083 &  0.0104 & -0.0076 &  0.3432 &  0.0404 & -1.0961 &  0.4267 &  1.1327 & -0.0044 \\\hline
  1 & -1.5722 & -4.7573 & -4.6618 & -0.0137 & -0.1573 & -0.2374 &  1.5457 & -0.2010  &  1.5569 & -0.0334 & -4.7501 & -0.5418 \\\hline
  2 &  1.5930  &  1.9531 &  0.8304 & -1.6783 & -1.5760  & -1.5841 & -1.5684 &  0.0604 & -0.4015 & -0.0202 &  1.5155 & -1.8278 \\\hline
  3 &  0.0228 &  1.5732 & -1.5732 & -1.5716 & -2.7976 & -1.5621 & -0.0520  &  0.0594 & -1.5664 &  1.7422 &  1.5781 & -0.0860  \\\hline
  4 & -1.9226 & -0.2589 &  2.7112 & -2.0992 &  0.0877 & -0.4603 & -1.2895 &  2.7726 & -0.1389 & -2.6832 & -2.0356 & -0.0042 \\\hline
  5 & -1.5758 &  4.6627 &  1.5745 &  3.3032 & -0.0037 & -0.7178 &  0.4848 & -1.4714 & -0.1673 &  1.6008 &  3.0597 &  1.2465 \\\hline
  6 &  0.0072 &  0.3859 & -0.2240  &  3.0196 & -0.4894 & -0.0041 &  0.0162 & -2.2549 &  0.8874 & -0.7982 &  0.3035 & -0.0108 \\\hline
  7 &  3.1959 &  1.5730  &  1.5746 & -0.0129 &  0.2335 &  0.0013 & -1.4587 &  1.5969 &  1.6607 & -0.0131 &  0.5083 &  0.0036 \\\hline
  8 & -1.8641 & -0.2125 &  2.7557 & -2.0107 &  3.1658 &  0.5005 & -1.4565 &  2.5604 &  0.0662 & -2.0192 & -1.5055 & -0.4916 \\\hline
  9 & -1.5729 &  4.6746 &  1.5732 &  3.2918 &  0.0080  & -0.7120  &  1.3067 & -1.5503 & -0.0798 &  1.5939 &  1.2361 &  1.2577 \\\hline
 10 &  0.0017 &  0.3691 & -0.2857 &  2.6193 & -0.1122 & -0.0086 &  0.4963 & -1.9117 &  0.8412 & -0.6049 &  0.8032 &  0.0027 \\\hline
 11 &  3.1826 &  1.5762 &  1.5861 & -0.0231 &  0.2348 &  0.0014 & -0.7771 &  1.5899 &  1.0078 &  0.0033 &  0.5246 & -0.0099 \\\hline
\end{tabularx}
\caption{The parameters $\theta_{ij}$ used to implement time evolution for $t=7$ in the free and interacting theories. The $\theta_{ij}$ correspond to the angles defined in the brickwall layer shown in Fig.~\ref{phi4:fig:circuit_elements}d: $i$ labels the layer, and $j$ labels the parameter within layer $i$. These parameters are determined using classical simulations and optimization.}
\label{phi4:tab:time_evolution_angles_7}
\end{table}

\begin{table}
\centering
\setlength{\tabcolsep}{3pt} 
\scriptsize
\begin{tabularx}{\linewidth}{|c||Y|Y|Y|Y|Y|Y||Y|Y|Y|Y|Y|Y|}
\hline
&  \multicolumn{6}{c||}{$\lambda=0$}  &  \multicolumn{6}{c|}{$\lambda=2$} \\\hline
$j$ & $i=0$ & $i=1$ & $i=2$ & $i=3$ & $i=4$ & $i=5$ & $i=0$ & $i=1$ & $i=2$ & $i=3$ & $i=4$ & $i=5$ \\
\hline\hline
  0 &  0.0017 &  0.1202 & -0.2994 &  0.2670  &  0.008  & -0.0089 &  2.0762 &  0.0757 & -1.9467 &  0.5305 &  1.2240  & -0.0050  \\\hline
  1 & -1.5720  & -4.7144 & -4.7091 &  0.0071 & -0.1362 & -0.1595 &  1.5288 & -0.7801 &  1.5855 & -0.1109 & -4.7125 & -0.4436 \\\hline
  2 &  1.5624 &  2.4975 &  0.7148 & -2.4083 & -1.5697 & -1.5723 & -1.5883 & -0.0486 & -0.6387 &  0.0399 &  1.5001 & -1.5977 \\\hline
  3 &  0.0355 &  1.5617 & -1.5725 & -1.5728 & -3.3672 & -1.5360  & -0.0729 &  0.8390  & -1.5779 & -0.1046 &  1.6059 & -0.0781 \\\hline
  4 & -2.9469 & -0.5245 &  2.2046 & -2.2773 &  0.0940  & -0.5461 & -1.6336 &  1.2492 &  0.6364 & -1.0464 & -1.9926 & -0.3007 \\\hline
  5 & -1.5805 &  4.7094 &  1.5692 &  3.2779 & -0.0022 & -1.0045 & -0.2026 & -1.4905 &  1.9586 &  1.5602 &  3.1292 &  1.2552 \\\hline
  6 &  0.0011 &  0.7103 & -0.2456 &  2.9032 &  0.0792 & -0.0064 & -0.1564 & -2.5584 &  1.6904 & -1.5897 & -0.2588 &  0.0067 \\\hline
  7 &  3.1444 &  1.5738 &  1.5574 & -0.0258 &  0.1597 &  0.0009 & -1.3099 &  1.5913 &  2.0877 & -0.0052 &  0.4419 & -0.0027 \\\hline
  8 & -2.8916 & -0.5562 &  2.2471 & -2.2285 &  3.0843 &  0.5374 & -1.6450  &  1.2560  &  0.5729 & -1.2107 & -1.8533 & -0.3861 \\\hline
  9 & -1.5770  &  4.7101 &  1.5731 &  3.2801 &  0.0088 & -0.9555 & -0.1530  & -1.5376 &  1.9699 &  1.5432 &  2.9588 &  1.2599 \\\hline
 10 & -0.0034 &  0.5683 & -0.3675 &  3.2224 & -0.2362 & -0.0142 & -0.0932 & -2.7019 &  1.7268 & -1.6101 & -0.4026 &  0.0064 \\\hline
 11 &  3.1386 &  1.5736 &  1.5699 & -0.0124 &  0.1559 & -0.0004 & -1.2813 &  1.5986 &  2.0409 & -0.0089 &  0.4543 & -0.0007 \\\hline
\end{tabularx}
\caption{The parameters $\theta_{ij}$ used to implement time evolution for $t=8$ in the free and interacting theories. The $\theta_{ij}$ correspond to the angles defined in the brickwall layer shown in Fig.~\ref{phi4:fig:circuit_elements}d: $i$ labels the layer, and $j$ labels the parameter within layer $i$. These parameters are determined using classical simulations and optimization.}
\label{phi4:tab:time_evolution_angles_8}
\end{table}

\begin{table}
\centering
\setlength{\tabcolsep}{3pt} 
\scriptsize
\begin{tabularx}{\linewidth}{|c||Y|Y|Y|Y|Y|Y||Y|Y|Y|Y|Y|Y|}
\hline
&  \multicolumn{6}{c||}{$\lambda=0$}  &  \multicolumn{6}{c|}{$\lambda=2$} \\\hline
$j$ & $i=0$ & $i=1$ & $i=2$ & $i=3$ & $i=4$ & $i=5$ & $i=0$ & $i=1$ & $i=2$ & $i=3$ & $i=4$ & $i=5$ \\
\hline\hline
  0 & -0.0008 &  0.2800   & -0.2194 &  0.3617 & -0.0039 &  0.1371 &  1.6507 &  0.0406 & -1.6259 &  0.6184 &  1.3403 & -0.0024 \\\hline
  1 & -1.5719 & -4.7144 & -4.7125 &  0.0032 &  0.1182 &  0.0038 &  1.2220  & -1.5253 &  1.5760  & -0.0568 & -4.7190  & -0.4047 \\\hline
  2 &  1.4799 &  2.7295 &  0.8456 & -2.5016 & -1.5347 & -1.5740  & -1.5872 & -0.0099 & -0.6879 &  0.0060  &  0.5950  & -1.6816 \\\hline
  3 &  0.0767 &  1.5663 & -1.5709 & -1.5705 & -3.1967 & -1.3495 & -0.0371 &  0.7480  & -1.5209 & -0.5640  &  1.5786 & -0.0048 \\\hline
  4 & -3.3481 & -1.0828 &  1.7347 & -2.9603 &  0.2967 &  0.5598 & -1.5525 &  1.5285 &  0.6970  & -1.5498 & -1.9947 & -0.3791 \\\hline
  5 & -1.5795 &  4.7124 &  1.5657 &  3.0253 & -0.0025 & -0.8377 & -0.2366 & -1.1993 &  2.1882 &  1.5553 &  3.1469 &  1.2310  \\\hline
  6 & -0.0013 &  0.7131 & -0.3294 &  3.1614 & -0.0275 &  0.0018 & -0.0657 & -2.5103 &  1.6019 & -1.6033 & -0.0107 & 0.0000      \\\hline
  7 &  3.1395 &  1.5733 &  1.5699 &  0.0879 & -0.0030  & -0.0001 & -0.6230  &  1.5356 &  2.0720  &  0.0043 &  0.4026 & -0.0003 \\\hline
  8 & -3.4245 & -0.9444 &  1.8114 & -2.9586 &  0.2792 &  0.5453 & -1.5796 &  1.4988 &  0.6103 & -1.5431 & -1.9636 & -0.3887 \\\hline
  9 & -1.5741 &  4.7115 &  1.5722 &  3.0286 & -0.0033 & -0.7946 & -0.1634 & -1.2214 &  2.1575 &  1.5569 &  3.1081 &  1.2298 \\\hline
 10 & -0.0042 &  0.6035 & -0.4230  &  3.1141 &  0.0180  & -0.0166 &  0.0585 & -2.6936 &  1.5923 & -1.5523 & -0.0084 & -0.0008 \\\hline
 11 &  3.1352 &  1.5725 &  1.5752 &  0.1049 & -0.0049 &  0.0013 & -0.5291 &  1.5413 &  1.9455 &  0.0013 &  0.4044 &  0.0008 \\\hline
\end{tabularx}
\caption{The parameters $\theta_{ij}$ used to implement time evolution for $t=9$ in the free and interacting theories. The $\theta_{ij}$ correspond to the angles defined in the brickwall layer shown in Fig.~\ref{phi4:fig:circuit_elements}d: $i$ labels the layer, and $j$ labels the parameter within layer $i$. These parameters are determined using classical simulations and optimization.}
\label{phi4:tab:time_evolution_angles_9}
\end{table}

\FloatBarrier

\section{Tables of results}
In this appendix, expectation values $\langle\phi_j^2\rangle_\text{2wp}-\langle\phi_j^2\rangle_\text{vac}$ are tabulated for $\lambda=0$ and $\lambda=2$. Error-mitigated results from {\tt ibm\_fez} are compared to classical data from a MPS circuit simulator. Larger uncertainties in the data from the quantum computer can be seen as a result of increasing circuit depth: the circuits for $t=1,2,3$ have a depth of 59 (Tables~\ref{phi4:tab:results_t_1}, ~\ref{phi4:tab:results_t_2}, ~\ref{phi4:tab:results_t_3}), those for $t=4,5,6$ have a depth of 81 (Tables ~\ref{phi4:tab:results_t_4}, ~\ref{phi4:tab:results_t_5}, ~\ref{phi4:tab:results_t_6}), and those for $t=7,8,9$ have a depth of 103 (Tables ~\ref{phi4:tab:results_t_7}, ~\ref{phi4:tab:results_t_8}, ~\ref{phi4:tab:results_t_9}).

\begin{table}
\centering
\scriptsize
\renewcommand{\arraystretch}{0.6}
\begin{tabularx}{\linewidth}{|c||Y|Y||Y|Y|}
\hline
\rule{0pt}{10pt} & \multicolumn{4}{c|}{\large$\langle\phi_j^2\rangle_\text{2wp}-\langle\phi_j^2\rangle_\text{vac}$}\\\hline\hline
\rule{0pt}{8pt} \multirow{2}{*}{Spatial site $j$} &  \multicolumn{2}{c||}{$\lambda=0$}  &  \multicolumn{2}{c|}{$\lambda=2$} \\\cline{2-5}
\rule{0pt}{8pt} & MPS & {\tt ibm\_fez} & MPS & {\tt ibm\_fez} \\
\hline\hline
 0 & 0.000 & 0.014(10)  & 0.000 & 0.018(23)  \\
  1 & 0.000 & 0.017(10)  & 0.000 & -0.008(09) \\
  2 & 0.000 & -0.005(11) & 0.000 & 0.009(10)  \\
  3 & 0.000 & -0.011(10) & 0.000 & 0.015(10)  \\
  4 & 0.000 & -0.015(10) & 0.000 & -0.005(12) \\
\hline
\end{tabularx}
\caption{Numerical values for the expectation value $\langle\phi_j^2\rangle_\text{2wp}-\langle\phi_j^2\rangle_\text{vac}$ at $t=1$ from MPS and {\tt ibm\_fez} for the free and interacting theories, as shown in Figs.~\ref{phi4:fig:scattering_heatmap} and~\ref{phi4:fig:results_by_time}.}
\label{phi4:tab:results_t_1}
\end{table}

\begin{table}[h]
\centering
\scriptsize
\renewcommand{\arraystretch}{0.6}
\begin{tabularx}{\linewidth}{|c||Y|Y||Y|Y|}
\hline
\rule{0pt}{10pt} & \multicolumn{4}{c|}{\large$\langle\phi_j^2\rangle_\text{2wp}-\langle\phi_j^2\rangle_\text{vac}$}\\\hline\hline
\rule{0pt}{8pt} \multirow{2}{*}{Spatial site $j$} &  \multicolumn{2}{c||}{$\lambda=0$}  &  \multicolumn{2}{c|}{$\lambda=2$} \\\cline{2-5}
\rule{0pt}{8pt} & MPS & {\tt ibm\_fez} & MPS & {\tt ibm\_fez} \\
\hline\hline
 0 &  0.000 & 0.007(10)  &  0.000 & 0.003(09)  \\
  1 &  0.000 & 0.013(08)  &  0.000 & -0.004(12) \\
  2 &  0.000 & 0.021(11)  &  0.000 & -0.003(11) \\
  3 &  0.000 & 0.008(08)  &  0.000 & 0.004(09)  \\
  4 &  0.000 & -0.013(11) &  0.000 & 0.007(11)  \\
\hline
\end{tabularx}
\caption{Numerical values for the expectation value $\langle\phi_j^2\rangle_\text{2wp}-\langle\phi_j^2\rangle_\text{vac}$ at $t=2$ from MPS and {\tt ibm\_fez} for the free and interacting theories, as shown in Figs.~\ref{phi4:fig:scattering_heatmap} and ~\ref{phi4:fig:results_by_time}.}
\label{phi4:tab:results_t_2}
\end{table}

\begin{table}[h]
\centering
\scriptsize
\renewcommand{\arraystretch}{0.6}
\begin{tabularx}{\linewidth}{|c||Y|Y||Y|Y|}
\hline
\rule{0pt}{10pt} & \multicolumn{4}{c|}{\large$\langle\phi_j^2\rangle_\text{2wp}-\langle\phi_j^2\rangle_\text{vac}$}\\\hline\hline
\rule{0pt}{8pt} \multirow{2}{*}{Spatial site $j$} &  \multicolumn{2}{c||}{$\lambda=0$}  &  \multicolumn{2}{c|}{$\lambda=2$} \\\cline{2-5}
\rule{0pt}{8pt} & MPS & {\tt ibm\_fez} & MPS & {\tt ibm\_fez} \\
\hline\hline
  0 &  0.000 & -0.013(11) &  0.000 & 0.007(09)  \\
  1 &  0.000 & -0.007(09) &  0.000 & 0.007(09)  \\
  2 &  0.000 & 0.002(11)  &  0.000 & 0.002(08)  \\
  3 &  0.000 & 0.010(08)  &  0.000 & 0.028(12)  \\
  4 &  0.000 & 0.004(08)  &  0.000 & 0.018(07)  \\
\hline
\end{tabularx}
\caption{Numerical values for the expectation value $\langle\phi_j^2\rangle_\text{2wp}-\langle\phi_j^2\rangle_\text{vac}$ at $t=3$ from MPS and {\tt ibm\_fez} for the free and interacting theories, as shown in Figs.~\ref{phi4:fig:scattering_heatmap} and~\ref{phi4:fig:results_by_time}.}
\label{phi4:tab:results_t_3}
\end{table}

\begin{table}[h]
\centering
\scriptsize
\renewcommand{\arraystretch}{0.6}
\begin{tabularx}{\linewidth}{|c||Y|Y||Y|Y|}
\hline
\rule{0pt}{10pt} & \multicolumn{4}{c|}{\large$\langle\phi_j^2\rangle_\text{2wp}-\langle\phi_j^2\rangle_\text{vac}$}\\\hline\hline
\rule{0pt}{8pt} \multirow{2}{*}{Spatial site $j$} &  \multicolumn{2}{c||}{$\lambda=0$}  &  \multicolumn{2}{c|}{$\lambda=2$} \\\cline{2-5}
\rule{0pt}{8pt} & MPS & {\tt ibm\_fez} & MPS & {\tt ibm\_fez} \\
\hline\hline
 0 &  0.000 & 0.023(12)  &  0.000 & 0.008(15)  \\
  1 &  0.000 & 0.012(11)  &  0.000 & 0.001(10)  \\
  2 &  0.000 & 0.004(11)  &  0.000 & 0.014(10)  \\
  3 &  0.000 & 0.005(11)  &  0.000 & 0.019(11)  \\
  4 &  0.000 & -0.020(13) &  0.000 & 0.003(12)  \\
\hline
\end{tabularx}
\caption{Numerical values for the expectation value $\langle\phi_j^2\rangle_\text{2wp}-\langle\phi_j^2\rangle_\text{vac}$ at $t=4$ from MPS and {\tt ibm\_fez} for the free and interacting theories, as shown in Figs.~\ref{phi4:fig:scattering_heatmap} and~\ref{phi4:fig:results_by_time}.}
\label{phi4:tab:results_t_4}
\end{table}

\begin{table}[h]
\centering
\scriptsize
\renewcommand{\arraystretch}{0.6}
\begin{tabularx}{\linewidth}{|c||Y|Y||Y|Y|}
\hline
\rule{0pt}{10pt} & \multicolumn{4}{c|}{\large$\langle\phi_j^2\rangle_\text{2wp}-\langle\phi_j^2\rangle_\text{vac}$}\\\hline\hline
\rule{0pt}{8pt} \multirow{2}{*}{Spatial site $j$} &  \multicolumn{2}{c||}{$\lambda=0$}  &  \multicolumn{2}{c|}{$\lambda=2$} \\\cline{2-5}
\rule{0pt}{8pt} & MPS & {\tt ibm\_fez} & MPS & {\tt ibm\_fez} \\
\hline\hline
 0 &  0.001 & 0.005(10)  &  0.000 & 0.008(13)  \\
  1 &  0.000 & -0.016(15) &  0.000 & -0.011(13) \\
  2 &  0.001 & 0.019(12)  &  0.000 & -0.021(14) \\
  3 &  0.000 & -0.030(14) &  0.000 & -0.010(14) \\
  4 &  0.001 & -0.038(15) &  0.000 & 0.003(11)  \\
\hline
\end{tabularx}
\caption{Numerical values for the expectation value $\langle\phi_j^2\rangle_\text{2wp}-\langle\phi_j^2\rangle_\text{vac}$ at $t=5$ from MPS and {\tt ibm\_fez} for the free and interacting theories, as shown in Figs.~\ref{phi4:fig:scattering_heatmap} and~\ref{phi4:fig:results_by_time}.}
\label{phi4:tab:results_t_5}
\end{table}

\begin{table}[h]
\centering
\scriptsize
\renewcommand{\arraystretch}{0.6}
\begin{tabularx}{\linewidth}{|c||Y|Y||Y|Y|}
\hline
\rule{0pt}{10pt} & \multicolumn{4}{c|}{\large$\langle\phi_j^2\rangle_\text{2wp}-\langle\phi_j^2\rangle_\text{vac}$}\\\hline\hline
\rule{0pt}{8pt} \multirow{2}{*}{Spatial site $j$} &  \multicolumn{2}{c||}{$\lambda=0$}  &  \multicolumn{2}{c|}{$\lambda=2$} \\\cline{2-5}
\rule{0pt}{8pt} & MPS & {\tt ibm\_fez} & MPS & {\tt ibm\_fez} \\
\hline\hline
 0 &  0.001 & 0.022(10)  &  0.000 & 0.025(18)  \\
  1 &  0.001 & -0.010(11) &  0.000 & -0.036(16) \\
  2 &  0.001 & 0.033(10)  &  0.000 & -0.020(17) \\
  3 &  0.001 & 0.010(10)  &  0.000 & -0.021(25) \\
  4 &  0.001 & 0.036(11)  &  0.000 & -0.003(17) \\
\hline
\end{tabularx}
\caption{Numerical values for the expectation value $\langle\phi_j^2\rangle_\text{2wp}-\langle\phi_j^2\rangle_\text{vac}$ at $t=6$ from MPS and {\tt ibm\_fez} for the free and interacting theories, as shown in Figs.~\ref{phi4:fig:scattering_heatmap} and~\ref{phi4:fig:results_by_time}.}
\label{phi4:tab:results_t_6}
\end{table}

\begin{table}[h]
\centering
\scriptsize
\renewcommand{\arraystretch}{0.6}
\begin{tabularx}{\linewidth}{|c||Y|Y||Y|Y|}
\hline
\rule{0pt}{10pt} & \multicolumn{4}{c|}{\large$\langle\phi_j^2\rangle_\text{2wp}-\langle\phi_j^2\rangle_\text{vac}$}\\\hline\hline
\rule{0pt}{8pt} \multirow{2}{*}{Spatial site $j$} &  \multicolumn{2}{c||}{$\lambda=0$}  &  \multicolumn{2}{c|}{$\lambda=2$} \\\cline{2-5}
\rule{0pt}{8pt} & MPS & {\tt ibm\_fez} & MPS & {\tt ibm\_fez} \\
\hline\hline
  0 &  0.000 & 0.023(14)  &  0.000 & -0.003(10) \\
  1 &  0.000 & 0.012(16)  &  0.000 & 0.008(15)  \\
  2 &  0.000 & 0.015(17)  &  0.000 & 0.014(12)  \\
  3 &  0.000 & 0.025(16)  &  0.000 & -0.001(13) \\
  4 &  0.000 & 0.002(16)  &  0.000 & -0.003(11) \\
\hline
\end{tabularx}
\caption{Numerical values for the expectation value $\langle\phi_j^2\rangle_\text{2wp}-\langle\phi_j^2\rangle_\text{vac}$ at $t=7$ from MPS and {\tt ibm\_fez} for the free and interacting theories, as shown in Figs.~\ref{phi4:fig:scattering_heatmap} and~\ref{phi4:fig:results_by_time}.}
\label{phi4:tab:results_t_7}
\end{table}

\begin{table}[h]
\centering
\scriptsize
\renewcommand{\arraystretch}{0.6}
\begin{tabularx}{\linewidth}{|c||Y|Y||Y|Y|}
\hline
\rule{0pt}{10pt} & \multicolumn{4}{c|}{\large$\langle\phi_j^2\rangle_\text{2wp}-\langle\phi_j^2\rangle_\text{vac}$}\\\hline\hline
\rule{0pt}{8pt} \multirow{2}{*}{Spatial site $j$} &  \multicolumn{2}{c||}{$\lambda=0$}  &  \multicolumn{2}{c|}{$\lambda=2$} \\\cline{2-5}
\rule{0pt}{8pt} & MPS & {\tt ibm\_fez} & MPS & {\tt ibm\_fez} \\
\hline\hline
 0 &  0.000 & -0.004(10) &  0.000 & -0.024(16) \\
  1 &  0.000 & 0.011(12)  &  0.000 & -0.014(15) \\
  2 &  0.000 & -0.007(13) &  0.000 & -0.013(10) \\
  3 &  0.000 & 0.012(10)  &  0.000 & -0.028(16) \\
  4 &  0.000 & 0.010(12)  &  0.000 & -0.019(13) \\
\hline
\end{tabularx}
\caption{Numerical values for the expectation value $\langle\phi_j^2\rangle_\text{2wp}-\langle\phi_j^2\rangle_\text{vac}$ at $t=8$ from MPS and {\tt ibm\_fez} for the free and interacting theories, as shown in Figs.~\ref{phi4:fig:scattering_heatmap} and~\ref{phi4:fig:results_by_time}.}
\label{phi4:tab:results_t_8}
\end{table}

\begin{table}[h]
\centering
\scriptsize
\renewcommand{\arraystretch}{0.6}
\begin{tabularx}{\linewidth}{|c||Y|Y||Y|Y|}
\hline
\rule{0pt}{10pt} & \multicolumn{4}{c|}{\large$\langle\phi_j^2\rangle_\text{2wp}-\langle\phi_j^2\rangle_\text{vac}$}\\\hline\hline
\rule{0pt}{8pt} \multirow{2}{*}{Spatial site $j$} &  \multicolumn{2}{c||}{$\lambda=0$}  &  \multicolumn{2}{c|}{$\lambda=2$} \\\cline{2-5}
\rule{0pt}{8pt} & MPS & {\tt ibm\_fez} & MPS & {\tt ibm\_fez} \\
\hline\hline
  0 &  0.000 & 0.007(12)  &  0.000 & -0.017(18) \\
  1 &  0.000 & 0.017(11)  &  0.000 & -0.003(15) \\
  2 &  0.000 & 0.033(11)  &  0.000 & 0.022(15)  \\
  3 &  0.000 & 0.020(14)  &  0.000 & 0.007(13)  \\
  4 &  0.000 & -0.001(13) &  0.000 & 0.036(14)  \\
\hline
\end{tabularx}
\caption{Numerical values for the expectation value $\langle\phi_j^2\rangle_\text{2wp}-\langle\phi_j^2\rangle_\text{vac}$ at $t=9$ from MPS and {\tt ibm\_fez} for the free and interacting theories, as shown in Figs.~\ref{phi4:fig:scattering_heatmap} and~\ref{phi4:fig:results_by_time}.}
\label{phi4:tab:results_t_9}
\end{table}

\end{subappendices}
\chapter{Digital quantum simulations of scattering in quantum field theories using W states}
\label{chap:ising_scattering}

\noindent
{\it This chapter is associated with Ref.~\cite{Farrell:2025nkx}: ``Digital quantum simulations of scattering in quantum field theories using W states'' by Roland C. Farrell, Nikita A. Zemlevskiy, Marc Illa, and John Preskill.}

\section{Introduction}
\label{i_s:sec:scattering_ising_intro}
\noindent
Quantum simulations of QFTs would enable first-principles studies of the behavior of matter present in extreme astrophysical environments, high-energy particle collisions and the early universe~\cite{Bauer:2022hpo,Catterall:2022wjq,Humble:2022klb,Banuls:2019bmf,Bauer:2023qgm,DiMeglio:2023nsa,Bauer:2025nzf}.
This pinnacle of ab initio scientific investigation, along with advances in quantum technologies, has motivated the first simulations of QFTs on quantum devices~\cite{Li:2024lrl,
Chai:2023qpq,
Cochran:2024rwe,
Schuster:2023klj,
Angelides:2023noe,
Guo:2024tnb,
Zhu:2024dvz,
Martinez:2016yna,
Kokail:2018eiw,
Meth:2023wzd,
Atas:2021ext,
ARahman:2021ktn,
Illa:2022jqb,
ARahman:2022tkr,
Atas:2022dqm,
Kavaki:2024ijd,
Than:2024zaj,
Lewis:2025xtu,
Nguyen:2021hyk,
Davoudi:2024wyv,
Mueller:2024mmk,
De:2024smi,
Klco:2018kyo,
Klco:2019evd,
Ciavarella:2021nmj,
Ciavarella:2021lel,
Ciavarella:2023mfc,
Turro:2024pxu,
Farrell:2022wyt,
Farrell:2022vyh,
Alexandrou:2025vaj,
Crippa:2024hso,
Klco:2019xro}.
Notably, quantum simulations of QFTs have been among the first to surpass the capabilities of exact statevector simulations~\cite{Farrell:2023fgd,Farrell:2024fit,Ciavarella:2024lsp,Ciavarella:2024fzw,Gonzalez-Cuadra:2024xul,Gyawali:2024hrz,Yang:2020yer,Hayata:2024fnh}, and are strong candidates for a quantum advantage.

Simulating QFT scattering on a quantum computer requires preparing single-particle wavepackets as the initial state.
Existing preparation methods incur either a classical computing overhead or a circuit depth that scales unfavorably with wavepacket size, limiting the lattice sizes accessible in practice.
This chapter introduces a wavepacket preparation algorithm whose circuit depth is independent of wavepacket size and depends only on the correlation length.
It is then used to perform a quantum simulation of inelastic scattering in which evidence for particle production is observed for the first time.
A key element of this simulation is a new algorithm for preparing the initial state, i.e., single-particle wavepackets. 
The algorithm has two steps:
\begin{itemize}
    \item[1.] Prepare a state $| W(k_0)\rangle$ that establishes the spatial profile, momentum content and quantum numbers of the target wavepacket, but has contributions from multi-particle states.
    \item[2.] Project $|W(k_0)\rangle$ onto single-particle eigenstates by minimizing the energy with symmetry-preserving quantum circuits. 
    The energy minimum corresponds to the target wavepacket.
\end{itemize}

The initial state $|W(k_0)\rangle$ is similar to the W state~\cite{Dur:2000zz} that has been well studied in quantum information science due to its robust entanglement structure and applications to quantum communication, sensing and optimization~\cite{Agrawal2006,Wang2007,Liu2011,Li_2007,Joo_2003,Wang_2020,9259949,Catalano:2024bdh}.
Recent work has shown that W states can be prepared in constant circuit depth using mid-circuit measurement and feedforward (MCM-FF) ~\cite{Piroli:2024ckr,Buhrman:2023rft,Piroli:2021fjn,Yu:2024szp}.
Inspired by techniques from Refs.~\cite{Smith:2022nbd, Cruz_2019}, we improve the W state preparation protocol in Ref.~\cite{Piroli:2024ckr} by removing the need for ancillas. 
To the best of our knowledge, this is the first ancilla-free protocol for preparing W states in constant depth.
A straightforward generalization of this method is used to efficiently initialize $|W(k_0)\rangle$ in the first step of the wavepacket preparation algorithm.

The symmetry-preserving circuits that minimize the energy in step 2 are found using the variational algorithm ADAPT-VQE~\cite{Grimsley:2018wnd} running on classical computers.
In the future, energy-minimizing circuits could be determined using ADAPT-VQE or a different algorithm with provable performance guarantees on a fault-tolerant quantum computer.
Our wavepacket preparation algorithm can be applied to a wide range of lattice models.
Furthermore, the required circuit depth is independent of wavepacket size and only scales with the correlation length.
This is a significant improvement over previous methods, which either incur an exponentially scaling classical computing overhead~\cite{Zemlevskiy:2024vxt,Farrell:2024fit} or require a circuit depth that scales polynomially with the wavepacket size~\cite{Davoudi:2024wyv,Chai:2023qpq,Jordan:2011ci,Hite:2025pvb,Turco:2023rmx,Turco:2025jot}.

We demonstrate the utility of this algorithm by constructing wavepacket preparation circuits in one-dimensional Ising field theory, scalar field theory, the Schwinger model, and in two-dimensional Ising field theory. 
In one-dimensional Ising field theory, the circuits that prepare wavepackets on lattices with 100+ qubits are determined using a Matrix Product State (MPS) simulator.
These wavepacket preparation circuits are then used to initialize a scattering simulation on 104 qubits of IBM's quantum computer {\tt ibm\_marrakesh}.
For current hardware capabilities, we find that linear-depth unitary circuits are more practical for preparing $|W(k_0)\rangle$ than the MCM-FF method.
The application of up to 45 steps of Trotterized time evolution (5,589 two-qubit gates) evolves the system well beyond the collision.
The paths of the particles throughout the scattering process are identified from measurements of the energy density.
The creation of a slow-moving heavy particle during the collision causes the energy density in the post-collision state to be skewed.
The skewness is extracted from our quantum simulations and is evidence for inelastic particle production.

Sections \ref{i_s:sec:WPsummary},~\ref{i_s:sec:qsim} and~\ref{i_s:sec:scattering_ising_discussion} constitute the main text of this chapter and are designed to be read sequentially.
Section~\ref{i_s:sec:WPsummary} provides a high-level overview of the wavepacket preparation algorithm.
Section~\ref{i_s:sec:qsim} presents results from using {\tt ibm\_marrakesh} to simulate inelastic scattering in one-dimensional Ising field theory.
Section~\ref{i_s:sec:scattering_ising_discussion} concludes and discusses prospects for a near-term quantum advantage in simulations of scattering.
The Appendices support the results presented in the main text.
In particular, the constant-depth circuit that prepares $|W(k_0)\rangle$ is given in App.~\ref{i_s:sec:WKprep}.
Circuits that prepare wavepackets in one-dimensional Ising field theory are determined using an exact statevector simulator in App.~\ref{i_s:sec:WPQFTcircs}.
In App.~\ref{i_s:sec:qcirc_scatt}, circuits that prepare wavepackets on large lattices in one-dimensional Ising field theory are constructed using a MPS circuit simulator, and in App.~\ref{i_s:sec:csimscatt} they are used to simulate scattering with MPS.
Details about the experiments performed on {\tt ibm\_marrakesh}, including the error mitigation techniques used, are given in App.~\ref{i_s:sec:qsimDetails}.
Appendix~\ref{i_s:sec:skew} shows how inelastic effects can be detected from the asymmetry in the outgoing energy density and explains our method for quantifying this with skewness.

\section{Overview of the wavepacket preparation algorithm}
\label{i_s:sec:WPsummary}
\noindent
A quantum simulation of particle scattering begins with the preparation of single-particle wavepackets.
Our method prepares wavepackets of the lightest particles with given quantum numbers by minimizing the energy starting from a suitable initial state.
The circuits that minimize the energy could be determined from running variational algorithms on a quantum computer.
For one-dimensional systems these variational algorithms can instead be implemented using a MPS circuit simulator.
Importantly, the energy minimization circuits can be made hardware-efficient by adapting the ansatze to the connectivity and native gate set of the target quantum computer.
Therefore, this method is particularly advantageous for simulations on near-term quantum computers that are limited by circuit depth.
For simplicity, this section focuses on preparing wavepackets in one-dimensional lattice QFTs.

\begin{figure*}
    \centering
     \includegraphics[width=0.85\linewidth]{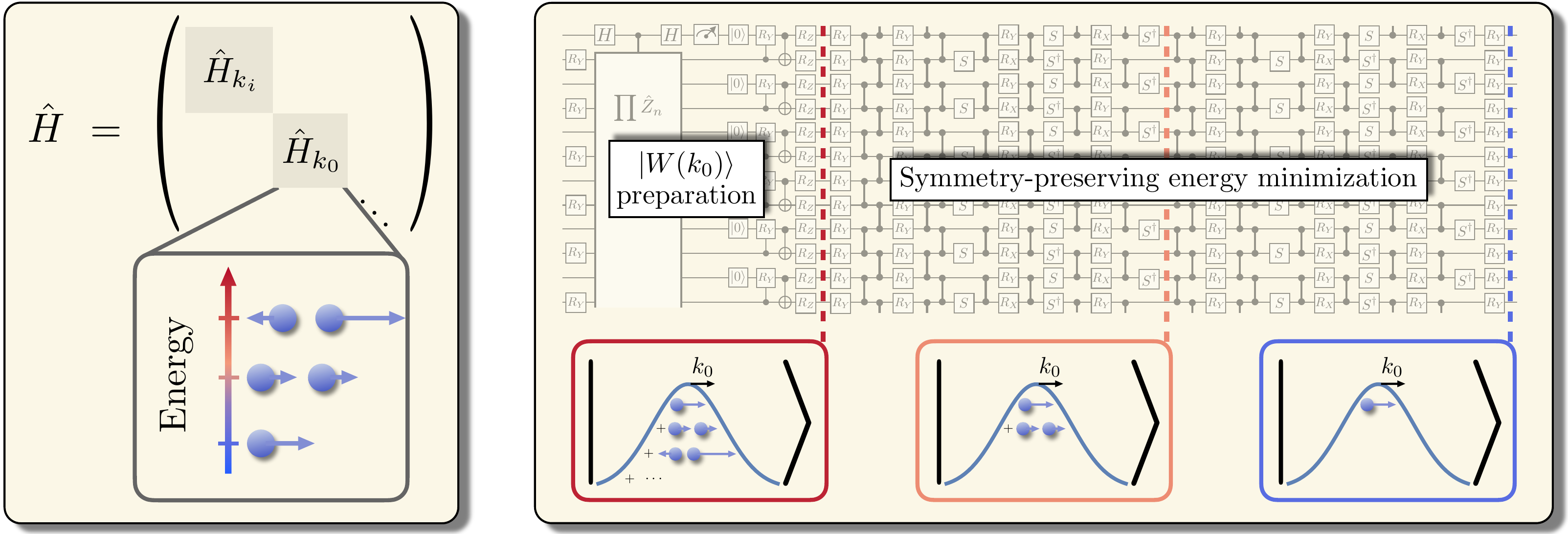}
    \caption{\textit{Wavepacket preparation using W states and symmetry-preserving energy minimization.} Left: due to translational invariance, the Hamiltonian decomposes into blocks $\hat{H}_{k}$, each with definite momentum $k$.
    The lowest-energy state in each block corresponds to the single-particle momentum eigenstate $|\psi_k\rangle$.
    Higher-energy states can be interpreted as multiple particles with total momentum $k$ (or heavier single-particle states not shown).
    Right: quantum circuits that prepare the target wavepacket $|\psi_{\text{wp}}\rangle$.
    First, the state $|W(k_0)\rangle$ is prepared with a constant-depth circuit using MCM-FF (step 1).
    This initial state has the momentum content of the target wavepacket but incorrectly contains higher-energy components.
    Next, the energy is minimized using circuits that are translationally invariant, real, and conserve other system-specific symmetries (step 2).
    These constraints preserve the momentum content of $|W(k_0)\rangle$ while projecting the wavefunction onto the desired single-particle states.
    The energy minimum corresponds to $|\psi_{\text{wp}}\rangle$.}
\label{i_s:fig:WPOverview}
\end{figure*}
Consider a system with a unique ground state that is invariant under time reversal and spatial translations, and has periodic boundary conditions (PBCs).
Its Hamiltonian $\hat{H}$ is block diagonal, with each block labeled by a momentum $k$.
If the Hamiltonian has a mass gap, the ground state and first excited state of the $k=0$ block correspond to the vacuum and the lightest particle at rest, respectively.
The lowest-energy states of the other $k\neq0$ blocks are the lightest single-particle eigenstates with momentum $k$ which we define as $|\psi_k\rangle$.
The block-diagonal structure of the Hamiltonian is illustrated in the left panel of Fig.~\ref{i_s:fig:WPOverview}.
These properties of the Hamiltonian motivate a variational quantum algorithm for preparing single-particle wavepackets. 
 
A wavepacket is a superposition of single-particle eigenstates that is localized around a position $x_0$ and a momentum $k_0$.
A wavepacket with a small spread in momentum space $\sigma_{}$ will have a large spread in position space $\sim\!\sigma_{}^{-1}$, and will propagate for a long time as a localized particle.
The wavepacket wavefunction is
\begin{equation}
\vert \psi_{\text{wp}} \rangle  \ = \ {\cal N}\sum_k e^{-i k x_0}\, e^{-(k_0 - k)^2/(4\sigma_{}^2)} \vert \psi_k \rangle \ ,
\label{i_s:eq:psiWPFull}
\end{equation}
where ${\cal N}$ is a normalization factor.\footnote{A consistent phase convention is necessary when defining the single-particle eigenstates $|\psi_k\rangle$, see App.~\ref{i_s:app:ExactWP}. Note also the difference in normalization to Chapter~\ref{chap:phi4_scattering}.} 
The sum runs over $k \in 2 \pi n/L$, where $L$ is the number of lattice sites and $n$ is an integer such that $k\in (-\pi, \pi]$.
For systems with a dispersion relation that is monotonic in $|k|$, a wavepacket that propagates to the right (left) will have the bulk of its amplitude in eigenstates with $k>0$ ($k<0$). 
As a warm up, first consider the task of preparing $|\psi_k\rangle$.

The state $|\psi_k\rangle$ with $k\neq 0$ can be prepared in two steps.
First, some state $|k\rangle$ that is in a definite momentum block of the Hamiltonian (i.e., an eigenstate under spatial translations, $e^{-i \hat{k} n}|k\rangle = e^{-i k n}|k\rangle$) is initialized.
One simple choice for $|k\rangle$ is the superposition of all states related by translation to $|00...001\rangle$ weighted by the appropriate phase. 
This assumes a basis where $|00...001\rangle$ has the quantum numbers of the target single-particle state.\footnote{For example, in scalar field theory, single-particle states are odd under the $Z_2$ symmetry that takes $\phi\to-\phi$.}
If one lattice site maps to one qubit, this state would be
\begin{align}
|k\rangle \ = \ \frac{1}{\sqrt{L}}\sum_{n=0}^{L-1} e^{i k n}|2^n\rangle \ ,
\label{i_s:eq:psik0}
\end{align}
where the state is specified by its binary value on $L$ bits, e.g., $|2^1\rangle = |00...010\rangle$.
This state is in the correct $k$ block of the Hamiltonian and has the long-range entanglement inherent to momentum eigenstates~\cite{Gioia:2021xtp}, but needs to be rotated to the lowest-energy state.
A circuit that approximately implements this rotation $|\psi_k\rangle \approx \hat{U}(\vec{\theta}_{\star}) |k\rangle$ can be determined by minimizing the energy,
\begin{align}
\hat{U}(\vec{\theta}_{\star}) \ = \ \text{argmin} \langle \psi_{\text{ansatz}} |\hat{U}(\vec{\theta})^{\dagger}\hat{H} \hat{U}(\vec{\theta})|\psi_{\text{ansatz}}\rangle \ ,
\label{i_s:eq:UThetastar}
\end{align}
where $|\psi_{\text{ansatz}}\rangle = |k\rangle$ and
$\hat{U}(\vec{\theta})$ is a translationally invariant unitary operator $e^{i \hat{k} n} \hat{U}(\vec{\theta}) e^{-i \hat{k} n} = \hat{U}(\vec{\theta})$ that depends on variational parameters $\vec{\theta}$.
This circuit must also preserve the other quantum numbers of $|\psi_k\rangle$, and translational invariance ensures that the resulting state remains in the correct $k$ block of the Hamiltonian.
This forms the basis for a variational algorithm, and circuits that prepare $|k\rangle$ and minimize the energy can readily be constructed.

The generalization of this strategy to prepare $|\psi_{\text{wp}}\rangle$ begins with the initialization of some state $|W(k_0)\rangle$, that has the correct amplitude and phase in each $k$ block of the Hamiltonian.
A simple choice is a wavepacket built from $|k\rangle$,
\begin{align}
\vert W(k_0) \rangle  \ &= \ {\cal N}\sum_k e^{-i k x_0}\, e^{-(k_0 - k)^2/(4\sigma_{}^2)} \vert k \rangle \ \nonumber \\ &=\ \sum_n e^{i\phi_n}c_n |2^n\rangle \ , 
\label{i_s:eq:psiWP0}
\end{align}
where the magnitude of $c_n$ follows a Gaussian centered at $n=x_0$.
The special case of a uniform superposition is the W state~\cite{Dur:2000zz}.\footnote{The use of W states for preparing wavepackets in analog quantum simulations was recently proposed in Ref.~\cite{Bennewitz:2024ixi}.}
Using techniques from Refs.~\cite{Cruz_2019,Smith:2022nbd} we improve the W state preparation circuits in Ref.~\cite{Piroli:2024ckr}, and generalize them to prepare $\vert W(k_0) \rangle$.
The circuit is given in App.~\ref{i_s:sec:WKprep} and utilizes MCM-FF to prepare $\vert W(k_0) \rangle$ in two-qubit gate depth 7. 
The depth is independent of lattice geometry, system size and target wavepacket size. 

A simplified description of the circuit is the following:
first, single-qubit rotations produce the state $|000...\rangle  + \sqrt{\delta}|W(k_0)\rangle \ + \ \delta|X\rangle $, where $|X\rangle$ contains states with two or more $1$s in their binary representations, and $\delta$ is an input parameter.
Next, the parity $\prod \hat{Z}_n$ (and only the parity) is measured using the constant-depth circuit from Ref.~\cite{Piroli:2024ckr}. 
Postselecting on odd parity prepares $|W(k_0)\rangle$ with infidelity ${\cal I} = {\cal O}(\delta^2)$ and succeeds with probability $p_{\text{success}} \geq 0.43\delta$ (assuming $\delta\le 1$).
For large wavepackets, the success probability and infidelity are independent of wavepacket size (see App.~\ref{i_s:sec:WKprep}).
The state $|W(k_0)\rangle$ has the momentum content of the target wavepacket but incorrectly contains components with two or more particles, as well as heavier single-particle components.
This is illustrated by the first step in the right panel of Fig.~\ref{i_s:fig:WPOverview}.

After preparing $|W(k_0)\rangle$, the wavefunction in each $k$ block is rotated to the lowest-energy state while preserving its amplitude ($e^{-(k_0 - k)^2/(4\sigma_{}^2)}$) and phase ($e^{-i k x_0}$).
The circuit that approximately performs this rotation $|\psi_{\text{wp}}\rangle \approx \hat{U}(\vec{\theta}_{\star}) |W(k_0)\rangle$ is again determined from minimizing the energy in Eq.~\eqref{i_s:eq:UThetastar}.
Now, $|\psi_{\text{ansatz}}\rangle = |W(k_0)\rangle$, and $\hat{U}(\vec{\theta})$ is translationally invariant, real, and conserves the other quantum numbers of $|W(k_0)\rangle$. 
Translational invariance preserves the amplitude of the initial state in each $k$ block and reality preserves the phase.
At a practical level, $\hat{U}(\vec{\theta})$ can be constructed from translationally invariant and imaginary operators $\hat{O}_j$ as $\hat{U}(\vec{\theta})= \prod_j e^{i \theta_j \hat{O}_j}$.
Each of the terms $e^{i \theta_j \hat{O}_j}$ is then converted to a quantum circuit via Trotterization.
Operators $\hat{O}_j$ generated from the Lie algebra of the Hamiltonian are particularly effective at minimizing the energy and guide the circuit design in this work.
The energy minimization effectively projects out all unwanted multi-particle and heavy single-particle components of the $|W(k_0)\rangle$ wavefunction, leaving only the target single-particle wavepacket $|\psi_{\text{wp}}\rangle$.
This is illustrated by the second step in the right panel of Fig.~\ref{i_s:fig:WPOverview}.

This wavepacket preparation method can be applied to a wide range of lattice models in any finite spatial dimension.
The required circuit depth is independent of wavepacket volume, provided that the qubit connectivity matches that of the target lattice.
Appendix~\ref{i_s:app:WP_efficient} discusses the limitations and possible extensions of the wavepacket preparation method.
In the next section, this algorithm is used to initialize simulations of scattering in one-dimensional Ising field theory on IBM's quantum computers.
In App.~\ref{i_s:app:moreCircuits}, it is applied to the preparation of wavepackets in one-dimensional scalar field theory, the Schwinger model, and two-dimensional Ising field theory.

\section{Quantum simulations of scattering in one-dimensional Ising field theory}
\label{i_s:sec:qsim}
\noindent
The physics of the Ising model illustrates the complexity of quantum phenomena that can emerge from simple interactions.
This has made it a target for quantum simulations, with recent demonstrations of
many-body localization~\cite{Shtanko:2023tjn}, discrete time crystals~\cite{Shinjo:2024vci}, Majorana edge modes~\cite{Mi:2022egw} and string breaking~\cite{De:2024smi}. 
Our work focuses on the field theory that emerges when the Ising model Hamiltonian,
\begin{align}
\hat{H}  \ &=\  -  \sum_{n=0}^{L-1}\left [ \frac{1}{2}\left (\hat{Z}_{n-1}\hat{Z}_{n}+\hat{Z}_n\hat{Z}_{n+1}\right ) +  g_x \hat{X}_n + g_z\hat{Z}_n \right ] \nonumber \\
&\equiv \ \sum_{n=0}^{L-1}\hat{H}_n \ , \label{i_s:eq:HIFT}
\end{align}
is tuned to criticality: $g_x\to1,\,g_z\to0$ and $L\to \infty$.
The energy density $\hat{H}_n$ will be an important observable and is defined in the second line.
The physics at the critical point is described by the conformal field theory (CFT) of a free, massless Majorana fermion.
A family of Ising QFTs is reached by tuning to criticality, but keeping the scaling invariant ratio
\begin{equation}
\eta_{\text{latt}} \ = \ \frac{g_x-1}{\vert g_z\vert^{\frac{D-\Delta_{\epsilon}}{D-\Delta_\sigma}}} \ = \ \frac{g_x-1}{\vert g_z\vert^{8/15}}
\end{equation}
held fixed.
The second equality uses the spacetime dimension $D=2$ and scaling dimensions of the relevant CFT deformations $\Delta_{\epsilon} = 1$ and $\Delta_{\sigma} = 1/8$.
For generic values of $\eta_{\text{latt}}$, the field theory describes massive interacting particles and is non-integrable.
The lattice spacing is set to 1 throughout this work and all lengths are expressed in lattice units. 
The Hamiltonian $\hat H$ is dimensionless; hence time is also dimensionless, and the operator $e^{-i\hat H t}$ describes evolution for dimensionless time $t$.
More information on Ising field theory can be found in Refs.~\cite{Zamolodchikov:1989hfa,Zamolodchikov:1989fp,Jha:2024jan}.

Dynamical simulations of scattering in Ising field theory were recently performed using MPS methods in Ref.~\cite{Jha:2024jan}.
By changing the energy of the initial state, clear distinctions between elastic scattering, scattering near a resonance, and particle production via inelastic scattering were observed.
This work showcased the wealth of information that can be accessed with real-time simulations of collisions.

Our quantum simulations use $g_x=1.25$ and $g_z=0.15$ where there are two stable particles: one light particle $|1\rangle$ with mass $m_1=1.59$ and one heavy particle $|2\rangle$ with mass $m_2=2.98$.
The initial state for our scattering simulations consists of two wavepackets of $|1\rangle$ particles that are separated in space.
As time goes on, they travel toward each other, collide, and separate. 
For low center-of-mass energies $E_{\text{tot}}$, only the elastic process $11\to 11$ is kinematically allowed. 
The lowest-energy inelastic process is $11\to 12$ where a single heavy $|2\rangle$ particle is produced during the collision. The observation of this process is the goal of our quantum simulations.
This inelastic channel opens up for energies above the particle production threshold, $E_{\text{tot}}>E_{\text{thr}}=m_1+m_2 = 4.57$, and can be detected from additional outgoing tracks in the energy density.
In the kinematic regime illustrated by the right panel, three different scattering outcomes occur in superposition: elastic scattering, inelastic scattering with the heavier particle moving right, and inelastic scattering with the heavier particle moving left. Hence four outgoing tracks are visible, even though there are only two outgoing particles for each outcome. 

For energies $E_{\text{tot}}\gg E_{\text{thr}}$ there are additional inelastic processes that produce many particles.
Such processes generate significant entanglement and are challenging to simulate with MPS methods.
Simulations of these ultra-high-energy collisions are also beyond the capabilities of current quantum hardware.
At the energies accessed in our quantum simulations all inelastic channels besides $11\to12$ can be neglected.\footnote{The quantum simulations in this work are at $E_{\text{tot}}=3.46m_1$; hence the three-body channel $11\to 111$ is kinematically allowed, but has a much smaller branching ratio than the $11\to 12$ channel~\cite{Jha:2024jan}.} 
See App.~\ref{i_s:app:kinematics} for details.

\subsection{Inelastic scattering on IBM's quantum computers}
\begin{figure}
    \centering
    \includegraphics[width=0.5\linewidth]{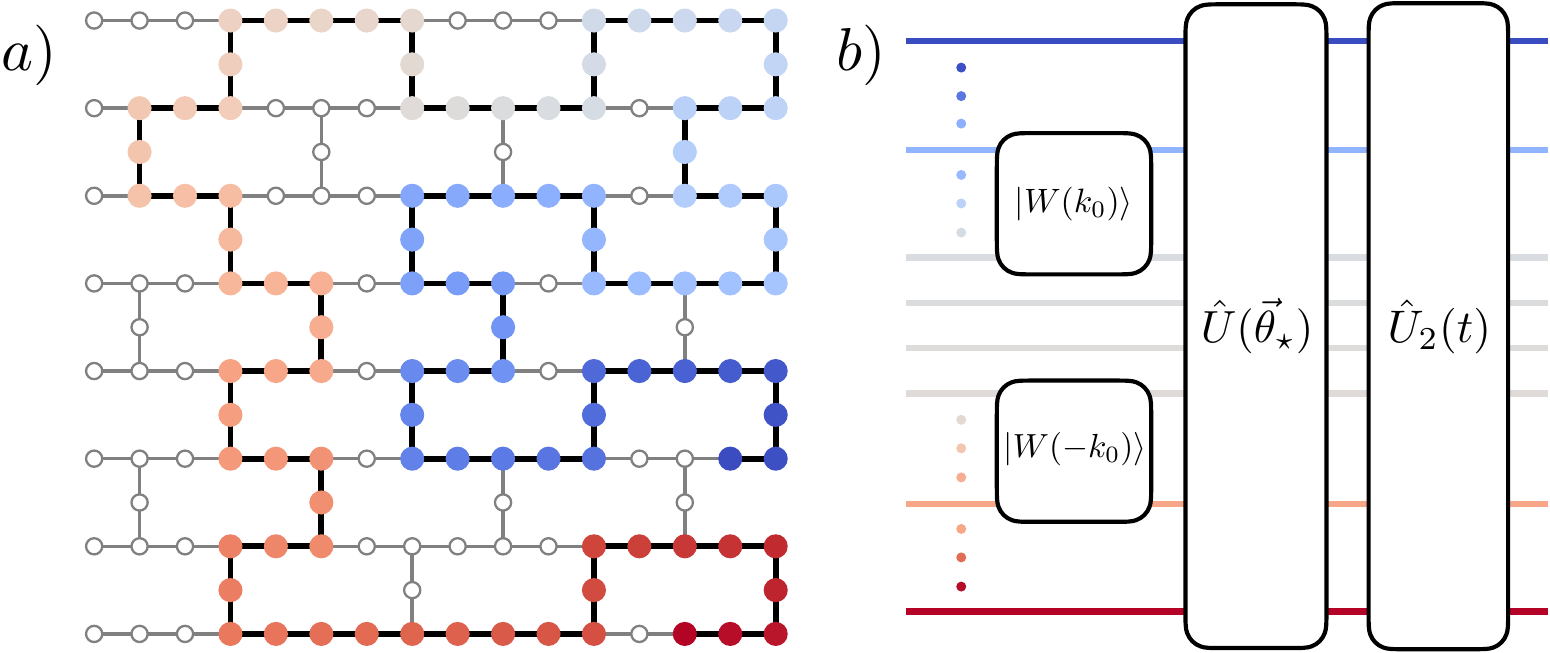}
    \caption{a) The $L=104$ qubit layout used on {\tt ibm\_marrakesh}. 
    b) The structure of the quantum circuits used to simulate scattering in Ising field theory. The colors indicate the qubits used in the lattice-to-device mapping.
    The energy minimization circuit $\hat{U}(\vec{\theta}_*)$ creates wavepackets when acting on $|W(\pm k_0)\rangle$.
    The circuit $\hat{U}_2(t)$ implements time evolution with second-order Trotterization.}
\label{i_s:fig:marrakesh_layout}
\end{figure}

\noindent 
In this section, inelastic scattering is simulated on $L=104$ qubits of {\tt ibm\_marrakesh} using the lattice-to-qubit mapping shown in Fig.~\ref{i_s:fig:marrakesh_layout}a).
A lattice with open boundary conditions (OBCs) is chosen to avoid noisy gates.
The boundary has exponentially small effects on wavepackets created in the bulk, but causes the wavefunction to deviate from the ground state near the lattice edges.
See App.~\ref{i_s:sec:qsimDetails} for a discussion, along with the modifications to the simulation protocols that are necessary to accommodate OBCs.
The first step of the wavepacket preparation algorithm outlined in Sec.~\ref{i_s:sec:WPsummary} is the initialization of $|W(k_0)\rangle$.
Appendix~\ref{i_s:sec:WKprep} shows that this can be achieved with a constant-depth circuit using MCM-FF.
This approach is currently not optimal on IBM's {\tt heron r2} quantum computers,\footnote{This is in part due to readout being $\sim\!3\times$ noisier and $\sim \!38\times$ slower than two-qubit gates (2584ns vs 68ns), and also the current incompatibility of MCM-FF with dynamical decoupling and native $R_{ZZ}(\theta)$-gates. Improvements are expected in future versions of the hardware and software stack.}
and instead $|W(k_0)\rangle$ is initialized with the unitary circuit in the left panel of Fig.~\ref{i_s:fig:IsingWPCircs}.
Wavepackets are then prepared by applying symmetry-preserving circuits $\hat{U}(\vec{\theta}_{\star})$ that minimize the energy.
These circuits are determined using ADAPT-VQE running on the MPS circuit simulator Quimb~\cite{gray2018quimb}.
Information on wavepacket preparation is given in App.~\ref{i_s:sec:WPQFTcircs} and App.~\ref{i_s:sec:qcirc_scatt}.
The structure of the circuit that prepares two wavepackets is shown in Fig.~\ref{i_s:fig:marrakesh_layout}b).
The initial state is chosen so the wavepackets have support on $d\gtrsim21$ sites, and the total energy is $E_{\text{tot}}/E_{\text{thr}}=1.2$. These parameters give the strongest signal of inelastic scattering in the least circuit depth (see App.~\ref{i_s:sec:qsimDetails}).
\begin{figure*}
    \centering
    \includegraphics[width=\linewidth]{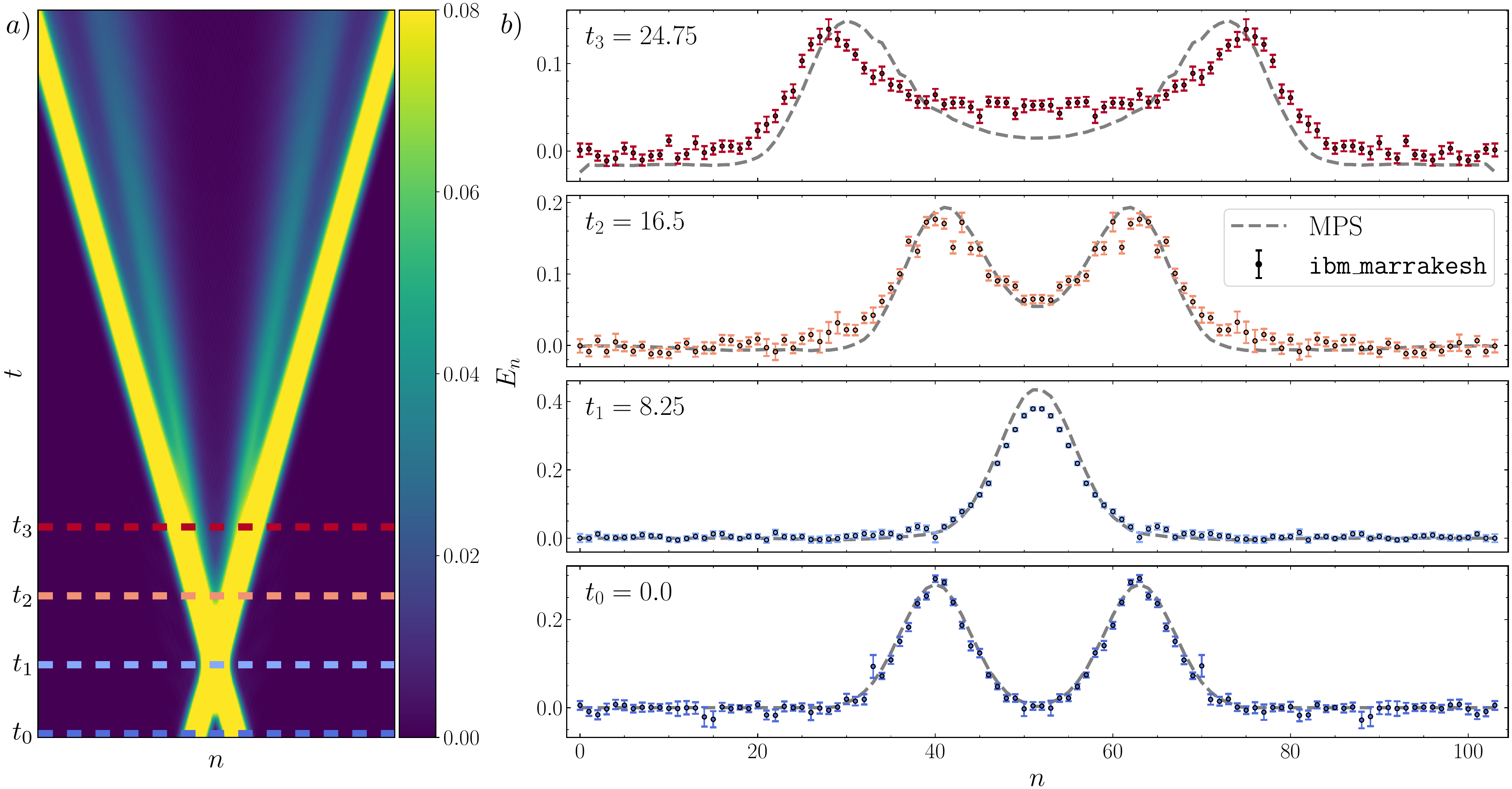}
    \caption{\textit{Simulations of inelastic particle production in one-dimensional Ising field theory.} a) The energy density $E_n$ throughout the scattering process obtained with a MPS circuit simulator on a $L=256$ lattice. 
    b) Results from simulations of scattering using $L=104$ qubits of {\tt ibm\_marrakesh} for a selection of times depicted in a).
    The wavepackets are initialized at $t_0=0$ and collide around $t_1=8.25$.
    Outgoing particles begin to form at $t_2=16.5$ and $t_3=24.75$. The asymmetry of the energy density in each ``bump'' signals the formation of the $|2\rangle$ particle produced in the inelastic process $11\to12$.
    The y-axis range at each time is different to clearly show the features of the energy density.
    The initial wavepacket parameters are $k_0=0.32\pi$ and $\sigma_{}=0.13$, and each wavepacket is supported on $d \gtrsim 21$ sites.
    A Trotter step size of $\delta t=1/16$ is used to evolve the system in a) and $\delta t=0.55$ is used in b).}
\label{i_s:fig:ibm_results}
\end{figure*}
Once established, the wavepackets are time evolved with a second-order Trotterized unitary $\hat{U}_2(t)$.
A Trotter step size of $\delta t=0.55$ is chosen to balance the Trotter error and circuit depth (see App.~\ref{i_s:app:systematics}).
For each simulation time, the vacuum-subtracted energy density
\begin{align}
E_n \ = \ \langle\psi_{\text{2wp}}|\hat{H}_n(t) |\psi_{\text{2wp}}\rangle \ - \ \langle\psi_{\text{vac}}|\hat{H}_n(t) |\psi_{\text{vac}}\rangle 
\label{i_s:eq:vacsubEn}
\end{align}
is measured. 
Here, $\hat{H}_n(t) = \hat{U}_2(t)^{\dagger}\hat{H}_n \hat{U}_2(t)$ with $\hat{H}_n$ defined in Eq.~\eqref{i_s:eq:HIFT}, $|\psi_{\text{2wp}}\rangle$ is the prepared two-wavepacket state and $|\psi_{\text{vac}}\rangle$ is the prepared vacuum.
The circuits that prepare $|\psi_{\text{2wp}}\rangle$ and $|\psi_{\text{vac}}\rangle$ have very similar structure and are explained in App.~\ref{i_s:sec:qsimDetails}.

MPS calculations of the energy density throughout the scattering process are shown in Fig.~\ref{i_s:fig:ibm_results}a).\footnote{MPS simulations are performed using CuPy \cite{cupy} for GPU acceleration.} 
Two pairs of particle tracks emerge from the collision region.
The majority of the energy is in the outer pair of tracks, which correspond to the trajectories of the light $|1\rangle$ particles.
The inner pair of tracks has a lower velocity compared to the elastic trajectories, and represents the heavy $|2\rangle$ particles produced in the inelastic process $11\to12$.
Schematically, this process is
\begin{align}
|1(k_0) \,  1(-k_0)\rangle \ \to \ |1(k_0')\,  2(-k_0')\rangle\ + \ 1\leftrightarrow2 \ ,
\label{i_s:eq:inelastic_scattering_wavefunction}
\end{align}
where $k_0'$ is the momentum of outgoing particles.
The kinematics are such that the trajectories of the outgoing $|1\rangle$ particles overlap with the elastic tracks and cannot be distinguished in Fig.~\ref{i_s:fig:ibm_results}a) (see App.~\ref{i_s:sec:csimscatt} for details).
It is shown in App.~\ref{i_s:app:kinematics} that the identification of the outgoing tracks with $|1\rangle$ and $|2\rangle$ particles is consistent with their velocities predicted from the single-particle dispersion relations.
We emphasize that the elastic and inelastic processes occur in superposition, and there are no components of the wavefunction with four particles.

The energy density measured on {\tt ibm\_marrakesh} is shown for a selection of simulation times in Fig.~\ref{i_s:fig:ibm_results}b).
Two distinct wavepackets initialized at $t_0=0$ propagate toward each other and collide around $t_1=8.25$.
Later simulation times probe the formation and evolution of the post-collision state.
At $t_2=16.5$ the outgoing wavepackets begin to separate, and by $t_3=24.75$ asymptotic particles begin to form.
Up to $t_2=16.5$, there is good agreement between the energy density determined from {\tt ibm\_marrakesh} (data) and MPS (gray dashed line).
Significant systematic errors develop in the quantum data at $t_3=24.75$ due to the larger circuit volume.
For $t>t_3$, the asymptotic $|2\rangle$ particle produced in the $11\to12$ process becomes identifiable as a separate ``bump'' in the energy density that propagates at a lower velocity.
Unfortunately, the cumulative effects of noise made simulations beyond $t=24.75$ impossible on {\tt ibm\_marrakesh}.

\begin{figure}
    \centering
    \includegraphics[width=0.5\linewidth]{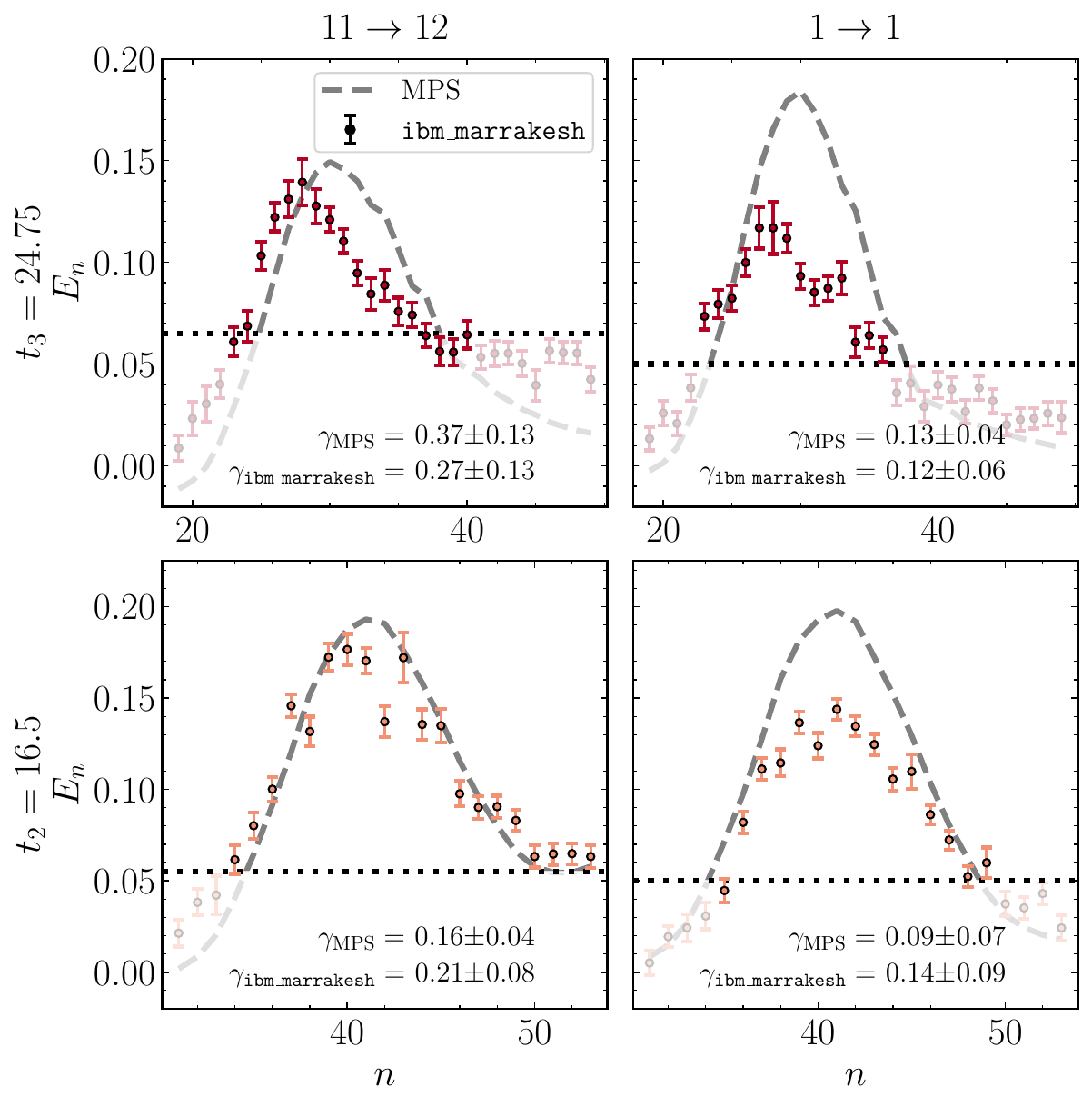}
    \caption{\textit{Asymmetry in the post-collision and single-particle energy densities.} 
    The regions of positive energy density in the left half of the lattice in the $11\to12$ scattering process (left column) and in the $1\to1$ process of single particle propagation (right column).
    Results at times $t_2$ and $t_3$ obtained from {\tt ibm\_marrakesh} and MPS are shown.
    The skewness of the energy density, $\gamma$, is computed by considering points in an interval determined by an energy cutoff (black dotted line).
    The uncertainty in $\gamma$ comes from varying the energy cutoff as described in App.~\ref{i_s:sec:skew}, and from statistical error.
    The skewness is increased in the $11\to12$ energy density compared to $1\to1$ due to the inelastic production of the heavy $|2\rangle$ particle.
    The simulation parameters are the same as in Fig.~\ref{i_s:fig:ibm_results}b).}
\label{i_s:fig:1wp_vs_2wp}
\end{figure}

Despite the device noise, we observe evidence for the inelastic production of the heavy $|2\rangle$ particle.
Long after the collision, the presence of a heavy particle could be confirmed by projecting onto exclusive scattering channels.
Before this, the energy density of the outgoing particles on each side of the collision becomes skewed toward the center of the lattice.
This is due to the presence of the $|2\rangle$ particle in the wavefunction that travels slower, and is absent for scattering below inelastic threshold, as shown in App.~\ref{i_s:sec:skew}.
We compute the skewness, $\gamma$, from the third moment of the energy density of each region of outgoing particles as described in App.~\ref{i_s:sec:skew}.
The left column of Fig.~\ref{i_s:fig:1wp_vs_2wp} shows the energy density of the bumps after the collision, and gives $\gamma$ obtained from both MPS and {\tt ibm\_marrakesh}.
The quantum results of the post-collision state at $t_2$ are right-skewed ($\gamma>0$) and in agreement with MPS.
The right-skewness increases at $t_3$ due to the emergence of the heavy $|2\rangle$ particle.
Some of this skewness is due to systematic errors in the quantum data that increase $E_n$ between the outgoing particles.
To isolate skewness due to inelastic effects from systematic errors, additional simulations of single wavepacket propagation are performed.
The energy density of a single wavepacket at times $t_2$ and $t_3$ is shown in the right column of Fig.~\ref{i_s:fig:1wp_vs_2wp}.
The MPS simulations of single wavepacket propagation still exhibit positive skewness due to boundary effects as discussed in App.~\ref{i_s:sec:skew}.
The skewness of the post-collision state is $0.6\sigma$ larger than the single-particle state at $t_2$, and $1\sigma$ larger at $t_3$.
This provides evidence that the post-collision skewness is, at least partially, due to inelastic particle production.

The quantum resources used for these simulations are detailed in App.~\ref{i_s:sec:qsimDetails}.
The latest simulation time corresponds to $n_T=45$ Trotter steps and requires 5,589 two-qubit gates with a two-qubit gate depth of 130. 
With such a large quantum volume, effective error mitigation is crucial for recovering reliable results.
A full description of our error mitigation strategy is provided in App.~\ref{i_s:sec:qsimDetails}, with certain aspects highlighted here.
Pauli twirling~\cite{Wallman:2015uzh} shapes the noise into a stochastic Pauli channel.
For each simulation time, circuits that evolve both $|\psi_{\text{2wp}}\rangle$ and $|\psi_{\text{vac}}\rangle $ are run.
The time evolution of the vacuum is used to learn how the Pauli noise channel affects local observables.
Noise-free expectation values in the scattering simulations are then estimated using the learned noise model.
This method, known as Operator Decoherence Renormalization (ODR)~\cite{Farrell:2023fgd}, is an extension of Refs.~\cite{Urbanek:2021oej,ARahman:2022tkr}.

Two unforeseen features of the device noise were encountered during the simulations.
First, the Pauli noise channel is very asymmetric, and $\langle\hat{Z}_n\rangle$ measurements are effectively $\sim\!3\times$ noisier than $\langle\hat{X}_n\rangle$ and $\langle\hat{Z}_n\hat{Z}_{n+1}\rangle$ measurements (see App.~\ref{i_s:sec:qsimDetails} for details on device noise characteristics).
The source of this asymmetry is a mystery, but could be due to incomplete twirling of the (non-Clifford) $R_{ZZ}$ gate.
Additionally, we observed that one faulty single- or two-qubit gate could degrade the whole simulation.
This was not noticed in previous simulations of wavepacket dynamics on IBM's quantum computers performed by the authors that utilized similar circuit depths~\cite{Zemlevskiy:2024vxt,Farrell:2024fit}.
We attribute this increased sensitivity to the larger simulation light cone of $c t_{\text{max}}\sim\!40$ qubits reached in this work, compared to $c t_{\text{max}}\sim\! 8$ qubits in our previous works.
This effectively allows the noise to contaminate a larger region of the device in the same circuit depth.
To reduce these effects, IBM's calibration data was used to carefully avoid high-error gates in the lattice-to-qubit mapping.
The simulation light cone also explains the increase in systematic errors at $t_3$ compared to $t_2$, despite the circuit depth only increasing by 30. 
This is because the simulation light cone at $t_3$ encapsulates approximately 2.5$\times$ more two-qubit gates than at $t_2$.
These observations highlight that the effects of circuit noise on the results of a quantum simulation depend heavily on the physical process being simulated.

\clearpage

\begin{subappendices}

\section{Discussion}
\label{i_s:sec:scattering_ising_discussion}
\noindent
The quantum simulations in this work are the first to probe the nonequilibrium QFT dynamics generated in the wake of particle collisions.
This is a significant improvement over previous quantum simulations of scattering that were constrained to early times and low energies.
Advancing beyond this regime required sophisticated strategies for managing device errors that are amplified by a large simulation light cone $ct_\text{max}$.
Algorithmic errors and quantum resource requirements must be balanced in an optimal simulation that operates within the correct regime of length scales.
Our simulations of Ising field theory used: system size $L=104$, particle propagation distance $vt_{\text{max}}\!\sim40$, wavepacket size $d\!\sim\!21$ and correlation length $\xi\propto1/m\sim0.6$ (where $m$ is the mass of the lightest particle).
Even with these minimal parameters, observing evidence of inelastic particle production required the full capability of state-of-the-art quantum hardware.

Our simulations are far from the continuum limit of $\xi\to\infty$ (in lattice units),
and are not designed to make precise predictions about the underlying QFT.
Instead, our work demonstrates that simulations far from the continuum can still be informative and capture features of nonequilibrium processes in QFTs.
The use of quantum computers to explore quantum many-body dynamics, without the goal of making precise quantitative predictions, aligns with the capabilities expected in the pre-fault-tolerant era. 
Precision simulations of scattering are possible in one-dimensional QFTs using MPS techniques~\cite{Pichler:2015yqa,Van_Damme_2021,Rigobello:2021fxw,Vovrosh:2022bpj,Belyansky:2023rgh,Papaefstathiou:2024zsu,Milsted:2020jmf,Su:2024uuc,Jha:2024jan,Barata:2025hgx}.
For example, the Ising field theory simulations in Ref.~\cite{Jha:2024jan} utilized an $L,\,vt_{\text{max}},\,d,\,\xi$, and inverse time step $1/\delta t$ that were about $10\times$ larger than are used in this work.
The smaller systematic errors in their simulations allowed individual inelastic channels to be isolated and enabled robust predictions of the underlying QFT.
Future quantum simulations, likely incorporating some form of error correction, may be able to access ultra-high-energy scattering where
the production of highly entangled, many-particle states renders MPS methods unreliable. 

The wavepacket preparation algorithm  developed in this work is particularly advantageous in higher dimensions.
By utilizing MCM-FF, wavepackets in any finite spatial dimension can be prepared with a circuit depth that is independent of the wavepacket size, provided that the qubits share the connectivity of the target lattice.
In contrast, existing methods have a circuit depth that scales polynomially in the spatial {\it volume} of the wavepacket.
This scaling is often worse if all-to-all connectivity is not assumed. 
The speedup from using MCM-FF arises because some long-range correlations in quantum states can be built 
with constant-depth circuits that are supplemented with local operations and classical communication.
Efficient algorithms allocate the minimal number of two-qubit gates needed to create the requisite entanglement, which is then distributed throughout the system with MCM-FF.
The key insight is that the long-range entanglement inherent to momentum plane waves~\cite{Gioia:2021xtp} is also present in W states~\cite{Gioia:2023adm}.
Therefore, wavepackets can be built from W states by only modifying correlations localized over $\sim\! \xi$ sites. 
Finding a basis where this observation is useful was guided by the symmetries and hierarchies in length scales characteristic of the target system.

The efficient preparation of wavepackets is a prerequisite for achieving a quantum advantage in simulations of scattering. 
An advantage could be realized in two dimensions, where no dynamical simulations of QFT scattering have ever been performed.
The underlying assumption of classical hardness is supported by empirical evidence that simulations of dynamics in two dimensions are challenging using classical computers~\cite{Haghshenas:2025euj,Park:2025vyf,Dziarmaga:2022via,PRXQuantum.6.020302}.
Such simulations would unlock the {\it ab initio} study of many phenomena absent in one-dimensional systems.
For example, in two dimensions, spatial rotations on the circle allow single particles to have spin, pairs of particles to have anyonic statistics, and multiple particles to have relative orbital angular momentum.
Dynamical studies of these effects in QFTs are now possible~\cite{Iqbal:2023wvm,Iqbal:2024drh,Xu:2024guv,Minev:2024pmj,Andersen:2022xmz} due to the emergence of flagship quantum computers with native two-dimensional connectivity~\cite{GoogleQuantumAIandCollaborators:2024efv,Gao:2024fik,DeCross:2024tmi,Chen:2023erd,Rodriguez:2024bhh,Muniz:2024dna}.
These advancements have the potential to open a new frontier of scientific exploration that is powered by quantum computing.

\section{A constant-depth circuit for initializing \texorpdfstring{$|W(k_0)\rangle$}{}}
\label{i_s:sec:WKprep}
\noindent
The efficient preparation of the state in Eq.~\eqref{i_s:eq:psiWP0} on $d$ sites,
\begin{equation}
\vert W(k_0) \rangle  \ =\ \sum_{n=0}^{d-1} e^{i\phi_n}c_n |2^n\rangle ,
\label{i_s:eq:psiWP02}
\end{equation}
is an important primitive for the preparation of wavepackets.
For $\phi_n = 0$ and $c_n=1/\sqrt{d}$  this reduces to the W state.
\begin{figure}
    \centering
    \includegraphics[width=\linewidth]{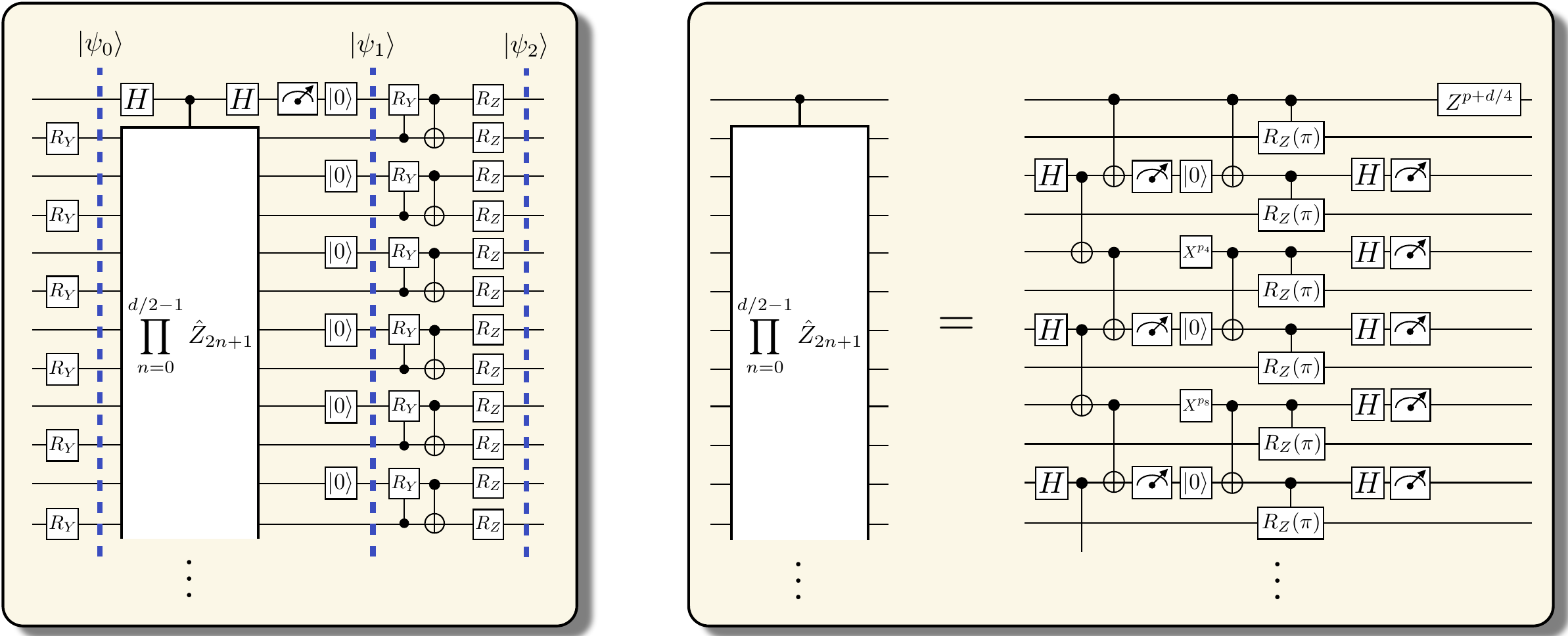}
    \caption{\textit{Circuits that prepare $|W(k_0)\rangle$ in constant depth using MCM-FF.} Left: a circuit that prepares $|W(k_0)\rangle$ in Eq.~\eqref{i_s:eq:psiWP02}.
    Right: the circuit from Ref.~\cite{Piroli:2024ckr} that implements the global controlled unitary in constant depth.}
    \label{i_s:fig:ConstantDepth}
\end{figure}
The authors of Ref.~\cite{Piroli:2024ckr} presented an ingenious circuit for preparing the W state in constant depth that requires $d$ ancilla qubits.
Their method begins by initializing $|\psi_0\rangle = (\sqrt{1-\delta/d}|0\rangle +\sqrt{\delta/d}|1\rangle )^{\otimes d}$ on the system qubits. Then the parity $\prod \hat{Z}_n$ is measured using a constant-depth circuit.
Odd parity occurs with $p_{\text{success}}\geq0.43\delta$ and prepares the W state with infidelity ${\cal I} ={\cal O}(\delta^2)$.
We generalize their method to prepare $|W(k_0)\rangle$ for arbitrary $c_n$ and $\phi_n$. 
Additionally, we remove the ancillas and provide exact expressions for $p_{\text{success}}$ and ${\cal I}$.

The circuit that prepares $|W(k_0)\rangle$ is shown in Fig.~\ref{i_s:fig:ConstantDepth}.
The states $|\psi_0\rangle,\,|\psi_1\rangle,\,|\psi_2\rangle$ are labeled at intermediate stages of the circuit and are described here.
First, $q_0$ is prepared in $|+\rangle$ and
\begin{align}
|\psi_0\rangle \ = \ \bigotimes_{n=0}^{d/2} \left [\cos{\left ( \frac{\theta_{2n+1}}{2}\right ) }|0\rangle \ + \ \sin{\left ( \frac{\theta_{2n+1}}{2}\right ) }|1\rangle \right ] \ , \ \ \sin{\left ( \frac{\theta_{2n+1}}{2}\right ) } \ = \ \sqrt\delta \sqrt{c_{2n}^2 + c_{2n+1}^2} \ ,
\label{i_s:eq:psi0}
\end{align}
is prepared on the odd-numbered qubits by applying single-qubit $R_Y$ rotations, 
\begin{align}
    \prod_n e^{-i \theta_{2n+1} \hat{Y}_{2n+1} /2} \ .
\end{align}
Then, the parity is kicked back to qubit $q_0$ by applying $\prod_{n=0}^{d-1}\hat{Z}_{2n+1}$ controlled on $q_0$.
This controlled-parity operation is implemented with the constant-depth circuit developed in Ref.~\cite{Piroli:2024ckr}, and shown in the right panel of Fig.~\ref{i_s:fig:ConstantDepth}.
Several details about this circuit are described here. The second layer of CNOTs acts on pairs of qubits $q_{n}q_{n+2}$ with $n=0,4,\ldots,d-6$.
The $p_n$ in the first round of MCM-FF are the sums of the outcomes on all measured qubits $q_m$ with $m<n$ and
the $p$ in the second round of MCM-FF is the sum of all measurement outcomes. 
After applying the controlled $\prod_{n=0}^{d-1}\hat{Z}_{2n+1}$ and a Hadamard on qubit $q_0$, the state of $q_0$ is correlated with the parity, i.e., $|\psi\rangle = |\text{even parity}\rangle|0\rangle + |\text{odd parity}\rangle|1\rangle$.

Next, $q_0$ is measured and outcomes are postselected on odd parity, which occurs with success probability
\begin{align}
p_{\text{success}}  &= \frac{1}{2} - \frac{1}{2}\prod_{n=0}^{d/2 - 1}\left [1-2\delta(c_{2n}^2+c_{2n+1}^2)\right ] \nonumber \\
&= \delta - \delta^2\left[1 - \sum_{n=0}^{d/2-1}(c_{2n}^2 + c_{2n+1}^2)^2\right] \nonumber\\
&\quad + \frac{2}{3}\delta^3\left[1 - 3\sum_{n=0}^{d/2-1}(c_{2n}^2 + c_{2n+1}^2)^2 + 2\sum_{n=0}^{d/2-1}(c_{2n}^2 + c_{2n+1}^2)^3\right] + {\cal O}(\delta^4) \nonumber\\
&\geq \frac{1}{2}\left (1-e^{-2\delta} \right ) \ \geq \ 0.43 \delta \ .
\label{i_s:eq:psuccess}
\end{align}
This expression has been simplified by using the normalization condition $\sum_n c_n^2 = 1$.
The lower bound of  $p_{\text{success}}$ is derived in App.~\ref{i_s:app:psuccess} for $\delta \leq 1$.
The even-numbered qubits are reset to $|0\rangle$ and the state after a successful measurement outcome is
\begin{align}
|\psi_1\rangle \ = \ \sum_{n=0}^{d/2-1}  \sqrt{c_{2n}^2+c_{2n+1}^2}\,  |2^{2n+1}\rangle  \ + \ {\cal O}(\delta)\ .
\end{align}
For a given pair of neighboring qubits, this state has concentrated the probability of finding either in $|1\rangle$ to the odd-numbered qubits.
Inspired by Ref.~\cite{Smith:2022nbd}, the final step spreads this probability to the even sites, previously used as ancillas.
This is done using the controlled-$R_Y$ CNOT sequence from Ref.~\cite{Cruz_2019}, with rotation angles
\begin{align}
\tan{\left (\frac{\theta_{2n}}{2} \right )} \ = \ \frac{c_{2n}}{c_{2n+1}} \ .
\end{align}
Lastly, the phases in $|W(k_0)\rangle$ are added with single-qubit $R_Z$ rotations, $\prod_n e^{-i \phi_n \hat{Z}_n /2}$.
The final state $|\psi_2\rangle$ is an approximation to $| W(k_0)\rangle$ with infidelity
\begin{align}
{\cal I} \ &= \ 1 \ - \ |\langle W(k_0)|\psi_2\rangle|^2 \ = \ 1 \ - \ \frac{1}{p_{\text{success}}} \sum_{n=0}^{d/2 - 1}\delta(c_{2n}^2 + c_{2n+1}^2)\prod_{\ell\neq n}\left[1 - \delta (c_{2\ell}^2 + c_{2\ell+1}^2)\right] \nonumber \\
&= \ \frac{\delta^2}{6}\left [1 -3\sum_{n=0}^{d/2-1}(c^2_{2n} + c_{2n+1}^2)^2 +2\sum_{n=0}^{d/2-1}(c^2_{2n} + c_{2n+1}^2)^3 \right ] \ + \ {\cal O}(\delta^3)  \nonumber \\
&\leq \  \frac{\delta^2}{6}\frac{(d-2)(d-4)}{d^2} \ + \ {\cal O}(\delta^3) \ . 
\label{i_s:eq:infidelity_full}
\end{align}
The upper bound on ${\cal I}$ is derived in App.~\ref{i_s:app:psuccess}.
Note that expanding ${\cal I}$ to the first nonzero order required $p_{\text{success}}$ in Eq.~\eqref{i_s:eq:psuccess} to third order.
Both $p_{\text{success}}$ and ${\cal I}$ are independent of $d$ for large $d$, and their worst case bounds are saturated for W states.

\section{Quantum circuits for preparing wavepackets in one-dimensional Ising field theory }
\label{i_s:sec:WPQFTcircs}
\noindent
Throughout this work the qubits are labeled $|q_{L-1}\dots q_2 q_1 q_0\rangle$ (little-endian notation) and the top wire in every circuit corresponds to $q_0$.
The system size is $L$ and all systems have PBCs (with $q_L=q_0$) unless otherwise stated.
The wavepacket size is $d$, the mass of the lightest particle is $m$, $|\psi_{\text{wp}}\rangle$ is the exact wavepacket, $|\psi_{\text{ansatz}}\rangle$ is the approximately prepared wavepacket and $|\psi_{\text{vac}}\rangle$ is the vacuum.

The algorithm outlined in Sec.~\ref{i_s:sec:WPsummary} can prepare wavepackets in a wide range of lattice QFTs.
This section provides the circuits for wavepackets in one-dimensional Ising field theory, with additional applications to $\lambda \hat{\phi}^4$ scalar field theory and the Schwinger model, as well as two-dimensional Ising field theory, given in App.~\ref{i_s:app:moreCircuits}.
Circuits that prepare wavepackets on system sizes with $\leq 28$ qubits are found using the {\tt qiskit} statevector simulator~\cite{Javadi-Abhari:2024kbf}, and the prepared state is benchmarked against the exact wavepacket determined from exact diagonalization.
Our exact diagonalization methods are explained in App.~\ref{i_s:app:ExactWP}.
In App.~\ref{i_s:sec:csimscatt}, the circuits that prepare wavepackets in one-dimensional Ising field theory on a $L=256$ lattice are determined using a MPS circuit simulator.

\begin{figure}
    \centering
    \includegraphics[width=\linewidth]{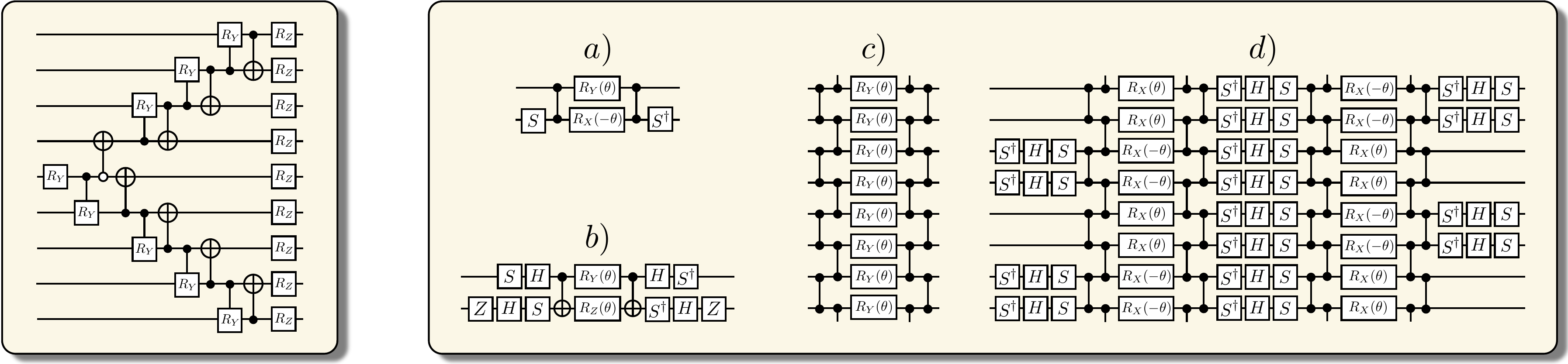}
    \caption{\textit{Circuits used to prepare wavepackets in one-dimensional Ising field theory. } Left: a unitary circuit that prepares $|W(k_0)\rangle$ in Eq.~\eqref{i_s:eq:psiWP02} across 9 sites.
    The rotation angles are given in Eq.~\eqref{i_s:eq:WP0angles}.
    Right: circuit elements used in ADAPT-VQE to prepare wavepackets in Ising field theory. They implement the unitary evolution of the operators in Eq.~\eqref{i_s:eq:opPool}.
    a) implements $\exp{-i\frac{\theta}{2}\left (\hat{Y}\hat{Z}+\hat{Z}\hat{Y} \right )}$ and 
    b) implements $\exp{-i\frac{\theta}{2}\left (\hat{Y}\hat{X}  + \hat{X}\hat{Y} \right )}$.
    Examples of circuits that implement  $\exp{-i\frac{\theta}{2}\sum \hat{Z}\hat{Y}\hat{Z}}$ and $\exp{-i\frac{\theta}{2}\sum \left (\hat{Z}\hat{X}\hat{Y} + \hat{Y}\hat{X}\hat{Z}\right )}$ across 8 qubits with PBCs are shown in c) and d), respectively.}
    \label{i_s:fig:IsingWPCircs}
\end{figure}

MCM-FF is inefficient on classical circuit simulators because each measurement outcome creates a new branch in the circuit that must be computed separately. 
In the worst case, this causes the simulation cost to scale exponentially with the number of measurements. 
Therefore, all simulations that use classical computing in this work initialize $|W(k_0)\rangle$ with the unitary circuit in the left panel of Fig.~\ref{i_s:fig:IsingWPCircs}.\footnote{The wavepackets prepared in this section using statevector simulators have the wavepacket size $d$ equal to the system size $L$.}
Additional circuits for preparing $|W(k_0)\rangle$ that reduce the circuit depth by utilizing beyond-linear connectivity and/or MCM-FF are given in App.~\ref{i_s:app:WP0prep}.
The approximate wavepacket is then prepared by variationally minimizing the energy using circuits that are translationally invariant, real and preserve the other system-specific symmetries.
We perform this optimization using ADAPT-VQE, which provides an efficient framework for symmetry-aware optimization. 
ADAPT-VQE is a greedy algorithm that builds the structure of the ansatz circuits layer by layer.
At each step in the algorithm, the circuit layer that is most effective at minimizing the energy is identified from a pool of symmetry-preserving parameterized circuits $\{ \hat{U}(\theta) \}$.
This circuit layer is appended to the ansatz circuit and the values of the variational parameters $\vec{\theta}$ are optimized to minimize the energy.
The key ingredient in ADAPT-VQE is the choice of circuit pool, or equivalently an operator pool $\{ \hat{O}\}$ from which the circuits are generated, i.e., $\hat{U}_j(\theta) = e^{i \theta_j \hat{O}_j}$.
Guided by symmetries, an operator pool inspired by the Lie algebra of the Hamiltonian will be used throughout this work.
Operator pools inspired by the Hamiltonian algebra have previously been used to great success in Refs.~\cite{Farrell:2023fgd,Farrell:2024fit,Zemlevskiy:2024vxt,Gustafson:2024bww,Ciavarella:2024lsp,VanDyke:2022ffj,Farrell:2024mgu}.
A full description of ADAPT-VQE is given in App.~\ref{i_s:app:57Adapt}.

In one-dimensional Ising field theory, one lattice site maps to one qubit and the initial state for preparing wavepackets is the $|W(k_0)\rangle$ in Eq.~\eqref{i_s:eq:psiWP02}.
A unitary circuit that prepares $|W(k_0)\rangle$ is shown in the left panel of Fig.~\ref{i_s:fig:IsingWPCircs}.
This circuit is most efficient for odd wavepacket sizes $d$. The $R_Y$ angles are given by recursively solving the equations,
\begin{align}
\left [\sin\left (\frac{\theta_{\eta}}{2}\right )\right ]^2 \ &= \ \sum_{i=\eta}^{d-1} \, c_i^2  \ , \nonumber \\[4pt]
\cos{\left (\frac{\theta_{\eta+j+1}}{2}\right )}\prod_{i=0}^{j} \sin{\left (\frac{\theta_{\eta+i} }{2}\right )} \ &= \ c_{\eta+j} \ \ , \ \ j\in[0,1,\ldots,\eta-1] \ , \nonumber \\[4pt] 
\cos{\left (\frac{\theta_{\eta}}{2}\right )}\cos{\left (\frac{\theta_{\eta-j-1}}{2}\right )}\prod_{i=2}^{j} \sin{\left (\frac{\theta_{\eta-i} }{2}\right )} \ &= \ c_{\eta-j} \ \ , \ \ j\in[1,2,\ldots,\eta-1] \ ,
\label{i_s:eq:WP0angles}
\end{align}
where $\eta\equiv(d-1)/2$. The $R_Z$ rotation angles are the $\phi_n$ from Eq.~\eqref{i_s:eq:psiWP02}.
The circuit depth scales linearly with wavepacket size,
\begin{equation}
\text{CNOT depth: } \ 2\lfloor d/2 \rfloor+2 \ ,
\end{equation}
and is likely optimal assuming linear connectivity and no MCM-FF.

The next step in the wavepacket preparation algorithm is to apply
translationally invariant and real circuits that minimize the energy.
This is achieved by using ADAPT-VQE to optimize
parameterized circuits $\{e^{i \theta \hat{O}}\}$ constructed from a symmetry-preserving pool of operators $\{\hat{O}\}_{1d \ \text{Ising}}$. 
Enforcing reality and hermiticity constrains all pool operators to be imaginary and anti-symmetric, and the union of all unique operators generated from $i\sum_n[\hat{H}, \hat{X}_n ]$ and $i\sum_n [ \hat{H}, [ \hat{H}, [\hat{H}, \hat{X}_n ] ] ]$ is found to be effective,
\begin{align}
\{ {\hat O}\}_{1d \ \text{Ising}} \  = \  \sum_{n=0}^{L-1} \bigg \{& \hat{Y}_n 
 ,   \hat{Z}_n\hat{Y}_{n+1}\hat{Z}_{n+2} , \left (\hat{Y}_n\hat{Z}_{n+1}+ \hat{Z}_n\hat{Y}_{n+1}\right )  ,   \left (\hat{Y}_n\hat{X}_{n+1}+ \hat{X}_n\hat{Y}_{n+1}\right )  ,\nonumber \\ 
 & \left (\hat{Z}_n\hat{X}_{n+1}\hat{Y}_{n+2}+ \hat{Y}_n\hat{X}_{n+1}\hat{Z}_{n+2}\right )  \bigg \} \ .
\label{i_s:eq:opPool}
\end{align}
Circuits implementing unitary evolution generated by the operators in the pool are given in the right panel of Fig.~\ref{i_s:fig:IsingWPCircs}.
The construction of the circuits is based on the methods in Ref.~\cite{Chernyshev:2025jyw}, and details can be found in App.~\ref{i_s:app:qcircs}.
The CNOT depth corresponding to the unitary evolution with respect to each pool operator in Eq.~\eqref{i_s:eq:opPool} is $\{0,4,4,4,8\}$

\begin{figure}
    \centering
    \includegraphics[width=\linewidth]{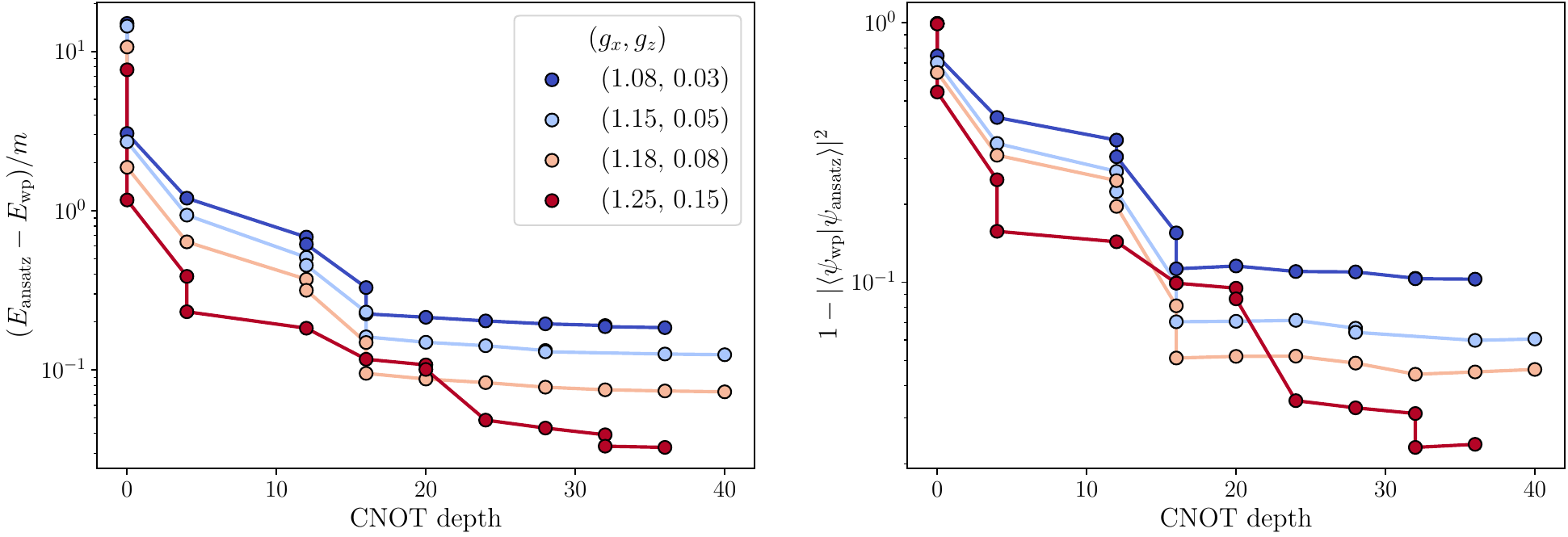}
    \caption{\textit{Quality of prepared wavepackets in one-dimensional Ising field theory.} 
    The deviation in energy (left) and infidelity (right) of the prepared wavepacket as a function of CNOT depth.
    The CNOT depth is monotonically related to the step of ADAPT-VQE, and corresponds to the additional depth after $|W(k_0)\rangle$ preparation.
    Results are shown for wavepacket parameters $\sigma_{}=0.13,\,k_0=0.36\pi$ and $L=28$ for various couplings $g_x,\,g_z$.}
    \label{i_s:fig:IsingADADPT_results}
\end{figure}

To assess the quality of the ADAPT-VQE-prepared wavepacket $|\psi_{\text{ansatz}}\rangle$, we compare it to the exact wavepacket  $|\psi_{\text{wp}}\rangle$ by computing the deviation of the energy $(E_{\text{wp}}-E_{\text{ansatz}})/m$ in units of the mass gap $m$, and the infidelity $1-|\langle\psi_{\text{wp}}|\psi_{\text{ansatz}}\rangle|^2$.
Figure~\ref{i_s:fig:IsingADADPT_results} shows these two quantities as functions of CNOT depth, obtained from up to 12 steps of ADAPT-VQE for $L=28$ and $\sigma_{}=0.13,\,k_0=0.36\pi$.
Results are shown for four sets of couplings that approach the field-theory limit of vanishing mass gap $\{(g_x,g_z,m)\}=\{(1.25,0.15,1.6),(1.18,0.08,1.1),(1.15,0.05,0.8),(1.08,0.03,0.7)\}$.
The circuits are optimized to minimize the energy (left plot), which improves the overlap with the target wavepacket (right plot).
An infidelity $<0.12$ is reached by depth 16 for all couplings considered.

A plateau in the convergence is observed around the 6th step (depth 16) of ADAPT-VQE for all couplings except $g_x=1.25,g_z=0.15$.
In App.~\ref{i_s:app:57Adapt} it is shown that these plateaus can be overcome by expanding the operator pool to include terms in the Lie algebra of the Hamiltonian that correspond to 5th- and 7th-order commutators.
In general, the convergence to the target wavepacket degrades with decreasing mass gap (increasing correlation length).
This is expected as deeper circuits are generically needed to build out longer correlations.
Recently, it has been demonstrated that variational quantum algorithms that utilize MCM-FF have a more favorable optimization landscape~\cite{Deshpande:2024kpt}, and are able to more efficiently prepare states with long-range correlations~\cite{Alam:2024mit,Niu:2024oxx,Yan:2024xev}.
This is an exciting direction where further improvements are expected.
The sequence of operators and variational parameters corresponding to the 8th step of ADAPT-VQE in Fig.~\ref{i_s:fig:IsingADADPT_results} are given in App.~\ref{i_s:app:ADAPTparam}.

\section{Quantum circuits for simulating scattering in one-dimensional Ising field theory}
\label{i_s:sec:qcirc_scatt}
\noindent
To reduce finite-size effects, the system size $L$ must be much larger than the spatial extent of the wavepackets $\sigma_{}^{-1}$.
Because of this, it is pragmatic to set amplitudes in the $|W(k_0)\rangle$ wavefunction below some threshold to zero (and then normalize the wavefunction).
This defines a wavepacket size $d$ that depends on $\sigma_{}$ but is independent of $L$.
The initial state is of the form
\begin{align}
|\psi_{\text{ansatz}}\rangle \ \sim \ |0\rangle^{\otimes (L-d)/2} \ \otimes \ |W(k_0)\rangle \ \otimes \ |0\rangle^{\otimes (L-d)/2} \ ,
\label{i_s:eq:psiinit}
\end{align}
and has errors that can be exponentially suppressed by increasing $d$. 
A wavepacket size of $d=22$ is shown to have negligible truncation errors for $\sigma_{}=0.13$ (see App.~\ref{i_s:app:systematics}), and will be used throughout this section. 
Roughly speaking, the ADAPT-VQE circuit $\hat{U}(\vec{\theta}_\star)$ that minimizes the energy acts as
\begin{align}
\hat{U}(\vec{\theta}_\star)\left (|0\rangle^{\otimes (L-d)/2}  \otimes |W(k_0)\rangle \otimes |0\rangle^{\otimes (L-d)/2}\right ) \ \rightarrow \ \left (|\psi_{\text{vac}}\rangle  \otimes |\psi_{\text{wp}}\rangle \otimes |\psi_{\text{vac}}\rangle\right ) \ .
\label{i_s:eq:psiinit2}
\end{align}
This expression is only schematic, as the wavefunction on the RHS does not factorize between vacuum and wavepacket components.
However, sufficiently far from the region where $|W(k_0)\rangle$ was constructed, local observables are indistinguishable from their vacuum expectation value.
This is due to the exponential decay of correlations inherent to one-dimensional gapped systems~\cite{Hastings:2005pr}. 
As a result, the vacuum can be approximately prepared from
\begin{align}
\hat{U}(\vec{\theta}_\star) |0\rangle^{\otimes L}  \ \approx \ |\psi_{\text{vac}}\rangle \ .
\label{i_s:eq:ADAPTWPvac}
\end{align}
In App.~\ref{i_s:app:57Adapt} it is shown that this approximately prepared vacuum is of even higher quality than the corresponding wavepacket.
All results presented have the energy density of this approximate vacuum subtracted, as defined in Eq.~\eqref{i_s:eq:vacsubEn}. 
This includes the results from scattering simulations, which have the energy density of the time-evolved approximate vacuum subtracted. 

\begin{figure}[t]
    \centering
    \includegraphics[width=0.75\linewidth]{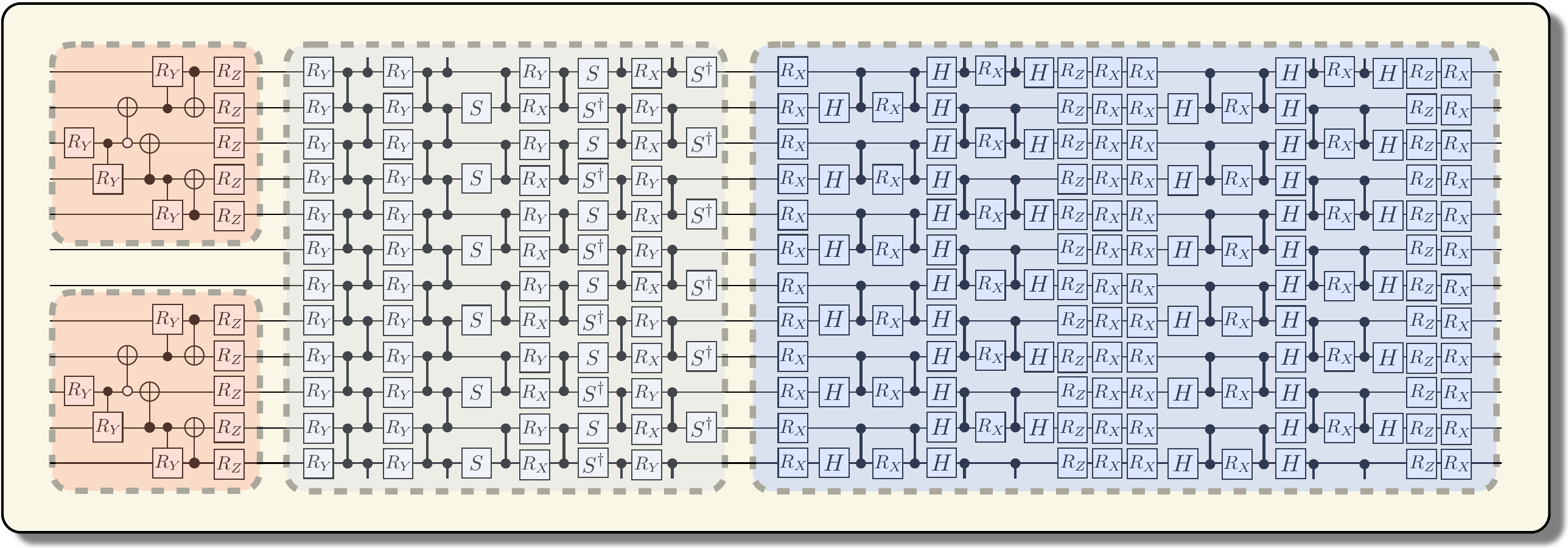}
    \caption{\textit{Circuits for simulating scattering in one-dimensional Ising field theory.}
    A circuit that simulates the scattering of two $d=5$ wavepackets on a $L=12$ lattice.
    The pieces of the circuit highlighted in light red prepare $|W(k_0)\rangle$ and $|W(-k_0)\rangle$, the layer in gray represents the symmetry-preserving energy minimization circuit $\hat{U}(\vec{\theta}_*)$ and the layer in blue is two steps of second-order Trotterized time evolution $\hat{U}_2(t)$.}
    \label{i_s:fig:scatcirc}
\end{figure}

The structure of the circuit that is used to simulate scattering in one-dimensional Ising field theory is shown in Fig.~\ref{i_s:fig:scatcirc}.
First, two wavepackets are prepared with opposite momenta.
This is done by initializing $|W(k_0)\rangle$ and $|W(-k_0)\rangle$ at the desired locations (light red), and then acting with the circuit that minimizes the energy, $\hat{U}(\vec{\theta}_\star)$ (gray).
Note that the $\hat{U}(\vec{\theta}_\star)$ for preparing  $|\psi_{\text{ansatz}}(k_0)\rangle$ works equally well for preparing $|\psi_{\text{ansatz}}(-k_0)\rangle$ since 
\begin{equation}
| \psi_{{\rm ansatz}}(-k_0)\rangle \ = \ \left (\hat{U}(\vec{\theta}_\star)|W(k_0)\rangle\right )^* \ = \ \hat{U}(\vec{\theta}_\star)|W(-k_0)\rangle \ . 
\end{equation}
The first equality follows from time-reversal symmetry, and the second equality follows from using real unitaries in ADAPT-VQE. Lastly, steps of second-order Trotterized time evolution are applied (blue).
%

\section{MPS simulations of scattering in one-dimensional Ising field theory}
\label{i_s:sec:csimscatt}
\noindent
In this section, the circuit elements in Fig.~\ref{i_s:fig:scatcirc} are used to simulate scattering in one-dimensional Ising field theory with MPS.
The simulation parameters are: $L=256$,\, $\sigma_{}=0.13$,\, $d=22$, and $g_x=1.25,\,g_z=0.15$, with corresponding mass gap $m=1.59$ determined from finite-size extrapolations (see App.~\ref{i_s:app:kinematics}). 
The circuits that minimize energy and prepare wavepackets are determined by implementing ADAPT-VQE on the Quimb MPS circuit simulator~\cite{gray2018quimb}.\footnote{It is important to set optimization bounds on the variational parameters so that the optimizer does not explore highly entangled states that are not well represented by MPS.}

\begin{figure}[t]
    \centering
    \includegraphics[width=0.5\linewidth]{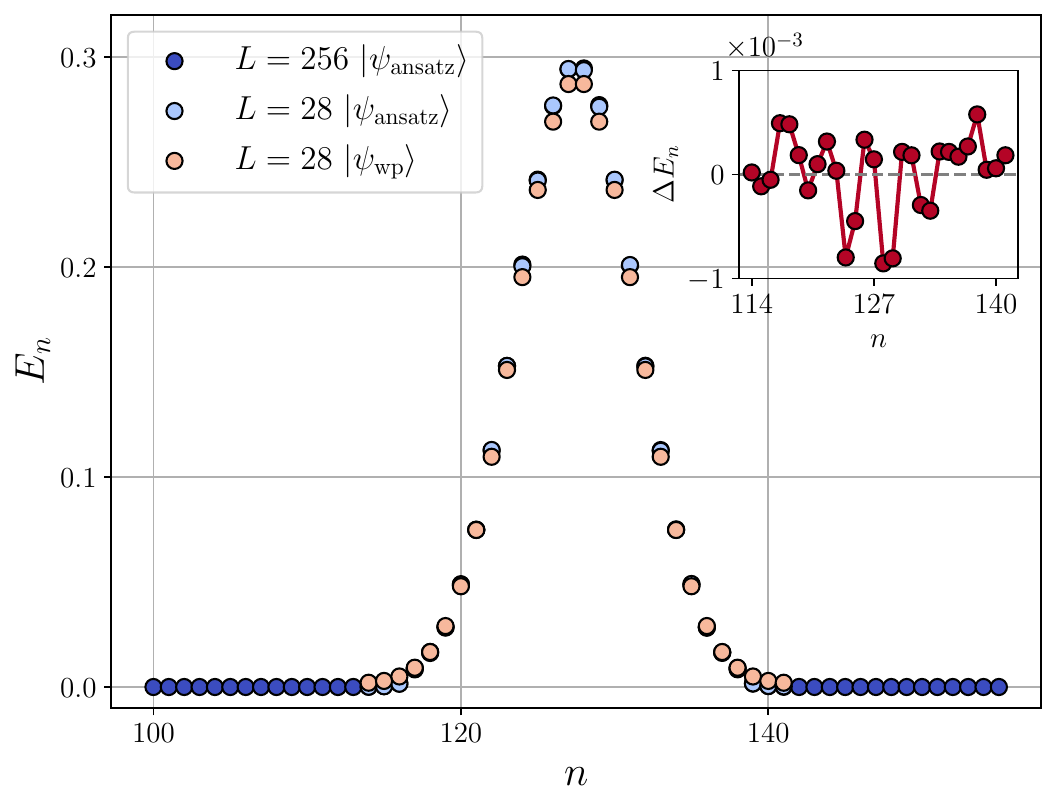}
    \caption{
    \textit{Comparison of wavepackets prepared on large and small lattices in one-dimensional Ising field theory.} 
    The vacuum-subtracted energy density $E_n$ of a wavepacket with $g_x=1.25$,\, $g_z=0.15$,\, $k_0=0.36\pi$,\, $\sigma_{}=0.13$ and $d=22$.
    The approximate wavepackets $|\psi_{\text{ansatz}}\rangle$ are prepared from 8 steps of ADAPT-VQE on a $L=256$ lattice (dark blue) and a $L=28$ lattice (light blue).
    The energy density of the exact $L=28$ wavepacket is also shown (tan).
    The inset shows the difference in the energy density $\Delta E_n$ between the approximate $L=28$ and $L=256$ wavepackets.
    Only the energy density of the 56 sites in the center of the lattice is shown, and the $L=28$ wavepackets are translated to align with the $L=256$ wavepacket.}
    \label{i_s:fig:L108L28WP}
\end{figure}
In App.~\ref{i_s:sec:WPQFTcircs} the prepared wavepackets were benchmarked against the exact wavepacket for $L=28$.
Since comparison with the exact wavepacket on a $L=256$ lattice is not possible, the energy densities of wavepackets on $L=256$ and $L=28$ lattices are compared instead.
Differences in the energy density of the exact wavepackets are expected to be small due to the exponential convergence of the single-particle spectrum with increasing system size (see App.~\ref{i_s:app:kinematics}).
The energy density of wavepackets with $k_0=0.36\pi$ prepared from 8 steps of ADAPT-VQE, 
as well as the energy density of the exact $L=28$ wavepacket are shown in Fig.~\ref{i_s:fig:L108L28WP}.\footnote{Due to the separation of length scales, $L\gg d$, special care must be taken with the energy minimization in ADAPT-VQE.
When $L\gg d$, the majority of $|\psi_{\text{ansatz}}\rangle$ in Eq.~\eqref{i_s:eq:psiinit} is $|00 \ldots 0\rangle$, 
and ADAPT-VQE prioritizes circuits that locally prepare the highest-quality vacuum.
We address this problem by instead minimizing the energy density summed across an interval that is a few correlation lengths larger than $d$.
This instance of ADAPT-VQE minimizes the energy density over $28=d+6 \approx d+10m^{-1}$ sites, and slight variations in the size of the interval have negligible effects.}
The inset shows that the differences in the energy density between the two prepared wavepackets are $\Delta E_n<10^{-3}$.
This is strong evidence that the $L=256$ and $L=28$ wavepackets are of similar quality.
Additionally, the energy density of the approximate vacuum on $L=256$ sites, prepared as in  Eq.~\eqref{i_s:eq:ADAPTWPvac}, agrees to 0.01\% with the vacuum on $L=28$.
The prepared wavepackets at other momenta exhibit similar behavior.
Further verification of the ADAPT-VQE circuits obtained with MPS comes from comparing the variational parameters determined for $L=28$ and $L=256$, which agree to 4 decimal places.
The variational parameters and operators that prepare all of the wavepackets used in this section are provided in App.~\ref{i_s:app:ADAPTparam}.

A goal of our simulations is to observe a clear distinction between scattering at energies above and below inelastic threshold.
There are two stable particles for $g_x=1.25,\,g_z=0.15$: $|1\rangle$ with mass $m_1=m = 1.59$ and $|2\rangle$ with mass $m_2=2.98$.\footnote{Away from $E_8$ theory~\cite{Zamolodchikov:1989hfa,Zamolodchikov:1989fp}, a stable particle is defined to have a mass less than $2m_1$ so that decay is kinematically forbidden. 
The existence of two stable particles at $\eta_{\text{latt}} = \frac{(g_x-1)}{|g_z|^{8/15}}\approx0.69$ is consistent with Ref.~\cite{Jha:2024jan}.}
The lowest-energy inelastic threshold corresponds to two $|1\rangle$ particles colliding and producing one $|1\rangle$ particle and one $|2\rangle$ particle, i.e., the process $11\to12$.
This threshold is at a center-of-mass energy of $E_{\text{thr}} = m_1 + m_2$.
The corresponding threshold momentum can be found from inverting the single-particle dispersion relation and solving $E(k_{\text{thr}}) = E_{\text{thr}}/2$.
In App.~\ref{i_s:app:kinematics}, the dispersion relation is computed exactly for $L=\{16,17,\ldots,28\}$ and shown to be exponentially converged in system size.
The predicted threshold momentum is $|k_{\text{thr}}|\approx0.24\pi$.
However, since our wavepackets are obtained via variational energy minimization, their energy is strictly higher than that of the exact wavepacket.
Because of this, the actual momentum threshold is a bit lower, closer to $|k_{\text{thr}}|\approx0.22\pi$.
Additionally, the inelastic threshold is not sharp due to the spread in momentum (or equivalently in energy) coming from a finite $\sigma_{}$.

\begin{figure}
    \centering
    \includegraphics[width=\linewidth]{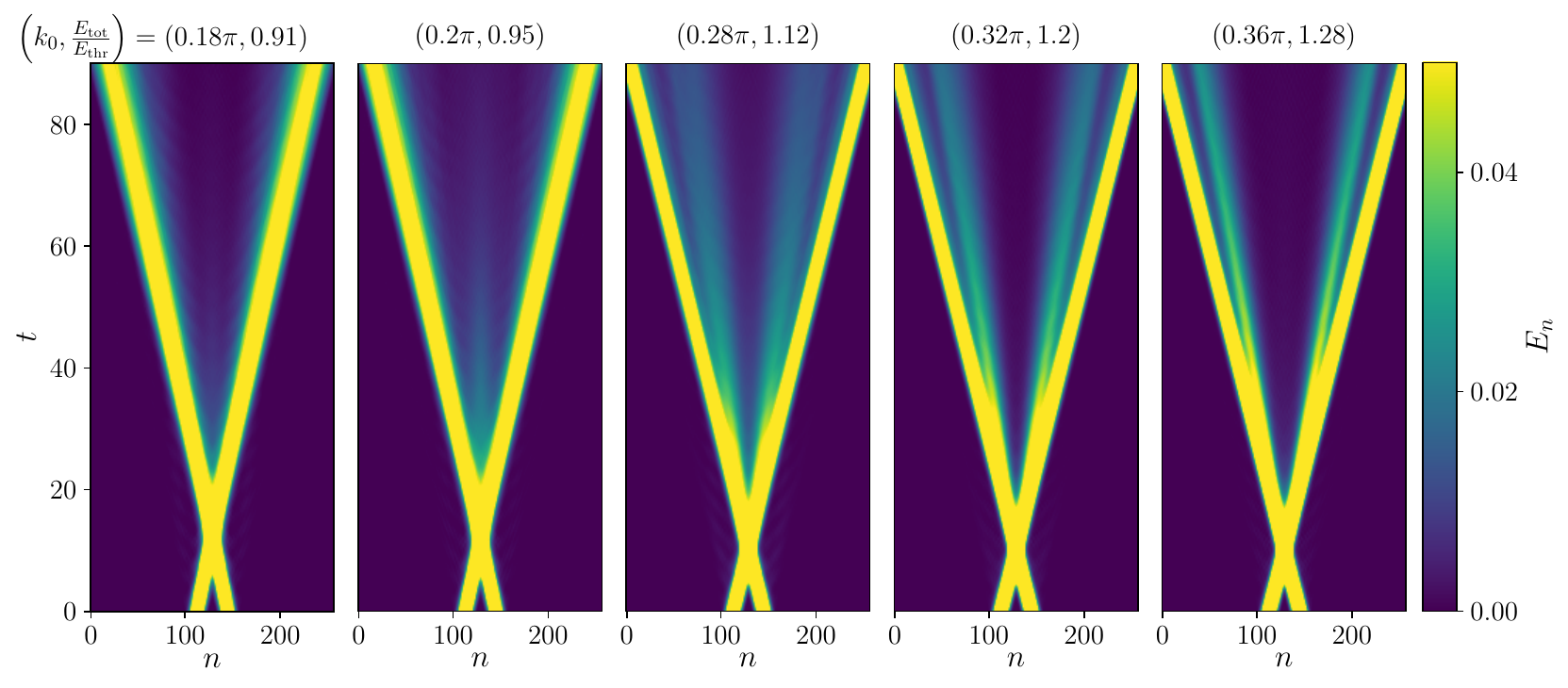}
    \caption{\textit{MPS circuit simulations of scattering in one-dimensional Ising field theory.}
    The vacuum-subtracted energy density $E_n$ throughout the scattering process for a selection of center-of-mass energies $E_{\text{tot}}$.
    The initial wavepackets are separated by 10 sites and evolved with a Trotter time step of $\delta t = 1/16$.
    The wavepackets are constructed from 8 steps of ADAPT-VQE except for $k_0=0.2\pi$ which is constructed from 10 steps.
    Results for $L=256$ and $\sigma_{}=0.13,\, d=22$,\, $g_x=1.25$,  $g_z=0.15$ and max bond dimension 350 are shown.}
    \label{i_s:fig:2WP_scattering_MPS}
\end{figure}
Results from MPS circuit simulations of scattering are shown in 
Fig.~\ref{i_s:fig:2WP_scattering_MPS}. 
These simulations compute the vacuum-subtracted energy density $E_n$ in Eq.~\eqref{i_s:eq:vacsubEn} as a function of lattice position and simulation time.
Each plot is labeled by the total energy $E_{\text{tot}}$ and momentum $k_0$ of the wavepackets.
The elastic process $11\to11$ is identified from outgoing particles with the same velocity as the incoming particles.
The inelastic process $11\to12$ converts some of the kinetic energy to mass to produce the heavier $|2\rangle$ particle.
As a result, the outgoing particles travel slower than the incoming ones.
The lowest momentum, $k_0=0.18\pi$, is well below inelastic threshold, and only elastic scattering is observed. 
Scattering at $k_0=0.2\pi$ reveals a very faint track in the middle of the lattice that does not propagate.
This corresponds to $11\to12$ where the post-collision particles are at rest.
Increasing the momentum further causes the left- and right-moving particle $|2\rangle$ tracks to be more identifiable, as shown in the plots for $k_0\geq0.28\pi$.
The kinematics are such that the particle $|1\rangle$ tracks in $11\to12$ are hidden behind the elastic scattering trajectories.
It is important to remember that expectations are averaged over the whole scattering wavefunction, which, in the case of $k_0\leq0.36\pi$, is a superposition of components with at most two particles.
This discussion has ignored three-particle production that can occur for $E_{\text{tot}}>3m_1=1.04 E_{\text{thr}}$, as it has a much smaller branching ratio compared to $11\to12$~\cite{Jha:2024jan}. See App.~\ref{i_s:app:kinematics} for a discussion of the other inelastic processes.
The MPS simulations used a maximum bond dimension of ${\tt max\_bond}=350$ and the energy density was converged to $.01\%$.

\section{Details on the simulations performed using IBM's quantum computers}
\label{i_s:sec:qsimDetails}
\noindent 
The simulations performed on IBM's quantum computers in Sec.~\ref{i_s:sec:qsim} use parameters that are chosen to maximize the clarity of the inelastic effects in the results. 
The interplay between the parameters is nontrivial, and our simulation choices were informed by the following considerations:
\begin{itemize}
    \item OBCs: OBCs are chosen instead of PBCs due to increased flexibility in the lattice-to-qubit mapping. See next bullet point.
    \item $L=104$: this was the largest lattice that could be mapped to {\tt ibm\_marrakesh} while  maintaining a maximum $CZ$ gate error of $ 7.5\times10^{-3}$.
    \item $\delta t=0.55$: this was the largest Trotter step size that did not have significant Trotter errors. See App.~\ref{i_s:app:systematics} for a quantification of the Trotter errors. 
    \item $t_{\text{max}}=24.75$ ($n_T=45$ Trotter steps): this was the maximum simulation time that could be reached before the results were overwhelmed by device errors.
    \item $k_0=0.32\pi$: informed by the results of MPS scattering simulations shown in Fig.~\ref{i_s:fig:2WP_scattering_MPS}, we determined that a wavepacket momentum of $k_0=0.32\pi$ has the strongest signal of inelastic scattering at $t_{\text{max}}=24.75$.
    \item $\sigma_{}=0.13$: the wavepacket extent in momentum space
    was chosen to minimize the wavepacket's spatial size and its spreading under time evolution. 
    Wavepacket spreading is most significant for low-energy, elastic scattering, e.g., $k_0=0.18\pi$ in Fig.~\ref{i_s:fig:2WP_scattering_MPS}.
    Clearly resolving elastic and inelastic scattering, i.e., distinguishing wavepacket delocalization and particle production,
    requires $\sigma_{}\leq0.13$.
    This is shown in App.~\ref{i_s:app:systematics}.
    \item $d=21$: this is the smallest spatial extent of $|W(k_0)\rangle$ that could be chosen for $\sigma_{}=0.13$ while keeping truncation effects small, see App.~\ref{i_s:app:systematics}. 
    \item No MCM-FF: the unitary $|W(k_0)\rangle$ preparation circuit in Fig.~\ref{i_s:fig:IsingWPCircs} was significantly less noisy when run on {\tt ibm\_marrakesh} than the constant-depth circuit in Fig.~\ref{i_s:fig:ConstantDepth} that utilizes MCM-FF.
    \item 7 steps of ADAPT-VQE: The quality of the prepared wavepacket significantly improves at 7 steps.
    \item 2 site wavepacket separation:
    this is the smallest separation that keeps interactions between the initial wavepackets negligible and preserves the $E_n = E_{L-1-n}$ reflection symmetry.

\end{itemize}

\begin{figure}
    \centering
    \includegraphics[width=0.8\linewidth]{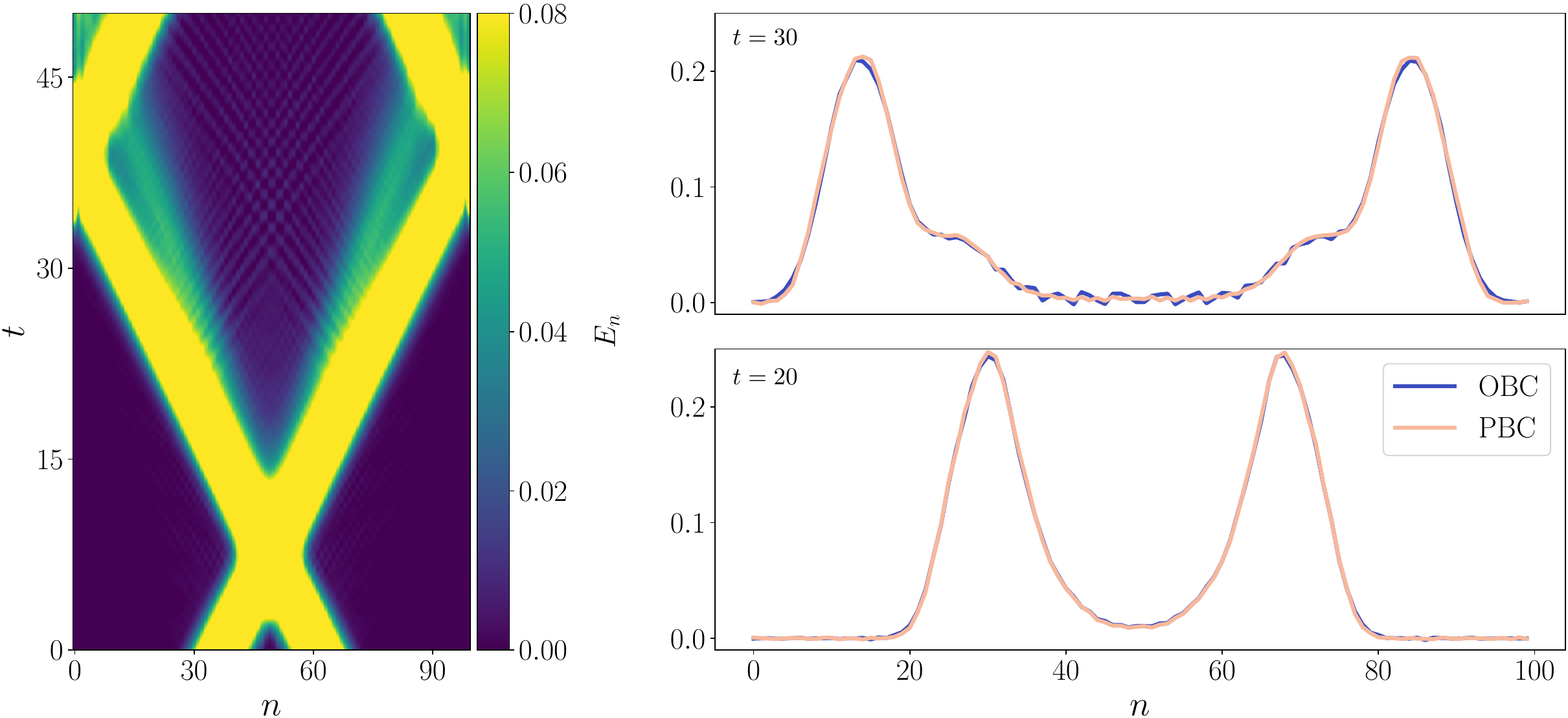}
    \caption{\textit{Comparison of OBCs and PBCs in one-dimensional Ising field theory.} Left: the vacuum-subtracted energy density $E_n$ throughout a MPS simulation of scattering on a $L=100$ lattice with OBCs.
    Wavepacket parameters $\sigma_{}=0.13,\,k_0=0.36\pi$ and a time step of $\delta t=1/16$ are used. 
    Right: the energy density at $t=20$ and $t=30$ for OBCs (blue) and PBCs (tan).}
    \label{i_s:fig:PBCOBC}
\end{figure}
The wavepacket preparation algorithm presented in this work relies on having a Hamiltonian that is block-diagonal in momentum space.
This is a consequence of translational symmetry, which is broken in a system with OBCs.
However, with OBCs, there is still an approximate translational symmetry in the bulk. 
For this reason, the quantum circuits that prepare PBC wavepackets can also be used to prepare wavepackets in a system with OBCs, if they act sufficiently far from the boundaries.
The initial state of the quantum simulations in Sec.~\ref{i_s:sec:qsim} has wavepackets localized $\gtrsim 30$ sites from boundary. 
This is much larger than the correlation length and therefore boundary effects may be ignored.
The circuits that minimize the energy and prepare wavepackets are determined in a system with PBCs to preserve the momentum content established by $|W(k_0)\rangle$.
The wavepacket with OBCs is prepared using the same circuit as for PBCs, but with all elements corresponding to Pauli strings that couple $q_0, q_{L-1}$ removed.
This causes the local vacuum near the boundaries to be of slightly lower quality than the rest of the state.
Time evolution is then implemented with the OBC Ising field theory Hamiltonian,  Eq.~\eqref{i_s:eq:HIFT} without the $\hat{Z}_0 \hat{Z}_{L-1}$ term.

The effects of OBCs on the quantum simulations in Sec.~\ref{i_s:sec:qsim} are illustrated in Fig.~\ref{i_s:fig:PBCOBC}.
The left plot shows a MPS simulation of the energy density throughout the inelastic scattering process in a $L=100$ system with OBCs.
The quality of the vacuum near the boundaries is lower than in the bulk, causing small perturbations to propagate inwards.
At $t\approx 20$, these perturbations interact with the post-collision state and generate small ripples in the energy density between the outgoing particles.
This is shown in more detail in the right plot of Fig.~\ref{i_s:fig:PBCOBC} which compares the energy density in a OBC and PBC simulation.
At $t=20$ the energy density is nearly identical and there are almost no boundary effects.
At $t=30$, fluctuations coming from the boundaries can be identified in the energy density between the particles.
These small fluctuations could be mitigated by adding boundary operators to the ADAPT-VQE operator pool to improve the vacuum near the edges~\cite{Farrell:2023fgd}.

The structure of the circuit that simulates scattering in Ising field theory is shown in Fig.~\ref{i_s:fig:scatcirc}. 
The initial state is prepared using the unitary circuit whose structure is shown in the left panel of Fig.~\ref{i_s:fig:IsingWPCircs}.
The symmetry-preserving circuits that minimize the energy and prepare wavepackets are determined by implementing ADAPT-VQE with a MPS circuit simulator.
This is described in App.~\ref{i_s:sec:csimscatt}, and the operator ordering and variational parameters that minimize the energy are given in App.~\ref{i_s:app:ADAPTparam}.
The corresponding circuits used in ADAPT-VQE are shown in the right panel of Fig.~\ref{i_s:fig:IsingWPCircs}.
Time evolution is implemented using $n_T$ second-order Trotter steps with ordering $\{R_X, R_{Z}, R_{ZZ}, R_X\}$.
The total circuit depth for the  simulations is
\begin{equation}
\text{two-qubit gate depth: }\ 22 \ + \ 18 \ + \ 2n_T \ ,
\label{i_s:eq:circuitDepth}
\end{equation}
with the terms corresponding to the preparation of $|W(k_0)\rangle$ with $d=21$, 7 steps of ADAPT-VQE and $n_T$ steps of Trotterized time evolution. The total circuit depth and number of two-qubit gates used in the quantum simulations are provided in Table~\ref{i_s:tab:device_run_params}.

IBM's {\tt heron} quantum computers support fractional gates \cite{IBM_Quantum_Computing_Blog_2024}, which include a native $R_{ZZ}(\theta)=e^{-i\theta\hat{Z}\hat{Z}/2}$ for $\theta \in [0,\pi/2)$.
Trotterized time evolution requires $e^{i \delta t \hat{Z}\hat{Z}}$ with the opposite sign of $\theta$.
This could be overcome by sandwiching the $R_{ZZ}$ gates with $\hat{X}$ to flip the sign of $\theta$ at the cost of two additional single-qubit gates per $R_{ZZ}$. 
To eliminate the single-qubit gate overhead, we instead swap the positions of the wavepackets and evolve backwards in time.
Time reversal symmetry ensures that this simulates exactly the same scattering process.
By using $R_{ZZ}$ instead of $CZ$ gates, the two-qubit gate depth of each Trotter step is reduced from 4 to 2.
At the beginning of each run, we execute a suite of $R_{ZZ}$ calibration circuits to tune the rotation angles and reduce coherent errors. 

At the end of the simulations, the energy density in Eq.~\eqref{i_s:eq:vacsubEn} is measured and compared to the expected, noise-free, value obtained by simulating the same circuits with the Quimb MPS simulator.\footnote{With OBCs, the energy density $\hat{H}_n$ in Eq.~\eqref{i_s:eq:HIFT} on sites 0 and $L-1$ contains only a single $\hat{Z}\hat{Z}$ term without the $1/2$ prefactor.}
A relatively large $\delta t=0.55$ causes the bond dimension to become quite large to maintain accuracy.
Our MPS simulations required a bond dimension up to ${\tt max\_bond}=2250$ to be well converged at $t=40$.\footnote{Increasing the maximum bond dimension from 2000 to 2250 only changes the energy density by $8\times 10^{-4}$.}
In comparison, ${\tt max\_bond}=350$ was sufficient for the MPS scattering simulations shown in Fig.~\ref{i_s:fig:2WP_scattering_MPS} because they used a much smaller Trotter step size of $\delta t=1/16$. 

\subsection{Error mitigation}
\begin{figure}
  \begin{minipage}[c]{.5\linewidth}
    \centering
    \includegraphics[width=0.4\linewidth]{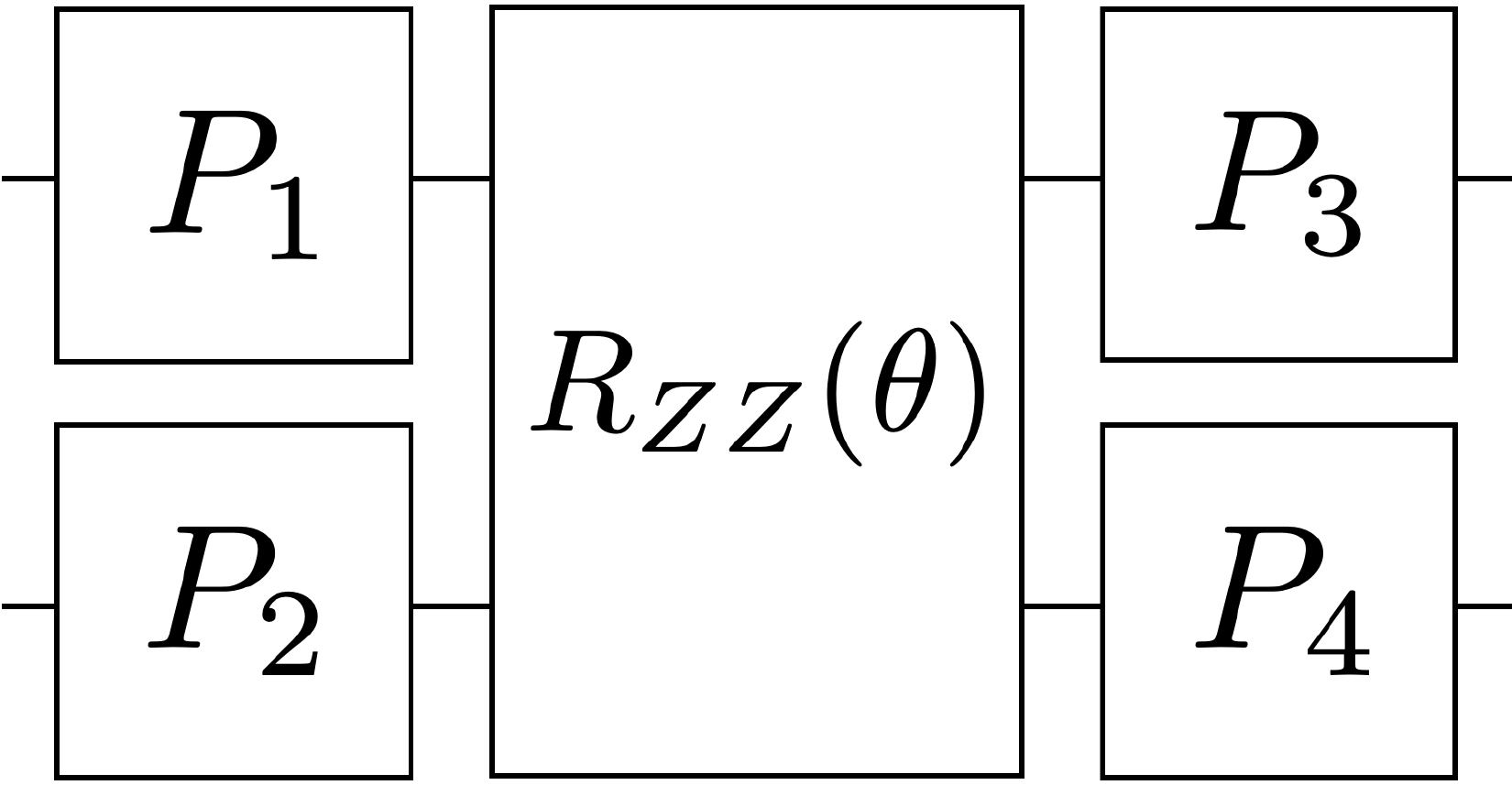}
  \end{minipage}\hfill
  \begin{minipage}[c]{.5\linewidth}
    \centering
    \begin{tabularx}{\linewidth}{|c||Y|Y|Y|Y|Y|Y|Y|Y|}
    \hline
    $P_1$ & $I$ & $X$ & $Y$ & $Z$ & $X$ & $Y$ & $Z$ & $I$ \\\hline
    $P_2$ & $I$ & $X$ & $Y$ & $Z$ & $Y$ & $X$ & $I$ & $Z$ \\\hline
    $P_3$ & $I$ & $X$ & $Y$ & $Z$ & $X$ & $Y$ & $Z$ & $I$ \\\hline
    $P_4$ & $I$ & $X$ & $Y$ & $Z$ & $Y$ & $X$ & $I$ & $Z$ \\\hline
    \end{tabularx}
    \renewcommand{\arraystretch}{1}
  \end{minipage}
  \caption{\textit{Pauli Twirling of the $R_{ZZ}(\theta)$ gate.} The Pauli gates $P_i$ are appended before and after $R_{ZZ}(\theta)$ as shown on the left. The table on the right shows the sets of $P_i$ that leave $R_{ZZ}(\theta)$ invariant for all $\theta$. }
  \label{i_s:fig:rzz_twirls}
\end{figure}

\noindent
A suite of error mitigation methods allows accurate predictions of observables to be made from noisy device data.  
Dynamical decoupling (DD) \cite{Viola:1998jx,Ezzell:2022uat} removes the effects of idle noise and crosstalk between qubits. 
Pauli Twirling (PT) \cite{Wallman:2015uzh} is applied to every two-qubit gate to convert coherent errors into stochastic noise. 
The 16 combinations of Paulis are used to twirl CZ gates, while the 8 combinations given in Fig.~\ref{i_s:fig:rzz_twirls} are used for $R_{ZZ}$ (a larger set compared to the one used in Ref.~\cite{Kim:2021gvc}).  
The complete Pauli basis for twirling $R_{ZZ}(\theta)$ is not available because it is a non-Clifford operation for arbitrary $\theta$ .
Twirled Readout Error eXtinction (TREX) \cite{Berg:2020ibi} is used to mitigate measurement errors.
In the limit of infinite PT and TREX, the noise is characterized by a Pauli channel~\cite{Wallman:2015uzh}, 
\begin{align}
    \rho \ \rightarrow \ \sum_i p_i \hat{P}_i \rho \hat{P}_i \ ,
    \label{i_s:eq:pauli_channel}
\end{align} 
where the sum runs over all $4^L$ Pauli operators $\hat{P}_i$ and $\sum_i p_i = 1$ with $p_i\geq0$.
Under this channel, expectation values of observables $\hat{O}$ are given by 
\begin{align}
    \langle \hat{O}\rangle_\text{meas} \ = \ \sum_i p_i \text{Tr}(\hat{P}_i\rho\hat{P}_i\hat{O}) \ ,
    \label{i_s:eq:pauli_channel_expectation_value}
\end{align}
For Pauli observables, $\hat{P}_i\hat{O}\hat{P}_i=\pm\hat{O}$, and this channel scales the measured values $\langle \hat{O}\rangle_\text{meas}$ relative to their predicted (noise-free) counterparts $\langle \hat{O}\rangle_\text{pred}$: $\langle \hat{O}\rangle_\text{meas} = p_{\hat{O}}\langle \hat{O}\rangle_\text{pred}$.
The signal strength $|p_{\hat{O}}|\leq1$ characterizes how much an observable is impacted by the Pauli noise channel.

Expectation values of local observables are then estimated using Operator Decoherence Renormalization (ODR) \cite{Farrell:2023fgd,Farrell:2024fit,Urbanek:2021oej,ARahman:2022tkr}.
For each observable $\hat{O}$, the signal strength $p_{\hat{O}}$ is determined using a ``mitigation'' circuit whose noise profile is similar to the original ``physics'' circuit, but whose output can be efficiently computed with classical computers. 
In this work, the Trotterized time evolution of the vacuum state, $\hat{U}_2(t)|\psi_\text{vac}\rangle$
is used for the mitigation circuit.
With exact state preparation and time evolution, $e^{-i\hat{H} t}|\psi_\text{vac}\rangle$ is trivial since $|\psi_\text{vac}\rangle$ is an eigenstate of $\hat{H}$.
In our simulations, errors coming from the approximate preparation of the vacuum in Eq.~\eqref{i_s:eq:ADAPTWPvac} are negligible.
However, Trotter errors are large and cause the energy density to fluctuate up to $15\%$ during time evolution (see App.~\ref{i_s:app:systematics}).
These fluctuations can be mitigated by enforcing energy conservation in post-processing, see App.~\ref{i_s:app:energy_rescale} for more details.

The observables evaluated in the time-evolved (approximate) vacuum are calculated via MPS.
For early times, this is efficient in one dimension due to the correspondence between ground states of gapped systems and area law entanglement~\cite{Hastings:2007iok}.
The signal strength is computed from  the mitigation circuit results, 
\begin{align}
    p_{\hat{O}} \ = \ \frac{\langle \hat{O}\rangle_\text{meas}}{\langle \hat{O}\rangle_\text{pred}} \ = \ \frac{\langle \psi_\text{vac}|\hat{U}^{\dagger}_2(t) \hat{O} \hat{U}_2(t)|\psi_\text{vac}\rangle_\text{meas}}{\langle \psi_\text{vac}|\hat{U}^{\dagger}_2(t) \hat{O} \hat{U}_2(t)|\psi_\text{vac}\rangle_\text{pred}} \ .
    \label{i_s:eq:scattering_ising_odr_p_j}
\end{align}
This $p_{\hat{O}}$ is then used to rescale $\langle \hat{O}\rangle_\text{meas}$ in the physics circuit and estimate its noise-free value. 
The $p_{\hat{O}}$ are also able to identify device runs with anomalously large amounts of noise. 
We filter out observables with $p_{\hat{O}}<0.01$~\cite{Farrell:2024fit}, and observables that involve a qubit with readout error $\epsilon_{\text{readout}}>0.035$.
Additionally, there is a parity symmetry that equates the energy density of sites that are related by reflection about the collision point.
For our simulations this implies $ E_n = E_{L-1-n}$.
By combining the uncorrelated measurements of observables related by this symmetry, our number of shots is effectively doubled.

\begin{table}[t]
\centering
\begin{tabularx}{0.9\linewidth}{|c||c|Y|Y|Y|Y|} \hline
$t$ & $n_T$ & \# two-qubit gates & Two-qubit gate depth & \# PTs  &\ \# shots \\\hline\hline
0 & 0 & 954 & 40 & 40 &$2.56\times10^6$ \\\hline
8.25 & 15 & 2,499 & 70 & 40 &$1.28\times10^6$ \\\hline
16.5 & 30 & 4,044 (4,022) & 100 & 40 (80) & $1.28\times10^6$ ($2.56\times10^6$)\\\hline
24.75 & 45 & 5,589 (5,567) & 130 & 80 (160) & $2.56\times10^6$ ($5.12\times10^6$)\\\hline
\end{tabularx}
\renewcommand{\arraystretch}{1}
\caption{\textit{Resources used in the quantum simulations performed on 104 qubits of {\tt ibm\_marrakesh}.} For a given simulation time $t$ (first column), the number of Trotter steps $n_T$ is given in the second column. 
The total number of two-qubit gates and corresponding two-qubit gate depth is given in columns three and four, respectively.
The total number of Pauli twirls and shots per simulation time, including all error mitigation overhead, are given in column five and six, respectively. The quantities in parenthesis correspond to the single wavepacket simulations.}
\label{i_s:tab:device_run_params}
\end{table}
A total of 16 TREX twirls, each with 500 shots, is used to mitigate the measurement errors for every PT, and the number of PTs run for each of the physics and mitigation circuits are given in Table~\ref{i_s:tab:device_run_params}. 
Additionally, measurements must be performed in both the $X$- and $Z$-bases to determine the energy density.
For the $t_3=24.75$, this gives a total of $500\times16\times80\times2\times2=2.56\times10^6$ shots. 
It is important that the PT and TREX twirls are applied identically for each pair of physics and mitigation circuits, so that their noise profiles are as similar as possible.\footnote{The mitigation circuits also work for TREX calibration, and no additional circuits are needed to mitigate readout errors.}
The initial state used for mitigation, $|\psi_\text{vac}\rangle$, is prepared by acting the wavepacket ADAPT-VQE circuit on the $|000...\rangle$ state, see the discussion around Eq.~\eqref{i_s:eq:ADAPTWPvac}.
It is desirable to have maximal similarity between the structure and noise profiles of the physics and mitigation circuits.
This is accomplished by preparing $|000...\rangle$ with the $|W(k_0)\rangle$ circuit in Fig.~\ref{i_s:fig:scatcirc}, but with the angle of the first $R_Y$-gate set to zero and the second CNOT changed from a control on $|0\rangle$ to a control on $|1\rangle$.
An average over TREX twirls is taken to compute an expectation value for each observable for each PT. 
This is then used as input to the rest of the error mitigation pipeline.
The expectation values $\langle\hat{Z}_n\rangle$, $\langle\hat{X}_n\rangle$, and $\langle\hat{Z}_n\hat{Z}_{n+1}\rangle$ are computed separately then added together to form the energy density.

\begin{figure}
    \centering
    \includegraphics[width=\linewidth]{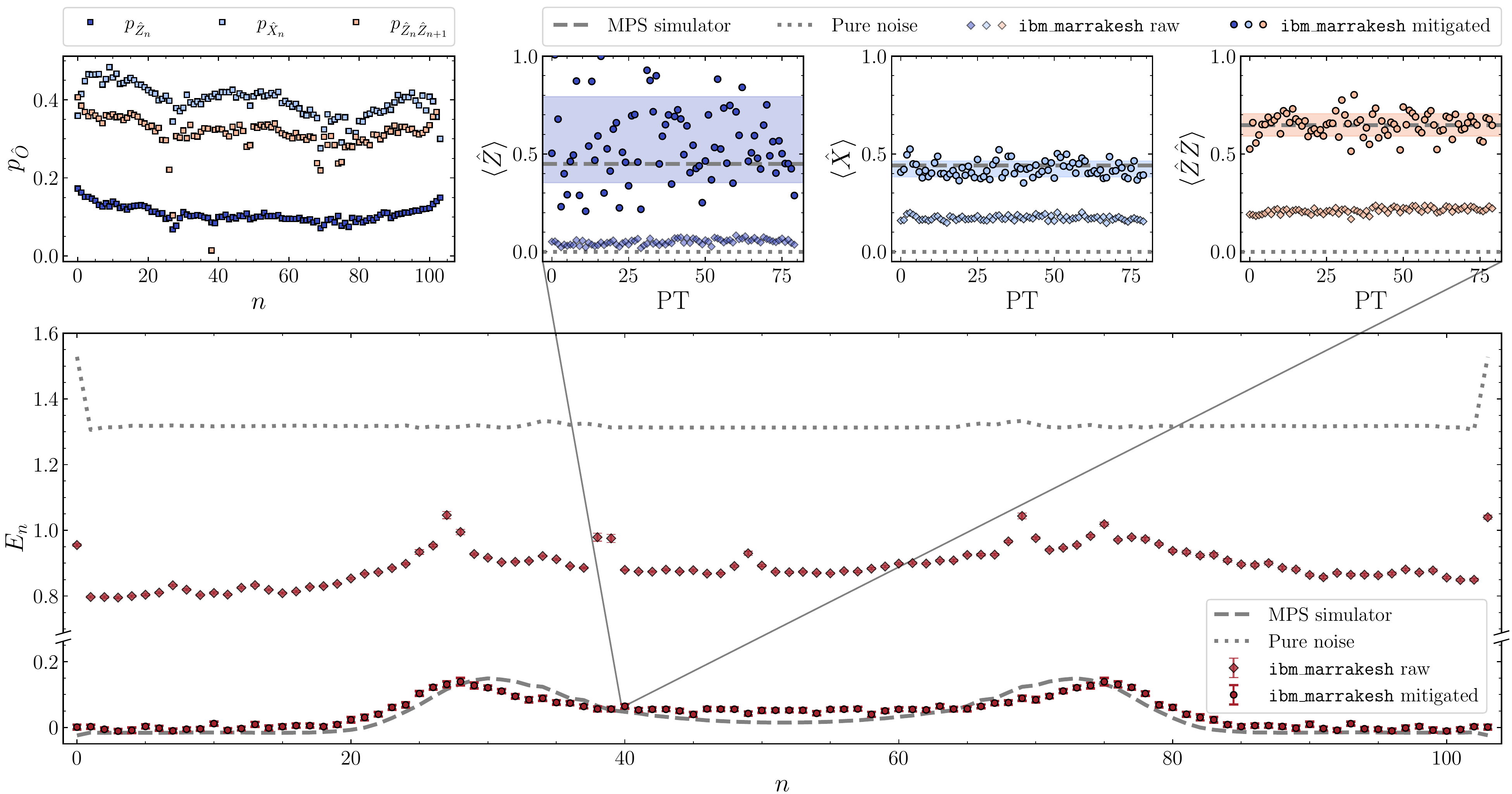}
    \caption{{\it The effect of error mitigation on the quantum simulation results from {\tt ibm\_marrakesh}.} 
    The data corresponds to the quantum simulation results in Fig.~\ref{i_s:fig:ibm_results} for $L=104$ and $t_3=24.75$ (130 two-qubit gate depth).
    Top left: the median signal strengths $p_{\hat{O}}$ defined in Eq.~\eqref{i_s:eq:scattering_ising_odr_p_j} as a function of lattice position $n$ for $\hat{O}=\{ \hat{Z}_n,\,\hat{X}_n,\,\hat{Z}_n\hat{Z}_{n+1} \}$. 
    The signal strength is computed for each $n$ by averaging over the TREX twirls, and then computing the median over the set of PTs. 
    Top right: the effect of ODR on the expectation values of the same $\hat{O}$ for one lattice site, $n=40$. The data points are determined by averaging over TREX twirls for each PT. The gray dotted lines represent the expectation value of a completely decohered state, while the gray dashed lines are the MPS predictions. The colored bands represent $\pm1$ standard deviation of the mitigated results.  
    Bottom: the vacuum-subtracted energy density $E_n$ of the raw device data (diamonds) and after all error mitigation (circles). 
    The raw device data represents the median over PTs (averaged over TREX twirls). 
    The error bars for both the raw and the mitigated data are obtained via bootstrap resampling.}
    \label{i_s:fig:error_mitigation}
\end{figure}

The raw and error-mitigated results are compared in Fig.~\ref{i_s:fig:error_mitigation} for $t_3=24.75$. 
The raw data from the device is close to the limit of pure depolarizing noise, which is far from the expected values determined from MPS.
The results after error mitigation are significantly improved, and lie close to MPS expectations.
There exist points in the bottom panel of Fig.~\ref{i_s:fig:error_mitigation} where the device results disagree with the MPS expectations by several standard deviations, despite all error mitigation techniques.
This is particularly noticeable in the regions containing the outgoing particles, highlighting a weakness of the chosen mitigation circuits for ODR. 
The time evolution of the vacuum used as the reference state in ODR, $U_2(t)|\psi_\text{vac}\rangle$, does not capture effects of state-dependent noise on the wavepackets. 

The signal strengths for $\langle\hat{Z}_n\rangle$, $\langle\hat{X}_n\rangle$, and $\langle\hat{Z}_n\hat{Z}_{n+1}\rangle$ at $t_3=24.75$ are shown in the top left panel of Fig.~\ref{i_s:fig:error_mitigation}.
The signal strength for $\hat{Z}$ is consistently $\sim\!3\times$ lower than for the other observables.
This indicates that the weights $p_i$ in the Pauli noise channel Eq.~\eqref{i_s:eq:pauli_channel}, are larger for the Paulis that anticommute with $\hat{Z}_n$.
The top right panels of Fig.~\ref{i_s:fig:error_mitigation} show both the raw and ODR-predicted estimates for the different observables at a single site $n=40$.
The extra noise in $\langle\hat{Z}_n\rangle$ leads to significantly larger variations in the ODR-predicted $\langle\hat{Z}_n\rangle$ across Pauli twirls compared to the other observables.
These large variations are a major cause of uncertainty in the final results, shown in the bottom panel of Fig.~\ref{i_s:fig:error_mitigation}. 
We additionally computed $\langle\hat{X}_n\hat{X}_{n+1}\rangle$ and found that it is as noisy as $\langle\hat{Z}_n\rangle$. 
More extensive simulations are needed to identify the origin of this asymmetry.
One cause could be the incomplete twirling of the $R_{ZZ}$ gate (only 8 combinations of the 16 are possible).
Another could be the imbalance of the couplings $g_x=1.25$, $g_z=0.15$.

\section{Calculating skewness}
\label{i_s:sec:skew}
\noindent
The inelastic process $11\to12$ probed by the quantum simulations presented in Sec.~\ref{i_s:sec:qsim} produces a heavy $|2\rangle$ particle that travels slower than the lighter $|1\rangle$ particle.
At late times, the $|2\rangle$ particle can be identified as a distinct bump in the energy density that travels slower.
Before this, but after the collision, the emergence of the heavy $|2\rangle$ particle skews the outgoing energy profile toward the point of the collision.
This skewness was identified in the data presented in Sec.~\ref{i_s:sec:qsim} and is evidence for inelastic particle production.

\begin{figure}
    \centering
    \includegraphics[width=\linewidth]{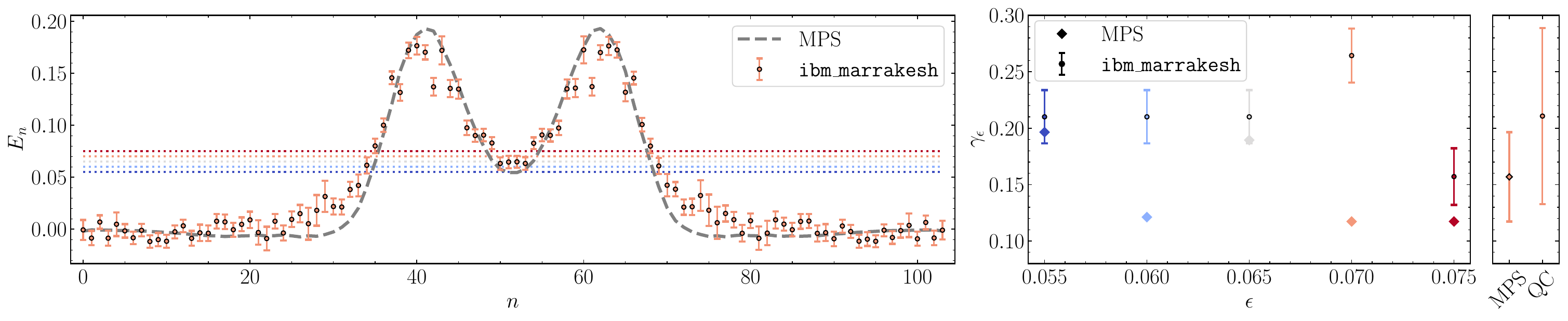}
    \caption{{\it Calculation of skewness for $t_2=16.5$.} 
    Left: the range of energy cutoffs $\epsilon$ (dotted lines) are overlaid on the energy density from Fig.~\ref{i_s:fig:ibm_results}. 
    Middle: the third moment of the energy density distribution  $\gamma_\epsilon$ is calculated for a contiguous interval of points where $E_n+\sigma_n>\epsilon$. 
    The error bars represent the standard deviation computed via bootstrap resampling. 
    Right: the reported value of $\gamma$ is computed via Eq.~\eqref{i_s:eq:gamma}, with error bars that cover the range of $\gamma_\epsilon$ coming from the variation of the energy cutoff and, for the quantum results, statistical errors from bootstrap resampling.}
    \label{i_s:fig:skewness_cutoffs}
\end{figure}

Our metric for the skewness is designed to account for the discrete nature of the data and incorporate statistical errors and device noise.
We compute the skewness of the energy density over an interval of sites $j\in [n_{\text{min}},n_{\text{max}}]$ that captures the region of positive energy density on one half of the lattice.\footnote{The energy density is symmetrized about the center of the collision as a part of error mitigation. See App.~\ref{i_s:sec:qsimDetails} for details.} 
Given an energy cutoff $\epsilon$, the window $[n_\text{min}, n_\text{max}]$ is chosen by contiguously including sites where $E_n+\sigma_n \geq \epsilon$, where $E_n$ and its standard deviation $\sigma_n$ are computed via bootstrap resampling as detailed in App.~\ref{i_s:sec:qsimDetails}.
The energy cutoffs are chosen to capture as much of the region of excitations as possible, while omitting outliers caused by noise. 
For each time, the energy cutoff is varied across a range given in Table~\ref{i_s:tab:skewness_cutoffs} to estimate systematic errors.
For each $\epsilon$, we compute the third moment $\gamma_\epsilon$ of the distribution of the energy density,
\begin{align}
    \gamma_\epsilon \ = \ \frac{\sum_j (j -\mu)^3  \tilde{E}_j}{\sigma_{\epsilon}^3} \ ,
\end{align}
where the normalized energy density $\tilde{E}$, mean $\mu$ and standard deviation $\sigma_{\epsilon}$ are defined as
\begin{align}
\tilde{E}_n \ = \ \frac{E_{n}}{\sum_j E_j} \ \ , \ \ \mu \ = \ \sum_{j} j\, \tilde{E}_j  \ \ , \  \ \sigma_{\epsilon}^2 \ = \ \sum_j (j -\mu)^2  \tilde{E}_j \ .
\end{align}
The sums run over all sites in the window $[n_{\text{min}},n_{\text{max}}]$.
The skewness metric $\gamma$ reported in Sec.~\ref{i_s:sec:qsim} is calculated from $\{\gamma_\epsilon\}$
\begin{align}
\gamma \ = \ \frac{\max \gamma_\epsilon + \min \gamma_\epsilon}{2}.
\label{i_s:eq:gamma}
\end{align}
A positive value of $\gamma$ corresponds to a right-skewed distribution, while a negative $\gamma$ corresponds to a left-skewed one.
Considerable variations are observed in the $\gamma_\epsilon$ values computed from different $\epsilon$. 
Reflecting this, the error bars for $\gamma$ include both the statistical errors from bootstrap resampling $\gamma_\epsilon$ and the variation of $\gamma_\epsilon$ with the cutoff $\epsilon$.
Figure~\ref{i_s:fig:skewness_cutoffs} shows an example of the choice of cutoffs for $t_2=16.5$, as well as the resulting values of $\gamma_\epsilon$ for each $\epsilon$, and the reported values of $\gamma$ along with the error bars. 

\begin{table}[t]
\centering
\begin{tabularx}{\linewidth}{|c|c||Y|Y||Y|Y|} \hline
Initial state & $t$ & $\epsilon_\text{MPS}$ & $\epsilon_{\tt ibm\_marrakesh}$ & $\gamma_{\text{MPS}}$ & $\gamma_{{\tt ibm\_marrakesh}}$ \\\hline\hline
$|\psi_\text{2wp}\rangle$ & $24.75$ & $0.02-0.04$ & $0.065-0.085$ & 0.37(13) & 0.27(13) \\\hline
$|\psi_\text{1wp}\rangle$ & $24.75$ & $0.03-0.05$ & $0.05-0.07$ & 0.13(4) & 0.12(6) \\\hline
$|\psi_\text{2wp}\rangle$ & $16.5$ & $0.055-0.075$ & $0.055-0.075$ & 0.16(4) & 0.21(8)\\\hline
$|\psi_\text{1wp}\rangle$ & $16.5$ & $0.03-0.05$ & $0.05-0.07$ & 0.09(7) & 0.14(9) \\\hline
$|\psi_\text{2wp}\rangle$ & $0.0$ & $0.02-0.04$ & $0.02-0.04$ & 0.0(0)& -0.09(7)\\\hline
\end{tabularx}
\renewcommand{\arraystretch}{1}
\caption{The ranges of energy cutoffs $\epsilon$ used for calculating $\gamma$ from the energy density obtained from MPS simulations and {\tt ibm\_marrakesh}. Cutoffs in increments of $0.005$ are used.}
\label{i_s:tab:skewness_cutoffs}
\end{table}

\begin{figure}
    \centering
    \includegraphics[width=\linewidth]{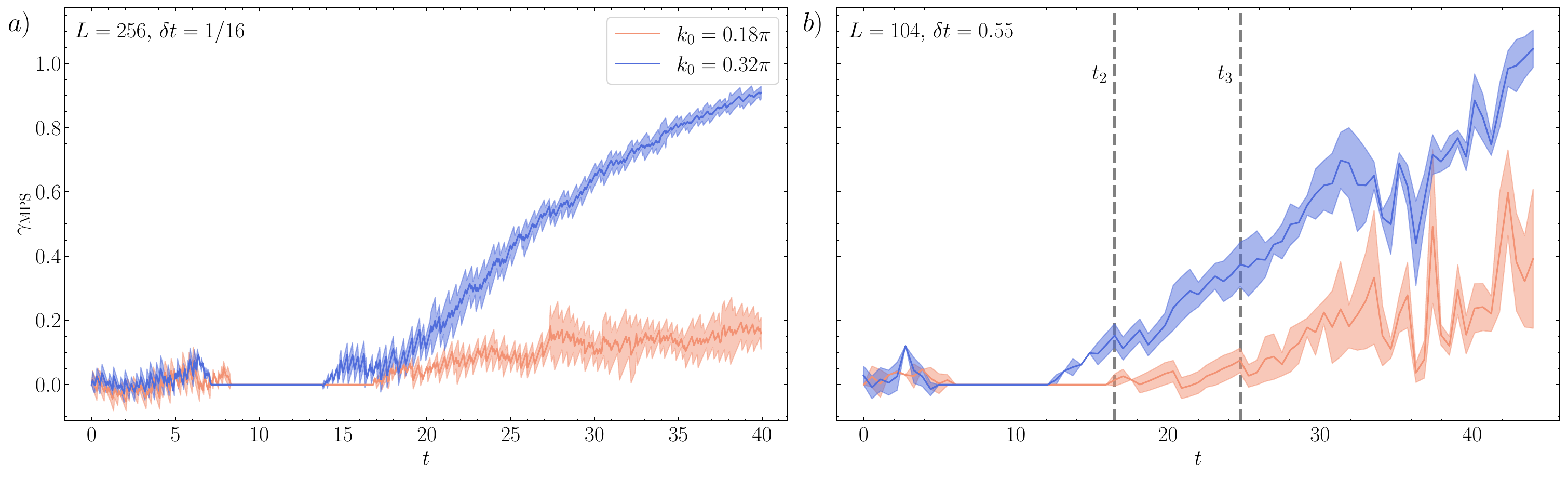}
    \caption{{\it Skewness as an indicator of inelasticity.} a) The skewness $\gamma_\text{MPS}$ calculated for MPS simulations of elastic $(k_0=0.18\pi)$ and inelastic $(k_0=0.32\pi)$ scattering with $L=256,\,\delta t=1/16,\,d=27,\,\sigma_{}=0.1$, PBCs, and max bond dimension 350.
    The energy cutoffs $\epsilon$ are chosen to be [$E_{L/2}+0.01,E_{L/2}+0.03$] in increments of $0.005$, and $\gamma_\text{MPS}$ is set to 0 if the maximum value of $E_n$ occurs at $L/2$ or if there are no points with $E_n>\epsilon$. 
    The shading represents the range of $\gamma_\text{MPS}$ coming from the variation of $\epsilon$. 
    b) The same but for $L=104,\,\delta t=0.55,\,d=21,\,\sigma_{}=0.13$, OBCs, and a max bond dimension 1500. 
    The vertical lines show the times $t_2=16.5$ and $t_3=24.75$ at which the skewness in Fig.~\ref{i_s:fig:1wp_vs_2wp} is calculated.}
    \label{i_s:fig:skewness_mps}
\end{figure}

Figure~\ref{i_s:fig:skewness_mps} compares the skewness in MPS simulations of elastic $(k_0=0.18\pi)$ and inelastic $(k_0=0.32\pi)$ scattering. 
Despite small fluctuations coming from the discrete nature of the data, a clear trend is identified distinguishing the elastic and inelastic channels.
The skewness is consistent with 0 before the collision for both momenta. After the collision $(t \approx 12-15)$, the skewness increases for inelastic scattering and stays close to zero for elastic scattering. 
This reinforces the use of skewness of the energy density as a quantitative metric to detect inelastic effects shortly after the collision.
Fig.~\ref{i_s:fig:skewness_mps}a) shows the skewness in a large system with PBCs and larger wavepackets, while b) replicates the simulations that were ran on {\tt ibm\_marrakesh} for Figs.~\ref{i_s:fig:ibm_results} and \ref{i_s:fig:1wp_vs_2wp}.
Differences between Fig.~\ref{i_s:fig:skewness_mps}a) and b) are primarily due to boundary effects and Trotter errors.

\section{Bounds on the success probability and infidelity when initializing \texorpdfstring{$|W(k_0)\rangle$}{}}
\label{i_s:app:psuccess}
\subsection{Success probability}
\noindent
The probability that the MCM-FF protocol in App.~\ref{i_s:sec:WKprep} succeeds is the probability of measuring odd parity in the state given in Eq.~\eqref{i_s:eq:psi0}.
This is,
\begin{align}
p_{\text{success}} \ &= \ \frac{1}{2}\left (1 - \langle(-1)^{\sum_{n=0}^{\frac{d}{2}-1} \hat{Z}_{2n+1}} \rangle\right ) \ = \ \frac{1}{2}\left (1 - \prod_{n=0}^{\frac{d}{2}-1}\langle(-1)^{ \hat{Z}_{2n+1}} \rangle\right ) \nonumber \\ 
&= \ \frac{1}{2} - \frac{1}{2}\prod_{n=0}^{\frac{d}{2}-1}\left [ 1-2\delta(c_{2n}^2 + c_{2n+1}^2) \right ] \ .
\label{i_s:eq:psuccess2}
\end{align}
The second equality follows from the state being a tensor product.
It is convenient to define $p_n = \delta(c_{2n}^2+c_{2n+1}^2)$ with $\sum_{n=0}^{\frac{d}{2}-1}p_n = \delta$.
We assume $\delta\leq 1$.
Lower bounding $p_{\text{success}}$ is equivalent to obtaining an upper bound for $\prod_{n=0}^{d/2 - 1}\left (1-2p_n\right )$.
Consider instead bounding the log,
\begin{equation}
\log\left [\prod_{n=0}^{d/2 - 1}\left (1-2p_n\right ) \right ] \ = \ 
\sum_{n=0}^{d/2 - 1}\log\left (1-2p_n\right ) .
\end{equation}
An upper bound can be obtained by noticing that $\log(1-2p)$ is concave and therefore Jensen's inequality can be applied.
Jensen's inequality states that for concave functions $f(p)$,
\begin{equation}
\sum_n w_nf(p_n)\ \leq \  f\left [\sum_n w_n p_n\right ] \ \ , \ \ \sum_n w_n = 1 \ .
\end{equation}
Identifying $w_n = 2/d$ and $f(p) = \log(1-2p)$ gives,
\begin{equation}
\frac{2}{d}\sum_{n=0}^{d/2 - 1}\log\left (1-2p_n\right ) \  \leq \ \log\left (1-\frac{4}{d}\sum_{n=0}^{d/2 - 1} p_n\right ) \ = \  \log\left (1-\frac{4 \delta }{d}\right ) \ .
\end{equation}
Exponentiating both sides gives,
\begin{equation}
\prod_{n=0}^{d/2 - 1}\left (1-2p_n\right ) \  \leq \   \left (1-\frac{4 \delta }{d}\right )^{d/2} \ .
\end{equation}
Inserting this into Eq.~\eqref{i_s:eq:psuccess2} it is easy to show that 
\begin{equation}
p_{\text{success}}  \ \geq  \ \frac{1}{2}\left (1-e^{-2\delta} \right )  \ \geq \ 0.43 \delta
\end{equation}
for $\delta \leq 1$.

\subsection{Infidelity}
\noindent
The infidelity of $|W(k_0)\rangle$ prepared with the constant depth MCM-FF circuit is given in Eq.~\ref{i_s:eq:infidelity_full}.
Defining $p_n = c_{2n}^2 + c_{2n+1}^2$ gives,
\begin{align}
{\cal I} \ &= \ \frac{\delta^2}{6}\left [1\ - \ 3\sum_{n=0}^{d/2-1}p_n^2  \ +  \ 2 \sum_{n=0}^{d/2-1}p_n^3\right ]  \ + \ {\cal O}(\delta^3) \nonumber \\
&= \ \frac{\delta^2}{6}\sum_n p_n(1-p_n)(1-2p_n) \ + \ {\cal O}(\delta^3) \nonumber \\
&\equiv \ \frac{\delta^2}{6}\sum_n f(p_n) \ + \ {\cal O}(\delta^3)
\ . 
\end{align}
The second line uses $\sum_n p_n=1$ and the third line defines $f(p_n)=p_n(1-p_n)(1-2p_n)$.
First, consider the case where all $p_n\leq 1/2$ and $f(p_n)$ is concave.
Applying Jensen's inequality gives,
\begin{align}
\sum_{n=0}^{d/2-1} \frac{2}{d} f(p_n) \ \leq \ f\left (\sum_{n=0}^{d/2-1}\frac{2}{d}p_n \right ) \ = \ f\left ( \frac{2}{d}\right ) \ .  
\end{align}
Rearranging gives the desired inequality
\begin{align}
\sum_{n=0}^{d/2-1} f(p_n) \ \leq \ \frac{d}{2} f\left ( \frac{2}{d}\right ) \ = \  \frac{(d-2)(d-4)}{d^2} \ .   
\end{align}
The remaining case is where one of the probabilities is $p_m> 1/2$. 
There can only be one since $\sum_n p_n = 1$.
Applying Jensen's inequality to the $p_{n\neq m}$ gives,
\begin{align}
\sum_{n=0}^{d/2-1} f(p_n) \ \leq \ f(p_m) \ + \  (d/2 - 1) f\left ( \frac{1-p_m}{d/2-1}\right )   \ \leq \  \frac{(d-2)(d-4)}{d^2}\ ,   
\end{align}
where the second inequality can be verified from direct computation.
Therefore, in both cases the upper bound on the infidelity is
\begin{align}
{\cal I} \ \leq \ \ \frac{\delta^2}{6}\frac{(d-2)(d-4)}{d^2} \ + \ {\cal O}(\delta^3) \ .
\
\end{align}
This bound is saturated for uniform $p_n$, i.e. for W states.

\section{Limitations and generalizations of the wavepacket preparation method}
\label{i_s:app:WP_efficient}
\noindent
The wavepacket preparation algorithm begins by initializing a state $|W(k_0)\rangle$ with the  momentum content and quantum numbers of the target wavepacket.
This establishes the correct amplitudes in each block of the Hamiltonian as illustrated in Fig.~\ref{i_s:fig:WPOverview}.
Minimizing the energy with symmetry preserving circuits then projects each of these blocks to its lowest energy state.
This protocol prepares wavepackets composed of the particles that have the lowest energy for their quantum numbers.
Due to the energy minimization, it cannot prepare higher energy particles that share the same quantum numbers.
For example, in $1+1$D Ising field theory, our wavepacket preparation algorithm cannot prepare a wavepacket of heavy $|2\rangle$ particles since there is no quantum number distinguishing them from the light $|1\rangle$ particles.
If there is access to the light-particle wavefunction $|\psi_k\rangle$ via e.g. MPS, then this limitation can be circumvented by minimizing the energy with the Hamiltonian modified by a chemical potential $\hat{H}\to \hat{H}+\mu|\psi_k\rangle\langle\psi_k|$.

In principle, there always exists a symmetry preserving unitary (circuit) that connects the W state to the target wavepacket.
One example is adiabatic time evolution if the target Hamiltonian is connected via an adiabatic path to a Hamiltonian with low-energy excitations that are W states (like the Ising Hamiltonian with zero transverse field or the 1D $XY$ Hamiltonian). 
However, there could be other circuits, distinct from adiabatic evolution, and
the generality of our method depends on the effectiveness of ADAPT-VQE at finding these circuits.
As ADAPT-VQE is a heuristic algorithm, this can likely only be addressed by testing different operator pools and looking at, e.g., the convergence in energy or, if the exact wavefunction is known, the fidelity.
A key design choice in ADAPT-VQE is the operator pool from which the ansatz is constructed.
The choice of pool operators is guided by the symmetries and hierarchies in correlations of the target wavefunction.
In this work, we use operator pools that are generated from the Hamiltonian Lie algebra as they already encode all of the relevant symmetries.
This approach performs well for Ising field theories and scalar field theory but runs into convergence problems in the Schwinger model, see App.~\ref{i_s:app:moreCircuits}.
From our statevector simulations of the Schwinger model, we believe this is due to a lack of expressivity of the chosen operator pool.

Going beyond the Schwinger model to higher-dimensional lattice gauge theories is essential for simulating scattering in systems that more closely resemble quantum chromodynamics.
In principle, we expect that our wavepacket preparation algorithm can be extended to this case with a few important changes.
First, the structure of the initial W state will need to be changed.
In the Schwinger model, we started from a wavepacket of strong coupling hadrons, which is connected to the W state by the circuit shown in Fig.~\ref{i_s:fig:SchwingerWP}.
The initial W state encodes the momentum information of the target state, and the rest of the circuit puts in the other quantum numbers of the massive photon (charge conjugation odd, electric charge zero).
In analogy to the Schwinger model, W-like states can still be used to establish the momentum space structure in a more complicated theory.
In higher-dimensional theories, the angular momentum of the target wavepacket will need to be encoded in the initial W state, and the energy-minimization circuits will need to conserve angular momentum.
In a non-abelian lattice gauge theory, the initial state will be in the singlet representation of the gauge group, e.g. $SU(3)$.
Establishing an initial non-abelian singlet could be much more complicated than in the $U(1)$ Schwinger model, since the constraints do not commute.

After initializing a state with the correct quantum numbers, the energy is minimized using a quantum circuit.
The ADAPT-VQE pool operators for a higher dimensional lattice gauge theory will be more complex than in the Schwinger model.
The pool operators chosen for the Schwinger model were commutators of the staggered mass terms and $e^{+}e^{-}$ creation/annihilation operators separated by different distances.
In higher dimensions, where dynamical gauge fields are present, it will be necessary to consider a variety of color singlet operators in the pool.
The pool will contain components with fermion creation/annihilation operators connected by a Wilson line, as well as components with just the gauge bosons, like Wilson loops.
In the long-term, adiabatic evolution from a W-like state may be preferred if an adiabatic path is known. 
Even if a shallower circuit exists, finding it with ADAPT-VQE may incur too large a sample overhead.
Adiabatic evolution would still have the benefits of creating long-range momentum entanglement in constant depth with the MCM-FF W state preparation circuits.
If an adiabatic path exists with a constant sized gap, then the complete circuit would be constant depth.

\section{Single-particle spectra, inelastic thresholds and wavepacket spreading}
\label{i_s:app:kinematics}
\begin{figure}[t!]
    \centering
    \includegraphics[width=0.75\linewidth]{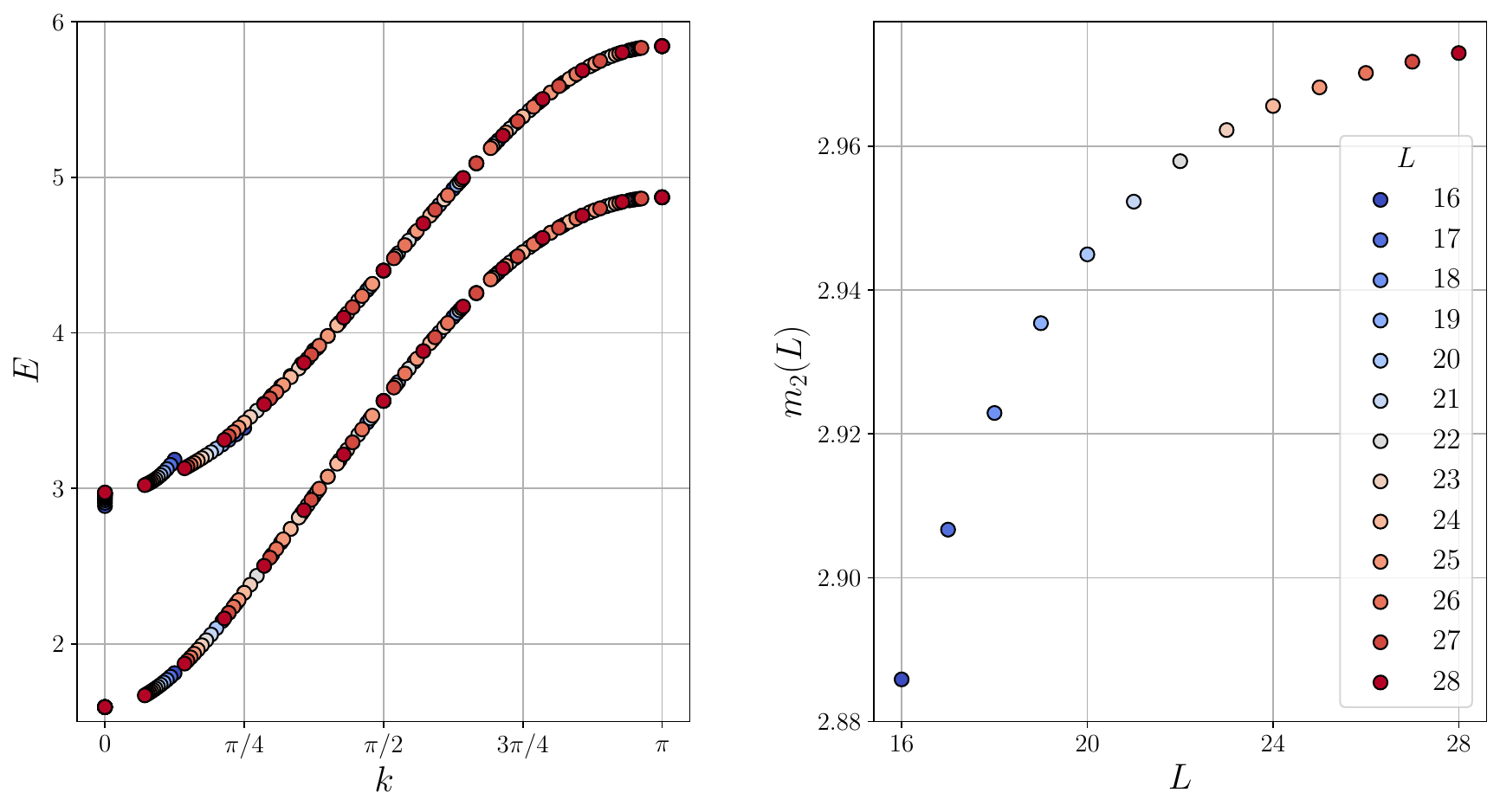}
    \caption{Left: the dispersion relation for particle $|1\rangle$ (lower) and $|2\rangle$ (upper) in one-dimensional Ising field theory determined from exact diagonalization for a range of system sizes.
    Right: the system size dependence of $m_2$.
    All results are for $g_x=1.25$ and $g_z=0.15$.}
    \label{i_s:fig:finite size}
\end{figure}

\noindent
For systems with a mass gap, the single-particle spectrum converges exponentially with increasing $L$ for sufficiently large $L$.
This is verified in the left plot of Fig.~\ref{i_s:fig:finite size} which shows the dispersion relation of particles $|1\rangle$ and $|2\rangle$ in one-dimensional Ising field theory for $g_x=1.25,\,g_z=0.15$ across a range of system sizes.
The smooth behavior of $E(k)$ for $L=\{16,17,\ldots,28\}$ provides strong evidence that $E(k)$ is well converged by $L=28$.
The finite-size effects, while small, are the most pronounced for particle $|2\rangle$ at low momentum.
The right plot of Fig.~\ref{i_s:fig:finite size} shows $m_2(L)$.
Both $m_{1,2}$ are fit well to an exponential of the form 
\begin{equation}
m_{1,2}(L) \ = \  m_{1,2}(\infty)\ + \ ae^{-b L} \ ,
\end{equation}
with $m_1(\infty) = 1.59377803(8)$ and $m_2(\infty) = 2.97682(4)$.  
The value of $m_1(\infty)$ has been confirmed by DMRG.

\begin{figure}[tb!]
    \centering
    \includegraphics[width=0.65\linewidth]{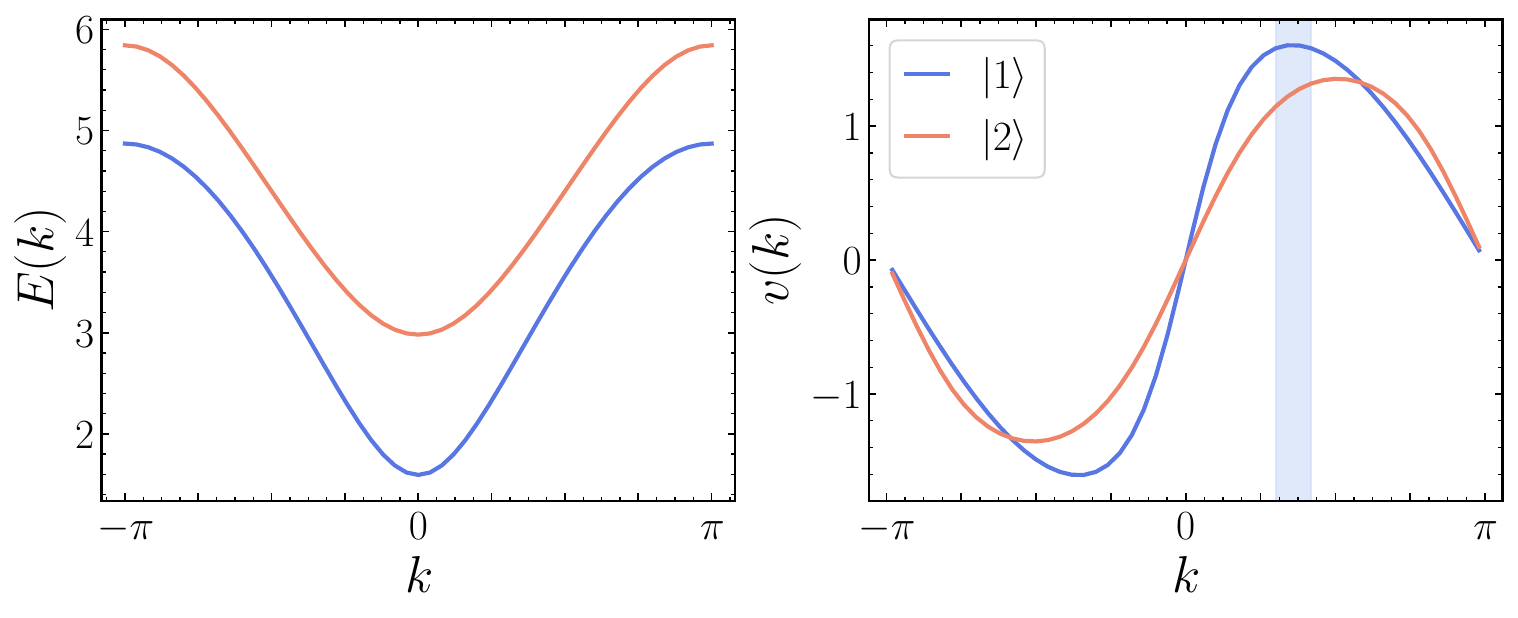}
    \caption{Left: the dispersion relation for particle $|1\rangle$ (blue) and $|2\rangle$ (tan) in one-dimensional Ising field theory with $g_x=1.25,g_z=0.15$. 
    The circles are obtained from exact diagonalization with $L=28$, and the dashed lines are from the approximation $E=\sqrt{m_{1,2}^2+4g_xk^2}$.
    Center: the corresponding group velocity $v(k)=dE/dk$. Right: the group velocity of the outgoing particles $v(k_0')$ as a function of the incoming momentum $k_0$ for the inelastic process $11\rightarrow12$.}
    \label{i_s:fig:dispersion}
\end{figure}

For arbitrary $g_x$ and $g_z$, the exact dispersion relation is unknown.
However, for $g_z=0$, the transverse field Ising model maps to a free massive Majorana fermion, and the exact dispersion relation is given by
\begin{align}
E(k) \ &= \ 2\sqrt{1+g_x^2 - 2g_x\cos{k}}  \nonumber\\
&= \ \sqrt{4(g_x-1)^2 + 4g_xk^2} \ + \ {\cal O}(k^4) \nonumber\\ &\equiv\ \sqrt{m^2 + 4g_xk^2}\ + \ {\cal O}(k^4) \ .
\label{i_s:eq:Isingdispersion}
\end{align} 
In the second equality, the low-energy field theory limit of $k\to 0$ has been taken.
A nonzero $g_z$ will alter the dispersion relation.
An approximation that works quite well for $|k|\lesssim\pi/4$ is the RHS of Eq.~\eqref{i_s:eq:Isingdispersion}  with $m=m_{1,2}$ determined from exact diagonalization.
This is compared to the exact dispersion relation for $L=28$ in the left plot of Fig.~\ref{i_s:fig:dispersion}.
This approximation predicts a group velocity of
\begin{align}
v(k) \ = \ \frac{dE}{dk} \ = \  \frac{4g_x k}{\sqrt{m^2 + 4g_xk^2}}\ ,
\end{align}
and is compared to the lattice group velocity in the center plot of Fig.~\ref{i_s:fig:dispersion}.\footnote{The lattice derivative is computed via the discrete symmetric finite difference $f'(x) \approx \frac{f(x+\epsilon) - f(x-\epsilon)}{2\epsilon}$.}
Low momentum corresponds to ``non-relativistic'' propagation with $v(k)$ increasing linearly with momentum.
At larger momentum there is a plateau in the group velocity corresponding to relativistic propagation with a ``speed of light'' of $c\approx1.6$.
Further increasing the momentum leads to a decrease in the group velocity due to lattice artifacts.

Energy and total momentum are conserved throughout scattering, and schematically the inelastic process $11 \to 12$ is
\begin{align}
|1(k_0)\,\,1(-k_0)\rangle \ \to \ |1(k_0')\,\,2(-k_0')\rangle \ + \ 1\leftrightarrow2 
\end{align}
with $|k_0'| < |k_0|$ and 
\begin{align}
2E_1(k_0) \ = \ E_1(k_0') + E_2(k_0')
\label{i_s:eq:energy_condition_inelastic}
\end{align}
where $E_{1,2}(k)$ is the dispersion relation for particles $|1\rangle$, $|2\rangle$.
Energy and momentum conservation uniquely fix the velocity of the outgoing particles in the $11\to12$ process.
The outgoing velocities as functions of the incoming momenta $k_0$ are shown in the right plot of Fig.~\ref{i_s:fig:dispersion}.
Near $k_\text{thr}$, there is a large difference in the group velocities of the outgoing $|1\rangle$ and $|2\rangle$ particles.
Selecting an incoming momentum $k_0$ in this region ensures that the trajectories of the $|1\rangle$ and $|2\rangle$ particles do not overlap and can be clearly distinguished.
This guides the selection of $k_0$ that has the largest signal of inelastic effects.
The predicted velocities agree well with the MPS simulations of inelastic scattering shown in Fig.~\ref{i_s:fig:2WP_scattering_MPS}.

\begin{figure}
    \centering
    \includegraphics[width=0.9\linewidth]{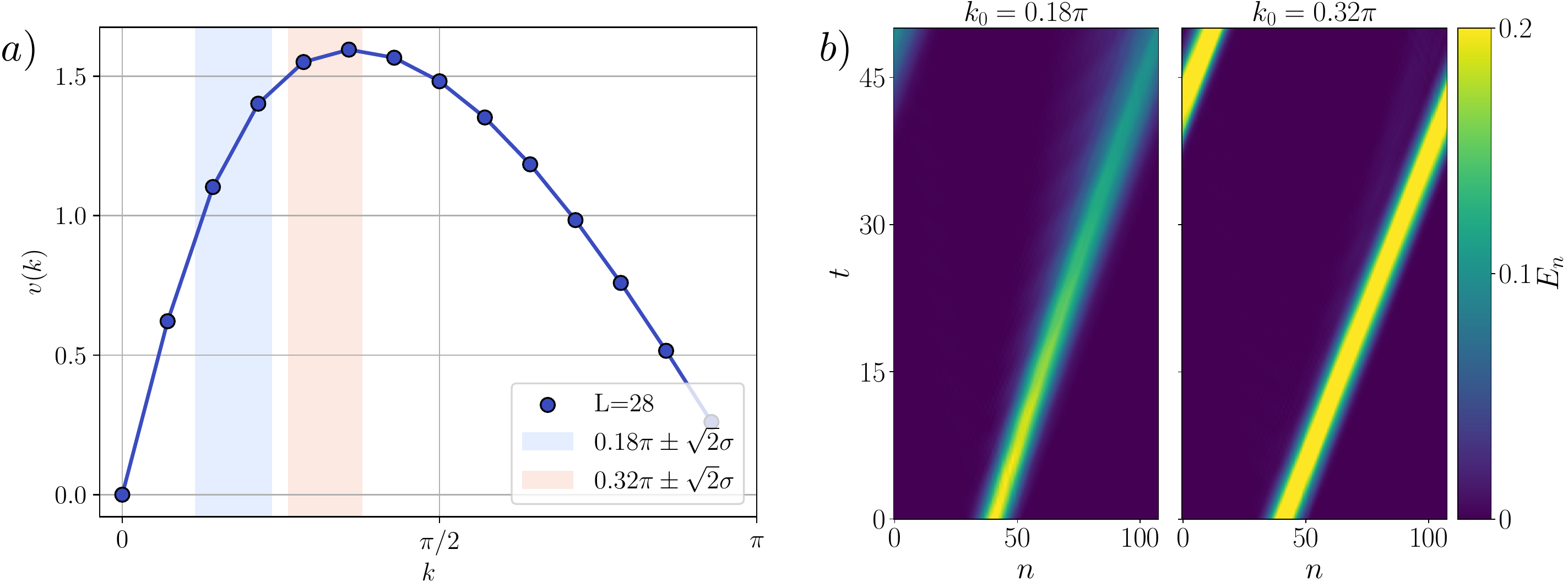}
    \caption{a) The group velocity $v(k)$ of particle $|1\rangle$ for $g_x=1.25,\, g_z=0.15$ and $L=28$ determined from exact diagonalization. 
    The shaded light blue (tan) regions correspond to the support ($\pm$ one standard deviation of the probability) of the wavepackets used to simulate elastic (inelastic) scattering.
    b) The time evolution of the vacuum-subtracted energy density $E_n$ for the propagation of a single wavepacket with $g_x=1.25$,\, $g_z=0.15$ and $L=108$.
    The wavepackets have $\sigma_{}=0.13,\, d=22$ and are constructed from 8 steps of ADAPT-VQE.}
    \label{i_s:fig:PBCWP}
\end{figure}

In App.~\ref{i_s:sec:csimscatt}, elastic and inelastic scattering in Ising field theory were simulated at momentum $k_0 = 0.18\pi$ and $k_0=0.32\pi$ respectively.
The group velocity of the momentum modes making up the two wavepackets is shown in Fig.~\ref{i_s:fig:PBCWP}a).
The shaded regions represent the support of 68\% of the wavefunction probability ($\pm 1$ standard deviation of the gaussian wavepacket).
In these regions, the group velocity varies between $0.9\lesssim v\lesssim1.45$ and $1.5\lesssim v\lesssim1.6$ for $k_0=0.18\pi$ and $k_0=0.32\pi$ respectively.
The large variance in velocity at small momentum causes the wavepacket to spread out as it propagates.
This is illustrated in the left and right plots of Fig.~\ref{i_s:fig:PBCWP}b) which show the propagation of a single wavepacket.
The energy density of the $k_0=0.18\pi$ wavepacket delocalizes under time evolution, whereas the energy density remains focused for $k_0=0.32\pi$.
The wavepacket velocities can also be determined from these plots, and are consistent with dispersion relation estimates of $v(0.18\pi)\approx1.25$ and $v(0.32\pi)\approx1.6$.


The wavepacket parameters $k_0=0.32\pi$ and $\sigma_{}=0.13$ were used in Sec.~\ref{i_s:sec:qsim} to simulate inelastic scattering on IBM's quantum computers.
As seen in Fig.~\ref{i_s:fig:PBCWP}a), one standard deviation in momentum space includes up to $k_0=0.38\pi$.
Scattering at $k_0=0.32\pi$ and $\sigma_{}=0.13$ accesses additional inelastic processes beyond $11\to 12$.
All inelastic processes $11\to X$ with $k_{\text{thr}}\leq0.42\pi$ are given in Table~\ref{i_s:tab:thresholds}.
The thresholds have been estimated from the dispersion relation shown in Fig.~\ref{i_s:fig:dispersion}.
All of these processes occur in superposition for simulations at $k_0=0.32\pi$.
However, the higher-energy processes are suppressed due to the kinematics of the initial state (see right column of Table~\ref{i_s:tab:thresholds}).
Additionally, the $11\to111$ process has a much smaller branching ratio than $11\to12$~\cite{Jha:2024jan}. 
For these reasons, $11\to12$ is the dominant inelastic process for the simulations in this work.

\begin{table}[t]
\centering
\begin{tabularx}{\linewidth}{|c||Y|Y|Y|} \hline
Process & $E_\text{thr}/m_1$ & $k_\text{thr}$ & $P(E_\text{tot}>E_\text{thr})$ \\\hline\hline
$\phantom{2}11\to12\phantom{121}$ & 2.87 & $0.24\pi$ & 0.9504 \\\hline
$\phantom{2}11\to111\phantom{21}$ & 3 & $0.26\pi$ & 0.8645 \\\hline
$\phantom{2}11\to22\phantom{121}$ & 3.75 & $0.38\pi$ & $5.92\times10^{-3}$ \\\hline
$\phantom{2}11\to112\phantom{21}$ & 3.87 & $0.4\pi$ & $7.90\times 10^{-4}$ \\\hline
$\phantom{2}11\to1111\phantom{1}$ & 4 & $0.42\pi$ & $6.99\times 10^{-5}$ \\\hline
\end{tabularx}
\renewcommand{\arraystretch}{1}
\caption{\textit{Kinematic thresholds of inelastic processes.} For each inelastic process (first column), the second column gives the threshold energy $E_\text{thr}$ in units of $m_1$, above which the process is kinematically allowed. 
The third column gives the corresponding momentum threshold $k_\text{thr}$ for a collision of two $|1\rangle$ particles. 
The last column shows the probability that the process is accessed in a collision of two $|1\rangle$ particles with $k_0=0.32\pi$ and $\sigma_{}=0.13$.}
\label{i_s:tab:thresholds}
\end{table}
%

\section{Additional circuits for preparing \texorpdfstring{$|W(k_0)\rangle$}{}}
\label{i_s:app:WP0prep}
\noindent
This appendix presents three additional ways of preparing the $\vert W(k_0) \rangle$ state in Eq.~\eqref{i_s:eq:psiWP02}.
These methods make use of beyond-linear connectivity and/or MCM-FF to reduce the CNOT depth relative to the circuit in the left panel of Fig.~\ref{i_s:fig:IsingWPCircs}.
Each circuit will focus on getting the amplitudes, $c_n$ in Eq.~\eqref{i_s:eq:psiWP02}, correct.
Afterwards, the phases are added with single qubit $R_Z$ rotations $\prod_n e^{-i \phi_n \hat{Z}_n /2}$.

\begin{figure}
    \centering
    \includegraphics[width=\linewidth]{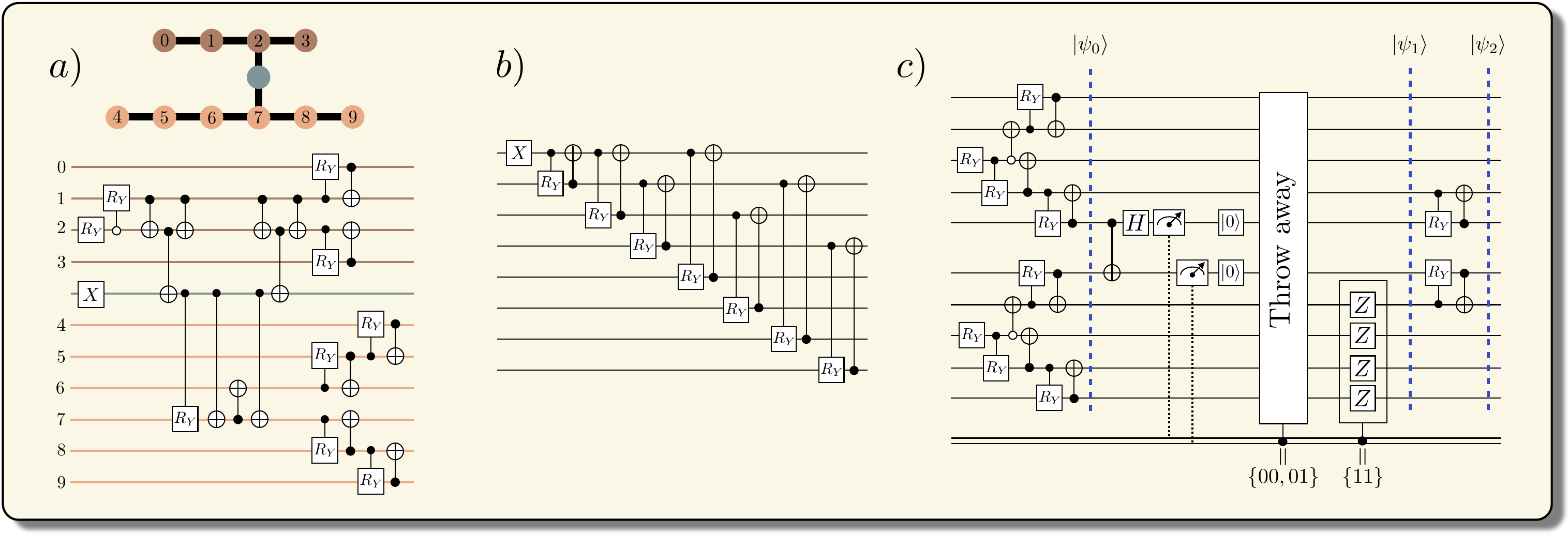}
    \caption{{\it Quantum circuits that prepare $|W(k_0)\rangle$}
    a) A circuit that uses an ancilla (gray) that is available with heavy-hex connectivity to simultaneously build $|W(k_0)\rangle$ on the dark and light brown-colored qubits.
    The qubit mapping is shown on top.
    b) A log-depth circuit that makes use of all-to-all connectivity.
    c) A circuit that makes use of MCM-FF to fuse together two smaller $|W(k_0)\rangle$ states.}
    \label{i_s:fig:WP0Prep}
\end{figure}

IBM's quantum computers natively support a heavy-hex connectivity.
A useful way to view heavy-hex is as a linear chain plus some connections between segments of the chain.
These connections can be used as ancillas to reduce the circuit depth.
A circuit illustrating this is shown in Fig.~\ref{i_s:fig:WP0Prep}a).
The idea is to split $|W(k_0)\rangle$ into two segments above and below the ancilla,
and then simultaneously build $|W(k_0)\rangle$ in both segments using the circuit in the left panel of Fig.~\ref{i_s:fig:IsingWPCircs}.
This strategy is the most efficient for $d = 2 + 4n$, with $n$ an integer. The angles for the $R_Y$ rotations are found by recursively solving the following equations,
\begin{align}
\left [\sin{\left ( \tfrac{\theta_{\eta}}{2}\right )} \right ]^2 \ &= \ \sum_{i=0}^{\eta-1}c^2_{\eta+i} \ , \nonumber \\
\left [\cos{\left ( \tfrac{\theta_{\eta}}{2}\right )}\sin{\left ( \tfrac{\theta_{\eta-1}}{2}\right )} \right ]^2 \ &= \ \sum_{i=0}^{\eta-1}c^2_{i} \ , \nonumber \\
\left [\sin{\left ( \tfrac{\theta_{3\eta+1}}{2}\right )}\cos{\left ( \tfrac{\theta_{\eta}}{2}\right )}\cos{\left ( \tfrac{\theta_{\eta-1}}{2}\right )} \right ]^2 \ &= \ \sum_{i=0}^{\eta}c^2_{3\eta+1+i} \ , \nonumber \\
\sin{\left ( \tfrac{\theta_{\eta}}{2}\right )}\cos{\left ( \tfrac{\theta_{\eta+j+1}}{2}\right )}\prod_{i=1}^j \sin{\left ( \tfrac{\theta_{\eta+i}}{2}\right )} \ &= \ c_{\eta+j} \ , \ j\in[0,1,\ldots,\eta-2] \ , \nonumber \\
\cos{\left ( \tfrac{\theta_{\eta}}{2}\right )}\sin{\left ( \tfrac{\theta_{\eta-1}}{2}\right )}\cos{\left ( \tfrac{\theta_{\eta-j-1}}{2}\right )}\prod_{i=2}^j \sin{\left ( \tfrac{\theta_{\eta-i}}{2}\right )} \ &= \ c_{\eta-j} \ , \ j\in[1,2,\ldots,\eta-1] \ , \nonumber \\
\cos{\left ( \tfrac{\theta_{3\eta+1}}{2}\right )}\cos{\left ( \tfrac{\theta_{\eta}}{2}\right )}\cos{\left ( \tfrac{\theta_{\eta-1}}{2}\right )}\cos{\left ( \tfrac{\theta_{3\eta-j-1}}{2}\right )}\prod_{i=1}^j \sin{\left ( \tfrac{\theta_{3\eta-i}}{2}\right )} \ &= \ c_{3\eta-j} \ , \ j\in[0,1,\ldots,\eta-1] \ , \nonumber \\
\sin{\left ( \tfrac{\theta_{3\eta+1}}{2}\right )}\cos{\left ( \tfrac{\theta_{\eta}}{2}\right )}\cos{\left ( \tfrac{\theta_{\eta-1}}{2}\right )}\cos{\left ( \tfrac{\theta_{3\eta+j+1}}{2}\right )}\prod_{i=2}^j \sin{\left ( \tfrac{\theta_{3\eta+i}}{2}\right )} \ &= \ c_{3\eta+j} \ , \ j\in[1,2,\ldots,\eta] \ ,
\end{align}
with $\eta \equiv (d-2)/4$.
The circuit depth is 
\begin{equation}
\text{CNOT depth: }\ 7+2\left \lceil \frac{d-2}{4}\right \rceil \ ,
\end{equation}
which is asymptotically half of the depth of the circuit in the left panel of Fig.~\ref{i_s:fig:IsingWPCircs}.

The state $|W(k_0)\rangle$ can be prepared with a log-depth circuit if the target connectivity is all-to-all.
This circuit is shown in Fig.~\ref{i_s:fig:WP0Prep}b) and generalizes the construction in Ref.~\cite{Cruz_2019}.
The set of equations relating the rotation angles $\{\theta\}$ of the controlled-$R_Y$ gates and the amplitudes $c_n$ can be determined from a $d\times d$ matrix $\Theta_{i,j}$.
The following pseudocode constructs this matrix:
\begin{lstlisting}[language=Python,mathescape=true]
$\Theta_{i,j}=1 \quad \forall \, i,j$
for $n$ in $[0,\ldots,d-1]$:
    for $j$ in $[0,\ldots,\min(2^n,d-2^n)-1]$:
        $\Theta_{j,2^n+j} = \cos{ \left( \frac{\theta_{2^n+j}}{2}\right) }$
        $\Theta_{2^n+j,2^n+j} = \sin{ \left( \frac{\theta_{2^n+j}}{2}\right) }$
for $n'$ in $[1,\ldots,d-1]$:        
    for $n$ in $[n',\ldots,d-1]$:
        for $j$ in $[0,\ldots,\min(2^n,d-2^n)-1]$:
            for $j'$ in $[j,j-2,\ldots,0]$:
                $\Theta_{2^n+j,j'} = \Theta_{2^{n-n'}+j-1,2^{n-n'}+j'-1}$
                $\Theta_{2^n+j,2^{n-n'}+j'} = \Theta_{j,2^{n-n'}+j'}$
\end{lstlisting}
The set of equations relating the $c_n$ and angles are found by multiplying each row, $c_i=\prod_{j=0}^{d-1}\Theta_{i,j}$.
This circuit has a circuit depth of,
\begin{equation}
\text{CNOT depth: }2\lceil \log_2(d)\rceil \ .
\end{equation}

Two mid-circuit measurements and one feedforward operation can be used to reduce the CNOT depth required to prepare $|W(k_0)\rangle$.
Inspired by the GHZ state preparation circuit in Ref.~\cite{Smith:2022nbd},
the idea is to split $|W(k_0)\rangle$ into two halves (assuming $d$ is even)
\begin{equation}
\sum_{n=0}^{d-1}  c_n |2^n\rangle  \ \rightarrow \ \left (\sum_{n=0}^{\frac{d}{2}-1}  a_n |2^n\rangle \right )\otimes\left (\sum_{n=0}^{\frac{d}{2}-1} b_n |2^n\rangle \right ) \ ,
\end{equation}
and fuse the two halves together with a Bell measurement followed by a feedforward operation.
One of the unitary constructions presented above can be used to prepare each half, and it will be shown that this method succeeds with probability $p_{\text{success}}=1/2$.

As an illustrative example, consider preparing a $d=6$ site wavepacket from fusing together two 3 site halves.
A circuit with the same structure extending this to the fusion of 5-site halves is shown in Fig.~\ref{i_s:fig:WP0Prep}c).
The input state $|\psi_0\rangle$ is
\begin{align}
\vert \psi_0\rangle\ = \  \left ( a_0|001\rangle + a_1|010\rangle + a_2|100\rangle \right )\otimes \left (b_0|001\rangle + b_1|010\rangle + b_2|100\rangle \right ) \ ,
\label{i_s:eq:mcminit}
\end{align}
with real coefficients that satisfy $a_0^2 + a_1^2 + a_2^2 = b_0^2 + b_1^2 + b_2^2 =1$.
Next, a Bell measurement is performed on qubits $q_3q_2$.
The states obtained conditioned on each possible Bell measurement outcome are given in Table~\ref{i_s:tab:Wmcm}.
\begin{table}[t]
\setlength{\tabcolsep}{1pt}
\begin{tabularx}{\linewidth}{|c||Y|c|}
\hline
     Meas. & State  & Probability \\
     \hline
     \hline
     $00$ & $a_0 b_2 |000\rangle |000\rangle  + \left ( a_1|010\rangle + a_2|100\rangle \right) \left (b_0|001\rangle + b_1|010\rangle \right )$ & $\frac{1}{2}\left (1-a_0^2-b_2^2+2a_0^2 b_2^2\right )$\\ 
     \hline 
     $01$ & $-a_0 b_2 |000\rangle |000\rangle  + \left ( a_1|010\rangle + a_2|100\rangle \right) \left (b_0|001\rangle + b_1|010\rangle \right )$ & $\frac{1}{2}\left (1-a_0^2-b_2^2+2a_0^2 b_2^2\right )$\\
     \hline 
     $10$ & $ a_0 |000\rangle \left (b_0|001\rangle + b_1 |010\rangle  \right ) + b_2\left (a_1|010\rangle + a_2 |100 \rangle \right ) |000\rangle$ & $\frac{1}{2}\left (a_0^2 + b_2^2 - 2a_0^2 b_2^2\right )$\\ 
     \hline 
     $11$ & $a_0 |000\rangle \left (b_0|001\rangle + b_1 |010\rangle  \right ) - b_2\left (a_1|010\rangle + a_2|100\rangle \right ) |000\rangle $ & $\frac{1}{2}\left (a_0^2 + b_2^2 - 2a_0^2 b_2^2\right )$\\ 
     \hline
\end{tabularx}
\caption{The (unnormalized) state (middle column) obtained after a particular measurement outcome of qubits $q_2$ and $q_3$ (left column).
The state being measured is given in Eq.~\eqref{i_s:eq:mcminit} and
qubits $q_2$ and $q_3$ are reset after measurement.
The probability of each measurement outcome is given in the right column.}
\label{i_s:tab:Wmcm}
\end{table}
After the measurement, the qubits $q_3q_2$ are reset to $|00\rangle$.
The protocol has succeeded for measurement outcomes $ q_3 q_2  = \{10,11 \}$ and failed for $ q_3 q_2  = \{00, 01 \}$.
A feedforward of $\hat{Z}_4 \hat{Z}_5$ is applied if $ q_3 q_2 = \{11 \}$, and the resulting state $|\psi_1\rangle$ has the structure of $|W(k_0)\rangle$ on the unmeasured qubits,
\begin{align}
|\psi_1\rangle \ = \  \frac{1}{\sqrt{a_0^2 + b_2^2 -2a_0^2 b_2^2}}\left (a_0b_0|000001\rangle + a_0b_1 |000010\rangle   + b_2a_1|010000\rangle + b_2a_2 |100000 \rangle  \right ) \
 .
\label{i_s:eq:psi1}
\end{align}
From Table~\ref{i_s:tab:Wmcm}, the probability of success is $p_{\text{success}} = p_{10} + p_{11} = a_0^2 + b_2^2 -2a_0^2 b_2^2$.
This has a maximum of $p_{\text{success}} =1/2$ when $a_0^2 = b_2^2 = 1/2$ which can be shown to correspond to a wavepacket where $\sum_{n=0}^{\frac{d}{2}-1} c_n^2 \ = \ \sum_{n=0}^{\frac{d}{2}-1} c_{\frac{d}{2}+n}^2 $.
This condition is always satisfied for even $d$ since the magnitudes of Gaussian wavepackets are symmetric about their midpoints.
Because of this, the remaining expressions will assume $a_0^2 = b_2^2 = 1/2$.
The last step is to perform a controlled-$R_Y$ CNOT sequence between qubits $q_1 q_2$ and $q_4 q_3$, giving
\begin{align}
|\psi_2\rangle \ =& \  b_0|000001\rangle + b_1\cos{\left (\frac{\theta_2}{2}\right )} |000010\rangle + b_1\sin{\left (\frac{\theta_2}{2}\right )}|000100\rangle  \nonumber \\
&+ a_1\sin{\left (\frac{\theta_3}{2}\right )}|001000\rangle + a_1\cos{\left (\frac{\theta_3}{2}\right )}|010000\rangle + a_2 |100000 \rangle  \ ,
\label{i_s:eq:psi2}
\end{align}
where $\theta_2$ and $\theta_3$ are the value of the $R_Y$ rotation angles.
$|W(k_0)\rangle$ is prepared by identifying
\begin{align}
&b_0=c_0 \ , \quad b_1^2 = c_1^2 + c_2^2 \ , \quad  \tan{\left ( \frac{\theta_2}{2}\right )} = \frac{c_2}{c_1} \ , \nonumber \\
&a_1^2 = c_3^2 + c_4^2 \ , \quad   \tan{\left ( \frac{\theta_3}{2}\right )} = \frac{c_3}{c_4} \ , \quad  a_2 = c_5 \ .
\end{align}

This method generalizes to any even $d$ and succeeds with a probability of success $p_{\text{success}}=1/2$.
The procedure is:
\begin{itemize}
    \item[1.] Prepare the initial state $ \left (\sum_{n=0}^{\frac{d}{2}-1}  a_n |2^n\rangle  \right )\otimes \left (\sum_{n=0}^{\frac{d}{2}-1} b_n |2^n\rangle \right )$ using one of the unitary $|W(k_0)\rangle$ preparation circuits presented previously.
    The coefficients are given by
    \begin{align}
        b_n \ &= \ c_n \ , \ \  n= \{0,1,\ldots,\frac{d}{2}-3\} \ , \nonumber \\ 
        a_n \ &= \ c_{\frac{d}{2}+n} \ , \ \  n= \{2,3,\ldots,\frac{d}{2}-1\} \ , \nonumber \\
        a_{0} \ &= \ b_{\frac{d}{2}-1} = \frac{1}{\sqrt{2}} \ , \nonumber \\
        a_1^2 \ &= \ c^2_{\frac{d}{2}} + c^2_{\frac{d}{2} + 1 } \ ,  \nonumber \\
        b_{\frac{d}{2}-2}^2\  &= \ c^2_{\frac{d}{2}-1} + c^2_{\frac{d}{2} -2 } \ .
    \end{align}
    \item[2.] Measure the qubits $q_{d/2} q_{d/2 - 1}$ in the Bell basis. 
    If the measurement outcome is $q_{d/2 }q_{d/2 - 1} =\{00,01 \}$, restart the procedure.
    If the measurement outcome is $q_{d/2 }q_{d/2 - 1} =\{11 \}$,
    feed forward $\prod_{i=d/2+1}^{d-1}\hat{Z}_{i}$. If the measurement outcome is $q_{d/2 }q_{d/2 - 1} =\{10 \}$, do nothing. 
    \item[3.] Reset $q_{d/2 }q_{d/2 - 1}$ to $|00\rangle$.
    \item[4.]
    Apply a controlled-$R_Y$ CNOT sequence between $q_{d/2-2}q_{d/2-1}$ and $q_{d/2+1}q_{d/2}$. 
    The $R_Y$ rotation angles are given by 
    \begin{equation}
        \tan\left (\frac{\theta_{\frac{d}{2}-1}}{2} \right ) =\frac{c_{\frac{d}{2}-1}}{c_{\frac{d}{2}-2}} \ \ , \ \ \tan\left (\frac{\theta_{\frac{d}{2}}}{2} \right ) =\frac{c_{\frac{d}{2}}}{c_{\frac{d}{2}+1}} \ .
    \end{equation}
\end{itemize}

Table~\ref{i_s:tab:summary_W} compares the scaling of the resources required to prepare W states using the methods in this chapter to other methods in the literature.

\begin{table}
\setlength{\tabcolsep}{1pt}
\footnotesize
\begin{tabularx}{\linewidth}{|c||Y|Y|Y|Y|Y|Y|Y|}
\hline
     Reference & Qubits & Connectivity & CNOT count & CNOT depth & MCM & $p_{\text{success}}$ & ${\cal I}$ \\
     \hline
     \hline
      Ref.~\cite{Cruz_2019} & $d$ & linear & $2d$ & $2d$ & 0 & 1 & 0 \\
     \hline
     Fig.~\ref{i_s:fig:IsingWPCircs} & $d$ & linear & $2d$ & $d$ & 0 & 1 & 0 \\ \hline
     Fig.~\ref{i_s:fig:WP0Prep}c) & $d$ & linear & $2d$& $d/2$ & 2 & 1/2 & 0 \\ \hline
     Ref.~\cite{Piroli:2024ckr} & $2d$ & linear & $13d/4$ & 11 & $3d/4$ & $\geq0.43 \,\delta$ & ${\cal O}(\delta^2)$ \\ 
     \hline
     Fig.~\ref{i_s:fig:ConstantDepth} & $d$ & linear & $17d/4$& $13$ & $3d/4$ & $\geq0.43\,\delta$ & ${\cal O}(\delta^2)$ \\ 
     \hline
    Fig.~\ref{i_s:fig:WP0Prep}a) & $d+1$ & heavy-hex & $2d$ &  $d/2$ & 0 & 1 & 0 \\ 
     \hline
    Ref.~\cite{Cruz_2019}, Fig.~\ref{i_s:fig:WP0Prep}b) & $d$ & all-to-all & $2d$ & $2\log_2(d)$ & 0 & 1 & 0 \\ \hline
\end{tabularx}
\caption{Summary of the scaling of resources (qubits, connectivity, CNOT gates, mid-circuit measurements, success probability $p_\text{success}$ and infidelity $\mathcal{I}$) needed to prepare the W state on $d$ qubits.
The values for the constant-depth circuits have been adjusted for a device with linear connectivity.}
\label{i_s:tab:summary_W}
\end{table}
%

\section{Quantum circuits for preparing wavepackets in other lattice quantum field theories}
\label{i_s:app:moreCircuits}

\subsection{One-dimensional scalar field theory}
\label{i_s:app:scalar}
\noindent
Historically, scalar field theory has served as a sandbox for understanding aspects of QFTs, with many applications to (beyond) Standard Model physics.
It describes the dynamics of the Higgs Boson~\cite{Higgs:1964pj}, as well as particles that emerge from the spontaneous breaking of exact and approximate symmetries, such as the pseudoscalar mesons~\cite{Gasser:1983yg} and the axion~\cite{PhysRevLett.40.279,PhysRevLett.40.223}.
Quantum simulations of scattering in scalar field theory were first laid out in two seminal papers~\cite{Jordan:2011ci,Jordan:2012xnu}.
These works gave protocols for simulating scattering in scalar field theory on quantum computers with an exponential quantum advantage.
Scattering in one-dimensional scalar field theory has also been shown to be BQP-complete~\cite{Jordan_2018}, meaning that any efficient computation on a quantum computer can be mapped to scattering in scalar field theory with a polynomial overhead.

There have been numerous efforts and proposals toward quantum simulations of scattering in scalar field theory \cite{Klco:2019xro,Klco:2019yrb,Klco:2020aud,Kurkcuoglu:2021dnw,Macridin:2021uwn,Liu:2021otn,Abel:2025zxb,Illa:2022jqb, Li:2022ped,Hardy:2024ric,Kreshchuk:2023btr,Briceno:2023xcm,Turco:2023rmx}.
Recently, the first such simulation was performed on 120 qubits of IBM's quantum computers~\cite{Zemlevskiy:2024vxt}.
That work used variational optimization in every stage of the quantum simulation to minimize the circuit depth.
Building off the techniques in Refs.~\cite{Farrell:2023fgd,Farrell:2024fit}, scalable circuits that prepared wavepackets were found by minimizing the infidelity with the exact wavepacket using classical computers.
One challenge was that there were no physics-informed constraints on the structure of the wavepacket preparation circuits, making the optimization difficult.
Additionally, because it relied on minimizing the infidelity, the wavepacket size was constrained by the limitations of exact statevector simulations.
The wavepacket preparation method described in this section overcomes both of these difficulties by:
1) relying on a minimization of a local observable (the energy), not the global infidelity, and 2) only having to optimize over quantum circuits whose structure is heavily constrained by symmetries.
This section begins with a background for scalar field theory on the lattice.

The Hamiltonian for $\lambda \hat{\phi}^4$ scalar field theory on a one-dimensional lattice is
\begin{align}
\hat{H} \ &= \ \sum_{n=0}^{L-1}\left (\frac{1}{2}m_0^2 \hat{\phi}_n^2 \ + \ \frac{1}{2}\hat{\Pi}_n^2 \ + \ \frac{1}{2}\left (  \hat{\phi}_{n+1}-\hat{\phi}_n \right )^2 \ +\ \frac{\lambda}{4!}\hat{\phi}_n^4\right ) \nonumber \\
&\equiv\ \sum_{n=0}^{L-1}\left (\hat{H}_{n}^{(\text{h.o.})} \ - \ \hat{\phi}_{n}\hat{\phi}_{n+1}  \ +\  \frac{\lambda}{4!}\hat{\phi}_n^4\right ) \ ,
\end{align}
where $m_0$ is the bare mass and $\lambda$ is the coupling strength.
For later use, the second line has separated out the single-site harmonic oscillator Hamiltonian,
\begin{align}
\hat{H}_n^{(\text{h.o.})} \ = \ \frac{1}{2}\left (m_0^2 + 2 \right )\hat{\phi}_n^2 + \frac{1}{2} \hat{\Pi}_n^2 \ .
\label{i_s:eq:Hscalar1}
\end{align}
The bosonic field operator $\hat{\phi}_n$ and its conjugate momentum $\hat{\Pi}_n$ satisfy the canonical commutation relations $[ \hat{\phi}_n, \hat{\Pi}_m ]=i\delta_{n,m}$.
This Hamiltonian has a $Z_2$ symmetry that takes $\hat{\phi}_n \to -\hat{\phi}_n$ (and $\hat{\Pi}_n \to -\hat{\Pi}_n$). The vacuum is even under $Z_2$, whereas the lowest-energy single-particle states are odd.

The bosonic nature of scalar particles means the Hilbert space on each lattice site is infinite-dimensional and identical to that of a quantum harmonic oscillator. To simulate this theory on a quantum computer, the local Hilbert space must be made finite.
This requires choosing a basis to represent the fields, and then truncating the number of basis states.
Two bases will be used in this work: the harmonic oscillator basis composed of eigenstates of $\hat{H}^{(\text{h.o.})}$, $\vert E_j\rangle^{(\text{h.o.})}$,
and the field basis where $\hat{\phi} \vert \phi_j\rangle = \phi_j\vert \phi_j\rangle$.
To minimize confusion, $n$ will be used as a position space index, and $j$ will be used as an index labeling states in the local Hilbert space at each lattice site $n$.
Once a basis is chosen, the local Hilbert space is truncated to $2^{n_q}$ states and the degrees of freedom of each site are mapped onto $n_q$ qubits. 

The prescription for digitizing scalar field theory in the field basis was described in Refs.~\cite{Jordan:2011ci, Jordan:2012xnu}.
The eigenvalues of the field operator are restricted to the interval $[-\phi_{\text{max}},\phi_{\text{max}}]$ and uniformly sampled at intervals $\delta_\phi$,
\begin{align}
\phi_j = -\phi_{\text{max}} + j \delta_{\phi} \ \ , \ \ \delta_{\phi} = \frac{2 \phi_{\text{max}}}{2^{n_q}-1} \ \ , \ \ j\in[0,2^{n_q}-1] \ .
\end{align}
The basis where $\hat{\Pi}$ is diagonal is related to the $\phi$-basis by a site-wise Fourier transform. 
By the Nyquist-Shannon sampling theorem, field-space digitization errors are minimized when the eigenvalues of $\hat{\Pi}$, $\Pi_j$ are also sampled uniformly,
\begin{align}
\Pi_j \ = \ -\frac{\pi}{\delta_{\phi}} + \left (j+\frac{1}{2} \right )\frac{2 \pi}{2^{n_q} \delta_{\phi}} \ \ , \ \ j\in[0,2^{n_q}-1] \ .
\label{i_s:eq:Pij}
\end{align}
In the field basis, there is a simple mapping from fields to spin operators:
\begin{align}
\hat{\phi}_n \ = \ - \frac{\phi_{\text{max}}}{2^{n_q} - 1} \sum_{j=0}^{n_q - 1}2^j \hat{Z}_{n_q  n + j} \ ,
\end{align}
and
\begin{align}
\hat{\Pi}_n \ = \ \hat{V}^{\dagger}\left (- \frac{\pi}{2^{n_q} \delta_{\phi}} \sum_{j=0}^{n_q - 1}2^j \hat{Z}_{n_q  n + j} \right )\hat{V} \ ,
\end{align}
where $\hat{V}$ is the $2^{n_q}\! \times \! 2^{n_q}$ symmetric Fourier transform~\cite{Klco:2018zqz} matrix $V_{kj} = \frac{1}{2^{n_q/2}}e^{i {\bf k} {\bf j}}$, with ${\bf k} = {\bf \Pi}\delta\phi$ and ${\bf \Pi}$ defined in Eq.~\eqref{i_s:eq:Pij}.
Importantly, the Fourier transform can be implemented with ${\cal O}(n_q)$ circuit depth~\cite{Klaver:2024vkw}, making this construction efficient for quantum simulation. 
There is one free parameter, $\phi_{\text{max}}$, which is chosen to minimize the field digitization errors.
This can be done by ensuring that the maximum energy of the $\hat{\phi}$ and $\hat{\Pi}$ terms in the single-site Hamiltonian $\hat{H}^{(\text{h.o.})}+\lambda/4! \,\hat{\phi}^4$, are equal.
This constraint gives an optimal $\phi_{\text{max}}$ that is the solution to the cubic equation\footnote{This was determined numerically in Refs.~\cite{Klco:2018zqz,Zemlevskiy:2024vxt}, and analytically for $\lambda=0$ in Ref.~\cite{Bauer:2021gek}.}
\begin{align}
\frac{\pi(2^{n_q}-1)^2}{2^{n_q +1}} \ = \ \phi_{\text{max}}^2\sqrt{(m_0^2+2)+\frac{\lambda}{12}\phi_{\text{max}}^2} \ .
\label{i_s:eq:phimax}
\end{align}
For $\lambda=0$ this solution also preserves the  symplectic symmetry of the harmonic oscillator that takes $\hat{\phi} \leftrightarrow  \hat{\Pi}/\sqrt{m_0^2 +2}$.
In the rest of this section, $n_q=3$ is used.

The first step in the wavepacket preparation algorithm outlined in Sec.~\ref{i_s:sec:WPsummary} is to initialize a state with the correct amplitude and phase in each momentum block of the Hamiltonian.
For a single-particle wavepacket in scalar field theory, this initial state should be odd under $Z_2$.
Therefore, it is convenient to work in the basis of harmonic oscillator eigenstates $|E_j\rangle^{(\text{h.o.})}$.
The vacuum $|E_0\rangle^{(\text{h.o.})} =  |0\rangle^{\otimes n_q}$ is $Z_2$-even and the first excited state $|E_1\rangle^{(\text{h.o.})} =  |0\rangle^{\otimes n_q -1}|1\rangle$ is $Z_2$-odd.
A $Z_2$-odd state with momentum $k$ is constructed using a straightforward generalization of Eq.~\eqref{i_s:eq:psik0}, 
\begin{align}
\vert k\rangle^{(\text{h.o.})} \ =\  \frac{1}{\sqrt{L}}\sum_{n=0}^{L-1}e^{i k n}\, \left (|E_0\rangle^{(\text{h.o.})}\right )^{\otimes L-n-1}\left (|E_1\rangle^{(\text{h.o.})}\right )\left (|E_0\rangle^{(\text{h.o.})}\right )^{\otimes n} \ ,
\label{i_s:eq:psik1}
\end{align}
and superpositions of $\vert k \rangle^{(\text{h.o.})}$ can be combined into a wavepacket,
\begin{align}
| W(k_0)\rangle^{(\text{h.o.})} \ &= \ {\cal N}\sum_ke^{-i k x_0}\, e^{-(k_0 - k)^2/(4\sigma_{}^2)} |k\rangle^{(\text{h.o.})} \nonumber\\ 
&= \ \sum_{n=0}^{L-1} e^{i\phi_n}c_n |000\rangle^{\otimes L-1-n}|001\rangle|000\rangle^{\otimes n}\ .
\label{i_s:eq:Wk0Scalar}
\end{align}
The second equality is for $n_q=3$ and has emphasized that $|W(k_0)\rangle^{(\text{h.o.})}$ is simply a generalized version of the initial state used to prepare wavepackets in Ising field theory, Eq.~\eqref{i_s:eq:psiWP02}. 
This state can be prepared using the $|W(k_0)\rangle$ preparation circuits constructed previously.

\begin{figure}
    \centering
    \includegraphics[width=\linewidth]{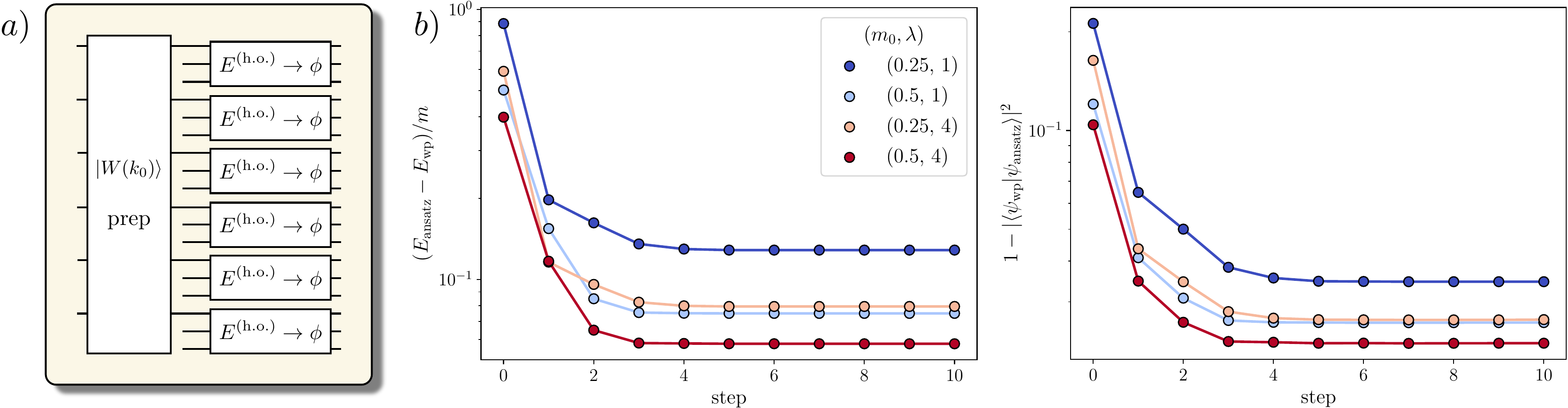}
    \caption{\textit{Wavepacket preparation in scalar field theory.} a) A circuit that prepares the initial state of the wavepacket preparation algorithm $|W(k_0)\rangle^{(\phi)}$. b) The deviation of the energy (left) and infidelity (right) of the wavepackets prepared with up to 10 steps of ADAPT-VQE. Results for $\sigma_{}=0.35,\,k_0=0.44\pi$ and $L=9$ are shown for $n_q=3$ (27 qubits).}
    \label{i_s:fig:scalarWP}
\end{figure}

The remainder of the wavepacket preparation is more efficient in the field basis.
While an efficient circuit that implements the $E^{(\text{h.o.})} \to \phi$ change of basis is not known, it can be found by exactly synthesizing the corresponding $2^{n_q}$-dimensional unitary into gates.
Since a modest $n_q$ is expected to be sufficient for many observables of interest~\cite{Klco:2018zqz}, this single-site change of basis during state preparation will likely not be the limiting factor in quantum simulations of scattering.
An example of a circuit preparing a wavepacket across 6 sites is shown in Fig.~\ref{i_s:fig:scalarWP}a).
First $|W(k_0)\rangle^{(\text{h.o.})}$ is prepared in the harmonic oscillator basis.
Then the circuits implementing the $E^{(\text{h.o.})}\to\phi$ change of basis are applied locally to each lattice site.
For $n_q=3$, the change of basis circuit found by BQSkit~\cite{osti_1785933} has CNOT depth 11 (compare to depth 6 for the symmetric Fourier transform). 
The result is $|W(k_0)\rangle^{(\phi)}$ in the field basis.
%


With $|W(k_0)\rangle^{(\phi)}$ initialized, the next step is to minimize the energy using ADAPT-VQE equipped with a pool of translationally invariant, real and $Z_2$-symmetric operators.
Inspired by the Lie algebra of the Hamiltonian, an operator pool similar to that used in Ref.~\cite{Zemlevskiy:2024vxt} is found to be effective,
\begin{align}
\{ \hat{O} \}_{\text{scalar}} \ &= \ \{ \hat{O}_{\phi\Pi} (s)\} \ = \ \left \{ -\frac{i}{2} \left [\sum_{n=0}^{L-1} \hat{\phi}_n \hat{\phi}_{n+s}\, , \  \sum_{n=0}^{L-1} \hat{\Pi}_{n}^2\right ] \right \} \nonumber \\
&\to \ \left \{ \sum_{n=0}^{L-1}\left (\hat{\phi}_n\hat{\Pi}_{n+s} + \hat{\Pi}_n \hat{\phi}_{n+s} \right )  \right \} \ .
\end{align}
The second line has assumed the fields satisfy the canonical commutation relations, which is only approximately true with digitized fields.
However, the pool operators still have the desired symmetry properties.
In the field basis, $\hat{\phi}_n$ is real and $\hat{\Pi}_n$ is imaginary, and therefore $\exp\left (i \theta \hat{O}_{\phi\Pi}\right )$ is real.
These operators build correlations over $s$ spatial sites with $s=0,1,2,\ldots,\lfloor L/2\rfloor$. 
The unitary evolution with respect to these operators is implemented by Trotterizing and switching from the $\phi$-basis to the $\Pi$-basis using the quantum Fourier transform.
For $\hat{O}_{\phi\Pi}(0)$, we instead find the matrix of eigenvectors $\hat{U}$ and eigenvalues $\hat{D}$ of $\left (\hat{\phi}_n\hat{\Pi}_{n} + \hat{\Pi}_n \hat{\phi}_{n}\right )$, and then implement $U^{\dagger} e^{i \theta \hat{D}} U$. 
BQSkit's unitary circuit synthesis finds a CNOT depth 10 implementation of $\hat{U}$ and $\hat{U}^{\dagger}$, and $e^{i\theta \hat{D}}$ is a product of diagonal rotations.

The ADAPT-VQE results for preparing wavepackets with $\sigma_{}=0.35$, $k_0=0.44\pi$ and $L=9$ (27 qubits) are shown in Fig.~\ref{i_s:fig:scalarWP}b).
A selection of couplings 
\begin{align}
    \{(m_0,\lambda,m)\} \ = \ \{(0.5,4,0.94),(0.25,4,0.85),(0.5,1,0.66),(0.25,1,0.52)\} \nonumber
\end{align}
are chosen, with $m$ the mass gap.
For these couplings, Eq.~\eqref{i_s:eq:phimax} gives 
\begin{align}
    \phi_{\text{max}} \ = \ \{2.21,2.23,2.41,2.45\} \ .
\end{align}.
Again, minimizing the energy rapidly decreases the infidelity with the target wavepacket.
It is surprising how well  $|W(k_0)\rangle^{(\phi)}$ (the state at step 0) approximates the exact wavepacket.
This state is a superposition over tensor products between different spatial sites, yet it still has high overlap with the exact wavepacket for all couplings tested.
Convergence is faster for larger mass gaps, and all couplings reach an infidelity $<0.05$ by the second step of ADAPT-VQE.

\subsection{The Schwinger model}
\label{i_s:app:SchwingerWP}
\noindent
The Schwinger model describes electrons, positrons and photons interacting in one  dimension, and is often studied as a toy model for Quantum Chromodynamics (QCD).
Like QCD, the Schwinger model is a lattice gauge theory that exhibits confinement and a chiral condensate.
It has emerged as a popular testbed for quantum simulation algorithms on quantum devices~\cite{Farrell:2023fgd,
Farrell:2024fit,
Klco:2018kyo,
Nguyen:2021hyk,
deJong:2021wsd,
Pomarico:2023png,
Zhou:2021kdl,
Zhang:2023hzr,
Mil:2019pbt,
Lu:2018pjk,
Mueller:2022xbg,
Riechert:2021ink,
Yang:2020yer,
Kokail:2018eiw,
Martinez:2016yna,
Guo:2024tnb,
Angelides:2023noe,
Schuster:2023klj}.
Recently, a wavepacket was prepared in the Schwinger model on 112 qubits of IBM's quantum computers~\cite{Farrell:2024fit}.
Wavepacket preparation circuits were first determined on small system sizes by minimizing the infidelity with the exact wavepacket using classical computers.
The SC-ADAPT-VQE algorithm~\cite{Farrell:2023fgd} was then used to scale these circuits up and prepare wavepackets on 112 qubits.
Similar to the discussion in the previous section on scalar field theory, this approach has limitations related to wavepacket scalability and circuit optimization.
The wavepacket preparation method described here overcomes both of these problems.

In this work, the Schwinger model is discretized onto a staggered lattice with fermions (antifermions) occupying even- (odd-) numbered sites, and the gauge fields occupying the links between them.
This maps $L$ spatial sites to $2L$ staggered sites.
Gauss's law constrains the allowed states of the gauge fields and fermions. 
With OBCs, explicit gauge field degrees of freedom can be completely removed from the theory, leaving a system of fermions interacting through a linear Coulomb potential.
With PBCs, there is a single mode of the gauge field whose dynamics is not constrained.
To keep translational invariance manifest, it is convenient to parameterize this mode by the average electric field $\hat{{\cal E}} = \sum_n \hat{{\cal E}}_n/(2L)$, and its conjugate parallel transporter $\hat{U}^{2L} = \prod_n \hat{U}_n$, such that $[\hat{{\cal E}},\hat{U}^{2L} ] = \hat{U}$~\cite{Dempsey:2022nys}.
After performing the Jordan-Wigner (JW) mapping from fermionic to spin operators the Hamiltonian is,
\begin{align}
\hat{H} \ = \ &\frac{m_0}{2}\sum_{n=0}^{2L-1}\left [ (-1)^n \hat{Z}_n + \hat{I}\right ] \  - \ \frac{g^2}{2}\sum_{n=0}^{2L-1}\sum_{s=1}^{L}\left (s - \frac{s^2}{2L} \right )\left (1-\frac{\delta_{s,L}}{2} \right )\hat{Q}_n\hat{Q}_{n+s} \nonumber \ + \ g^2 L \hat{{\cal E}}^2 \\
&+ \ \frac{1}{2}\sum_{n=0}^{2L-2}\left (\hat{\sigma}_n^+ \hat{U} \hat{\sigma}_{n+1}^- + {\rm h.c.} \right ) \ + \ \frac{1}{2}(-1)^{L + 1}\left (\hat{\sigma}^+_{2L-1}\hat{U}\hat{\sigma}^-_{0} +{\rm h.c.} \right ) \ .
\end{align}
The bare fermion mass is $m_0$, the electric charge is $g$ and the staggered electric charge operator is $\hat{Q}_n = -\frac{1}{2}[(-1)^n +\hat{Z}_n ]$. 
The $(-1)^{L +1}$ in the hopping term between sites $2L-1$ and $0$ comes from the JW mapping.\footnote{The extra sign can be verified by comparing the spectrum of a JW-mapped free fermion with PBCs to the exact result.} 

The zero mode of the gauge field is bosonic and its Hilbert space is formally infinite.
Like for scalar field theory, this Hilbert space is made finite by choosing a basis and truncating to some number of states.
In this section, we truncate the Hilbert space to one state, i.e., will not consider gauge field dynamics.
This can be improved by allocating some number of qubits to represent the gauge field.
With this truncation, the Hamiltonian only contains operators acting on staggered fermion sites,\footnote{The Hamiltonian in Eq.~\eqref{i_s:eq:SchwingerH} is equivalent to that used in Ref.~\cite{Zache:2018cqq} with the right choice of initial conditions.}
\begin{align}
\hat{H} \ \to \ &\frac{m_0}{2}\sum_{n=0}^{2L-1}\left [ (-1)^n \hat{Z}_n + \hat{I}\right ] \  - \ \frac{g^2}{2}\sum_{n=0}^{2L-1}\sum_{s=1}^{L}\left (s - \frac{s^2}{2L} \right )\left (1-\frac{\delta_{s,L}}{2} \right )\hat{Q}_n\hat{Q}_{n+s} \nonumber \\
&+ \ \frac{1}{2}\sum_{n=0}^{2L-2}\left (\hat{\sigma}_n^+  \hat{\sigma}_{n+1}^- + {\rm h.c.} \right ) \ + \ \frac{1}{2}(-1)^{L + 1}\left (\hat{\sigma}^+_{2L-1}\hat{\sigma}^-_{0} +{\rm h.c.} \right ) \ .
\label{i_s:eq:SchwingerH}
\end{align}
This is the Hamiltonian that will be used in the rest of this section.

The first step for preparing wavepackets is to initialize a state with the momentum content and quantum numbers of the target wavepacket. 
In the Schwinger model, due to confinement, the single-particle eigenstates are charge-neutral ``hadrons''.
As a result, the initial state must also be charge-neutral, not composed of individual electron ($e^{-}$) or positron ($e^+$) excitations.
One place to start is at strong coupling (SC), $g\to\infty$, where hadrons are tightly bound pairs of $e^+ e^-$ on neighboring staggered sites.
The wavefunction for a plane wave of hadrons at SC is,
\begin{align}
|k\rangle^{(\text{SC})} \ = \ &\frac{1}{\sqrt{2L}}\Bigg [ \sum_{\ell=0}^{L-1}e^{i k \ell}\, |01\rangle^{\otimes L-\ell-1}|10\rangle |01\rangle^{\otimes \ell} \nonumber \\
&- e^{ik/2} \left ( \sum_{\ell=0}^{L-2}e^{i k \ell}\, |01\rangle^{\otimes L-\ell-2}|00\rangle|11\rangle|01\rangle^{\otimes \ell}  +   (-1)^{L+1}e^{ik \left (L-1\right )}|11\rangle |01\rangle^{\otimes L-2}\vert 00\rangle\right ) \Bigg ] \ .
\end{align}
The sum index $\ell$ emphasizes that these sums are over spatial sites.
The state in the first line represents a plane wave of $|e^+ e^{-}\rangle$, with the $e^+$ on the higher-numbered staggered site.
The states in the second line are related by charge conjugation, and are plane waves of $|e^- e^+\rangle$ with the $e^-$ on the higher-numbered staggered site.
At all couplings, the lightest hadron corresponds to a massive photon and is therefore odd under charge conjugation.
This is why there is a $(-1)$ between the sums in the first and second lines.
On a staggered lattice, charge conjugation is realized by a global spin flip followed by a translation by one staggered site.
One staggered site is half of a spatial site, and this translation gives the $e^{i k/2}$ relative phase between the sums in the first and second line.
Finally, when the $|e^-e^+\rangle$ is translated across sites $0$ and $2L-1$, there is an additional factor of $(-1)^{L+1}$ due to the kinetic term in the JW mapping.

\begin{figure}
    \centering
    \includegraphics[width=\linewidth]{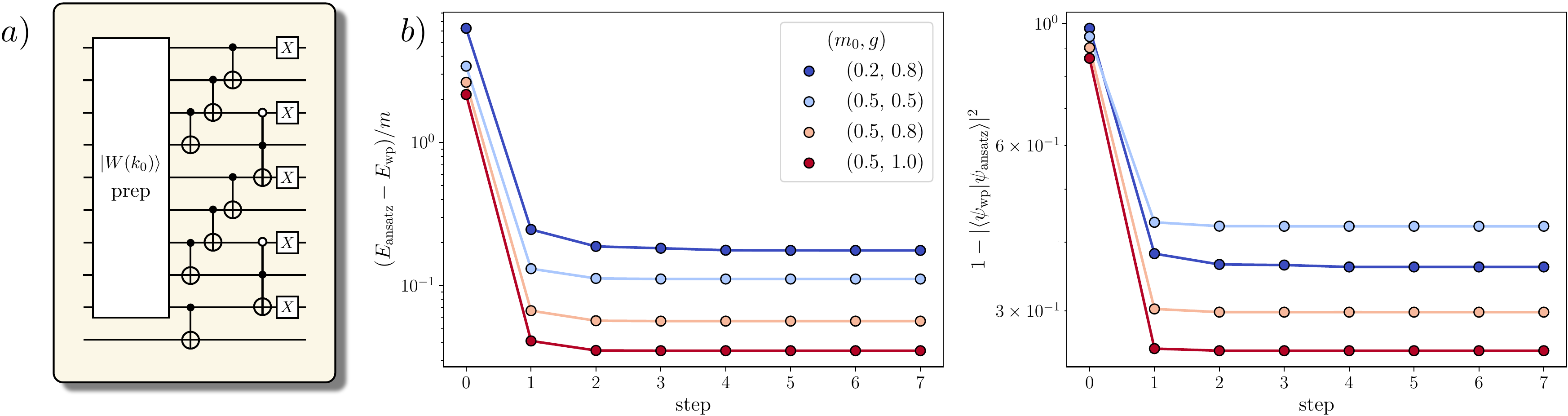}
    \caption{\textit{Wavepacket preparation in the Schwinger model.} a) A circuit that prepares the initial wavepacket of hadrons at strong coupling $|W(k_0)\rangle^{(\text{SC})}$ on $2L=10$ staggered sites.
    b) The deviation of the energy (left) and infidelity (right) of a wavepacket prepared with up to 7 steps of ADAPT-VQE.
    Results for $\sigma_{}=0.25,\,k_0=0.36\pi$ and $L=14$ (28 staggered sites) are shown.}
    \label{i_s:fig:SchwingerWP}
\end{figure}

An initial wavepacket of hadrons at SC is written as,\footnote{The SC hadron that crosses sites $2L-1$ and 0 has been omitted since, in practice, wavepackets will be localized to a spatial region away from site 0.}
\begin{align}
| W(k_0)\rangle^{(\text{SC})} \ &= \ {\cal N}\sum_{k}e^{-i k x_0}\, e^{-(k_0 - k)^2/(4\sigma_{}^2)}  |k\rangle^{(\text{SC})} \ \nonumber \\
&\equiv \ \sum_{n=0}^{L-1} c_{2n}e^{i\phi_{2n}}  |01\rangle^{\otimes L-n-1}|10\rangle |01\rangle^{\otimes n} \nonumber \\
& \ \ + \ \sum_{n=0}^{L-2} c_{2n+1}e^{i\phi_{2n+1}} |01\rangle^{\otimes L-n-2}|00\rangle|11\rangle|01\rangle^{\otimes n} \  .
\label{i_s:eq:psiWPSC}
\end{align}
Hadron wavepackets at SC have previously been used in MPS simulations of scattering~\cite{Papaefstathiou:2024zsu,Barata:2025hgx}.
An example of a circuit that prepares $| W(k_0)\rangle^{(\text{SC})}$ over ten staggered sites is given in Fig.~\ref{i_s:fig:SchwingerWP}a).
First, $|W(k_0)\rangle$ in Eq.~\eqref{i_s:eq:psiWP02} is prepared over all but the last staggered site using one of the $|W(k_0)\rangle$ preparation circuits previously developed.
The CNOT and CCNOT sequence that follows turns each ``$1$'' in $|W(k_0)\rangle$ into ``$11$''.
The desired wavepacket of SC hadrons is produced by applying $\hat{X}_n$ to all even-numbered staggered sites.
The circuits that prepare $|W(k_0)\rangle^{(\text{SC})}$ over an arbitrary number of spatial sites are determined by a straightforward generalization.

The next step in wavepacket preparation is to perform ADAPT-VQE using a pool of operators that are translationally invariant, real, charge conjugation and parity symmetric, and conserve electric charge.
We choose an operator pool generated by taking all nested commutators of the mass and kinetic terms (the Lie algebra of the free Hamiltonian),
\begin{align}
\{\hat{O} \}_{\text{Schwinger}} \ &= \ \{\hat{O}_{mh}(s)\} \ , \nonumber \\[4pt]
\hat{O}_{mh}(s) \ &=\ i \left [\, \sum_{n=0}^{2L-1} (-1)^n \hat{Z}_n   \ , \ \sum_{n=0}^{2L-1} v(n,s)(1-\frac{1}{2}\delta_{s,L})\left (\hat{\sigma}^+_{n} \hat{Z}^{s-1}\hat{\sigma}^-_{n+s} + {\rm h.c.}\right ) \right ]
\label{i_s:eq:QCDpool}
\end{align}
where $v(n,s) = (\pm 1)^{L+1}$ with $(+)$ if $n+s \leq 2L-1$ and $(-)$ if $n+s > 2L-1$. 
This sign is needed to account for the minus sign in the kinetic term for even $L$.
The range of $s$ is $s\in \{1,3,\ldots,L\}$, and only odd $s$ is generated as a consequence of charge conjugation symmetry.
This operator pool is the PBC version of the pool used in Refs.~\cite{Farrell:2023fgd,Farrell:2024fit,Farrell:2024mgu} to prepare the Schwinger model vacuum.
Quantum circuits corresponding to the unitary evolution of these operators are a straightforward extension of those in Ref.~\cite{Farrell:2023fgd}.
For convenience, the {\tt PauliEvolutionGate} method in {\tt qiskit} is used to generate these circuits.
This method converts a sum of Pauli strings into circuits using a first order Trotterization.

The results from using ADAPT-VQE to prepare wavepackets are shown in Fig.~\ref{i_s:fig:SchwingerWP}b) for $L=14$ and wavepacket parameters $k_0=0.36\pi$ and $\sigma_{}=0.25$.
Four sets of couplings were chosen
$\{(m_0,g,m)\}=\{(0.5,1.0,1.61),(0.5,0.8,1.44),(0.5,0.5,1.23),(0.2,0.8,0.82)\}$ that approach the continuum limit of vanishing mass gap $m\to0$. 
Convergence to the target wavepacket mostly plateaus after the second step of ADAPT-VQE.
The convergence is better with larger $m_0$ and $g$ likely because the mass gap is larger and/or it is closer to the SC limit.
These fidelities will likely be sufficient for near-term demonstrations of scattering.
However, higher-quality wavepackets will eventually be needed for simulations that approach the continuum limit.
Preliminary attempts to expand the operator pool by including commutators between the kinetic term and $\hat{Q}_n\hat{Q}_{n+s}$ did not improve convergence.
It is possible that an initial state with hadrons of different lengths, i.e., not only $e^-e^+$ on neighboring staggered sites, would perform better.
This would allow for the valence fermion and anti-fermion to have different relative momentum, like in Refs.~\cite{Davoudi:2024wyv,Rigobello:2021fxw}.

\subsection{Two-dimensional Ising field theory}
\label{i_s:app:2DIsing}
The wavepacket preparation method described in Sec.~\ref{i_s:sec:WPsummary} readily generalizes to lattice QFTs beyond one dimension.
Consider the two-dimensional tilted-field Ising model on a PBC square lattice defined by the Hamiltonian,
\begin{align}
\hat{H} \ = \ -\sum_{\langle ij\rangle}\hat{Z}_i \hat{Z}_j \ - \sum_i \left (g_x \hat{X}_i \ + \ g_z \hat{Z}_i \right ) \ .
\label{i_s:eq:2DIsingH}
\end{align}
The first sum is over nearest neighbors $\langle i,j\rangle$ and the second sum is over every lattice site.
This Hamiltonian has a translational symmetry in both the $x$- and $y$-direction, as well as a symmetry under $\pi/2$ rotations.\footnote{The recovery of the $SO(2)$ rotational symmetry in the continuum limit is subtle, and may require lattice and operator smearing.
See, e.g., Refs.~\cite{HadronSpectrum:2009krc,Davoudi:2012ya}.}
On an infinite lattice and at the critical point, $g_c=3.04438(2)$~\cite{PhysRevE.66.066110} and $g_z = 0$, this system is described by the 3D Ising CFT.
A massive interacting QFT is obtained by taking $g_x \to g_c$ and $g_z\to0$ but keeping fixed
the scaling invariant ratio,
\begin{align}
\eta_{\text{latt}}^{2d} \ = \ 
\frac{g_x-g_c}{|g_z|^{\frac{D-\Delta_{\epsilon}}{D-\Delta_{\sigma}}}} \ = \ 
\frac{g_x-g_c}{|g_z|^{0.63959303(29)}} \ .
\label{i_s:eq:eta2d}
\end{align}
The second equality has used $D=3$ spacetime dimensions and the scaling dimensions of the relevant deformations, $\Delta_{\epsilon} = 1.41262528(29)$ and $\Delta_{\sigma} = 0.518148806(24)$, determined using conformal bootstrap~\cite{Chang:2024whx}.
The universal physics of the 3D Ising CFT also describes the Wilson-Fisher fixed point in two-dimensional scalar field theory~\cite{PhysRevLett.28.240}.
As a result, the continuum limit for arbitrary $\eta_{\text{latt}}^{2d}$ corresponds to a massive interacting scalar field theory in two dimensions.

A wavepacket on a $L\times L$ lattice is,
\begin{equation}
\vert \psi_{\text{wp}} \rangle  \ = \ {\cal N}\sum_{\vec{k}} e^{-i \vec{k}\cdot \vec{x}_0}\, e^{-|\vec{k}_0 - \vec{k}|^2/(4\sigma_{}^2)} \vert \psi_{\vec{k}} \rangle \ ,
\label{i_s:eq:psiWPFull2D}
\end{equation}
where, e.g., $\vec{k} = (k_x,k_y)$ and $\vert \psi_{\vec{k}} \rangle$ are the lowest-energy single-particle states with momentum $\vec{k}$.\footnote{For simplicity the same spread in momentum $\sigma_{}$ is used for $k_x$ and $k_y$.}
The sum over $\vec{k}$ runs over $L^2$ terms where $k_x=2\pi n_x/L$ and $k_y=2\pi n_y/L$ with $n_x,n_y$ integers and $k_x,k_y \in (-\pi,\pi]$.
The Hamiltonian can be block diagonalized into $L^2$ different blocks labeled by $\vec{k}$.
A simple initial state with the correct amplitude and phase in each momentum block is
\begin{equation}
\vert W(k_0)\rangle  \ = \ {\cal N}\sum_{\vec{k}} e^{-i \vec{k}\cdot \vec{x}_0}\, e^{-|\vec{k}_0 - \vec{k}|^2/(4\sigma_{}^2)} \vert \vec{k} \rangle  \ = \  \sum_{n_x=0}^{L-1}\sum_{n_y=0}^{L-1} e^{i \phi_{n_x,n_y}}c_{n_x,n_y}|2^{L n_y + n_x}\rangle \ ,
\label{i_s:eq:psiWP02d}
\end{equation}
where
\begin{align}
|\vec k\rangle \ = \ \frac{1}L\sum_{n_x=0}^{L-1}\sum_{n_y=0}^{L-1} e^{i \vec{k}\cdot \vec{n}}|2^{L n_y + n_x}\rangle \ .
\end{align}
These are the two-dimensional generalizations of the states given in Eqs.~\eqref{i_s:eq:psik0} and~\eqref{i_s:eq:psiWP0}, and $\vert W(k_0)\rangle$ can be prepared with the circuit in the left panel of Fig.~\ref{i_s:fig:IsingWPCircs}.

\begin{figure}
    \centering
    \includegraphics[width=\linewidth]{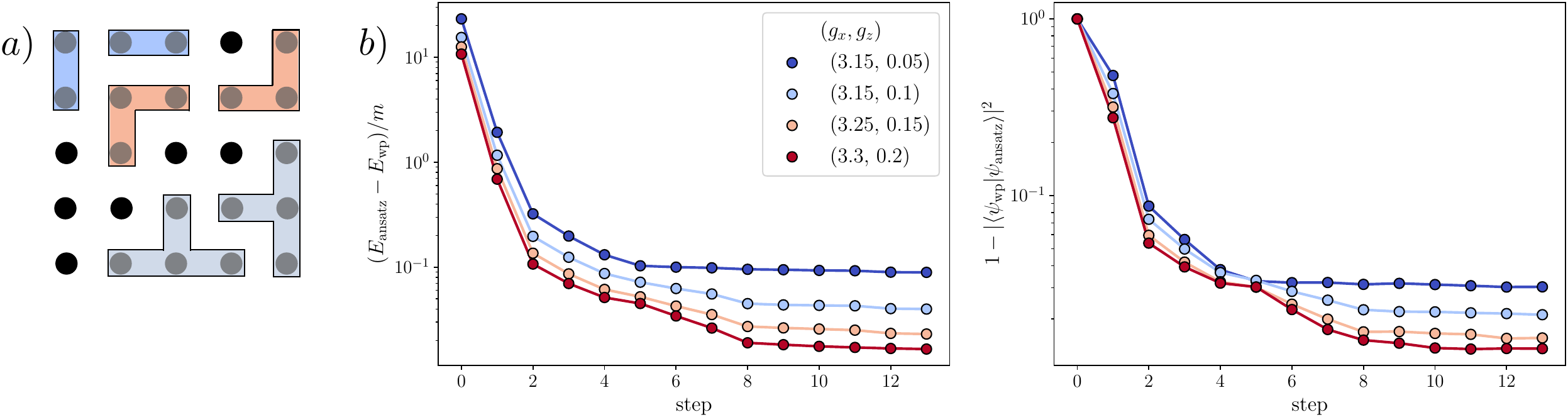}
    \caption{\textit{Wavepacket preparation in two-dimensional Ising field theory.} a) Operators in the ADAPT-VQE pool Eq.~\eqref{i_s:eq:opPool2d}, that are related by rotations.
    The blue operators span two sites, and the vertical strip is related to the horizontal strip by a $\pi/2$ rotation.
    The tan operators span three sites in an `L' shape and are related by a $\pi$ rotation.
    The gray operators span four sites in a `T' shape and are related by a $\pi/2$ rotation.
    b) The deviation of the energy (left) and infidelity (right) of a wavepacket prepared with up to 13 steps of ADAPT-VQE. 
    Results for a $5\times 5$ lattice with $\vec{k}_0=(0.4\pi,0.4\pi)$ and $\sigma_{}=0.5$ are shown.}
    \label{i_s:fig:2DWP}
\end{figure}
The next step in wavepacket preparation is to minimize the energy using translationally invariant and real circuits starting from $\vert W(k_0)\rangle$.
This is done using ADAPT-VQE and an operator pool inspired by the Lie algebra of the Hamiltonian is found to be effective.
Keeping all unique operators up to third-order commutators, $i\sum_n [\hat{H}, \hat{X}_n ]$ and $i\sum_n [ \hat{H}, [ \hat{H}, [\hat{H}, \hat{X}_n ] ]  ]$, gives an operator pool
{\allowdisplaybreaks
\begin{align}
&\{ {\hat O}\}_{2d \ \text{Ising}}\ = \ \sum_{n_x=0}^{L-1}\sum_{n_y=0}^{L-1} \bigg \{ \hat{Y}_{n_x,n_y} \,  
 , \nonumber \\
 &\Big (\hat{Y}_{n_x,n_y}\hat{Z}_{n_x+1,n_y}+ \hat{Z}_{n_x,n_y}\hat{Y}_{n_x+1,n_y}+\hat{Y}_{n_x,n_y}\hat{Z}_{n_x,n_y+1}+ \hat{Z}_{n_x,n_y}\hat{Y}_{n_x,n_y+1}\Big )\, , \nonumber \\
 &\Big (\hat{Y}_{n_x,n_y}\hat{X}_{n_x+1,n_y}+ \hat{X}_{n_x,n_y}\hat{Y}_{n_x+1,n_y}+\hat{Y}_{n_x,n_y}\hat{X}_{n_x,n_y+1}+ \hat{X}_{n_x,n_y}\hat{Y}_{n_x,n_y+1}\Big )\, , \nonumber \\
 &\Big ( \hat{Z}_{n_x,n_y}\hat{Y}_{n_x+1,n_y}\hat{Z}_{n_x+2,n_y}+\hat{Z}_{n_x,n_y}\hat{Y}_{n_x,n_y+1}\hat{Z}_{n_x,n_y+2} \Big ) \, , \nonumber \\
 &\Big (\hat{Z}_{n_x,n_y}\hat{X}_{n_x+1,n_y}\hat{Y}_{n_x+2,n_y}+ \hat{Y}_{n_x,n_y}\hat{X}_{n_x+1,n_y}\hat{Z}_{n_x+2,n_y}+\hat{Z}_{n_x,n_y}\hat{X}_{n_x,n_y+1}\hat{Y}_{n_x,n_y+2}\nonumber\\ &+\hat{Y}_{n_x,n_y}\hat{X}_{n_x,n_y+1}\hat{Z}_{n_x,n_y+2}\Big ) \, , \nonumber \\
 &\Big(\hat{Z}_{n_x,n_y}\hat{Y}_{n_x+1,n_y}\hat{Z}_{n_x+1,n_y+1}+\hat{Z}_{n_x,n_y}\hat{Y}_{n_x,n_y+1}\hat{Z}_{n_x-1,n_y+1}+\hat{Z}_{n_x,n_y}\hat{Y}_{n_x-1,n_y}\hat{Z}_{n_x-1,n_y-1} \nonumber\\ 
 &+\hat{Z}_{n_x,n_y}\hat{Y}_{n_x,n_y-1}\hat{Z}_{n_x+1,n_y-1} \Big ) \, , \nonumber \\
 &\Big ( \hat{Z}_{n_x,n_y}\hat{X}_{n_x+1,n_y}\hat{Y}_{n_x+1,n_y+1}+\hat{Z}_{n_x,n_y}\hat{X}_{n_x,n_y+1}\hat{Y}_{n_x-1,n_y+1}+\hat{Z}_{n_x,n_y}\hat{X}_{n_x-1,n_y}\hat{Y}_{n_x-1,n_y-1}\nonumber\\
 &+\hat{Z}_{n_x,n_y}\hat{X}_{n_x,n_y-1}\hat{Y}_{n_x+1,n_y-1} +\hat{Y}_{n_x,n_y}\hat{X}_{n_x+1,n_y}\hat{Z}_{n_x+1,n_y+1}+\hat{Y}_{n_x,n_y}\hat{X}_{n_x,n_y+1}\hat{Z}_{n_x-1,n_y+1}\nonumber\\
 &+\hat{Y}_{n_x,n_y}\hat{X}_{n_x-1,n_y}\hat{Z}_{n_x-1,n_y-1}+\hat{Y}_{n_x,n_y}\hat{X}_{n_x,n_y-1}\hat{Z}_{n_x+1,n_y-1} \Big ) \ , 
\nonumber \\
&\Big ( \hat{Y}_{n_x,n_y}\hat{Z}_{n_x+1,n_y}\hat{Z}_{n_x-1,n_y}\hat{Z}_{n_x,n_y+1} + \hat{Y}_{n_x,n_y}\hat{Z}_{n_x+1,n_y}\hat{Z}_{n_x-1,n_y}\hat{Z}_{n_x,n_y-1}\nonumber\\
&+\hat{Y}_{n_x,n_y}\hat{Z}_{n_x+1,n_y}\hat{Z}_{n_x,n_y+1}\hat{Z}_{n_x,n_y-1} + \hat{Y}_{n_x,n_y}\hat{Z}_{n_x-1,n_y}\hat{Z}_{n_x,n_y+1}\hat{Z}_{n_x,n_y-1} \Big )  \bigg \} \ .
\label{i_s:eq:opPool2d}
\end{align}
}
The rotational symmetry of the Hamiltonian on a square lattice groups together all operators related by $\pi/2$ rotations.
Lines 2-5 combine together horizontal and vertical strips of operators.
Operators spanning three sites can additionally be arranged in an `L' shape,
and lines 6-8 group together the four rotations of `L'. 
The four operators in lines 9 and 10 have a `T' shape and are related by rotations.
See Fig.~\ref{i_s:fig:2DWP}a) for an illustration.
The {\tt PauliEvolutionGate} method in {\tt qiskit} is used to convert the exponential of the operators in Eq.~\eqref{i_s:eq:opPool2d} into quantum circuits.

The results from preparing wavepackets  on a $5\times 5$ lattice using ADAPT-VQE are shown in Fig.~\ref{i_s:fig:2DWP}b).
Wavepacket parameters $\sigma_{}=0.5$ and $\vec{k}_0=(0.4\pi,0.4\pi)$, and four sets of couplings
$\{ (g_x,g_z,m) \} = \{ (3.3,0.2,3.27) , (3.25,0.15,2.77) , (3.15,0.1,2.17) , (3.15,0.05,1.48) \}$ with decreasing mass gap $m$ are chosen.
By minimizing the energy, the solution converges to the desired wavepacket, as seen by the decreasing infidelity with the exact wavepacket.
Like in all the other examples, convergence is slower for smaller mass gaps.
The initial operators chosen by ADAPT-VQE are ordered: one site, two sites nearest-neighbor, three sites in an `L' configuration and three sites in a line configuration.
This is a result of the hierarchy in correlations (short-distance correlations are more significant than long-distance ones) present in gapped systems.

\section{More details on using ADAPT-VQE to prepare wavepackets and vacua}
\label{i_s:app:57Adapt}
\noindent
The algorithm developed in this work for preparing wavepackets uses ADAPT-VQE to minimize the energy.
Our implementation of ADAPT-VQE is the following:
\begin{itemize}
    \item[1.] Define a pool of operators $\{ \hat{O} \}$ that are translationally invariant and imaginary (so that $e^{i \theta \hat{O}}$ is real).
    They should also respect the other symmetries of the Hamiltonian.
    An operator pool inspired by the Lie algebra of the Hamiltonian is found to be effective.
    \item[2.] Initialize a state with the correct amplitude and phase in each momentum block of the Hamiltonian.
    This state should also have the other relevant quantum numbers of the desired wavepacket, e.g., $Z_2$ parity in scalar field theory or electric charge in the Schwinger model.
    \item[3.] For each operator in the pool $\hat{O}_i$, optimize the variational parameters in $ e^{i \theta_i \hat{O}_i}\lvert \psi_{{\rm ansatz}} \rangle$ to minimize the energy.
    The previously optimized values for $\theta_{1,\ldots,i-1}$ and $\theta_i=0$, are used as initial conditions. 
    \item[4.] Identify the operator $\hat{O}_n$ and variational parameters $\vec{\theta}$ that had the lowest energy in the previous step.
    Update the ansatz $\lvert \psi_{{\rm ansatz}} \rangle \to e^{i \theta_n \hat{O}_n}\lvert \psi_{{\rm ansatz}} \rangle$ using the optimal parameters.
    \item[5.] Return to step 3 until the desired tolerance is achieved.
\end{itemize}
Previous applications of ADAPT-VQE have selected the operator $\hat{O}_n$ in step 4 with the largest energy gradient~\cite{Grimsley:2018wnd,Farrell:2023fgd,Gustafson:2024bww}.
We found that using the gradient leads to worse convergence compared to optimizing every operator in the pool, and choosing the operator that produces the lowest energy. The resulting algorithm is less greedy as it explores more circuits, and is seen to perform better.
This has a computational overhead that scales with the number of operators in the operator pool.
However, our operator pools are very small due to symmetry constraints, and this overhead was not a limitation.

\begin{figure}
    \centering
    \includegraphics[width=0.6\linewidth]{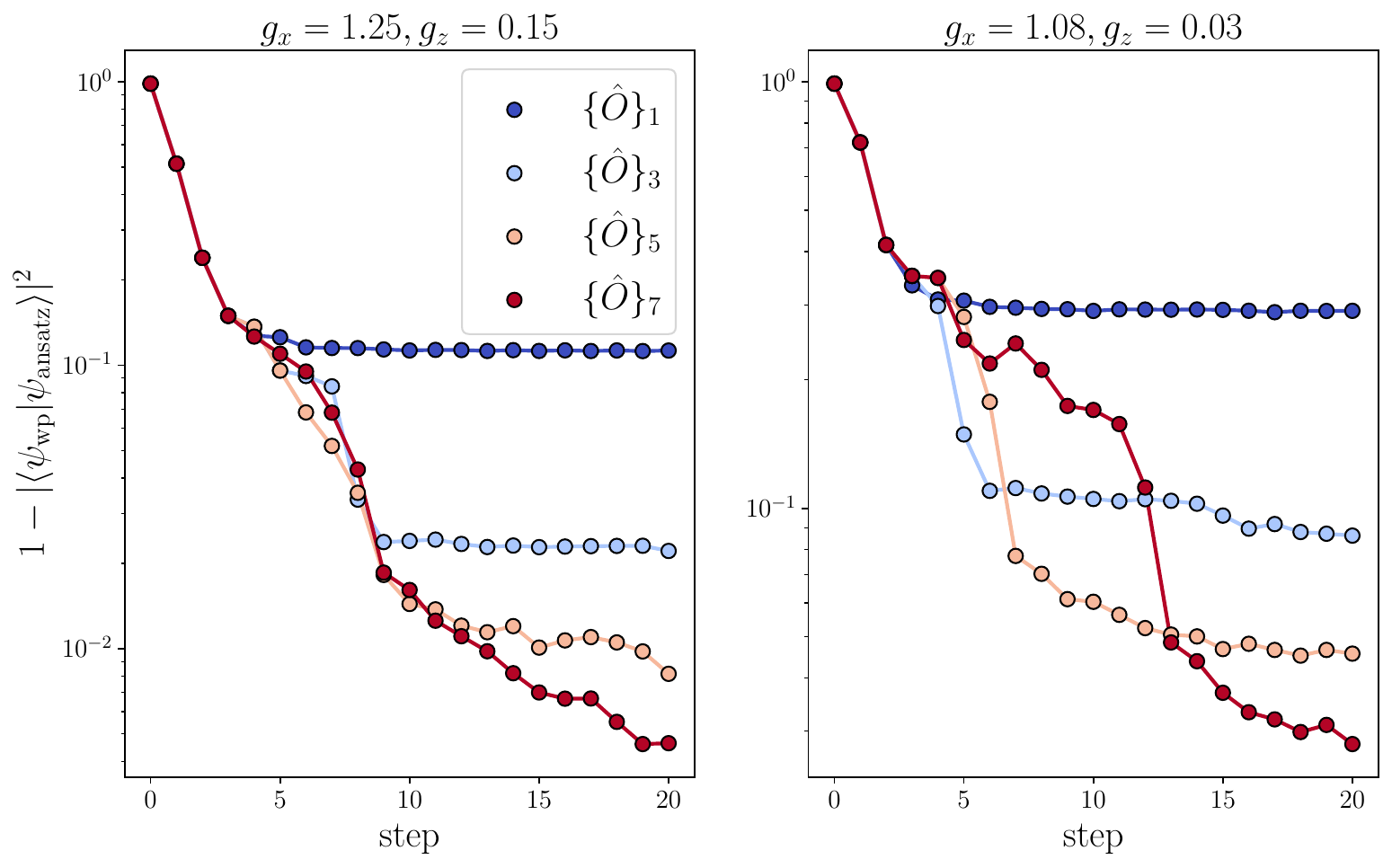}
    \caption{{\it The effect of larger ADAPT-VQE operator pools on wavepacket quality in one-dimensional Ising field theory.}
    The infidelity of wavepackets prepared with up to 20 steps of ADAPT-VQE and parameters $\sigma_{}=0.15,\,k_0=0.36\pi$.
    Results are shown for $L=24$ and using four different operators pools $\{\hat{O}\}_1,\{\hat{O}\}_3,\{\hat{O}\}_5,\{\hat{O}\}_7$, defined in Eq.~\eqref{i_s:eq:opPool1357}.
    The left (right) plot is for $g_x=1.25,\,g_z=0.15$ ($g_x=1.08,\,g_z=0.03$).}
    \label{i_s:fig:ADAPT1357}
\end{figure}
A successful application of ADAPT-VQE requires using an effective operator pool.
A pool that is too large will be difficult to optimize, partly due to the presence of many local minima in the optimization landscape.
On the other hand, a pool that is too small does not have sufficient expressivity to prepare the desired wavefunction.\footnote{This tradeoff was discussed in the context of brickwall circuits in Ref.~\cite{Zemlevskiy:2024vxt}.}
This tradeoff is explored by preparing wavepackets in one-dimensional Ising field theory using ADAPT-VQE equipped with an operator pool consisting of the unique operators in the Hamiltonian algebra up to 1st, 3rd (used in the main text), 5th and 7th order commutators,
\begin{align}
&\{ {\hat O}\}_1  =  \sum_{n=0}^{L-1} \bigg \{ \hat{Y}_n \,  , \  \left (\hat{Y}_n\hat{Z}_{n+1}+ \hat{Z}_n\hat{Y}_{n+1}\right )   \bigg \} \ , \nonumber \\
&\{ {\hat O}\}_3  =  \{ {\hat O}\}_1 \ \bigcup \ \sum_{n=0}^{L-1} \bigg \{ \hat{Z}_n\hat{Y}_{n+1}\hat{Z}_{n+2}  ,  \  \left (\hat{Y}_n\hat{X}_{n+1}+ \hat{X}_n\hat{Y}_{n+1}\right )  ,  \nonumber \\ &\left (\hat{Z}_n\hat{X}_{n+1}\hat{Y}_{n+2}+ \hat{Y}_n\hat{X}_{n+1}\hat{Z}_{n+2}\right )  \bigg \} \ , \nonumber \\
 &\{ {\hat O}\}_5 \ = \  \{\hat{O}\}_3 \ \bigcup \ \sum_{n=0}^{L-1}\Bigg \{ \left (\hat{Z}_n\hat{Y}_{n+1}\hat{X}_{n+2}+\hat{X}_n\hat{Y}_{n+1}\hat{Z}_{n+2} \right ),
\left (\hat{Y}_n\hat{X}_{n+1}\hat{X}_{n+2}+\hat{X}_n\hat{X}_{n+1}\hat{Y}_{n+2} \right ),&\nonumber\\  
&\hat{Y}_n\hat{Y}_{n+1}\hat{Y}_{n+2} , \left(\hat{Z}_n\hat{X}_{n+1}\hat{Y}_{n+2}\hat{Z}_{n+3}+\hat{Z}_n\hat{Y}_{n+1}\hat{X}_{n+2}\hat{Z}_{n+3} \right ), 
\left (\hat{Y}_n\hat{Z}_{n+1}\hat{Z}_{n+2}+\hat{Z}_n\hat{Z}_{n+1}\hat{Y}_{n+2} \right ), \nonumber \\
&\left (\hat{Z}_n\hat{X}_{n+1}\hat{X}_{n+2}\hat{Y}_{n+3}+\hat{Y}_n\hat{X}_{n+1}\hat{X}_{n+2}\hat{Z}_{n+3} \right ) \Bigg \} \ , \nonumber \displaybreak \\
&\{ {\hat O}\}_7 \ = \  \{\hat{O}\}_5 \ \bigcup \ \sum_{n=0}^{L-1} \Bigg \{ \left (\hat{Z}_n\hat{I}_{n+1}\hat{Y}_{n+2}+\hat{Y}_n\hat{I}_{n+1}\hat{Z}_{n+2} \right ),
\left (\hat{Y}_n\hat{Z}_{n+1}\hat{X}_{n+2}+\hat{X}_n\hat{Z}_{n+1}\hat{Y}_{n+2} \right )\ , \nonumber \\
&\left (\hat{Z}_n\hat{Z}_{n+1}\hat{Y}_{n+2}\hat{Z}_{n+3}+\hat{Z}_n\hat{Y}_{n+1}\hat{Z}_{n+2}\hat{Z}_{n+3} \right ), \left (\hat{Z}_n\hat{X}_{n+1}\hat{Y}_{n+2}\hat{X}_{n+3}+\hat{X}_n\hat{Y}_{n+1}\hat{X}_{n+2}\hat{Z}_{n+3} \right ),\nonumber\\ 
&\hat{X}_n\hat{Y}_{n+1}\hat{X}_{n+2}, \left (\hat{Y}_n\hat{Y}_{n+1}\hat{Y}_{n+2}\hat{Z}_{n+3}+\hat{Z}_n\hat{Y}_{n+1}\hat{Y}_{n+2}\hat{Y}_{n+3} \right ), \nonumber \\
&\left (\hat{Z}_n\hat{Y}_{n+1}\hat{X}_{n+2}\hat{X}_{n+3}+\hat{X}_n\hat{X}_{n+1}\hat{Y}_{n+2}\hat{Z}_{n+3} \right ),
\left (\hat{Y}_n\hat{X}_{n+1}\hat{X}_{n+2}\hat{X}_{n+3}+\hat{X}_n\hat{X}_{n+1}\hat{X}_{n+2}\hat{Y}_{n+3} \right ),\nonumber \\
&\left (\hat{Z}_n\hat{X}_{n+1}\hat{X}_{n+2}\hat{Y}_{n+3}\hat{Z}_{n+4}+\hat{Z}_n\hat{Y}_{n+1}\hat{X}_{n+2}\hat{X}_{n+3}\hat{Z}_{n+4} \right ), \nonumber \\
&\left (\hat{Y}_n\hat{X}_{n+1}\hat{Z}_{n+2}\hat{Z}_{n+3}+\hat{Z}_n\hat{Z}_{n+1}\hat{X}_{n+2}\hat{Y}_{n+3} \right ),
\left (\hat{Y}_n\hat{X}_{n+1}\hat{Y}_{n+2}\hat{Y}_{n+3}+\hat{Y}_n\hat{Y}_{n+1}\hat{X}_{n+2}\hat{Y}_{n+3} \right ), \nonumber \\
&\left (\hat{Z}_n\hat{Y}_{n+1}\hat{I}_{n+2}\hat{Z}_{n+3}+\hat{Z}_n\hat{I}_{n+1}\hat{Y}_{n+2}\hat{Z}_{n+3} \right ),\left (\hat{Z}_n\hat{X}_{n+1}\hat{Z}_{n+2}\hat{Y}_{n+3}+\hat{Y}_n\hat{Z}_{n+1}\hat{X}_{n+2}\hat{Z}_{n+3} \right ), \nonumber \\
&\hat{Z}_n\hat{X}_{n+1}\hat{Y}_{n+2}\hat{X}_{n+3}\hat{Z}_{n+4} , \left (\hat{X}_n\hat{I}_{n+1}\hat{Y}_{n+2}+\hat{Y}_n\hat{I}_{n+1}\hat{X}_{n+2} \right ), \nonumber \\ 
&\left (\hat{Z}_n\hat{X}_{n+1}\hat{X}_{n+2}\hat{X}_{n+3}\hat{Y}_{n+4}+\hat{Y}_n\hat{X}_{n+1}\hat{X}_{n+2}\hat{X}_{n+3}\hat{Z}_{n+4} \right )\Bigg \} \ .
\label{i_s:eq:opPool1357}
\end{align}

The performance of ADAPT-VQE using these different operator pools is shown in Figure~\ref{i_s:fig:ADAPT1357}.
The infidelity is given as a function of ADAPT-VQE step for $L=24$, wavepacket parameters $\sigma_{}=0.15,\,k_0=0.36\pi$ and two sets of couplings, $g_x=1.25,\,g_z=0.15$ and $g_x=1.08,\,g_z=1.03$, with mass gaps of $m=1.6$ and $m=0.7$, respectively.
For early ADAPT-VQE steps, the best performing operator pool changes from step to step.
However, for later ADAPT-VQE steps, the largest operator pool $\{\hat{O}\}_7$ always performs the best.
This is more pronounced for the couplings with a smaller mass gap, $g_x=1.08,\,g_z=0.03$, likely because the longer Pauli strings in $\{\hat{O}\}_7$ are able to more effectively build out long-range correlations.
Note that these results do not show the circuit depths, 
which will generally be larger for the higher-order commutator operator pools due to the longer Pauli strings.
We identified $\{\hat{O}\}_3$ as the best balance between convergence and circuit depth, and it was therefore used for the quantum simulations performed in this work.

\begin{figure}
    \centering
    \includegraphics[width=\linewidth]{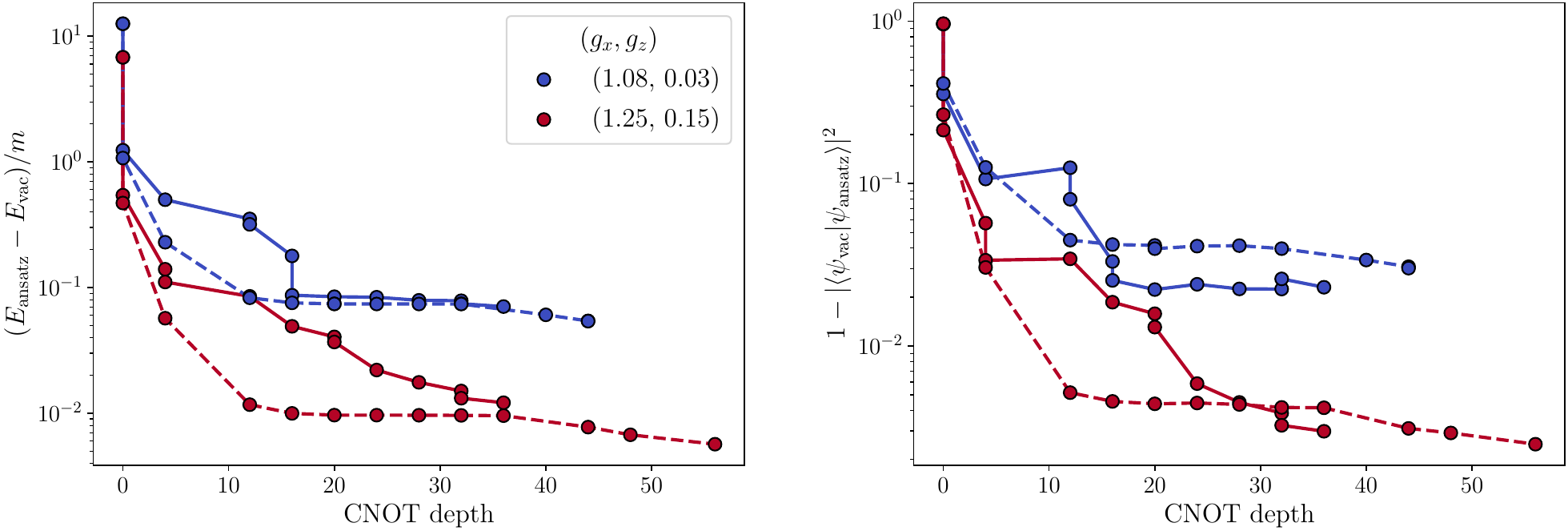}
    \caption{{\it The quality of the one-dimensional Ising field theory vacuum prepared using ADAPT-VQE.}
    The deviation in the energy (left) and infidelity (right) are shown for $L=28$ and up to 12 steps of ADAPT-VQE.
    The dashed line corresponds to the vacuum prepared using $\hat{U}_{\text{vac}}(\vec{\theta}_\star)$ in Eq.~\eqref{i_s:eq:ADAPTvac} and the solid line corresponds to $\hat{U}(\vec{\theta}_\star)$  in Eq.~\eqref{i_s:eq:ADAPTWPvac2}.
    The $\hat{U}(\vec{\theta}_\star)$ circuits are the same ones that were used in Fig.~\ref{i_s:fig:IsingADADPT_results} to prepare wavepackets with $k_0=0.36\pi,\,\sigma_{}=0.13$.}
\label{i_s:fig:IsingADADPT_VACresults}
\end{figure}

In the main text, vacuum-subtracted quantities were computed by time evolving an approximate vacuum constructed from the {\it wavepacket} ADAPT-VQE circuit,
\begin{align}
\hat{U}(\vec{\theta}_\star) |0\rangle^{\otimes L}  \ \approx \ |\psi_{\text{vac}}\rangle \ .
\label{i_s:eq:ADAPTWPvac2}
\end{align}
Alternatively, ADAPT-VQE can be performed with the goal of preparing the vacuum by starting from the state $|\psi_{\text{ansatz}}\rangle=|00 \ldots 0\rangle$.
This defines the {\it vacuum} ADAPT-VQE circuit,
\begin{align}
\hat{U}_{\text{vac}}(\vec{\theta}_\star) |0\rangle^{\otimes L}  \ \approx \ |\psi_{\text{vac}}\rangle \ .
\label{i_s:eq:ADAPTvac}
\end{align}
The results from preparing the vacuum in these two ways are shown in Fig.~\ref{i_s:fig:IsingADADPT_VACresults}.
The vacuum ADAPT-VQE circuits $\hat{U}_{\text{vac}}(\vec{\theta})$ are constructed from the operator pool in Eq.~\eqref{i_s:eq:opPool}.
Comparing the infidelities with Fig.~\ref{i_s:fig:IsingADADPT_results}, it is seen that the vacuum prepared by $\hat{U}(\vec{\theta}_\star)$ (solid line) is of higher quality than the corresponding wavepacket.
The vacuum prepared with $\hat{U}_{\text{vac}}(\vec{\theta}_\star)$ (dashed line)  initially converges quite quickly, but then plateaus after a few steps.
In comparison, the vacuum prepared with $\hat{U}(\vec{\theta}_\star)$ converges slower, but does not feature as prominent of a plateau.
Notice that the vacuum prepared with $\hat{U}(\vec{\theta}_\star)$ is, in some cases, better than with $\hat{U}_{\text{vac}}(\vec{\theta}_\star)$.
This is a result of the greedy nature of ADAPT-VQE that does not always pick the optimal operator ordering.
Comparing the two sets of couplings $(g_x,g_z,m)=\{(1.25,0.15,1.6),(1.08,0.03,0.7)\}$, it is seen that convergence to the vacuum is slower for a smaller mass gap, as expected.
The ADAPT-VQE operator ordering and variational parameters that prepare the vacuum are given in App.~\ref{i_s:app:ADAPTparam}.
%

\section{State preparation and Trotter errors}
\label{i_s:app:systematics}
\begin{figure}
    \centering
    \includegraphics[width=0.5\linewidth]{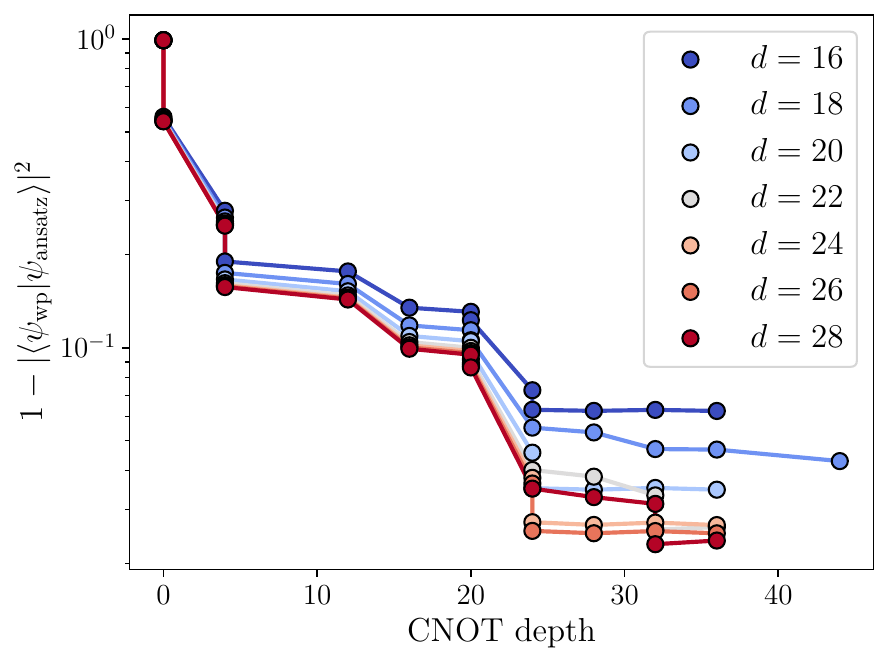}
    \caption{{\it The effect of truncating the spatial extent of $|W(k_0)\rangle$ on the quality of the prepared wavepacket in one-dimensional Ising field theory.
    }
    The infidelity of the prepared wavepacket for a selection of truncations $d$, defined in Eq.~\eqref{i_s:eq:WPd}.
    Results are shown for up to 12 steps of ADAPT-VQE with parameters $g_x=1.25,\,g_z=0.15,\,L=28,\,k_0=0.36\pi,\,\sigma_{}=0.13$.}
    \label{i_s:fig:dError}
\end{figure}
\noindent
As explained in App.~\ref{i_s:sec:qcirc_scatt}, when the lattice size is much larger than the wavepacket width, $L\gg \sigma_{}^{-1}$, the spatial support of the
initial wavepacket $|W(k_0)\rangle$ is truncated to a spatial interval $d$.
Explicitly, for a wavepacket centered at $x_0$,
\begin{align}
|W(k_0)\rangle \ = \ \sum_{n=0}^{L-1} e^{i\phi_n}c_n |2^n\rangle \ \to  \ {\cal N}\sum_{|n-x_0|\leq d/2} e^{i\phi_n}c_n |2^n\rangle \ . 
\label{i_s:eq:WPd}
\end{align}
Results from preparing wavepackets in Ising field theory with different $d$ are shown in Fig.~\ref{i_s:fig:dError} for $L=28,\,g_x=1.25,\,g_z=0.15$ and wavepacket parameters $\sigma_{}=0.13,\,k_0=0.36\pi$.
Increasing $d$ decreases the infidelity, as expected, and for these wavepacket parameters, there are essentially no truncation effects for $d\geq20$.
This informs the choice of $d=21$ used in the quantum simulations presented in Sec.~\ref{i_s:sec:qsim}.

\begin{figure}
    \centering
    \includegraphics[width=\linewidth]{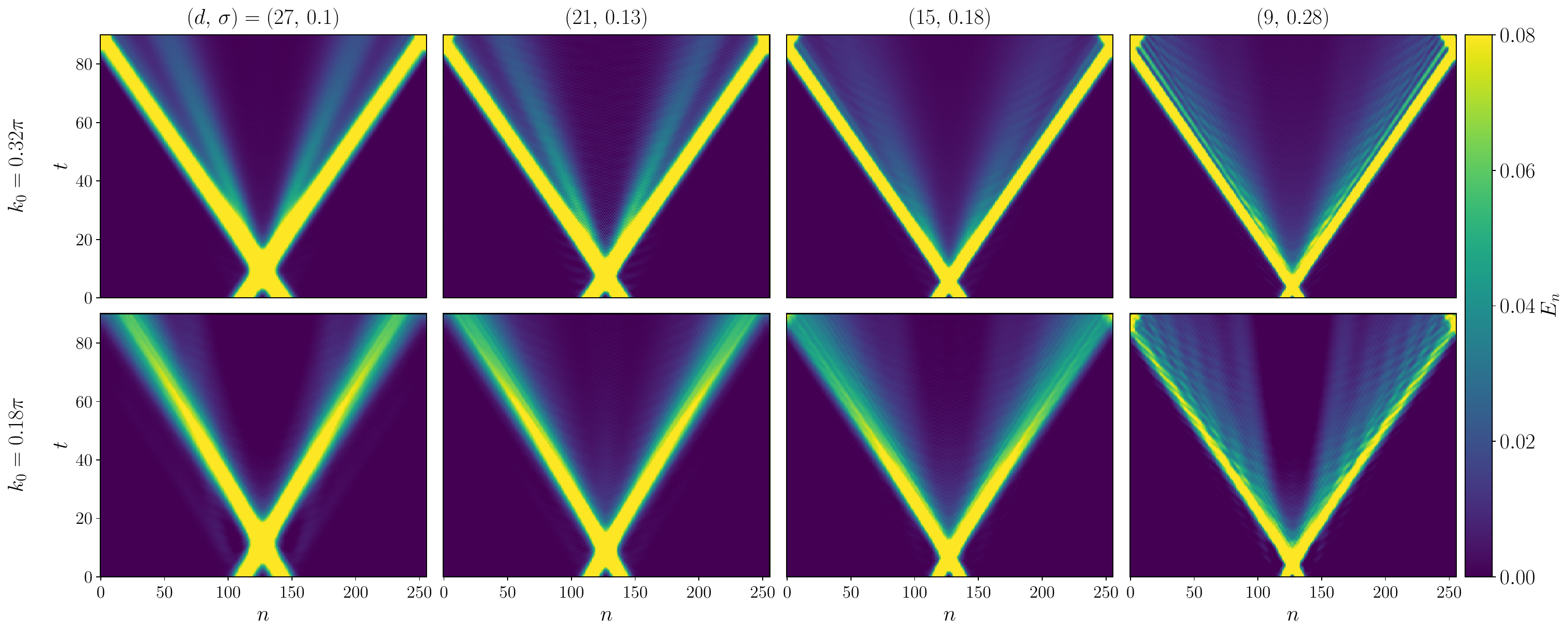}
    \caption{\textit{The effect of wavepacket size on scattering in one-dimensional Ising field theory.}
    The vacuum-subtracted energy density $E_n$ throughout MPS simulations of inelastic (top row) and elastic (bottom row) scattering is shown for an $L=256$ system with PBCs and a time step of $\delta t=1/16$. 
    The spread in momentum $\sigma_{}$ is chosen such that the wavepacket quality is the same after truncation to $d$ sites.}
\label{i_s:fig:wp_size_comparison}
\end{figure}

Larger wavepackets take longer to overlap and scatter but spread less under time evolution.
The latter is a result of variations in the group velocity, e.g., see the middle plot in Fig.~\ref{i_s:fig:dispersion}. 
At a fixed wavepacket quality, increasing $d$ allows the spread in momentum space $\sigma_{}$ to decrease.
The effect of varying $d$ on scattering dynamics is shown in Fig.~\ref{i_s:fig:wp_size_comparison}.
In all plots, $\sigma_{}$ is chosen to maintain a constant wavepacket amplitude after truncating $|W(k_0)\rangle$ to $d$ lattice sites as in Eq.~\eqref{i_s:eq:WPd}.
The top row shows the gradual washing out of the inelastic signal for scattering at $k_0=0.32\pi$ as $d$ is decreased from 27 to 9, where they are completely absent.
The bottom row shows elastic scattering for wavepackets with $k_0=0.18\pi$. 
For small $d$, interference effects from negative momenta and higher energies become significant, resulting in a checkerboard pattern in the energy density.
This prevents the elastic and inelastic channels from being reliably distinguished.
The wavepacket size $d=21$ is chosen for the quantum simulations in Sec.~\ref{i_s:sec:qsim} as a balance between resolution in momentum space and time to collision.
It is surprising that such a large hierarchy between $d$ and the correlation length $\xi\propto1/m=0.6$ is required to distinguish elastic and inelastic scattering.

The Trotter step size $\delta t$ is chosen to balance Trotter errors and circuit depth.
Smaller $\delta t$ approximates the exact time evolution more closely, but requires deeper circuits to reach a target $t_{\text{max}}$, thus incurring more device errors. 
In principle, if the device noise is characterized well, there exists an optimal $\delta t$ for each target simulation time $t$.
Examples of such exploration can be found for the digital case in Refs.~\cite{Knee:2015,Clinton:2021,Haghshenas:2025euj} and for the analog setting in Ref.~\cite{Zemlevskiy:2023eyw}.
When such knowledge is not readily available, as in this work, $\delta t$ must be chosen heuristically.
\begin{figure}
    \centering
    \includegraphics[width=\linewidth]{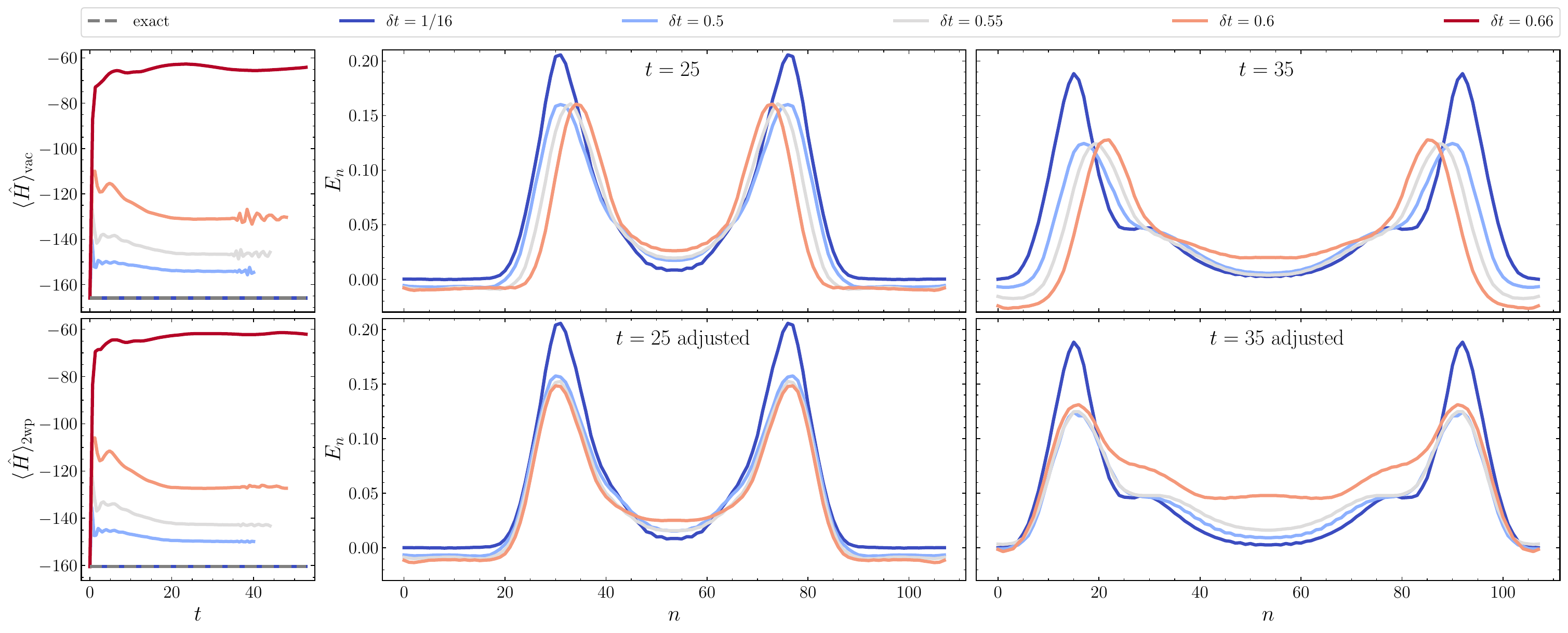}
    \caption{\textit{The effect of Trotter step size $\delta t$ on scattering in one-dimensional Ising field theory.} 
    Results are shown for a $L=108$ system with PBCs simulated using MPS with $k_0=0.32\pi$ and $\sigma_{}=0.13$. 
    Left column: the total energy of $|\psi_\text{vac}\rangle$ (top) and $|\psi_\text{2wp}\rangle$ (bottom) as a function of time $t$ for various $\delta t$. 
    Right: the top plots show the vacuum-subtracted energy density $E_n$ at $t=25$ and $t=35$. 
    The bottom plots show $E_n$ for $\delta t=1/16$  at $t=25,35$, as well as larger $\delta t$ with $t$ chosen so that the wavepacket peaks are aligned.}
\label{i_s:fig:trotter_errors}
\end{figure}

Figure~\ref{i_s:fig:trotter_errors} shows the effects of Trotter errors on the scattering process through MPS simulations.
The plots in the left column give the total energy of the vacuum and the two-wavepacket state as a function of time $t$ for various $\delta t$. 
A large $\delta t$ effectively quenches the state at $t=0$, causing a sharp change in the energy.
The system is heated to the chaotic regime for $\delta t=0.66$, where the dynamics no longer resemble those of the exact time evolution.
A sharp transition to chaotic dynamics is a generic feature of Trotterized time evolution, e.g., see Refs.~\cite{Heyl_2019,Sieberer_2019,Kargi:2021rww}.
The quench effects are also seen in the four plots on the right, where the vacuum-subtracted energy density $E_n$ is shown for $t=25,35$. 
The height of the wavepacket peaks is reduced for $\delta t > 1/16$ suggesting evolution toward a thermal state~\cite{DAlessio:2014rzv,Lazarides:2014loq,Yang:2023nak}. 

The Trotters errors cause the group velocity to be modified, and the wavepackets travel slower for larger $\delta t$.
This is seen in the top right plots of Fig.~\ref{i_s:fig:trotter_errors} for $t=25,35$, where the wavepacket peaks are closer together for larger $\delta t$, indicating slower propagation.
To compare the state when the wavepackets have propagated equal distances, the simulation time is adjusted so that the wavepacket peaks align (bottom right plots of Fig.~\ref{i_s:fig:trotter_errors}). 
The best approximation to the exact time evolution is given by the smallest Trotter step $\delta t=1/16$.
A larger $\delta t$ tends to make the energy density more uniform, and washes out the small ``bumps'' that are the signatures of inelastic particle production.
The bottom right plots of Fig.~\ref{i_s:fig:trotter_errors} show the bump from the emergence of a second pair of tracks vanishing as $\delta t$ is increased.
In addition, $E_n$ near the point of the collision does not return to the vacuum after the wavepackets have left the region.
Furthermore, a systematic offset from $E_n=0$ is seen to develop in the regions outside of the wavepackets for $\delta t>1/16$. 
This effect increases with Trotter step size and is persistent across all bond dimensions.
We attribute this to the quench from Trotterized time evolution creating a ``negative energy excitation'' that travels in the opposite direction from each wavepacket starting at $t=0$.
A Trotter step size of $\delta t=0.55$ is chosen for the quantum simulations presented in Sec.~\ref{i_s:sec:qsim} as it retains signatures of inelasticities while minimizing circuit depth and Trotter errors.

\section{Determining wavepackets with exact diagonalization}
\label{i_s:app:ExactWP}
\noindent
In App.~\ref{i_s:sec:WPQFTcircs}, the wavepackets prepared with ADAPT-VQE are compared to the exact wavepacket determined from exact diagonalization.
The exact wavepacket is the superposition of single particle eigenstates $|\psi_k\rangle$ as in Eq.~\eqref{i_s:eq:psiWPFull}.
These $|\psi_k\rangle$ are obtained by finding the lowest energy state of the Hamiltonian projected onto the desired $k$ block.
This appendix outlines how this projection is done.

The columns of the projection matrix consist of orthonormal basis vectors that span a given $k$ block. 
The task is to construct a basis for a given $k$ block.
A generic state with momentum $\tilde{k}$, $|\tilde{k}\rangle$, is an eigenstate under translations, $e^{i \hat{k} \Delta}|\tilde{k}\rangle \ = \ e^{i \tilde{k} \Delta}|\tilde{k}\rangle$.
A basis for these states can be built by grouping together all states that are related by translation with the appropriate phase.
As an example, consider a $L=4$ lattice in Ising field theory.
A basis for each momentum sector is,
\begin{align}
\{|k=0\rangle \} \ = \ & \Big \{ |0000\rangle \ , \ \frac{1}{2}\left (|0001\rangle +|0010\rangle + |0100\rangle +|1000\rangle \right ) \ , \nonumber \\
&\frac{1}{2}\left (|0011\rangle +|0110\rangle + |1100\rangle +|1001\rangle \right ) \ , \ \frac{1}{\sqrt{2}}\left (|0101\rangle +|1010\rangle \right )\ ,\nonumber \\ 
&\frac{1}{2}\left (|0111\rangle +|1110\rangle + |1101\rangle +|1011\rangle \right )\ , \ |1111\rangle \Big \} \ , \nonumber \\[4pt]
\{|k=\pi/2\rangle \} \ = \ 
& \Big \{ \frac{1}{2}\left (\vert0001 \rangle + i \vert0010 \rangle - \vert0100 \rangle -i \vert1000 \rangle\right ) \ , \nonumber\\  
&\frac{1}{2}\left (\vert0011 \rangle + i \vert0110 \rangle - \vert1100 \rangle -i \vert1001 \rangle\right ) \ , \nonumber \\
&\frac{1}{2}\left (\vert1110 \rangle + i \vert1101 \rangle - \vert1011 \rangle -i \vert0111 \rangle\right ) \Big \} \ , \nonumber \\[4pt]
\{|k=-\pi/2\rangle \} \ = \ 
& \Big \{ \frac{1}{2}\left (\vert0001 \rangle - i \vert0010 \rangle - \vert0100 \rangle +i \vert1000 \rangle\right ) \ , \nonumber\\  
&\frac{1}{2}\left (\vert0011 \rangle - i \vert0110 \rangle - \vert1100 \rangle +i \vert1001 \rangle\right ) \ , \nonumber \\
&\frac{1}{2}\left (\vert1110 \rangle - i \vert1101 \rangle - \vert1011 \rangle +i \vert0111 \rangle\right ) \Big \} \ , \nonumber \\[4pt]
\{|k=\pi\rangle \} \ = \ 
& \Big \{ \frac{1}{2}\left (\vert0001 \rangle - \vert0010 \rangle + \vert0100 \rangle - \vert1000 \rangle\right ) \ , \  \frac{1}{2}\left (\vert0011 \rangle - \vert0110 \rangle + \vert1100 \rangle - \vert1001 \rangle\right ) \ , \nonumber \\
&\frac{1}{2}\left (\vert1110 \rangle -  \vert1101 \rangle + \vert1011 \rangle - \vert0111 \rangle\right ) \ , \  \frac{1}{\sqrt{2}}\left (|0101\rangle -|1010\rangle \right )\Big \} \ .
\label{i_s:eq:L4kbasis}
\end{align}
All the above states are orthonormal and together span the $2^4$ dimensional Hilbert space.
Note that the state $\vert 0101 \rangle$ is not included in the $k=\pm\pi/2$ set as it maps onto itself with a minus sign under a translation by two sites.
Also notice that the states in $k=+\pi/2$ and $k=-\pi/2$ are related by complex conjugation as required by time reversal symmetry.

The basis vectors are then collected into the columns of the projection matrix $\hat{V}_k$, and the Hamiltonian is projected into a $k$ block $\hat{H}_k$ by
\begin{equation}
\hat{H}_k \ = \ \hat{V}^{\dagger}_{k} . \hat{H} . \hat{V}_k \ .
\end{equation}
For example, projecting onto $k=0$ using the basis in Eq.~\eqref{i_s:eq:L4kbasis} gives a $\hat{V}_0$ that is a $16\times 6$ matrix and a $\hat{H}_0$ that is a $6\times 6$ matrix.
Once the Hamiltonian has been projected into the desired $k$ block, the single particle eigenstates $|\psi_k\rangle$ are determined by diagonalizing $\hat{H}_k$.

The $|\psi_k\rangle$ can now be added in superposition to form the desired wavepacket (after using $\hat{V}_k$ to map them back to the full Hilbert space).
Here, a subtlety occurs. 
The eigenstates of a matrix are only determined up to an overall complex phase.
Therefore, the $|\psi_k\rangle$ that are obtained will generically have an overall complex phase that changes every time $\hat{H}_k$ is diagonalized.
This is a problem when constructing $|\psi_{\text{wp}}\rangle$ because the relative phases are set by the $e^{-i kx_0}$ in Eq.~\eqref{i_s:eq:psiWPFull}.
A consistent phase convention can be established by choosing a basis state that has a nonzero amplitude in every $|\psi_k\rangle$, and then multiplying each $|\psi_k\rangle$ by an overall phase to make that amplitude real.
For example, in Ising field theory, every $|\psi_k\rangle$ will contain the state $|0\rangle^{\otimes L-1}|1\rangle$, and $|\psi_k\rangle$ can be multiplied by an overall phase that makes the amplitude of $|0\rangle^{\otimes L-1}|1\rangle$ real.

\section{Useful circuit identities}
\label{i_s:app:qcircs}
\begin{figure}
    \centering
    \includegraphics[width=0.8\linewidth]{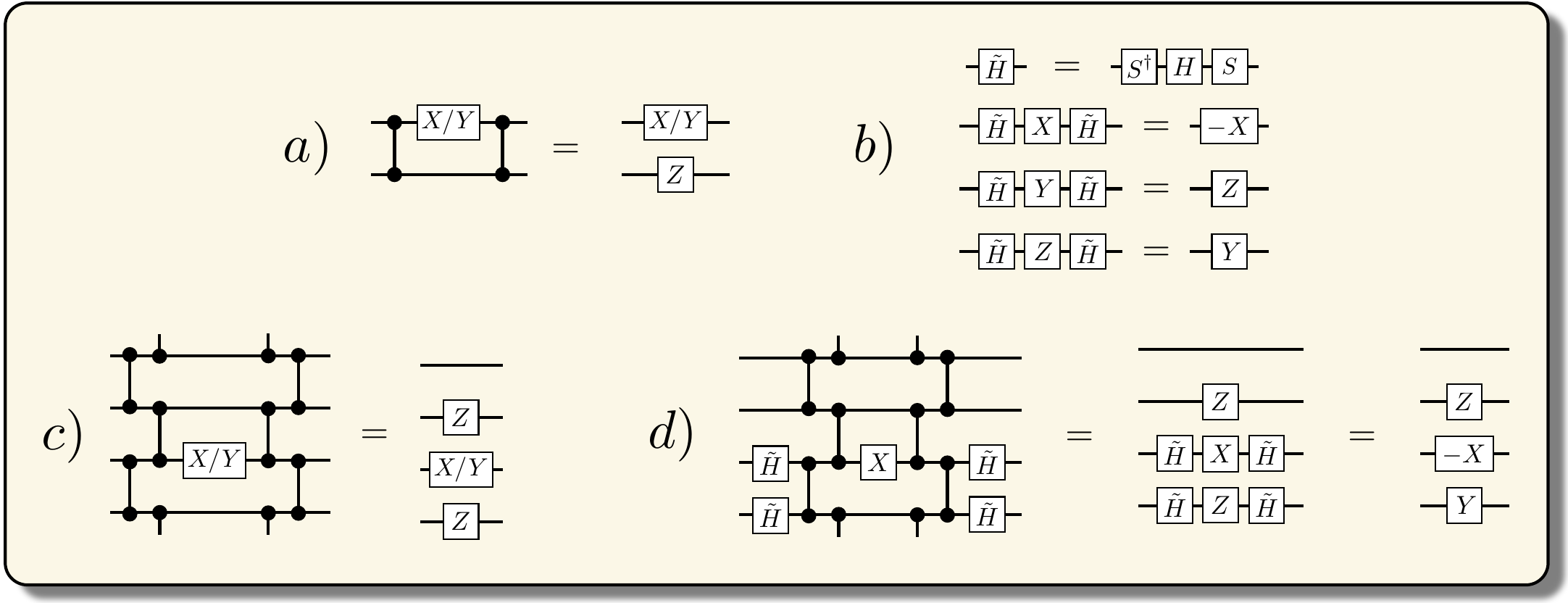}
    \caption{Circuit identities used to construct the circuits in Fig.~\ref{i_s:fig:IsingWPCircs} that prepare wavepackets in one-dimensional Ising field theory.}
    \label{i_s:fig:circIdentities}
\end{figure}

\noindent
The quantum circuits used to implement the unitary evolution with respect to the operator pool in Eq.~\eqref{i_s:eq:opPool} are inspired by the techniques in Ref.~\cite{Chernyshev:2025jyw}.
First, consider the circuit in Fig.~\ref{i_s:fig:IsingWPCircs}c) that implements unitary evolution of $\sum_n\hat{Z}_n\hat{Y}_{n+1}\hat{Z}_{n+2}$.
The building block for this circuit is shown in Fig.~\ref{i_s:fig:circIdentities}c).
Defining a brickwall sequence of $CZ$ as $\overline{CZ}$, this circuit implements the following unitary,
\begin{equation}
\left (\overline{CZ}\right )\hat{X}_n\left (\overline{CZ}\right ) \to \hat{Z}_{n-1}\hat{X}_n\hat{Z}_{n+1} \ \ , \ \ \left (\overline{CZ}\right )\hat{Y}_n \left (\overline{CZ}\right )\to \hat{Z}_{n-1}\hat{Y}_n\hat{Z}_{n+1}
\end{equation}
These relations can be derived by applying the identity shown in Fig.~\ref{i_s:fig:circIdentities}a).
The circuit in Fig.~\ref{i_s:fig:IsingWPCircs}c) immediately follows 
\begin{equation}
\left (\overline{CZ}\right )\prod_{n=0}^{L-1} R_Y(\theta)_n \left (\overline{CZ}\right )\to \exp \left [-i \frac{\theta}{2}\left (\sum_{n=0}^{L-1}\hat{Z}_{n-1}\hat{Y}_n\hat{Z}_{n+1}\right )\right ] \ . 
\end{equation}

Next, consider the circuit in Fig.~\ref{i_s:fig:IsingWPCircs}d) that implements unitary evolution of 
\begin{align}
    \sum_n\left (\hat{Z}_n\hat{X}_{n+1}\hat{Y}_{n+2} + \hat{Y}_n\hat{X}_{n+1}\hat{Z}_{n+2}\right )\nonumber
\end{align}
A useful gate is $\tilde{H} \equiv S.H.S^{\dagger}$, which acts like a Hadamard between the $Y$- and $Z$-bases, as shown in Fig.~\ref{i_s:fig:circIdentities}b). 
In Fig.~\ref{i_s:fig:circIdentities}d) this change of basis is combined with Fig.~\ref{i_s:fig:circIdentities}c) to show that
\begin{equation}
\tilde{H}_n\tilde{H}_{n+1}\left (\overline{CZ}\right ) \hat{X}_n\left (\overline{CZ}\right )\tilde{H}_n\tilde{H}_{n+1} \ = \ -\hat{Z}_{n-1}\hat{X}_n \hat{Y}_{n+1} \ . 
\end{equation}
Similarly it can be shown that 
\begin{align}
&\tilde{H}_n\tilde{H}_{n+1}\left (\overline{CZ}\right ) \hat{X}_{n+1}\left (\overline{CZ}\right )\tilde{H}_n\tilde{H}_{n+1} \ = \ -\hat{Y}_{n}\hat{X}_{n+1} \hat{Z}_{n+2} \ ,\nonumber \\[4pt]
&\tilde{H}_n\tilde{H}_{n+1}\left (\overline{CZ}\right ) \hat{X}_{n-1}\left (\overline{CZ}\right )\tilde{H}_n\tilde{H}_{n+1} \ = \ \hat{Z}_{n-2}\hat{X}_{n-1}\hat{Y}_n \ .
\end{align}
Next, split the Pauli strings in $\sum_n\left (\hat{Z}_n\hat{X}_{n+1}\hat{Y}_{n+2} + \hat{Y}_n\hat{X}_{n+1}\hat{Z}_{n+2}\right )$ into two commuting sets, 
\begin{align}
&\{\hat{O}_1\} \ = \ \left \{\sum_{n=0}^{\lfloor (L-1)/3\rfloor}\left (\hat{Z}_{3n}\hat{X}_{3n+1}\hat{Y}_{3n+2}  +\delta_{3n+1<L}\hat{Z}_{3n+1}\hat{X}_{3n+2}\hat{Y}_{3n+3} +\delta_{3n+2<L}\hat{Y}_{3n+2}\hat{X}_{3n+3}\hat{Z}_{3n+4} \right ) \right \} \ , \nonumber \\[4pt]
&\{\hat{O}_2\} \ = \ \left \{\sum_{n=0}^{\lfloor (L-1)/3\rfloor}\left (\hat{Y}_{3n}\hat{X}_{3n+1}\hat{Z}_{3n+2}  +\delta_{3n+1<L}\hat{Y}_{3n+1}\hat{X}_{3n+2}\hat{Z}_{3n+3} +\delta_{3n+2<L}\hat{Z}_{3n+2}\hat{X}_{3n+3}\hat{Y}_{3n+4}  \right ) \right \} \ .
\end{align}
The unitary evolution of the exponential of each set is implemented separately using the construction in Fig.~\ref{i_s:fig:circIdentities}d).
The circuit in Fig.~\ref{i_s:fig:IsingWPCircs}d) first implements the unitary of evolution of $\{ \hat{O}_1\}$ and then $\{\hat{O}_2\}$.
Half of the rotations are $R_X(-\theta)$ due to the minus sign in $\tilde{H}\hat{X}\tilde{H}=-\hat{X}$.

%

\section{Operators and variational parameters for preparing wavepackets in one-dimensional Ising field theory}
\label{i_s:app:ADAPTparam}
\noindent
The variational parameters and operator ordering used to prepare wavepackets in one-dimensional Ising field theory with ADAPT-VQE for the quantum simulations performed in Sec.~\ref{i_s:sec:qsim} are given in Table~\ref{i_s:tab:qsimIsingAdaptAng},
for the {\tt qiskit} statevector simulations performed in App.~\ref{i_s:sec:WPQFTcircs} are given in Table~\ref{i_s:tab:IsingAdaptAngL28},
for the MPS simulations performed in App.~\ref{i_s:sec:csimscatt} are given in  Tables~\ref{i_s:tab:IsingAdaptAngL256} and~\ref{i_s:tab:IsingAdaptAngL256_2}
and for the vacuum preparation in App.~\ref{i_s:app:57Adapt} are given in Table~\ref{i_s:tab:VacIsingAdaptAng}.

As in Refs.~\cite{Farrell:2024fit,Zemlevskiy:2024vxt},
the variational parameters that prepare wavepackets converge exponentially with increasing system size.
For example, the parameters used to prepare wavepackets with $g_x=1.25,\,g_z=0.15$ and $k_0=0.36\pi$ for $L=28$, Table~\ref{i_s:tab:IsingAdaptAngL28}, and for $L=256$, Table~\ref{i_s:tab:IsingAdaptAngL256} agree to 4 decimal places. 
This is due to the exponential convergence of the mass gap with increasing system size, and the associated exponential suppression of finite size effects.
This convergence motivated the SC-ADAPT-VQE workflow~\cite{Farrell:2023fgd} (see also Refs.~\cite{Klco:2019yrb,Klco:2020aud}) where quantum circuits are optimized on small lattices and then systematically extrapolated to prepare states on large lattices.
For preparing wavepackets, this approach has the limitation that it can only prepare wavepackets of a fixed size, i.e., $d/L$ is a constant.
Indeed, in this work we found that the operators and variational parameters that prepare small wavepackets were not optimal for preparing larger wavepackets.
In this work no parameter extrapolations were necessary as our state preparation circuits for large $L$ could be directly simulated using MPS.
However, circuit extrapolations and the SC-ADAPT-VQE workflow will likely be helpful for preparing wavepackets in higher dimensions where classical computing techniques are not able to faithfully simulate large circuits.
\begin{table}[!t]
\begin{tabularx}{\textwidth}{|c || Y | Y | Y | Y | Y | Y  | Y |}
  \hline
 \diagbox{$k_0$}{$\theta_i$} & 
 $\hat{Y}$&
 $\hat{Y}\hat{Z}$ & 
 $\hat{Y}$& $\hat{Z}\hat{X}\hat{Y}$& $\hat{Y}\hat{Z}$&$\hat{Z}\hat{Y}\hat{Z}$&
 $\hat{Y}$

 \\
 \hline\hline
$0.32\pi$ &  0.0191&  0.0276& -0.4497&  0.0226&  0.0618&  0.0900&   -0.2238\\
 \hline
\end{tabularx}
\caption{
Operators/parameters used to prepare wavepackets in the quantum simulations of scattering in one-dimensional Ising field theory presented in Sec.~\ref{i_s:sec:qsim}.
These parameters correspond to the 7th step of ADAPT-VQE for $L=104$, $g_x=1.25,\,g_z=0.15$ and $k_0=0.32\pi,\, \sigma_{}=0.13$.
For clarity, each operator is labeled by one term, e.g., $\hat{Y}\hat{Z}$ corresponds to $\hat{Y}\hat{Z} + \hat{Z}\hat{Y}$ in Eq.~\eqref{i_s:eq:opPool}. }
 \label{i_s:tab:qsimIsingAdaptAng}
\end{table}
\begin{table}[!t]
\footnotesize
\setlength{\tabcolsep}{1pt}
\begin{tabularx}{\textwidth}{|c || Y | Y | Y | Y | Y | Y | Y | Y | Y |Y|Y|Y|}
  \hline
 \diagbox{$(g_x,g_z)$}{$\theta_i$} & 
 $\hat{Y}$&
 $\hat{Y}\hat{Z}$ &  $\hat{Z}\hat{X}\hat{Y}$ & 
 $\hat{Y}$ &  
 $\hat{Z}\hat{X}\hat{Y}$&  $\hat{Z}\hat{Y}\hat{Z}$&  
 $\hat{Y}\hat{Z}$&  
 $\hat{Y}$&
 $\hat{Y}\hat{Z}$
 &
 $\hat{Y}$
&
 $\hat{Y}\hat{X}$
 &
 $\hat{Z}\hat{Y}\hat{Z}$

 \\
 \hline\hline
$(1.25,0.15)$ &  0.1212 & 0.0185 & -- & -0.5452 &  0.0397 & --&0.0599& --&0.0556 &-0.2637& --&0.0566\\
 \hline
 $(1.18,0.08)$ & -0.3517 & 0.0610 &  0.0477 & -0.1425 &--& 0.1107&--& -0.2030 &  0.0310&--&   0.0176&--\\
 \hline
  $(1.15,0.05)$ &-0.3866 & 0.0700&    0.0473&  0.0201&--&  0.0705&--& -0.3268 & 0.0288&--&  --&0.0368\\
   \hline
  $(1.08,0.03)$ & -0.3653 & 0.0754 & 0.0539 &-0.1910&--&   0.1234&--& -0.1772&  0.0313&--&  0.0266&--\\
 \hline
\end{tabularx}
\caption{
Operators/parameters used to prepare wavepackets in one-dimensional Ising field theory for $L=28$ and $k_0=0.36\pi,\, \sigma_{}=0.13$.
These parameters correspond to the 8th step in Fig.~\ref{i_s:fig:IsingADADPT_results}.
For clarity, each operator is labeled by one term, e.g., $\hat{Y}\hat{Z}$ corresponds to $\hat{Y}\hat{Z} + \hat{Z}\hat{Y}$ in Eq.~\eqref{i_s:eq:opPool}.}
 \label{i_s:tab:IsingAdaptAngL28}
\end{table}
\begin{table}[!t]
\footnotesize
\setlength{\tabcolsep}{1pt}
\begin{tabularx}{\textwidth}{|c || Y | Y | Y | Y | Y | Y | Y | Y | Y |Y|Y|Y|Y|Y|}
  \hline
 \diagbox{$k_0$}{$\theta_i$} & 
 $\hat{Y}$&
 $\hat{Y}\hat{Z}$ & $\hat{Y}$& 
 $\hat{Z}\hat{X}\hat{Y}$&  $\hat{Y}$&$\hat{Y}\hat{Z}$
 &$\hat{Z}\hat{Y}\hat{Z}$&  
 $\hat{Y}\hat{Z}$&  
 $\hat{Y}$&
 $\hat{Y}\hat{Z}$
 &$\hat{Z}\hat{Y}\hat{Z}$&
 $\hat{Y}$
 &
 $\hat{Z}\hat{X}\hat{Y}$

 \\
 \hline\hline
$0.36\pi$ &  0.1212&  0.0185& -0.5452&  0.0397&--&  0.0599&--&  0.0556& -0.2637&--&0.0566&--&--\\
\hline
$0.32\pi$ &  0.0505&  0.0006&-0.3983&  0.0316&--&  0.0750 & 0.0868& --&-0.3029&--& 0.0349&--&--\\
\hline
$0.28\pi$ &  0.0499&  0.0314& -0.5092&  0.0231&--&  0.0401&--&  0.0325&--&--&  0.0739&-0.2064&--\\
\hline
$0.18\pi$ &  -0.2663&  0.0566&--&  0.0470& -0.1871&--&  0.0806&--& -0.2214& 0.0524&--&--&0.0127\\
 \hline
\end{tabularx}
\caption{
Operators/parameters used to prepare wavepackets in one-dimensional Ising field theory for $L=256$,\, $g_x=1.25,\,g_z=0.15$ and $\sigma_{}=0.13$.
These parameters were used to prepare the initial state of the scattering simulations in Fig.~\ref{i_s:fig:2WP_scattering_MPS}.
For clarity, each operator is labeled by one term, e.g., $\hat{Y}\hat{Z}$ corresponds to $\hat{Y}\hat{Z} + \hat{Z}\hat{Y}$ in Eq.~\eqref{i_s:eq:opPool}.}
 \label{i_s:tab:IsingAdaptAngL256}
\end{table}
\begin{table}[!t]
\footnotesize
\setlength{\tabcolsep}{1pt}
\begin{tabularx}{\textwidth}{|c || Y | Y | Y | Y | Y | Y  | Y |Y|Y|Y|Y|Y|}
  \hline
 \diagbox{$k_0$}{$\theta_i$} & 
 $\hat{Y}$&
 $\hat{Y}\hat{Z}$ & 
 $\hat{Z}\hat{X}\hat{Y}$& $\hat{Y}$& $\hat{Z}\hat{Y}\hat{Z}$&$\hat{Y}$&
 $\hat{Y}\hat{Z}$&  
 $\hat{Z}\hat{Y}\hat{Z}$&
 $\hat{Y}$
 &
 $\hat{Z}\hat{X}\hat{Y}$

 \\
 \hline\hline
$0.20\pi$ &  -0.3124&  0.0412&  0.0511&  0.1360 &  0.0235& -0.3669&  0.0579&
        0.0828& -0.1367&  0.0009\\
 \hline
\end{tabularx}
\caption{
Same as Table~\ref{i_s:tab:IsingAdaptAngL256} but for $k_0=0.20\pi$ and with 10 ADAPT-VQE steps.
Operators/parameters used to prepare wavepackets in one-dimensional Ising field theory for $L=256$,\, $g_x=1.25,\,g_z=0.15$ and $\sigma_{}=0.13$.
These parameters were used to prepare the initial state of the scattering simulations in Fig.~\ref{i_s:fig:2WP_scattering_MPS}.
For clarity, each operator is labeled by one term, e.g., $\hat{Y}\hat{Z}$ corresponds to $\hat{Y}\hat{Z} + \hat{Z}\hat{Y}$ in Eq.~\eqref{i_s:eq:opPool}.}
 \label{i_s:tab:IsingAdaptAngL256_2}
\end{table}
\begin{table}[!t]
\footnotesize
\setlength{\tabcolsep}{1pt}
\begin{tabularx}{\textwidth}{|c || Y | Y | Y | Y | Y | Y | Y |Y|Y|Y|}
  \hline
 \diagbox{$(g_x,g_z)$}{$\theta_i$} & 
 $\hat{Y}$&
 $\hat{Y}\hat{Z}$ &  $\hat{Z}\hat{X}\hat{Y}$ &  
 $\hat{Y}\hat{Z}$&
 $\hat{Y}\hat{Z}$&
 $\hat{Z}\hat{Y}\hat{Z}$&
 $\hat{Y}$&$\hat{Y}\hat{X}$
 &$\hat{Z}\hat{Y}\hat{Z}$&
 $\hat{Y}\hat{X}$
 \\
 \hline\hline
$(1.25,0.15)$ & -0.4735 & 0.0244 & 0.0145 & 0.0209 & 0.0195 & 0.0008 &--&-0.0024& -0.0029&-- \\
   \hline
  $(1.08,0.03)$ &-0.4885 & 0.0288 & 0.0194 & 0.0248 & 0.0235 & --&-0.0048 & 0.0021&--& -0.0026 \\
 \hline
\end{tabularx}
\caption{
Operators/parameters used to prepare the vacuum in one-dimensional Ising field theory for $L=28$.
These parameters correspond to the 8th step in Fig.~\ref{i_s:fig:IsingADADPT_VACresults}.
For clarity, each operator is labeled by one term, e.g., $\hat{Y}\hat{Z}$ corresponds to $\hat{Y}\hat{Z} + \hat{Z}\hat{Y}$ in Eq.~\eqref{i_s:eq:opPool}.}
 \label{i_s:tab:VacIsingAdaptAng}
\end{table}
%

\section{Error mitigation with energy rescaling}
\label{i_s:app:energy_rescale}
\noindent
An ideal simulation would conserve the total energy $E_{\text{tot}}=\sum_nE_n$.
In a quantum simulation, the combination of Trotter errors and the imperfect mitigation of device noise leads to a violation of energy conservation.
We find that a new error mitigation step that restores energy conservation in post-processing can reduce systematic errors.
This is done by applying a multiplicative rescaling of the vacuum-subtracted energy density predicted by ODR,
\begin{align}
    E_n = E_n^{(\text{ODR})} \frac{E_{\text{tot}}}{\sum_{j=0}^{L-1} E_j^{(\text{ODR})}}\ ,
\end{align}
where $E_{\text{tot}}$ is the energy of the initial state and $E_j^{(\text{ODR})}$  is the energy density predicted after ODR.
This rescaling corrects Trotter and device errors that lead to violations of energy conservation.
It also reduces the classical computing overhead required in ODR.
Our implementation of ODR, described in App.~\ref{i_s:sec:qsimDetails}, assumed that observables could be efficiently computed in the time-evolved vacuum $U_2(t) |\psi_{\text{vac}}\rangle$.
This observable $\langle \hat{O}\rangle_\text{pred}$ is used to compute the signal strength in Eq.~\eqref{i_s:eq:scattering_ising_odr_p_j}.
Efficient classical computation of observables in the time-evolved vacuum assumes that the time-evolved state remains close to the true vacuum, which can be efficiently represented by a MPS in one dimension.
However, the large Trotter step size of $\delta t = 0.55$ used in this work effectively quenches the vacuum, and the bond dimension required to compute observables in $U_2(t) |\psi_{\text{vac}}\rangle$ can be quite large (although not as large as in the scattering simulations).
This challenge is overcome if the energy conservation is enforced in post-processing because the {\it unevolved} vacuum can be used to compute $\langle \hat{O}\rangle_\text{pred}$ and determine the signal strength,
\begin{align}
    p_{\hat{O}} \ = \ \frac{\langle \hat{O}\rangle_\text{meas}}{\langle \hat{O}\rangle_\text{pred}} \ = \ \frac{\langle \psi_\text{vac}|\hat{U}^{\dagger}_2(t) \hat{O} \hat{U}_2(t)|\psi_\text{vac}\rangle_\text{meas}}{\langle \psi_\text{vac}| \hat{O} |\psi_\text{vac}\rangle_\text{pred}} \ .
    \label{i_s:eq:scattering_ising_odr_p_j_vac}
\end{align}
The $\sim\!15\%$ ``error'' incurred by using the static vacuum to compute $\langle \hat{O}\rangle_\text{pred}$ instead of the time-evolved vacuum is fixed when energy conservation is enforced.

\begin{figure}[t!]
    \centering
    \includegraphics[width=0.75\linewidth]{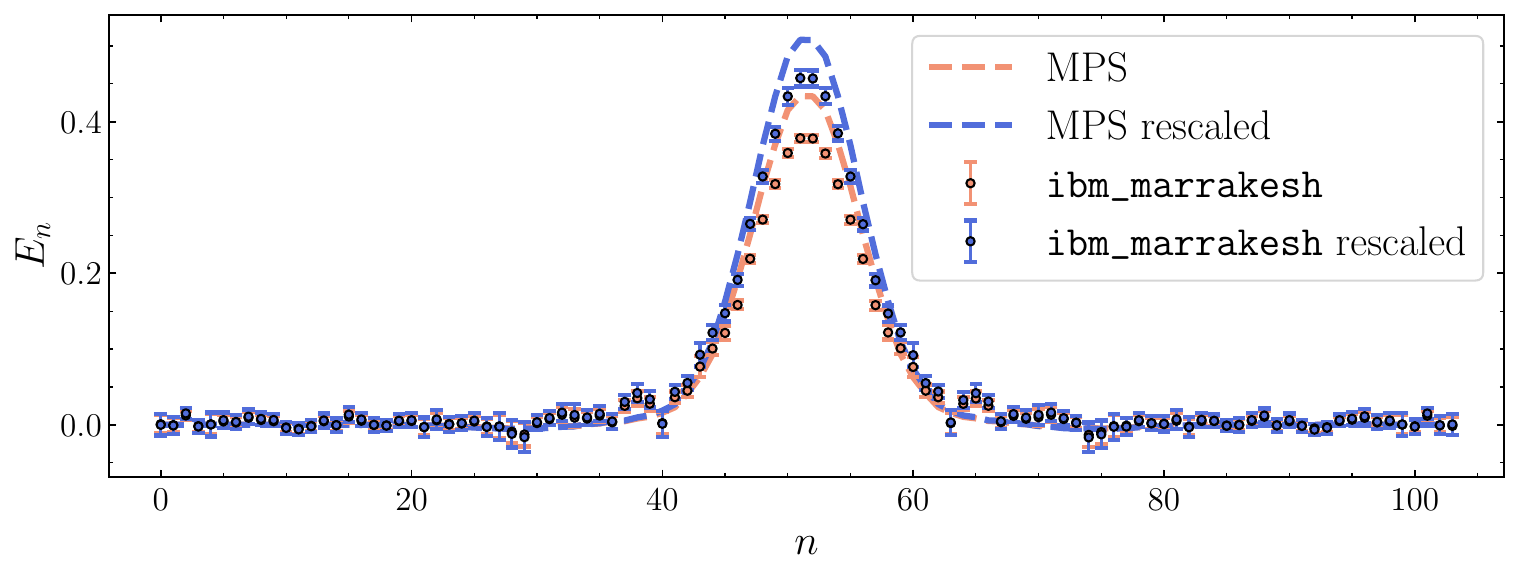}
    \caption{{\it The effect of energy rescaling.}
    MPS predictions (dashed lines) are compared with data obtained from {\tt ibm\_marrakesh} (points) for $t_1=8.25$.
    The results without energy rescaling presented in Fig.~\ref{i_s:fig:ibm_results}b) are shown in orange while the results with energy conservation enforced are shown in blue.}
    \label{i_s:fig:Energy_rescale}
\end{figure}

The effect of enforcing energy conservation on the results is shown in Fig.~\ref{i_s:fig:Energy_rescale} for $t_1 = 8.25$.
Without energy conservation, the energy density determined from {\tt ibm\_marrakesh} (orange points) is significantly lower than the MPS predictions in the middle of the collision (orange dashed line).
Enforcing energy conservation has the largest impact on the energy density in the middle of the lattice, and the rescaled energy density (blue points) perfectly aligns with MPS.
However, for a fair a comparison, the rescaled energy density should be compared to the rescaled MPS predictions (blue dashed line).
The energy in the MPS simulations changes because of Trotter errors.
The systematic error in the energy density near the point of collision is seen to be reduced, although it is still present.
This is primarily because there is a slight positive-energy bias compared with MPS in the quantum results away from the middle of the lattice. 
When summed over many sites, this small bias becomes significant and reduces the energy rescale value.
For later simulations times, the positive energy bias in the quantum results becomes more significant and the energy rescaling becomes ineffective.
Techniques that use the light cone to zero the energy density outside of the scattering region would remove the positive bias, and could improve the efficacy of this error mitigation strategy.
%

\section{Tables of results}
\noindent
Table~\ref{i_s:tab:results_t} gives data for the vacuum-subtracted energy density $E_n$ throughout the quantum simulations of inelastic scattering presented in Sec.~\ref{i_s:sec:qsim}, and Table~\ref{i_s:tab:results_t_1wp} gives data for $E_n$ for the single wavepacket simulations.
An $L=104$ system with OBCs and $g_x=1.25,\,g_z=1.5$ is simulated using a Trotter step size $\delta t=0.55$.
The wavepacket parameters $k_0=0.32\pi,\,\sigma_{}=0.13$ are used. 
For each time $t$, the left column gives the MPS data and the right column shows the results from {\tt ibm\_marrakesh} after error mitigation.
Parity symmetry about the collision point equates $E_n$ and $E_{L-1-n}$, and the energy density is not shown for sites related by this symmetry. 
The MPS values of $E_n$ are not exactly 0 outside of the simulation light cone due to finite Trotter step size. 
\begin{table}[h]
\renewcommand{\arraystretch}{0.6}
\scriptsize
\setlength{\tabcolsep}{1pt}
\begin{tabularx}{\linewidth}{|c||Y|Y||Y|Y||Y|Y||Y|Y|}
\hline
\rule{0pt}{10pt} & \multicolumn{8}{c|}{\large$E_n$}\\\hline\hline
\rule{0pt}{10pt} \multirow{2}{*}{Spatial site $n$} &  \multicolumn{2}{c||}{$t_0=0.0$}  &  \multicolumn{2}{c||}{$t_1=8.25$} & \multicolumn{2}{c||}{$t_2=16.5$} & \multicolumn{2}{c|}{$t_3=24.75$} \\\cline{2-9}
\rule{0pt}{10pt} & MPS & QC & MPS & QC & MPS & QC & MPS & QC\\
\hline\hline
  0 & 0.000 & 0.005(10)  &  0.000 & 0.001(12)  & -0.002 & -0.001(10) & -0.023 & 0.001(08)  \\
  1 & 0.000 & -0.009(10) &  0.000 & -0.000(09) & -0.001 & -0.009(07) & -0.014 & 0.002(06)  \\
  2 & 0.000 & -0.016(11) &  0.000 & 0.012(06)  & -0.001 & 0.007(06)  & -0.015 & -0.006(05) \\
  3 & 0.000 & -0.003(12) &  0.000 & -0.002(08) & -0.001 & -0.009(07) & -0.015 & -0.011(06) \\
  4 & 0.000 & 0.007(11)  &  0.000 & 0.001(13)  & -0.001 & 0.005(11)  & -0.015 & -0.008(08) \\
  5 & 0.000 & 0.006(11)  &  0.000 & 0.004(09)  & -0.001 & -0.002(07) & -0.015 & 0.003(06)  \\
  6 & 0.000 & -0.002(09) &  0.000 & 0.003(07)  & -0.001 & -0.008(06) & -0.015 & -0.002(07) \\
  7 & 0.000 & 0.001(10)  &  0.000 & 0.009(08)  & -0.002 & -0.001(06) & -0.015 & -0.010(06) \\
  8 & 0.000 & 0.001(09)  &  0.000 & 0.006(07)  & -0.002 & -0.012(06) & -0.014 & -0.006(06) \\
  9 & 0.000 & 0.000(08)  &  0.000 & 0.005(06)  & -0.003 & -0.010(06) & -0.014 & -0.004(05) \\
 10 & 0.000 & -0.001(11) &  0.000 & -0.003(06) & -0.003 & -0.012(06) & -0.014 & 0.012(06)  \\
 11 & 0.000 & -0.002(16) &  0.000 & -0.005(06) & -0.004 & -0.002(06) & -0.014 & -0.008(06) \\
 12 & 0.000 & -0.000(15) &  0.000 & -0.001(07) & -0.004 & 0.003(06)  & -0.014 & -0.003(06) \\
 13 & 0.000 & -0.001(13) &  0.000 & 0.005(07)  & -0.005 & -0.009(07) & -0.014 & 0.009(08)  \\
 14 & 0.000 & -0.021(22) &  0.000 & -0.000(07) & -0.006 & -0.004(08) & -0.014 & -0.002(07) \\
 15 & 0.000 & -0.026(18) &  0.000 & 0.011(08)  & -0.006 & -0.004(08) & -0.015 & 0.003(06)  \\
 16 & 0.000 & 0.002(10)  &  0.000 & 0.005(07)  & -0.006 & 0.008(07)  & -0.014 & 0.005(06)  \\
 17 & 0.000 & -0.002(09) &  0.000 & -0.000(07) & -0.007 & 0.007(07)  & -0.014 & 0.006(06)  \\
 18 & 0.000 & 0.000(05)  &  0.000 & -0.001(06) & -0.007 & -0.000(07) & -0.012 & 0.003(06)  \\
 19 & 0.000 & -0.002(08) &  0.000 & 0.005(07)  & -0.007 & 0.005(07)  & -0.010 & 0.008(06)  \\
 20 & 0.000 & 0.007(08)  &  0.000 & 0.005(08)  & -0.006 & 0.009(08)  & -0.005 & 0.023(08)  \\
 21 & 0.000 & -0.017(10) & -0.001 & -0.003(11) & -0.006 & -0.003(12) &  0.002 & 0.030(09)  \\
 22 & 0.000 & -0.017(13) & -0.001 & 0.006(10)  & -0.006 & -0.009(11) &  0.014 & 0.039(07)  \\
 23 & 0.000 & 0.003(12)  & -0.001 & 0.001(08)  & -0.006 & 0.008(07)  &  0.031 & 0.061(07)  \\
 24 & 0.000 & -0.000(07) & -0.002 & 0.002(08)  & -0.005 & -0.004(07) &  0.050 & 0.068(08)  \\
 25 & 0.000 & 0.001(09)  & -0.003 & 0.004(08)  & -0.006 & 0.009(07)  &  0.070 & 0.103(07)  \\
 26 & 0.000 & -0.011(17) & -0.004 & -0.002(10) & -0.007 & 0.015(08)  &  0.094 & 0.121(07)  \\
 27 & 0.000 & -0.001(12) & -0.004 & -0.002(15) & -0.006 & 0.005(15)  &  0.117 & 0.130(09)  \\
 28 & 0.001 & -0.006(08) & -0.005 & -0.009(15) & -0.007 & 0.018(15)  &  0.132 & 0.139(12)  \\
 29 & 0.002 & 0.001(09)  & -0.005 & -0.013(16) & -0.005 & 0.031(15)  &  0.143 & 0.127(08)  \\
 30 & 0.012 & 0.019(12)  & -0.004 & 0.002(08)  & -0.003 & 0.022(08)  &  0.148 & 0.120(06)  \\
 31 & 0.020 & 0.015(10)  & -0.003 & 0.008(08)  &  0.002 & 0.021(07)  &  0.144 & 0.110(06)  \\
 32 & 0.036 & 0.019(12)  & -0.002 & 0.013(10)  &  0.008 & 0.038(07)  &  0.139 & 0.095(06)  \\
 33 & 0.057 & 0.094(26)  & -0.002 & 0.010(12)  &  0.020 & 0.042(11)  &  0.127 & 0.084(08)  \\
 34 & 0.088 & 0.072(07)  &  0.001 & 0.008(08)  &  0.041 & 0.062(08)  &  0.123 & 0.088(08)  \\
 35 & 0.124 & 0.108(09)  &  0.002 & 0.012(08)  &  0.065 & 0.080(07)  &  0.106 & 0.075(06)  \\
 36 & 0.168 & 0.150(11)  &  0.005 & 0.003(09)  &  0.092 & 0.100(07)  &  0.088 & 0.074(06)  \\
 37 & 0.209 & 0.183(09)  &  0.005 & 0.025(08)  &  0.120 & 0.146(06)  &  0.083 & 0.064(06)  \\
 38 & 0.246 & 0.236(09)  &  0.008 & 0.035(10)  &  0.152 & 0.132(08)  &  0.066 & 0.056(07)  \\
 39 & 0.271 & 0.253(08)  &  0.011 & 0.028(09)  &  0.174 & 0.172(08)  &  0.054 & 0.056(06)  \\
 40 & 0.280 & 0.292(08)  &  0.016 & 0.002(15)  &  0.187 & 0.177(09)  &  0.049 & 0.064(06)  \\
 41 & 0.271 & 0.284(06)  &  0.023 & 0.036(07)  &  0.193 & 0.170(07)  &  0.044 & 0.053(06)  \\
 42 & 0.245 & 0.239(07)  &  0.041 & 0.045(08)  &  0.191 & 0.137(08)  &  0.038 & 0.055(06)  \\
 43 & 0.210 & 0.187(07)  &  0.063 & 0.077(14)  &  0.176 & 0.172(14)  &  0.034 & 0.055(06)  \\
 44 & 0.166 & 0.140(10)  &  0.093 & 0.101(08)  &  0.158 & 0.136(08)  &  0.029 & 0.050(06)  \\
 45 & 0.126 & 0.124(09)  &  0.139 & 0.121(09)  &  0.138 & 0.135(09)  &  0.027 & 0.040(08)  \\
 46 & 0.087 & 0.074(06)  &  0.192 & 0.158(06)  &  0.116 & 0.098(07)  &  0.023 & 0.057(05)  \\
 47 & 0.058 & 0.048(06)  &  0.251 & 0.219(05)  &  0.096 & 0.090(06)  &  0.021 & 0.055(05)  \\
 48 & 0.035 & 0.021(07)  &  0.315 & 0.271(05)  &  0.080 & 0.091(06)  &  0.019 & 0.055(05)  \\
 49 & 0.021 & 0.022(09)  &  0.370 & 0.318(05)  &  0.066 & 0.083(06)  &  0.018 & 0.042(06)  \\
 50 & 0.012 & -0.003(16) &  0.415 & 0.359(05)  &  0.058 & 0.063(06)  &  0.017 & 0.052(07)  \\
 51 & 0.003 & 0.004(09)  &  0.434 & 0.378(05)  &  0.055 & 0.065(06)  &  0.017 & 0.052(05)  \\
 \hline
\end{tabularx}
\caption{The vacuum-subtracted energy density $E_n$ for $t=0,\,8.25,\,16.5,\,24.75$ corresponding to Fig.~\ref{i_s:fig:ibm_results}. 
Results are shown from MPS simulations and from quantum simulations using {\tt ibm\_marrakesh} after error mitigation (columns labeled QC). 
The energy density of the sites not shown can be obtained by the parity symmetry that relates $E_n = E_{L-1-n}$.}
\label{i_s:tab:results_t}
\end{table}

\begin{table}[h]
\renewcommand{\arraystretch}{0.6}
\scriptsize
\setlength{\tabcolsep}{0pt}
\begin{tabularx}{\linewidth}{|c||Y|Y||Y|Y||c||Y|Y||Y|Y|}
\hline
\rule{0pt}{10pt} & \multicolumn{4}{c||}{\large$E_n$} & \rule{0pt}{10pt} & \multicolumn{4}{c|}{\large$E_n$}\\\hline\hline
\rule{0pt}{10pt} \multirow{2}{*}{Spatial site $n$} &  \multicolumn{2}{c||}{$t_2=16.5$} & \multicolumn{2}{c||}{$t_3=24.75$} & \rule{0pt}{10pt} \multirow{2}{*}{Spatial site $n$} & \multicolumn{2}{c||}{$t_2=16.5$} & \multicolumn{2}{c|}{$t_3=24.75$} 
\\\cline{2-5}
\cline{7-10}
\rule{0pt}{10pt} & MPS & QC & MPS & QC & \rule{0pt}{10pt} & MPS & QC & MPS & QC\\
\hline\hline
0 & -0.001 & 0.008(11)  & -0.021 & -0.002(09) &  52 &  0.020 & 0.043(06)  &  0.001 & 0.027(06)  \\
  1 & -0.001 & 0.006(09)  & -0.015 & 0.008(07)  &  53 &  0.018 & 0.024(07)  &  0.000 & -      \\
  2 & -0.001 & -0.013(07) & -0.016 & -0.002(06) &  54 &  0.015 & 0.034(08)  & -0.001 & 0.025(07)  \\
  3 & -0.001 & -0.003(07) & -0.016 & -0.005(06) &  55 &  0.013 & 0.041(05)  & -0.002 & 0.036(05)  \\
  4 & -0.001 & 0.005(09)  & -0.016 & -0.002(07) &  56 &  0.010 & 0.025(06)  & -0.003 & 0.028(05)  \\
  5 & -0.001 & 0.004(07)  & -0.016 & 0.010(07)  &  57 &  0.009 & 0.030(07)  & -0.004 & 0.019(06)  \\
  6 & -0.001 & 0.004(06)  & -0.016 & -0.006(07) &  58 &  0.007 & 0.011(09)  & -0.004 & 0.031(08)  \\
  7 & -0.001 & 0.005(05)  & -0.016 & -0.001(06) &  59 &  0.006 & 0.036(06)  & -0.005 & 0.027(06)  \\
  8 & -0.001 & -0.000(06) & -0.015 & -0.002(06) &  60 &  0.004 & 0.031(06)  & -0.005 & 0.021(06)  \\
  9 & -0.001 & 0.014(06)  & -0.015 & -0.001(06) &  61 &  0.004 & 0.036(06)  & -0.006 & 0.030(06)  \\
 10 & -0.001 & 0.012(06)  & -0.015 & 0.004(06)  &  62 &  0.003 & 0.028(05)  & -0.006 & 0.015(06)  \\
 11 & -0.001 & 0.008(06)  & -0.015 & -0.002(06) &  63 &  0.002 & 0.028(08)  & -0.007 & 0.029(06)  \\
 12 & -0.001 & -0.001(07) & -0.015 & -0.002(07) &  64 &  0.001 & 0.015(06)  & -0.006 & 0.021(06)  \\
 13 & -0.001 & 0.006(07)  & -0.015 & -0.004(08) &  65 &  0.001 & 0.005(08)  & -0.007 & 0.028(07)  \\
 14 & -0.001 & 0.000(06)  & -0.015 & 0.003(08)  &  66 & -0.001 & 0.015(06)  & -0.008 & 0.019(06)  \\
 15 & -0.002 & 0.007(07)  & -0.014 & 0.004(07)  &  67 & -0.001 & 0.012(07)  & -0.006 & 0.010(06)  \\
 16 & -0.002 & 0.009(06)  & -0.013 & -0.001(07) &  68 & -0.003 & 0.008(06)  & -0.005 & 0.005(06)  \\
 17 & -0.002 & 0.014(05)  & -0.012 & 0.001(06)  &  69 & -0.003 & 0.005(07)  & -0.009 & 0.001(07)  \\
 18 & -0.002 & -0.001(06) & -0.010 & 0.011(06)  &  70 & -0.004 & -0.005(07) & -0.006 & 0.008(07)  \\
 19 & -0.002 & 0.001(05)  & -0.007 & 0.013(06)  &  71 & -0.004 & 0.015(06)  & -0.005 & 0.013(06)  \\
 20 & -0.002 & 0.006(06)  & -0.003 & 0.026(06)  &  72 & -0.004 & 0.021(06)  & -0.007 & 0.015(05)  \\
 21 & -0.002 & 0.013(08)  &  0.004 & 0.021(06)  &  73 & -0.005 & 0.009(07)  & -0.006 & 0.011(07)  \\
 22 & -0.002 & -0.000(07) &  0.018 & 0.038(06)  &  74 & -0.004 & 0.011(08)  & -0.007 & -0.001(12) \\
 23 & -0.002 & -0.016(07) &  0.037 & 0.073(06)  &  75 & -0.004 & -      & -0.007 & -      \\
 24 & -0.002 & -0.021(08) &  0.059 & 0.079(07)  &  76 & -0.003 & 0.019(12)  & -0.006 & -0.000(10) \\
 25 & -0.002 & -0.006(07) &  0.082 & 0.082(06)  &  77 & -0.004 & 0.019(08)  & -0.008 & 0.008(07)  \\
 26 & -0.002 & -0.011(08) &  0.113 & 0.100(07)  &  78 & -0.003 & 0.009(07)  & -0.008 & 0.018(06)  \\
 27 & -0.002 & 0.005(10)  &  0.142 & 0.117(10)  &  79 & -0.002 & 0.008(08)  & -0.009 & -0.010(07) \\
 28 & -0.001 & -      &  0.161 & 0.117(13)  &  80 & -0.003 & 0.013(07)  & -0.010 & 0.006(07)  \\
 29 &  0.001 & 0.007(09)  &  0.175 & 0.112(07)  &  81 & -0.003 & 0.014(08)  & -0.010 & 0.007(07)  \\
 30 &  0.004 & 0.010(07)  &  0.179 & 0.093(06)  &  82 & -0.003 & 0.023(09)  & -0.011 & 0.014(10)  \\
 31 &  0.008 & 0.005(07)  &  0.169 & 0.085(06)  &  83 & -0.003 & 0.002(07)  & -0.011 & 0.006(11)  \\
 32 &  0.015 & 0.019(06)  &  0.155 & 0.087(06)  &  84 & -0.004 & 0.001(06)  & -0.011 & 0.012(06)  \\
 33 &  0.027 & 0.024(07)  &  0.133 & 0.092(08)  &  85 & -0.004 & -0.011(06) & -0.011 & 0.003(06)  \\
 34 &  0.047 & 0.031(07)  &  0.120 & 0.061(07)  &  86 & -0.004 & -0.005(06) & -0.011 & 0.010(06)  \\
 35 &  0.072 & 0.045(06)  &  0.094 & 0.064(06)  &  87 & -0.004 & -0.003(06) & -0.010 & 0.014(06)  \\
 36 &  0.098 & 0.082(06)  &  0.069 & 0.057(06)  &  88 & -0.004 & 0.004(06)  & -0.011 & 0.008(07)  \\
 37 &  0.128 & 0.111(06)  &  0.060 & 0.036(06)  &  89 & -0.003 & -0.003(07) & -0.010 & 0.019(06)  \\
 38 &  0.160 & 0.114(07)  &  0.040 & 0.041(08)  &  90 & -0.003 & -0.003(07) & -0.010 & -0.002(08) \\
 39 &  0.182 & 0.136(06)  &  0.028 & 0.029(08)  &  91 & -0.002 & 0.011(06)  & -0.009 & 0.000(06)  \\
 40 &  0.193 & 0.124(07)  &  0.025 & 0.040(06)  &  92 & -0.002 & -0.003(06) & -0.009 & 0.003(05)  \\
 41 &  0.198 & 0.144(06)  &  0.022 & 0.038(06)  &  93 & -0.001 & -0.008(06) & -0.010 & 0.009(05)  \\
 42 &  0.192 & 0.135(06)  &  0.018 & 0.027(06)  &  94 & -0.001 & -0.006(06) & -0.009 & 0.012(05)  \\
 43 &  0.172 & 0.125(06)  &  0.015 & 0.038(06)  &  95 & -0.001 & -0.002(06) & -0.010 & 0.006(05)  \\
 44 &  0.152 & 0.106(06)  &  0.013 & 0.032(06)  &  96 &  0.000 & 0.006(06)  & -0.010 & 0.008(05)  \\
 45 &  0.129 & 0.110(09)  &  0.011 & 0.020(05)  &  97 &  0.000 & 0.009(07)  & -0.011 & 0.008(05)  \\
 46 &  0.103 & 0.086(05)  &  0.009 & 0.023(06)  &  98 &  0.000 & 0.006(07)  & -0.010 & 0.010(06)  \\
 47 &  0.081 & 0.072(05)  &  0.007 & 0.023(05)  &  99 &  0.000 & 0.002(09)  & -0.010 & 0.003(07)  \\
 48 &  0.062 & 0.052(06)  &  0.006 & 0.026(06)  & 100 &  0.000 & -0.002(07) & -0.010 & 0.006(06)  \\
 49 &  0.044 & 0.060(08)  &  0.004 & 0.024(08)  & 101 &  0.000 & 0.003(07)  & -0.010 & 0.007(06)  \\
 50 &  0.031 & 0.037(07)  &  0.003 & 0.029(06)  & 102 &  0.000 & 0.001(10)  & -0.009 & 0.011(06)  \\
 51 &  0.024 & 0.035(06)  &  0.002 & 0.033(05)  & 103 &  0.000 & 0.007(10)  & -0.016 & -0.002(08) \\
 \hline
\end{tabularx}
\caption{The vacuum-subtracted single wavepacket energy density $E_n$ for  $t=16.5,\,24.75$ corresponding to Fig.~\ref{i_s:fig:1wp_vs_2wp}. 
Results are shown from MPS simulations and from quantum simulations using {\tt ibm\_marrakesh} after error mitigation (columns labeled QC). 
Several entries are missing data as a result of filtering out qubits with noisy readout.}
\label{i_s:tab:results_t_1wp}
\end{table}

\end{subappendices}

\part{Quantum information in physical systems}
\chapter{Exclusive scattering channels from entanglement structure in real-time simulations}
\label{chap:exclusive_channels}

\noindent
{\it This chapter is associated with Ref.~\cite{Zemlevskiy:2026kpc}: ``Exclusive scattering channels from entanglement structure in real-time simulations'' by Nikita A. Zemlevskiy.}

\section{Introduction}
\label{i_m:sec:intro}
\noindent
The late-time wavefunction after a scattering event encodes the possible products of the collision and their respective amplitudes.
Decomposing the full superposition into individual channels, i.e., resolving the exclusive final states rather than summing inclusively over all configurations, is a prerequisite for extracting branching ratios and reconstructing channel-specific kinematics.
Extracting comparable information from real-time simulations requires methods to resolve exclusive final states.

This chapter shows that the entanglement structure, revealed through the Schmidt decomposition, provides a natural way to resolve these channels.
Individual channels are identified through the tensor product factorization at bipartitions separating components from different channels.
This method leverages spatial separations in the state to dissect the full superposition into orthogonal components, each corresponding to a distinct scattering outcome, with Schmidt coefficients that directly yield the channel probabilities.
Physical meaning has been ascribed to Schmidt vectors in various contexts, for instance in string breaking and hadronization~\cite{Grieninger:2026bdq,Florio:2024aix}, in characterizing topological features~\cite{Li:2008kda,Fidkowski:2010nhf}, and identifying quasiparticle excitations~\cite{Zauner-Stauber:2018gqr,Cocchiarella:2025mtv}.
In the context of collisions, entanglement and quantum complexity measures provide a powerful lens on scattering and particle production dynamics~\cite{MISHIMA2004371,Robin:2025ymq,Martin:2025hzm,Cheng:2025zaw,Mendez:2024wqn,Quinta:2023ink,Zagirdinova:2022awg,Hentschinski:2022rsa,Cervera-Lierta:2017tdt,Hida:2009jdg,Harshman:2007omw,Giorgi:2006lse,Yuasa:2006faz,Lamata:2006au,Lombardi:2006tzp,Harshman:2006sec,Peschanski:2019yah,Bai:2022hfv,Kowalska:2024kbs,Hales:2022osm,PhysRevA.73.052313,Kouzakov:2019hbt,Karlsson_2021,Sou:2025tyf,Tkachev:2024iuj,Gargalionis:2025iqs,Beane:2018oxh,Robin:2024oqc}.
Recent work has shown that the entanglement at a spatial bipartition characterizes the nature of the scattering process~\cite{Jha:2024jan,Milsted:2020jmf,Pavesic:2025nwm,Papaefstathiou:2024zsu,Belyansky:2023rgh,Barata:2025rjb,Barata:2025hgx,Rigobello:2021fxw,Van_Damme_2021,Bennewitz:2024ixi}. 
In particular, it is known that the late-time state after an inelastic collision is more entangled than the elastic case.
This work reveals a correspondence between the presence of additional reaction channels and the entanglement structure of the post-scattering state.

This correspondence is used to analyze the final state of scattering simulations by isolating channels based on the spatial structure of the late-time wavefunction.
This method is demonstrated in simulations of inelastic collisions in the Ising field theory, where channel isolation enables the extraction of particle masses from individual scattering outcomes and provides confirmation of heavy particle production.
While the present work focuses on scattering, the approach extends naturally to other settings where distinct physical processes occur in superposition.

\section{Isolating scattering channels}
\label{i_m:sec:method}
\begin{figure}
    \centering
    \includegraphics[width=0.6\linewidth]{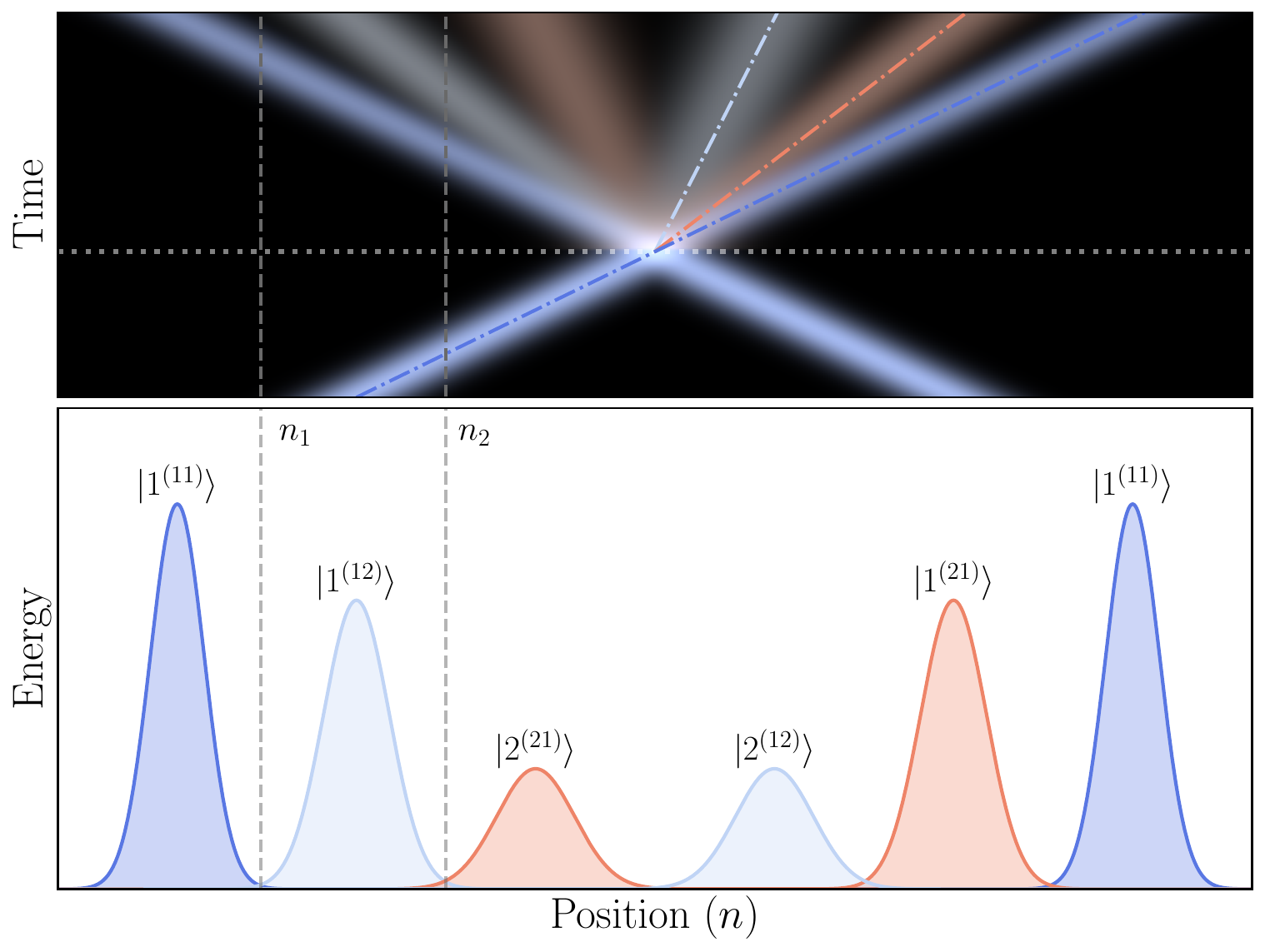}
    \caption{{\it Channel isolation in the post-inelastic scattering state.}
    Top: the energy density as a function of position and time with the $|11\rangle$ (dark blue), $|12\rangle$ (light blue), and $|21\rangle$ (orange) states distinguished.
    The dotted line marks the collision time and the dot-dashed lines show the trajectories of the outgoing particles in each state.
    Bottom: the energy density as a function of position for the final state in the heatmap in the top panel. 
    The superscripts on the individual particles label the states they belong to.
    The dashed lines show the positions $n_1$ and $n_2$ where bipartitions can be made to isolate $|11\rangle$, $|12\rangle$, and $|21\rangle$.
    }
    \label{i_m:fig:late_time_schematic}
\end{figure}

\noindent
In regimes where both elastic and inelastic two-particle scattering are possible, these processes happen in superposition. 
The method to isolate these channels is general and only requires spatial separation of collinear outgoing particles.
For concreteness, consider a theory in one spatial dimension with two stable particles, $|1\rangle$ and $|2\rangle$, such that $m_1<m_2$.
In a high-energy collision, kinetic energy can be converted into mass, forming a heavier outgoing particle.
Such particle production is possible in a collision of two $|1\rangle$ with momentum $k_i>k_\text{thr}$, where $k_\text{thr}$ is the threshold momentum at which the total energy is equal to $m_1+m_2$.\footnote{Working in the center-of-momentum frame, this threshold corresponds to $\sqrt{s} \geq m_1 + m_2$ in terms of the Lorentz-invariant Mandelstam variable $s = (E_1 + E_2)^2 - (k_1 + k_2)^2$. For simplicity, higher-order processes such as 2-3 scattering are not considered.} 
The scattering process of two $|1\rangle$ particles with $k_i>k_\text{thr}$ is
\begin{equation}
\begin{aligned}
   |1(k_i)\rangle |1(-k_i)\rangle \ \to \ \alpha &|1^{(11)}(-k_i)\rangle |1^{(11)}(k_i)\rangle \\ + \ \beta &|1^{(12)}(-k_f)\rangle |2^{(12)}(k_f)\rangle \\ + \ \gamma &|2^{(21)}(-k_f)\rangle |1^{(21)}(k_f)\rangle \\ 
   \equiv \ \ &|f\rangle \ ,
   \label{i_m:eq:production}
\end{aligned}
\end{equation}
where the first entry represents an outgoing particle traveling left, and the second entry corresponds to a particle moving right.
The superscripts label the state each particle belongs to and $\beta=\gamma$ by parity symmetry.
Throughout this work, 11 and 12 refer to the elastic and inelastic channels, and $|11\rangle$, $|12\rangle$, and $|21\rangle$ represent the states that contribute to the late-time wavefunction.
For theories where energy depends nontrivially on momentum, excitations generally have different group velocities $v_{1,2}(k)=dE_{1,2}(k)/dk$.
Wavepackets corresponding to different particle species therefore separate in space over time, allowing different scattering channels to be distinguished by the presence or absence of excitations in particular spatial regions.
An example of the late-time energy density corresponding to Eq.~\eqref{i_m:eq:production} is shown in Fig.~\ref{i_m:fig:late_time_schematic}.

Previous work has determined particle content by constructing projectors onto localized states with given quantum numbers~\cite{Jha:2024jan,Van_Damme_2021}. 
Entanglement considerations are central to MPS simulations and allow leveraging spatial separation between channels with different outgoing velocities.
The key ingredient of the method in this work is the Schmidt decomposition across a bipartition between collinear components of different channels, which forces the wavefunction into a sum of tensor product states. 
By finding the basis that diagonalizes correlations between two regions, the Schmidt decomposition naturally yields the spatially separated channels. 
When all wavepackets are spatially separated, the states in Eq.~\eqref{i_m:eq:production} occupy distinct regions of space.
Since only $|1^{(11)}\rangle$ has support to the left of $n_1$ in Fig.~\ref{i_m:fig:late_time_schematic}, it is orthogonal to $|1^{(12)}\rangle$ and $|2^{(21)}\rangle$.
The Schmidt decomposition at $n_1$ therefore isolates the fast-moving elastic components from the slower inelastic ones,
\begin{equation}
    |f\rangle \ = \ \alpha|1^{(11)}\rangle\otimes|1^{(11)}\rangle + \beta|\text{vac}\rangle \otimes|\psi_{12}\rangle \ ,
    \label{i_m:eq:proj_n1}
\end{equation}
where $|\text{vac}\rangle$ schematically denotes the vacuum to the left of the cut and $|\psi_{12}\rangle =|1^{(12)}\rangle |2^{(12)}\rangle + |2^{(21)}\rangle |1^{(21)}\rangle$.
The tensor product structure at $n_l$ is made explicit through $\otimes$.
Retaining the second term in this decomposition projects onto the wavefunction of the inelastic channel.\footnote{This projection is implemented in MPS through the singular value decomposition (SVD) at the bipartition. 
The state of interest is obtained by setting the singular values corresponding to all other Schmidt vectors to zero.}
Since only the left-moving $|1^{(12)}\rangle$ has support between $n_1$ and $n_2$, a cut at $n_2$ on the second term in Eq.~\eqref{i_m:eq:proj_n1} separates the two branches of the inelastic channel, 
\begin{equation}
    |\psi_{12}\rangle \ = \ |1^{(12)}\rangle \otimes |2^{(12)}\rangle + |\text{vac}\rangle \otimes |2^{(21)}\rangle|1^{(21)}\rangle \ .
\end{equation}
The exclusive states corresponding to this decomposition are shown by different colors in the bottom panel of Fig.~\ref{i_m:fig:late_time_schematic}.

From the separation into the elastic and inelastic channels it is straightforward to extract the branching ratio for particle production.
The Schmidt coefficients give the probabilities for each process:
\begin{equation}
    P(11) \ = \ |\alpha|^2 \ , \quad
    P(12) \ = \ 2|\beta|^2 \ ,
    \label{i_m:eq:branching_ratios}
\end{equation}
from which exclusive cross sections can be extracted.
This result is obvious from Eq.~\eqref{i_m:eq:production} but easily obtained from this decomposition.

The presence of a heavy particle in $|f\rangle$ is confirmed by computing the masses of excitations in the exclusive final states through the dispersion relation together with the momenta of the excitations.
Momentum may be measured with several different techniques and is discussed in App.~\ref{i_m:app:momentum}.
In this work, the velocity of local observables is matched to the group velocities of different excitations to determine their momenta, and solve for their masses from the dispersion relations.

This method is exact in the limit where the outgoing particles belonging to different channels traveling in the same direction are well-separated. 
This requires initial states of sufficiently large, collimated wavepackets, as well as sufficiently late times.
In practice, the combined effects of wavepacket spreading and finite simulation time lead to small residual mixing between channels and corresponding uncertainties in the extracted probabilities.
These approximations can be quantified and systematically reduced by increasing the momentum-space resolution of the simulation.

\section{Channel isolation after particle production in Ising field theory }
\label{i_m:sec:ising_scattering_momentum_results}
\noindent
This work investigates particle production in an inelastic scattering process in the Ising model,
\begin{align}
\hat{H}  \ &=\  -  \sum_{n=0}^{L-1}\left [ \frac{1}{2}\left (\hat{Z}_{n-1}\hat{Z}_{n}+\hat{Z}_n\hat{Z}_{n+1}\right ) +  g_x \hat{X}_n + g_z\hat{Z}_n \right ] \nonumber \\
&\equiv \ \sum_{n=0}^{L-1}\hat{H}_n \ , \label{i_m:eq:h_ising}
\end{align}
which corresponds to a quantum field theory (QFT) when tuned near criticality (i.e., $g_x\approx1,\ g_z\approx0,\ L\to\infty$).
Throughout this work, the couplings $g_x=1.25,\ g_z=0.15$ and system size $L=400$ are used.
With these couplings, the theory has two stable particles, $|1\rangle$ and $|2\rangle$.
Trotter time evolution is used with MPS to simulate the scattering of two wavepackets of $|1\rangle$ particles, with $k_i=0.36\pi>k_\text{thr}$ and width in momentum space $\sigma_k=0.059\pi$.
See App.~\ref{i_m:app:sim_details} for details on the simulation techniques.

\begin{figure}
    \centering
    \includegraphics[width=0.6\linewidth]{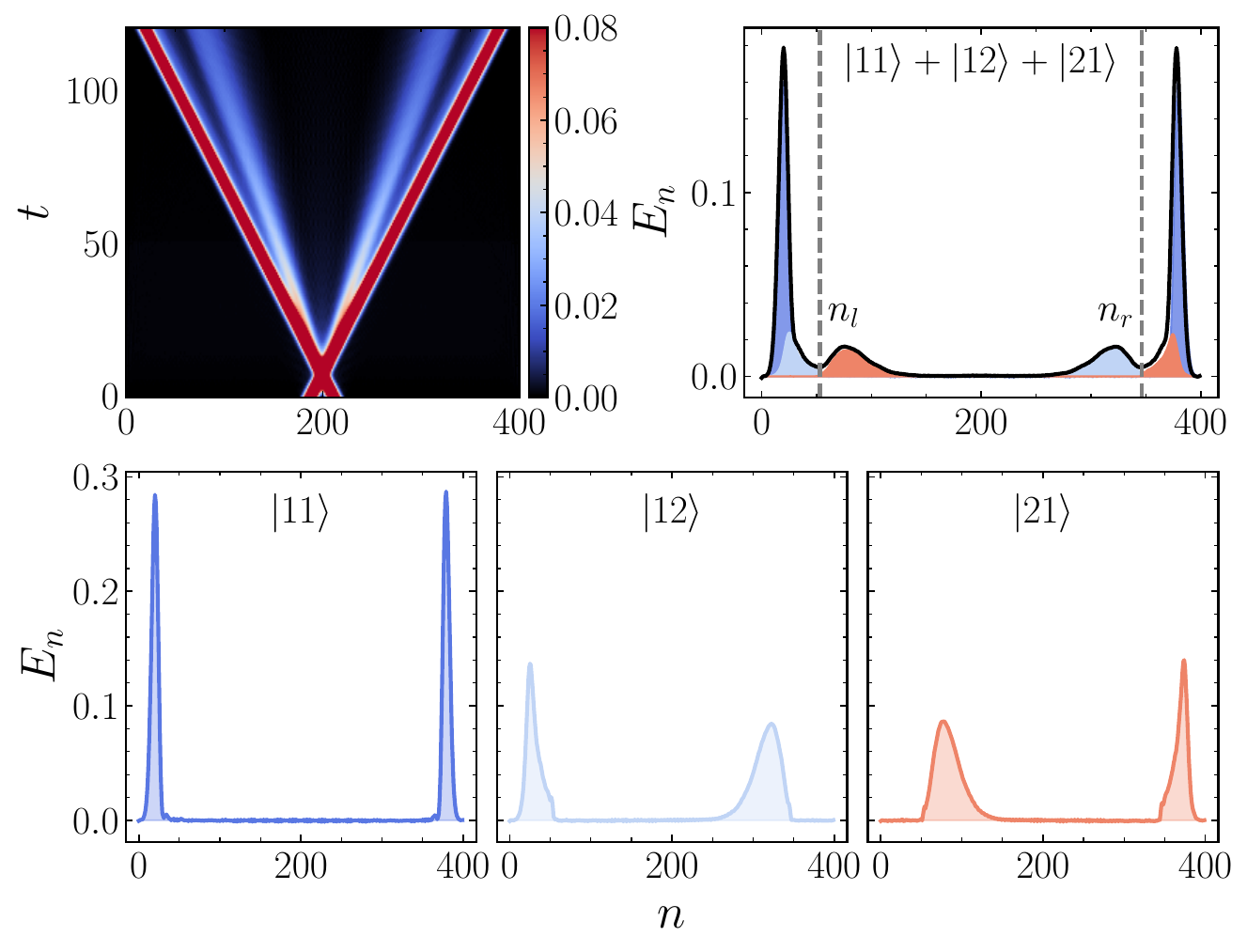}
    \caption{{\it Leading-order channel isolation after particle production.}
    Top left: the energy density $E_n$ as a function of position $n$ and time $t$ for the collision of two $|1\rangle$ particles with $k_i=\pm0.36\pi$. 
    Top right: $E_n$ corresponding to $t=120$ in the top left panel. 
    The energy density of the inclusive post-scattering state $|f\rangle$ is shown in black.
    The color shading shows the contributions of the individual states: $|11\rangle$ (dark blue), $|12\rangle$ (light blue), and $|21\rangle$ (orange).
    The positions $n_l$ and $n_r$ of the bipartitions used to isolate the states composing the elastic and inelastic channels are marked by the dashed lines.
    Bottom: $E_n$ of the exclusive states contributing to $|f\rangle$ in the top right panel.
    }
    \label{i_m:fig:channel}
\end{figure}

The top left panel of Fig.~\ref{i_m:fig:channel} shows the vacuum-subtracted energy density, 
\begin{align}
E_n \ = \ \langle f(t)|\hat{H}_n |f(t)\rangle - \langle\text{vac}|\hat{H}_n |\text{vac}\rangle\ ,
\label{i_m:eq:e_n}
\end{align}
throughout the scattering process.
As in Eq.~\eqref{i_m:eq:production}, the simulation begins with the $|1\rangle$ particles moving toward each other. 
At $t\sim8$ they collide and travel away from each other. 
After the collision, the faint tracks toward the center of the lattice represent the contributions of the slower $|2\rangle$ particle produced in the inelastic channel.

Entanglement and quantum complexity are known to be powerful indicators of dynamics in physical processes. 
For instance, entanglement entropy has been shown to track the onset of inelastic thresholds and the creation of new particles in meson collisions~\cite{Rigobello:2021fxw,Papaefstathiou:2024zsu}, the transition from fermionic Fock states to meson-like bound states during hadronization is revealed by properties of the entanglement spectrum~\cite{Florio:2023dke,Florio:2024aix}, and measures of quantum complexity expose nonlocal correlations during string breaking~\cite{Grieninger:2026bdq}.
Increased levels of entanglement have been observed in late-time states after inelastic scattering.
This work identifies the reason for this: elastic collisions contain a single dominant Schmidt vector, whereas the number of significant Schmidt components increases with the inelasticity and the number of possible distinct processes.

\begin{figure}
    \centering
    \includegraphics[width=0.6\linewidth]{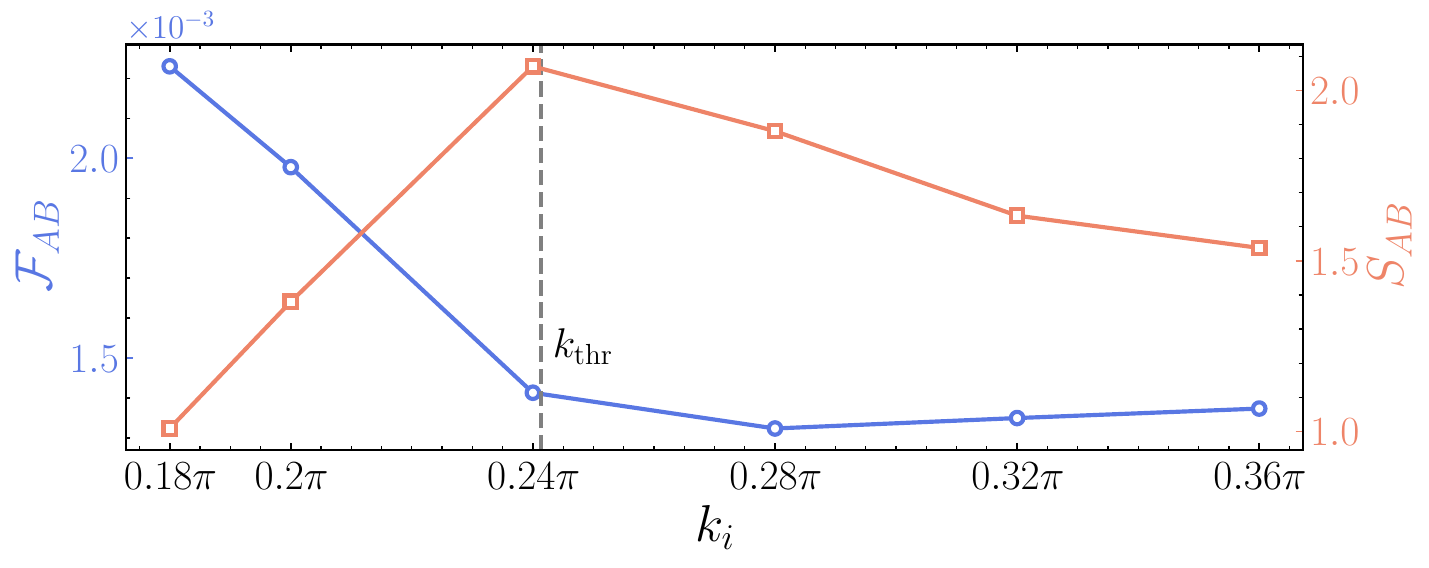}
    \caption{{\it Entanglement structure of the post-scattering state.}
    The antiflatness ${\cal F}_{AB}$ (blue) and the entanglement entropy $S_{AB}$ (orange) as a function of initial momentum $k_i$.
    The dashed line marks the threshold momentum $k_\text{thr}$ above which the process $11\to12$ is kinematically allowed.
    }
    \label{i_m:fig:af_ee}
\end{figure}

The antiflatness of the entanglement spectrum $\{\lambda_i\}$ (Schmidt coefficients squared) is measured by its variance,\footnote{Similar measures are known to bound nonstabilizerness~\cite{Tirrito:2023fnw}.} 
\begin{align}
    {\cal F}_{AB} \ = \ \sum_i \lambda_i^2 - \left(\sum_i \lambda_i\right)^2 \ ,
    \label{i_m:eq:af}
\end{align}
characterizes the distribution of entanglement at a bipartition $AB$ in the system. 
High values of ${\cal F}_{AB}$ correspond to fewer significant components contributing to the entanglement distribution, whereas low ${\cal F}_{AB}$ indicates a more uniform entanglement structure.
Figure~\ref{i_m:fig:af_ee} shows ${\cal F}_{AB}$ and the entanglement entropy $S_{AB}$ at $n=L/2$ in late-time states of collision simulations as a function of initial momentum $k_i$.
The decrease in ${\cal F}_{AB}$ and increase in $S_{AB}$ past $k_i=k_\text{thr}$ indicates that there are more significant Schmidt components with increasing momentum.\footnote{The changes in ${\cal F}_{AB}$ and $S_{AB}$ are not sharp because the wavepackets have $\sigma_k\neq0$.}
This corresponds to the inelastic channel opening up.
Entanglement is maximized around $k_\text{thr}$ because there is significant probability to produce a particle $|2\rangle$ at rest or moving slowly, whose wavefunction is bisected by a bipartition at $L/2$. 
This relation of entanglement with the physical processes that drive the structure of the state can also be seen explicitly in Table~\ref{i_m:tab:schmidt_vals}, where the significant Schmidt components are shown for each $k_i$.
\begin{table}
\centering
\begin{tabularx}{\linewidth}{|c||Y|Y|Y|Y|Y|Y|} \hline
$k_i$ & $0.18\pi$ & $0.2\pi$ & $0.24\pi$ & $0.28\pi$ &
$0.32\pi$ & $0.36\pi$ \\\hline\hline
& & & & 0.6595, & 0.6574, & 0.6620,\\ 
$\lambda_i>10^{-1}$ & 0.8844 & 0.8325 & 0.6983 & 0.1296, & 0.1545, & 0.1570, \\
& & & & 0.1170 & 0.1376 & 0.1431 \\\hline
\end{tabularx}
\renewcommand{\arraystretch}{1}
\caption{{\it Significant components of the post-scattering state.} 
The dominant values of the $t=120$ entanglement spectrum $\{\lambda_i\}$ (Schmidt spectrum squared) at $L/2$ of $|f\rangle$ for various momenta $k_i$. 
}
\label{i_m:tab:schmidt_vals}
\end{table}

The entanglement structure at the center of the lattice reveals the composition of the late-time wavefunction, but does not distinguish the individual channels.
As explained in Sec.~\ref{i_m:sec:method}, a series of spatial Schmidt decompositions is used to distinguish the elastic from the inelastic channels. 
The cut locations are shown schematically in the top right panel of Fig.~\ref{i_m:fig:channel} and are chosen to maximize the separation of the elastic and inelastic components.

In general, $|1^{(12)}\rangle$ travels at a different velocity than $|1^{(11)}\rangle$ because some of its kinetic energy is converted to $|2^{(12)}\rangle$. 
However, an ultra-high-energy collision, such as in the top left panel of Fig.~\ref{i_m:fig:channel}, leaves the velocity of $|1^{(12)}\rangle$ almost unchanged and its path largely overlaps with $|1^{(11)}\rangle$.
Compared to Fig.~\ref{i_m:fig:late_time_schematic} where there are three distinct regions on each half of the lattice, in Fig.~\ref{i_m:fig:channel} there are only two: ``fast'' and ``slow''.
The elastic channel is distinguished by having a particle in both fast regions, whereas $|12\rangle$ ($|21\rangle$) has a fast (slow) particle on the left and a slow (fast) particle on the right.
Since $\langle1^{(11)}|1^{(12)}\rangle\neq0$, the light particle in the inelastic channel can be decomposed as $|1^{(12)}\rangle = a|1^{(11)}\rangle + b|\delta\rangle$ where $\langle1^{(11)}|\delta\rangle=0$.
The Schmidt decomposition at $n_l$ diagonalizes the reduced density matrix of the left side giving three orthogonal left Schmidt vectors: $|1^{(11)}\rangle$, $|\delta\rangle$, and $|2^{(21)}\rangle$.
The late-time state in Eq.~\eqref{i_m:eq:production} becomes
\begin{equation}
\begin{aligned}
    |f \rangle \
&= \ | 1^{(11)} \rangle \otimes
\left(
\alpha | 1^{(11)} \rangle + a\beta | 2^{(12)} \rangle
\right) \\
&+ \ b \beta |\delta\rangle \otimes |2^{(12)}\rangle \\
&+ \ \beta |\text{vac}\rangle \otimes | 2^{(21)} \rangle | 1^{(21)} \rangle  \ .
\label{i_m:eq:cut_1}
\end{aligned}
\end{equation}
This separates the light left-moving components belonging to both channels from the heavy left-moving particle, which is identified as the last term in Eq.~\eqref{i_m:eq:cut_1}.

The right-moving $|1^{(11)}\rangle$ and $|2^{(12)}\rangle$ are orthogonal since they are different particle types at different positions. 
Since the $a$ and $b$ components of $|1^{(12)}\rangle$ share the same right-moving state $|2^{(12)}\rangle$, a second Schmidt decomposition at $n_r$ groups them back together.
This distinguishes the remaining $|11\rangle$ and $|12\rangle$ states.\footnote{Components containing the tails of the wavepackets due to finite particle separation and wavepacket spreading are suppressed, as are components corresponding to higher-order inelastic processes. See App.~\ref{i_m:app:channel_details} for details.}

The energy densities of the exclusive final states are shown in the lower panels of Fig.~\ref{i_m:fig:channel}.
In this figure, the difference between $v_1(k_i)$ and $v_1(k_f)$ is seen by comparing the peaks of the fast-moving components in the elastic and inelastic channels.
The broader profile of outgoing $|2\rangle$ particles results from larger variations in $v_2$ than $v_1$ around $k_f$.
Sharp features in the energy density are the result of bipartitions taken between wavepackets that have not fully separated.

The entanglement is found to be spread roughly equally between $|11\rangle$, $|12\rangle$, and $|21\rangle$. 
Each state has a single dominant Schmidt component (${\cal F}_{AB}\sim0$), consistent with unentangled asymptotic outgoing particles.\footnote{This split is similar in nature to ``configuration entropy'' and entropy within symmetry sectors in a gauge theory~\cite{Ghosh:2015iwa,Turkeshi:2020yxd}.}
Numerical values for ${\cal F}_{AB}$, $S_{AB}$, and the entanglement spectrum are given in App.~\ref{i_m:app:channel_details}.

\begin{figure}
    \centering
    \includegraphics[width=0.6\linewidth]{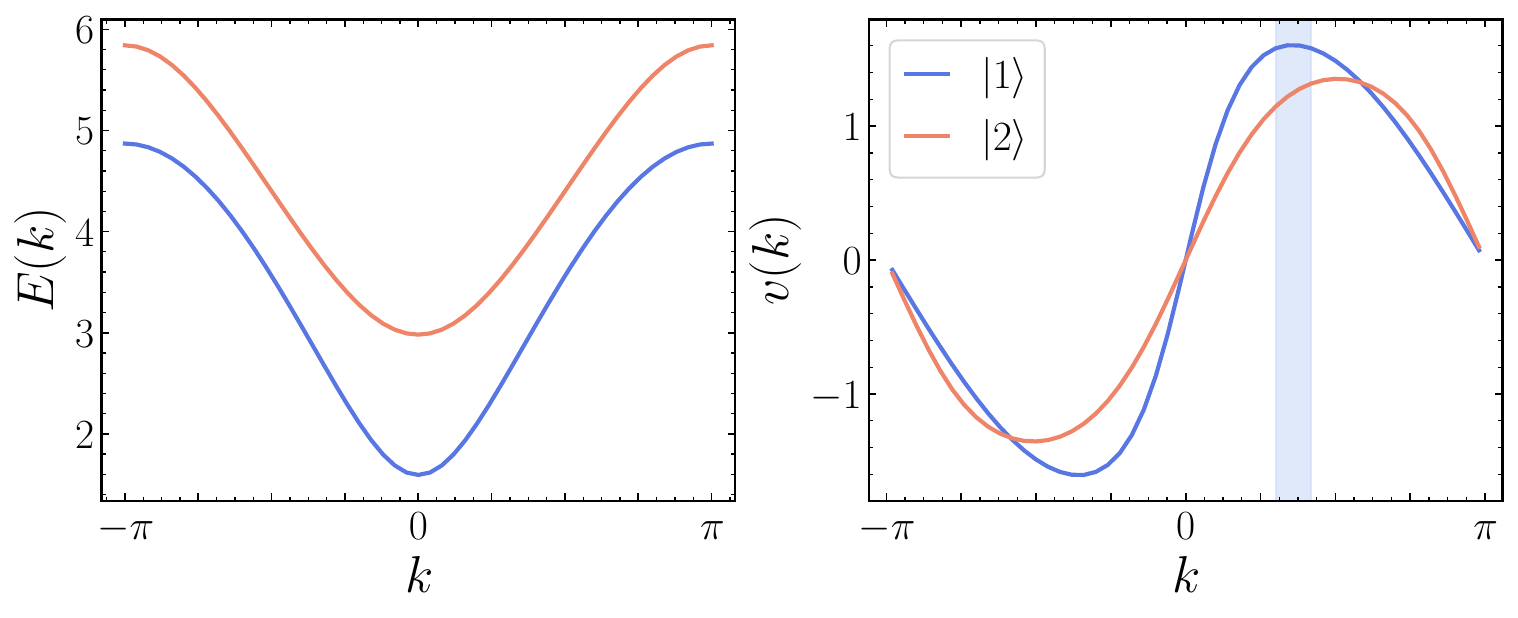}
    \caption{
    Dispersion relations $E(k)$ (left) and the group velocities $v(k)=dE(k)/dk$ (right) for particles $|1\rangle$ (blue) and $|2\rangle$ (orange) calculated in a $L=\infty$ system using the quasiparticle excitation ansatz.
    The light blue band on the right plot shows $k_i\pm\sigma_k$ used for the scattering simulations in this work.
    }
    \label{i_m:fig:ising_momentum_dispersion}
\end{figure}

The particles within each state are then classified by mass using dispersion relations and group velocities calculated in the $L=\infty$ system, confirming the presence of a $|2\rangle$ particle (as opposed to, e.g., a slow $|1\rangle$ particle).
The group velocities $v(k)$ shown in the right panel of Fig.~\ref{i_m:fig:ising_momentum_dispersion} extract the momenta $k(v)$ of particles traveling with speed $v$ measured by tracking peaks in $E_n$ over time.
The momentum is then used in the dispersion relation (left panel of Fig.~\ref{i_m:fig:ising_momentum_dispersion}) to compute the energy $E(k(v))$ that particles $|1\rangle$ and $|2\rangle$ traveling with velocity $v$ would have.
Masses are determined by comparing $E(k(v))$ to the energy of the individual excitations $E_\text{wp}$ measured in each exclusive state and searching for (approximate) equality.\footnote{In addition to approximations discussed in Sec.~\ref{i_m:sec:method}, the use of infinite-system dispersion relations also introduces (exponentially small) errors.}
These calculations are detailed in App.~\ref{i_m:app:particle_classification}.
Applying this self-consistent classification shows $E(k(v))$ and $E_\text{wp}$ are in agreement up to 8\% error, which can likely be improved with larger wavepackets and better spatial separation. 
This classification confirms the absence of $|2\rangle$ particles in the elastic channel and the presence of exactly one $|2\rangle$ particle in each inelastic state.

The branching ratios for $k_i=0.36\pi$ calculated from the decomposition are
\begin{align}
    P(11) \ = \ 0.5638\ , \quad P(12) \ = \ 0.3395 \ .
\end{align} 
Based on the identifications of the Schmidt components with the channels, 2-3 scattering ($11\to111$) and other higher-order inelastic processes have probabilities $O(10^{-2})$.
Up to approximations discussed in Sec.~\ref{i_m:sec:method}, these branching ratios are in agreement with previous results~\cite{Jha:2024jan} and could be confirmed by perturbative calculations near the integrable points of the theory. 
These calculations, together with the classification of particles in individual channels, confirm that a heavy $|2\rangle$ particle is produced in this collision event with probability 0.3395.

\section{Discussion}
\label{i_m:sec:conclusion}
\noindent
This technique provides a way to measure inelastic cross sections in simulations of scattering and is directly motivated by experimental detection protocols at collider facilities.
Akin to path reconstruction from layered detectors in collision experiments, spatial bipartitions separate wavefunction amplitudes that are then combined to form channel amplitudes, with particle velocities serving the role of time-of-flight measurements.
Notably, this method shows that a limited number of detectors, informed only by kinematics and symmetries, is enough to classify inelastic channels.
A promising direction for quantum simulations is to make this experimental analogy direct, implementing channel tagging by placing ``detectors'' throughout the lattice.
On each {\it shot} of the simulation, detector readings would indicate the presence or absence of a (certain type of) particle, allowing shot results to be directly categorized into channels.
Detectors in such simulations would parallel the role of bipartitions of MPS wavefunctions in this work.

While the utility of successive bipartitions is clear for cases where the kinematics constrains the outgoing momenta, higher-order events such as $11\to111$ will produce a distribution of states with different momenta, precluding the construction of, e.g., Dalitz plots~\cite{Dalitz:1953cp,Dalitz:1954cq} from spatial information alone.
This fundamental limitation of spatial cuts is encountered in 2-2 scattering if attempting to distinguish momentum modes within a wavepacket.
In such cases, the Schmidt spectrum still reflects the number of contributing channels, but isolating individual wavefunctions requires spatial separation of the outgoing particles.
A similar challenge exists in simulations of higher-dimensional systems, where event-shape observables such as Fox-Wolfram moments~\cite{Fox:1978vw} could provide a basis for channel classification beyond spatial bipartitions.
Addressing this limitation likely requires the incorporation of detectors both in classical and quantum simulations, with projections onto detector outcomes providing channel classification.

Spatial entanglement structure plays a crucial role in the channel isolation method and reveals the structure of the inclusive post-collision state. 
The direct correspondence of the entanglement spectrum with scattering channels points toward a potential broader relationship between entanglement and the S matrix~\cite{Vanderstraeten:2013xda,Beane:2018oxh,Beane:2021zvo,Beane:2020wjl}, which could be explored through the entanglement Hamiltonian~\cite{https://doi.org/10.1002/andp.202200064}.
While spatial bipartitions are natural for MPS, quantum simulations do not carry the restrictions associated with a spatial representation. 
As a result, this observable-driven measurement design generalizes to any quantum number that distinguishes outgoing channels, providing a framework for exclusive measurements in quantum simulations of scattering.

\clearpage
\begin{subappendices}

\section{Channel isolation details}
\subsection{Entanglement spectrum and higher-order channels}
\label{i_m:app:channel_details}
\noindent
To accommodate finite simulation time and wavepacket spreading, as well as a small difference in the outgoing velocities of $|1^{(11)}\rangle$ and $|1^{(12)}\rangle$, the general method described in Sec.~\ref{i_m:sec:method} is modified.
At the bipartition of the system at lattice site $n_l=53$ (see Fig.~\ref{i_m:fig:channel}), the system has three significant Schmidt components, $\lambda_0,\ \lambda_1,$ and $ \lambda_2$.
Here $\lambda_{l,r}$ labels the index of the Schmidt component at successive cuts $n_l,\ n_r$.
The first row of Table~\ref{i_m:tab:schmidt_vals_nl_nr} shows the values of the Schmidt coefficients at this bipartition.
\begin{table}
\centering
\begin{tabularx}{\linewidth}{|c|c|c||Y|} \hline
$|\psi\rangle$ & $\langle\psi|\psi\rangle$ & Cut location & $\lambda_i>10^{-2}$
\\\hline\hline
$|f\rangle$ & 1 & $n_l$ & $\lambda_0=0.6129$, $\lambda_1=0.2074$, $\lambda_2=0.1463$\\\hline
$|f_{02}\rangle$ & 0.7592 & $n_r $ & $\lambda_{02,0}=0.5638$, $\lambda_{02,1}=0.1747$, $\lambda_{02,2}=0.0259$ \\\hline
$|f_{1}\rangle$ & 0.2074 & $n_r $ & $\lambda_{1,0}=0.1648$, $\lambda_{1,1}=0.0300$  \\\hline
$|\text{vac}\rangle$ & 1 & $n_l$ & 0.9786, 0.0105 \\\hline
$|\text{vac}_0\rangle$ & 0.9786 & $n_r$ & 0.9665, 0.0105\\\hline
\end{tabularx}
\renewcommand{\arraystretch}{1}
\caption{The entanglement spectrum (Schmidt coefficients squared) $\{\lambda_i\}$ in the post-scattering state with $k_i=0.36\pi$ after successive cuts at locations $n_l$, $n_r$.
The state is shown in the first column. 
Where present, $l$ in $|\psi_l\rangle$ corresponds to the components that are kept at the first bipartition. 
The second column shows the norm of the state in the first column. 
The fourth column gives the significant values of the entanglement spectrum obtained after cutting the state in the first column at the position given in third column.
}
\label{i_m:tab:schmidt_vals_nl_nr}
\end{table}
This cut separates out the heavy particles from the light particles moving left, but does not orthogonalize the light particles.
The wavefunction of each of these states is obtained by projecting onto the corresponding Schmidt component as described in Sec.~\ref{i_m:sec:method}.
Since $\langle1^{(11)}|1^{(21)}\rangle\neq0$, some mixing between $|11\rangle$ and $|12\rangle$ is present in the decomposition.
By inspecting the energy density of each wavefunction, it is determined that $|f_0\rangle$ and $|f_2\rangle$ mix the $|11\rangle$ and $|12\rangle$ states, while $|f_1\rangle$ contains $|21\rangle$. 

\begin{figure}
    \centering
\includegraphics[width=0.6\linewidth]{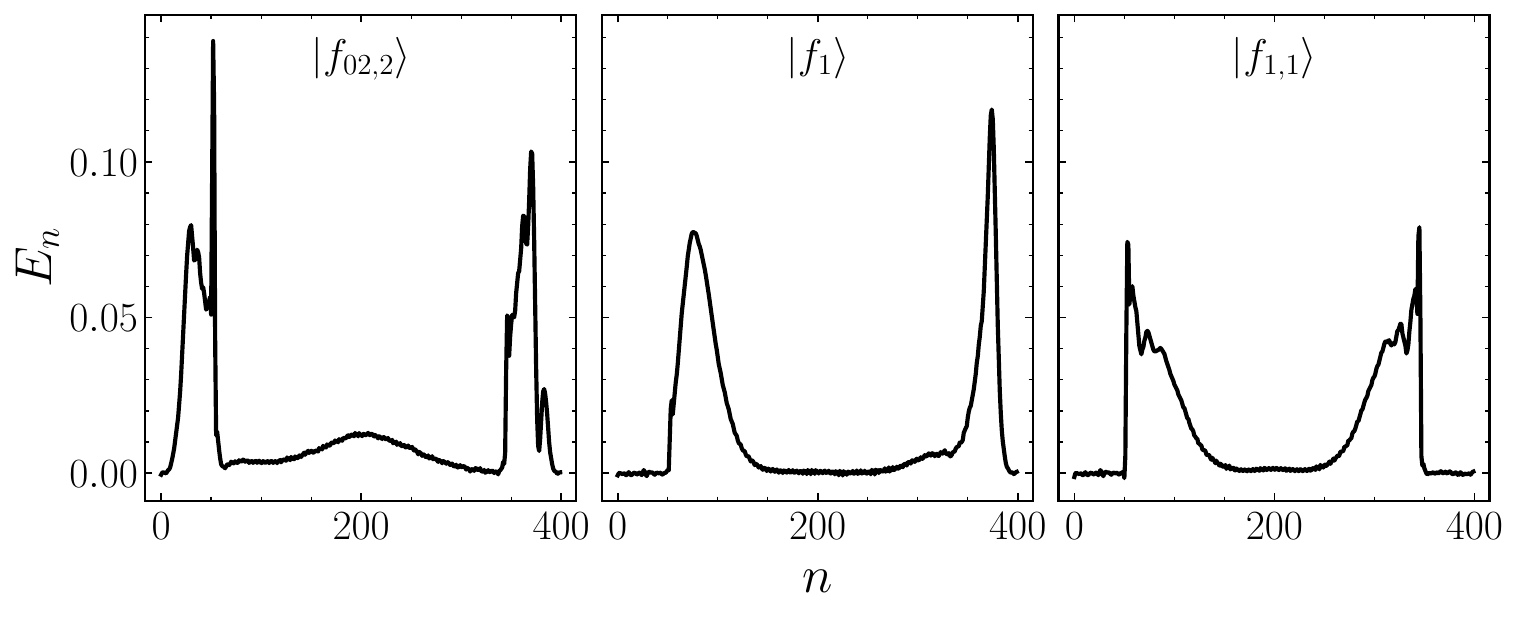}
\caption{
Energy densities $E_n$ for the Schmidt components used in the channel isolation process. 
The left panel shows $|f_{02,2}\rangle$, which is attributed to higher-order processes.
The middle panel shows $|f_1\rangle$, which is the combination of $|21\rangle$ and $|f_{1,1}\rangle$ shown on the right.
The state $|f_{1,1}\rangle$ is also attributed to higher-order processes.
}
\label{i_m:fig:other_channels}
\end{figure}

Projecting out the $|f_1\rangle$ retains the $|11\rangle$ and $|12\rangle$ states ($|f_{02}\rangle$), which are distinguished by a second bipartition at $n_r=L-n_l-1=346$.
Within $|f_{02}\rangle$, the second cut at $n_r$ also has three significant Schmidt coefficients: $\lambda_{02,0}$, which is identified with $|11\rangle$, $\lambda_{02,1}$, corresponding to $|12\rangle$, and $\lambda_{02,2}$.
The numerical values of these are given in the second row of Table~\ref{i_m:tab:schmidt_vals_nl_nr}.
The energy density in the last component, $|f_{02,2}\rangle$, is shown in the left panel of Fig.~\ref{i_m:fig:other_channels}, and is classified as a contribution from higher-order inelastic processes such as $11\to111$, as well as some potential mixing with the elastic channel.\footnote{Sharp features in the energy density around $n_l$ and $n_r$ are caused by forcing tensor product states at these locations.}
Parity symmetry requires the $|12\rangle$ and $|21\rangle$ components to be equal in amplitude, and rough equality is observed as expected ($\lambda_{02,1}\sim\lambda_{1,0}$).
Further decomposition of this wavefunction is only possible by cuts where excitations have significant support, which would separate different momentum modes belonging to the same channel, and are not done in this work.

The middle panel of Fig.~\ref{i_m:fig:other_channels} shows the energy density of the wavefunction $|f_1\rangle$ obtained by retaining the $\lambda_{1}$ component at the bipartition $n_l$, where a tail is seen on the left side of the $|1^{(21)}\rangle$ particle.
Performing a cut at $n_r$ on this wavefunction gives two significant Schmidt components $\lambda_{1,0}$ and $\lambda_{1,1}$, which are shown in the third row of Table~\ref{i_m:tab:schmidt_vals_nl_nr}.
The energy density of $|f_{1,0}\rangle$ is classified as $|21\rangle$ and is shown in the bottom right panel of Fig.~\ref{i_m:fig:channel}. 
The second component $|f_{1,1}\rangle$ is shown in the right panel of Fig.~\ref{i_m:fig:other_channels}, and is deemed to similarly arise from higher-order processes.
As a result, $\lambda_{02,2}+\lambda_{1,1}$ is reported as the contribution from higher-order channels.
Schmidt decompositions of the vacuum are dominated by a single vector across all cuts, shown in the bottom two rows of Table~\ref{i_m:tab:schmidt_vals_nl_nr}.
As a result of translation invariance, $|\text{vac}\rangle$ has the same entanglement structure throughout the lattice.

In summary, the states contributing to the elastic and inelastic channels are isolated by selecting the following Schmidt components:
\begin{equation}
    |f_{02,0}\rangle \ = \ |11\rangle \ , \ 
    |f_{02,1}\rangle \ = \ |12\rangle \ , \ 
    |f_{1,0}\rangle \ = \ |21\rangle
\end{equation}

Table~\ref{i_m:tab:ee_af_channel} shows the dominant Schmidt components, ${\cal F}_{AB}$ and $S_{AB}$ in the exclusive final states across a bipartition at the center of the lattice.
The remaining components are $O(10^{-3})$, suppressed by two orders of magnitude compared to the dominant states.
Together with the near-zero values of ${\cal F}_{AB}$, this indicates that the exclusive states $|11\rangle$, $|12\rangle$, and $|21\rangle$ are well described by product states, consistent with unentangled asymptotic outgoing particles.
The entanglement appears to be shared roughly evenly between the states, with variations likely caused by the differences in the cutting procedure described earlier.
Compared to $|12\rangle$ and $|21\rangle$, $|11\rangle$ is seen to have a larger value of ${\cal F}_{AB}$.
This is caused by the elastic channel having a single dominant component with more weight than the main dominant components in the inelastic channel. 
\begin{table}
\centering
\begin{tabularx}{\linewidth}{|c||Y|Y|Y|} \hline
State & $\lambda_i>10^{-2}$ & ${\cal F}_{AB}$ & $S_{AB}$
\\\hline\hline
$|f\rangle = |11\rangle + |12\rangle + |21\rangle$ & 0.6620,0.1570,0.1431 & $1.373\times 10^{-3}$ & 1.5380 \\ \hline\hline
$|f_{02,0}\rangle = |11\rangle$ & 0.5553 & $8.7843\times10^{-4}$ & 0.5419 \\ \hline
$|f_{02,1}\rangle = |12\rangle$ & 0.1621 & $7.4977\times10^{-5}$ & 0.5347 \\ \hline
$|f_{1,0}\rangle = |21\rangle$ & 0.1607 & $7.3546\times10^{-5}$ & 0.4646 \\ \hline
\end{tabularx}
\renewcommand{\arraystretch}{1}
\caption{Values of entanglement measures of the final states (first column) across a bipartition at $L/2$.
The second column gives significant values in the entanglement spectrum (Schmidt spectrum squared) at $t=120$.
The antiflatness ${\cal F}_{AB}$ and entanglement entropy $S_{AB}$ are given in the third and fourth columns respectively. 
The bottom three rows give the exclusive states determined from the inclusive state $|f\rangle$ (first row).
The exclusive states are not normalized so that their probabilities and entropies sum to the values in the first row.
}
\label{i_m:tab:ee_af_channel}
\end{table}
%

\subsection{Particle classification}
\label{i_m:app:particle_classification}
\noindent
After the states contributing to $|f\rangle$ are isolated, the kinematics of the process can be used to classify the particles by type within each channel. 
Table~\ref{i_m:tab:disp_calc} shows the details of the classification calculation. 
The velocity $v$ of the excitations in each state is given in the second column.
It may be computed by isolating exclusive states at several times and comparing positions of peaks in $E_n$.
Another method to determine $v$ involves finding the collision time $t_0$ by maximizing $E_{L/2}(t)$ and finding the velocity starting of each excitation from $t_0$ to $t=120$.
Equivalently, $t_0$ can be determined by a calibration to the 11 channel where the outgoing speed is known, which gives $t_0=8.2$ (compared with $t_0=8$ obtained from maximizing $E_n$).
The velocity computed with the latter method is reported in Table~\ref{i_m:tab:disp_calc}.
The group velocity $v_{1,2}(k)$ is then inverted to determine $k_{1,2}(v)$, which is plugged into the dispersion relation to determine $E_{1,2}(k_{1,2}(v))$.
Where present, the several entries in a single column correspond to cases where multiple solutions are possible.
A $-$ entry in Table~\ref{i_m:tab:disp_calc} indicates there is no solution, eliminating the possibility to classify the given excitation as that type of particle.
A classification between $|1\rangle$ and $|2\rangle$ is made by comparing $E(k(v))$ to the energy of the individual excitations measured in each exclusive state $|\psi_\text{chan.}\rangle$, ${E_{\text{wp}} = \sum_{n\in\text{wp}} \langle \psi_\text{chan.}|\hat H_n |\psi_\text{chan.}\rangle -\langle \text{vac}|\hat H_n |\text{vac}\rangle}$ and selecting the closest match.
In all cases, the error $|E_\text{wp}-E(k(v))|/E_\text{wp}<8\%$ and the classification identifies the absence of $|2\rangle$ in the elastic channel and the presence of a single $|2\rangle$ in the inelastic states.
Making a misclassification requires a 28\% error in the calculation.
Mismatches in the energies determined in this process stem from the approximate nature of the isolation process, as well as approximate state preparation.
\begin{table}
\centering
\tiny
\setlength{\tabcolsep}{1pt} 
\begin{tabularx}{\linewidth}{|c||Y|Y|Y|Y|Y|Y||c|} \hline
State & $v$ & $k_1(v)$ & $k_2(v)$ & $E_1(k_1(v))$ & $E_2(k_2(v))$ & $E_\text{wp}$ & Classification
\\\hline\hline
$|1^{(11)}\rangle$ & 1.6057 & 0.3551,\textbf{0.3600} & -- & 2.8489,\textbf{2.8735} & -- & 2.8512 & $|1\rangle$ \\\hline
$|1^{(12)}\rangle$ & -1.5610 & -0.4428,\textbf{-0.2817} & -- & 3.2877,\textbf{2.4816} & -- & 2.3616 & $|1\rangle$ \\\hline
$|2^{(12)}\rangle$ & 1.0958 & 0.1361,0.6771 & \textbf{0.2779},0.7338 & 1.8491,4.2918 & \textbf{3.5154},5.3364 & 3.3464 & $|2\rangle$ \\\hline
$|2^{(21)}\rangle$ & -1.0869 & -0.6803,-0.1346 & -0.7374,\textbf{-0.2742} & 4.3027,1.8437 & 5.3486,\textbf{3.5026} & 3.3527 & $|2\rangle$ \\\hline
$|1^{(21)}\rangle$ & 1.5610 & \textbf{0.2817},0.4428 & -- & \textbf{2.4816},3.2877 & -- & 2.3088 & $|1\rangle$ \\\hline
\end{tabularx}
\renewcommand{\arraystretch}{1}
\caption{{\it Particle classification based on measured energy and velocity of local observables.}
For a given individual excitation (first column) in the exclusive states, the second column gives the velocity of the peaks in $E_n$ over time.
The third and fourth columns give the results of extracting the possible momenta from the group velocities of particles $|1\rangle$ and $|2\rangle$ respectively. 
A $-$ entry indicates no solution is found, and multiple entries in a single column correspond to cases where multiple solutions are possible.
The fifth and sixth columns show the energy calculated from the dispersion relation using the momenta from the third and fourth columns.
The seventh column gives the energy $E_\text{wp}$ of the excitations calculated from the area under $E_n$ in the bottom panel of Fig.~\ref{i_m:fig:channel}.
The last column shows the classification of each excitation into particle type.
Bold entries correspond to the closest match between $E(k(v))$ (columns five and six) and $E_\text{wp}$ (column seven).
Since the left-moving $|1^{(11)}\rangle$ only differs from the right-moving $|1^{(11)}\rangle$ by the sign of its velocity, its calculations are not included in this table.
}
\label{i_m:tab:disp_calc}
\end{table}
%

\section{Momentum measurement in classical and quantum simulations}
\label{i_m:app:momentum}
\noindent
The simplest way to estimate the momentum of excitations traveling on the lattice is by recording the velocity of local observables sensitive to the excitations, such as $E_n$.
This is the method that is used in Sec.~\ref{i_m:sec:ising_scattering_momentum_results} and has no extra overhead in quantum or classical simulations.
When $E(k)$ is not known, several other techniques are possible both in classical and quantum simulations.

In the context of MPS, momentum can be measured in real-time by overlapping with plane-wave states, such as those prepared by the quasiparticle excitation ansatz described in App.~\ref{i_m:app:sim_details}. 
In principle, any basis of states with well-defined momentum, such as those in Eq.~\eqref{i_m:eq:spatial_v}, can be used to perform this measurement.
This is efficient with MPS, and may be done by sampling a subset of momenta instead of measuring all $2^{n_Q}$ modes.
This approach has also been used to detect particles by overlapping with localized wavepackets~\cite{Jha:2024jan}.

On a quantum computer, the momentum could be measured in a local region of the lattice by applying a unitary Fourier transform $V$ on the lattice indices.
This is the equivalent method of computing overlaps with plane-wave states in MPS described above.
On a 4-qubit region of the lattice, $V$ carries out the following transformation.

\begin{equation}
\begin{aligned}
    &\text{Hamming weight 0} \, \left\{\,
    \begin{aligned}
        |0000\rangle \ &\to \ k=0: \ &&|0000\rangle & 
    \end{aligned}
    \right.&
    \\[10pt]
    &\text{Hamming weight 1} \left\{
    \begin{aligned}
        |0001\rangle \ &\to \ k=0: &&\frac{1}{2} \left( |0001\rangle + |0010\rangle + |0100\rangle + |1000\rangle \right) \\
        |0010\rangle \ &\to \ k=\frac{\pi}{2}: &&\frac{1}{2} \left( |0001\rangle + i|0010\rangle - |0100\rangle - i|1000\rangle \right) \\
        |0100\rangle \ &\to k=\pi: &&\frac{1}{2} \left( |0001\rangle - |0010\rangle + |0100\rangle - |1000\rangle \right) \\
        |1000\rangle \ &\to \ k=-\frac{\pi}{2}: &&\frac{1}{2} \left( |0001\rangle - i|0010\rangle - |0100\rangle + i|1000\rangle \right)
    \end{aligned}
    \right.&
    \\[10pt]
    &\text{Hamming weight 2} \left\{
    \begin{aligned}
        |0011\rangle \ &\to \ k=0: &&\frac{1}{2} \left( |0011\rangle + |0110\rangle + |1100\rangle + |1001\rangle \right) \\
        |0110\rangle \ &\to \ k=\frac{\pi}{2}: &&\frac{1}{2} \left( |0011\rangle + i|0110\rangle - |1100\rangle - i|1001\rangle \right) \\
        |1100\rangle \ &\to \ k=\pi: &&\frac{1}{2} \left( |0011\rangle - |0110\rangle + |1100\rangle - |1001\rangle \right) \\
        |1001\rangle \ &\to \ k=-\frac{\pi}{2}: &&\frac{1}{2} \left( |0011\rangle - i|0110\rangle - |1100\rangle + i|1001\rangle \right) \\
        |0101\rangle \ &\to \ k=0: &&\frac{1}{\sqrt 2} \left( |0101\rangle + |1010\rangle \right) \\
        |1010\rangle \ &\to \ k=\pi: &&\frac{1}{\sqrt 2} \left( |0101\rangle - |1010\rangle \right)
    \end{aligned}
    \right.&
    \\[10pt]
    &\text{Hamming weight 3} \left\{
    \begin{aligned}
        |0111\rangle \ &\to \ k=0: &&\frac{1}{2} \left( |0111\rangle + |1110\rangle + |1101\rangle + |1011\rangle \right) \\
        |1110\rangle \ &\to \ k=\frac{\pi}{2}: &&\frac{1}{2} \left( |0111\rangle + i|1110\rangle - |1101\rangle - i|1011\rangle \right) \\
        |1101\rangle \ &\to \ k=\pi: &&\frac{1}{2} \left( |0111\rangle - |1110\rangle + |1101\rangle - |1011\rangle \right) \\
        |1011\rangle \ &\to \ k=-\frac{\pi}{2}: &&\frac{1}{2} \left( |0111\rangle - i|1110\rangle - |1101\rangle + i|1011\rangle \right)
    \end{aligned}
    \right.&
    \\[10pt]
    &\text{Hamming weight 4} \,\left\{\,
    \begin{aligned}
        |1111\rangle \ &\to \ &k=0: \ &&|1111\rangle
    \end{aligned}
    \right.&
\end{aligned}
\label{i_m:eq:spatial_v}
\end{equation}
While finding a circuit for this transformation is challenging, its structure would mirror the classical Cooley-Tukey Fast Fourier Transform (FFT)~\cite{Cooley:1965zz} on $n_Q=2^n$ qubits.
Similar circuits have been developed for the fermionic FFT~\cite{Ferris:2014jmb,Barak:07,Verstraete:2008qpa}, and have been used in the context of neutral atom platforms~\cite{Maskara:2025oab} to implement fermion permutations.
Following the transformation to (spatial) momentum space (in contrast to Hilbert space momentum), measurements on the qubits within the ``detector'' region would correspond to the momentum modes present in that region.
This would be an approximate measurement of momentum with the resolution set by the size of the detector $L_d$, yielding a momentum uncertainty $\Delta k\sim 2\pi/L_d$ at the cost of a circuit depth overhead.

\begin{figure}
    \centering
\includegraphics[width=0.9\linewidth]{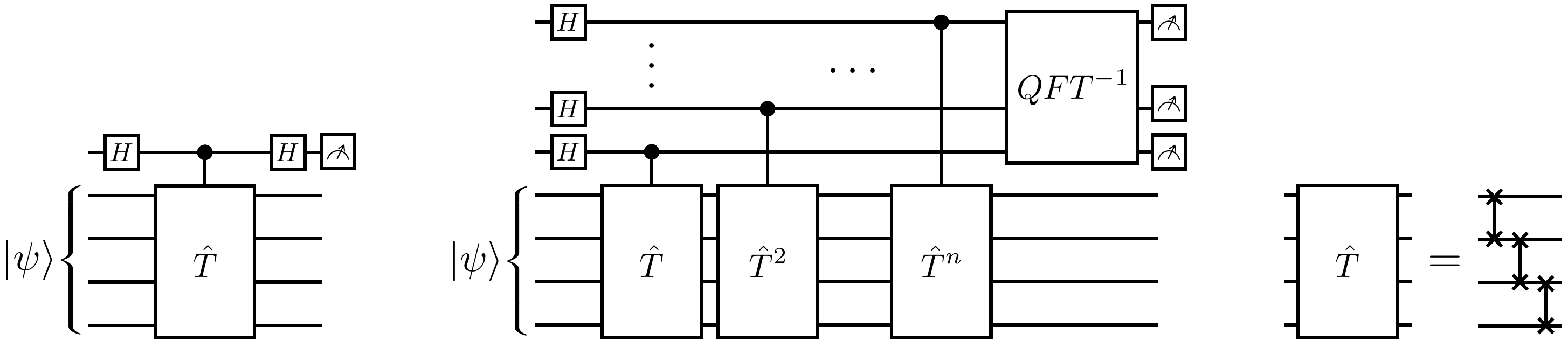}
\caption{
Circuits that could be used to measure momentum using the translation operator $\hat T$.
Left: the circuit that measures $\langle \psi|\hat T|\psi\rangle$. 
Center: the quantum phase estimation circuit that can be used for momentum measurement.
Right: the implementation of the translation operator in a spin system.
}
\label{i_m:fig:measure_t}
\end{figure}

Another approach involves measuring the expectation value of the translation operator $\hat T$. 
On a 4-qubit state $\hat T$ acts as $\hat T|0001\rangle \ = \ |0010\rangle$.
On a state with well-defined momentum $|\psi_k\rangle$, $\hat T$ shifts the state by one lattice site and returns a phase 
\begin{align}
    \hat T|k\rangle \ = \ e^{ik}|k\rangle \ .
    \label{i_m:eq:T_action}
\end{align}
The expectation value $\langle \hat T\rangle$ can be measured using the Hadamard test~\cite{10.1098/rspa.1998.0164}, requiring a single application of a controlled-$\hat T$ unitary.
The circuit implementing this is shown in the left panel of Fig.~\ref{i_m:fig:measure_t}.
The probability of measuring 0 on the ancilla qubit is related to $\langle \hat T\rangle$ by $P(0) = \frac{1+Re\langle \hat T\rangle}{2}$.
The imaginary component of the phase is similarly extracted by measuring in the $Y$ basis.
For spin systems, this circuit requires a network of controlled-swap gates, and is long-range in nature.
This approach may also be extended to full phase estimation, requiring additional ancilla qubits and applications of the controlled-$\hat T$ unitary. 
This is shown in the central panel of Fig.~\ref{i_m:fig:measure_t}, and would give information regarding the amplitude present in each momentum sector, up to the resolution set by the number of ancilla qubits.\footnote{Recently, there have been optimizations to the phase estimation algorithm developed which allow full phase estimation with a single ancilla qubit at the expense of sample overhead, and removing controlled-unitary dependence altogether~\cite{Clinton:2024sij}.}
This approach carries an overhead in shots and circuit depth.
Since proper momentum states only have the phase relation of Eq.~\eqref{i_m:eq:T_action} when translated along the whole lattice, this only yields approximate measurements of momentum when applied locally, and is thus of limited use.
However, it could be used to test for local translation invariance.

\section{Computational methods}
\label{i_m:app:sim_details}
\subsection{State preparation}
\label{i_m:app:state_prep}
\noindent
Wavepackets are prepared using the state preparation algorithm described in Chapter~\ref{chap:ising_scattering}.
This state preparation algorithm initializes the localized wavepackets and the vacuum of the periodic boundary conditions (PBC) system elsewhere.
Although MPS are much more efficient for open boundary conditions (OBCs)~\cite{PhysRevLett.93.227205}, PBCs are used in this work to avoid propagating excitations from imperfect state preparation near the boundary in OBCs.

\begin{figure}
    \centering
\includegraphics[width=0.5\linewidth]{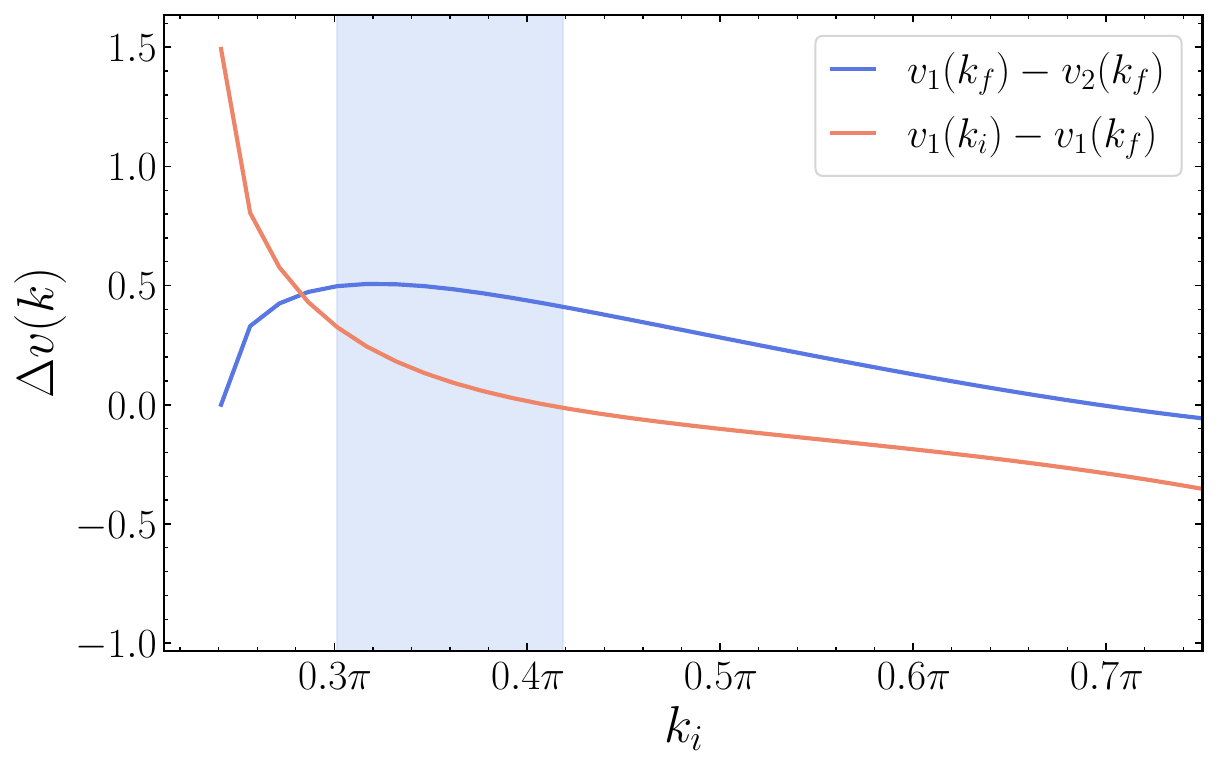}
\caption{
Velocity differences $\Delta v(k)$ of the outgoing particles as a function of incoming momentum $k_i$ of the $|1\rangle$ particles. 
The difference in velocity of $|1^{(12)}\rangle$ and $|2^{(21)}\rangle$ is shown in blue, while the difference between the velocities of the two light particles $|1^{(11)}\rangle$ and $|1^{(12)}\rangle$ is shown in orange.
The light blue shading shows $k_i\pm\sigma_k$ used in scattering simulations in Sec.~\ref{i_m:sec:ising_scattering_momentum_results}.
}
\label{i_m:fig:outgoing_group_velocities}
\end{figure}

The initial momentum of the wavepackets $k_i=\pm0.36\pi$ is chosen for the simulations producing the final state shown in Fig.~\ref{i_m:fig:channel}.
Figure~\ref{i_m:fig:outgoing_group_velocities} shows the differences between the velocities of the outgoing particles in the elastic and inelastic channels, which set the spacing between the particles in Fig.~\ref{i_m:fig:late_time_schematic}.
While the separation between $|1^{(11)}\rangle$ and $|1^{(12)}\rangle$ is small at $k_i=0.36\pi$ and $t=120$, the difference in the outgoing velocities of the light and heavy particles is close to maximal.
A smaller $k_i$ would increase the difference $v_1(k_i)-v_1(k_f)$, but would decrease the velocities of all outgoing particles (see left panel of Fig.~\ref{i_m:fig:ising_momentum_dispersion}).
The combined effects of wavepacket spreading and slower propagation result in the best spatial separation between $|1\rangle$ and $|2\rangle$ being near $k_i=0.36\pi$.

The wavepacket width is set to $\sigma_k=0.059\pi$ and 99.54\% of the $|\psi_\text{wp}\rangle$ norm is retained by truncating the wavepacket in position space to $d=21$ sites.

\subsection{MPS details}
\noindent 
The MPS time evolution simulations are done with a bond dimension of ${\tt max\_bond}=600$ and cutoff of $10^{-9}$ to $t=40$, well past the collision time ($t\sim8$).
After this time, the MPS is truncated to ${\tt max\_bond}=350$ with the same cutoff to increase the speed of the simulation, which introduces truncation errors of $O(10^{-3})$.
The norm of the state is used to estimate the truncation effects, and is shown for various times in the simulation in Table~\ref{i_m:tab:trunc_errors}.
\begin{table}
\centering
\footnotesize
\begin{tabularx}{\linewidth}{|c||Y|Y|Y|Y|Y|Y|} \hline
& \multicolumn{6}{c|}{$\langle f|f\rangle$} \\ \hline
$k_i$ & $0.18\pi$ & $0.2\pi$ & $0.24\pi$ & $0.28\pi$ &
$0.32\pi$ & $0.36\pi$ \\\hline\hline
$t=40$, ${\tt max\_bond}=600$ & 0.9938 & 0.9945 & 0.9926 & 0.9914 & 0.9911 & 0.9928 \\\hline
$t=40$, ${\tt max\_bond}=350$ & 0.9908 & 0.9920 & 0.9886 & 0.9870 & 0.9868 & 0.9897 \\\hline
$t=120$, ${\tt max\_bond}=350$ & 0.9331 & 0.9546 & 0.9379 & 0.9313 & 0.9355 & 0.9488 \\\hline
\end{tabularx}
\renewcommand{\arraystretch}{1}
\caption{Truncation errors given by the norms of $|f\rangle$ at various times and initial momenta $k_i$ in the MPS simulation. The first row shows the norm well after the scattering events at $t=40$ at ${\tt max\_bond}=600$. 
At this point, {\tt max\_bond} is truncated to 350 (second row).
The third row shows the norms at the final time $t=120$.
All rows use a cutoff of $10^{-9}$.
}
\label{i_m:tab:trunc_errors}
\end{table}
The final states at $t=120$ have truncation errors of $<7\%$ and are normalized to unity for purposes of channel isolation in Sec.~\ref{i_m:sec:ising_scattering_momentum_results}.
While this error is non-negligible, it is assumed that the most physically important components are retained by the SVD truncation, since the $t=40$ post-scattering state with ${\tt max\_bond}=600$ has norm near unity and no additional interactions take place after this time.
A Trotter step size of $\delta t=1/32$ is used to simulate the scattering process to $t=120$. 
This could be improved by using an optimized higher-order Trotter formula~\cite{Barthel:2019kch}.\footnote{It is interesting to consider the possibility of simulating scattering events in momentum space, where the states are much simpler at the expense of having highly nonlocal interactions~\cite{VanDamme:2022lax,Corbett:2025flm}.}

For one-dimensional gapped theories with isolated single-particle excitations, dispersion relations can be efficiently computed in MPS using the quasiparticle excitation ansatz~\cite{Haegeman:2013xcv}. 
Where this is not possible, they could be computed with exact diagonalization for small system sizes and extrapolated to suitable $L$~\cite{Farrell:2023fgd,Zemlevskiy:2024vxt}.
Dispersion relations could be evaluated on a quantum computer with knowledge of the single-particle excitation structure. 
This can be done by preparing single-particle plane waves (e.g., Ref.~\cite{Farrell:2025nkx}) and measuring their energy, or by preparing wavepackets and recording their speed.
The dispersion relations in Sec.\ref{i_m:sec:ising_scattering_momentum_results} were constructed using the quasiparticle excitation ansatz in the thermodynamic limit of the Hamiltonian in Eq.~\eqref{i_m:eq:h_ising}.
This algorithm first finds the (translationally invariant) vacuum of the $L=\infty$ system using Variational Uniform Matrix Product States (VUMPS)~\cite{Zauner-Stauber:2017eqw} specified by the tensor $A$,
\begin{equation}
|\text{vac}\rangle \ = \ \dots
\begin{tikzpicture}[baseline=-0.5ex, scale=0.8,
    tensor/.style={draw, rounded corners=3pt, minimum size=0.7cm, fill=white, inner sep=2pt}
]
    \node[tensor] (A0) at (0, 0) {$A$};
    \node[tensor] (A1) at (1.4, 0) {$A$};
    \node[tensor] (A)  at (2.8, 0) {$A$};
    \node[tensor] (A2) at (4.2, 0) {$A$};
    \node[tensor] (A3) at (5.6, 0) {$A$};

    \draw (A0.west) -- ++(-0.35,0);
    \draw (A0.east) -- (A1.west);
    \draw (A1.east) -- (A.west);
    \draw (A.east) -- (A2.west);
    \draw (A2.east) -- (A3.west);
    \draw (A3.east) -- ++(0.35,0);

    \foreach \t in {A0,A1,A,A2,A3} {
        \draw (\t.south) -- ++(0,-0.25);
    }

    \node[below] at ($(A0.south)+(0,-0.25)$) {\small $\cdots$};
    \node[below] at ($(A1.south)+(0,-0.25)$) {\small $s_{n-1}$};
    \node[below] at ($(A.south)+(0,-0.25)$)  {\small $s_n$};
    \node[below] at ($(A2.south)+(0,-0.25)$) {\small $s_{n+1}$};
    \node[below] at ($(A3.south)+(0,-0.25)$) {\small $\cdots$};
\end{tikzpicture}
\dots \ ,
\end{equation}
where the tensor legs below the tensors $A$ correspond to the physical indices of the state.
Low-lying excitations of momentum $k$ are constructed as momentum superpositions of a local operator $B_{k;j}$ acting on the vacuum,
\begin{equation}
|\psi_{k;j}\rangle \ = \ \sum_n e^{ikn} \dots
\begin{tikzpicture}[baseline=-0.5ex, scale=0.8,
    tensor/.style={draw, rounded corners=3pt, minimum size=0.7cm, fill=white, inner sep=2pt}
]
    \node[tensor] (A0) at (0, 0) {$A$};
    \node[tensor] (A1) at (1.4, 0) {$A$};
    \node[tensor] (B)  at (2.8, 0) {$B_{k;j}$};
    \node[tensor] (A2) at (4.2, 0) {$A$};
    \node[tensor] (A3) at (5.6, 0) {$A$};

    \draw (A0.west) -- ++(-0.35,0);
    \draw (A0.east) -- (A1.west);
    \draw (A1.east) -- (B.west);
    \draw (B.east) -- (A2.west);
    \draw (A2.east) -- (A3.west);
    \draw (A3.east) -- ++(0.35,0);

    \foreach \t in {A0,A1,B,A2,A3} {
        \draw (\t.south) -- ++(0,-0.25);
    }

    \node[below] at ($(A0.south)+(0,-0.25)$) {\small $\cdots$};
    \node[below] at ($(A1.south)+(0,-0.25)$) {\small $s_{n-1}$};
    \node[below] at ($(B.south)+(0,-0.25)$)  {\small $s_n$};
    \node[below] at ($(A2.south)+(0,-0.25)$) {\small $s_{n+1}$};
    \node[below] at ($(A3.south)+(0,-0.25)$) {\small $\cdots$};
\end{tikzpicture}
\dots \ .
\label{i_m:eq:psi_k_mps}
\end{equation}
Here $j$ specifies the excitation number, corresponding to particle $|1\rangle$, $|2\rangle$ or higher excitations. 
For the simulations in this work, the states $|\psi_{k;1}\rangle$ correspond to the single-particle plane waves discussed in App.~\ref{i_m:app:state_prep}.
This ansatz provides an approximation to the true momentum states of the theory with an error that is exponentially small in the support of the local operator $B_{k;j}$. 
The vacuum of the $L=\infty$ system is well-represented by a bond dimension of ${\tt max\_bond}=12$, and the dispersion relation computations require ${\tt max\_bond}=24$.

\end{subappendices}

\chapter{The quantum complexity of string breaking in the Schwinger model}
\label{chap:schwinger_magic}

\noindent
{\it This chapter is associated with Ref.~\cite{Grieninger:2026bdq}: ``The quantum complexity of string breaking in the Schwinger model'' by Sebastian Grieninger, Martin J. Savage, and Nikita A. Zemlevskiy.}

\section{Introduction}
\label{m_s:sec:magic_schwinger_intro}
\noindent
Measures of quantum complexity are sensitive probes of emergent phenomena~\cite{Haferkamp:2021uxo, Chitambar:2018rnj,Brown:2017jil,Eisert:2008ur,Leone:2021rzd,Robin:2020aeh,Haug:2023hcs,Tarabunga:2023hau,Hengstenberg:2023ryt,Haug:2024ptu,Emerson:2013zse,Howard:2017maw,Hamaguchi:2023zpb,Tirrito:2023fnw,Chernyshev:2024pqy,Cao:2024nrx,Robin:2024oqc,brokemeier2025quantum,Robin:2025ymq,Jiang:2025wpj}.
This chapter investigates multipartite entanglement and magic in the ground state of the Schwinger model, a well-established testbed for quantum simulations of quantum field theories~\cite{PhysRevA.90.042305,Muschik:2016tws,Klco:2018kyo,Farrell:2023fgd,Farrell:2024fit,Nguyen:2021hyk}, as the system evolves from a confining string to isolated bound states (hadrons).
We describe minimal simulation requirements for lattice studies of string breaking, 
and perform simulations to study the inherent quantum complexity 
(quantum correlations) in the wavefunction, revealing previously unknown structure.
Measures of quantum complexity are found to change rapidly in the vicinity of 
string breaking,
offering a complementary view of mechanisms involved in the formation of hadrons and the associated vacuum rearrangement.

\section{Simulating the lattice Schwinger model}
\label{m_s:sec:LSM}

\subsection{The Hamiltonian and Gauss's law}
\label{m_s:sec:Hami}
\noindent
In 1+1D with open boundary conditions (OBCs), 
the distribution of fermion charges completely constrains the gauge field through Gauss's law.
Further, explicit gauge degrees of freedom are absent in axial gauge,
leaving a nonlocal fermionic interaction in the Hamiltonian induced by the Coulomb potential.
In the Kogut-Susskind formulation  \cite{Kogut:1974ag,Banks:1975gq}, with staggered fermion discretization \cite{Susskind:1976jm} and the Jordan-Wigner transformation to spin degrees of freedom~\cite{Jordan:1928wi,Lieb:1961fr}, the lattice Hamiltonian of the Schwinger model is (for a derivation, see Ref.~\cite{Banks:1975gq})
\begin{eqnarray}
	&& \hat{H} (d) \ =\   \frac{1}{4a}  \sum_{n=1}^{N-1}\left( \hat{X}_n \hat{X}_{n+1} + \hat{Y}_n \hat{Y}_{n+1} \right)  +  \frac{m_\text{lat}}{2} \sum_{n=1}^{N} (-1)^n  \hat{Z}_n 
    +\frac{a}{2} \sum_{n=1}^{N-1} \left(\hat{E}_n +E_{\text{ext},n}(d)\right)^2
    \ ,
    \label{m_s:eq:h_obc}
\end{eqnarray}
where the lattice sites are labeled from $n=1$ to $N$, and the link indices range from $n=1$ to $N-1$.
Here, ${m_\text{lat}=m-\frac{g^2a}{8}}$~\cite{Dempsey:2022nys} is the chirally improved (bare) fermion mass, $m$ and $g$ are the bare mass and coupling respectively and $N$ is the number of staggered sites (corresponding to 
$N_\text{phys}=N/2$
physical sites), and $a$ is the 
(staggered) lattice spacing.\footnote{We have not included constant terms in Eq.~\eqref{m_s:eq:h_obc}.
The physical lattice spacing is twice the staggered spacing, $a_{\rm phys}=2 a$,
and the number of physical lattice sites is half the number of staggered sites, $N_{\rm phys}=N/2$.
The physical length of the lattice is $L = aN = a_{\rm phys}N_{\rm phys}$.}
The last term specifies the nonlocal fermionic interaction, where 
\begin{align}
 \hat{E}_n \ =\  E_{0}+g \sum_{k=1}^{n} \hat{Q}_k \   , \quad 
 \hat{Q}_k \ = \ \frac{1}{2}\left((-1)^k+\hat{Z}_k\right)
 \ ,
 \label{m_s:eq:e_field_charge}
\end{align}
are the electric field between sites $n,n+1$, and the charge on site $k$, respectively.
By convention, we set the background field $E_0=0$.
A flux tube connecting two static external charges is represented by the external electric field $E_{\text{ext},n}(d)$~\cite{Grieninger:2025rdi}.\footnote{Instead of writing the electric field in terms of $\hat E_n$ and $E_\text{ext,n}(d)$ one could also modify Gauss's law as explained in~\cite{Buyens:2015tea}.}
A string of length $d$ centered on the middle of the lattice is implemented by applying the external field
\begin{equation}
	E_{\textrm{ext,}n}(d) \ = \ g\ \text{sign}(E)\cdot \Theta\!  \left( \frac{d/a - 1}{2} - \left| n - \frac{N}{2}\right|+\epsilon \right) \ ,
    \label{m_s:eq:L_ext}
\end{equation}
where $\Theta$ is the Heaviside step function and $\epsilon$ is a small number. 
This represents a string created by a static external fermion at $n=(N+1-d/a)/2$ and an antifermion at $n=(N+1+d/a)/2$. 

In the absence of dynamical screening and vacuum rearrangement, the ground state of the Hamiltonian~\eqref{m_s:eq:h_obc} encodes the rearrangement of the degrees of freedom in the vacuum in response to the insertion of static external charges.
We refer to the ground state of $H(d=0)$ as the ``vacuum'', 
and study the behavior of the ground states of $H(d)$ as a function of $d$.

\subsection{Minimal lattice requirements for studies of string breaking}
\label{m_s:sec:lattice_requirements}
\noindent
In lattice simulations, physical states must be well-contained within the volume of the lattice, so that boundary effects can be quantified and systematically removed. 
The Compton wavelength $l$ of excited states relevant to the physics of the process must be much smaller than the lattice size $L$. 
Furthermore, the resolution of low-lying state wavefunctions must be fine enough to resolve quantities of interest perturbatively close to their continuum values, meaning that $l/a_{\rm phys}\gg1$.
Therefore, the lattice parameters must satisfy 
$\frac{1}{L} \ll \Delta \ll \frac{1}{a_{\rm phys}}$ 
for a lattice simulation to reliably capture the physics of the process. 
Here $\Delta \sim 1/l$ is the gap to the first excited state.\footnote{As an example, for lattice QCD calculations at the physical values of the quark masses, 
with $m_\pi\sim 140~{\rm MeV}$,
this condition implies that $m_\pi L\gg 1$ and $m_\pi a\ll 1$. 
With a lattice spacing of $a=0.1~{\rm fm}$, this gives $m_\pi a\sim 0.07 \ll 1$, 
and 64 lattice sites in each spatial direction gives $m_\pi L\sim 4.5 \gg 1$.
}
When simulating confining systems with static background charges separated by a distance $d$ (or multiparticle states), boundary effects on the perturbations must be exponentially suppressed.
This requires $L\gg d$.\footnote{For a simple system of charges on a 1D lattice, the potential experienced by a static charge is the sum over image charge contributions~\cite{Huang:1957im,Hamber:1983vu,Luscher:1986pf,Luscher:1990ux,Luscher:1990ck,Detmold:2007wk,Lu:2018pjk}. 
For PBCs, ${V^{\rm eff}(d)  =  \sum_{n=-\infty}^{+\infty}\  V(|d+nL|) }$.
Twisted boundary conditions can be helpful in minimizing the effects of the boundary, see e.g., Ref.~\cite{Briceno:2013hya}, particularly i-PBCs for which the leading contributions to single-particle observables vanish.
Boundary effects are discussed in App.~\ref{m_s:app:obc_vs_pbc}.
}

In this work, we select parameters for the spectrum and interactions to satisfy these conditions.
As a starting point, we chose 
$g=0.09$, 
$m_\text{lat}=0.045$, $a=1$ and $N=220$ staggered lattice sites ($L=110$)
in Eq.~(\ref{m_s:eq:h_obc}).
With these parameters, the vector meson mass is found to be $M_v=0.13647$, so that $M_v a_{\rm phys} \sim 0.2729$ and $M_v L \sim 15.0117$, 
satisfying the conditions for perturbative finite volume and lattice spacing corrections.
Further, 
the finite volume corrections are suppressed by 
${\cal O}(e^{-M_v L/2}) \sim 2\times 10^{-4}$ for $d=L/2$.
Continuum physics is identified from the $a\rightarrow 0$, $N\rightarrow\infty$
limit of the Hamiltonian in
Eq.~(\ref{m_s:eq:h_obc}) with fixed $m$ and $g$.
To approach the continuum with this physical parameter set, we perform simulations with 
$\{N,a\} = \{220,1\}, \{440,\frac{1}{2}\}$ and $\{880,\frac{1}{4}\}$ so that $L=110$.

\section{Measures of quantum complexity and gauge invariance}
\label{m_s:sec:QCGI}
\noindent
The distribution of quantum information in a state provides insight into its underlying structure.
In gauge theories, physically meaningful measures of entanglement and quantum complexity must be gauge invariant, which restricts their definition to operations that preserve global charges.
The ground state wavefunction $|\psi\rangle$  of the Schwinger model has vanishing total electric charge, $Q=0$.
However, reduced states have contributions from all charge sectors.
For a bipartition of the lattice into regions $A$ and $B$, the reduced density matrix $\hat{\rho}_A$ has a block diagonal structure, with the blocks characterized by the charge within region $A$,
\begin{eqnarray}
    |\psi\rangle & \ = \ & \sum_{Q,i}\ c^{(Q)}_i |\psi^{(Q,i)}_A\rangle \otimes |\psi^{(-Q,i)}_B\rangle
    \ \ , \nonumber\\ 
    \hat{\rho}_A & \ =\  &   {\rm Tr}_B \left[ \hat{\rho}_{AB} \right]\ =\ \sum_Q \ p_Q \ \hat{\rho}^{(Q)}_A
    \ ,
\label{m_s:eq:ABQ}
\end{eqnarray}
where 
$\hat{\rho}_{AB}=|\psi\rangle\langle\psi|$.
The block-diagonal form of $\hat{\rho}_A$ allows the entanglement entropy to be decomposed 
into contributions from the charge sectors~\cite{Ghosh:2015iwa,Turkeshi:2020yxd,Buividovich:2008gq,Donnelly:2011hn,Casini:2013rba,Radicevic:2014kqa,Aoki:2015bsa,Soni:2016ogt,Goldstein:2017bua,Nishioka:2018khk,Amorosso:2024leg,Amorosso:2024glf}
\begin{eqnarray}
S(\hat{\rho}_A) & \  = \ & 
\sum_Q p_Q \ S(\hat{\rho}^{(Q)}_A) - \sum_Q p_Q \log_2 p_Q
    \ ,
\label{m_s:eq:SEQ}
\end{eqnarray}
where $S(\hat{\rho})=-{\rm Tr} \left[ \hat{\rho}\log_2\hat{\rho} \right]$ is the von Neumann entanglement entropy.
By treating each charge sector separately, this quantity is naturally gauge invariant.
This definition extends to other quantities, including mutual information  (MI) and antiflatness.
The MI is defined by
\begin{eqnarray}
I(A:B) & \ = \ & S(\hat{\rho}_A) + S(\hat{\rho}_B) - S(\hat{\rho}_{AB})
    \ ,
\label{m_s:eq:IAB}
\end{eqnarray}
using the definition in Eq.~(\ref{m_s:eq:SEQ}).
The antiflatness for the same bipartition, 
$ {\cal F}(\hat{\rho}_A)$, 
which provides a lower bound to the nonlocal magic~\cite{Cao:2024nrx},\footnote{
Results from small-qubit systems suggest the nonlocal magic is related to its lower bound by ${\cal M}_2^{(\text{NL})}=4{\cal F}$.} becomes 
\begin{eqnarray}
{\cal F}(\hat{\rho}_A) & \ = \ & \sum_Q p_Q^2\  {\cal F}_A(\hat{\rho}_A^{(Q)})
\ ,
\nonumber\\
{\cal F}_A(\hat{\rho}) & \ = \ &  {\rm Tr}\hat{\rho}^3 -\ \left({\rm Tr}\hat{\rho}^2\right)^2 \ =\ {\rm Var}(\hat{\rho}^2)
    \ .
\label{m_s:eq:AFQ}
\end{eqnarray}

The recently identified Stabilizer Rényi Entropies (SREs) 
are calculable measures of the quantum magic 
in both pure and mixed states but they are not monotones for mixed states.
The density matrix of a pure state $|\psi\rangle$ may be written in the Pauli basis,
\begin{equation}
    \hat{\rho} \ = \ \ket{\psi} \bra{\psi} \ = \ 
    \frac{1}{\mathbf d} \sum_{\hat P } \langle \psi |\hat{P} | \psi \rangle \, \hat{P} 
    \ =\ 
    \frac{1}{\mathbf d} \sum_{\hat P } c_P \, \hat{P} 
    \; ,
\end{equation}
where $c_{\hat{P}} \equiv \langle \psi |\hat{P} | \psi \rangle$,
and ${\mathbf d}=2^{n_Q}$ for $n_Q$  qubits.
For pure states,
the quantity $\Xi_{\hat{P}} \equiv   c_{\hat{P}}^2/{\mathbf d}$
is a probability distribution~\cite{Leone:2021rzd}.
For a stabilizer state $\ket{\psi}$,
$c_{\hat{P}} = \pm 1$ for ${\mathbf d}$ commuting Pauli strings
and $c_{\hat{P}} = 0$ for the remaining  ${\mathbf d}^2-{\mathbf d}$ strings~\cite{zhu2016clifford}.
The SREs measure the deviation from stabilizer states, 
\begin{equation} 
\mathcal{M}_{\alpha}(\ket{\psi}) \ = \ -\log_2 {\mathbf d} + \frac{1}{1-\alpha} \log_2 
\left( \sum_{\hat{P} } \Xi_{\hat{P}}^{\alpha} \right) \; ,
\label{m_s:eq:Renyi_entropy_def1}
\end{equation}
with ${\cal M}_2$ used in this work:
\begin{eqnarray}
    {\cal M}_2 & \ = \ &    -
    \log_2 \ {\mathbf d} \sum_{\hat{P} }  \Xi_{\hat{P}}^2
\ .
    \label{m_s:eq:MlinM1M2_def}
\end{eqnarray}

Nonlocal magic of a bipartition of a system into regions A and B is the amount of magic that cannot be removed by local unitary transformations.
Operationally, it is defined by 
\begin{eqnarray} 
\mathcal{M}^{(\text{NL})}_{\alpha}(\ket{\psi}) & \ = \ & 
\min_{ {\bm\theta}_A,  {\bm\theta}_B} 
\mathcal{M}_{\alpha}(U_A({\bm\theta}_A) U_B({\bm\theta}_B)\ket{\psi}) 
\ ,
\label{m_s:eq:NLMAGIC}
\end{eqnarray}
for charge-preserving transformations to maintain gauge invariance.
An upper bound for the nonlocal magic ${\cal M}_2$ is found by writing the 
bipartitioned $Q=0$ ground state wavefunction as
in Eq.~(\ref{m_s:eq:ABQ})~\cite{Falcao:2024msg},
 \begin{eqnarray}
&& e^{-{\cal M}_2}\ =\   
 \sum_{ {\bm\sigma}^{(1)} , {\bm\sigma}^{(2)} , {\bm\sigma}^{(3)} , {\bm\sigma}^{(4)}}
 c_{{\bm\sigma}^{(1)}}\ 
 c_{{\bm\sigma}^{(2)}}\ 
 c_{{\bm\sigma}^{(3)}}\ 
 c_{   {\bm\sigma}^{(1)}   {\bm\sigma}^{(2)}   {\bm\sigma}^{(3)}   }\ 
 \nonumber\\
 &&
 \qquad\times
 c^*_{   {\bm\sigma}^{(1)}   {\bm\sigma}^{(2)}   {\bm\sigma}^{(4)}   }\ 
 c^*_{   {\bm\sigma}^{(1)}   {\bm\sigma}^{(3)}   {\bm\sigma}^{(4)}   }\ 
 c^*_{   {\bm\sigma}^{(2)}   {\bm\sigma}^{(3)}   {\bm\sigma}^{(4)}   }\ 
 c^*_{{\bm\sigma}^{(4)}}
 \ ,
 \label{m_s:eq:Tara}
 \end{eqnarray}
where 
$c_i= c^{(Q)}_i$
and products of ${\bm\sigma}^{(s)} $ denote elementwise multiplication. 

The Robustness of Magic  (RoM)~\cite{Howard:2017maw,Pashayan:2015cos,Heinrich:2019aei} is defined by the minimum distance to the surface of stabilizer states,
\begin{eqnarray}
    R(\hat{\rho})  & \ = \ &  \min_{\bm x} \left\{
    \ ||\bm x||_1 \ \ \  \bigg\rvert \ \ \ \hat{\rho} = \sum_{i}
     x_i\hat{\rho}_{s_i}\right\} 
     \ ,
\label{m_s:eq:RoMdef}
\end{eqnarray}
given by the 1-norm of the coefficients of stabilizer density matrices $\hat{\rho}_{s_i}$.
States that are elements of the stabilizer polytope will have $x_i>0$ and $R(\hat{\rho})=1$, while nonstabilizer states will have one or more $x_i<0$ and thus $R(\hat{\rho})>1$.
As the trace of the density matrix is unity, $R(\hat{\rho})$ measures the amount of negativity in the 
expansion coefficients,
\begin{eqnarray}
||\bm x||_1 \ = \ 1 + 2 \sum\limits_{i, x_i<0} |x_i|
\ ,
\end{eqnarray}
and hence provides a measure of the difficulty for classical simulation
(more specifically, $R(\hat{\rho})-1$).
This is a faithful magic monotone for both pure and mixed states.
The nonlocal RoM (NL RoM) associated with subsystems $A$ and $B$ is defined by a further minimization with respect to charge-preserving 
local unitary transformations in each region,
\begin{eqnarray}
     R_{AB}^{({\rm NL})}(\hat{\rho}_{AB})  \ =\   \min_{{\bm x}, {\bm\theta}_A,  {\bm\theta}_B} 
    \left\{
    \ ||{\bm x}||_1 \ \ \  \bigg\rvert \ \ \ 
     \hat{U}_A({\bm\theta}_A) \hat{U}_B({\bm\theta}_B) \hat{\rho}_{AB} \hat{U}_A^\dagger ({\bm\theta}_A) \hat{U}_B^\dagger ({\bm\theta}_B)  = \sum_{i}
     x_i\hat{\rho}_{s_i}
     \right\} 
     \ .
\label{m_s:eq:NLRoMdef}
\end{eqnarray}
The transformations are constrained to preserve charge in each region
for a gauge-invariant definition of $R_{AB}^{({\rm NL})}$.
In the Schwinger model with the mapping to qubits 
as in Eq.~\eqref{m_s:eq:h_obc}, 
for regions $A$ and $B$ each having support on two qubits, the transformations are each parameterized by four angles 
\begin{eqnarray}
    \hat{U}({\bm \theta}) & \ = \ & e^{-i \left( \theta_0 \hat{Z}\hat{I} + \theta_1 \hat{I}\hat{Z} + \theta_2\hat{Z}\hat{Z} + \theta_3 (\hat{X}\hat{X}+\hat{Y}\hat{Y}) \right)}
    \ \ ,
\label{m_s:eq:QpresTrans}
\end{eqnarray}
as opposed to the 15 angles for an arbitrary SU(4) transformation.

\section{Results obtained with classical simulations}
\label{m_s:sec:magic_schwinger_results}
\begin{figure}
     \centering
     \includegraphics[width=0.6\linewidth]{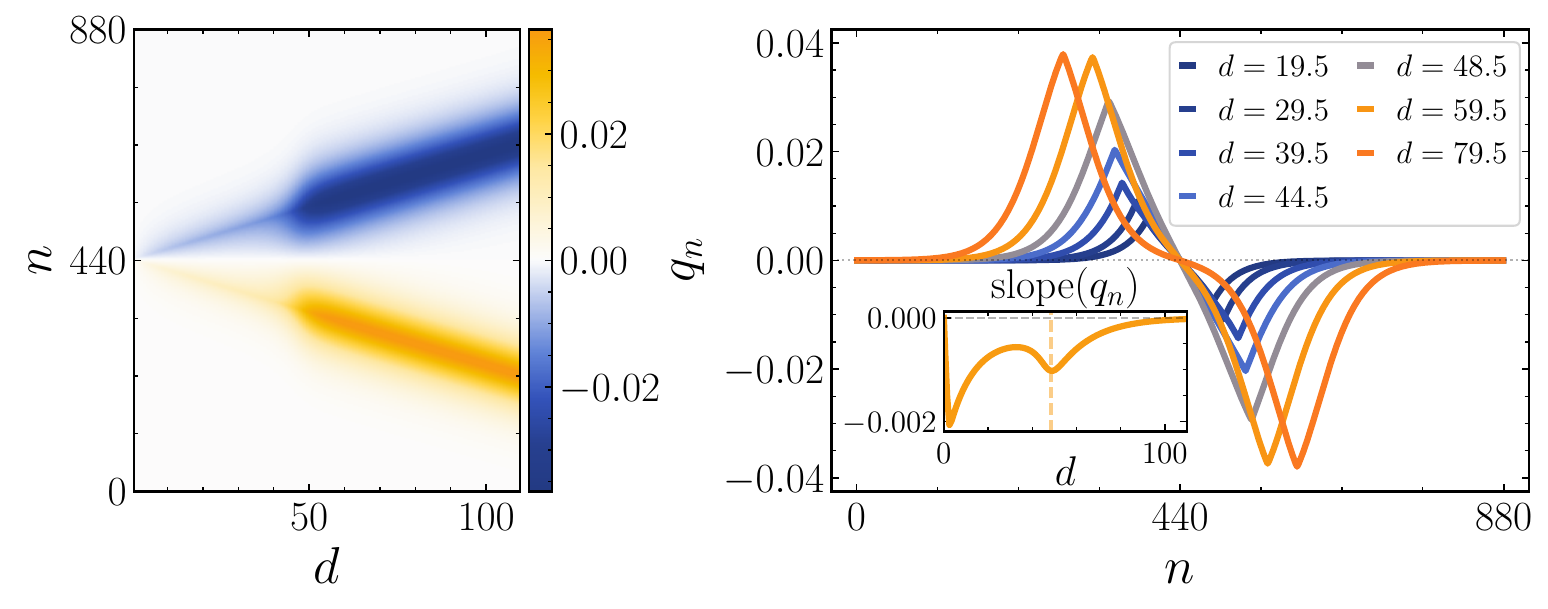} 
     \caption{{\it The charge density $q_n$ through the string breaking process.}
     Left: the vacuum-subtracted $q_n$ obtained with simulation parameters described in the text using $N=880$ and $a=1/4$, as a function of position $n$ and external charge separation $d$. 
     Right: cross sections of $q_n$ for a selection of $d$'s. 
     The inset shows the slope of $q_n$ at the center of the lattice, which has a local extremum at $d=48.5$.
     }
     \label{m_s:fig:chargespatial}
\end{figure}

\noindent 
We first show results obtained for classical local observables, conventionally studied in the context of string breaking.
The left panel of Fig.~\ref{m_s:fig:chargespatial} shows the vacuum-subtracted charge density $q_n = \langle \hat{Q}_n\rangle/a$ as a function of the static charge separation $d$. 
With the simulation parameters described in the previous section, a small range of $d$ captures significant changes in the structure of the ground state wavefunction. 
Dramatic changes in the charge distribution around $d=45-50$ reflect the transition from a single quarkonium-type state with a string between the static charges to a molecular-type state~\cite{Guo:2017jvc} of two mesons. 
As the external charges separate beyond this point, the mesons become isolated and the string between them vanishes.

The charge distribution between the static charges 
(right panel of Fig.~\ref{m_s:fig:chargespatial}) 
is observed to transition from linear to curved with increasing $d$,
consistent with charge localization due to confinement.
The inset shows the profile of the charge distribution.
The width of this peak provides a measure of the separations over which the ground state structure changes from string-dominated to meson-dominated.
The accompanying energy-momentum tensor components, 
chiral condensate and electric field 
are discussed in App.~\ref{m_s:app:classical}. 
These classical quantities have been well-studied in the context of string breaking (e.g.,~\cite{Buyens:2015tea,Grieninger:2025rdi}) and our results agree with existing literature.

Quantum correlations across a bipartition of the system at the string center constitute an informative class of complexity measures.
The entanglement entropy of such bipartitions has been studied previously (e.g., Refs.~\cite{Grieninger:2025rdi,Florio:2025hoc}).
For this half-lattice bipartition, 
we examine the (gauge-invariant) entanglement entropy using Eq.~\eqref{m_s:eq:SEQ}, the lower bound of the nonlocal magic using the antiflatness (Eq.~\eqref{m_s:eq:AFQ})
and the upper bound of the nonlocal magic using Eq.~\eqref{m_s:eq:Tara}.
\begin{figure}
     \centering
     \includegraphics[width=0.6\linewidth]{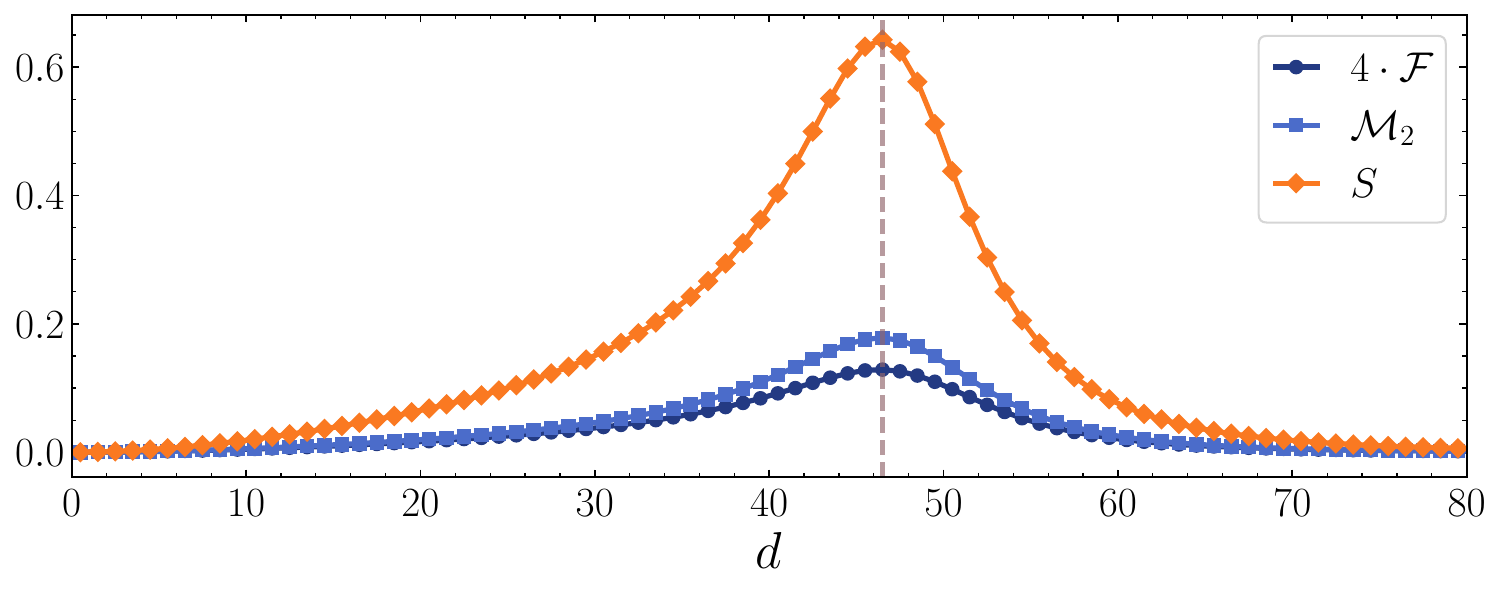} 
     \caption{{\it Bipartite measures of entanglement and quantum complexity in string breaking.}
     The symmetric bipartition entanglement entropy, antiflatness and the upper bound on the nonlocal ${\cal M}_2$ using simulation parameters $N=880, a=1/4, g=0.09, m=0.04601$, as a function of separation between static sources in terms of physical spatial sites.
     These quantities all peak at $d=46.5$ (see Table~\ref{m_s:tab:peaks}).
     }
     \label{m_s:fig:EEandAF}
\end{figure}
Figure~\ref{m_s:fig:EEandAF} shows these quantities as a function of $d$.  
The previously identified entanglement entropy peak in the vicinity of string breaking is also observed in both the lower and upper bounds of nonlocal magic. 
This result is consistent with the nonlocal magic being driven by entanglement between the regions.\footnote{
The bounds provide all the accessible information about bipartition nonlocal magic because the required minimizations are impractical for the lattice sizes considered in this work.}

We conclude that the nonlocal magic increases during string breaking before returning to its vacuum value as the mesons separate.
This is typical of a transition from long-range to short-range correlations.
Moreover, this is consistent with the bipartition surface being (exponentially) insensitive to the presence of one or more mesons located in the bulk of the lattice, far from boundaries.
This can be considered a magic barrier, analogous to those observed in the time evolution of nonequilibrium systems~\cite{Ebner:2025pdm}.

We probe quasi-local quantum correlations with the $n$-tangle $\tau^{(n)}$~\cite{PhysRevA.63.044301}
\begin{eqnarray}
    \tau^{(n)}_{(i_1 ... i_n)} \ & = & \ |\langle \psi | 
    \hat Y_{i_1}\otimes \dots \otimes \hat Y_{i_n}
    | \psi^* \rangle|^2 \ ,
\label{m_s:eq:n-tangle}
\end{eqnarray}
where $\hat{Y}_{i_k}$ is the Pauli matrix acting on qubit $i_k$. 
Specifically, the 2-tangle $\tau^{(2)}_{(i,i+1)}$ provides a measure of nearest-neighbor entanglement.
\begin{figure}
     \centering
   \includegraphics[width=0.6\linewidth]{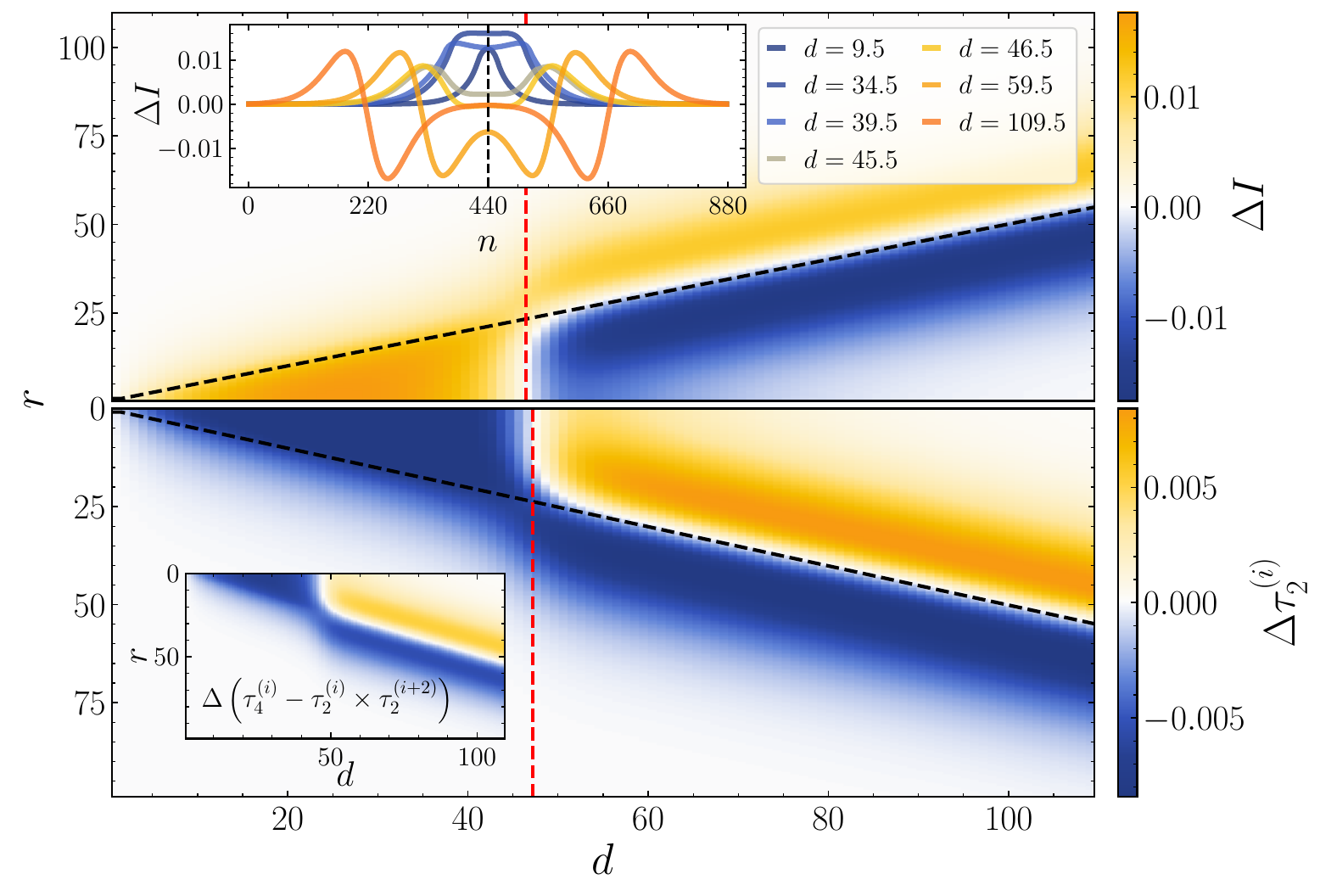}  
     \caption{{\it Local and multipartite entanglement in string breaking.} 
     Top: the local vacuum-subtracted MI as a function of distance from the center $r$ and external charge separation $d$ for $N=880, a=1/4, g=0.09, m=0.04601$.
     The inset shows cross sections of the MI for a selection of $d$'s.
     Bottom: the vacuum-subtracted 2-tangle for the same parameters. 
     The inset on the bottom panel shows the vacuum-subtracted 4-tangle. The black dashed lines indicate the peak of the charge distribution shown in Fig.~\ref{m_s:fig:chargespatial}. The red dashed lines show $d$ where the respective quantity changes sign for the smallest $r$.
}
     \label{m_s:fig:tau2}
\end{figure}
The lower panel of Fig.~\ref{m_s:fig:tau2} shows the vacuum-subtracted $\tau^{(2)}_{(i,i+1)}$ 
across the lattice as a function of $d$.
The nearest-neighbor entanglement is modified along the length of the string, which rapidly vanishes, changes sign in the vicinity of string breaking, then becomes confined into the mesons as they become isolated.
This behavior is also evident in the vacuum-subtracted 4-tangle, 
$\tau^{(4)}_{(i,i+1,i+2,i+3)} - \tau^{(2)}_{(i,i+1)} \tau^{(2)}_{(i+2,i+3)}$, 
(inset of lower panel in Fig.~\ref{m_s:fig:tau2})
and in the vacuum-subtracted MI 
$\Delta I$,
(upper panel of Fig.~\ref{m_s:fig:tau2}),
both between adjacent spatial sites. 
The shape looks similar to $\tau^{(2)}_{(i,i+1)}$,
indicating that much of the information is contained in the 2-tangle.\footnote{
We find the quantity 
${\tau^{(4)}_{(i,i+1,i+2,i+3)} -c\, \tau^{(2)}_{(i,i+1)} \tau^{(2)}_{(i+2,i+3)}\rightarrow 0}$ 
for $c\approx 1.78$. 
Similarly, the nonlocal 4-tangle ${\tau^{(4)}_{(N/2,N/2+1,i,i+1)} - \tau^{(2)}_{(N/2,N/2+1)}\tau^{(2)}_{(i,i+1)} }\to 0$ for $i$ away from the center.}
This suggests that (local) entanglement structure is limited to physical sites.
There is a separation for which $\tau^{(2)}_{(i,i+1)}$ vanishes along much of the string (indicated by the red dashed line), coinciding with the point of string breaking.
Notably, the zero that develops in $\tau^{(2)}$ and $\Delta I$ tracks the maximum of the charge distribution (shown with the black dashed line) after the string breaking. 
Deviations of $\Delta I$ from 0 show the localization of correlations as a function of $d$, reflecting the underlying structure of string breaking and meson formation.
The $\Delta I$ in the central region after string breaking arises from the nuclear force between the mesons, which vanishes when the mesons are sufficiently separated. 
It is interesting to note, that both the maximum and minimum of $\Delta I$ are within the width of the meson defined by the charge profile peak (Fig.~\ref{m_s:fig:chargespatial}).

We use the RoM, defined in Eq.~(\ref{m_s:eq:RoMdef}), as a measure of both local and nonlocal magic.
\begin{figure}
     \centering
     \includegraphics[width=0.6\linewidth]{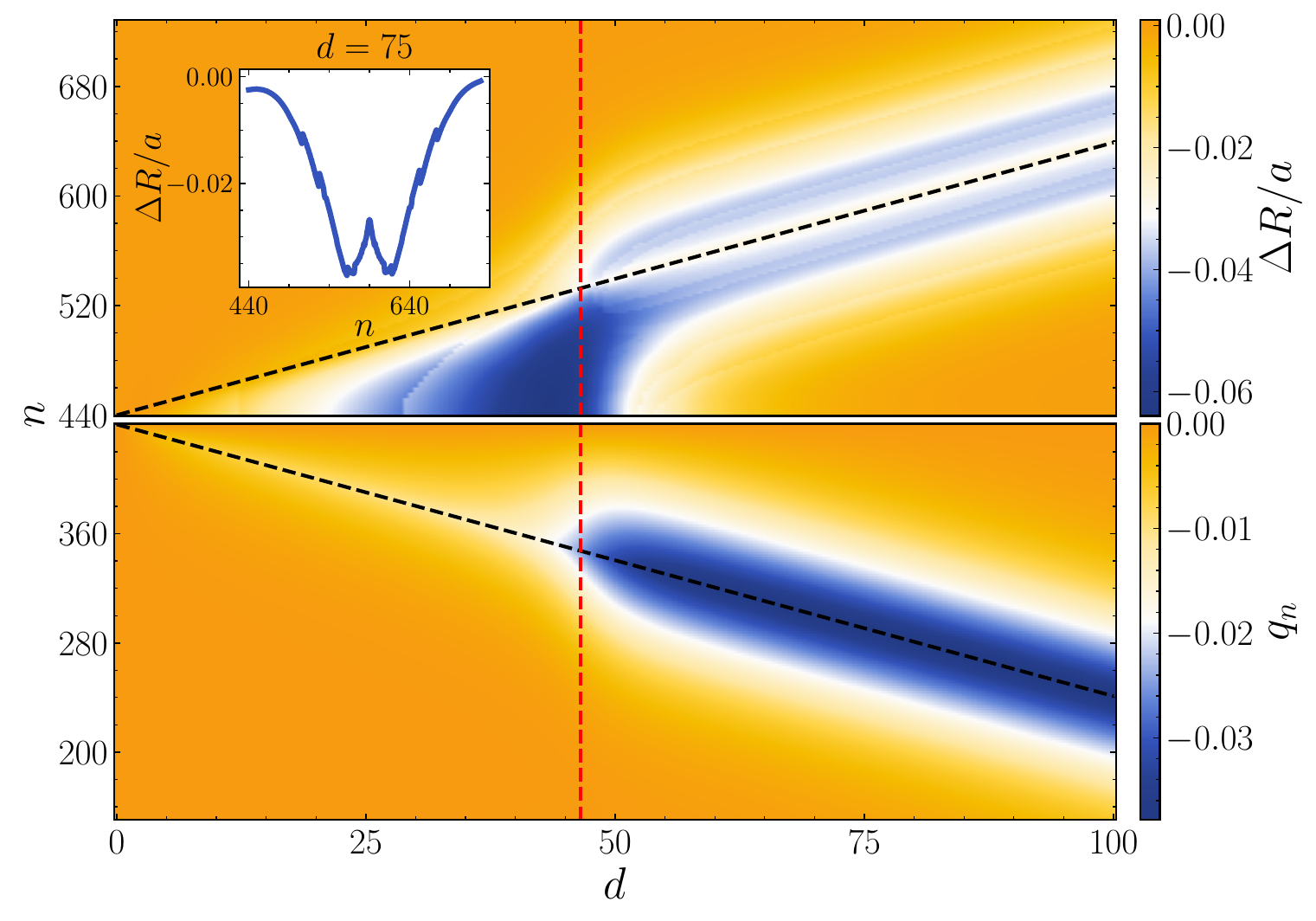}
     \caption{{\it Internal structure of the outgoing meson revealed by the RoM.} 
     Top: the vacuum-subtracted RoM of adjacent sites, $\Delta R$, in units of $a$, shown
      for $n \geq 440$ as a function of $d$, for $N=880,a=1/4,g=0.09, m=0.04601$.
     The inset shows $\Delta R$ inside the meson region for $d=75$.
     Bottom: the charge density $q_n$ for $n\leq440$ for the same system.
     The black dashed lines show the external charge positions, and the red dashed lines mark the peaks of the bipartite complexity measures shown in Fig.~\ref{m_s:fig:EEandAF}.}
     \label{m_s:fig:ROMadj}
\end{figure}
Figure~\ref{m_s:fig:ROMadj} shows the vacuum-subtracted RoM,
$\Delta R$, defined via Eq.~(\ref{m_s:eq:RoMdef}) as
\begin{eqnarray}
    \Delta R(i) \ & = & \ R(i:i+3) - R_{\Omega}(i:i+3)
    \ ,
    \label{m_s:eq:R4}
\end{eqnarray}
of four adjacent lattice sites, $i, \dots,  i+3$,
as a function of position and the separation of the external charges.
Here $R_{\Omega}$ denotes the value at $d=0$.
With increasing $d$, the deviation of $\Delta R$ from zero increases along the string, 
becoming maximal at the point where the string breaks, then decreasing rapidly as the mesons separate.
The (local) RoM reveals structure in the string state before it breaks, and in the outgoing mesons.

The inset of Fig.~\ref{m_s:fig:ROMadj} shows $\Delta R$ within the meson at $d=75$. 
This suggests that moments of RoM distributions in hadronic structure calculations could be complementary probes of structure, in the same way that moments of the charge or axial distributions are computed (and measured experimentally). 
This shares similarities with the behavior of $\Delta I$ within the meson (inset of top panel in Fig.~\ref{m_s:fig:tau2}).
The dependence of $\Delta R$ on lattice spacing and $m/g$ is discussed in App.~\ref{m_s:app:supplemental_results}.

A novel class of quantum correlations in the context of string breaking are those between 
spatially separated subregions~\cite{Qian:2025oit}.
The mutual information $\Delta I(A:B)$ provides a measure of both classical and quantum correlations between regions $A$ and $B$.
\begin{figure}
     \centering
     \includegraphics[width=0.6\linewidth]{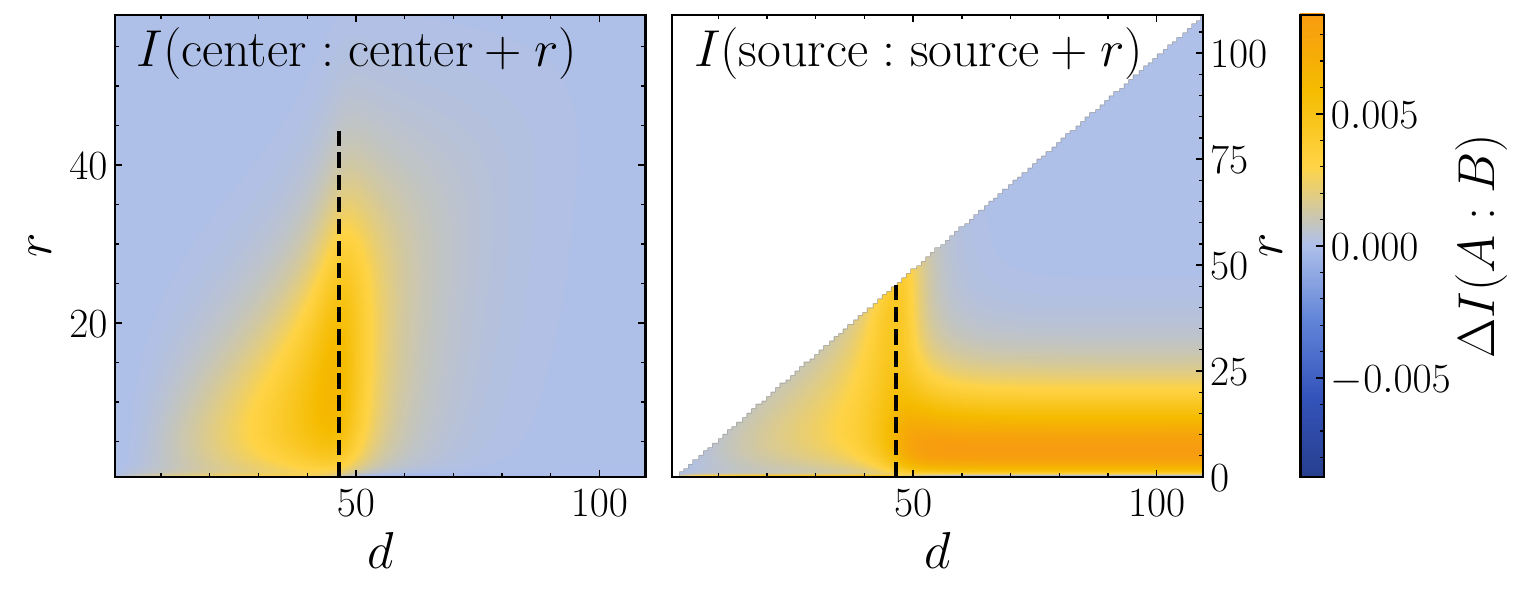} 
     \caption{{\it Mutual information in string breaking.} 
     Left: The vacuum-subtracted MI,
      $\Delta I(A:B)$, where region $A$ is at the center of the lattice, $A=\left(\frac{N}{2}+1,\frac{N}{2}+2\right)$ and $B$ is a distance $r$ away, $B=A+(r,r+1)$.
      The parameters $N=880,a=1/4,g=0.09, m=0.04601$ are used.
      Right: $\Delta I(A:B)$ where $A$ is at the left external charge, $A=\left(\frac{N}{2}-\frac{d}{2a}-1,\frac{N}{2}-\frac{d}{2a}\right)$, and $B$ is a distance $r$ away, $B=A+(r,r+1)$.
      The dashed lines mark the peak of the bipartite complexity measures shown in Fig.~\ref{m_s:fig:EEandAF}.}
\label{m_s:fig:MI}
\end{figure}
The left panel of Fig.~\ref{m_s:fig:MI} shows  $\Delta I(A:B)$ between spatial sites along the string with respect to the center of the system.
The MI 
is continuous along the string, peaking at the point of string breaking, and rapidly returning to  vacuum values thereafter.
Perhaps it is not surprising that the string supports long-distance correlations (classical or quantum) along the length of the string, which vanish as the system transitions to isolated mesons.
However, the transition region extends (at least)
over the range of the nuclear force between the mesons.\footnote{
The behavior of $\Delta I$ and $\Delta R^{\text{(NL)}}_{AB}$ for a system with unnatural scattering parameters may determine whether the string-breaking scale, the nuclear force or the scattering parameters drive changes in quantum complexity.}
The right panel of Fig.~\ref{m_s:fig:MI} shows $\Delta I$ over the interval between the static charges, where region $A$ is fixed at the position of the left charge. 
Its behavior along the string is analogous to the left panel, and it is nonzero over the extent of the meson for larger $d$.

\begin{figure}
     \centering
     \includegraphics[width=0.6\linewidth]{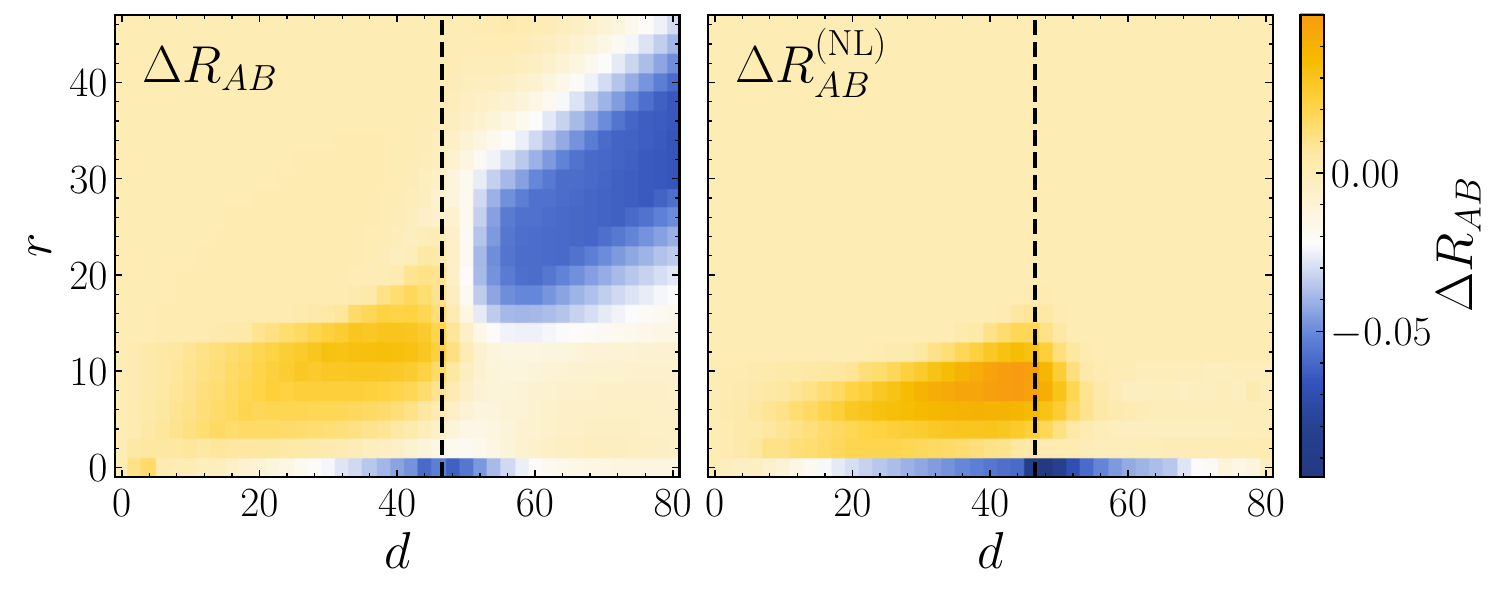}
     \caption{{\it Nonlocal quantum correlations in string breaking as measured by the NL RoM.}
     Left: the RoM of disjoint regions, $\Delta R_{AB}$, where region $A$ is at the center of the lattice and region $B$ is a distance $r$ away. 
     The vacuum value ($d=0$) and the values as $r\to\infty$ are subtracted.
     The system parameters $N=220,a=1,m_\text{lat}=0.045,g=0.09$ are used.
     Right: the NL RoM, $\Delta R^\text{(NL)}_{AB}$ for the same system parameters.
     The dashed lines mark the peak of the bipartite complexity measures shown in Fig.~\ref{m_s:fig:EEandAF}.
     }
     \label{m_s:fig:RoM}
\end{figure}
The left panel of Fig.~\ref{m_s:fig:RoM} shows the RoM $\Delta R_{AB}$ between disjoint subsystems on the lattice as a function of the distance between the regions $r$.
As in the left panel of Fig.~\ref{m_s:fig:MI}, region $A$ is at the center of the lattice, ${A=(N/2+1,N/2+2)}$, and region $B$ is distance $r$ away, 
${B=A+(r,r+1)}$.
We define this quantity with a different vacuum subtraction to that in Eq.~(\ref{m_s:eq:R4}),
\begin{eqnarray}
    \Delta R_{AB} \ & = & \ R(A,B) - R_{\infty}(A,B) - R_{\Omega}(A,B)
    \ ,
    \label{m_s:eq:R4AB}
\end{eqnarray}
where $R_{\infty}$ 
denotes the quantity evaluated as $r\rightarrow\infty$ 
and $R_{\Omega}$ denotes the value at $d=0$ (as before).
The reason for this is that the wavefunction is modified at the center of the lattice 
before string breaking, providing a modification in $R$ for all values of $r$.
Unlike $\Delta I$, $\Delta R_{AB}$ exhibits a localized rapid change of sign at the point of string breaking.
This is consistent with the center being the location of the most ``restructuring'' of the wavefunction.
This increase in charge density seen in Fig.~\ref{m_s:fig:chargespatial} 
coincides with the ``hot spot'' in the left panel of Fig.~\ref{m_s:fig:RoM}.
The behavior of $\Delta R_{AB}$ in the transition region immediately after string breaking may reveal new insights into the strong nuclear force.
Interestingly, the NL RoM (defined in Eq.~\eqref{m_s:eq:NLRoMdef}) shows similar, but more localized behavior to the MI (cf. Fig.~\ref{m_s:fig:MI}). 
The outgoing meson ``track'' vanishes in the NL components of the RoM (right panel of Fig.~\ref{m_s:fig:RoM}), suggesting that  NL quantum information is only encoded in the string state.
The presence of NL RoM establishes a link between spatially extended physical objects (strings) and nonlocal quantum complexity independent of basis choice.
These purely quantum correlations, 
absent from traditional models of hadronization, 
could imprint themselves into final states in high-energy collisions.

\section{Summary and discussion}
\label{m_s:sec:Conc}
\noindent
Continuing along the path of scientific discovery in high-energy and nuclear physics, techniques from quantum information science are now being used to understand confinement and develop predictive capabilities for the resulting dynamical phenomena.
We have performed classical simulations of string formation and breaking in the (Abelian) Schwinger model, finding that quantum complexity varies rapidly during string breaking, and provides distinct probes complementary to classical quantities like energy density.
Importantly, nonlocal 
quantum complexity (as measured by both the MI and the NL RoM) is found along the string.
This shows that both classical and quantum correlations develop during string breaking, and suggests that the longer distance 
correlations are predominately classical.
Our nonlocal complexity computations are limited to small systems due to the resources required for subregion minimizations and stabilizer polytope considerations.
For these reasons, we have not computed higher-body nonlocal magic measures and cannot rule them out.

The quantum complexity exhibits sharp changes in the transition from flux tubes to isolated hadronic bound states, akin to phase transitions~\cite{Grieninger:2025rdi}. 
Dynamical simulations of string breaking, via wavepacket evolution~\cite{Farrell:2024fit,Zemlevskiy:2024vxt,Farrell:2025nkx} or time-dependent Hamiltonians~\cite{Farrell:2024mgu,Li:2025sgo}, would clarify mechanisms of charge extraction and confinement that are difficult to investigate with statics alone. 
Applying our techniques to non-Abelian and higher-dimensional systems will be illuminating.
We also anticipate that the quantum complexity in string breaking will be imprinted in final states of 
collisions of hadrons and nuclei.  
Analyses quantifying entanglement and magic in top-anti-top production events at the LHC are already underway~\cite{White:2024nuc,Yazgan:2025pah}.
If detected in experiment, the hierarchies of these correlations and their (non)locality may affect modeling of fragmentation and hadronization.

\clearpage

\begin{subappendices}

\section{Analysis of lattice spacing artifacts}
\label{m_s:app:lattice_artifacts}
\begin{figure}[ht!]
    \centering
     \includegraphics[width=1.0\linewidth]{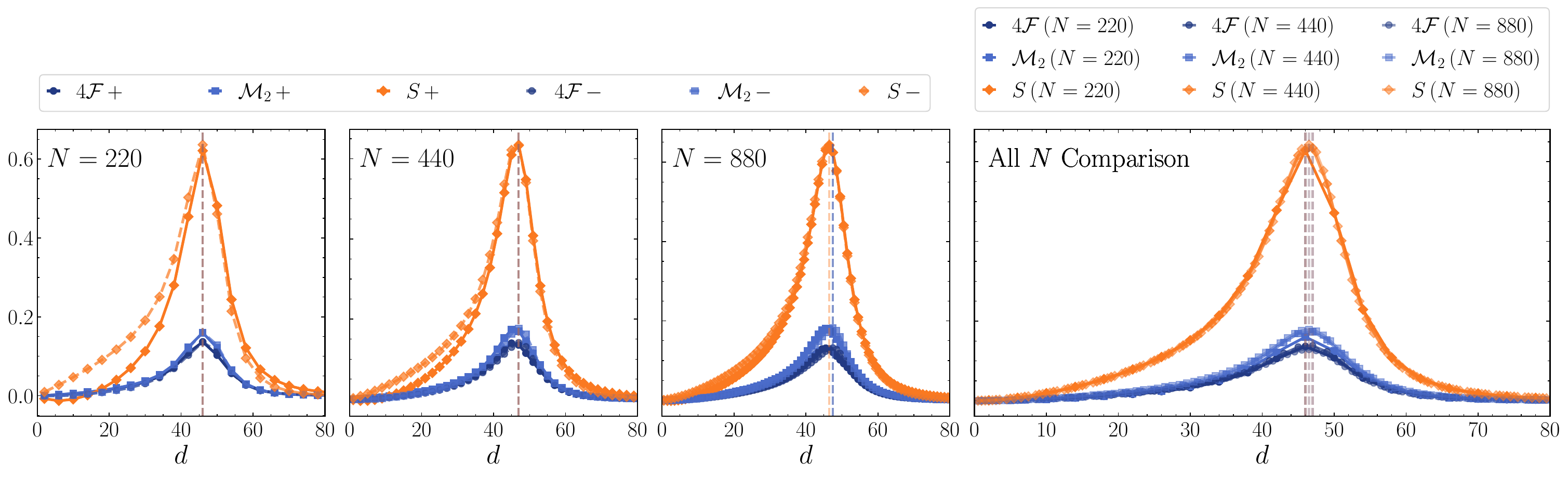}
     \caption{Bipartite measures of entanglement and quantum complexity as a function of $N,a$ for $(N,a)=(220,1), (440,1/2)$, and $(880,1/4)$.
     The antiflatness $4{\cal F}$, the upper bound to the nonlocal magic ${\cal M}_2$, and the entanglement entropy $S$ are shown for both orientations of the external field $E\text{ext}=-$ and $E_\text{ext}=+$.
     The right panel shows the average over the orientations of $E_\text{ext}$ for all $N,a$.
     The locations of the peaks for each measure are given in Table~\ref{m_s:tab:peaks}.}
     \label{m_s:fig:quantsN}
\end{figure}

\noindent
On a staggered lattice, the electric field points from antifermion to fermion sites. For our choice of parameters, odd sites correspond to fermions and even sites to antifermions. Hence, for $N/2$ even the middle link is between an antifermion and fermion site and the external electric field should point to the right (i.e., larger indices in the chain). This corresponds to our $d=a$ configuration. For $d=3a$, the opposite configuration is supported. Note that both the $d=a$ and $d=3a$ configuration belong to the same physical site. This implies that we should average over neighboring configurations to obtain physical sites, i.e., we average $d=a$ and $d=3a$, $d=5a$ and $d=7a$, etc. The averaged configurations then correspond to physical separations of $d=2a, 6a,$ etc. In the continuum, there is no distinction between fermion and antifermion sites and the results obtained for either direction of the electric field should be equal. Hence, a good proxy for the impact of lattice artifacts is to compare the results obtained for both directions of the external electric field. This is exemplified in Fig~\ref{m_s:fig:quantsN}. In the first three panels, 
we keep $L=N a$ fixed and decrease the lattice spacing from $a=1$ to $a=1/4$. For $a=1$ there is a clear difference in the entanglement entropy for the two different directions of the $E_\text{ext}$. In fact, the $S_+$ curve even turns negative at small $d$. The two curves converge in the $a\to 0$ limit. To account for the difference at finite lattice spacing, we average $S,\, \cal{F}$ and ${\cal M}_2$ over the two different directions of the electric field as shown in the fourth plot in Fig~\ref{m_s:fig:quantsN}.

\begin{table}[h]
\footnotesize
\setlength{\tabcolsep}{1pt} 
\begin{tabularx}{\linewidth}{|c||Y|Y|Y||Y|Y|Y||Y|Y|Y|}
\hline
\rule{0pt}{10pt} & \multicolumn{9}{c|}{ Peak position}\\\hline\hline
\rule{0pt}{10pt} \multirow{2}{*}{Measure} &  \multicolumn{3}{c||}{$N=220,a=1$}  &  \multicolumn{3}{c||}{$N=440,a=1/2$} & \multicolumn{3}{c|}{$N=880,a=1/4$}  \\\cline{2-10}
& $E_\text{ext}=-$ &  $E_\text{ext}=+$ & avg. & $E_\text{ext}=-$ & $E_\text{ext}=+$ & avg. & $E_\text{ext}=-$ & $E_\text{ext}=+$ & avg. \\
\hline\hline
${\cal F}$ & 46.0 & 46.0 & 46.0  & 47.0 & 45.5 & 46.25 & 45.5 & 47.5 & 46.5 \\\hline
${\cal M}_2$ & 46.0 & 46.0 & 46.0 &  47.0 & 45.0 & 46.0 & 45.5 & 47.5 & 46.5 \\\hline
$S$ & 46.0 & 46.0 & 46.0 & 47.0 & 47.0 & 47.0 & 46.5 & 46.5 & 46.5 \\
 \hline
\end{tabularx}
\caption{The values of $d$ for which the measures (left column) reach their maximum value, for parameters in the main text and both orientations of the external field $E_\text{ext}=-$ and $E_\text{ext}=+$, and the average of the two orientations.}
\label{m_s:tab:peaks}
\end{table}

\section{OBC vs PBC}
\label{m_s:app:obc_vs_pbc}
\begin{figure}[ht!]
    \centering
     \includegraphics[width=0.75\linewidth]{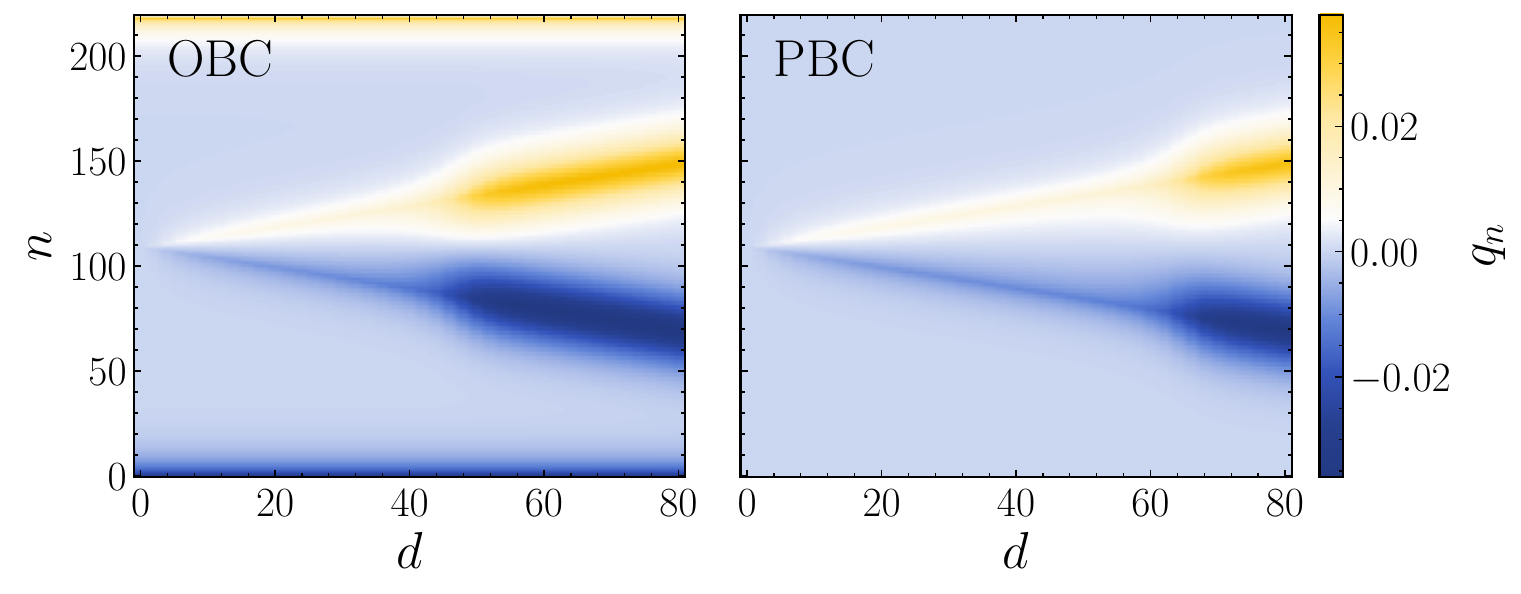}
     \caption{The charge density for OBC (left) and PBC (right) $N=220,a=1,m_\text{lat}=0.045,g=0.09$.}
     \label{m_s:fig:obc_vs_pbc}
\end{figure}

\noindent
With the parameters chosen as described in the main text, the effect of boundaries and finite volume on the physics of the simulation is exponentially suppressed.
As a result, the choice of boundary conditions should not impact the results of the computations.\footnote{OBCs are highly preferred due to the inefficiency of standard MPS in representing periodic states~\cite{PhysRevLett.93.227205}.} 
With OBCs, dynamical gauge field degrees of freedom can be eliminated from the theory, resulting in long range fermionic interactions through a linear potential. 
In PBCs, there is a single gauge field degree of freedom, the ``zero mode'', whose dynamics is not constrained by Gauss's law.
We follow the approach of Refs.~\cite{Farrell:2025nkx,Zache:2018cqq,Zache:2020qny}, where the gauge field zero mode is truncated to a single state and zero mode dynamics is not considered.
This approximation is valid for low-energy dynamics, and corrections are suppressed by $L$.
The Hamiltonian is modified from Eq.~\eqref{m_s:eq:h_obc}:
\begin{align}
    \hat{H}_\text{PBC}\,(d) \ &= \ \frac{m_\text{lat}}{2}\sum_{n=1}^{N}(-1)^n \hat{Z}_n \nonumber \\
&\quad- \frac{g^2 a}{2}\sum_{n=1}^{N}\left\{\sum_{s=1}^{N_\text{phys}}
\left(s - \frac{s^2}{N}\right)
\left(1 - \frac{\delta_{s,N_\text{phys}}}{2}\right)
\hat{Q}_n\hat{Q}_{n+s} + \frac{2}{g}\sum_{k=1}^{N-1}\frac{N-k}{N}E_{\text{ext},n}(d)\hat{Q}_{n+k}\right\} \nonumber \\
&\quad
+ \frac{1}{4a}\sum_{n=1}^{N-1}
\left(\hat{X}_n \hat{X}_{n+1} + \hat{Y}_n \hat{Y}_{n+1}\right)
+ \frac{1}{4a}(-1)^{N_\text{phys}+1}
\left(\hat{X}_{N} \hat{X}_{1} + \hat{Y}_{N} \hat{Y}_{1}\right) \ .
\end{align}
With the lattice parameters chosen as in the main text, we observe a difference in the string breaking process between PBCs and OBCs with the same system size $N$.
As shown in Fig.~\ref{m_s:fig:obc_vs_pbc}, the string breaks at a larger separation $d$ for PBCs. 
This is attributed to image charge effects in PBCs that are absent in OBCs,\footnote{This is verified by considering a larger lattice with the same lattice spacing, where the image charges are further away and the string breaks at an $d$ closer to OBCs.} indicating that these calculations are not far into regime prescribed in the main texts.
The boundary effects, both from the lattice boundary in OBCs and image charges in PBCs, extend into the lattice volume by a length scale set by the lightest hadron mass.
As expected, the charge density of the OBC and PBC vacua determined by DMRG agree in the bulk.

\section{Charge rearrangement during string breaking}
\label{m_s:app:charge_rearrangement}
\noindent
During string breaking, charge is extracted from the vacuum to screen the heavy charges, so it is natural to study the rearrangement of charge. 
Considering a bipartition of the system at the center of the lattice, the ground state wavefunction may be written as in Eq.~\eqref{m_s:eq:ABQ}.
The charge rearrangement may be examined through the effect that screening has on the sector weights $p_Q$.
At $d=0$, the ground state is the lowest-lying state with total $Q=0$, so that only charge-neutral terms appear in Eq.~\eqref{m_s:eq:ABQ}.
Introducing background charges causes the weights $p_Q$ to change, as the vacuum rearranges to the new lowest-energy state.
When the background charges are far apart, effects near the bipartition are exponentially suppressed by confinement, so only the light degrees of freedom that do not participate in screening contribute to $p_Q$.
The half-lattice charge changes by one. 
In the following, $|Q^+\rangle$ denotes an external charge of $+1$, and $|q\rangle$ denotes the (reduced) state of the half-lattice. The changes for that the lowest several charge sectors experience are
\begin{eqnarray}
|Q^+\rangle\otimes |q=0\rangle &\rightarrow & |Q^+e^-\rangle \otimes |q=+1\rangle
 \ , \nonumber\\
|Q^+\rangle\otimes |q=+1\rangle &\rightarrow & |Q^+e^-\rangle \otimes |q=+2\rangle
 \ , \nonumber\\
|Q^+\rangle\otimes |q=-1\rangle &\rightarrow & |Q^+e^-\rangle \otimes |q=0\rangle
 \ ,  \label{m_s:eq:charge_transitions}
\\
 |Q^+\rangle\otimes |q=+2\rangle &\rightarrow & |Q^+e^-\rangle \otimes |q=+3\rangle
 \ , \nonumber\\
|Q^+\rangle\otimes |q=-2\rangle &\rightarrow & |Q^+e^-\rangle \otimes |q=-1\rangle
 \ , \nonumber
\end{eqnarray}
and similarly for the $|Q^-\rangle$ region.
In other words, for the half of the lattice with the $|Q^+\rangle$ external charge, ${p_{q=0}\to p_{q=+1}}$, ${p_{q=+1}\to p_{q=+2}}$, etc.

\begin{figure}[ht!]
     \centering
     \includegraphics[width=0.75\linewidth]{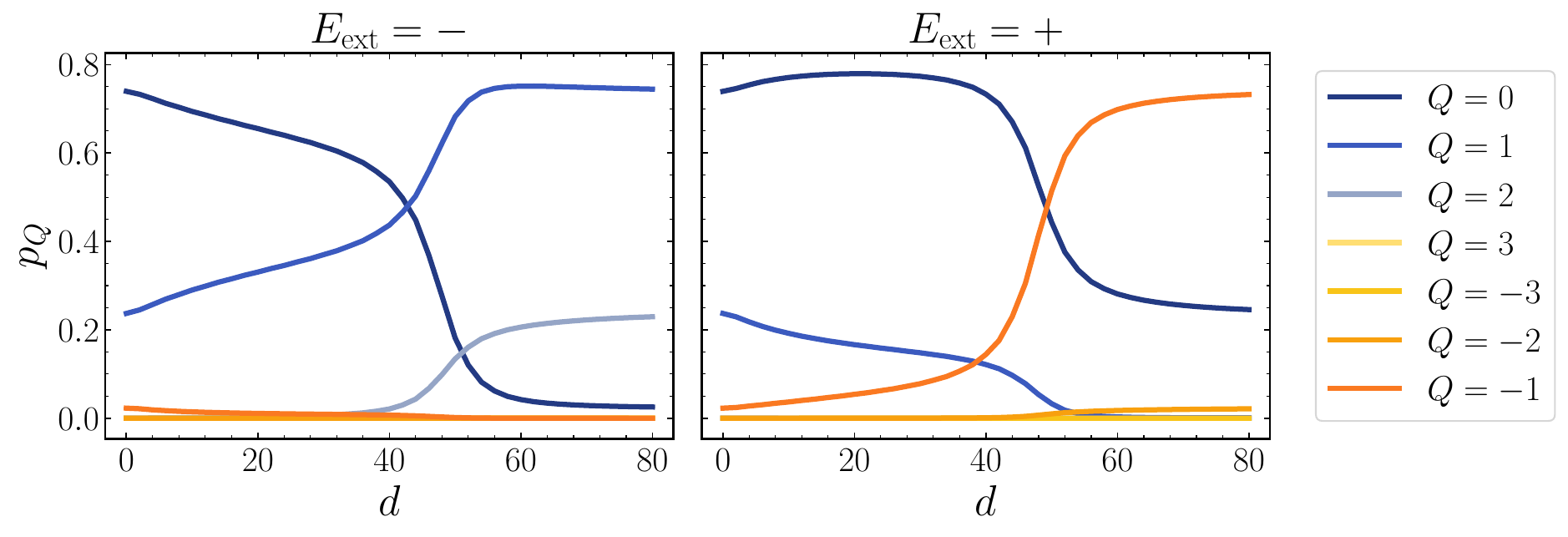}
     \caption{The charge sector weights $p_Q$ as a function of external charge separation $d$ for the two different orientations of the external field, $E_\text{ext}=-$ and $E_\text{ext}=+$. The system parameters $N=220,a=1,m_\text{lat}=0.045,g=0.09$ are used.}
     \label{m_s:fig:charge_spectrum}
\end{figure}

In our formalism, string breaking may be observed in two equivalent ways, with the external field pointing toward $n=0$, $E_\text{ext}=-$, and toward $n=N-1$, $E_\text{ext}=+$.\footnote{The configuration $E_\text{ext}=-$ corresponds to an external fermion on the left half of the lattice, and an antifermion on the right, and vice-versa for $E_\text{ext}=+$.}
Fig.~\ref{m_s:fig:charge_spectrum} shows the rearrangement of $p_Q$ for the two choices of $E_\text{ext}$.

Physical quantities should not depend on the choice of $E_\text{ext}$ orientation. 
However, we observe that the $p_Q$ transitions do not follow Eq.~\eqref{m_s:eq:charge_transitions} exactly, and behave differently for the different orientations of the electric field.
OBCs generate a charge asymmetry along the lattice from the ordering of fermions and antifermions (seen Fig.~\ref{m_s:fig:obc_vs_pbc}), created by configurations localized around the boundaries.
These boundary charge densities present a challenge for interpreting global measures of entanglement and quantum complexity, that have sensitivity to the entire lattice or half-lattices.
This effect is responsible for the differences with Eq.~\eqref{m_s:eq:charge_transitions} and the asymmetry of the left and right panels of Fig.~\ref{m_s:fig:charge_spectrum}.

Attempts to compensate for boundary effects by inserting image charges did not reduce the asymmetry between the two panels of Fig.~\ref{m_s:fig:charge_spectrum}.
Switching to PBCs as described in App.~\ref{m_s:app:obc_vs_pbc} introduces a second entangling surface, and makes Eq.~\eqref{m_s:eq:charge_transitions} superficially true due to symmetry. 
Considering a subregion of the bipartition that does not include the external boundary has similar issues. 
As seen in the left panels of Fig.~\ref{m_s:fig:quantsN} and Table~\ref{m_s:tab:peaks}, these effects do not go away with reduced lattice spacing, as naively expected. 
For this reason, it is important to consider measures of quantum complexity that are insensitive to boundary effects.

\section{Additional classical observables}
\label{m_s:app:classical}
\begin{figure}[ht!]
     \centering
     \includegraphics[width=0.6\linewidth]{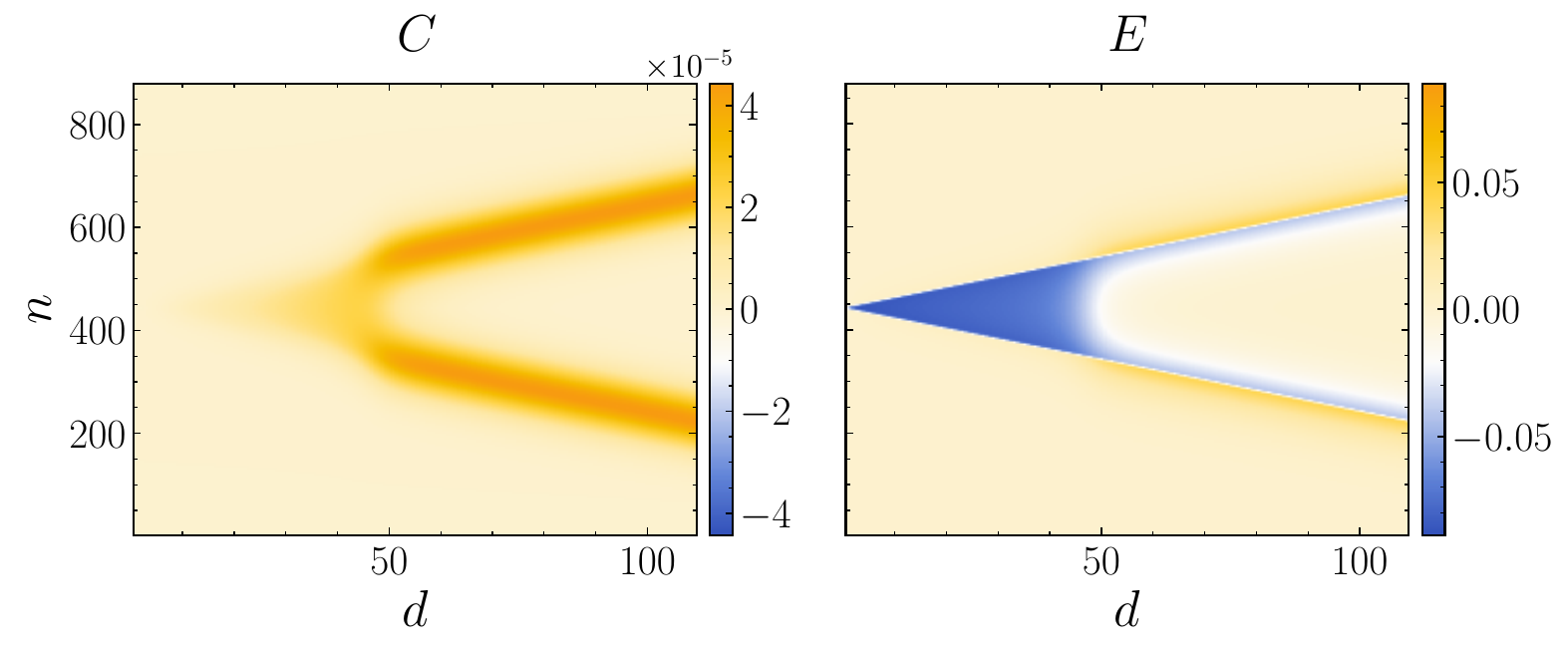}      
     \caption{The vacuum-subtracted chiral condensate (left panel) and electric field (right panel)
     obtained with simulation parameters described in the text using $N=880$ and $a=1/4$.
}
     \label{m_s:fig:cond}
\end{figure}

\noindent
Figure~\ref{m_s:fig:cond} shows the chiral condensate and the electric field as functions of position $n$ and separation of external charges $d$. The chiral condensate $C$ is defined as
\begin{align}
    C_n \ &= \ \frac{(-1)^n}{2a}\langle Z_n\rangle  \ ,
\end{align}
and the electric field $E_n$ is given in Eq.~\eqref{m_s:eq:e_field_charge}.
The relevant components of the energy-momentum tensor are given by~\cite{Grieninger:2025mbm}
\begin{align}
    \hat T^{00}_n\ &=\
      \frac{1}{2a}\left( \hat K_n+\hat K_{n-1}+m(-1)^n \hat Z_n 
     \,+ \frac{a}{2}\left(\hat E_{\text{tot},n}^2+\hat E_{\text{tot},n-1}^2\right) \right),\\
         \hat T^{11}_n\ &=\
      \frac{1}{2a}\left( \hat K_n+\hat K_{n-1}- \frac{a}{2}\left(\hat E_{\text{tot},n}^2+\hat E_{\text{tot},n-1}^2\right) \right),
\end{align}
Here, $\hat K_n=\frac{1}{4a} (\hat X_{n+1}\hat X_n+\hat Y_{n+1}\hat Y_n)$ is the kinetic term, $\hat E_{\text{tot},n}=\hat E_n+E_{\text{ext}, n}$ and all indices that are smaller than 1 or larger than $N$ are zero with open boundary conditions. 
The third component $\hat T^{01}_n$ vanishes in the ground state.
The energy density $\varepsilon$ and the pressure p are related to the energy-momentum tensor components by
\begin{align}
\hat\varepsilon_n\ &=\ \frac{1}{2}\left(\hat T^{00}_n -\hat  T^{11}_n \pm \sqrt{(\hat T^{00}_n + \hat T^{11}_n)^2 - 4(\hat T^{01}_n)^2}\right), \\
\hat p_n\ &=\ \hat \varepsilon - (\hat T^{00}_n - \hat T^{11}_n),
\end{align}
where the positive sign is chosen for $\hat T^{00}_n +\hat  T^{11}_n\ge0$ and the negative sign otherwise.

\begin{figure}[ht!]
     \centering
     \includegraphics[width=0.6\linewidth]{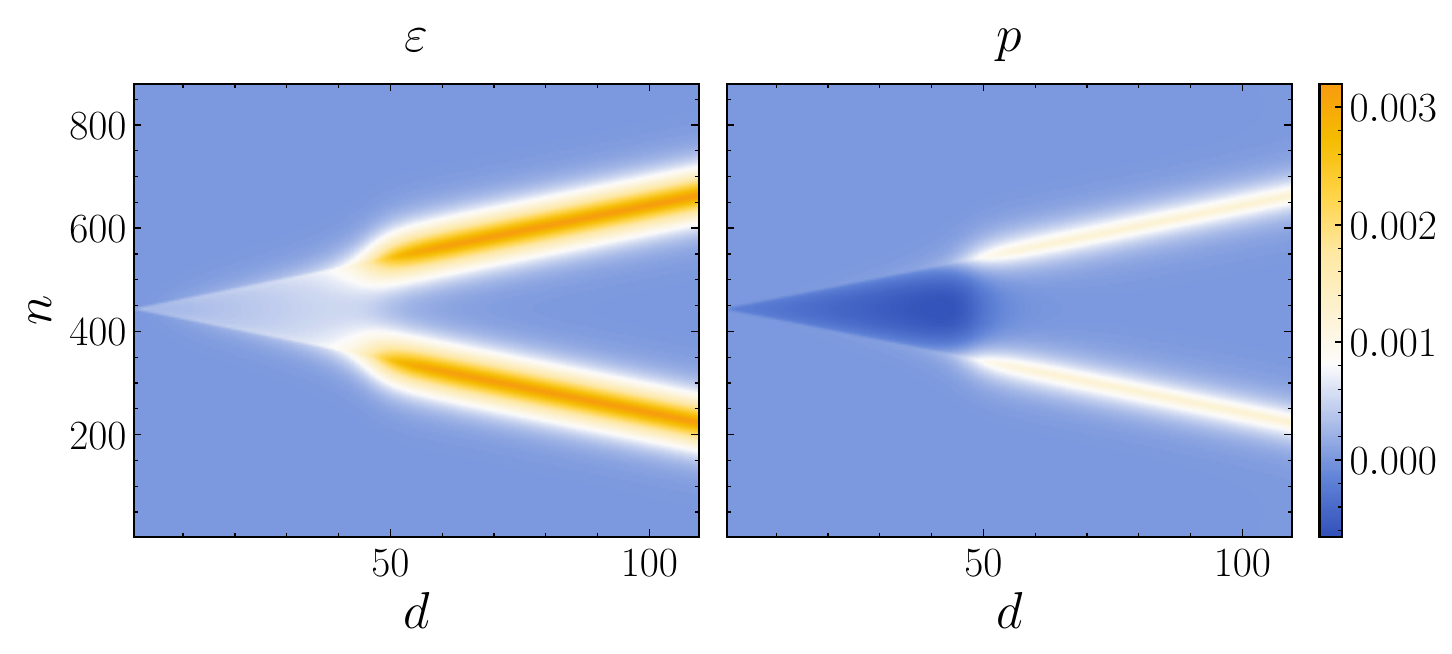}
     \caption{Components of the vacuum-subtracted energy-momentum tensor obtained with simulation parameters described in the text using $N=880$ and $a=1/4$.
     The left panel shows the energy density $\varepsilon$, while the right panel shows the pressure $p$.
     }
     \label{m_s:fig:emten}
\end{figure}

We computed the energy density and pressure, which are displayed in Fig.~\ref{m_s:fig:emten}.
The results are consistent with those given in Ref.~\cite{Grieninger:2025rdi}, but with finer resolution of the string-breaking process.  
The pressure distribution is particularly interesting.
The pressure is seen to become increasingly negative with increasing static-charge separation, and then revert rapidly to the vacuum value after string breaking.  In contrast, the pressure within the mesons rapidly increases from its vacuum  value after string breaking.

\section{Supplemental quantum complexity results}
\label{m_s:app:supplemental_results}
\noindent 

\begin{figure}[ht!]
     \centering
     \includegraphics[width=0.8\linewidth]{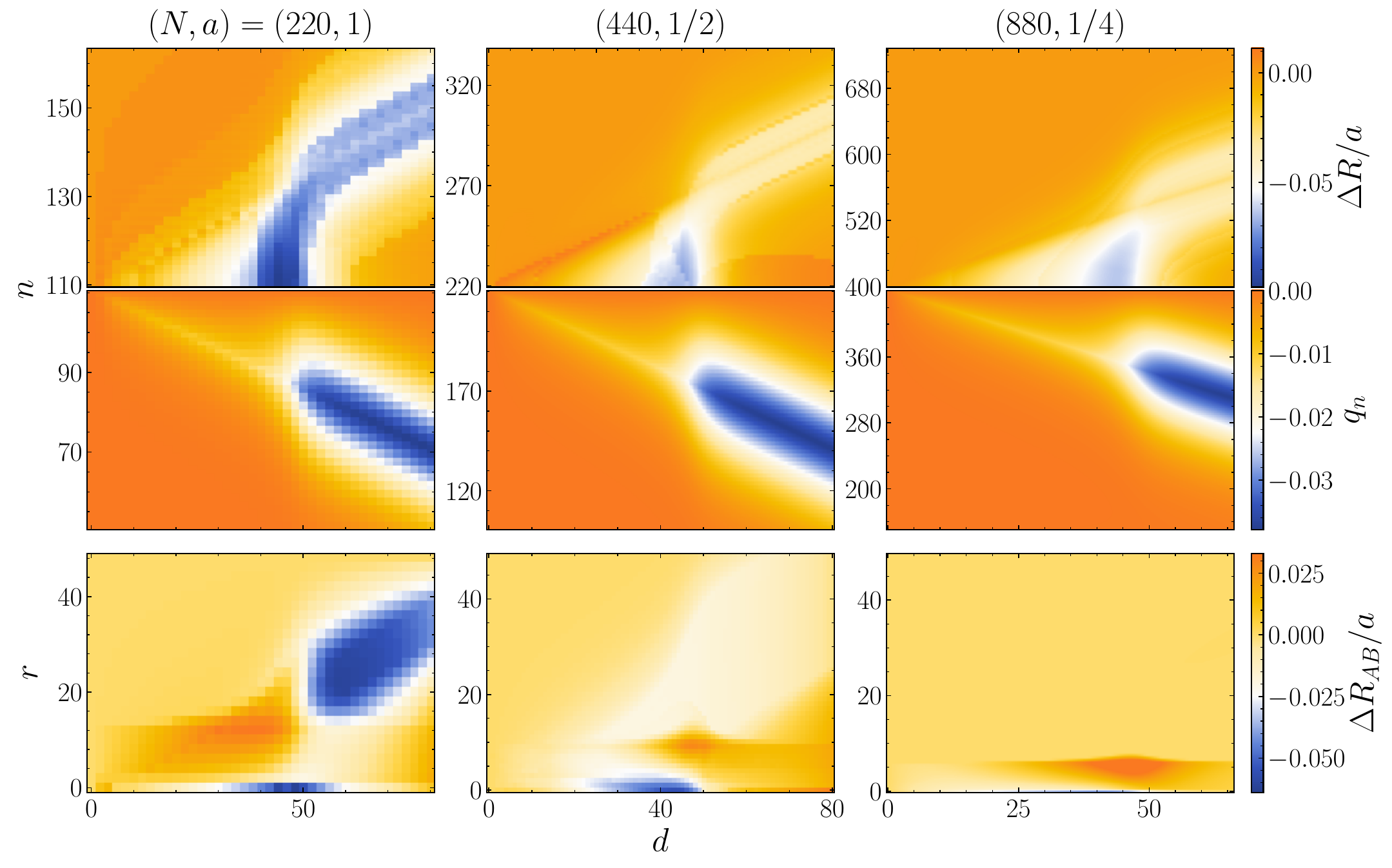}
     \caption{The dependence of the RoM on system size and lattice spacing.
     Top: The vacuum-subtracted RoM of adjacent sites, $\Delta R$ for a selection of system sizes and lattice spacings $(N,a)=(220,1)$ (left), $(440,1/2)$ (center), and $(880,1/4)$ (right). 
     The RoM is plotted in units of $a$ for half of the lattice $n\geq N/2$ as a function of external charge separation $d$.
     The system parameters from the main text are used.
     Middle: the associated charge densities for half of the lattice $n\leq N/2$.
     Bottom: the RoM of disjoint regions, $\Delta R_{AB}$, with region $A$ at the center of the lattice and region $B$ a distance $r$ away. 
     The vacuum value ($d=0$) and the value as $r\to\infty$ are subtracted.
     The RoM is plotted in units of $a$ for the same system parameters.
     }
     \label{m_s:fig:ROMadjALL}
\end{figure}

\noindent A natural quantity to consider is the behavior of magic in the continuum as $a\to0$. In principle, this could be studied by computing the RoM of a region of fixed physical volume, (i.e., $L=N_\text{phys}a = const.$).
However, the number of stabilizer states over which the RoM must be computed grows superexponentially~\cite{Aaronson:2004xuh}, and is intractable for more than 8 qubits~\cite{Hamaguchi:2023zpb}.
Instead of this, we compute the RoM for a subregion of fixed size, independent of $a$. 
While this does not probe the continuum limit of RoM in a physical region, it probes finer components of the wavefunction that cannot be seen with smaller system sizes.
Figure~\ref{m_s:fig:ROMadjALL} shows the RoM of adjacent physical sites (4 qubits) $\Delta R$ as a function of the position $n$ and separation of external charges $d$.
The plots with smaller $a$ provide a finer, ``zoomed in'' resolution view into the structure of the states. 
At $N=440,a=1/2$, additional structure is seen in $\Delta R$ between the mesons after the string has broken.
Interestingly, this structure resembles the structure in the pressure shown in Fig.~\ref{m_s:fig:emten}. 
The bottom row of Fig.~\ref{m_s:fig:ROMadjALL} shows the RoM between disjoint regions of the lattice as a function of separation (defined in Eq.~\eqref{m_s:eq:R4AB}) $\Delta R_{AB}$, scaled by $a$.
The structure initially present in the $N=220, a=1$ data is seen to vanish as $a$ is decreased.
The fixed-size RoM calculation (4 qubits) shows less information for finer $a$ because the wavefunction is spread over more lattice sites.
Instead, only quasi-local structure (small $r$) remains. 
\begin{figure}[ht!]
     \centering
     \includegraphics[width=\linewidth]{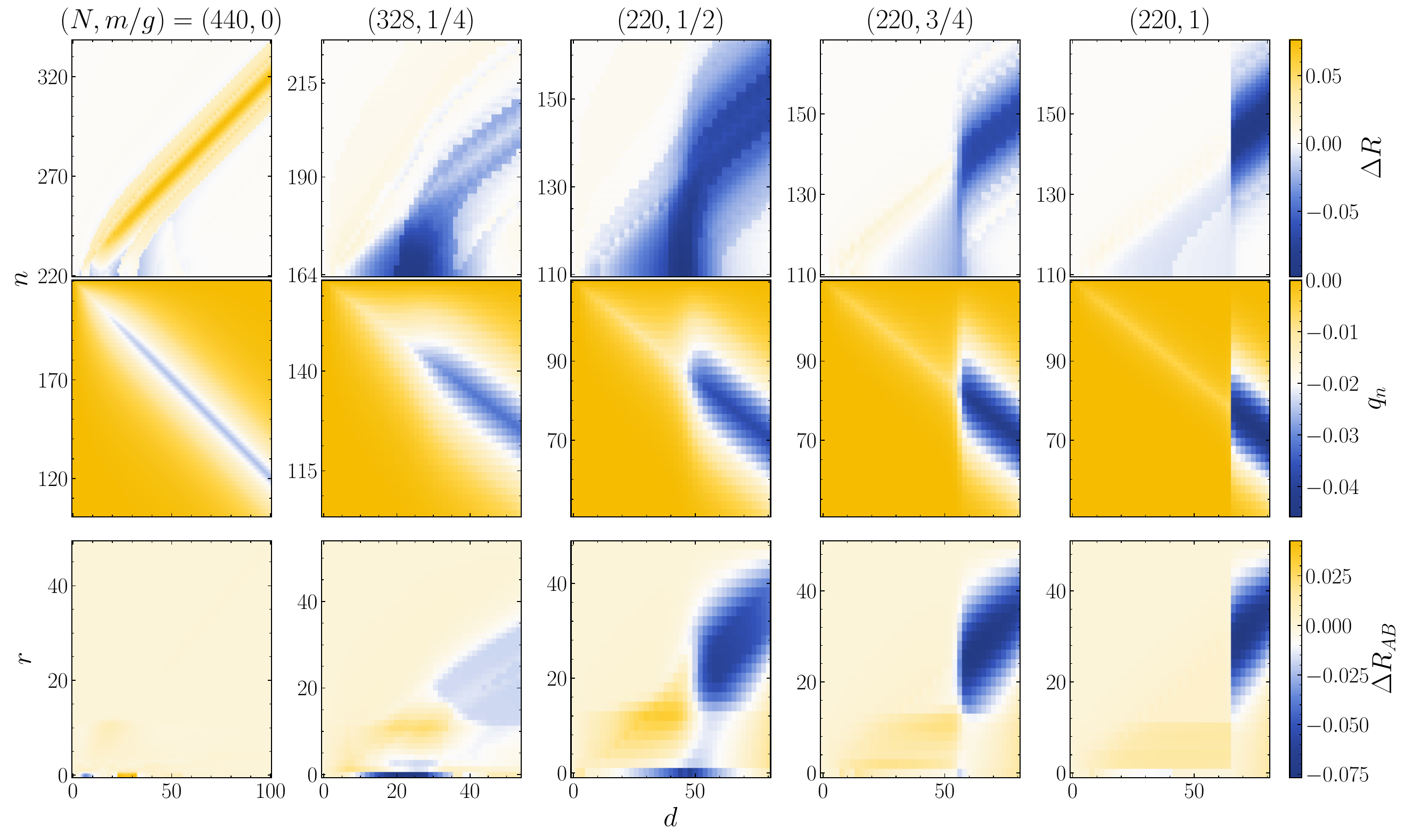}    
     \caption{The dependence of the RoM breaking on fermion mass for a selection of masses, $m/g=0$ (first column), $m/g=1/4$ (second column), $m/g=1/2$ (third column), $m/g=3/4$ (fourth column) and $m/g=1$ (fifth column).
     Top: the local vacuum-subtracted RoM $R$ as a function of $d$ for half of the lattice $n\geq N/2$ with $a=1$.
     Middle: the associated charge densities for half of the lattice $n\leq N/2$ for the same parameters.
     Bottom: the RoM of disjoint regions, $\Delta R_{AB}$, with region $A$ at the center of the lattice and region $B$ a distance $r$ away.
     The vacuum value ($d=0$) and the value as $r\to\infty$ are subtracted.
     The RoM is plotted in units of $a$ for the same system parameters.
     }
     \label{m_s:fig:m_g_scan}
\end{figure}

Figure~\ref{m_s:fig:m_g_scan} details the behavior of the quantum complexity as $m/g$ is varied.
We consider several values of $m$ holding $g$ fixed as in the main text, and expand the lattice to account for the increased correlation length with smaller mass.
The top and middle rows display the RoM between adjacent sites, $\Delta R$, and the charge density $q_n$ as a function of position $n$ and separation $d$. 
The transition from strings to mesons becomes increasingly sharper with larger $m/g$, which is seen in both $\Delta R$ and $q_n$.
The string also appears to break at a larger separation $d$ for larger $m/g$.
This is an expected result of fermions being more ``expensive'' to excite from the vacuum to screen the external charges.
The bottom row, as in Fig.~\ref{m_s:fig:ROMadjALL}, shows the RoM between disjoint $\Delta R_{AB}$ subsystems on the lattice as a function of the distance between the regions $r$.
Region $A$ is at the center of the lattice, and region $B$ is distance $r$ away.
The vacuum-subtracted $\Delta R_{AB}$ only shows structure at small $r$ for the massless case, consistent with the system only having short-range quantum correlations before string breaking. 
At intermediate $m/g$, the correlations before string breaking are longest-range, and then die off again at $m/g=1$.
The propagation of the bound hadrons is seen in all parameters besides the massless case.

\section{Computational methods}
\label{m_s:app:mps_methods}
\noindent
The reduced density matrix $\hat{\rho}_{AB}$ is required for the nonlocal measures of complexity such as MI and RoM, where regions $A$, $B$ may be separated on the lattice.
To avoid exponentially scaling resource requirements for creating a long-distance reduced density matrix, we apply a swap network to the MPS to move region $B$ next to region $A$, and then trace the regions outside $A$, $B$. 
This introduces long-range entanglement in the system and has the same cost as exactly contracting the region between A and B.
We approximate this operation by introducing a bond dimension and cutoff for the swap network operation.

We find ground states and low-lying excited states in MPS with DMRG. 
We use a bond dimension of 400, a cutoff of $10^{-12}$ and 40 sweeps. 
The convergence of our MPS computations, both for DMRG and for computing nonlocal reduced states is verified by examining the results as the precision is increased.

The RoM is calculated using the approach of Ref.~\cite{Hamaguchi:2023zpb}.
The minimization problem of Eq.~\eqref{m_s:eq:RoMdef} is reframed as a sparse linear programming problem,
\begin{align}
    R(\hat{\rho}) \ = \ \min_{\bm x} \left\{
    \ ||\bm x||_1 \ \ \  \bigg\rvert \ \ \  \ \hat{A} \bm x = \bm b\right\} \ .
    \label{m_s:eq:rom_axb}
\end{align}
Here the matrix $\hat{A}$ is defined as $ \hat{A}_{ij}=\text{Tr}(\hat{P}_i \hat{\rho}_{s_j})$, and $\bm b$ is the expansion of the state $\hat{\rho}$ in the Pauli basis: $b_i=\text{Tr}(\hat{P}_i\hat{\rho})$.
In practice, this may be readily solved in the following form:
\begin{align}
    \underset{\bm u} {\rm minimize} \ \sum_i u_i \ , \ \text{s.t.} \ \left( \hat{A} \quad - \hat{A}\right)\bm u=\bm b \ , \  u_i \geq 0 \ ,
\end{align}
A further simplification developed in Ref.~\cite{Hamaguchi:2023zpb} is that only a small fraction of stabilizer states have $x_i>0$, and these are correlated with having large values of the overlap $|2^{n_Q}\text{Tr}\left(\hat{\rho}_{s_i}\hat{\rho}\right)|$.
This enables further sparsification of the problem, only retaining a fraction $K$ of the states $\hat{\rho}_{s_i}$.
In this work, $K=0.05$ of states with the largest overlaps are kept for the optimization problem.

The NL RoM is calculated as the minimum of 1000 optimizations as in Eq.~\eqref{m_s:eq:NLRoMdef}, each with randomly chosen initial conditions.

\end{subappendices}

\part{Errors and fault-tolerance in quantum simulations}
\chapter{Optimization of algorithmic errors in analog quantum simulations}
\label{chap:spiral}

\noindent
{\it This chapter is associated with Ref.~\cite{Zemlevskiy:2023eyw}: ``Optimization of algorithmic errors in analog quantum simulations'' by Nikita A. Zemlevskiy, Henry F. Froland, and Stephan Caspar.}

\section{Introduction}
\noindent
Many analog quantum simulators are naturally described by an Ising-type Hamiltonian, but a broader class of dynamics, such as Heisenberg-type evolution, is often of greater physical interest~\cite{gong_zhu_sheng_2014, ma_dakic_naylor_zeilinger_walther_2011,Nachman_2021, Bauer:2022hpo, Caspar:2022llo, 2a,maldacena2023simple,Florio:2023dke,PhysRevD.105.083020}. 
Realizing this evolution requires engineering the native Hamiltonian through pulse sequences or digital decompositions — a general strategy known as Hamiltonian engineering. 
Continuously driven fields have also been shown to implement Heisenberg-type evolution in Ising systems~\cite{1a}.
This chapter studies the errors that accompany such engineering on realistic analog devices.

Major limitations of all these devices are the numerous sources of error that accompany any particular simulation. 
Several error correction schemes have been proposed for analog quantum computation~\cite{PhysRevLett.80.4088, PhysRevLett.119.180507}, but they have not yet been implemented on devices available today. 
Because of this, the ability to quantify these errors is crucial to make precise statements about a simulation. 
The three main sources of error that are important to consider for the purposes of simulation are the encoding error, the algorithmic error, and the hardware error~\cite{Klco:2018zqz}. 
The encoding and algorithmic error come from the approximations that are made in translating physics to a form best suited for a particular device (e.g., Hilbert space truncations, commutator error for product formulas, etc.). 
The hardware error comes from experimental imperfections and noise on a device, which is usually uncontrolled and needs to be dealt with using error mitigation. 
Much work has been done examining Trotter errors~\cite{Wiebe_2010,Childs_2019,childs_wiebe_2021,Cai:2022rnq,Endo_2018} and errors induced by analog devices~\cite{Shaffer_Megidish_Broz_Chen_Haffner_2021, trivedi2022quantum, PRXQuantum.1.020308, PhysRevX.12.021049} individually, but little attention has been paid to the interplay of these different sources of error. 
As it turns out, there are particular choices of parameters that will minimize the cumulative error within and across these different categories.

\begin{figure}
\centering
\includegraphics[trim={1cm 3cm 2cm 0.5cm}, clip, width=\linewidth]{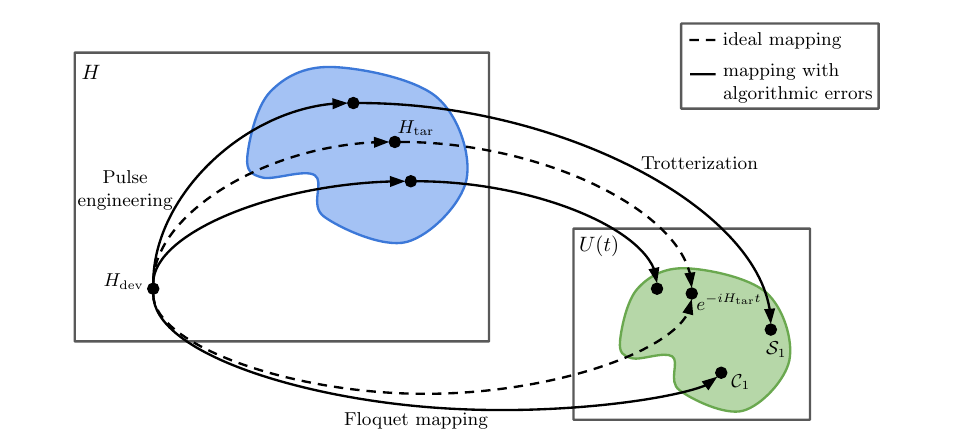}
\caption{{\it Hamiltonian engineering methods.} The device Hamiltonian $H_\text{dev}$ can be used to emulate a target system of interest with Hamiltonian $H_\text{tar}$ using Trotter-like methods (top path) or continuous driving methods (bottom path). In the limit of perfect pulses (equivalently, infinite magnetic field), the dashed curves implement the target time evolution exactly. The solid curves represent the mappings that are approximate due to algorithmic errors. These errors increase the overall error when mapping between Hamiltonians due to idle errors and when implementing the time evolution operator $U(t)$ (Trotter errors). Algorithmic errors affect constant-field time evolution engineering methods as well. The constant-field method $\mathcal{C}_1$ and one of the Trotter methods $\mathcal{S}_1$ are labeled as resulting time evolution methods within the space of possible $U(t)$ operators.}
\label{spiral:fig:error_schematic}
\end{figure}

With the advent of analog quantum simulation, an understanding of uncertainties stemming from mapping physical problems and implementations of simulation algorithms is now necessary. 
This work investigates the interplay of two types of algorithmic errors occurring in digital quantum computations on analog systems, Trotter errors and idle errors. 
Trotter errors arise due to the decomposition of the evolution operator into a finite product formula of non-commuting unitaries. 
Idle errors, on the other hand, stem from the inability of analog systems to turn off interactions while applying local pulses (gates). 
The decomposition of the algorithmic error into Trotter and idle errors is shown in Fig.~\ref{spiral:fig:error_schematic}.

On a perfect device, improving precision is straightforward if these two error sources are analyzed independently. In the absence of realistic device constraints, decreasing the Trotter step size is guaranteed to decrease the Trotter error, and decreasing the pulse width on the device lowers the idle error contribution. However, these error sources play against each other on real-world machines. For example, taking many short Trotter steps will incur a large error if the idle errors are high in the device. Furthermore, taking very small Trotter steps may be impossible due to limited pulse width. Therefore, it is necessary to analyze these effects collectively to find the parameters that maximize the precision of the output from simulations on a real device. In this work, a general framework for analyzing the error scaling of engineered time evolution methods is presented and is used to compare the performance of the aforementioned methods. We compare with numerical results for an analog simulator with (quasi-) local Ising interactions and global longitudinal and transverse-field addressing. This analysis gives the optimal choice of simulation parameters as a function of device parameters. In Sec.~\ref{spiral:sec:engineering_methods}, the general Hamiltonian engineering methods are described, Sec.~\ref{spiral:sec:error_analysis} introduces the types of algorithmic error studied, Sec.~\ref{spiral:sec:spiral_results} presents the results of the error analysis, and Sec.~\ref{spiral:sec:spiral_discussion} discusses the implications of this work's findings.

\section{Hamiltonian engineering methods} \label{spiral:sec:engineering_methods}
Consider a general Hamiltonian that may be implemented in various experimental systems: 
\begin{align}\label{spiral:eq:h_general}
    H_\text{dev}(t) \ &= \ H_\text{idle} + H_\text{drive}(t)\ , \nonumber\\
    H_\text{idle} \ &= \ \sum_{i<j} J_{ij} \left(c_xX_iX_j + c_y Y_iY_j + c_z Z_iZ_j\right)\ , \\ 
    H_\text{drive}(t) \ &= \ \vec{B}(t) \cdot \left( \sum_i \vec{S_i} \right)\ . \nonumber 
\end{align}
Using carefully engineered pulse sequences of the global magnetic field $\vec{B}(t)$, the coefficients $c_{x,y,z}$ in $H_\text{idle}$ can be modified into any target Hamiltonian $H_\text{tar} = H_\text{idle}'$ in the interaction picture, obeying $c_x+c_y+c_z = c_x'+c_y'+c_z'$ and $\min c_{x,y,z} \leq c_{x,y,z}' \leq \max c_{x,y,z}$~\cite{masanes2002timeoptimal}. The targeted Hamiltonian $H_\text{tar}$ arises through dynamical decoupling of the strong magnetic fields $|\vec{B}(t)| \leq \Omega \gg \mathop{\max}J_{ij}$
\begin{align}
    H_\text{tar} \ &= \ \frac{1}{\tau} \int_0^\tau d t ~ U_\text{drive}(t)^\dag ~ H_\text{idle} ~ U_\text{drive}(t) \ , \\
    i\partial_t U_\text{drive}(t) \ &= \ H_\text{drive}(t) U_\text{drive}(t) \ , \\
    U_\text{drive}(0) \ &= \ \mathbb{1}\ ,
    \label{spiral:eq:intpicture}
\end{align}
where an average is taken over a period $\tau$ of the pulse sequence with $\vec{B}(t+\tau)=\vec{B}(t)$ and $\Omega$ is the maximum field strength that can be implemented on the device. The rest of the chapter focuses on the concrete case of emulating the isotropic Heisenberg $c_{x,y,z}'=\frac{1}{3}$ evolution from Ising interactions $c_z=1,c_{x,y}=0$
\begin{align}
    H_\text{idle} \ = \ H_\text{Ising} \ &= \ \sum_{i < j} J_{ij} Z_i Z_j
    \quad \to \quad
    H_\text{tar} \ = \ H_{XXX} \ = \ \sum_{i < j} \frac{J_{ij}}{3} \left(X_i X_j + Y_i Y_j + Z_i Z_j\right)\ . \label{spiral:eq:h_heisenberg}
\end{align}

Consider two types of pulse engineering methods, intermittent and continuous driving. Intermittent driving is inspired by Trotterized time-evolution on digital quantum computers. The strong magnetic field allows approximate global $X$, $Y$, or $Z$ $\frac{\pi}{2}$ -rotations  through short pulses $\epsilon = \frac{\pi}{2\Omega} \ll J_{ij}^{-1}$ at the maximal field strength $\Omega$
\begin{align} 
\label{spiral:eq:AnalogRotations}
\begin{split}
    R^\pm_X(\epsilon) \ & = \ \exp{\left(-i\epsilon \sum_{i<j} J_{i,j} \hat{Z}_i \hat{Z}_j \mp \frac{i \pi}{4} \sum_j \hat{X}_j\right)} , \\
    R^\pm_Y(\epsilon) \ & = \ \exp{\left(-i \epsilon \sum_{i<j} J_{i,j} \hat{Z}_i \hat{Z}_j \mp \frac{i \pi}{4} \sum_j \hat{Y}_j\right)} , \\
    R^\pm_Z(\epsilon) \ & = \ \exp{\left(-i \epsilon \sum_{i<j} J_{i,j} \hat{Z}_i \hat{Z}_j \mp \frac{i \pi}{4} \sum_j \hat{Z}_j\right)}.
\end{split}
\end{align}
Note that at infinite $\Omega$ these are exact rotation gates. These pulses effectively permute the coefficients $c_{x,y,z}$, yielding a set of global gates which can be composed in the usual Trotter-like fashion. There have been several recent developments showing Trotter-like methods of Hamiltonian engineering where similar target systems are recovered~\cite{Martin:2022wyl, lukin_metrology_2020, tyler2023higherorder, zhou2023robust, Zhou:2023xnx, Geier:2021uxg, PRXQuantum.3.020303, choi_zhou_knowles_landig_choi_lukin_2020, richerme_gong_lee_senko_smith_foss-feig_michalakis_gorshkov_monroe_2014, jurcevic_lanyon_hauke_hempel_zoller_blatt_roos_2014,PhysRevA.95.013602, PhysRevA.97.023611}. For the native Ising interactions of Eq.~\eqref{spiral:eq:h_heisenberg}, there are global evolutions generated by native $XX$, $YY$ and $ZZ$ gates
\begin{align} \label{spiral:eq:AnalogGateSet}
\begin{split}
    R_{ZZ}(t) \ & = \ \exp{\left(-i t H_\text{Ising}\right)}, \\
    R_{XX}^{\pm}(t,\epsilon) \ &= \ R^{\mp}_{Y}(\epsilon)R_{ZZ}(t)R^{\pm}_{Y}(\epsilon), \\
    R_{YY}^{\pm}(t,\epsilon) \ &= \ R^{\mp}_{X}(\epsilon)R_{ZZ}(t)R^{\pm}_{X}(\epsilon).
\end{split}
\end{align}
Taking the product of these three gates defines a sequence which is referred to as ``pseudo-first order''
\begin{align} \label{spiral:eq:firstpseudo}
    \mathcal{S}_{1/2}(\vec{t},\epsilon) \ &= \ R_{ZZ}(t_z) R^{+}_{YY}(t_y,\epsilon) R^{+}_{XX}(t_x,\epsilon).
\end{align}
$\mathcal{S}$ stands for pulse sequence and the subscript indicates the error scaling with $\epsilon$. As in digital product formulas, order is defined by the scaling of the error with the parameters of the sequence. For instance, a first-order sequence has errors that only scale like $\mathcal{O}(t^2, \epsilon t, \epsilon^2)$. The ``pseudo''-modifier indicates that this sequence generates unwanted errors at order $\mathcal{O}(\epsilon)$ which contribute to $H_\text{tar}$ in Eq.~\eqref{spiral:eq:intpicture}. Such errors are avoidable if the alternative gate set
\begin{align} \label{spiral:eq:AltAnalogGateSet}
\begin{split}
    \tilde{R}_{XX}^{\pm}(t,\epsilon) &= R^{\pm}_{Y}(\epsilon)R_{ZZ}(t)R^{\pm}_{Y}(\epsilon), \\
    \tilde{R}_{YY}^{\pm}(t,\epsilon) &= R^{\pm}_{X}(\epsilon)R_{ZZ}(t)R^{\pm}_{X}(\epsilon),
\end{split}
\end{align}
is used to define a ``true-first order'' method
\begin{align} \label{spiral:eq:firsttrue}
    \mathcal{S}_1(\vec{t},\epsilon) &= R_{ZZ}(t_z)\tilde{R}^{+}_{YY}(t_y,\epsilon)\tilde{R}^{+}_{XX}(t_x,\epsilon).
\end{align}
Applying $\frac{\pi}{2}$-pulses in the same direction twice leads to destructive interference of the error terms plaguing $\mathcal{S}_{1/2}$. This comes at the cost of a residual global $Z$ rotation by $\pi$. This overall rotation is easy to keep track of and correct for in subsequent stages of a simulation. To extend this Trotter-like approach, we present a ``pseudo-second order'' sequence, $\tilde{\mathcal{S}}_1$, which carefully rearranges the gates in $\mathcal{S}_1$ to remove errors at order $\mathcal{O}(t^2)$
\begin{align} \label{spiral:eq:secondpseudo}
    \tilde{\mathcal{S}}_1(\vec{t},\epsilon) \ &= \ R_{ZZ}(t_z/2)R^{+}_{X}(\epsilon)R_{ZZ}(t_y/2,\epsilon)\tilde{R}^{+}_{XX}(t_x, \epsilon) R_{ZZ}(t_y/2) R^{-}_{X}(\epsilon)R_{ZZ}(t_z/2).
\end{align}
Finally, there is the fully symmetrized version of $\mathcal{S}_1$ according to the formula~\cite{Suzuki:1991jtk}
\begin{equation}
\mathcal{S}_2(\vec{t},\epsilon) \ = \ \mathcal{S}_1(\vec{t},\epsilon) \mathcal{S}_1(-\vec{t},-\epsilon)^\dagger \ . \nonumber
\end{equation}
As one would expect, all errors $\mathcal{O}(\epsilon^2,\epsilon t,t^2)$ vanish due to the symmetry, making this a ``true-second order'' sequence
\begin{align} \label{spiral:eq:secondtrue}
    \mathcal{S}_2(\vec{t},\epsilon) \ &= \ R_{ZZ}(t_z/2) \tilde{R}^{+}_{YY}(t_y/2,\epsilon) \tilde{R}^{+}_{XX}(t_x/2,\epsilon) \tilde{R}^{-}_{XX}(t_x/2,\epsilon) \tilde{R}^{-}_{YY}(t_y/2,\epsilon) R_{ZZ}(t_z/2).
\end{align}

\begin{figure}
\centering
\includegraphics[width=0.5\linewidth]{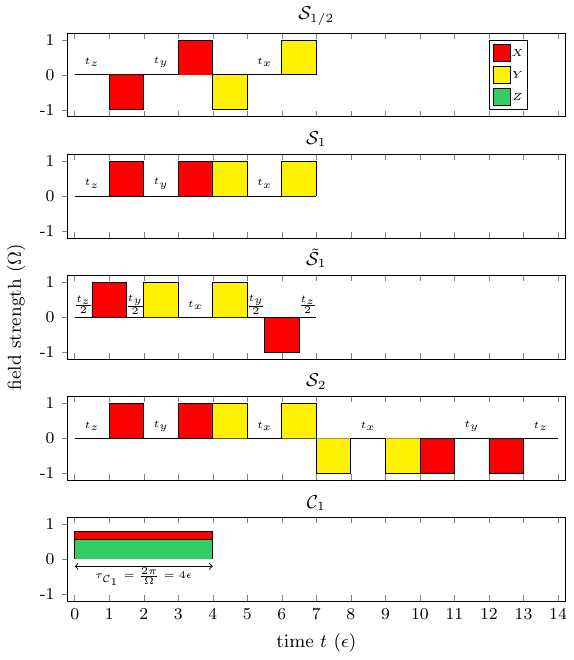}
\caption{{\it Pulse sequences for generating Heisenberg evolution from $ZZ$ interactions.} The pulse sequences correspond to Eqs.~\eqref{spiral:eq:firstpseudo}, ~\eqref{spiral:eq:firsttrue} -~\eqref{spiral:eq:bspiral}. For each pulse sequence, the magnetic field strength (in units of $\Omega$) is given as a function of time (in units of $\epsilon = \frac{\pi}{2 \Omega}$). The length of each sequence $U$ is denoted by $\tau_U$. The idling times $t_i$ are shown in the graphics for each of the Trotter sequences.}
\label{spiral:fig:zz_pulse_sequences}
\end{figure}

These pulse sequences are illustrated in Fig.~\ref{spiral:fig:zz_pulse_sequences}. The parameter $\vec{t} = (t_x,t_y,t_z)$ specifies the amount of time to evolve without a magnetic field in the respective frames $XX,YY,ZZ$. They obey $\tau c_i' = t_i + \mathcal{O}(\epsilon)$ since the finite pulse width also contributes to these coefficients, and the $t_i$ need to be adjusted accordingly. It is important to reiterate that the sequences $\mathcal{S}_1$ and $\tilde{\mathcal{S}}_1$ reproduce evolution under $H_{XXX}$ up to a global $Z$ $\pi$-rotation, which is accounted for in this analysis. 

The constant drive field method of Refs.~\cite{1a,1c} is an example of continuous driving. Similar methods have been explored in various contexts within simulations~\cite{eckstein2023largescale}. This approach involves the constant driving field
\begin{equation} \label{spiral:eq:bspiral}
    \vec{B}(t) \ = \ \frac{\Omega}{\sqrt{3}}
    \begin{pmatrix}
    \sqrt{2}\\
    0\\
    1
    \end{pmatrix}.
\end{equation}
In contrast with finite pulses, this method has the effect of gradually rotating the interaction term around the Bloch sphere. It is straightforward to show that after a full period $\tau_{C_1} =  4 \epsilon = 2\pi/\Omega$, the generated $H_\text{tar}$ reproduces $H_{XXX}$ at leading-order in $\epsilon$. The error analysis for the constant field method is less involved than for the pulse sequences, since there is only one parameter $\epsilon$. The engineered Hamiltonian in Eq.~\eqref{spiral:eq:intpicture} is nothing but the leading-order term in the Magnus expansion
\begin{align}
    U_\text{int}(t) \ &= \ U_\text{drive}(t)^\dag U(t) \ = \ e^{\sum_k \Omega_k(t)}, &
    \tau_{\mathcal{C}_1} H_\text{tar} \ &= \ i~\Omega_1(\tau_{\mathcal{C}_1}).
\end{align}
Due to the scaling $\Omega_k(\tau) = \mathcal{O}(\tau^k)$, the expected error is of order $\mathcal{O}(\tau^2)=\mathcal{O}(\epsilon^2)$, which is equivalent to a first-order Trotter formula. This method is referred to as $\mathcal{C}_1$ with $\mathcal{C}$ standing for constant-field, and the subscript again indicating the error scaling with $\epsilon$. $\mathcal{C}_1$ is shown graphically in the bottom panel of Fig.~\ref{spiral:fig:zz_pulse_sequences}.

The methods presented in this chapter for approximating desired Hamiltonians using pulse engineering, as well as the methods described for analyzing the errors incurred, are general and can be applied to any case of Eq.~\eqref{spiral:eq:h_general}. The next section describes how to analyze and quantify the errors associated with different time evolution methods.

\section{Characterization of error type and error rate}\label{spiral:sec:error_analysis}
Consider the problem of simulating evolution for a total time $T$, which is broken up into steps of length $\tau$ in physical time. As seen in Fig~\ref{spiral:fig:zz_pulse_sequences}, the time it takes to implement a single step on the device varies for each sequence; these are labeled by $\tau_U$ for a pulse sequence $U$. This section analyzes the error accrued during a single step of the sequences described and shows how to optimize the length of this step to minimize the error. The contributions of different sources of error are isolated by considering a time evolution operator $U$ that is a smooth function of two parameters $t$ and $\epsilon$ admits the following Taylor expansion:
\begin{align}\label{spiral:eq:pftaylor}
    U(t, \epsilon) \ &= \ \sum_{k_1=0}^{\infty}\sum_{k_2=0}^{\infty}t^{k_1}\epsilon^{k_2}U_{t^{k_1}\epsilon^{k_2}}, & U_{t^{k_1}\epsilon^{k_2}} \ = \ \frac{1}{k_1!k_2!}\partial_t^{k_1}\partial_{\epsilon}^{k_2}U(t,\epsilon)\big|_{t,\epsilon=0}.
\end{align}
The figure of merit that will be used to compare the error scaling of the different evolution methods is the error rate
\begin{align}\label{spiral:eq:ErrorRate}
    ER_{U}(t,\epsilon) \ &= \ \frac{1}{\tau_U (t,\epsilon)}\specnorm{U(t,\epsilon)-e^{-i \tau_U (t,\epsilon) H_{tar}}}
    \leq \frac{1}{\tau_U (t,\epsilon)}\sum_{k_1=0}^{\infty}\sum_{k_2=0}^{\infty}t^{k_1}\epsilon^{k_2}\specnorm{\mathcal{E}_{U;t^{k_1}\epsilon^{k_2}}},
\end{align}
where 
\begin{equation}\label{spiral:eq:deriv_terms}
    \mathcal{E}_{U;t^{k_1}\epsilon^{k_2}} = U_{t^{k_1}\epsilon^{k_2}}-\frac{1}{k_1!k_2!}\partial_t^{k_1}\partial_{\epsilon}^{k_2}e^{-i\tau_U (t,\epsilon)H_{tar}}\big|_{t,\epsilon=0}
\end{equation}
are defined to be the error terms that contribute at $\mathcal{O}(t^{k_1}, \epsilon^{k_2})$. As before, $H_\text{tar}$ describes the system to be simulated (in this work it is $H_{XXX}$), and $\tau_U(t, \epsilon)$ is the optimal step size for a given time evolution $U$ and parameters $t$ and $\epsilon$. In principle, any norm can be used in Eq.~\eqref{spiral:eq:ErrorRate}; throughout this work the spectral norm induced by the Hilbert space 2-norm is used, which is defined by: $||A|| = \mathop{\sup}\limits_{v}\frac{||Av||_2}{||v||_2}$. While the error for a single application of $U$ will always decrease with step size $\tau_U$, multiple applications may cause this error to grow unfavorably. The error accrued per step, the error rate $ER_U$, is used to account for varying step size.

The different terms $\mathcal{E}_{U;t^{k_1}\epsilon^{k_2}}$ can be characterized as follows. Terms with $k_2=0$ are the standard Trotter error, as these come strictly from the algorithm used to simulate time evolution (i.e. commutator error from Trotter-Suzuki formulae). The magnitude of this type of error is controlled by $t$. Terms with $k_1=0$ are called ``idle errors'' and come strictly from the inability to implement ideal single qubit rotations. For the device considered in this work, the source of idle error is due to the persistent Ising interaction during $\frac{\pi}{2}$-pulses. The magnitude of this type of error is controlled by $\epsilon$, which is inversely related to the maximum magnetic field strength. Terms that correspond to $k_1,k_2\neq 0$ are referred to as ``mixed errors'', these can be thought of as $\mathcal{O}(\epsilon^{k_2})$ idle error that is propagated through the simulation by the $\mathcal{O}(t^{k_1})$ action of the time evolution algorithm. 

In the Trotter-like digital methods there is freedom in the choice of $\tau_U$. The aforementioned sources of error can balance against one another to give overly pessimistic simulation results. This means that not all choices of $\tau_U$ are equally good, and some choices will give more error than others. To be precise, for a given time evolution method $U$, the optimal step size $\tau_U$ is split into 
\begin{align}
    \tau_U(t, \epsilon) \ &= \ 3(t_\text{rot} + t_\text{comp} + t), \label{spiral:eq:tau_u}
\end{align}
where $t_\text{rot}$ corresponds to the evolution due to the rotation gates, $t_\text{comp}$ is the compensation for unwanted evolution due to those gates, and $t$ is optional evolution time. The factor of 3 comes from mapping $H_\text{Ising} \rightarrow H_{XXX}$ in Eq.~\eqref{spiral:eq:h_heisenberg}. The optimization of $\tau_U$ proceeds as follows. $t_\text{rot}$ is fixed by the sequence, and so $t_\text{comp}$ must be applied to match $H_\text{tar}$ evolution at leading-order in the expansion of Eq.~\eqref{spiral:eq:pftaylor}. $t$ is then chosen to minimize the contribution of terms not cancelled by $t_\text{comp}$ to $ER_U$; the optimal choice of $t$ is labelled $t_*$. Here no assumptions are made about the relative smallness of $t$ and $\epsilon$, only that they are small enough to give systematically improvable expansions for each type of error such that Eq.~\eqref{spiral:eq:pftaylor} may be truncated at finite-order.

The freedom in choosing $\tau_U$ for the Trotter sequences lies in the choice of $\vec{t}$ appearing in Eqs.~\eqref{spiral:eq:firstpseudo}-~\eqref{spiral:eq:secondtrue}. Specifically, $\vec{t}$ is chosen so that $t_i = t_{\text{comp},i} + t$\, where $t$ is the optional evolution time of Eq.~\eqref{spiral:eq:tau_u} (this is expanded upon in App.~\ref{spiral:app:first_order_error_analytics}). An important upshot of this relation is that all product formulas have a minimum value of $\tau_U$ where $t_i=0$ for some $i$ (e.g., corresponding to $\tau_{\mathcal{S}_1} = 6\epsilon$). In other words, for given experimental parameters and a given sequence, the step size $\tau_U$ cannot be made arbitrarily small. Therefore the optimization of $t$ must be constrained such that $\tau_U$ is always larger than its minimal value.

The optimal choice of $t$ (assuming this is greater than the minimal simulation time) is then given by
\begin{equation}
    t_{U*} \ = \ \mathop{\arg \min}\limits_{t_i>0} \left[ER_U(t)\right]. \label{spiral:eq:step_size_formula}
\end{equation}
For the low-order product formulas considered in this work, the solution becomes analytically tractable. Explicit expressions for these bounds are presented in subsequent sections, with derivations in App.~\ref{spiral:app:first_order_error_analytics}. For the constant-field method $\mathcal{C}_1$, the step size is completely set by the Floquet period $\tau_{\mathcal{C}_1} = 4\epsilon$ and so this method only incurs (algorithmic) error as a function of $\epsilon$.

\begin{figure}
    \centering
    \includegraphics[scale=1]{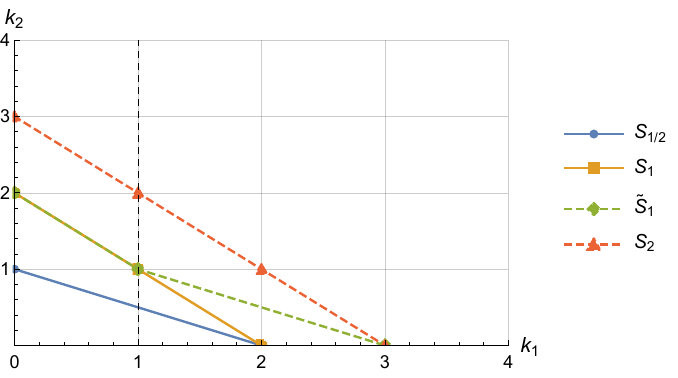}
    \caption{{\it Leading-order contributions to the error rate.}  The optimal scaling is given by the negative slope of each line at $k_1=1$ and the corresponding error rate scaling is given by the $k_1=1$ intercept.}
    \label{spiral:fig:graphicMethod}
\end{figure}

There is a useful graphical view of the relevant error terms that makes the optimal scaling of $t$ and error rate with $\epsilon$ manifest. The error rate in Eq.~\eqref{spiral:eq:ErrorRate} is expanded using Eq.~\eqref{spiral:eq:pftaylor} and truncated based on the leading-order Trotter and idle errors (i.e. the lowest powers of strictly $k_2$ or $k_1$ respectively). This leaves a finite number of terms contributing to the error rate in unique powers of $t$ and $\epsilon$, which can be represented as a curve $k_2=\mathcal{B}(k_1)$ in the $k_1-k_2$ plane, as in Fig.~\ref{spiral:fig:graphicMethod}. Specifically, the error rate series becomes

\begin{equation}
    ER_U(t) \ \leq \ \frac{1}{\tau_U(t,\epsilon)}\left(t^{n+1}\specnorm{\mathcal{E}_{U;t^{n+1}}}+\epsilon^m\specnorm{\mathcal{E}_{U;\epsilon^{m}}}+\sum_{k=1}^{m-1}t^{\kappa}\epsilon^k\specnorm{\mathcal{E}_{U;t^{\kappa}\epsilon^{k}}}\right),\label{spiral:eq:errRateBound}
\end{equation}
where $\kappa$ is defined such that
\begin{equation}
    \specnorm{\mathcal{E}_{U;t^{\kappa}\epsilon^{k}}}  \ \neq \ 0,\qquad \text{inf}\left(\specnorm{\mathcal{E}_{U;t^{\kappa-1}\epsilon^{k}}},\specnorm{\mathcal{E}_{U;t^{\kappa}\epsilon^{k-1}}}\right) \ = \ 0
\end{equation}
for some fixed $k$. By working in the asymptotic limit, which is defined as $\epsilon \ll 1$ with $t\propto\epsilon^{\alpha}$ for some number $\alpha$, the error rate scaling may be analyzed in terms of the individual powers of each of the contributing error terms. The leading-order $k_1$ error is $\alpha (k_1-1)+\mathcal{B}(k_1)$, where 1 is subtracted because $\tau_U(t,\epsilon)=\mathcal{O}(t,\epsilon)$. The optimal error rate, for a given choice of $\alpha$, then scales with $\epsilon$ as 
\begin{equation}
    \mathop{\sup}\limits_{k_1} \alpha (k_1-1)+\mathcal{B}(k_1).
\end{equation}
This quantity then needs to be minimized over $\alpha$. This is a modified Legendre transform of $\mathcal{B}$, where $\alpha$ and $k_1$ form a conjugate pair. Because of this, the optimal scaling of the error rate is given by $\mathcal{B}(k_1=1)$ and the optimal scaling $\alpha$ is given by $-\left.\frac{d\mathcal{B}}{dk_1}\right|_{k_1=1}$. Comparing Fig.~\ref{spiral:fig:graphicMethod} to the results below demonstrates these properties. An upshot of this analysis is that for curves whose slope at $k_1=1$ is undefined (e.g. $\tilde{\mathcal{S}}_1$), there are actually a range of optimal scalings between the slopes from the left and right side that give rise to the same optimal error rate.

\section{Results} \label{spiral:sec:spiral_results}
\subsection{Analytic results} \label{spiral:sec:analytic_results}
This section gives the results of applying the analysis presented in the previous section to the Hamiltonian engineering methods of Sec.~\ref{spiral:sec:engineering_methods}. Specifically, $U=\mathcal{S}_{1/2}, \mathcal{S}_1, \mathcal{C}_1$ are used to demonstrate the analysis described above, with details of the calculations contained in Apps.~\ref{spiral:app:constant_field_drive_appendix} and~\ref{spiral:app:first_order_error_analytics}. To isolate the contributions of different sources of error $\mathcal{E}$ to the error rate, the expansion of Eq.~\eqref{spiral:eq:pftaylor} is applied to the two Trotter sequences. A bound on these contributions is then placed by separating the part of the expression depending on the geometry of the system from the operator content. The leading-order contributions to the error are found to be $\mathcal{O}(N)$. These are given in Eqs.~\eqref{spiral:eq:E1_tilde_eps_norm} -~\eqref{spiral:eq:E1_t2_norm}.

As described in the previous section, the optimal step size $\tau_U$ is found by first choosing $t_\text{comp}$ to cancel leading-order error terms and optimizing $t$ to minimize the remaining terms at leading or next-leading order, the result of which is shown in Eqs.~\eqref{spiral:eq:topt_S1t} and~\eqref{spiral:eq:topt_S1}. $\tau_U$ is expected to vary for different sequences since the error made depends on the sequence used. The magnitudes of the different leading-order contributions to the error are used to find $t_*$. Subject to the constraints described in the previous section, the optimal $t_*$ values are (up to $\mathcal{O}(1)$ factors determined by system geometry, see Eqs.~\eqref{spiral:eq:topt_S1t_with_constraint} and~\eqref{spiral:eq:topt_S1_with_constraint})
\begin{align}
    &t_{\mathcal{S}_{1/2}*} \ = \ \mathcal{O}(\epsilon^{\frac{1}{2}}),
    &t_{\mathcal{S}_1*} \ = \ 0.
\label{spiral:eq:approximate_optimal_step_sizes}
\end{align}
The step size for $\mathcal{C}_1$ is fixed by $\tau_{\mathcal{C}_1}$, so there is no optimization to be done for the constant-field method. 

The error rate (Eq.~\eqref{spiral:eq:ErrorRate}) is the main point of comparison between different pulse sequences. It is computed by using the individual error source contributions $\mathcal{E}_{U;\{\dots\}}$ and the optimal step sizes $t_*$ to find the error rate scaling as a function of $\epsilon$ for the sequences $\mathcal{S}_{1/2}$ and $\mathcal{S}_1$ 
\begin{align} \label{spiral:eq:approximate_Trotter_error_rates}
\begin{split}
    ER_{\mathcal{S}_{1/2}} \ &\lesssim \ \epsilon^{\frac{1}{2}}\sqrt{\left(\sum_{i\neq j\neq k}J_{ik}J_{kj}\right)\left(\sum_{i\neq j}J_{ij}\right)} \ = \ \mathcal{O}(\epsilon^{\frac{1}{2}} N),\\
    ER_{\mathcal{S}_1} \ &\lesssim \ \epsilon \left(\sum_{i\neq j\neq k}J_{ik}J_{kj} + \sum_{i\neq j}J_{ij}^2\right)=\mathcal{O}(\epsilon N).
\end{split}
\end{align}
Here $\mathcal{O}(1)$ factors are again neglected for clarity. The full expressions can be found in Eqs.~\eqref{spiral:eq:ER_S1t_with_constraint} and~\eqref{spiral:eq:ER_S1_with_constraint}. To find the error rate for the constant-field method, the evolution implemented on the device is expanded in a Magnus series and the Heisenberg terms are subtracted off at leading-order to get a bound for $ER_{\mathcal{C}_1}$. Dropping $\mathcal{O}(1)$ factors, the error rate for this method is given by:
\begin{align} \label{spiral:eq:approximate_spiral_error_rate}
    ER_{\mathcal{C}_1} \ 
    &\lesssim \ \epsilon\left(\sum_{i\neq j\neq k}J_{ij}J_{jk}+\sum_{i\neq j}J_{ij}^2\right)=\mathcal{O}(\epsilon N).
\end{align}
The error rates are found to be extensive quantities, which is not surprising: the number of terms in the Hamiltonian increases with the system size. For comparing systems of different sizes and analyzing local observables, the error rate density is useful: $\frac{ER_U}{N}$.

The methods described in this work can be used to map between device and target Hamiltonians obeying a simple relation described in Sec.~\ref{spiral:sec:engineering_methods}. As an example of this, the setup of Ref.~\cite{PRXQuantum.3.020303} can accommodate time evolution methods similar to those presented in previous sections. Their work assumes an experimental system with dipole $XX + YY$ interactions, and tunable $X$ and $Y$ magnetic fields. With this setup, the pulse sequences of Eqs.~\eqref{spiral:eq:firstpseudo}, ~\eqref{spiral:eq:firsttrue} -~\eqref{spiral:eq:secondtrue} implementing Trotterized time evolution carry over without change to emulate an isotropic Heisenberg system. In addition, if a tunable $Z$ magnetic field is assumed, the constant-field method of Eq.~\eqref{spiral:eq:bspiral} works as well. Following the analysis for $ER_U$, the method proposed in Ref.~\cite{PRXQuantum.3.020303} has a $\mathcal{O}(\sqrt{\epsilon})$ error rate scaling, similar to $\mathcal{S}_{1/2}$, with a slightly different prefactor.

There exist many sequences that have the same error scaling. Examples of this are the sequences $\mathcal{S}_1$ and $\tilde{\mathcal{S}}_1$, which are essentially reordered versions of each other. The method of Ref.~\cite{PRXQuantum.3.020303} and $\mathcal{S}_{1/2}$ also both have $ER_U = \mathcal{O}(\sqrt{\epsilon})$. This non-uniqueness comes from the fact that different combinations of $\pm\frac{\pi}{2}$-rotations may yield the same evolution and similar scaling but with different prefactors due to the combination of the errors incurred (see App.~\ref{spiral:app:pi_2_pulse_error}). This work shows low-lying sequences up to $\mathcal{O}(\epsilon^2)$ in the error scaling. It is possible to extend these using the standard Trotter methods~\cite{Suzuki:1991jtk} to stitch them together to create higher-order sequences. Furthermore, as the subscript on $\mathcal{C}_1$ suggests, it is possible to create higher-order variants of constant-field methods by stitching together lower-order methods in the appropriate manner, e.g., $\mathcal{C}_2(2\epsilon) = \mathcal{C}_1(\epsilon) \mathcal{C}_1(-\epsilon)$.

\subsection{Numerical results} \label{spiral:sec:numerical_results}
The reliance of this analysis on inequalities rooted in the triangle inequality is a source of inflation of the derived upper bounds. In practice, cancellations between different terms give rise to a smaller error overall, compared to the worst-case bounds. Realistic values for the error rate can be found by evaluating Eq.~\eqref{spiral:eq:ErrorRate} numerically to understand the extent to which our bounds ignore these cancellations.

For concreteness, the physical Hamiltonian underlying experimental systems of Rydberg atoms~\cite{RevModPhys.82.2313, Wu_2021} is used for the rest of the chapter, corresponding to $c_x = c_y = 0, c_z = 1$ in Eq.~\eqref{spiral:eq:h_general}. Rydberg atoms have an all-to-all coupling mediated by a Van der Waals interaction (i.e. $J_{ij} = \frac{C_6}{|\vec{r}_i - \vec{r}_j|^6}$). This interaction only depends on the geometry of the Rydberg array which is assumed to be fixed for the duration of the simulation. In this work, the values of $C_6 = \mathcal{O}(10^6)$ \unit{MHz . \um^6 } and the lattice spacing $a = \mathcal{O}(1)$ \unit{\um} are fixed to realistic values based on current hardware parameters~\cite{wurtz2023aquila}. Changing these parameters is equivalent to rescaling the maximum field strength $\Omega$, so numerical calculations are performed with these constants fixed. In reality, devices require some time to change the magnetic fields; this time is determined by the slew rate. This analysis assumes the ideal case of an infinite slew rate for simplicity, although including realistic slew rates does not change the results qualitatively.

\begin{figure*}
\centering
\includegraphics[width=0.75\linewidth]{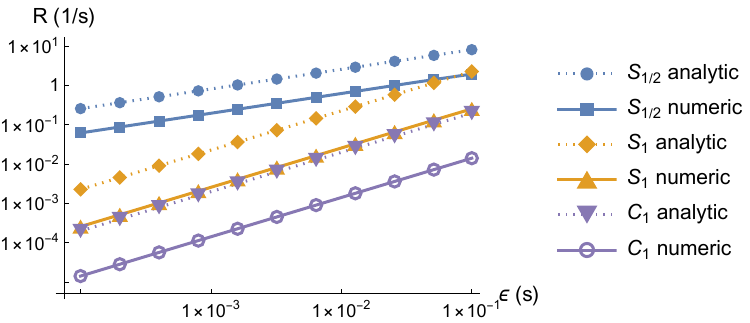}
\caption{{Bounds for a single engineered Trotter step.} Bounds for the error rate for a single Trotter step from Eqs.~\eqref{spiral:eq:approximate_Trotter_error_rates} and~\eqref{spiral:eq:approximate_spiral_error_rate} are compared with numerical results as a function of the device parameter $\epsilon$ for $2\times2$ systems. The analytic bounds consistently exceed the numerical results by factors of $\mathcal{O}(1)-\mathcal{O}(10)$, which is an expected consequence of using the triangle inequality to evaluate the bounds~\cite{childs_wiebe_2021}.}
\label{spiral:fig:error_rates}
\end{figure*}

The individual contributions $\mathcal{E}$ are compared to the numerical values (obtained by directly evaluating the $\mathcal{E}$ terms in Eq.~\eqref{spiral:eq:deriv_terms}) in Fig.~\ref{spiral:fig:individual_term_scalings} as a function of system configuration. The results for the error rates of Eqs.~\eqref{spiral:eq:approximate_Trotter_error_rates} and~\eqref{spiral:eq:approximate_spiral_error_rate}, along with their numerical counterparts, can be seen in Fig.~\ref{spiral:fig:error_rates}. In both Fig.~\ref{spiral:fig:error_rates} and Fig.~\ref{spiral:fig:individual_term_scalings}, we see the analytic bounds exceed the numerical values. This happens because the inequalities applied to derive the bounds effectively ignore the cancellations between various terms that happen in practice.

\begin{figure*}
\centering
\includegraphics[width=0.6\linewidth]{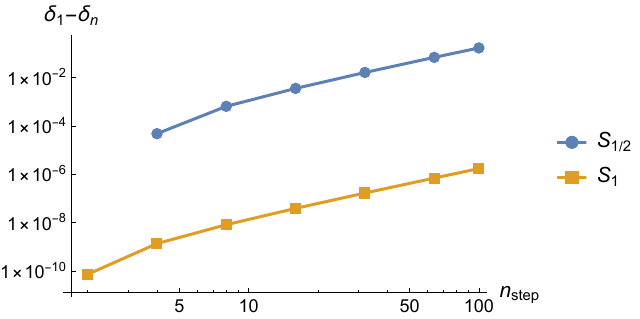}
\caption{{\it The effect of multiple Trotter steps on errors for $2\times2$ systems with $\epsilon \sim 10^{-5}$ and pulse sequences $\mathcal{S}_{1/2}$ and $\mathcal{S}_1$}. The quantity $\delta_1 - \delta_n = \specnorm{U_\text{trot}(T) - e^{-i H_{XXX} T}} - \specnorm{\prod_{n}U_\text{trot}(T/n) - e^{-i H_{XXX} T}}$ is positive when taking many small steps of $U_\text{trot}(T/n)$ replicates $e^{-iH_{XXX}T}$ better than taking one large step $U_\text{trot}(T)$. Here, $T = n \tau_U$ and $\tau_U$ is chosen optimally according to the Eq.~\eqref{spiral:eq:approximate_optimal_step_sizes}. Taking many steps to reach a total time $T$ is beneficial for $\mathcal{S}_{1/2}$ and $\mathcal{S}_1$ compared to taking a single large step when the step size is chosen optimally. This demonstrates the importance of choosing the step size correctly to take advantage of the Trotter error scaling.}
\label{spiral:fig:crossovers}
\end{figure*}

So far, the error for a single Trotter step has been considered. Fig.~\ref{spiral:fig:crossovers} investigates whether it is better to take many Trotter steps or a single large step by considering the quantity 
\begin{equation}
    \delta_1 - \delta_n \ = \ \specnorm{U_\text{trot}(T) - e^{-i H_{XXX} T}} - \specnorm{\prod_{n}U_\text{trot}(T/n) - e^{-i H_{XXX} T}} \ . \nonumber
\end{equation}
For a given Trotter time evolution method $U_\text{trot}$, measures how much closer $U_\text{trot}$ with $n$ steps is to the true time evolution than $U_\text{trot}$ with one large step. Considering this figure, the effect of the different error contributions on the overall error for each pulse sequence becomes evident: while decreasing $\epsilon$ will always decrease the error, the choice of the step size $\tau_U$ is of crucial importance to be able to take advantage of the scaling properties of Trotter formulas. In other words, a poorly chosen $\tau_U$ may inhibit successive application of the sequence and incur more error than a single large application as a result of constructive interference of different $\mathcal{E}$ contributions. Figs.~\ref{spiral:fig:error_rates} and~\ref{spiral:fig:crossovers} also demonstrate the orders of magnitude difference between the rate at which errors accrue for $\mathcal{S}_{1/2}$ and $\mathcal{S}_1$.

\section{Discussion}\label{spiral:sec:spiral_discussion}

\begin{figure}
    \centering
\includegraphics[width=0.6\linewidth]{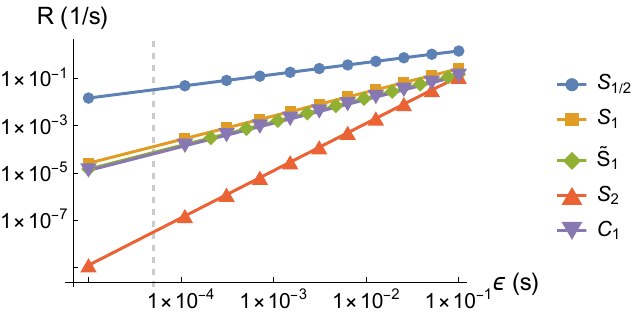}
\caption{{\it Numerical error rates for engineered time evolution.}
The numerical error rates are shown corresponding to Eq.~\eqref{spiral:eq:ErrorRate} for a single Trotter step (Floquet period for $\mathcal{C}_1$) for $2\times2$ systems. The slopes of lines on these log-log plots indicate the scaling of the error rate for each method (see Table~\ref{spiral:tab:method_comparison}) and the y-intercepts give the scaling prefactor. The lines for $\mathcal{S}_{1/2}, \mathcal{S}_1$, $\mathcal{C}_1$ are the same as the numerical ones in Fig.~\ref{spiral:fig:error_rates}. Current device capabilities report pulse control to the ns level~\cite{wurtz2023aquila}, meaning pulse sequences with $\epsilon \sim \mathcal{O}(10^{-5})$ can be applied, indicated by the dashed vertical line.
}
\label{spiral:fig:zz_norms}
\end{figure}

While simulation of real-life physical systems remains an outstanding goal, the onset of devices capable of supporting analog quantum simulation provides a path forward in the near-term. Because of the level of control and the large number of qubits compared to digital quantum computers, these devices are attractive platforms for attempts at simulating classically inaccessible physics. Control and understanding of errors are important for any quantitative study. As demonstrated in this work for the case of analog simulations, the error on real-life devices is not always minimized by trivially minimizing the sources of error individually. Instead, parameters of the simulation must be chosen specifically to maximize the precision of the simulation metrics of interest given the physical limitations of the device. This work has outlined and shown examples of the analysis necessary to quantify the error incurred by methods of simulating Hamiltonians using analog quantum devices. Different sources of error and the interplay between them have been investigated, and the growth of these contributions with system parameters has been described. The analytic results provide worst-case guarantees for the methods presented; in practice, numerical simulations of the methods give more realistic reflections of the error. The error rate as a function of minimal $\frac{\pi}{2}$-pulse duration $\epsilon$ is used to compare the effectiveness of several time evolution methods. For methods emulating time evolution under $H_{XXX}$ starting from $H_\text{Ising}$, error rates are shown in Fig.~\ref{spiral:fig:zz_norms}, and relevant features and parameters are given in Table~\ref{spiral:tab:method_comparison}. Extensions of this analysis method may include using different norms, since typically only a small subset of the Hilbert space is of interest, such as low-energy effective field theory spaces~\cite{_ahino_lu_2021}. Furthermore, physically relevant observables may have additional protection from errors due to their locality~\cite{Heyl_2019}, and properties of the physical system being simulated may combine with algorithmic and device errors to produce unexpectedly different results~\cite{Chinni_2022}. Overall, the considerations presented in this study are important for many analog quantum simulations on devices whose evolution is naturally described by Ising-type Hamiltonians such as $H_\text{dev}$. The discussion presented is an important step toward a complete quantification of uncertainties for quantum simulation on analog devices.

\begin{table}
\centering
\begin{tabular}{||l|l|l|l|l|l||}
\hline
 Method & \makecell{Error \\rate}& 
 \makecell{$\tau_U$} & \makecell{$\vec{t}_\text{rot}$} & \makecell{$\vec{t}_\text{comp}$} & \makecell{$t_*$} \\ 
\hline
\hline
$\mathcal{S}_{1/2}$ & $\mathcal{O}\left(\sqrt{\epsilon}\right)$ & $\mathcal{O}\left(\sqrt{\epsilon}\right)$ & $(\epsilon, \epsilon, 2\epsilon)$ & $(\epsilon, \epsilon, 0)$ & $\mathcal{O}\left(\sqrt{\epsilon}\right)$ \\ 
\hline
$\mathcal{S}_1$ & $\mathcal{O}(\epsilon$) & $6\epsilon$ & $(\epsilon, \epsilon, 2\epsilon)$ & $(\epsilon, \epsilon, 0)$ & 0 \\ 
\hline
$\tilde{\mathcal{S}}_1$ & $\mathcal{O}(\epsilon)$ & $6\epsilon$ & $(\epsilon, 2\epsilon, \epsilon)$ & $(\epsilon, 0, \epsilon)$ & 0 \\ 
\hline
$\mathcal{S}_2$ & $\mathcal{O}(\epsilon^2)$ & $12\epsilon$ & $(2\epsilon, 2\epsilon, 4\epsilon)$ & $(2\epsilon, 2\epsilon, 0)$ & 0 \\ 
\hline
$\mathcal{C}_1$ & $\mathcal{O}(\epsilon)$ & $4\epsilon$ & - & - & - \\ 
\hline
\end{tabular}
\renewcommand{\arraystretch}{1}
\caption{{\it Comparison of time evolution methods.} The error rates, step size, and step parameters for each method to simulate $H_{XXX}$ evolution are shown. The three rightmost columns show the breakdown of $\tau_U$ into the evolution due to the rotations $t_\text{rot}$, the evolution to compensate for the rotations $t_\text{comp}$, and the optimal optional evolution time $t_*$. $t_*$ is given by Eqs.~\eqref{spiral:eq:step_size_formula} and~\eqref{spiral:eq:approximate_optimal_step_sizes}, and following similar analysis for the other Trotter sequences. For $\mathcal{C}_1$ the step size corresponds to the Floquet period $\tau_{\mathcal{C}_1} = 4\epsilon$. The essentially identical properties of $\mathcal{S}_1$ and $\tilde{\mathcal{S}}_1$ and the better error rate scaling of $\mathcal{S}_2$ at the cost of longer device time can be seen.}
\label{spiral:tab:method_comparison}
\end{table}

Current device precision capabilities may limit the ability to implement the time evolution methods that have been presented. To implement a sequence whose error grows as $\mathcal{O(\epsilon)}$, the device should have control of the pulse width at $\mathcal{O}(\epsilon^2)$ or better precision. If such control is not possible, then the errors will wash out any algorithmic scaling assumed by this analysis. Because of this, it is important to consider the level of control that current devices have. Recently, nanosecond level control has been demonstrated on a Rydberg system~\cite{wurtz2023aquila}. This corresponds to $\epsilon^2 \sim \mathcal{O}(10^{-9})$, meaning sequences using $\epsilon \sim \mathcal{O}(10^{-5})$ can be supported. This value of $\epsilon$ is shown by the dashed vertical line in Fig.~\ref{spiral:fig:zz_norms}. Although constant-field methods such as $\mathcal{C}_1$ do not require rapid switching of the magnetic field, decreasing $\epsilon$ shortens the Floquet period, which makes the engineered time evolution more accurate. In other words, the level of control of the pulse width $\epsilon$ limits both types of time evolution engineering methods considered in this work. From Fig.~\ref{spiral:fig:zz_norms}, it can be seen that at the level of control available today, $\mathcal{S}_2$ outperforms all other considered engineering methods. Unless the minimal pulse width $\epsilon \sim \mathcal{O}(10^{-1})$, there is benefit to using higher-order formulas at the cost of rapid magnetic-field switching.

Many devices that implement Hamiltonians of the form of $H_\text{dev}$ are also able to modify the geometry of the system, such as allowing arbitrary placement of sites in the experimental plane of the apparatus using optical tweezers~\cite{schlosser_ohl_demello_schaffner_preuschoff_kohfahl_birkl_2020}. The ability to run experiments with such configurations has been proposed to simulate interesting physical phenomena such as nontrivial $\theta$ dependence in non-linear $\sigma$-models~\cite{Caspar:2022llo}, dynamical quantum phase transitions~\cite{Zache:2018cqq}, systems with topological order~\cite{Samajdar_2021}, and optimizations for 2+1-d gauge theory simulations~\cite{Muller:2023nnk}. These changes to the geometry would enter the error analysis in the form of modifying the sums in the bounds of Eqs.~\eqref{spiral:eq:approximate_Trotter_error_rates} and~\eqref{spiral:eq:approximate_spiral_error_rate}.

The interactions that can natively be supported on devices described by $H_\text{dev}$ are limited. Because of this, Hamiltonian engineering is an attractive way to emulate other models. While this work constitutes an important step towards quantifying the errors involved in mapping a physical problem to an analog device, a collective study including other sources of error such as device imperfections and other limitations would be necessary for complete control of uncertainties of an analog quantum simulation. The dynamics of Heisenberg evolution are of broad interest in physics, and analog simulations are enabling their study today. Therefore, a full understanding of the impact of errors and approximations on analog quantum simulations of the Heisenberg model is vital in a path toward scientifically practical results. With the errors under control, data from simulations of systems beyond the reach of classical computers may be used to make predictions about nature.

\clearpage
\begin{subappendices}

\section{Constant-field method error rate derivation}
\label{spiral:app:constant_field_drive_appendix}

For the choice of $\vec{B}(t)$ given by~\eqref{spiral:eq:bspiral}, time evolution in the interaction picture is given by 
\begin{equation*}
    U_\text{Ising}(t,0) \ = \ \text{exp}_{\mathcal{T}}\left\{-i\int_0^tdt' H_\text{int}(t')\right\},
\end{equation*}
where $H_\text{int}(t)=\frac{1}{2}\sum_{i\neq j}J_{ij}Z_i(t)Z_{j}(t)$, $Z_i(t)=2\vec{e}(t)\cdot\vec{S}_i$, and
\begin{equation}
    \vec{e}(t) \ = \ \frac{1}{3}
    \begin{pmatrix}
        2\sqrt{2}\sin^2\left(\frac{\Omega t}{2}\right)\\
        -\sqrt{6}\sin{\Omega t}\\
        1+2\cos{\Omega t}\\
    \end{pmatrix}.
\end{equation}
The error rate at Floquet periods $\tau_{\mathcal{C}_1}=\frac{2\pi}{\Omega}$ (i.e., the period of $\vec{e}(t)$) is given by
\begin{equation}
    ER_{\mathcal{C}_1}(\tau_{\mathcal{C}_1}) \ = \ \frac{1}{\tau_{\mathcal{C}_1}}\specnorm{\text{exp}_{\mathcal{T}}\left\{-i\int_0^{\tau_{\mathcal{C}_1}}dt H_\text{int}(t)\right\} - \text{exp}\left\{-i\frac{\tau_{\mathcal{C}_1}}{3}H_{XXX}\right\}}.
\end{equation}
To get the tightest possible bounds on $ER_{\mathcal{C}_1}$ with minimal application of triangle inequalities, the two propagators are organized in powers of $\Omega$ (equivalently powers of $\tau_{\mathcal{C}_1}$). For the target evolution,
\begin{equation*}
    \text{exp}\left\{-i\frac{\tau_{\mathcal{C}_1}}{3}H_{XXX}\right\} \ = \ 1-i\frac{\tau_{\mathcal{C}_1}}{3}H_{XXX}-\frac{\tau_{\mathcal{C}_1}^2}{18}H_{XXX}^2+\mathcal{O}(\tau_{\mathcal{C}_1}^3).
\end{equation*}
Working up to second-order in a Magnus expansion, the Floquet evolution gives:
\begin{align}
    \text{exp}_{\mathcal{T}}\left\{-i\int_0^{\tau_{\mathcal{C}_1}} dt H_\text{int}(t)\right\} \ &= \ e^{-i\sum_{k=1}^{\infty}\Omega_k(\tau_{\mathcal{C}_1})}\\
    &= \ 1-\left(i\Omega_1(\tau_{\mathcal{C}_1})-i\Omega_2(\tau_{\mathcal{C}_1}) + \frac{1}{2}\Omega^2_1(\tau_{\mathcal{C}_1})\right)+\mathcal{O}(\tau_{\mathcal{C}_1}^3),\label{spiral:eq:magConst}
\end{align}
where
\begin{align}
    \Omega_1(\tau_{\mathcal{C}_1}) \ &= \ \int_0^{\tau_{\mathcal{C}_1}}dt H_\text{int}(t)=\frac{\tau_{\mathcal{C}_1}}{3}H_{XXX},\\
    \Omega_2(\tau_{\mathcal{C}_1}) \ &= \ \frac{1}{2}\int_0^{\tau_{\mathcal{C}_1}}dt_1\int_0^{t_1}dt_2 [H_\text{int}(t_1),H_\text{int}(t_2)].\label{spiral:eq:magConstSecondOrder}
\end{align}

Evaluating Eq.~\eqref{spiral:eq:magConstSecondOrder} and defining $\vec{e}_i=\vec{e}(t_i)$
\begin{align}
    \Omega_2(\tau_{\mathcal{C}_1}) \ =&- \ i\int_0^{\tau_{\mathcal{C}_1}}dt_1\int_0^{t_1}dt_2 \Bigg(8\sum_{i\neq j\neq k}J_{ij}J_{jk}\left(\vec{e}_1\cdot\vec{S}_i\right)\left((\vec{e}_1\times\vec{e}_2)\cdot\vec{S}_j\right)\left(\vec{e}_2\cdot\vec{S}_m\right) \nonumber \\ 
   &\ +2\sum_{i\neq j}J_{ij}^2\left(\vec{e}_1\cdot\vec{e}_2\right)\left(\vec{e}_1\times\vec{e}_2\right)\cdot\vec{S}_j\Bigg),
\end{align}
and the following definitions to simplify expressions,
\begin{align}
    \Lambda_{\alpha\beta\gamma}(t_1,t_2) \ &= \ \left(\vec{e}_1\right)_{\alpha}\left(\vec{e}_1\times\vec{e}_2\right)_{\beta}\left(\vec{e}_2\right)_{\gamma},\\
    \tilde{\Lambda}_{\alpha}(t_1,t_2) \ &= \ \left(\vec{e}_1\cdot\vec{e}_2\right)\left(\vec{e}_1\times\vec{e}_2\right)_{\alpha},
\end{align}
the $\Omega_1$ terms exactly cancel the target Heisenberg evolution, giving the error rate at leading-order in $\tau_{\mathcal{C}_1}$ as:
\begin{align}
    ER_{\mathcal{C}_1}(\tau_{\mathcal{C}_1}) \ &= \ \frac{1}{\tau_{\mathcal{C}_1}}\specnorm{\Omega_2(\tau_{\mathcal{C}_1})} + \mathcal{O}(\tau_{\mathcal{C}_1}^3)\\
    &\ \leq \frac{1}{2 \pi \Omega} \Biggg[\specnorm{\int_0^1\int_0^{t_1}dt_1dt_2\Lambda(t_1,t_2)}\left(\sum_{i\neq j\neq k}J_{ij}J_{jk}\right) \nonumber \\ 
    &\ \qquad+\specnorm{\int_0^1\int_0^{t_1}dt_1dt_2\tilde{\Lambda}(t_1,t_2)}\left(\sum_{i\neq j}J_{ij}^2\right)\Biggg]\\
    &= \ \frac{2\epsilon}{\pi}\Biggg[\frac{\sqrt{142+24\pi^2}}{18}\left(\sum_{i\neq j\neq k}J_{ij}J_{jk}\right)+\frac{1}{3}\left(\sum_{i\neq j}J_{ij}^2\right)\Biggg]\label{spiral:eq:e_mag}\\
    &= \ \mathcal{O}(\epsilon N).
\end{align}

\section{Derivation of the \texorpdfstring{$\frac{\pi}{2}$}{pi/2}-pulse error}\label{spiral:app:pi_2_pulse_error}
As noted in the main body of the text, the global rotation gates cannot be perfectly implemented due to the persistent Ising interaction. The deviation from the ideal behavior is found using a Magnus expansion, Following App. A of~\cite{1c}:
\begin{align}
    \Upsilon^{\pm}_X(\epsilon) \ &= \ R_X\left(\mp\frac{\pi}{2}\right)R_X^{\pm}(\epsilon) \ = \ \text{exp}\left(-i\sum_{k=1}{\chi^{\pm}_k\epsilon^k}\right)\label{spiral:secondordergates1},\\
    \Upsilon^{\pm}_Y(\epsilon) \ &= \ R_Y\left(\mp\frac{\pi}{2}\right)R_Y^{\pm}(\epsilon) \ = \ \text{exp}\left(-i\sum_{k=1}{\nu^{\pm}_k\epsilon^k}\right)\label{spiral:secondordergates2}.
\end{align}
To derive the error rate bounds, it is only necessary to work to $\mathcal{O}(\epsilon^2)$ in $ \Upsilon^{\pm}_{X,Y}$. The expressions for $\chi_{1,2}^{\pm}$ and $\nu_{1,2}^{\pm}$ are:
\begin{align}
    \chi^{\pm}_1 &= \int_0^1dt e^{\pm\frac{i\pi t}{4}\sum_iX_i}\left(\sum_{i<j}J_{ij}Z_iZ_j\right)e^{\mp\frac{i\pi t}{4}\sum_iX_i}\nonumber\\
    &= \sum_{i<j}J_{ij}\left(\frac{Y_iY_j+Z_iZ_j}{2}\pm\frac{Y_iZ_j+Z_iY_j}{\pi}\right),\\
    \nu^{\pm}_1 &= \sum_{i<j}J_{ij}\left(\frac{X_iX_j+Z_iZ_j}{2}\mp\frac{X_iZ_j+Z_iX_j}{\pi}\right),
\end{align}
and
\begin{align}
    \chi^{\pm}_2 \ &= \ \frac{-i}{2}\int_0^1\int_0^{t_1}dt_1dt_2 \nonumber \\ 
    &\ \qquad \left[e^{\pm\frac{i\pi t_1}{4}\sum_iX_i}\left(\sum_{i<j}J_{ij}Z_iZ_j\right)e^{\mp\frac{i\pi t_1}{4}\sum_iX_i},e^{\pm\frac{i\pi t_2}{4}\sum_iX_i}\left(\sum_{i<j}J_{ij}Z_iZ_j\right)e^{\mp\frac{i\pi t_2}{4}\sum_iX_i}\right] \nonumber\\
    &= \ -i\sum_{i<j,m<n}J_{ij}J_{mn}\left(\frac{1}{\pi^2}\left[Y_iY_j,Z_mZ_n\right]\pm\frac{1}{8\pi}\left[Y_iY_j-Z_iZ_j,Z_mY_n+Y_mZ_n\right]\right)\nonumber\\
    &= \ \frac{1}{\pi}\sum_{i,m\neq j}J_{ij}J_{jm}\frac{Y_iX_jZ_m+X_jZ_mY_i}{\pi}\pm\frac{Y_iY_mX_j+Z_iX_jZ_m+Y_mX_jY_i+X_jZ_mZ_i}{8},\\
    \nu_2^{\pm} \ &= \ -\frac{1}{\pi}\sum_{i,m\neq j}J_{ij}J_{jm}\frac{X_iY_jZ_m+Y_jZ_mX_i}{\pi}\mp\frac{X_iX_mY_j+Z_iY_jZ_m+X_mY_jX_i+Y_jZ_mZ_i}{8}.
\end{align}

\section{First-order product formula error}
\label{spiral:app:first_order_error_analytics}
As an example, the analysis of Sec.~\ref{spiral:sec:error_analysis} is applied to the product formulas $\mathcal{S}_{1/2}$ and $\mathcal{S}_1$, with the specific choice $H_\text{tar}=H_{XXX}$. The result of this analysis will be Eqs.~\eqref{spiral:eq:approximate_optimal_step_sizes} and~\eqref{spiral:eq:approximate_Trotter_error_rates}. The error rate formulas are
\begin{align}
    ER_{\mathcal{S}_{1/2}} \ &\leq \ \frac{1}{\tau_{\mathcal{S}_{1/2}}(t,\epsilon)}\left(t^2\specnorm{\mathcal{E}_{\mathcal{S}_{1/2};t^2}}+\epsilon\specnorm{\mathcal{E}_{\mathcal{S}_{1/2};\epsilon}}\right)\label{spiral:eq:ER_S1t},\\
    ER_{\mathcal{S}_1} \ &\leq \ \frac{1}{\tau_{\mathcal{S}_1}(t,\epsilon)}\left(t^2\specnorm{\mathcal{E}_{\mathcal{S}_1;t^2}}+\epsilon^2\specnorm{\mathcal{E}_{\mathcal{S}_1;\epsilon^2}}+t\epsilon\specnorm{\mathcal{E}_{\mathcal{S}_1;t\epsilon}}\right).\label{spiral:eq:ER_S1}
\end{align}
The choices of $t$ that optimizes the error rates are
\begin{align}
    t_{\mathcal{S}_{1/2}*} \ &= \ \epsilon^{\frac{1}{2}}\sqrt{\frac{\specnorm{\mathcal{E}_{\mathcal{S}_{1/2};\epsilon}}}{\specnorm{\mathcal{E}_{\mathcal{S}_{1/2};t^2}}}}+\mathcal{O}(\epsilon),\label{spiral:eq:topt_S1t}\\
    t_{\mathcal{S}_1*} \ &= \ \epsilon\left(\sqrt{4-\frac{2\specnorm{\mathcal{E}_{\mathcal{S}_1;t\epsilon}}-\specnorm{\mathcal{E}_{\mathcal{S}_1;\epsilon^2}}}{\specnorm{\mathcal{E}_{\mathcal{S}_1;t^2}}}}-2\right).\label{spiral:eq:topt_S1}
\end{align}
The corresponding error rate bounds are (valid for $t_{\mathcal{S}_{1/2}*},t_{\mathcal{S}_1*}>0$) are given by
\begin{align}
    ER_{\mathcal{S}_{1/2}} \ &\leq \ 2\sqrt{\specnorm{\mathcal{E}_{\mathcal{S}_{1/2};t^2}}\specnorm{\mathcal{E}_{\mathcal{S}_{1/2};\epsilon}}}\epsilon^{\frac{1}{2}}+\mathcal{O}(\epsilon),\\
    ER_{\mathcal{S}_1} \ &\leq \ \left(2\sqrt{\specnorm{\mathcal{E}_{\mathcal{S}_1;t^2}}\left(4\specnorm{\mathcal{E}_{\mathcal{S}_1;t^2}}+\specnorm{\mathcal{E}_{\mathcal{S}_1;\epsilon^2}}-2\specnorm{\mathcal{E}_{\mathcal{S}_1;t\epsilon}}\right)}+\specnorm{\mathcal{E}_{\mathcal{S}_1;t\epsilon}}-4\specnorm{\mathcal{E}_{\mathcal{S}_1;t^2}}\right)\epsilon.
\end{align}

An important point to consider is that Eq.~\eqref{spiral:eq:topt_S1} suggests that $t_{\mathcal{S}_1*}$ may be negative, while in both Eq.~\eqref{spiral:eq:ER_S1t} and~\eqref{spiral:eq:ER_S1}, $t_{\mathcal{S}_{1/2}*}$ and $t_{\mathcal{S}_1*}$ were assumed to be positive numbers. This apparent contradiction simply reflects the fact that this is an optimization problem whose solution is on the boundary $t_{\mathcal{S}_1*}=0$. It is now clear which error terms must be calculated to recover the optimal step sizes and the error rates, which is done below.

\subsection{First-order error in \texorpdfstring{$\epsilon$}{epsilon}}
At leading-order in $\epsilon$ and $t$, Eqs.~\eqref{spiral:eq:firstpseudo} and~\eqref{spiral:eq:firsttrue} take the form
\begin{align}
    \mathcal{S}_{1/2} \ &= \ \mathbbm{1} - i\sum_{i<j}J_{ij}\left(t_xX_iX_j+t_yY_iY_j+t_zZ_iZ_j\right) \nonumber \\ 
    &\ \qquad - i\epsilon\sum_{i<j}J_{ij}\left(X_iX_j+Y_iY_j+2Z_iZ_j+\frac{2}{\pi}\left(Y_iZ_j+Z_iY_j-X_iZ_j-Z_iX_j\right)\right),\\
    \mathcal{S}_1 \ &= \ R_{Z}(\pi)\nonumber \\ &\ \times \ \left(\mathbbm{1} - i\sum_{i<j}J_{ij}\left(t_xX_iX_j+t_yY_iY_j+t_zZ_iZ_j\right)- i\epsilon\sum_{i<j}J_{ij}\left(X_iX_j+Y_iY_j+2Z_iZ_j\right)\right).
\end{align}
The leading-order target evolution is $\mathbbm{1}-i\tau_U(t,\epsilon)H_{XXX}$, and matching these evolutions requires $t_x,t_y=\epsilon+t$ and $t_z=t$. These choices imply that $\tau_U(t,\epsilon)=2\epsilon+t$. With these choices, the leading-order Heisenberg evolution is the same for $\mathcal{S}_{1/2}$ and $\mathcal{S}_1$, implying that $\tau_{\mathcal{S}_{1/2}}=\tau_{\mathcal{S}_1}$. The individual terms in the expansion~\eqref{spiral:eq:pftaylor} are
\begin{align}  
    \mathcal{S}_{1/2;t} \ &= \ -i\sum_{i<j}J_{ij}\left(X_iX_j+Y_iY_j+Z_iZ_j\right),\\
    \mathcal{S}_{1;t} \ &= \ R_{Z}(\pi)\left(-i\sum_{i<j}J_{ij}\left(X_iX_j+Y_iY_j+Z_iZ_j\right)\right),\\
    \mathcal{S}_{1/2;\epsilon} &=\ -2i\sum_{i<j}J_{ij}\left(X_iX_j+Y_iY_j+Z_iZ_j+\frac{Y_iZ_j+Z_iY_j-X_iZ_j-Z_iX_j}{\pi}\right),\\
    \mathcal{S}_{1;\epsilon} \ &= \ R_{Z}(\pi)\left(-2i\sum_{i<j}J_{ij}\left(X_iX_j+Y_iY_j+Z_iZ_j\right)\right).
\end{align}
The error terms that contribute to the error rates at this order are
\begin{align}
    \mathcal{E}_{\mathcal{S}_1;t},\mathcal{E}_{\mathcal{S}_1;\epsilon} \ &= \ 0,\\
    \mathcal{E}_{\mathcal{S}_{1/2};t} \ &= \ 0,\\
    \mathcal{E}_{\mathcal{S}_{1/2};\epsilon} \ &= \ -\frac{2i}{\pi}\sum_{i<j}J_{ij}\left((Y_i-X_i)Z_j+Z_i(Y_j-X_j)\right).
\end{align}
There is $\mathcal{O}(\epsilon)$ error appearing for $\mathcal{S}_{1/2}$ which causes the $t\propto\sqrt{\epsilon}$ scaling. To find the optimal $t_*$ for $\mathcal{S}_1$, higher-order terms must be calculated.

\subsection{Higher-order error in \texorpdfstring{$\epsilon, t$}{epsilon, t}}
The error terms that contribute to the error rates are (up to a factor of $R_{Z}(\pi)$)
\begin{align}
    \mathcal{E}_{\mathcal{S}_1;t^2} \ &= \ 8i\sum_{i\neq j\neq k}J_{ij}J_{jk}\left(X_iY_jZ_k-Y_iX_jZ_k-Y_iZ_jX_k\right),\\
    \mathcal{E}_{\mathcal{S}_1;\epsilon^2} \ &= \ 2i\sum_{i\neq j\neq k} \Bigg\{ J_{ij}J_{jk}\Bigg(8Y_iZ_jX_k-4(Y_iX_j+X_iY_j)Z_k \nonumber \\ 
    & \ \ + \ \frac{7(Y_iX_jY_k-X_iY_jX_k)+Z_i(Y_j-X_j)Z_k}{\pi}\Bigg)\Bigg\} \ +\frac{12 i}{\pi}\sum_{i\neq j}J_{ij}^2(X_i-Y_i),\\
    \mathcal{E}_{\mathcal{S}_1;t\epsilon} \ &= \ 8i\sum_{i\neq j\neq k}\Bigg[J_{ij}J_{jk}\left(2Y_iZ_jX_k-\frac{3}{2}X_iY_jZ_k+\frac{1}{2}Y_iX_jZ_k+\frac{Y_iX_jY_k-X_iY_jX_k}{\pi}\right)\nonumber \\ 
    & \ \qquad+\frac{8i}{\pi}\sum_{i\neq j}J_{ij}^2(X_i-Y_i)\Bigg].
\end{align}

\subsection{Evaluation of prefactors}
\label{spiral:app:prefactor_eval}
\begin{figure*}
\centering
\begin{minipage}{0.84\linewidth}
\subfloat[]{
\includegraphics[width=0.49\linewidth]{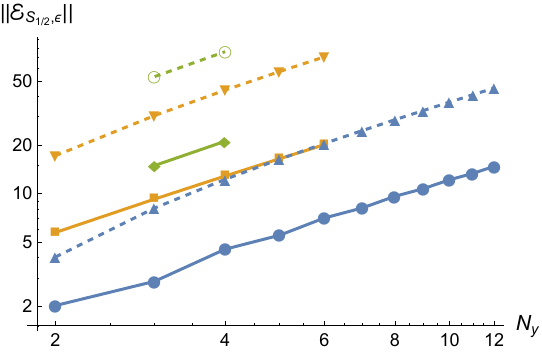}
}
\subfloat[]{
\includegraphics[width=0.49\linewidth]{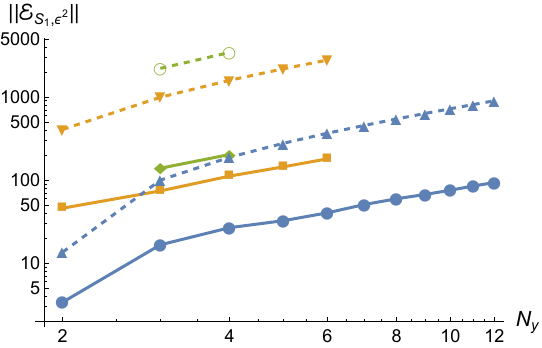}
}
\hspace{0mm}
\subfloat[]{
\includegraphics[width=0.49\linewidth]{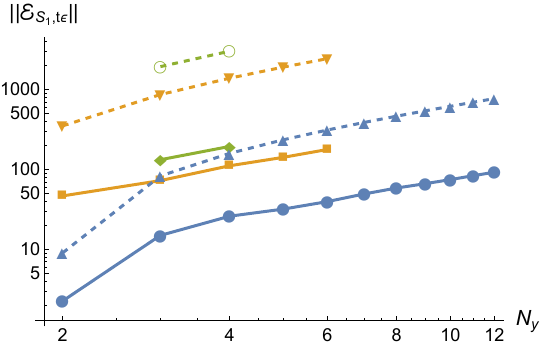}
}
\subfloat[]{
\includegraphics[width=0.49\linewidth]{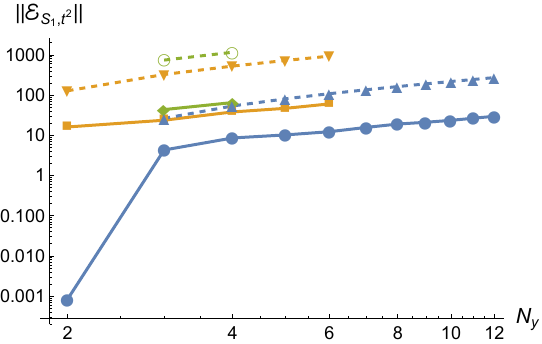}
}
\end{minipage}
\begin{minipage}{0.14\linewidth}

\subfloat{
\includegraphics[width=\linewidth]{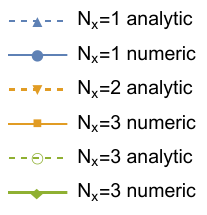}
}
\end{minipage}
\caption{{Contributions to the error rate.}
The scaling of the contributions $\mathcal{E}$ to the error from different sources in a single Trotter step of the pulse sequences $\mathcal{S}_{1/2}$ and $\mathcal{S}_1$ is shown as a function of system geometry. $N_y$ is varied on the $x$ axis and each line represents the contribution to the error with a different value of $N_x$. Here the analytic bounds of Eqs.~\eqref{spiral:eq:E1_tilde_eps_norm} -~\eqref{spiral:eq:E1_t2_norm} are compared to values obtained by numerically evaluating the norm of Eq.~\eqref{spiral:eq:deriv_terms} for each type of contribution. Note that in the latter three panels, the $N_x=1, N_y=1$ points are much lower than the rest of the $N_x=1$ lines because of the fact that there are three-body contributions to the error for these terms; when the system has less than three sites, these contributions are absent, leading to a much smaller error.}
\label{spiral:fig:individual_term_scalings}
\end{figure*}

The scaling prefactors in Eqs.~\eqref{spiral:eq:topt_S1t} and~\eqref{spiral:eq:topt_S1} require evaluation of the following quantities:
\begin{align}
    \specnorm{\mathcal{E}_{\mathcal{S}_{1/2};\epsilon}} \ &\leq \ \frac{4\sqrt{2}}{\pi}J_1,\label{spiral:eq:E1_tilde_eps_norm}\\
    \specnorm{\mathcal{E}_{\mathcal{S}_1;\epsilon^2}} \ &\leq \ 28J_3+\frac{12\sqrt{2}}{\pi}J_2,\label{spiral:eq:E1_eps2_norm}\\
    \specnorm{\mathcal{E}_{\mathcal{S}_1;t\epsilon}} \ &\leq \ 25J_3+\frac{8\sqrt{2}}{\pi}J_2,\label{spiral:eq:E1_teps_norm}\\
    \specnorm{\mathcal{E}_{\mathcal{S}_{1/2};t^2}} \ = \ \specnorm{\mathcal{E}_{\mathcal{S}_1;t^2}} \ &\leq \ 8\sqrt{3+2\sqrt{3}}J_3,\label{spiral:eq:E1_t2_norm}
\end{align}
where the factors $J_1,J_2,J_3$ are defined in Eqs.~\eqref{spiral:eq:jsums1} -~\eqref{spiral:eq:jsums3}. As explained in Sec.~\ref{spiral:sec:spiral_results}, numerics are used to see the values of the different $\mathcal{E}$ terms in practice. Eqs.~\eqref{spiral:eq:E1_tilde_eps_norm} -~\eqref{spiral:eq:E1_t2_norm} are compared to values obtained numerically as a function of system size in Fig.~\ref{spiral:fig:individual_term_scalings}. 

With these expressions, calculating Eqs.~\eqref{spiral:eq:approximate_optimal_step_sizes} and~\eqref{spiral:eq:approximate_Trotter_error_rates} is now straightforward.
\begin{align}
    t_{\mathcal{S}_{1/2}*} \ &\approx \ .3\sqrt{\frac{J_1}{J_3}}\epsilon^{\frac{1}{2}}+\mathcal{O}(\epsilon)\label{spiral:eq:topt_S1t_with_constraint},\\
    t_{\mathcal{S}_1*} \ &= \ 0\label{spiral:eq:topt_S1_with_constraint}.
\end{align}
Note that calculating $t_{\mathcal{S}_1*}$ using the bounds derived gives a negative number, which is why Eq.~\eqref{spiral:eq:topt_S1_with_constraint} is $0$; this conclusion is supported by numerics.
The corresponding optimal error rate bounds are
\begin{align}
    ER_{\mathcal{S}_{1/2}} \ &\leq \ 12.2\sqrt{J_1J_3}\epsilon^{\frac{1}{2}}+\mathcal{O}(\epsilon),\label{spiral:eq:ER_S1t_with_constraint}\\
    ER_{\mathcal{S}_1}\  &\leq \ (14J_3+2.7J_2)\epsilon.\label{spiral:eq:ER_S1_with_constraint}
\end{align}

\section{Analysis of the interaction sums}
Throughout this work, various sums over the Ising interaction terms appear. These sums encapsulate the geometry of the device used. $N_x$ is the horizontal extent of the atoms and $N_y$ is the vertical extent. These sums depend very cleanly on total number of atoms $N=N_xN_y$. In the following analysis it is assumed that the configuration of atoms in the experimental apparatus is two-dimensional and rectangular (i.e. $N_x,N_y\geq 2$). The sums analyzed are:
\begin{align}
    J_1(N) \ &= \ \sum_{i\neq j}J_{ij} = \tilde{J}_{11}N-\tilde{J}_{12},\label{spiral:eq:jsums1}\\
    J_2(N) \ &= \ \sum_{i\neq j}J_{ij}^2 = \tilde{J}_{21}N-\tilde{J}_{22},\label{spiral:eq:jsums2}\\
    J_3(N) \ &= \ \sum_{i\neq j\neq k}J_{ij}J_{jk} = \tilde{J}_{31}N-\tilde{J}_{32}.\label{spiral:eq:jsums3}
\end{align}
By explicit numerical evaluation of the sums for the considered system sizes, the following fits are found, which are plotted in Fig.~\ref{spiral:fig:jplots}:

\begin{figure*}
\centering
\subfloat{
\includegraphics[width=0.49\linewidth]{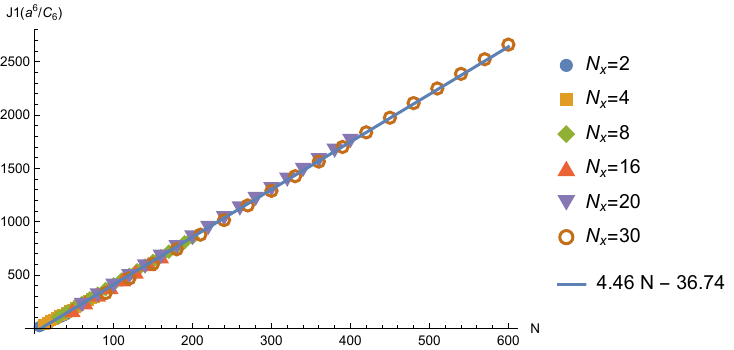}
\label{spiral:fig:appE1}
}
\subfloat{
\includegraphics[width=0.49\linewidth]{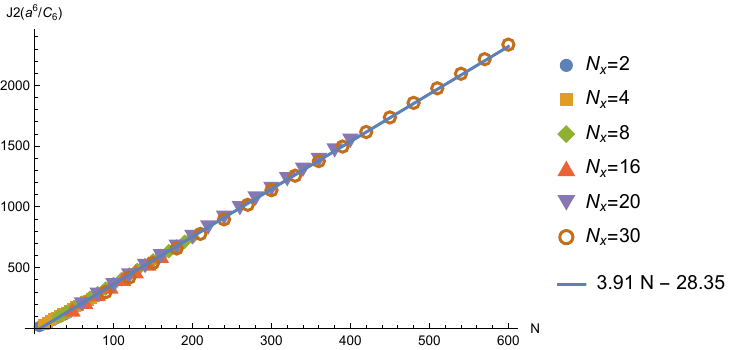}
\label{spiral:fig:appE2}
}
\hspace{0mm}
\subfloat{
\includegraphics[width=0.49\linewidth]{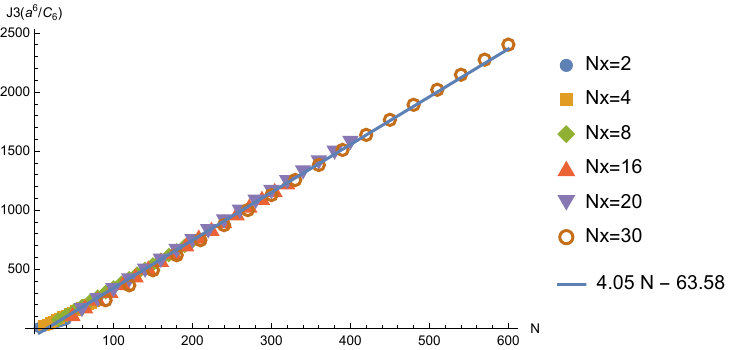}
\label{spiral:fig:appE3}
}
\caption{{\it The $J_1,J_2,J_3$ sums evaluated for explicit array configurations with fits.} The extent of the array in the $x$ direction is $N_x=20$ atoms. While there is a deviation from the for smaller array sizes, a linear scaling is still seen.}
\label{spiral:fig:jplots}
\end{figure*}

\begin{align*}
    \tilde{J}_{11} \ &= \ 4.46,\qquad\tilde{J}_{12} \ = \ -36.74,\\
    \tilde{J}_{21} \  &= \ 3.91,\qquad\tilde{J}_{22} \ = \ 28.35,\\
    \tilde{J}_{31} \ &= \ 4.05,\qquad\tilde{J}_{32} \ = \ -63.58.\\
\end{align*}

To see how this scaling arises, consider the $J_1$ sum, which may be written as:
\begin{align*}
    J_1 \ &= \ \frac{C_6}{a^6}\sum^{N_x,N_y}_{\substack{x_1=1 \\ y_1=1}}\sum^{N_x,N_y}_{\substack{x_2=x_1+1\\y_2=y_1+1}}\frac{1}{\left((x_2-x_1)^2+(y_2-y_1)^2\right)^3}\\
    &= \ \frac{C_6}{a^6}\sum^{N_x,N_y}_{\substack{x_1=1 \\ y_1=1}}\sum^{\substack{N_x-x_1 \\ N_y-y_1}}_{\substack{x'_2=1\\y'_2=1}}\frac{1}{\left((x'_2)^2+(y'_2)^2\right)^3}\\
    &\leq \ \frac{C_6}{a^6}\sum^{N_x,N_y}_{\substack{x_1=1 \\ y_1=1}}\int_1^{\infty}\frac{dx'_2dy'_2}{\left((x'_2)^2+(y'_2)^2\right)^3}.
\end{align*}
The integral converges to some constant, so we see (for $N=N_xN_y$):
\begin{equation*}
    J_1 \ = \ \mathcal{O}(Na^{-6}).
\end{equation*}
Scalings for $J_2$ and $J_3$ may be shown in a similar fashion. This scaling may be understood intuitively in the following manner. The power law interactions are rapidly decaying and so the dominant contribution to these sums will be given by a single sum over an effective nearest neighbor interaction. For a square array, this implies a $\sim 4N$ scaling, which is seen in Fig.~\ref{spiral:fig:jplots}. The deviation in the fit values from the ideal $4N$ scaling stem from the fact that the fits include values from geometrically thin arrays (e.g., $2\times 20$), where boundary interactions play a larger role.

\end{subappendices}

\chapter{Realizing error suppression in partially fault-tolerant quantum simulations with IBM quantum computers}
\label{chap:dev_422}

\noindent
{\it This chapter is associated with Ref.~\cite{Froland:2026rzt}: ``Realizing error suppression in partially fault-tolerant quantum simulations with IBM quantum computers'' by Henry Froland, Dorota M. Grabowska, Sebastian Grieninger, Jeremy Hartse, Anne L. Lashbrook, Zhiyao Li, Ziyuan Li, Sarah J.M. Powell, Martin J. Savage, Xiaojun Yao, and Nikita A. Zemlevskiy.}

\section{Introduction} 
\noindent
Performing quantum simulations with sufficient precision to compute many scientifically relevant observables requires levels of error suppression attainable only through quantum error correction (QEC)~\cite{Steane:1996ghp,Steane:1996va,Gottesman:1997zz,Gottesman:1997qd,Aliferis:2005ftz}.
The qubit and gate overhead of fault-tolerant (FT) quantum computation has so far kept large-scale QEC out of reach for practical quantum simulations.
An approach to computing with encoded qubits that is compatible with currently available hardware is to supplement FT components with non-FT (nFT) ones, 
allowing some physical errors to go undetected in exchange for lower circuit overhead~\cite{Preskill:2025cbl,Gerhard:2024peb,Reichardt:2026xbk,Akahoshi:2023xck,Yamamoto:2025iyx,Zhong:2025jox}.
In this partially FT approach, for a physical error rate $p$,\footnote{In practice many processes contribute to $p$. 
In this work $p$ characterizes the combined error rate of all of these processes, which is typically dominated by two-qubit gate errors.}
nFT components contribute a factor of $Ap$ to the logical error rate (LER), whereas FT components that remove $\mathcal O(p)$ errors provide a suppression of $Bp^2$, giving a total logical error rate of $p_L(p)=Ap+Bp^2$.
By carefully designing the nFT components such that $A\ll B\ll 1$, the computation can go below the ``pseudothreshold'', i.e., $p$ such that $p_L(p)<p$.
The nFT components generally consist of logical rotations, for which a fully FT implementation would require magic state injection~\cite{Bravyi:2004isx} or code switching~\cite{Paetznick:2013kwr,Daguerre:2025boq}. 
Although topological codes such as the surface code~\cite{Kitaev:1997wr,Dennis:2001nw} and the heavy-hex code~\cite{Chamberland:2019zev} are naturally suited to limited-connectivity devices~\cite{Benito:2024mll}, the measurement and decoding overhead of their lattice-surgery operations places them out of reach for present-day simulations~\cite{Horsman:2011hyt}.

Using multiple interacting code blocks of the $[[4,2,2]]$ Iceberg code, we perform Hamiltonian simulations on IBM's heavy-hexagonal quantum computers and achieve beyond break-even performance on the calculation of local observables.
This is done through the development of shallow syndrome extraction circuits designed for the device's native qubit connectivity.
We implement nFT logical rotations, minimizing their depth by selecting specific code block layouts and gate scheduling.
One unavoidable feature of QED codes is the exponential loss of ensemble size as the number of syndrome extraction rounds increases. 
For a computation with many code blocks, this loss prohibits running even a single round of syndrome extraction if all shots record an error and are discarded.
This work proposes a selective-filtering strategy,  Observable-Ranked Postselection (ORP), that discards detection events based on their probability of causing a logical error, determined through correlations between syndrome flips and logical observable values.

We demonstrate the benefits of these QED methods through simulations of false-vacuum decay in the 1+1D and 2+1D Mixed-Field Ising model (MFIM).
Despite a two to six times increase in circuit depth, encoded 1D simulations outperform their unencoded counterparts in estimating local observables by $2$-$6\%$ in the regimes considered in this work. 
As the logical degrees of freedom are spread across a block rather than being localized to a single physical qubit, the encoding relaxes the connectivity constraints of a logical computation. 
As a result, encoded simulations on a 2D square lattice are therefore possible at a 1.5 times lower overall circuit depth than a direct unencoded implementation. 
There, the encoding improvement grows with circuit depth, exceeding $200\%$ at late times.
Importantly, we compare the performance of encoded and unencoded circuits \emph{without any} error mitigation
schemes that utilize auxiliary circuits to ``learn'' the noise~\cite{Urbanek:2021oej,ARahman:2022tkr,Farrell:2023fgd,ZNE1,Froland:2026aff,Temme:2016vkz,PhysRevLett.120.210501,Klco:2018kyo,Kandala:2018kwe,Berg:2022ugn}.
This choice isolates the improvements due to QED from those introduced by mitigation schemes.
In both 1D and 2D, we find that infrequent error detection in systems mapped to many Iceberg-code logical qubits occupies a crossover regime between NISQ-era heuristics and fully FT computation.

Sections~\ref{dev_422:sec:422_code}--\ref{dev_422:sec:discussion} constitute the main text of this chapter and are designed to be read sequentially. 
Section~\ref{dev_422:sec:422_code} describes the implementation of multiple blocks of the $[[4,2,2]]$ code on heavy-hex connectivity, as well as the details of our detector filtering technique, ORP.
Section~\ref{dev_422:sec:results} presents results of encoded and unencoded simulations of the Ising model on {\tt ibm\_boston}.
Section~\ref{dev_422:sec:discussion} concludes with a discussion of the results and the role of encoded computations in the early FT era of quantum simulation for scientific discovery.
The Appendices contain information supporting the results in the main text.
In particular, \ref{dev_422:sec:filtering} elaborates on ORP.
Appendix~\ref{dev_422:sec:layouts} explains the procedure used to determine optimal device layouts.
Logical circuit scheduling is discussed in App.~\ref{dev_422:sec:scheduling}.
Appendices~\ref{dev_422:sec:quantum_sim} and~\ref{dev_422:sec:classical_sim}
provide the details and design choices behind our quantum and classical simulations, respectively.

\section{Fault tolerance in the \texorpdfstring{$[[4,2,2]]$}{[[4,2,2]]} code}
\label{dev_422:sec:422_code}
\noindent
The key consideration behind the FT design of logical circuits is that the spread of errors should be controlled and minimized. 
Namely, such circuits should not introduce more errors than they can detect, 
and they should not propagate detectable errors into undetectable ones. 
This work implements a version of FT that is inspired by level-1 FT (1-FT)~\cite{Aliferis:2005ftz} by choosing ``gadgets'', i.e. subcircuits that implement an encoded operation, that satisfy the following conditions:
\begin{enumerate}
    \item A gadget with no errors during operation detects one input error and outputs at most one error.
    
    \item A gadget with one error during operation and no input errors outputs at most one error.
\end{enumerate}
For modest-sized circuit gadgets, such as those used in this work, these conditions can be verified by classical simulations.
Two-qubit gates are the operations that typically suffer from the highest levels of noise and therefore set the dominant physical error rate. 
While faulty two-qubit gates can in principle produce undetectable errors at the same rate as single-qubit errors, this work finds that using these conditions as criteria for FT is sufficient for practical error suppression.

\begin{figure*}
    \centering
    \includegraphics[width=\linewidth]{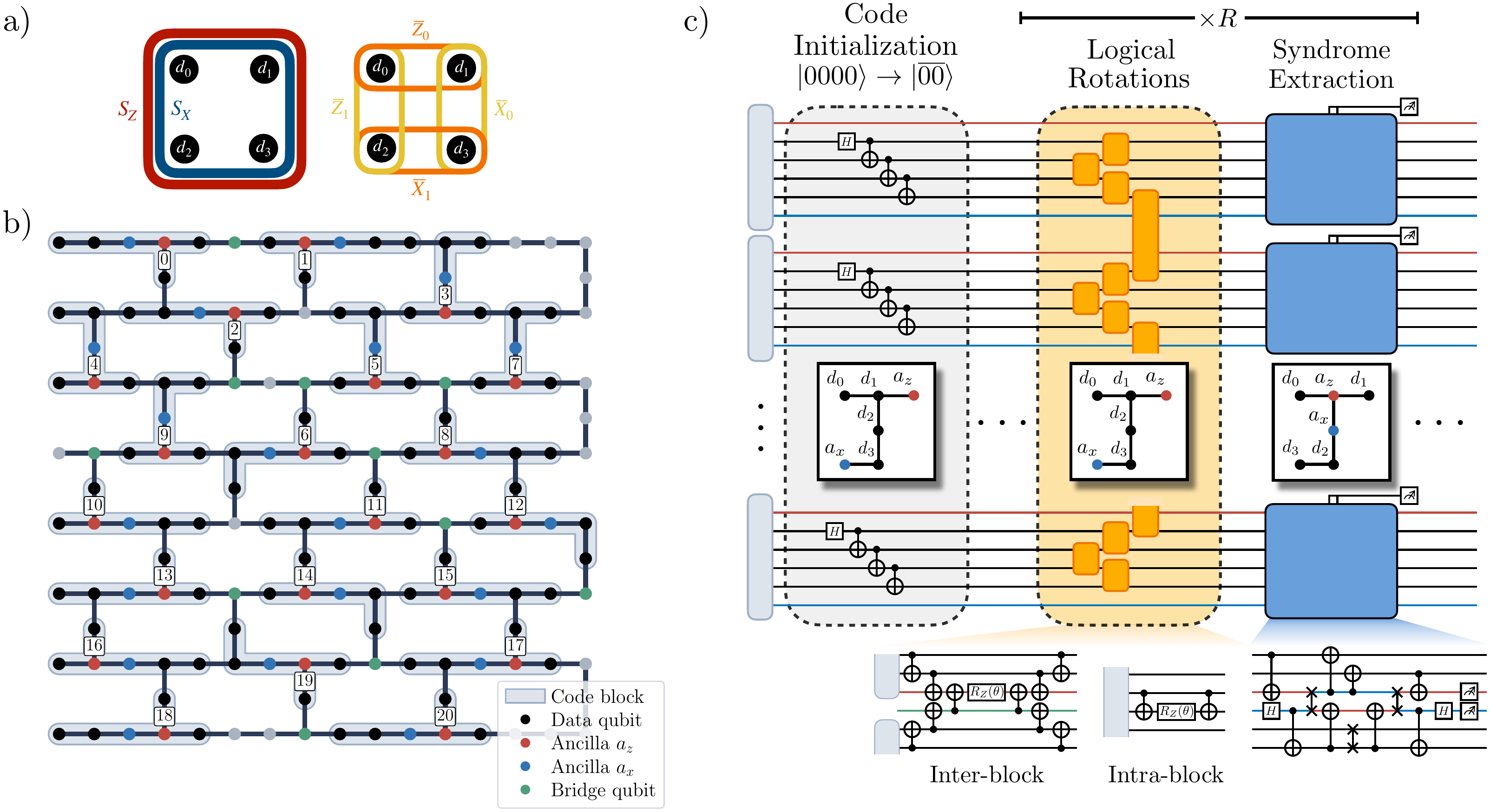}
    \caption{{\it Error detection in quantum simulations on heavy-hex topology.}
    a)~The stabilizers and logical operators that act on data qubits $d_0$-$d_3$ in the $[[4,2,2]]$ error-detecting code. 
    The $Z$-stabilizer is shown in red and the $X$-stabilizer is shown in blue. Logical operators are grouped into commuting sets represented by the color.
    b)~The placement of code blocks (grey) onto the {\tt ibm\_boston} quantum processor.
    Each code block has four data qubits (black) and two ancillas for $Z$- and $X$-stabilizer extraction (red and blue respectively).
    Bridge qubits between blocks used for routing are shown in green.
    c)~Circuit operation for encoded quantum simulations.
    Dashed outlines indicate nFT operations, while solid outlines indicate FT gadgets.
    Each code block starts in the ``compute'' configuration (shown in the left box in the center).
    Within each code block, the logical $|\overline{00}\rangle$ state is prepared with a nFT circuit. 
    Non-FT logical operations (orange boxes) are performed in the compute configuration, example circuits of which are shown at the bottom.
    The blue boxes represent FT syndrome extraction done in the ``syndrome'' configuration within each block (right two boxes), which is implemented with the circuit shown in the bottom.
    Applications of logical operations and syndrome extraction are repeated $R$ times in any given computation.
    }
    \label{dev_422:fig:overview}
\end{figure*}

The $[[4,2,2]]$ code is a Calderbank-Shor-Steane (CSS)
code~\cite{Calderbank:1995dw} that forms two logical qubits from four physical qubits, and is the smallest code in the family of Iceberg codes.
It is a distance $d=2$ code and so is capable of detecting single-qubit errors.
The stabilizers of this code, shown on the left side of Fig.~\ref{dev_422:fig:overview}a), are given by 
\begin{equation}
    S_X \ = \ X_0X_1X_2X_3,\quad S_Z \ = \ Z_0Z_1Z_2Z_3 \ .
\end{equation}
These operators define the codespace as the joint $+1$-eigenspace of $S_X$ and $S_Z$, i.e., the even-parity sector of the four-qubit Hilbert space. 
The logical $\ket{\overline{00}}$ state is a four-qubit GHZ state, ${\ket{GHZ(4)}=1/\sqrt{2}(|0000\rangle + |1111\rangle)}$,
and the logical operators on the codespace are given by 
\begin{equation}
\overline X_i \ = \ X_{i+1} X_3\ , \quad \overline Z_i \ = \ Z_0 Z_{i+1}
\label{dev_422:eq:logical_ops}
\end{equation}
for $i=0,1$ and are shown on the right side of Fig.~\ref{dev_422:fig:overview}a).
Any operator that contains only an even number of physical $X$s and $Z$s preserves the codespace, so errors of this form correspond to undetectable logical errors. 
Further, since operators on the logical space are equivalent up to multiplication by the stabilizers, the only errors this code fails to detect must contain two $X$s or two $Z$s.

\subsection{The \texorpdfstring{$[[4,2,2]]$}{[[4,2,2]]} code on heavy-hex quantum computers}
\label{dev_422:sec:422_device}
\noindent
To perform encoded computations on a quantum device, it is necessary to choose a hardware embedding that allows for efficient FT syndrome extraction and shallow implementation of logical operations. 
Six physical qubits are used to represent each $[[4,2,2]]$ code block: four data qubits $d_0$-$d_3$ and two ancilla qubits $a_x,\,a_z$ used for syndrome readout and logical operations. 
The embedding of a single code block into the heavy-hex connectivity requires one degree-3 vertex and two degree-2 vertices.
Within this single-block topology, logical operations and syndrome extraction have different optimal arrangements of qubits.
We call these the ``compute'' and ``syndrome'' configurations, which are shown in Fig.~\ref{dev_422:fig:overview}c) in the center of the circuits. 
The FT syndrome extraction circuit shown at the bottom right of Fig.~\ref{dev_422:fig:overview}c) measures $S_X$ and $S_Z$ simultaneously and requires all code blocks to have the same topology.
During syndrome extraction, the ancillas $a_x$ and $a_z$ also act as mutual error flags, controlling the proliferation of ``hook'' errors that propagate from ancilla to data qubits~\cite{Chao:2017wck,Chao:2017owu,Chamberland:2018ken,Chao:2019leh}.
This circuit is derived from Ref.~\cite{Reichardt:2018kqi} and is verified to be 1-FT through classical simulations.
The circuits required to switch between the compute and syndrome configurations (and the gate overhead) are given in~\ref{dev_422:sec:quantum_sim}.

To maintain scalability and locality of logical operations, we represent logical qubits with multiple $[[4,2,2]]$ code blocks.
The optimal block placement is chosen in several steps which are explained in App.~\ref{dev_422:sec:layouts}.
Avoiding noisy two-qubit gates and qubits with high measurement error rates, the maximum number of code blocks we find is 21, corresponding to 42 logical qubits.
Logical qubits are labeled by their block and qubit number within the block: 
\begin{align}
    (b,l_b) \ 
     {\rm with}\ \  l_b\in\{0,1\}
     \ .
     \label{dev_422:eq:blb}
\end{align}
Any given placement of code blocks can realize a number of distinct logical connectivity graphs.
This work selects the placement that has the most square-grid-like logical connectivity graph from all possible 21-block placements, which is relevant for the 2D simulations discussed in Sec.~\ref{dev_422:sec:results}.
See \ref{dev_422:sec:layouts} for more details on the layout selection process. 
The chosen block placement on {\tt ibm\_boston} has four missing internal edges and is shown in Fig.~\ref{dev_422:fig:overview}b).
This block placement has both directly adjacent blocks, and blocks that have a ``bridge'' qubit between them, shown in green.

\subsection{Observable-ranked postselection (ORP)}
\label{dev_422:sec:filtering_main_text}
\noindent
Postselection is the task of identifying and discarding shots corrupted by errors. 
Syndrome measurements serve as imperfect proxies for identifying these affected shots, both due to the partially FT approach used in this work and because 
multiple errors can incorrectly return the logical state into the codespace.
As a result, errors affecting logical information will not always trigger a syndrome.
Further, a logical observable will be unaffected by errors that are causally disconnected from it or otherwise commute with it.
We introduce ORP to sidestep the exponential shot loss associated with complete postselection.\footnote{The form of the exponential shot loss is a constant plus an exponential in the number of rounds of stabilizer measurements.
The constant is determined by the ratio of the size of the codespace to the physical-qubit space.
It is typically small enough to be neglected for practical ensemble sizes.}
This method ranks error detection events by their probability of causing a logical error and uses this ranking to iteratively clean the measured dataset.

Active reset of the ancilla qubits between rounds consumes device coherence time, and so resets are omitted in this work.
Each measured outcome therefore reflects the accumulated parity of syndrome measurements in the computation up to that point~\cite{Geher:2024lkc}.
The measurement outcome for syndrome type-$g$ on block $b$ and round $r$ is denoted as $s_j=s_{(g,b,r)}$.
These are combined into ``detectors'' by comparing each stabilizer's outcome $s_j$ to its value in a previous round,
\begin{align}
    v_{j+1} \ = \ s_{j-1} \oplus s_{j+1} \ , 
    \label{dev_422:eq:detectors}
\end{align}
where $j\pm1$ is shorthand for $(g,b,r\pm1)$ and $\oplus$ is the XOR operation implementing modulo-$2$ addition.
The definition~\eqref{dev_422:eq:detectors} implies a detector only fires ($v_j=1$) when this comparison is violated. 
Crucially, in the absence of errors, all detectors will take the deterministic value $v_j=0$.

A detector $v_j$'s impact on a given observable $\overline O$ is inferred by conditioning $\langle\overline O\rangle$ on the value of $v_j$ by
\begin{align}\label{dev_422:eq:delta_def}
    \Delta_j \ = \ \big|\langle \overline O \rangle_{v_j=0} - \langle \overline O\rangle_{v_j=1} \big| \ ,  
\end{align}
which measures the correlation between detection events and logical errors that bias $\langle \overline O\rangle$. 
For the $j$th detector, $\Delta_j$ can be interpreted as quantifying the effective probability that, given a detection event, a logical error occurred~\cite{Chen:2021num,Reichardt:2026xbk,Blume-Kohout:2025kvx}.\footnote{For a local observable $\overline O$, the detectors selected by ORP are found to coincide with the backwards lightcone of $\overline O$, such that detectors causally disconnected from $\overline O$ have values of $\Delta_j$ parametrically smaller than those within the lightcone.
This is discussed in~\ref{dev_422:app:pp_lightcone}.}
Detectors are then ranked by $\Delta_j$, and ORP uses this ranking to postselect on detection events with high $\Delta_j$. 
A cut at level $k$ keeps only shots in which none of the top-$k$ detectors fire, resulting in a sequence of increasingly strict cuts with smaller ensemble sizes and progressively cleaner estimates of $\langle \overline O\rangle_k$.
Each time the cutoff $k$ is increased, the proportion of shots containing harmful errors relative to the remaining ensemble size decreases, and so the expectation value of a logical observable progressively approaches the noiseless value until it plateaus at a level set by the undetectable errors.
However, the ensemble size also decreases with increasing $k$, inflating the statistical uncertainty on $\langle O\rangle_k$. 
The optimal cut level therefore balances the bias from residual device noise against the variance from shot loss.

Since the residual bias is bounded below by the undetectable errors, once the shots carrying the dominant detectable errors have been removed $\langle \overline O\rangle_k$ stops moving and its distribution is consistent with statistical fluctuations alone.
We identify plateaus in $k$ over which this criterion holds, indicating that the estimate has converged. 
The optimal cut level $k^*$ is chosen to be at the lower edge of the most stable plateau, and ORP uses this value to compute $\langle \overline O\rangle_{k^*}$.
Technical details of ORP and the plateau-finding algorithm are given in App.~\ref{dev_422:sec:filtering} and alternatives considered are discussed in App.~\ref{dev_422:app:filtering_metric_details}.

\section{Error-detected quantum simulations of the Ising model}
\label{dev_422:sec:results}

\begin{figure*}
    \centering
    \includegraphics[width=\linewidth]{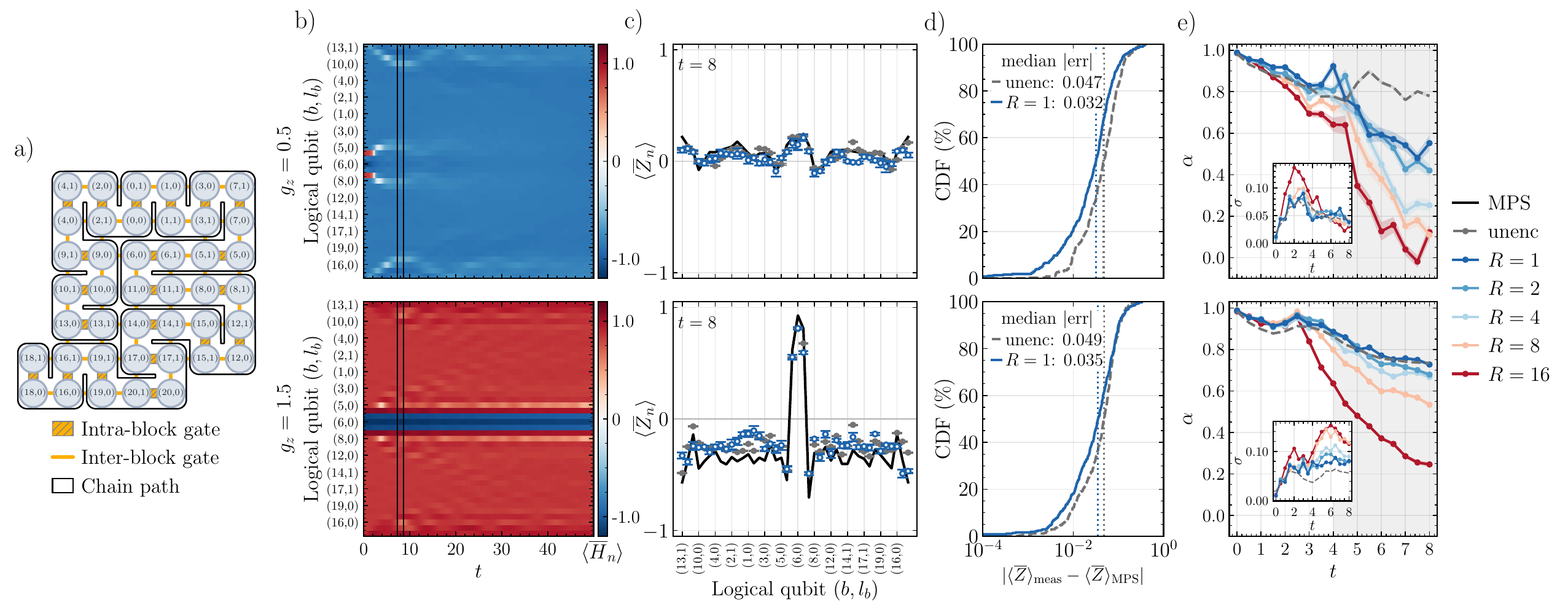}
    \caption{{\it Encoded simulations of dynamics in the 1D Ising model.} 
    a)~The realized logical connectivity grid by the block placement in Fig.~\ref{dev_422:fig:overview}b).
    Logical qubits $(b,l_b)$ (grey) are joined by intra-block gates (hatched orange) and inter-block gates (orange). 
    The black outline traces the gates used to simulate a 1D chain.
    b)~The energy density $\langle \overline{H}_n\rangle$ computed from MPS simulations as a function of time $t$ and logical qubit $n=(b,l_b)$ for $g_z=0.5$ (top) and $g_z=1.5$ (bottom).
    A size-three true-vacuum bubble located at the center is evolved with $n_T=500$ Trotter steps of size $\delta t=0.1$. 
    The black rectangle shows the $t$ for which device results are compared in c).
    c)~Comparison of encoded (blue) and unencoded (grey) magnetization $\langle \overline{Z}_n\rangle$ from {\tt ibm\_boston} to MPS simulations (black) at $t=8$, corresponding to 16 Trotter steps of $\delta t=0.5$ for $g_z=0.5$ (top) and $g_z=1.5$ (bottom).
    The encoded simulations shown have $R=1$ syndrome extraction rounds. 
    d)~The cumulative distribution function (CDF) of the per-qubit, per-time absolute error $|\langle \overline{Z}\rangle_\text{meas} - \langle \overline{Z}\rangle_\text{MPS}|$ for each value of $g_z$.
    The median error (dotted vertical lines) and CDFs corresponding to one layer ($R=1$) of syndrome measurements are shown compared to unencoded results (solid and dashed lines respectively).
    Results up to $t=4$ are used, where an encoding improvement is observed.
    e)~The signal-survival factor $\alpha$, defined in Eq.~(\ref{dev_422:eq:Zslopedef}),
    as a function of $t$ for various $R$ (colored lines) and each value of $g_z$ compared to unencoded results (dashed grey line).
    The shading for $t=4$-$8$ represents times excluded for the CDFs shown in d). 
    The insets show the residual error $\sigma$ in the fit.
    The statistical uncertainty is determined through bootstrap resampling.
    }
    \label{dev_422:fig:chain_results}
\end{figure*}
\noindent
This work focuses on simulations of false-vacuum decay under quench dynamics in the MFIM in 1D and 2D. 
False-vacuum decay has been well studied in both 1D and 2D using 
classical~\cite{Kormos:2016osj,Milsted:2020jmf,Lagnese:2021grb,Yin:2024hjm,Pavesic:2024ryc,Pavesic:2025nwm,Borla:2026fdb,Balducci:2022kvd,Pavesic:2024ryc,Pavesic:2026yiz,Lerose:2019jrs,Lagnese:2021hjt} and quantum methods~\cite{Darbha:2024srr,Chao:2025rhr,Luo:2025qlg,Humar:2026gbs}.
This work extends the tensor network study of Ref.~\cite{Pavesic:2024ryc} to nonzero longitudinal field.
The MFIM Hamiltonian with open boundary conditions (OBCs) is
\begin{align}
\overline{H} \ = \ -J\sum_{\langle ij\rangle} \overline{Z}_i \overline{Z}_j \ - \sum_i \left (g_x \overline{X}_i \ + \ g_z \overline{Z}_i \right ) \ .
\label{dev_422:eq:h_ising}
\end{align}
The sum over $\langle ij\rangle$ denotes nearest-neighbor coupling in 1D and couples adjacent logical qubits on a 2D lattice.

False-vacuum decay phenomenology models the nucleation of true-vacuum bubbles from a false-vacuum background and their subsequent evolution~\cite{Coleman:1977py,Callan:1977pt}.
This work studies the quench dynamics of product-state bubbles, i.e., the evolution of bubbles after their nucleation.
Our simulations begin in a product state ${|\psi(t=0)\rangle = \prod_{i\in B}\overline X_i |\overline1\rangle^{\otimes N}}$ on $N$ qubits, where $B$ is the initial true-vacuum bubble region. 
Recent work has predicted that bubbles melt for $|g_z|/|g_x|<1$ and remain localized for $|g_z|/|g_x|>1$ in the limit of large $J$~\cite{Balducci:2022zym}, respectively indicating the presence and absence of false-vacuum decay.
We work with the couplings $J=1,\,g_x=0.75,$ and identify a bubble-melting regime at $g_z=0.5$ and a localized regime at $g_z=1.5$, where the bubble's extent is fixed by the quench dynamics.
This phenomenon is driven by the linear potential between bubble walls that causes the momentum of excitations to wrap around the Brillouin zone, resulting in bubble oscillations known as Bloch oscillations~\cite{Pomponio:2021ltz}.
In 1D, the oscillation of the bubble width (equivalently, Stark localization of the interface~\cite{Kormos:2016osj,Lerose:2019jrs}) is predicted to have an amplitude set by the ratio $r_\text{max} \sim |g_x|/|g_z|$.

\begin{figure*}
    \centering
    \includegraphics[width=\linewidth]{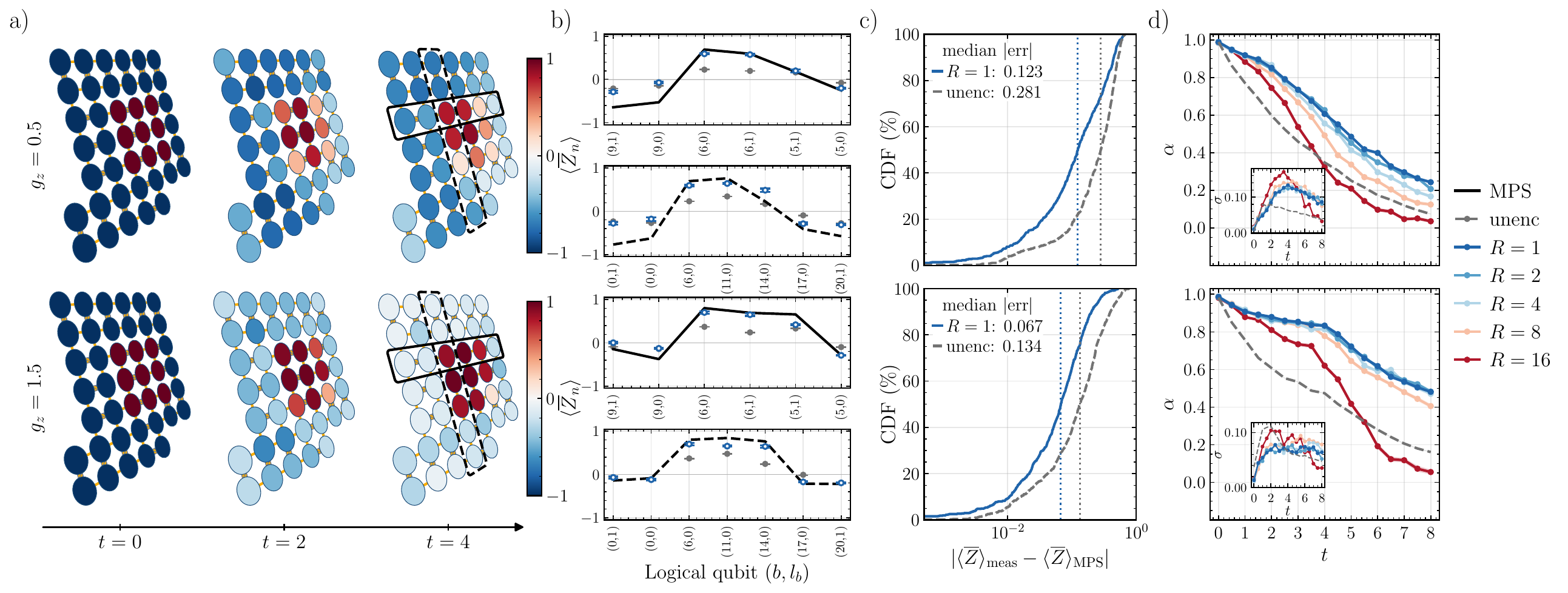}
    \caption{{\it Encoded simulations of the 2D Ising model.} 
    a)~The magnetization $\langle \overline{Z}_n\rangle$ of a $3\times3$ true-vacuum bubble as a function of time $t$ for $g_z=0.5$ (top row) and $g_z=1.5$ (bottom row),  computed from MPS simulations with $\delta t=0.1$.
    The black solid and dashed rectangles show cross-sections of the grid at $t=4$ for which data is plotted in b). 
    b)~Encoded (blue) and unencoded (grey) results from {\tt ibm\_boston} compared to MPS results (black) for the horizontal (solid) and vertical (dashed) cross-sections outlined in a), as a function of logical qubit number $n=(b,l_b)$.
    The encoded data has $R=1$ syndrome measurement rounds.
    c)~Cumulative distribution functions (CDFs) of the per-qubit, per-time absolute error $|\langle \overline{Z}\rangle_\text{meas} - \langle \overline{Z}\rangle_\text{MPS}|$ for both values of $g_z$.
    The median error and CDFs corresponding to  encoded runs with $R=1$ syndrome measurements and unencoded runs are shown (solid and dashed lines respectively).
    d)~The signal-survival factor $\alpha$ as a function of $t$ for various $R$ (colored lines) and each value of $g_z$, compared to unencoded results (dashed grey line).
    The insets show the residual error $\sigma$ in the fit.
    The statistical uncertainty is determined through bootstrap resampling.
    }
    \label{dev_422:fig:grid_results}
\end{figure*}

In quantum simulations using the encoding discussed in the previous section, logical operations between qubits $(b,0)$ and $(b,1)$ within a block $b$ are straightforwardly implemented in the compute configuration, shown in Fig.~\ref{dev_422:fig:overview}c). 
Depending on how adjacent blocks are placed relative to one another, some logical inter-block gates require routing through ancilla and bridge qubits.
These inter-block gates constitute the bulk of the depth in our time evolution circuits.
The MFIM Hamiltonian~\eqref{dev_422:eq:h_ising} is realized by the operations $\overline Z_i,\, \overline X_i,\, \overline{Z}_i \overline{Z}_j$ available on our logical connectivity graph.

\subsection{Results from IBM's quantum computers}
\noindent
We simulate the dynamics of a 1D chain under the Hamiltonian in Eq.~(\ref{dev_422:eq:h_ising}) using the logical qubit grid realized by the block placement shown in Fig.~\ref{dev_422:fig:overview}b).
By retaining all intra-block connections and removing specific inter-block edges from the grid, we implement a chain of 42 logical qubits shown in Fig.~\ref{dev_422:fig:chain_results}a).
All results shown in this section are obtained using dynamical decoupling (DD)~\cite{Viola:1998jx,Ezzell:2022uat}.

The expectation value of the energy density ${ \overline{H}_n=-J\overline Z_n\overline Z_{n+1} -(g_x \overline X_n + g_z \overline Z_n)}$ as a function of time and logical qubit $n=(b,l_b)$ determined from MPS simulations is shown in Fig.~\ref{dev_422:fig:chain_results}b).
For $g_z=0.5$ the bubble oscillates about its center, while for $g_z=1.5$ it remains localized.
This behavior is expected from the Bloch-oscillation picture of~\ref{dev_422:app:physics}, in which $r_\text{max}$ falls from well above the bubble size to comparable with it, and is consistent with the melting/localization transition reported in Ref.~\cite{Balducci:2022zym}.
The magnetization $\langle \overline Z_n\rangle$ from encoded results determined through ORP and unencoded runs on {\tt ibm\_boston} are compared to MPS results for $t=8$ in Fig.~\ref{dev_422:fig:chain_results}c).
Deviations from MPS expectations are quantified through the absolute error $|\langle \overline{Z}\rangle_\text{meas} - \langle \overline{Z}\rangle_\text{MPS}|$.
The cumulative distribution function (CDF) of the absolute errors over all qubits and $t\leq4$ is shown in Fig.~\ref{dev_422:fig:chain_results}d) for both values of $g_z$. 
The largest encoding improvement in median absolute error is found to be $ 36.1 \% \pm3.3\%$ up to $t=4$ for $g_z=0.5$ and $R=1$ (corresponding to a single syndrome extraction round at the end of the circuit).
The median absolute error is lower for encoded runs and the improvement is consistent for both $g_z$. 
Further, their CDFs lie predominantly to the left of unencoded CDFs, overlapping only at large absolute errors.
Error detection is found to concentrate the bulk of the distribution at small errors but retains a tail of poorly-performing samples. 
This is consistent with most qubits and gates operating just below the pseudothreshold, with several outliers.

To quantify how faithfully the encoded circuit reproduces the ideal dynamics independent of the overall signal magnitude, 
$\langle \overline Z\rangle_\text{meas}$ is fit to $\langle \overline Z\rangle_\text{MPS}$
with a line through the origin.
The slope $\alpha$ is the signal-survival factor and $\sigma$ is the root-mean-square of the fit residuals 
\begin{align}
    \langle \overline Z\rangle_\text{meas}\ = \ \alpha \langle \overline Z\rangle_\text{MPS} + \epsilon \ , \ \sigma \ = \ \sqrt{\frac{1}{n}\sum_i\epsilon_i^2} 
    \ .
    \label{dev_422:eq:Zslopedef}
\end{align}
Here the sums run over all $n$ samples of the observable (for a given qubit and time).
The signal-survival $\alpha=1$ indicates perfect retention, while $\alpha<1$ implies signal loss due to noise, so higher values of $\alpha$ at comparable $\sigma$ indicate more faithful results.
Figure~\ref{dev_422:fig:chain_results}e) shows $\alpha$ and $\sigma$ as a function of $t$ for both couplings and various numbers of syndrome extraction rounds $R$.
Here $R$ denotes the application of up to $R$ syndrome extraction rounds, with at most one syndrome round per Trotter step. 
Encoded runs for all values of $R$ are found to outperform their unencoded counterparts up to intermediate times, after which only small-$R$ encoded results for $g_z=1.5$ show an improvement in $\alpha$.
As expected, $\alpha$ trends lower with $t$ due to the accumulation of undetected errors.
The typical encoding improvement in the localized regime compared to the melting regime over all $t$ is attributed to a narrower backwards lightcone at $g_z=1.5$, which restricts error propagation to a smaller spatial region and lets the encoding detect a greater fraction of errors. 
Several points in the melting regime break the monotonic decay of $\alpha$; this is again attributed to the faster propagation of errors at $g_z=0.5$.
Since $\alpha$ is similar for unencoded runs at $g_z=0.5$ and $1.5$, this suggests that encoding is more effective when dynamics are slow and error propagation is limited.

An unforeseen feature of the encoding improvement is its dependence on $R$, represented by the colors in Fig.~\ref{dev_422:fig:chain_results}e). 
Early-time results show a modest advantage of increased syndrome extraction frequency (larger $R$), but beyond $t\sim4$ this reverses and fewer extraction rounds perform better.
This indicates that conserving coherence outweighs the benefit of additional syndrome extractions, and is due to the interplay between ORP and error behavior in the absence of ancilla resets. 
Without resets, an ancilla that detects an error remains in $|1\rangle$, and subsequent logical operations that couple to it inject effective errors that propagate through the system.
ORP removes some of this contamination, but the degradation with extraction frequency reveals its limitation: error detection events that ORP does not choose leave coherent contamination in the retained shots. 
Notably, adding resets does not significantly improve $\alpha$, indicating that the coherence-time cost of resets roughly cancels the benefit of eliminating syndrome-induced error propagation.
See App.~\ref{dev_422:app:more_analysis} for a detailed analysis of this effect.

Two-dimensional lattice simulations are implemented by using all inter- and intra-block edges in the logical connectivity graph shown in Fig.~\ref{dev_422:fig:chain_results}a).
Figure~\ref{dev_422:fig:grid_results}a) shows the magnetization $\langle \overline Z_n\rangle$ on the grid for several selected times determined from MPS simulations.
Although the dynamics in 2D is found to be slower than in 1D, an initial $3\times3$ bubble is still found to spread for $g_z=0.5$ and remain localized for $g_z=1.5$. 

Figure~\ref{dev_422:fig:grid_results}b) shows $\langle \overline Z_n\rangle$ for horizontal and vertical cross-sections on the grids in a) at $t=4$ (solid and dashed outlines respectively). 
While unencoded results show significant decoherence toward $\langle \overline Z_n\rangle=0$, encoded results computed with ORP are closer to MPS expectations. 
The encoding largely removes the geometric overhead associated with representing a square lattice on heavy-hex hardware, 
and only the depth overhead from Trotter step serialization is present. 
The retained geometric overhead is reflected in large unencoded circuit depths, and the encoding improvement is expected to be larger than in 1D simulations.
The CDFs of the absolute error over all logical qubits and times are shown in Fig.~\ref{dev_422:fig:grid_results}c), with encoded CDFs lying entirely to the left of their unencoded counterparts. 
Further, the median absolute error for encoded runs is seen to be significantly smaller than unencoded runs. 
The largest encoding improvement in median absolute error is found to be $ 56.9 \% \pm0.8\%$ for $g_z=0.5$ and $R=1$.
Both the unencoded and encoded distributions show higher median error in Fig.~\ref{dev_422:fig:grid_results}c) compared to Fig.~\ref{dev_422:fig:chain_results}d) due to increased circuit depth.

The encoded signal-survival factor $\alpha$ is found to exceed unencoded values for all times and all $R$ with the exception of $R=16$ past $t\sim4$ (Fig.~\ref{dev_422:fig:grid_results}d)). 
Similar to the CDFs, the $\alpha$ values are observed to decrease faster in 2D than in 1D. 
Further, the improvement from encoding is significantly larger than in 1D.
~\ref{dev_422:app:more_analysis} shows that $\alpha$ drops significantly and the improvement disappears for $R=4,8,16$ when postselection is removed.
This indicates that the improvement in $\alpha$ seen in Fig.~\ref{dev_422:fig:grid_results} is largely due to postselection as opposed to reduced gate depth. 
In contrast to 1D, the decay of $\alpha$ is observed to be monotonic,
which we attribute to the slower proliferation of undetectable errors.
Runs with $R=1,2$ are found to produce comparable results,
indicating that the gain from more syndrome measurement rounds is balanced with the added errors from longer circuits.

\begin{figure}
    \centering
    \includegraphics[width=0.6\linewidth]{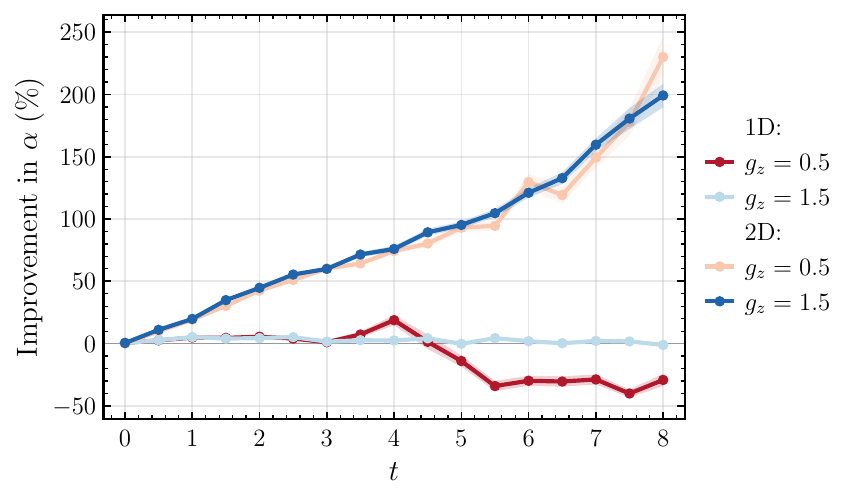}
    \caption{{\it Encoding improvement in $\alpha$.} 
    Percent improvement in $\alpha$ of encoded over unencoded results as a function of time $t$ for 1D and 2D simulations. 
    Results with $R=1$ syndrome extraction rounds for the melting regime ($g_z=0.5$) are shown in shades of red, and the localized regime ($g_z=1.5$) is shown in shades of blue.
    The shading represents statistical uncertainty determined through bootstrap resampling.
    }
    \label{dev_422:fig:improvement}
\end{figure}

Figure~\ref{dev_422:fig:improvement} shows the percentage improvement in $\alpha$ over time for $R=1$ in 1D and 2D simulations.
The improvement in the results from runs with encoding grows approximately linearly in 2D, with an average slope of $23.5\pm0.8\%$ in the melting regime and $22.8\pm0.5\%$ in the localized regime.
Because each caught error prevents corruption over a lightcone volume scaling as $\mathcal{O}(t^2)$ in 2D compared to $\mathcal{O}(t)$ in 1D, detection has greater leverage in 2D and the encoding improvement grows with $t$.
The greater than $200\%$ improvement in 2D is also driven by rapidly decohering unencoded results, as is similarly seen in Fig.~\ref{dev_422:fig:grid_results}b). 
Dynamics at $g_z=0.5$ and $g_z=1.5$ are more similar in 2D than in 1D, explaining their close trajectories in Fig.~\ref{dev_422:fig:improvement}.
Beyond $t\sim4.5$, the melting regime in 1D shows a disadvantage from encoding because of the proliferation of errors enabled by fast dynamics. 
In 1D, the encoding improvement is approximately constant: for $g_z=0.5$ an improvement is seen up to $t=4.5$ (average improvement of $5.99\pm0.99\%$), and for $g_z=1.5$ the average improvement is $2.5\pm0.23\%$ over all $t$.
Additional analysis of these results is shown in~\ref{dev_422:app:more_analysis} and data, including acceptance rates, is given in~\ref{dev_422:app:tables}.
Together, these results show that both the density of nFT gates and the dynamics being simulated affect the encoding improvement.

\section{Discussion}
\label{dev_422:sec:discussion}
\noindent
The partially FT quantum simulations performed in this work using IBM's quantum computers provide a concrete demonstration of the benefit of encoded error detection on contemporary hardware.
Despite significant overhead associated with encoding and the shot loss resulting from postselection, results obtained using our scheme show improvements for both 1D and 2D simulations. 
These results reside in a crossover regime between NISQ heuristics and fully FT device operation~\cite{Preskill:2025cbl}, which shapes our circuit design.
Our circuits combine FT syndrome extraction with nFT logical operations, resulting in jointly optimized circuit overhead and error resilience properties.
Increased frequency of syndrome extraction is seen to improve results at early times, as expected from a fully-FT computation. 
Late-time performance instead benefits from the nFT reduction in overhead provided by sparser error detection.
This balance reflects a central tradeoff in the crossover regime: more syndrome measurements aid filtering but consume some of the coherence time of the quantum computer. 
A similar tradeoff exists for the addition of ancilla resets following syndrome readout. 
Designed to operate inside this tradeoff, ORP postselects on a ranked subset of observable-relevant checks and recovers much of the benefit of full postselection without an exponential loss in statistics.
The implementation of a QEC code would avoid the filtering machinery built to manage the postselection overhead.
Our circuit design choices and error-detection scheme are tailored to the specific simulation and observables of interest.
We expect this kind of customization to be a persistent feature of the crossover regime.

The potential for encoded advantage in quantum simulation is underscored by its sensitivity to device performance. 
The absence of an advantage on devices with slightly worse error rates ($\mathcal{O}(10^{-3})$) and coherence times ($\mathcal{O}(100\mu \text s)$) than {\tt ibm\_boston} shows that even modest improvement in device characteristics is enough to cross the pseudothreshold.
Since our figure of merit is local observable estimation, our results do not directly imply a favorable logical error-rate scaling.
However, the improvement in observable estimation suggests that state-of-the-art devices are in the vicinity of the pseudothreshold for encoding procedures with a small footprint.
This reinforces the understanding that any improvement in the crossover regime is tied to specifics of code choice, overhead, and hardware operating point rather than being a generic feature of the encoding, and should become more robust and widespread as physical fidelities improve.

While the encoded improvements found in this work hold across dynamics and dimensionality, both encoded and unencoded results would likely benefit further from error mitigation.
Studying FT constructions that remove leading-order errors, alongside mitigation techniques that suppress higher-order contributions, is a very promising direction and should be pursued. 
This includes both standard error mitigation techniques based on physical-noise learning~\cite{Wallman:2015uzh,Farrell:2023fgd,Urbanek:2021oej,Froland:2026aff,ARahman:2022tkr,Berg:2020ibi,Temme:2016vkz,Berg:2022ugn}, as well as potential techniques operating directly on syndrome information.
The incorporation of such techniques involves circuit and observable-dependent design choices, and the implications for the encoded advantage remain to be seen.

Beyond the direct advantage in observable estimation, encoding expands the versatility of logical computations where no shallow unencoded analog exists. 
It enables simulations whose direct unencoded implementation carries substantially higher overhead from connectivity constraints alone. 
The 2D lattice simulations in this work illustrate this: the realized connectivity graph is costly to embed directly on heavy-hex hardware. 
This approach extends to more complex logical connectivities, such as three-dimensional lattices required for lattice gauge theories, e.g. Refs.~\cite{Klco:2021lap,Bauer:2022hpo,Beck:2023xhh,Bauer:2023qgm,DiMeglio:2023nsa} and the Fermi-Hubbard model, e.g. Ref.~\cite{Esslinger:2010rek}.
As these constraints relax, we expect a more diverse range of simulations to become feasible in the crossover regime, where the dominant constraints shift to coherence time and circuit depth.
The benefit of encoding is therefore not limited to solely improving the accuracy of a given simulation, but also extends to making entirely new classes of simulations accessible.

\clearpage
\begin{subappendices}

\section{ORP details}
\label{dev_422:sec:filtering}

\subsection{Detector construction from syndromes}
\label{dev_422:sec:detector_construction}
\noindent
Syndrome measurements record the eigenvalue of a stabilizer at the time of measurement, and connecting this value to the underlying error mechanisms is done through the construction of detectors.
A detector $v_j$ is a parity of various syndrome measurements that has a deterministic value for a noiseless circuit~\cite{Derks:2024jyw}, where $j=(g,b,r)$ is a ``space-time'' label denoting the stabilizer $g$, block $b$, and round $r$ that the detector measures.
The same construction applies to both $X$-type and $Z$-type detectors: the generator index $g$ runs over the $X$ stabilizers to build the $X$ detectors and over the $Z$ stabilizers to build the $Z$ detectors, with each detector formed only from syndrome bits of a single stabilizer within a single block.
The two stabilizer types are therefore structurally identical, and differ only through the ``temporal boundaries'', or how detectors are constructed at initialization and when terminal measurements on the data qubits are made.
In this case detectors are set by the logical state in which each block is prepared and the basis in which it is measured out.

\medskip
\noindent\textit{No-reset circuits.} For the no-reset circuits used in this work, the ancilla measuring a given stabilizer is not reinitialized between cycles, so its outcome records the running parity of that stabilizer's eigenvalue history rather than its instantaneous eigenvalue.
The deterministic combination is then the parity of two syndrome measurements in the same code block two cycles apart~\cite{Geher:2024lkc},
\begin{equation}
    v_{j+1} \ = \ s_{j+1}\oplus s_{j-1} \ ,
\end{equation}
where $j\pm 1$ is shorthand for $(g,b,r\pm 1)$, so that the intervening cycle is skipped.
Taking the parity across two cycles cancels the accumulated history carried by the unreset ancilla and isolates the change in the stabilizer eigenvalue, ensuring that an isolated error flips only a small, bounded set of detectors. 
For the first detector with $j=1$, the syndrome measurements are padded with a reference value of $0$ (so no error has occurred at the beginning of the circuit). 
This padding renders the first detector deterministic for every stabilizer whose eigenvalue is fixed by the state preparation. 
In this work, each block of the $[[4,2,2]]$ code is initialized in $\ket{\overline{00}}$, which is a simultaneous $+1$ eigenstate of both $S_X$ and $S_Z$.
Both stabilizer eigenvalues are therefore fixed at preparation, and the padding yields a deterministic first detector for the $X$-type and $Z$-type stabilizers alike.
Errors during initialization will go undetected unless the preparation circuits are themselves FT, however since $\ket{GHZ(4)}$ circuits are exceptionally shallow errors during this stage do not contribute very much to the overall logical error rate.
Had the block instead been prepared in a bare computational-basis product state, only the $Z$-type stabilizers would be fixed by the preparation. 
The $X$-type eigenvalue on the other hand would be random and its first detector would form only once two in-circuit syndrome measurements are available and their relative parities are deterministic.
At the final cycle the data-qubit readout reconstructs the stabilizers compatible with the measurement basis, and this reconstructed value plays the role of $s_{j+1}$ in the boundary detector, closing the space-time volume.
For $R$ rounds, the final detector in the measurement basis reads
\begin{equation}
    v_{R+1} \ = \ s_{R-1}\oplus s_{R}\oplus\left(\bigoplus_{d\in\text{supp}(g)}d\right) \ ,
\end{equation}
where $\text{supp}(g)$ are the data qubits on which stabilizer $g$ is defined; in this work it is always the $S_Z$ stabilizer as measurements are always made in the $Z$ basis.
For the simulations with only a single stabilizer-measurement round, the two-cycle rule has no earlier in-circuit round to reference, and the $0$-padding convention makes each detector coincide with its stabilizer measurement, $v_{1}=s_{1}$. 
Because the GHZ initialization fixes both eigenvalues, this holds for the $X$-type and $Z$-type stabilizers alike, supplemented by the measurement-basis boundary detector,
\begin{align}
    v_{2} \ & = \  s_{1}\oplus\left(\bigoplus_{d\in\text{supp}(S_Z)}d\right)
    \ .
\end{align}

\medskip
\noindent\textit{Reset circuits.}
When the ancilla is instead reset to $\ket{0}$ after each syndrome measurement, its outcome reports the instantaneous stabilizer eigenvalue rather than a running parity. 
The detector then reduces to the parity of syndrome measurements in consecutive cycles,
\begin{equation}
    v_j \ = \ s_{j-1} \oplus s_{j} \ ,
\end{equation}
where the index convention is the same as in the no-reset case and the same padding convention supplies the reference value for the first detector. Similarly, the construction of the $X$ and $Z$ detectors and the initialization boundary are unchanged, and both stabilizer types again acquire a deterministic first detector from the GHZ preparation. 
The only modification at the terminal boundary is that the reconstructed data value is compared against a single syndrome measurement round,
\begin{equation}
    v_{R+1} \ = \ s_R\oplus \left( \bigoplus_{d\in\text{supp}(g)}d \right) \ ,
\end{equation}
since $s_{R}$ is already the instantaneous eigenvalue. The remaining practical difference is in which detectors are flipped by an error: a measurement fault flips a pair of \emph{adjacent} detectors, $v_j$ and $v_{j+1}$, rather than the next-nearest pair $v_j$ and $v_{j+2}$ produced in the no-reset case, so the time-like edges connect neighboring rather than next-neighboring cycles.

\subsection{Detector selection}
\label{dev_422:sec:det_selec}
\begin{figure}
    \centering
    \includegraphics[width=0.5\linewidth]{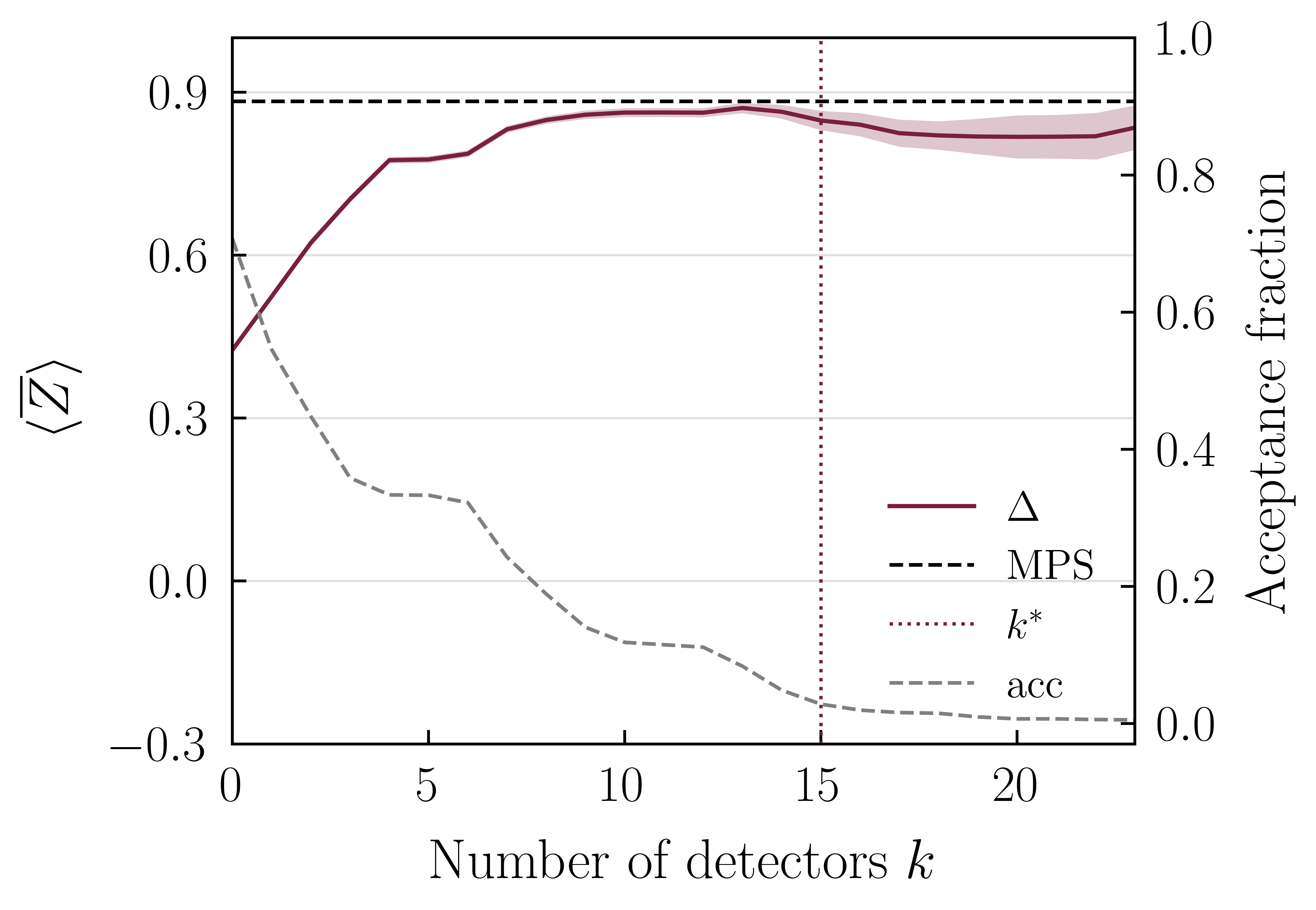}
    \caption{\textit{Convergence of ORP on {\tt ibm\_boston}}. 
    Four Trotter steps of size $\delta t=0.4$ are applied to four code blocks, each initialized in the $|\overline{00}\rangle$ state, with couplings $g_x=g_z=1$.
    The solid line shows $\langle \overline Z_{(0,0)}\rangle$ computed through ORP as a function of the number of detectors $k$ used in the postselection criterion, with the noiseless MPS prediction given by the horizontal dashed line. 
    The corresponding acceptance fraction is shown by the dotted line (right axis). 
    The vertical red dotted line marks the value of $k^*$ chosen by the plateau-finding algorithm.
    \label{dev_422:fig:nkeep_methods}}
\end{figure}

\noindent
The following plateau-finding algorithm is used with ORP to determine which detectors should be used for postselection.
\begin{enumerate}
    \item Given an observable $\overline{O}$, detectors $v_j$ are ranked\footnote{Since $\overline O$ is defined up to conjugation by $S_X$ and $S_Z$, the calculation of $\Delta_j$ depends on which data qubits are used to compute $\langle \overline O\rangle$.} in order of decreasing 
    \begin{align}
        \Delta_j \ = \ |\langle \overline{O} \rangle_{v_j=0} - \langle \overline{O}\rangle_{v_j=1} | \ .
    \end{align}
    \item A sweep over cutoff levels $k$ is done, where each level $k$ only retains shots for which none of the top-$k$ detectors record an error.
    Final blockwise parity postselection is always applied regardless of $k$.
    This produces a sequence of postselected expectation values $\langle \overline{O}\rangle_k$ with bootstrapped uncertainties $\sigma_k$.
    The acceptance fraction among all blockwise parity-even shots $f_k$ is recorded for each $k$.
    \item A reference value is built by combining the most heavily filtered levels using an inverse-variance weighted average, 
    \begin{align}
        O_\text{ref} \ = \ \frac{\sum_{k\in\{k_\text{ref}\}}\langle \overline{O}\rangle_k/\sigma_k^2}{\sum_{k\in\{k_\text{ref}\}}1/\sigma_k^2} \ ,
    \end{align}
    where $\{k_\text{ref}\}$ includes levels $k$ that satisfy ${f_\text{min}\leq f_k \leq f_\text{ref}}$.
    \item Starting from large $k$ and moving down, a plateau is found by looking for the longest run of $k$ where ${|\langle \overline{O}\rangle_k-O_\text{ref}|\leq z\sigma_k}$. 
    \item The selected value $k^*$ is the most permissive $k$ in this run.
    If a sufficiently long plateau isn't found, the cut level is set to be ${k^* = \argmin\limits_k \, [(\langle \overline{O}\rangle_k - O_\text{ref})^2 + \sigma_k^2]}$.
\end{enumerate}
This procedure is carried out on the held-out half of the results, and the resulting $k^*$ is applied to the complementary half to report $\langle O\rangle_{k^*}$.
This work uses the parameters $f_\text{min}=0.005,\,f_\text{ref}=0.05,\,z=1$, although the quality of $\langle \overline{O}\rangle_k$ is observed to be similar for a range of parameters.
Observables for all times $t$ and logical qubits $(b,l_b)$ are calculated using this algorithm. An example of the convergence of local expectation values for four blocks on {\tt ibm\_boston} as a function of the number of detectors used in the postselection criteria is shown in Fig.~\ref{dev_422:fig:nkeep_methods}. The plateau-finding algorithm successfully chooses a point within the vicinity of the noiseless MPS prediction, predicting $\langle \overline Z\rangle = 0.85(2)$ compared to an ideal value of $0.88$. While the associated acceptance fraction decreases exponentially, choosing to filter only on $k^*$ detectors does not leave the total ensemble exponentially small in system size.

\section{Block placement selection}
\label{dev_422:sec:layouts}
\noindent
The goal of layout selection is to arrange the logical qubits so that they closely resemble some target geometry, in our case a regular square lattice that we call a ``logical grid''.
The heavy-hex quantum processor's physical connectivity, combined with the choice of logical connectivity, creates a large search space of candidate logical grids. 
This can be formulated as a rectangle placement problem, since each $[[4,2,2]]$ block encodes two logical qubits, we represent it as a $2\times1$ rectangle occupying two sites of a square lattice.
Two logical qubits couple when their blocks occupy adjacent lattice sites such that they are directly connected by a physical gate.
We apply this formulation to the heavy-hex connectivity, though it generalizes to arbitrary device geometries.

We first enumerate all placements $P_i$ of blocks onto the device that maximize the number of blocks used, subject to the heavy-hex connectivity and excluding a set of qubits with poor performance. 
For {\tt ibm\_boston} (avoiding four noisy qubits), we find 42 different embeddings $P_i$ supporting 21 code blocks each, i.e., 42 logical qubits per embedding.

\begin{figure}
    \centering
    \includegraphics[width=0.4\linewidth]{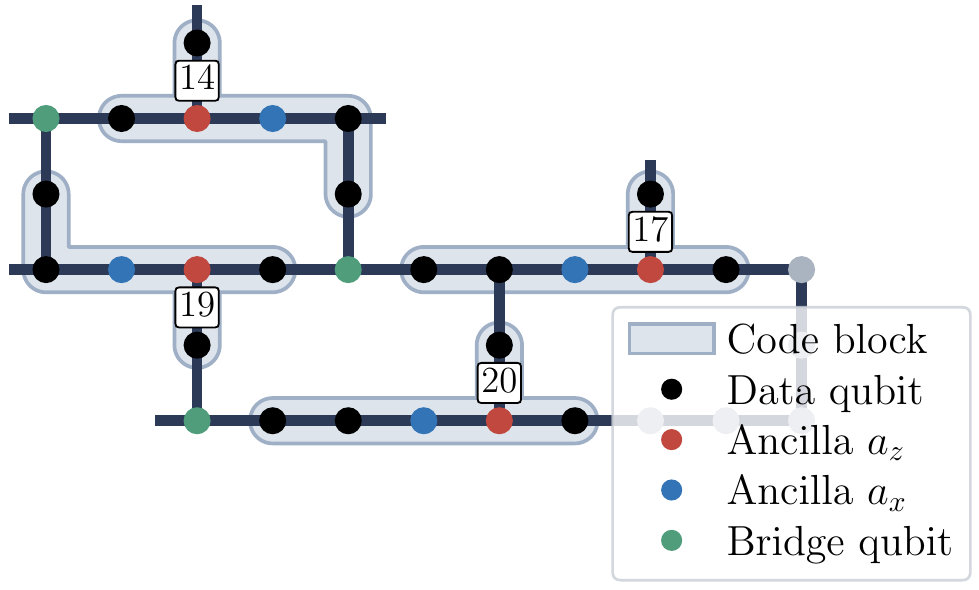}
    \caption{{\it Example $[[4,2,2]]$ code block placement onto heavy-hex connectivity}.
    Several blocks (grey) are shown placed onto a heavy-hex connected quantum processor.
    Data qubits are given in black, $Z$- and $X$-type ancillas are shown in red and blue respectively, and a bridge qubit is shown in green. }
    \label{dev_422:fig:block_placement_example}
\end{figure}

Within a given $P_i$, logical qubits within two blocks are allowed to couple if their blocks are adjacent or share a single ``bridge'' qubit on the device.
An example of this is shown in Fig.~\ref{dev_422:fig:block_placement_example}, where blocks 20 and 17 are directly adjacent, while blocks 17 and 14 are considered adjacent through the green bridge qubit.
Relabeling logical qubits $(b,0)\leftrightarrow(b,1)$ or ancillas $a_x\leftrightarrow a_z$ within a block leaves the intra-block topology unchanged, so we treat these configurations as equivalent.
For a given placement $P_i$, the realized edge set is
\begin{align}
    E(P_i) \ =\ &\underbrace{\{(b,0),(b,1)\}_{b\in P_i}}_{\text{intra-block, always present}} \nonumber\\
    &\cup\ \bigl\{\{(a,l_a),(b,l_b)\} : \nonumber\\
    &\qquad (a,l_a)\ \text{adjacent to}\ (b,l_b) \;\forall b\in P_i\bigr\}
    \ ,
\end{align}
i.e., the set of intra-block edges and available adjacent inter-block edges.
The degree of a given vertex $v$ is $\text{deg}(v) = 1+\#\{\text{inter-block edges}\}$, and we require $2\leq \text{deg}(v) \leq 4$ for all $v$.

Finding the best logical grid is a combinatorial optimization over the space of edge configurations, which is too large to search exhaustively.
Instead, we use simulated annealing~\cite{Kirkpatrick:1983zz} to find the approximate solution.
Each candidate configuration is scored as ${S(E) = |E| - \lambda (\kappa(E)-1)}$.
Here $\kappa(E)$ is the number of connected components and $\lambda$ is a large positive coefficient that penalizes disconnected graphs.
This form allows the algorithm to briefly pass through disconnected configurations to reach higher-scoring connected ones.
In the annealing search, modifications $E'$ to the current layout $E$ are accepted with the probability $p_\text{accept} = \min(1, e^{(S(E') - S(E))/T})$, where ${T(t) = T_0 \left(T_\text{end}/T_0\right)^{t/(N_\text{it}-1)}}$ is the cooling schedule which specifies the search rate.
A periodic reheating every $R_h$ steps, $T\leftarrow \max(T,T_\text{reheat})$ to escape local minima is included. 
We use the parameters ${T_0=4},{T_\text{end}=0.01},{T_\text{reheat}=2},{R_h=6000},{N_\text{it}=25000}$.

The set of allowed updates are single-rectangle relocations, single rectangle rotations, two-rectangle swaps, and three-rectangle cyclic swaps.
In addition, with probability $0.15$ an under-realized block (one that is not yet coupled to all its available neighbors) is relocated to a spot that maximizes its realized edges. 
This prevents the algorithm from getting stuck on grids with thin chains attached.
This process is repeated 500 times for every placement $P_i$ with random seeds.
The parameters for the simulated annealing are heuristically chosen to determine a suitable logical grid.
With this algorithm, we find the grid shown in Fig.~\ref{dev_422:fig:overview}b), which has 66 logical edges (45 inter-block and 21 intra-block).\footnote{In principle, the Trotter step depth could be optimized jointly with the layout. 
This work does not do so and instead fixes it in advance. Trotter step optimization is discussed in App.~\ref{dev_422:sec:scheduling}.} 
\ref{dev_422:app:other_layouts} discusses other block placements that may be more favorable on similar heavy-hex quantum processors.

\section{Circuit scheduling}
\label{dev_422:sec:scheduling}
\noindent
Once the optimal logical grid is determined, a Trotter step of the Hamiltonian in Eq.~\eqref{dev_422:eq:h_ising} must be built. 
Naive placement of inter-block $ZZ$ rotations results in deep circuits and long idle times.
As shown later in this section, finding the optimal Trotter step for a given grid reduces to determining the best placement of inter-block gates $U_e(\delta t) = e^{-i \delta t/2 \overline{Z}_{(a,l_a)} \overline{Z}_{(b,l_b)}}$.
The logical operations within a single block are defined in Eq.~\eqref{dev_422:eq:logical_ops}.
It is assumed that blocks start in the compute configuration shown in the center of Fig.~\ref{dev_422:fig:overview}c).
In each block $b$, 
$\overline{Z}_{(b,0)}$ and $\overline{X}_{(b,1)}$ can be implemented directly in this configuration while $\overline{Z}_{(b,1)}$ and $\overline{X}_{(b,0)}$ require SWAP($d_1,d_2$). 
A block's frame $f\in\{0,1\}$ records whether $d_1$ and $d_2$ have been swapped relative to their starting position.
Together with intra-block gates, inter-block gates on both logical qubits within a block require at least one frame toggle per Trotter step.

\begin{figure*}
    \centering
    \includegraphics[width=0.9\linewidth]{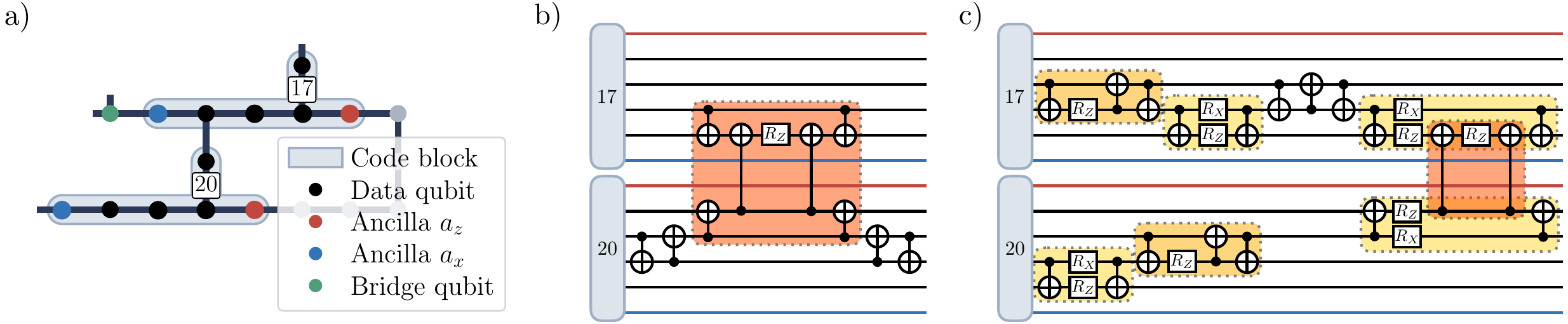}
    \caption{{\it Encoded Trotter step circuits.}
    a)~An example placement of two blocks in the compute configuration. 
    Data qubits are given in black, $Z$- and $X$-type ancillas are shown in red and blue respectively, and a bridge qubit is shown in green.
    b)~The CNOT fanout necessary for implementing a $\overline R_{ZZ}$ between logical qubits $(17,0)$ and $(20,1)$ (dark orange), with the associated SWAP gate converted to CNOTs.
    c)~A Trotter step involving logical qubits in blocks 17 and 20, with the inter-block gate between qubits $(17,0)$ and $(20,1)$.
    Single-logical-qubit gates (yellow) and intra-block $\overline R_{ZZ}$ (light orange) are folded into the CNOT fanout required for the inter-block $\overline R_{ZZ}$ gate, the last part of which is shown in dark orange. 
    Block 20 exits the Trotter step in frame $f=1$.}
    \label{dev_422:fig:trot_step_circs}
\end{figure*}

Given two blocks $a$ and $b$, and a single connection point between them (a direct connection or through a bridge qubit), the shallowest implementation of the unitary $U_e$ applies CNOT fanout on both blocks (to accumulate the required parity), applies a $Z$ rotation on the central qubit, and uncomputes the accumulated parities by applying the fanouts in reverse.\footnote{In the case where two adjacent blocks have two connections, such as blocks 14 and 19 in Fig.~\ref{dev_422:fig:block_placement_example}, the algorithm selects a schedule that may use both connections to minimize overall circuit depth.} 
An example of such a circuit between qubits $(17,0)$ and $(20,1)$ is shown in Fig.~\ref{dev_422:fig:trot_step_circs}b). 
The depths of the inter-block $ZZ$ gadgets between all neighboring pairs $(a,l_a),(b,l_b)$ are shown in Table~\ref{dev_422:tab:cost}.
\begin{table*}
\centering
\begin{tabularx}{\linewidth}{|c||Y|Y|Y|Y||c||Y|Y|Y|Y|}
\hline
\multirow{3}{*}{\makecell{Blocks\\$(a,b)$}} & \multicolumn{4}{c||}{Two-qubit gate depth} & \multirow{3}{*}{\makecell{Blocks\\$(a,b)$}} & \multicolumn{4}{c|}{Two-qubit gate depth}  \\ \cline{2-5}\cline{7-10}
 & \multicolumn{4}{c||}{$(l_a,l_b)$} &  & \multicolumn{4}{c|}{$(l_a,l_b)$} \\\cline{2-5}\cline{7-10}

 & (0,0) & (0,1) & (1,0) & (1,1) &  & (0,0) & (0,1) & (1,0) & (1,1) \\
\hline\hline
(0, 1) & 6 & 8 & 8 & 10 & (10, 13) & 4 & 8 & 6 & 8 \\
\hline
(0, 2) & 4 & 8 & 6 & 8 & (11, 12) & 6 & 8 & 8 & 10 \\
\hline
(1, 3) & 6 & 8 & 8 & 10 & (11, 14) & 4 & 8 & 6 & 8 \\
\hline
(2, 4) & 6 & 8 & 8 & 10 & (11, 15) & 6 & 8 & 8 & 10 \\
\hline
(2, 9) & 6 & 10 & 8 & 10 & (12, 15) & 6 & 8 & 8 & 10 \\
\hline
(3, 5) & 4 & 8 & 6 & 8 & (13, 14) & 6 & 8 & 8 & 10 \\
\hline
(3, 7) & 4 & 8 & 6 & 8 & (13, 16) & 4 & 8 & 6 & 8 \\
\hline
(4, 9) & 4 & 8 & 6 & 8 & (13, 19) & 6 & 10 & 8 & 10 \\
\hline
(5, 6) & 6 & 8 & 8 & 10 & (14, 15) & 4 & 8 & 6 & 8 \\
\hline
(5, 7) & 6 & 8 & 8 & 10 & (14, 17) & 8 & 10 & 10 & 12 \\
\hline
(5, 8) & 6 & 8 & 8 & 10 & (14, 19) & 6 & 10 & 8 & 10 \\
\hline
(6, 8) & 6 & 8 & 8 & 10 & (15, 17) & 4 & 8 & 6 & 8 \\
\hline
(6, 9) & 4 & 8 & 6 & 8 & (16, 18) & 4 & 8 & 6 & 8 \\
\hline
(6, 11) & 6 & 8 & 8 & 10 & (16, 19) & 6 & 8 & 8 & 10 \\
\hline
(7, 8) & 6 & 8 & 8 & 10 & (17, 19) & 6 & 10 & 8 & 10 \\
\hline
(8, 11) & 6 & 8 & 8 & 10 & (17, 20) & 4 & 8 & 6 & 8 \\
\hline
(8, 12) & 4 & 8 & 6 & 8 & (19, 20) & 6 & 10 & 8 & 10 \\
\hline
(9, 10) & 6 & 8 & 8 & 10 &  &  &  &  &  \\
\hline
    \end{tabularx}
    \caption{The inter-block $R_{ZZ}$ CNOT fanout depth for pairs of logical qubits $(l_a,l_b)$ between pairs of code blocks $(a,b)$ corresponding to the block placement shown in Fig.~\ref{dev_422:fig:overview}b).
    The depth includes CNOTs used for SWAPs and bridge qubits where necessary.}
    \label{dev_422:tab:cost}
\end{table*}
Three choices leave the logical circuit invariant, 
but may lower the depth: the orientation of block $o_b\in\{0,1\}$ (whether $a_x\leftrightarrow a_z$ is applied), $l_b\in \{0,1\}$ which specifies whether the $(b,0)\leftrightarrow (b,1)$ relabeling is applied, and $w_e$ controlling which connection to use between two blocks when multiple exist. 
We parameterize the schedule with $A = \{o,l,w\}$ for all blocks and edges in the logical grid.
Within a block, couplings in the same frame share the same accumulated $Z$-parity, so this parity is computed once per frame segment via a single fanout.

Scheduling for a fixed $A$ is itself a combinatorial optimization problem: even with the gate placements fixed, the ordering and layering of two-qubit gates into a minimum-depth circuit has a search space too large to explore exhaustively. 
The schedule for a given $A$ is determined using the constrained-satisfaction solver CP-SAT from the package {\tt ortools}~\cite{cpsatlp,ortools}.
This is possible because of commutation between the terms of each Trotter step. 
Together with fixing $f=0$ at the start of each Trotter step and placement of single qubit gates described below, this constrains the Trotter step to a single consistent ordering on all blocks, leaving the depth as the only free objective.
The objective minimizes the total depth with the additional degree of freedom that the exit frame need not be $f=0$. 
The parameters $A$ are determined by coordinate descent.

A full Trotter step of the Hamiltonian~\eqref{dev_422:eq:h_ising} can be implemented at no extra depth using the schedule of inter-block $\overline R_{ZZ}$ gates, provided the logical grid satisfies the conditions in App.~\ref{dev_422:sec:layouts}.
The single-logical qubit gates and the intra-block two-logical qubit gates can be folded into the inter-block $\overline R_{ZZ}$ gates.  
These fanouts already collect the $XX$ and $ZZ$ parities required for single qubit $\overline X$- and $\overline Z$-rotations.
Since each vertex $v$ on the selected lattice has $2\leq \text{deg}(v) \leq 4$, all four single-logical-qubit rotations can be implemented in all blocks with the addition of only single-physical-qubit rotations.
The parity required for the intra-block $\overline R_{ZZ}$ is also collected in each fanout at the time of the SWAP($d_1,d_2$) application, so the addition of this interaction also carries no extra two-qubit depth cost.
The specific location at which each single-logical-qubit gate and the intra-block $\overline R_{ZZ}$ is inserted depends on both the block's current frame and the structure of the particular inter-block $ZZ$ gadget being used.
An example of a Trotter step spanning two adjacent blocks is shown in Fig.~\ref{dev_422:fig:trot_step_circs}c), with the inter-qubit and intra-qubit gates highlighted.
Since a Trotter step does not necessarily return every block to $f=0$, consecutive odd-numbered Trotter steps are appended in reverse order, so that frames of all blocks are restored to $f=0$ after even $n_T$.

\section{Quantum simulation details}
\label{dev_422:sec:quantum_sim}
\noindent
While more efficient time evolution techniques and Trotter step implementations are possible for unencoded circuits and MPS simulations,
we execute circuits with the same Trotter order to hold observable expectation values fixed.
Further, results from encoded and unencoded runs with the same number of shots are compared, resulting in biased unencoded results with small error bars.
The ``extra'' shots in the unencoded runs could be instead used to efficiently mitigate errors in the results through noise-learning methods~\cite{Wallman:2015uzh,Farrell:2023fgd,Urbanek:2021oej,ARahman:2022tkr,Berg:2020ibi,Temme:2016vkz,Berg:2022ugn}.
Ideally, syndromes would be measured after every operation that could spread errors to multiple qubits.
In practice this is prohibitively costly due to limited device coherence time.
Tables~\ref{dev_422:tab:chain_circ_nums} and~\ref{dev_422:tab:grid_circ_nums} show the overhead in terms of depth, gate counts, and coherence time for encoded and unencoded runs in 1D and 2D, respectively.
\begin{table*}
\begin{tabularx}{\linewidth}{|c||Y|Y|Y|Y|Y|Y|}
\hline
& Unenc. & $R=1$ & $R=2$ & $R=4$ & $R=8$ & $R=16$ \\
\hline\hline
\# measurements & 42 & 126 & 168 & 252 & 420 & 756\\\hline
Measurement duration ($dt$) & 446 & 446 & 671 & 1121 & 2021 & 3821\\\hline
Total duration ($dt$) & 6017 & 10171 & 11139 & 13075 & 16947 & 26451\\\hline
Depth & 116 & 272 & 297 & 347 & 447 & 719 \\\hline
\# 2-qubit gates & 1126 & 3266 & 4112 & 5804 & 9188 & 16404\\\hline
\# shots & 32000 & 32000 & 32000 & 32000 & 32000 & 32000 \\\hline
\end{tabularx}
\caption{{\it Details of 1D chain simulations on {\tt ibm\_boston} at $t=8$}.
The columns compare unencoded counts to counts in encoded runs with various numbers of syndrome extraction rounds $R$.
The second row shows the total number of measurements (both mid-circuit and terminal).
The maximum coherence time used by measurements on any qubit in units of $dt=4\times10^{-9}$s is given in the third row.
The fourth row gives the total duration of the circuit.
Depths, total numbers of two-qubit gates, and numbers of shots executed are given in the fifth, sixth and seventh rows respectively}
\label{dev_422:tab:chain_circ_nums}
\renewcommand{\arraystretch}{1.0}
\end{table*}
\begin{table*}
\begin{tabularx}{\linewidth}{|c||Y|Y|Y|Y|Y|Y|}
\hline
& Unenc. & $R=1$ & $R=2$ & $R=4$ & $R=8$ & $R=16$ \\
\hline\hline
\# measurements & 42 & 126 & 168 & 252 & 420 & 756\\\hline
Measurement duration ($dt$) & 446 & 446 & 671 & 1121 & 2021 & 3821\\\hline
Total duration ($dt$) & 19135 & 13931 & 14935 & 16943 & 20959 & 30591 \\\hline
Depth & 626 & 443 & 470 & 524 & 632 & 912 \\\hline
\# 2-qubit gates & 4286 & 5540 & 6414 & 8162 & 11658 & 19418 \\\hline
\# shots & 32000 & 32000 & 32000 & 32000 & 32000 & 32000 \\\hline
\end{tabularx}
\caption{{\it Details of 2D grid simulations on {\tt ibm\_boston} at $t=8$}.
The same rows and columns as Table~\ref{dev_422:tab:chain_circ_nums} but for the 2D simulations presented in Sec.~\ref{dev_422:sec:results}.}
\label{dev_422:tab:grid_circ_nums}
\renewcommand{\arraystretch}{1.0}
\end{table*}
Given that the two-qubit gate duration is 
$17\ dt$ to $22\ dt$ on {\tt ibm\_boston}, 
both the encoded and unencoded circuits have significant idle time. 
The circuits in this work do not exceed the median coherence times on {\tt ibm\_boston}.\footnote{The median T1 and T2 times on {\tt ibm\_boston} were $234\mu$s and $275\mu$s when the data in this work was taken.}
However, degradation with depth is still observed, dominated by qubits with low T1/T2 times, high measurement error rates, or regions with faulty two-qubit gates.
The 1D encoded circuits are two to six times deeper than their unencoded counterparts.
This is similarly reflected in the durations and total gate counts. 
Table~\ref{dev_422:tab:grid_circ_nums} shows the geometric overhead of embedding a square lattice onto heavy-hex connectivity for the unencoded circuits, which are $\sim1.5$ times deeper than the shallowest encoded circuit. 
This is one of the driving factors of their degrading performance as seen in Fig.~\ref{dev_422:fig:grid_results}, and is also reflected in inflated gate counts and durations. 

\begin{figure}
    \centering
    \includegraphics[width=0.6\linewidth]{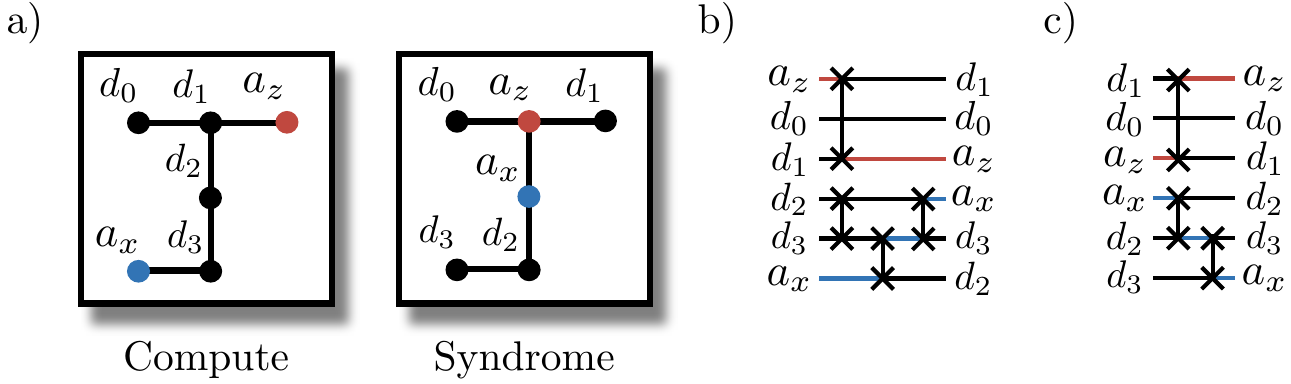}
    \caption{{\it Circuits for switching between compute and syndrome block configurations.}
    a)~The compute and syndrome block configurations as shown in the center of Fig.~\ref{dev_422:fig:overview}c). 
    Data qubits are shown in black, $Z$-type ancillas are shown in red, and $X$-type ancillas are shown in blue.
    b)~The circuit to switch from the compute to the syndrome configuration.
    c)~The corresponding circuit to switch from the syndrome to the compute configuration.
    The syndrome extraction circuit applies a SWAP$(d_2,d_3)$, which is taken into account in the configuration switching circuit.
    }
    \label{dev_422:fig:switch_layouts_circs}
\end{figure}

Switching between the compute and syndrome block configuration introduces a depth $\sim15$ overhead at each syndrome measurement.\footnote{Some of this depth is reduced by CNOT cancellations with the Trotter steps in practice.}
Figure~\ref{dev_422:fig:switch_layouts_circs} shows the SWAP circuits necessary to switch between the compute and syndrome configuration necessary for the syndrome extraction circuit, which is shown at the bottom right of Fig.~\ref{dev_422:fig:overview}c).

\section{Classical simulation details}
\label{dev_422:sec:classical_sim}
\noindent
Matrix product state simulations are used as a classical comparison to the results from {\tt ibm\_boston}. 
Despite the quench dynamics, for the 1D chain this is efficient because of the small system size, while in 2D the cost grows sharply.
Mapping an $L\times L$ grid onto a 1D MPS introduces long-range couplings, and the entanglement across a cut scales with the linear cut length.
Faithfully representing a general state on a 2D lattice requires a {\tt max\_bond} growing roughly exponentially in $L$, with per-step cost $\mathcal{O}(L^2\ {\tt max\_bond}^3)$. 
The required {\tt max\_bond} is set by the entanglement, which grows in time as the quench spreads correlations.
While more efficient classical simulation techniques for 2D lattices exist, 
such as tree tensor networks, belief propagation, and sparse Pauli dynamics, MPS is used for simplicity.

Time evolution is simulated by applying Trotter steps with ${\tt max\_bond}=1500$ and singular-value cutoff of ${\tt cutoff}=10^{-8}$.
Two sets of simulations are run. 
Late-time simulations with $\delta t=0.1$ with up to $n_T=500$ (corresponding to Fig.~\ref{dev_422:fig:mps_2d_gz_comparison}) are used to confirm the presence of physical effects absent at early times and washed out by large $\delta t$. 
While truncation errors in 1D simulations are negligible, significant error accumulates for a 2D lattice.
The missing links of the logical connectivity grid shown in Fig.~\ref{dev_422:fig:chain_results}a) modify the conventional ``snake'' MPS ordering, which likely modifies the convergence.
The error, quantified by the norm of the MPS wavefunction at $n_T=100$ (the latest time in Fig.~\ref{dev_422:fig:mps_2d_gz_comparison}) is $\mathcal{O}(10^{-3})$.
While this value is relatively large, changing {\tt max\_bond} from 1200 to 1500 only modifies observables at $\mathcal{O}(10^{-2})$ which is sufficient for this qualitative study.
The latter simulations execute the exact unencoded circuits run on ${\tt ibm\_boston}$ with $\delta t=0.5$ and up to $n_T=20$ Trotter steps for a noiseless comparison to device results.
The truncation errors in these simulations, both for 1D and 2D lattices, are negligible.
Larger values of $\delta t$ deform the dispersion relation and add an additional quench to the dynamics~\cite{Yang:2023nak,Milsted:2020jmf,Farrell:2025nkx}.
The localized regime ($g_z=1.5$) is observed to spread entanglement less than the melting regime ($g_z=0.5$).
The results of the $5\times5$ simulations displayed in Fig.~\ref{dev_422:fig:mps_2d_gz_uniform_comparison} 
were computed by {\tt qiskit} statevector simulation.

\section{Details on filtering metrics}
\label{dev_422:app:filtering_metric_details}
\noindent
This appendix presents a detailed analysis of the filtering metric $\Delta_j$ used in ORP. 
It also introduces $\beta_j$, an alternative metric based on linear regression.
First, both of these quantities are defined in~\ref{dev_422:app:metric_defs}. 
Then, a worked example on a small system of one code block and two rounds of syndrome measurements is given in~\ref{dev_422:app:worked_example} to provide a concrete implementation of these metrics.
After,~\ref{dev_422:app:correlation_description} introduces a general, high-level description of errors in terms of the correlations among detector measurements and correlations between detector and observable measurements. 
The quantities $\Delta_j$ and $\beta_j$ are then expressed in terms of the general model parameters in~\ref{dev_422:app:metric_details}, demonstrating the types of errors that they filter. 
These expressions simplify substantially when considered in a weak-noise limit, and can be seen to directly connect to the probability of an error detection event propagating to a logical error on an observable.
Finally,~\ref{dev_422:app:pp_lightcone} connects the detectors chosen based on the $\Delta_j$ ranking, as done in ORP, to the backwards lightcone of a target local observable using Pauli Propagation.

In what follows, the observables considered are Pauli strings and are denoted by $\overline{O}$ when considered as operators, with expectation values denoted as $\langle \overline{O} \rangle$. 
The result of a single measurement of such an observable will be denoted as $Q$ ($Q_i$ when the measurement is a member of an ensemble, where $i$ is the shot index), which can take values $Q\in\{-1,1\}$.
Expectation values and conditional expectation values of $Q$ over an ensemble are denoted as $\mathbb{E}[\cdot]$ and $\mathbb{E}[\cdot|\cdot]$. 
More general observables can be decomposed into a weighted sum of Pauli strings, and many of the following results can be extended to this context.

\subsection{Definitions of \texorpdfstring{$\Delta_j$}{Delta\_j} and \texorpdfstring{$\beta_j$}{Beta\_j}}
\label{dev_422:app:metric_defs}
\noindent
Recall the definition of $\Delta_j$ from Eq.~\eqref{dev_422:eq:delta_def}, rewritten as an expectation value over shots
\begin{align}\label{dev_422:eq:delta_exp}
    \Delta_j \ = \ \big|\mathbb{E}[Q | v_j=0] - \mathbb{E}[Q | v_j=1] \big| \ .
\end{align}
This quantity is calculated by taking the difference in the expectation values of observable $Q$ over shots postselected on the detector $v_j$ not detecting/detecting an error respectively. 
To define $\beta_j$, let $X_i$ be a length $N_d$ vector of detector outcomes for a single shot, so that $X$ is an $N_{s}\times (N_d+1)$ matrix where $N_d$ is the number of detectors and $N_s$ is the number of shots.
The first column of $X$ is all $1$s, and captures detector-independent behavior.
The relationship between a particular detector outcome and the value of the logical observable can be modeled as
\begin{equation}
    Q_i \ = \ \beta_0 \ +\ \sum_{j=1}^{N_{d}}\ \beta_j\ X_{ij} \ + \ \varepsilon_i
    \ ,
\end{equation}
where $\beta_0$ is a shot-independent bias term 
and $\varepsilon_i$ is an error term that accounts for observable behavior that is not explainable by the detector values. 
This relation is represented as a matrix equation $\vec{Q}=X\vec\beta+\vec \varepsilon$ where $\vec{Q}$ collects all $Q_i$ into a vector.
To minimize the cumulative error across all shots, we seek $\beta^*$ which is the solution of the following minimization
\begin{equation}\label{dev_422:eq:beta_min}
    \beta^* \ = \ \operatorname*{arg\,min}_{\beta} L(\vec{Q},X,\beta)
    \ ,
\end{equation}
where 
$L(\vec{Q},X,\beta)=||\vec{Q}-X\beta||^2_2$ is the $2$-norm of the error vector $\varepsilon$. 
The solution to Eq.~\eqref{dev_422:eq:beta_min} satisfies the \emph{normal equation}
\begin{equation}\label{dev_422:eq:normal_equations}
    X^TX\beta^* \ = \ X^T\vec{Q} \ .
\end{equation}
The metric $\beta_j$ is then the magnitude of the $j$-th element of the vector $\beta^*=(X^TX)^{-1}X^T\vec{Q}$.

\subsection{A worked example of calculating \texorpdfstring{$\Delta_j$}{Delta\_j} and \texorpdfstring{$\beta_j$}{Beta\_j}}
\label{dev_422:app:worked_example}
\noindent
As an explicit example, consider a circuit with one $[[4,2,2]]$ code block and two rounds of $S_Z$ syndrome measurements and no $S_X$ measurements (using one ancilla without resets, with measurements recorded to classical bits) corresponding to three detectors. 
The initial state is $\ket{\overline{00}}$ and the observable of interest is $\overline{O}=\overline{Z}_{0}$. 
For each shot, we consider the different times during syndrome extraction that an $X$ error can occur. 
For the set of $N_s=4$ measurements in this example, the outcomes on code block $A$ are:
\begin{equation}
\Bigl\{\ [\text{syndrome}\  |\  \text{data}]\ \Bigr\}
\ = \ 
\Bigl\{\ 
    [\,0\;0 \mid\;0\;0\;0\;0]\ ,\
    [\,1\;0 \mid\;1\;0\;0\;0]\ ,\
    [\,0\;1 \mid\;0\;0\;0\;1]\ ,\
    [\,0\;0 \mid\;1\;0\;0\;0]
    \ \Bigr\}
    \ ,
\end{equation}
where the notation $[\text{syndrome}\  |\  \text{data}]$ describes a measurement outcome. 
In the first shot, no error occurred. 
In the second shot, an error occurred before the first syndrome measurement. 
In the third shot, an error occurred between the first and second round of extraction. 
In the fourth shot, an error occurred after the second round of extraction. 
The detectors and observable outcomes (with notation $[v_1\;v_2\;v_3\;|\;Q_i]$) using the construction in App.~\ref{dev_422:sec:detector_construction} are then given by,
\begin{equation}
    [\,0\;0\;0 \mid +1\,]\ ,\quad
    [\,1\;0\;0 \mid -1\,]\ ,\quad
    [\,0\;1\;0 \mid +1\,]\ ,\quad
    [\,0\;0\;1 \mid -1\,]
    \ .
\end{equation}
The values of $\Delta_j$ can then be calculated as,\footnote{This example is a pure memory experiment on a single block, and so each detector heralds errors equally impacting the logical information causing all $\Delta_j$ to be equal. Adding more blocks or nFT logical rotations would introduce a hierarchy in the $\Delta_j$, which is the case for the simulations presented in the main text.}
\begin{equation}
    \Delta_1 \ = \ 4/3\ , \quad\Delta_2 \ = \ 4/3\ , \quad\Delta_3 \ = \ 4/3
    \ .
\end{equation}
To calculate the $\beta_j$, the detector matrix $X$ and observable vector $\vec{Q}$ are constructed as,
\begin{align}
    X \ & = \
    \begin{bmatrix}
        1 & 0 & 0 & 0 \\
1 & 1 & 0 & 0 \\
1 & 0 & 1 & 0 \\
1 & 0 & 0 & 1
    \end{bmatrix}
    \ \ ,\ \ 
\vec{Q} \ = \
\begin{bmatrix}
    +1 \\ -1 \\ +1 \\ -1
\end{bmatrix}
\ ,
\end{align}
and the vector of $\beta_j$ is given by $\beta^*=\left[(X^TX)^{-1}X^TQ\right]_{j>0}=-[2,0,2]$.

\subsection{Model of error mechanisms}
\label{dev_422:app:correlation_description}
\noindent
This section presents a higher-level description of the noise on a quantum processor, inspired by the treatments in Ref.~\cite{Chen:2021num, Kishony:2026qtg}.
This model stands in contrast to a circuit-level noise model, which describes the probabilities of a specific error occurring, where the probabilities of flipping a detector or causing a logical error are derived quantities.
Instead, this model treats error mechanisms that flip detectors as the fundamental objects and directly assigns probabilities to these events, which is favorable when compared to a circuit-level noise model because the probability of a detector flip can be directly calculated from the raw device data.
We will refer to the underlying causes of detector flips as `mechanisms' to emphasize that detector flips are due to the aggregate effect of a number of different underlying physical errors~\cite{Blume-Kohout:2025kvx}. 
In general there is no single underlying process that can be identified with the flip of a detector on a quantum computer. 

Let $G=(V,E)$ be a graph describing the errors, where $V$ is a set of nodes and $E$ are the edges. 
Each node $v_j\in V$ corresponds to a detector and can be flipped by error mechanisms $e_j,e_{ij}\in E$ with $e_j/e_{ij}=\text{Bernoulli}(p_j/p_{ij})$ respectively. 
A single index $j$ corresponds to a flip of detector $v_j$ if $e_j=1$ and the two indices $(i,j)$ correspond to different ``pair'' processes flipping two detectors $v_i,v_j$ if $e_{ij}=1$. 
Triplet processes $(i,j,k)$ and higher could be included to enhance the expressivity of this model, adding hyperedges to the error graph $G$, but analyzing these higher-order structures is beyond the scope of the current work. 
For brevity, both single node $j$ and pair processes $(i,j)$ will be denoted by a single index $m$. 
The support $\text{supp}(e_m)$ of an error mechanism $e_m$ is defined by the set of detectors that are flipped when that mechanism occurs.

If an error mechanism occurs, it flips the value of an observable $Q$ with probability $q_m$.
For an individual physical error in a Clifford circuit with Pauli noise, the observable flip probability is deterministic, $q_m \in\{0,1\}$, corresponding to whether the resulting error flips the operator.
The mechanisms of the present model are cumulative: each $(p_m, q_m)$ aggregates all microscopic faults sharing the same detector signature, with $p_m$ the combined rate of the class and $q_m$ its occurrence-weighted average flip probability. 
An intermediate value of $q_m$ therefore arises whenever the errors identified by a detector induce faults with different logical actions, and is expected even in Clifford experiments, for instance when a Pauli error coincides in detector signature with a leakage or crosstalk process. 
Intermediate values likewise arise for individual faults under non-Clifford gates.

The error mechanisms are taken to be mutually independent. 
The measured value of a logical observable $\overline{O}$ for a single shot is then represented as
\begin{equation}\label{dev_422:eq:O}
    Q \ = \ Q_I\prod_{m}(-1)^{f_m} \ ,
\end{equation}
where $Q_I\in\{-1,1\}$ is the noiseless value of $Q$\footnote{$\mathbb{E}[Q_I]$ approaches the noiseless value of the observable $\langle \overline{O}\rangle$ for a large ensemble size.} and $f_m\in\{0,1\}$ correspond to parity flips of the measured observable so that $Q\in\{-1,1\}$. 
The noiseless value $Q_I=-1$ with probability $\lambda$ depends on the encoded wavefunction and $f_m=1$ indicates process $m$ has occurred and flipped the logical observable, i.e. $P(f_m=1)=q_mp_m$. 
Separating the measured $Q$ into its ideal part $Q_I$ and associated parity flips $f_m$ allows one to treat the effect of noise independently of the encoded information in the state. 
This is an approximation that is only strictly true for Clifford circuits and restricted types of noise, e.g. stochastic Pauli noise. Therefore the circuits presented in the main text will exhibit correlations between detector and observable values that are not captured by this model.
However, its simplicity affords both analytical tractability and insight, and it is empirically found to capture certain features of device noise like dependence on the magnitude of $\langle\overline{O}\rangle$.

\subsection{Expressing \texorpdfstring{$\Delta_j$}{Delta\_j} and \texorpdfstring{$\beta_j$}{Beta\_j} in terms of detector probabilities}
\label{dev_422:app:metric_details}
\noindent
\subsubsection{Calculations for \texorpdfstring{$\Delta_j$}{Delta\_j}}
\noindent
This section expresses Eq.~\eqref{dev_422:eq:delta_exp} in terms of the probabilities $p_m,q_m$.
A detector $v_j$ can be written in terms of the error mechanisms $e_m$ that it detects as
\begin{equation}
    v_j \ = \ \bigoplus_{e_m\in E | j\in \text{supp}(e_m)} e_m
    \ ,
\end{equation}
where the condition $e_m\in E \,|\, j\in \text{supp}(e_m)$ will be denoted as $m\ni j$. 
In words, $v_j$ is defined the relative parity of the error mechanisms whose support contains it. 
First, we seek to solve for the expectation value $\mathbb{E}[Q | v_j=b]$ where $b\in\{0,1\}$. 
We begin by separating the product over $m$ in Eq.~\eqref{dev_422:eq:O}
into those whose support $\text{supp}(e_m)$ contains $j$ and those where it does not
\begin{align}\label{dev_422:eq:Qexpt}
    \mathbb{E}[Q | v_j=b] \ &= \ \mathbb{E}\left[Q_I\prod_{m\not\ni j}(-1)^{f_m}\right]\mathbb{E}\left[\prod_{m\ni j}(-1)^{f_m} \bigg|\ v_j=b\right]
    \nonumber\\
    \ &= \ \langle \overline{O}\rangle B_j\times\mathbb{E}\left[\prod_{m\ni j}(-1)^{f_m}\bigg|\ v_j=b\right] \ .
\end{align}
Here, $\langle \overline{O}\rangle=1-2\lambda$ and $B_j=\prod_{m\not\ni j}(1-2p_mq_m)$ denotes the background effect of the error mechanisms that do not touch detector $j$.
Thus, we must simplify the remaining expectation value in Eq.~\eqref{dev_422:eq:Qexpt}. Consider a single mechanism $e_n$ whose support contains $j$. 
The expectation value is written in terms of the conditional outcomes of $e_n$,
\begin{align}\label{dev_422:eq:remove_n}
    \mathbb{E}\!\left[(-1)^{\sum\limits_{m\ni j}f_m}\big|\ v_j=b\right] \ 
    &= \ \sum_{c\in\{0,1\}} \Bigg\{P(e_n=c \mid v_j=b)\,(1-2q_n)^{c}\,\nonumber \\
    &\ \ \ \  \times \ \mathbb{E}\!\left[(-1)^{\sum\limits_{m\ni j,\, m\neq n}f_m}\Big|\ 
    \bigoplus\limits_{m\ni j,\, m\neq n}e_m=b\oplus c\right]\Bigg\}
    \nonumber\\[4pt]
    &= \ P_0(e_n\mid v_j=b)\,
    \mathbb{E}\!\left[(-1)^{\sum\limits_{m\ni j,\, m\neq n}f_m}\Big|\ 
    \bigoplus\limits_{m\ni j,\, m\neq n}e_m=b\right]
    \nonumber\\
    &\hspace{30pt}
    + P_1(e_n\mid v_j=b)\,(1-2q_n)\,\nonumber \\ 
    & \hspace{60pt} \times  \ \mathbb{E}\!\left[(-1)^{\sum\limits_{m\ni j,\, m\neq n}f_m}\Big|\ 
    \bigoplus\limits_{m\ni j,\, m\neq n}e_m=\bar{b}\right] \ ,
\end{align}
where $P_b(v)=P(v=b)$. In the first line, the total parity has been conditioned on the value $e_n=c$, which shifts the target parity to $b\oplus c$ and introduces a factor $(1-2q_n)^c$; here $\oplus$ denotes addition modulo $2$. In the second line, $\bar{b}$ denotes the flipped value of $b$. The conditional probabilities are solved for by using
\begin{equation}
    P_c(e_n|v_j=b) \ = \ \frac{P(e_n=c,v_j=b)}{P(v_j=b)} \ = \ (1-c-p_n+2cp_n)\frac{P_{b\oplus c}(v_j')}{P_b(v_j)}
    \ ,
\end{equation}
where the notation $v'_j$ denotes the detector with mechanism $n$ removed. 
Both these quantities can be solved for by observing that
\begin{align}
    P_b(v_j) + P_{\bar{b}}(v_j) \ &= \ 1
    \ \ , \ \  
    P_b(v_j) - P_{\bar{b}}(v_j) \ = \ (-1)^b\ \mathbb{E}[(-1)^{v_j}]
    \ ,
\end{align}
with
\begin{equation}
    \mathbb{E}[(-1)^{v_j}] \ = \ \prod_{m\ni j}\mathbb{E}[(-1)^{e_m}] \ =\ \prod_{m\ni j}(1-2p_m)
    \ ,
\end{equation}
implying that
\begin{equation}
    P_b(v_j) \ = \ \frac{1}{2}\left(1+(-1)^b\prod_{m\ni j} (1-2p_m)\right)
    \ .
\end{equation}

Returning to evaluating Eq.~\eqref{dev_422:eq:remove_n}, the process of removing a node is now repeated. 
Denote the detector with $k$ mechanisms removed as $v_j^{(k)}=\bigoplus\limits_{m\ni j-k}e_m$, with $v_j^{(0)}=v_j$, and the expectation value with $k$ mechanisms removed as
\begin{equation}
    E_b^{(k)} \ = \ \mathbb{E}\left[(-1)^{\sum\limits_{m\ni j-k}f_m} \big|\ v_j^{(k)}=b\right]
    \ ,
\end{equation}
where Eq.~\eqref{dev_422:eq:remove_n} denotes the first step in this iteration. 
Generalizing to all nodes, the iterative equation reads
\begin{equation}\label{dev_422:eq:iteration}
    P_{b}(v_j^{(k)}) \ E_b^{(k)} \ = \ \left(1-p_{n_k}\right)\ 
    P_b\left(v_j^{(k+1)}\right)\ 
    E^{(k+1)}_b+p_{n_k}\left(1-2q_{n_k}\right)\ 
    P_{\bar{b}}\left(v_j^{(k+1)}\right)\ E_{\bar{b}}^{(k+1)}
    \ {\rm for}\ \ \ k\geq 0
    \ ,
\end{equation}
where $n_k$ denotes the removed mechanism. 
Note that the R.H.S of this equation only refers to the mechanisms that have $j$ in their support, without the removed mechanism $n_k$.
For both values of $b$, Eq.~\eqref{dev_422:eq:iteration} can be written as a matrix equation,
\begin{equation}
    V^{(k)} \ = \ M_{n_k}V^{(k+1)},\quad 
    V^{(k)} \ = \  
    \begin{bmatrix}
    P_0(v_j^{(k)})E_0^{(k)} \\
    P_1(v_j^{(k)})E_1^{(k)}
    \end{bmatrix},\quad
    M_{n_k}
     \ = \
    \begin{bmatrix}
    (1-p_{n_k}) & (1-2q_{n_k})p_{n_k} \\
    (1-2q_{n_k})p_{n_k} & (1-p_{n_k})
    \end{bmatrix}
    \ .
\end{equation}
After all $M$ mechanisms containing $j$ have been removed, the parity is even and the sum vanishes, giving the base case $V^{(M)}=(1,0)^{T}$. 
The matrix $M_{n_k}$ can be diagonalized by the basis 
$E_{\pm}^{(k)}=P_0(v_j^{(k)})\ E_0^{(k)}\pm P_1(v_j^{(k)})\ E_1^{(k)}$, 
with $E_{\pm}^{(M)}=1$, allowing the iteration relation to be solved exactly,
\begin{equation}
    E_{\pm}^{(0)} \ = \ \prod_{m\ni j} \Big[(1-p_m)\pm(1-2q_m)p_m\Big]
    \ .
\end{equation}
Rewriting this in terms of the original $E_{0}^{(0)}$ and $E_{1}^{(0)}$, the desired expectation value is then given by,
\begin{equation}
    \mathbb{E}[Q|v_j=b] \ = \ 
    \langle \overline{O}\rangle \ B_j \ 
    \frac{\prod\limits_{m\ni j}(1-2p_mq_m)+(-1)^b\prod\limits_{m\ni j}(1-2p_m(1-q_m))}{1+(-1)^b\prod\limits_{m\ni j}1-2p_m}
    \ .
\end{equation}
Finally, the filtering metric is obtained by subtracting the two conditional expectation values. Denote the three products over mechanisms containing $j$ as,
\begin{align}
    A_j \ &= \ \prod_{m\ni j}(1-2p_mq_m)\ ,\ \ 
    C_j \ = \ \prod_{m\ni j}\big(1-2p_m(1-q_m)\big)\ ,\ \ 
    D_j \ = \ \prod_{m\ni j}(1-2p_m)\ ,
\end{align}
so that 
\begin{align}
\mathbb{E}[Q|v_j=b] \ & = \ \langle \overline{O}\rangle\  B_j\ 
\frac{(A_j+(-1)^bC_j)}{(1+(-1)^bD_j)}
\ .
\end{align}
Subtracting the $b=0$ case from the $b=1$ case gives 
\begin{align}
    \Delta_j
     \ &= \ \Bigg|2\langle \overline{O}\rangle B_j\,\frac{A_jD_j-C_j}{1-D_j^2}\Bigg|
     \ .
\end{align}
The denominator is directly measurable: comparing with the parity probability derived above, ${D_j = 1-2P_1(v_j)}$, so that $1-D_j^2 = 4P_1(v_j)\big(1-P_1(v_j)\big)$. 
Expanding to leading order in the mechanism probabilities, ${A_jD_j-C_j\approx -4\sum_{m\ni j}p_mq_m}$ and ${1-D_j^2\approx 4\sum_{m\ni j}p_m}$, giving
\begin{equation}
    \Delta_j \ \approx \ \Bigg|2\langle \overline{O}\rangle \,\left(\sum\limits_{m\ni j}p_mq_m\right)\Bigg/\left(\sum\limits_{m\ni j}p_m\right)\Bigg|
    \label{dev_422:eq:DeltajestA}
    \ .
\end{equation}
The denominator can be viewed as the cumulative ``firing'' rate for all processes whose support contains detector $v_j$, while the numerator is this same quantity weighted by the probability of that process causing a logical error. 
Thus, $\Delta_j$ takes on the interpretation of an effective probability of all processes affecting $v_j$ that cause a logical error, weighted by $\langle \overline{O}\rangle$. 
As such, detectors that take on higher values of $\Delta_j$ should be postselected on first, as these are the most detrimental to the encoded information. This forms the basis of ORP.

This analysis reveals that the scale of $\Delta_j$ is set by the value of the logical observable and the background factor $B_j$. 
If $\langle \overline{O} \rangle\ll 1$ or if the background attenuation is large, i.e., $B_j\ll 1$, and all the individual firing rates $p_m$ are close to $1$, then every value of $\Delta_j$ will take on parametrically small values. 
Due to finite measurement statistics, it can become challenging to resolve differences at this scale, reducing the effectiveness of $\Delta_j$ as a quality metric distinguishing different detectors.

\subsubsection{Calculations for \texorpdfstring{$\beta_j$}{Beta\_j}}
\noindent
In the limit of many shots, $\frac{1}{N_s}X^TX$ and $\frac{1}{N_s}X^T\vec{Q}$ converge to matrices of moments.
The normal equation (Eq.~\eqref{dev_422:eq:normal_equations})
can be expressed component-wise explicitly in terms of the moments of detectors $v_j$ and observable $Q$
\begin{equation}
    \begin{bmatrix}
        1 & \mathbb{E}[v_k]^{T} \\
        \mathbb{E}[v_j] & \mathbb{E}[v_jv_k]
    \end{bmatrix}
    \begin{bmatrix}
        \beta_0^* \\ \beta_k^*
    \end{bmatrix}
    \ = \ 
    \begin{bmatrix}
        \mathbb{E}[Q] \\ \mathbb{E}[Qv_j]
    \end{bmatrix}\ ,
    \qquad j,k \ = \ 1,\dots,N_d
    \ .
\end{equation}
Eliminating $\beta_0^* = \mathbb{E}[Q]-\sum_k\mathbb{E}[v_k]\beta_k^*$ from
the remaining rows reduces this to
\begin{equation}
    G\beta^* \ = \ a
    \ ,
\end{equation}
for the connected two point function $G_{jk}=\mathbb{E}[v_jv_k]-\mathbb{E}[v_j]\mathbb{E}[v_k]$ and $a_j=\mathbb{E}[Qv_j]-\mathbb{E}[Q]\mathbb{E}[v_j]$, with the bias recovered from the eliminated row. 
Each moment is evaluated using the products derived above. 
Rewriting the detectors in terms of parities, $v_j=\tfrac{1}{2}\big(1-(-1)^{v_j}\big)$, the one-point functions are
\begin{equation}
    \mathbb{E}[v_j] \ = \ \frac{1-D_j}{2} \ ,\qquad
    \mathbb{E}[Q] \ = \ \langle \overline{O}\rangle B_j A_j
    \ ,
\end{equation}
where the second expression is independent of the choice of $j$. 
For the two-point function, the shared mechanism $e_{jk}$ cancels from the parity $v_j\oplus v_k$, so that $\mathbb{E}[(-1)^{v_j\oplus v_k}]=D_jD_k/(1-2p_{jk})^2$, and
\begin{equation}
    \mathbb{E}[v_jv_k] \ = \ \frac{1}{4}\left(1 - D_j - D_k + \frac{D_jD_k}{(1-2p_{jk})^2}\right)\ ,\qquad
    \mathbb{E}[v_j^2] \ = \ \mathbb{E}[v_j] \ = \ \frac{1-D_j}{2}
    \ ,
\end{equation}
where the diagonal follows from $v_j^2=v_j$ for a detector. 
The cross moment follows from the conditional expectation derived above, or directly from $\mathbb{E}[Q(-1)^{v_j}]=\langle \overline{O}\rangle B_jC_j$,
\begin{equation}
    \mathbb{E}[Qv_j] \ = \ \frac{\langle \overline{O}\rangle B_j}{2}\left(A_j-C_j\right)
    \ .
\end{equation}
The connected components are then given as,
\begin{equation}
    G_{jk} \ = \ \frac{D_jD_k}{4}\left(\frac{1}{(1-2p_{jk})^2}-1\right)\quad(j\neq k)\ , \ 
    G_{jj} \ = \ \frac{1-D_j^2}{4}\ , \
    a_j \ = \ \frac{\langle \overline{O}\rangle B_j}{2}\left(A_jD_j-C_j\right)
    \ .
\end{equation}
The limit of weak noise, $p_m\ll 1$, is used to simplify the expressions. 
Expanding to leading order, $D_j\approx 1-2\sum_{m\ni j}p_m$
and $(1-2p_{jk})^{-2}-1\approx 4p_{jk}$, so that
\begin{equation}
    G_{jk} \ \approx \ p_{jk}\quad(j\neq k)\ ,\qquad
    G_{jj} \ \approx \ p_j + \sum_{k\neq j}p_{jk}\ ,\qquad
    a_j \ \approx \ -2\langle \overline{O}\rangle\Big(p_jq_j+\sum_{k\neq j}p_{jk}q_{jk}\Big)
    \ ,
\end{equation}
where in $a_j$ the terms carrying rates alone cancel between $A_jD_j$ and
$C_j$, leaving the flip-weighted sums. 
The normal equation then reads
\begin{equation}
    \Big(p_j+\sum_{k\neq j}p_{jk}\Big)\beta^*_j
    + \sum_{k\neq j}p_{jk}\,\beta^*_k
    \ = \ -2\langle \overline{O}\rangle\Big(p_jq_j+\sum_{k\neq j}p_{jk}q_{jk}\Big)
    \ .
\end{equation}
Here, the RHS is the same magnitude as the numerator of $\Delta_j$ for the same detector.
Furthermore, omitting the second term on the LHS yields $\beta_j^*=\Delta_j$, showing that the correlations that cause these two metrics to differ are the pair processes $p_{ij}$ that connect two detectors. 
Specifically, $\Delta_j$ attributes the logical error caused by pair mechanisms $e_{ij}$ to \emph{both} detectors $v_i$ \emph{and} $v_j$. 
The regression, on the other hand, splits the contribution of $e_{ij}$ between $v_i$ and $v_j$ in proportion to all the other rates of mechanisms that have $v_j$ in their support. 
\begin{figure}
    \centering
    \includegraphics[width=0.45\linewidth]{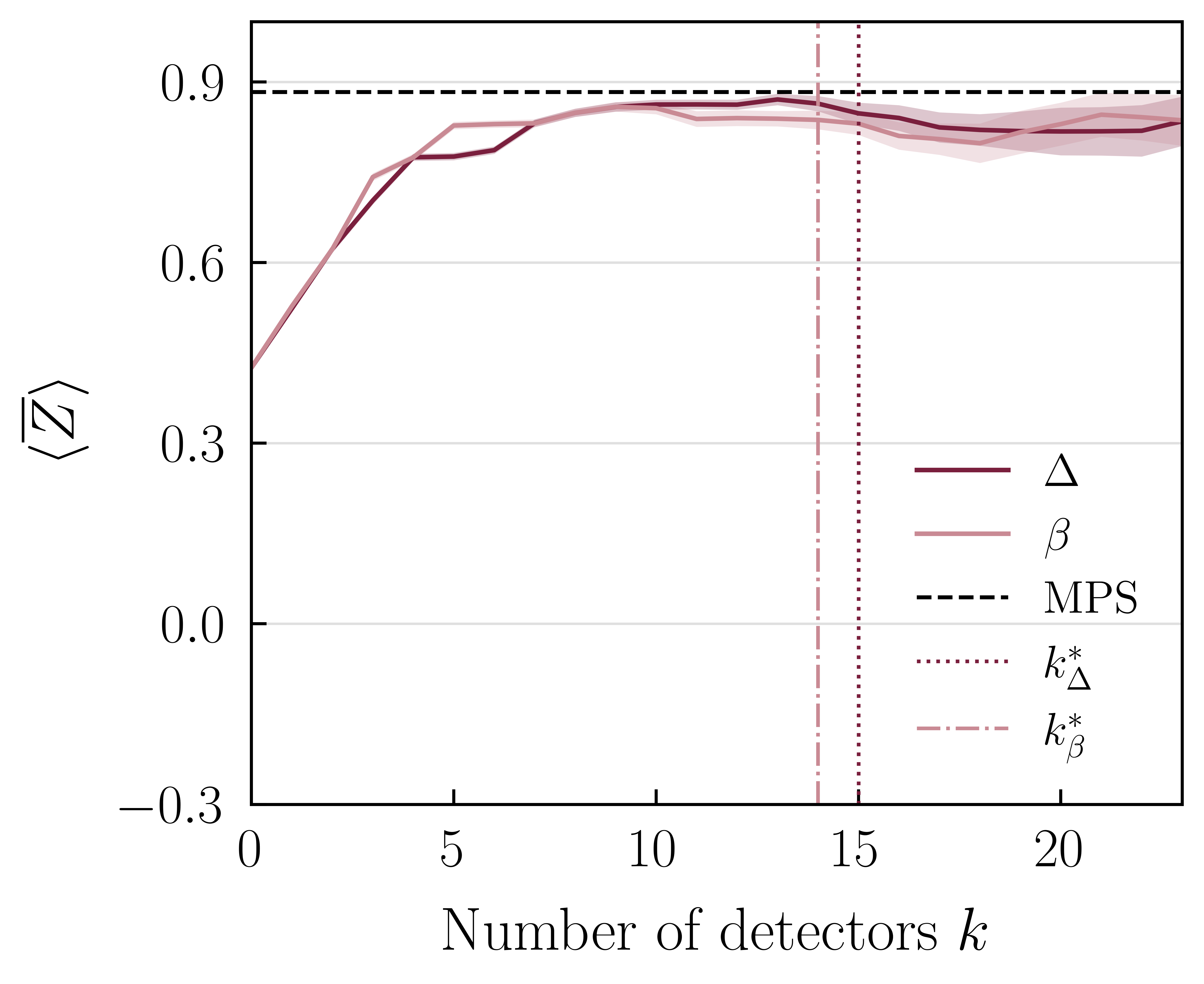}
    \caption{\textit{Comparison of $\Delta_j$ and $\beta_j$ predictions on }{\tt ibm\_boston}. The metrics $\Delta_j$ and $\beta_j$ are calculated using Eq.~(\ref{dev_422:eq:delta_exp}) and Eq.~(\ref{dev_422:eq:normal_equations}) respectively.
    The two solid lines show the predicted value of $\langle \overline{Z}\rangle$ as a function of the number $k$ of detectors used in postselection for $\Delta_j$ (dark) and $\beta_j$ (light). The horizontal dashed line is the noiseless MPS prediction. The vertical dashed lines correspond to the value of $k^*$ returned from the plateau finding algorithm using $\Delta_j$ (dark) and $\beta_j$ (light). The expectation value is from four Trotter steps on four code blocks.}
    \label{dev_422:fig:nkeep_appendices}
\end{figure}
Therefore, a ranking based on $\Delta_j$ quantifies the probability that detection events at $v_j$ result in a logical error, whereas a ranking based on $\beta_j$ quantifies the degree to which the logical error rate is uniquely attributed to the detector $v_j$. 
This distinction suggests that $\Delta_j$ is a more effective measure for identifying the noisiest shots,  and thus $\Delta_j$ is used in ORP to produce the results in the main text.
However, in practice, filtering based on $\beta_j$ or $\Delta_j$ is found to give comparable results, see e.g. Fig.~\ref{dev_422:fig:nkeep_appendices}.

\subsection{The lightcone of the filtered hierarchy}
\label{dev_422:app:pp_lightcone}
\noindent
To connect the detector ranking chosen by $\Delta_j$ to the properties of a partially FT circuit $U$, we Pauli-propagate the observable $\overline{O}$ on which $\Delta_j$ is conditioned backward through $U$ to the location of each detector $v_j$. 
The propagated observable can be represented as
\begin{align}\label{dev_422:eq:pp_obs}
    \tilde{O} \ &= \ U\overline{O}U^{\dagger}\nonumber\\
     \ &= \ \sum_P c_p P \ ,
\end{align}
where the sum is over all Pauli strings $P$. 
For a detector $v_{j=(g,b,r)}$, located at code block $b$ in round $r$, its ranking, as determined by Pauli-propagation, is given by
\begin{equation}\label{dev_422:pp_ranking}
    \gamma_j \ = \ \frac{1}{\mathcal{N}}\sum_{P : \text{supp}(P)\in b} |c_P|
\end{equation}
where the sum is over all Pauli strings that have support (i.e. non-identity components) contained in code block $b$, and $\mathcal{N} = \sum_P |c_P|$ is a normalization that ensures $0\leq \gamma_j \leq 1$.
The ranking based on $\Delta_j$ is set by the magnitude of $\Delta_j$.

\begin{figure}
    \centering
    \includegraphics[width=0.75\linewidth]{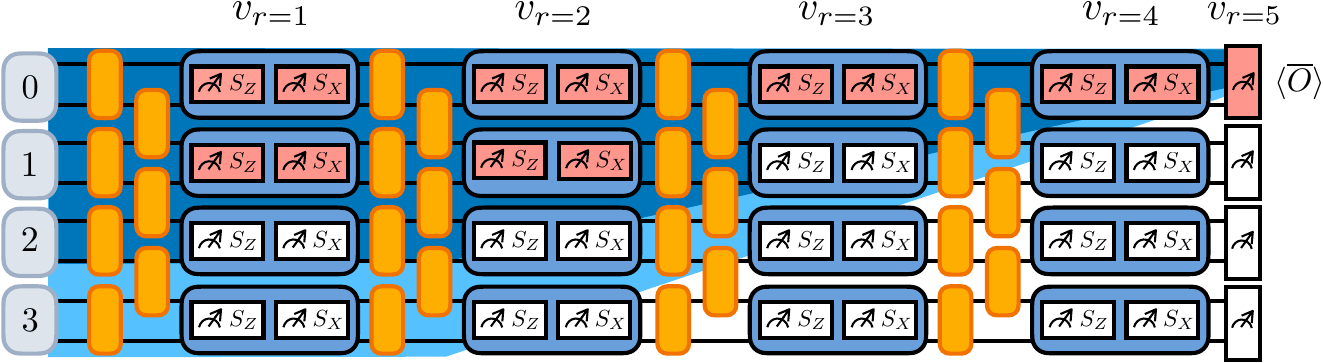}
    \caption{{\it depiction of most important detectors.} Calculated for the logical observable $\overline{O}=\overline{Z}_{(b=0,0)}$, as determined by the rankings calculated in Table~\ref{dev_422:tab:filtering_comparison}. 
    Highlighted in red are the locations of the seven most important detectors, as determined by both ORP and Pauli-propagation. 
    The grey region shows the observable's circuit backwards lightcone, containing all of the detectors that are connected to the observable through gates. 
    The dark blue region captures the physical reverse-lightcone of $\overline O$, showing the detector locations where the back-propagated observable has the largest Pauli weights.
    }
    \label{dev_422:fig:detector_lightcone_ranking}
\end{figure}

As an example, Table~\ref{dev_422:tab:filtering_comparison} shows a comparison between the detector ranking determined by $\gamma_j$ and the detector ranking determined by $\Delta_j$ using simulated results.
\begin{table}
\begin{tabularx}{\linewidth}{|c||Y|Y|Y|Y|Y|Y|Y|Y|Y|Y|}
\hline
Detector ranking & 1 & 2 & 3 & 4 & 5 & 6 & 7 & 8 & 9 & 10 \\
\hline\hline
$\Delta_j$ & $v_{(0,1)}$ & $v_{(0,2)}$ & $v_{(0,4)}$ & $v_{(0,3)}$ & $v_{(0,5)}$ & $v_{(1,1)}$ & $v_{(1,2)}$ & $v_{(3,1)}$ & $v_{(3, 3)}$ & $v_{(3,4)}$ \\\hline
$\gamma_j$ & $v_{(0,5)}$ & $v_{(0,4)}$ & $v_{(0,3)}$ & $v_{(0,2)}$ & $v_{(0,1)}$ & $v_{(1,1)}$ & $v_{(1,2)}$ & $v_{(1,3)}$ & $v_{(2,1)}$ & $v_{(3,5)}$\\\hline
\end{tabularx}
\caption{{\it Comparison between detector ranking using Pauli-propagation and ORP.} 
The top row shows the position in the ranking, where $1$ is the highest ranking and $10$ is the lowest ranking among the detectors shown.
The middle row shows the detectors $v_{(b,r)}$ ranked by their $\Delta_j$ value and the bottom row shows them ranked by their $\gamma_j$ value.
The rankings are computed for the observable $\overline{Z}_{(b=0,0)}$ using a four-block circuit consisting of $n_T=4$ Trotter steps with step size $\delta t=0.1$, and one syndrome extraction round per Trotter step, starting from an all $\ket{\overline{00}}$ initial state. The rankings are computed for $g_x=g_z=1$.
}
\label{dev_422:tab:filtering_comparison}
\renewcommand{\arraystretch}{1.0}
\end{table}
The rankings are computed for the observable $\overline{Z}_{(b=0,0)}$ on four code blocks $b=0,1,2,3$ with $n_T = 4$ Trotter steps and a syndrome extraction rate of one round per Trotter step, as illustrated in Fig.~\ref{dev_422:fig:detector_lightcone_ranking}.
\noindent
The $\Delta$-rankings are computed using classical simulations subject to two-qubit depolarizing noise with error rate $p=0.003$, which is comparable with error rates on current IBM hardware. 
These rankings are found to be consistent with those computed from hardware results.
Although the two orderings do not match exactly, ORP is consistent with Pauli-propagation, since the set of the top-$k$ detectors selected typically are the same, with the most important detectors residing in the code block containing the observable being measured. 
Figure~\ref{dev_422:fig:detector_lightcone_ranking} shows the locations of the most important detectors chosen by the two methods and illustrates how the highest ranked detectors selected by ORP coincide with the local observable's backwards lightcone.

\section{Fault tolerant benchmarking of circuit elements}
\label{dev_422:app:circuit_elts_benchmark}

\subsection{Logical error rates of circuit gadgets}
\label{dev_422:sec:ler_cg}
\noindent
This appendix numerically studies the logical performance in the simulations by calculating the logical error rate (LER) as a function of the device error rate.
The noise model assumes a two-qubit depolarizing channel, defined as
\begin{equation}
    \mathcal{E}(\rho) = (1-p)\rho + \frac{p}{15}\sum_{P\neq I} P\rho P
\end{equation}
where $p\in[0,1]$ is the physical error rate and the sum is over all two-qubit Paulis $P$, excluding the identity. 
Additionally, single-qubit gates are followed by a single-qubit depolarizing channel with error rate $p/10$.
In the numerical experiments, the ideal state is a logical $\ket{\overline{00}}$, which is a GHZ state on four qubits. 
A logical error occurs when there is a physical error that keeps the state in the codespace $\mathcal{C}$ and goes undetected by the syndrome measurements. Removing shots where a nontrivial syndrome is detected is equivalent to projecting the noisy state back onto the codespace, defined by the projector
\begin{equation}\label{dev_422:eq:code_space_proj}
    \Pi_{\mathcal{C}} = \frac{1}{4}(1+S_Z)(1+S_X)
    \ ,
\end{equation}
where errors take the state out of the simultaneous $+1$ eigenspace of $S_Z$ and $S_X$.
For a state $\rho$ whose noisy evolution is given by $\tilde{\rho}$, the LER is defined by
\begin{equation}
    p_L \ = \ 1-F_L,\qquad F_L \ = \ \text{Tr}\,\rho\tilde{\rho}\, \Pi_{\mathcal{C}}
    \ .
\end{equation}

\begin{figure}
    \centering
    \includegraphics[trim = {0 120 0 120}, clip, width=\linewidth]{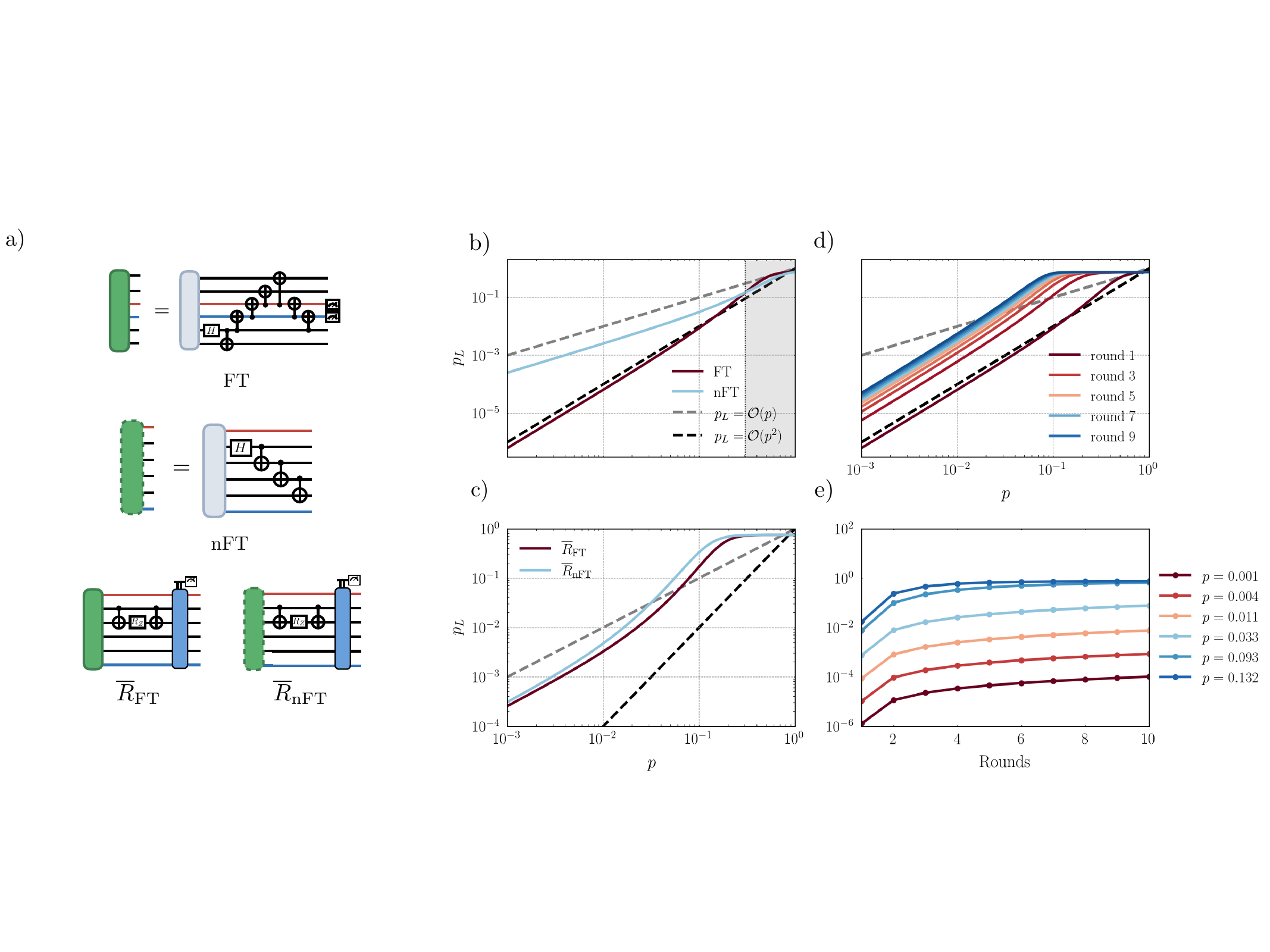}
    \caption{\textit{Logical error rate of different circuit components} 
    a)~The different FT and nFT circuit elements whose logical error rates are considered in ~\ref{dev_422:sec:ler_cg}.
    The green elements denote initialization gadgets, labeled FT (solid) and nFT (dashed). 
    The bottom two circuits $\overline R_\text{FT}/\overline R_\text{nFT}$ are FT/nFT initialization gadgets followed by a nFT logical rotation $\overline{R}_{ZZ}$.
    b)~The LER of the two different initialization gadgets. 
    The grey region is where nFT crosses to a lower LER than the FT procedure. The grey and black dashed lines represent $\mathcal{O}(p)$ and $\mathcal{O}(p^2)$ scaling for the logical error rate $p_L$.
    c)~ The LER for the FT and nFT initialization followed by a logical rotation.
    d)~The LER of the syndrome extraction circuits for multiple rounds.
    e)~The LER of syndrome extraction as a function of number of rounds for different $p$.
    }
    \label{dev_422:fig:p_vs_pl_gadget}
\end{figure}

The LER $p_L$ can be expressed as a function of $p$. For $p$ small, the leading order contribution is a good approximation for how $p_L$ scales with $p$.
For level-$0$ FT (no encoding) the leading order scaling is $\mathcal{O}(p)$.
Gadgets that are 1-FT have leading order behavior $\mathcal{O}(p^2)$, and so mixing both types results in an LER of the form $p_L=Ap+Bp^2$, where the constants $A,B$ encode the geometric features of a circuit. 
In particular, $A$ represents nFT circuit features that can cause undetectable errors while $B$ represents the extra circuit overhead required to detect errors. 
The relative magnitudes of $A$ and $B$ can have meaningful impacts on the error rates, and in some cases imply that a nFT circuit gadget can give superior performance to its encoded counterpart.
An example of this is seen by comparing the nFT rotations used in this work to a FT approach like magic state injection.

The clearest examples of the improved performance of including nFT elements are the $\ket{\overline{00}}$ initialization circuits that prepare a four-qubit GHZ state.
A simple nFT way of preparing this state requires a CNOT depth of four, whereas a 1-FT technique for preparing such a state on heavy-hex topology requires a CNOT depth of seven followed by two measurements.
Both of these circuits are shown in Fig.~\ref{dev_422:fig:p_vs_pl_gadget}a), with their respective LERs determined from classical simulations shown in Fig.~\ref{dev_422:fig:p_vs_pl_gadget}b) over a range of $p$.
When considering the LER of both of these circuit gadgets, the FT circuit has error suppression quadratic in $p$ compared to the lack of suppression in the nFT case.
For devices with long measurement times, idle errors can occur on the data qubits being measured, amplifying the FT overhead $B$ by almost an order of magnitude and increasing the number of errors.
Together with a two-qubit gate depth almost twice that of the nFT gadget, these extra circuit elements degrade the performance at large enough $p$.
This is reflected in the LER curves, where the FT $p_L$ passes the nFT value at a physical error rate of $p\sim 0.2$.

Considering both FT and nFT elements together in the same circuit qualitatively modifies this picture. 
In particular, examining the leading-order expansion for $p_L$ reveals a crossover point at $p^*\approx A/B$ such that for $p>p^*$ there is a region of quadratic suppression of errors, which then degrades to a linear decay with no suppression for $p\lesssim p^*$.
Examples of such behavior are shown in Fig.~\ref{dev_422:fig:p_vs_pl_gadget}c), which compares the effect of both FT and nFT state preparation, followed by a logical rotation (nFT) and syndrome extraction (FT). 
The extra circuit depth pushes the LER high enough such that the nFT state prep is no longer able to outperform its FT counterpart. 
Further, even though the asymptotic error dependence is linear, the addition of FT components can push the LER below that of nFT rotations. 
The suppression at intermediate $p>p^*$ indicates that for physical error rates that are small but not yet below threshold, there are still advantages in error suppression gained by incorporating FT elements into circuit design. 
Finally, Fig.~\ref{dev_422:fig:p_vs_pl_gadget}d) analyzes the LER of the syndrome extraction circuits with FT initialization. 
Here, all of the circuit elements are 1-FT and so $p_L = \mathcal{O}(p^2)$. 
The overhead factor $B$ increases with the number of syndrome measurement rounds, but then quickly plateaus to an $\mathcal{O}(1)$ constant as shown in Fig.~\ref{dev_422:fig:p_vs_pl_gadget}e).

\subsection{Memory experiments}
\label{dev_422:sec:memory}
\begin{figure}
    \centering
    \includegraphics[width=\linewidth, trim = 0 250 0 200, clip]{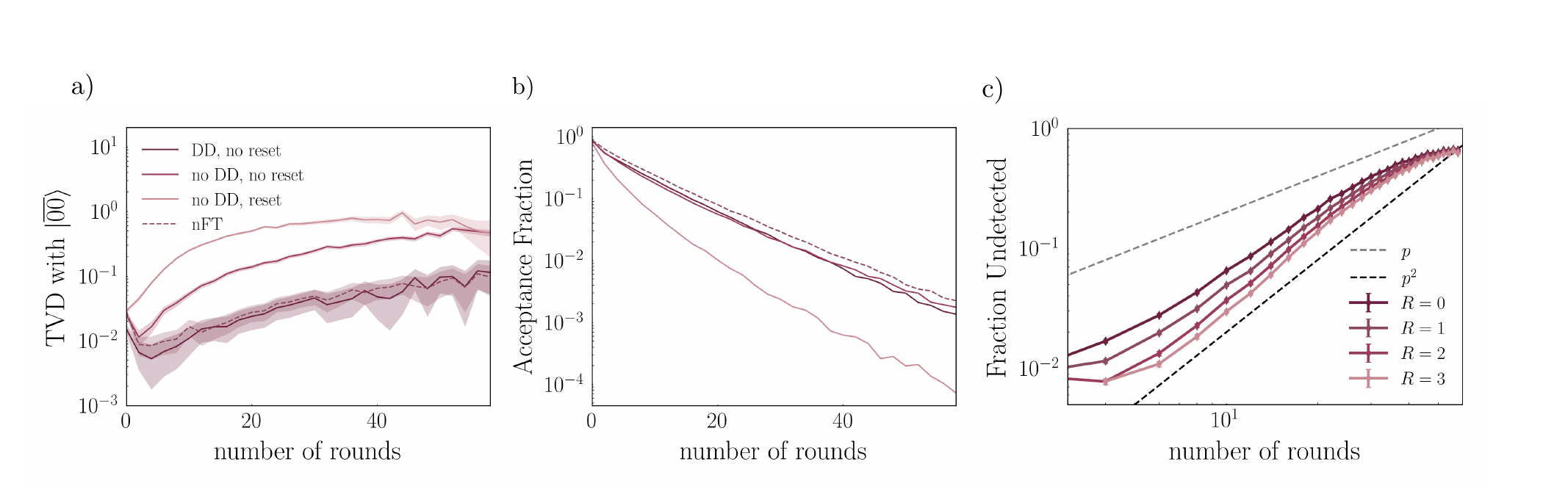}
    \caption{\textit{Memory experiment performance of multiple rounds on {\tt ibm\_boston}} 
    a)~The Total Variation Distance (TVD) from the logical $\ket{\overline{00}}$ state 
    over $60$ rounds of syndrome extraction.
    b)~The acceptance fraction over $60$ rounds. 
    c)~Suppression of errors at leading order as a function of number of syndromes used for postselection.}
    \label{dev_422:fig:memory_figs}
\end{figure}

\noindent
To assess the effectiveness of the syndrome extraction circuits used in this work, memory experiments are performed on {\tt ibm\_boston} where both FT and nFT state prep are followed by many rounds of syndrome extraction. 
Assuming noiseless circuit operation, the system will remain in the logical $\ket{\overline{00}}=\frac{1}{\sqrt{2}}(\ket{0000}+\ket{1111})$ state, and measurements in the $Z$-basis result in either ``0000'' or ``1111'' bitstrings.
Noise populates other bitstrings with a nonzero probability, with some being detected by syndrome extraction and others causing undetectable errors. 
The undetected errors will appear as even parity bitstrings with Hamming-weight two in the final bitstring distribution. 
After postselecting all shots whose syndrome measurements flagged an error, the filtered probability distribution is compared to that of an ideal GHZ state using the Total Variation Distance (TVD) defined by
\begin{equation}
    \text{TVD}(q,q') \ = \ \frac{1}{2}\sum_s|q(s)-q'(s)|
    \ ,
\end{equation}
where the sum is over all bit strings, $q$ is the measured distribution and $q'$ is the GHZ distribution, with $q'(\text{``0000''})=q'(\text{``1111''})=0.5$.
Figure~\ref{dev_422:fig:memory_figs}a) shows the TVD of the postselected distribution as a function of the number of syndrome extraction rounds.
The four curves compare 
(1) no DD and reset, 
(2) no DD and no reset, 
(3) DD and no reset and 
(4) DD and no reset 
with nFT state prep over the course of 60 rounds of syndrome extraction (all with a circuit depth of 761 CNOTs).
Incrementally adding in DD and removing resets have dramatic impacts on the error, lowering it by over an order of magnitude. 
The nFT state preparation contains more errors for a few rounds of syndrome extraction, but remains comparable to the FT state over many rounds.
Figure~\ref{dev_422:fig:memory_figs}b) shows the corresponding acceptance fraction for these cases, defined as the number of measurements that record no errors. 
This fraction decreases exponentially with the number of rounds, with DD having the most profound effect on preserving shot lifetime.

The $[[4,2,2]]$ code is a distance-2 code, and so all events contributing to the LER at first order in the physical error rate $p$ are flagged, and $p_L$ will scale as $\mathcal{O}(p^2)$.
Observing this scaling in quantum hardware runs requires tuning the physical error rate, which is not possible to do directly. 
We leverage the fact that syndrome extraction acts as the logical identity in the codespace, and so multiple rounds of extraction can serve to amplify the effective physical error rate.
As such, the physical error rate is increased by modifying which rounds of syndrome extraction are postselected against. 
Specifically, instead of considering the results of syndrome measurements on every round, we only consider postselecting on the final round of extraction. To account for the errors from initializing the code state $\ket{\overline{00}}$, a constant number of syndromes are additionally postselected on.
As a proxy for the LER $p_L$, we count the fraction of even parity Hamming-weight 2 bitstrings that survive the postselection, which at leading order is proportional to the number of undetectable errors that occurred. 
Figure~\ref{dev_422:fig:memory_figs}c) shows the fraction of undetected errors as a function of the number of rounds with an increasing number of syndromes postselected at the beginning. 
This fraction grows quadratically with the number of rounds, and this number is interpreted as the scaling factor for the physical error rate. 
At low amplification, 
this fraction plateaus due to nFT state prep errors and finite measurement statistics, whereas at high amplification there is a plateau due to exceeding the pseudothreshold.
Therefore, the scaling should be understood as appearing in an intermediate regime, in the plot approximately between rounds $8$-$30$.

\subsection{Rotations with ancilla qubits}
\label{dev_422:app:analog_ancilla_rots}
\begin{figure}
    \centering
    \includegraphics[width=\linewidth]{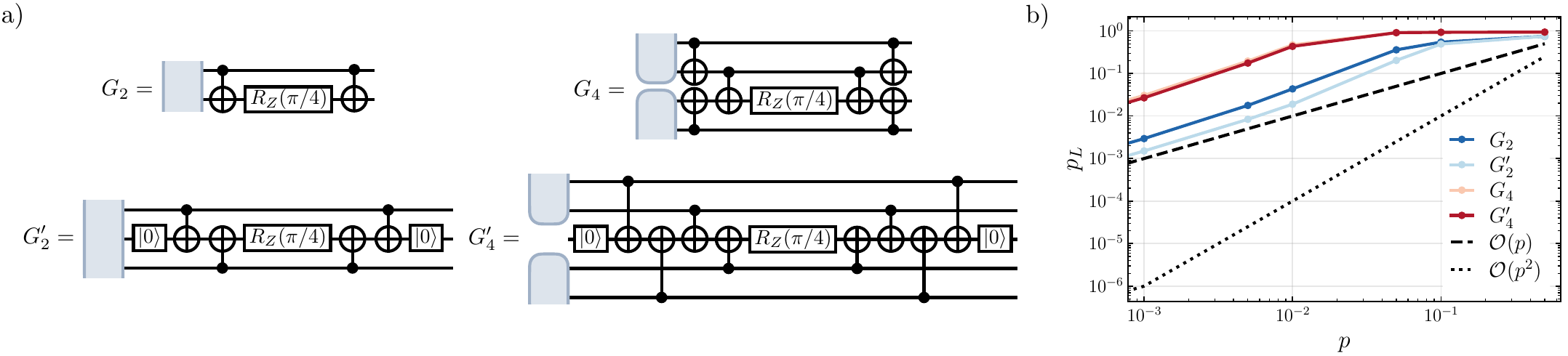}
    \caption{{\it Ancilla vs. no-ancilla rotations.}
    a)~The top row shows ancilla-free circuits $G_2$ and $G_4$ implementing $e^{-i\pi/8 \overline{Z}_{(b,0)}\overline{Z}_{(b,1)}}$ within a single code block (left) $b$ and $e^{-i\pi/8 \overline{Z}_{(a,0)}\overline{Z}_{(b,0)}}$ between two code blocks $a$ and $b$ (right).
    The bottom row shows circuits $G_2'$,\, $G_4'$ with ancilla qubits that implement the same rotations.
    b)~The LER $p_L$ as a function of the physical error rate $p$ for the circuits in a). 
    The logical error rate is computed by taking the average over Haar-random states under repeated rotations with depolarizing noise.}
    \label{dev_422:fig:ancilla_rotations}
\end{figure}

\noindent
A method to reduce the error spread by nFT rotations, at the cost of ancillas and measurements, was studied in Refs.~\cite{Zhong:2026jps,Gerhard:2024peb}.
This method couples an ancilla qubit prepared in $|0\rangle$ to the rotation circuit in such a way that the bulk of single-qubit $X$ errors are concentrated on the ancilla qubit.
As an example, let $P=ZZ$ and $P'=ZZZ$ where the first two qubits are data and the last one is an ancilla.
Then $e^{-i\theta P'}=e^{-i\theta P}\ket{0}\bra{0}+e^{i\theta P}\ket{1}\bra{1}$ and the correct rotation is recovered by postselection on ancilla state $|0\rangle$.
Figure~\ref{dev_422:fig:ancilla_rotations}a) shows such circuit pairs $G_2/G_2'$ and $G_4/G_4'$, implementing $ZZ$-rotations within a block and between logical qubits of neighboring blocks.
In these circuits, the only errors that spread to become high-weight errors are $Z$-errors indistinguishable from logical rotations.
Ancilla measurements can be deferred to the end of the circuit, provided later errors do not affect them.
Since $X$ errors are accumulated on the ancilla qubit, 
an undetectable (logical) error can occur when two physical errors cancel
\footnote{This cancellation can be studied using an error model with $X$ errors only on the ancilla qubit.}.
Note that applying this gadget with an ancilla qubit in state $|1\rangle$ implements a rotation with the opposite angle, which is a logical error that is detectable via postselection on ancilla measurement.
The implications of this are discussed in~\ref{dev_422:app:more_analysis}.

While this method enables detecting more errors, its gate and measurement overhead introduces additional errors.
To determine which approach is best-suited for currently available hardware, 
we compare implementations of $e^{-i\pi/8 \overline{Z}_{(b,0)}\overline{Z}_{(b,1)}}$ within a single code block $b$ and $e^{-i\pi/8 \overline{Z}_{(a,0)}\overline{Z}_{(b,0)}}$ between two code blocks $a$ and $b$.
The data qubits are positioned within code blocks so that SWAP gates are not required to implement these rotations.
The circuits in  Fig.~\ref{dev_422:fig:ancilla_rotations}a) are repeatedly applied such that a net $\theta=2\pi$ rotation is implemented.
The logical error rate $p_L$ is determined from classical simulations with depolarizing noise by computing the fidelity of the initial state $\rho$ with the noisy final state $\tilde\rho$, $p_L = 1-\text{Tr}(\rho\tilde\rho\Pi_{\mathcal{C}})$.
Since the effect of rotations depends on the initial state, we compute the Haar average of $p_L$ over all initial states.
A two-design is necessary to compute the Haar average of the fidelity~\cite{Dankert:2009yux}.
The set of stabilizer states forms a three-design~\cite{Zhu:2017psv}, so one- and two-qubit stabilizer states are used as initial states for the circuits in Fig.~\ref{dev_422:fig:ancilla_rotations}a). 
The LER as a function of physical error rate $p$ for eight repeated $\pi/4$ rotations of $G_2,\,G_2',\,G_4,\,G_4'$ is shown in Fig.~\ref{dev_422:fig:ancilla_rotations}b).
Both implementations exhibit $p_L=\mathcal{O}(p)$ scaling. 
The ancilla method $G_2'$ is found to have a two to three times smaller prefactor than $G_2$, and $G_4$ is found to be roughly equivalent to $G_4'$ on average when accounting for errors detected by the $[[4,2,2]]$ code.
This suggests that although the ancilla method can offer an advantage in specific cases, consistent with Refs.~\cite{Zhong:2026jps,Gerhard:2024peb}, the two approaches perform similarly on average under depolarizing noise.
Further, this examination assumes noiseless (and instantaneous) measurements.
On realistic hardware, a measurement consumes significant coherence time and introduces measurement errors. 
For these reasons the simulations in this work are carried out with the standard NISQ logical rotation gadget.\footnote{Ancillas are used to implement rotations between blocks through bridge qubits where necessary, see App.~\ref{dev_422:sec:scheduling}.}

\section{Additional analysis of encoding improvement}
\label{dev_422:app:more_analysis}

\subsection{Qubit dependence}
\begin{figure}
    \centering
    \includegraphics[width=\linewidth]{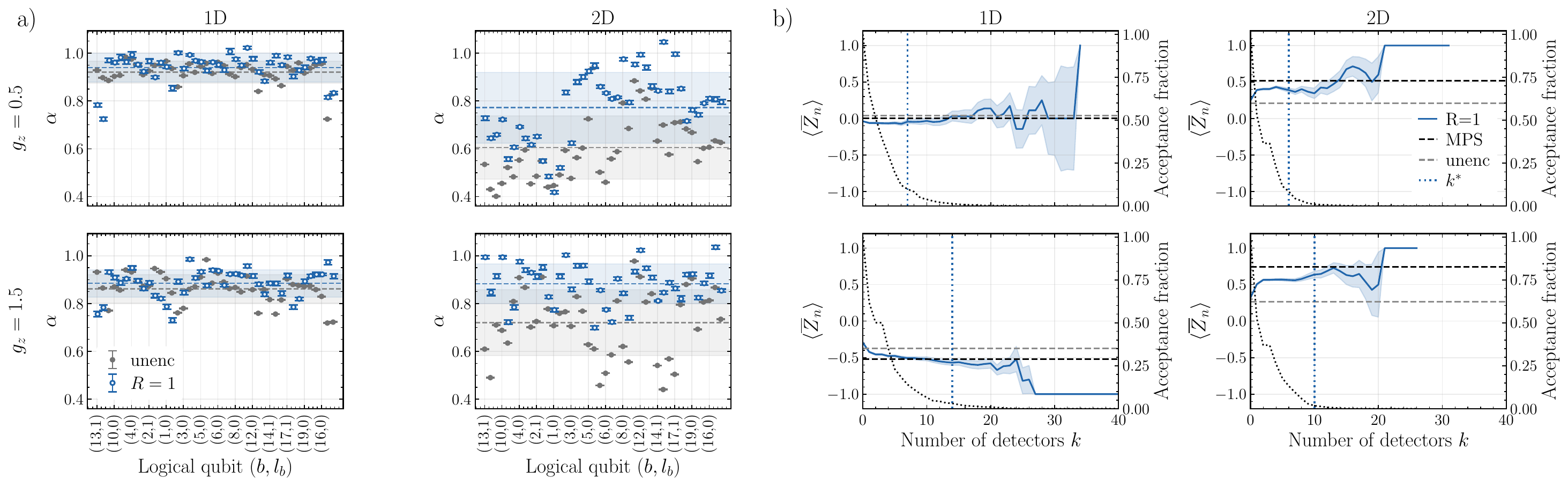}
    \caption{{\it Postselection performance and signal survival in encoded simulations.}
    a)~The signal-survival factor $\alpha$ as a function of logical qubit over all $t$ for $g_z=0.5$ (top) and $g_z=1.5$ (bottom).
    Encoded results with $R=1$ are given in blue, and unencoded results are shown in grey. 
    Results for the 1D chain are shown in the left column, and 2D grid results are shown in the right column.
    The shading represents one standard deviation of the $\alpha$ distribution over qubits.
    Error bars on individual points represent uncertainty determined through bootstrap resampling.
    b)~Convergence of an example observable on qubit $(14,1)$ at $t=5$ as a function of the number of detectors kept for postselection $k$. 
    The shading represents uncertainty from bootstrap resampling.
    Results from the unencoded runs and from MPS are given by grey and black dashed horizontal lines.
    The vertical black dotted line shows the $k^*$ selected by the plateau finding algorithm.
    Acceptance rates for each $k$ are shown by the dotted grey line.
    }
    \label{dev_422:fig:chain_grid_alpha_convergence}
\end{figure}

\noindent
Figure~\ref{dev_422:fig:chain_grid_alpha_convergence} shows additional metrics comparing encoded to unencoded runs presented in Sec.~\ref{dev_422:sec:results}.
As seen in Fig.~\ref{dev_422:fig:improvement}, the encoding improvement in 1D is most pronounced at early and intermediate times, and decreases at late times.
In 2D, the improvement grows with time.
Figure~\ref{dev_422:fig:chain_grid_alpha_convergence}a) shows $\alpha$ per logical qubit across all $t$.
For both 1D and 2D simulations, the encoded $\alpha$ values are on average higher than the unencoded, but are broadly distributed.
Several outliers are present in both encoded and unencoded results; 
these can be matched to locations on the quantum processor with faulty gates or qubits with high measurement error rate or low T1/T2 times.
Pauli twirling of the coherent and amplitude-damping/dephasing components of the noise could convert part of this structured error into a stochastic Pauli channel~\cite{Wallman:2015uzh} more amenable to detection and postselection.
Omitting these blocks in the analysis improves results considerably, consistent with most regions of the quantum processor having an operating error rate just below the pseudothreshold.
However, this exclusion is not done in the results reported in the main text.
Further, the per-qubit encoded points carry visibly larger bootstrap uncertainties than the unencoded points, which is an expected feature of shot loss from postselection and increased circuit depth due to encoding. The results obtained from the simulations in 2D have a significantly larger spread than those from simulations in 1D, and are generally noisier (lower $\alpha$). 
This again is a consequence of the deeper circuits required to implement a dense logical connectivity. 

Figure~\ref{dev_422:fig:chain_grid_alpha_convergence}b) shows the effects of the plateau-finding algorithm explained in App.~\ref{dev_422:sec:filtering} for qubit $(14,1)$ at $t=5$, 
as an example.
As detectors are added, the acceptance fraction (grey dotted line) falls steeply while the estimate settles onto a plateau in approaching the MPS value (black dashed line) and away from the biased unencoded result (grey dashed line).
The plateau-finding algorithm selects $k^*$ (blue dotted vertical line) at the onset of this stable region.
If such a value is not found, the algorithm selects $k^*$ to minimize the squared error, as explained in step five of the plateau-finding algorithm in App.~\ref{dev_422:sec:det_selec}, and examples of this are shown in both 2D plots in Fig.~\ref{dev_422:fig:chain_grid_alpha_convergence}b) and in the $g_z=1.5$ 1D plot.
Across all times and qubits, most show similar convergence behavior with the exception of especially faulty device regions.

\begin{figure}
    \centering
    \includegraphics[width=0.75\linewidth]{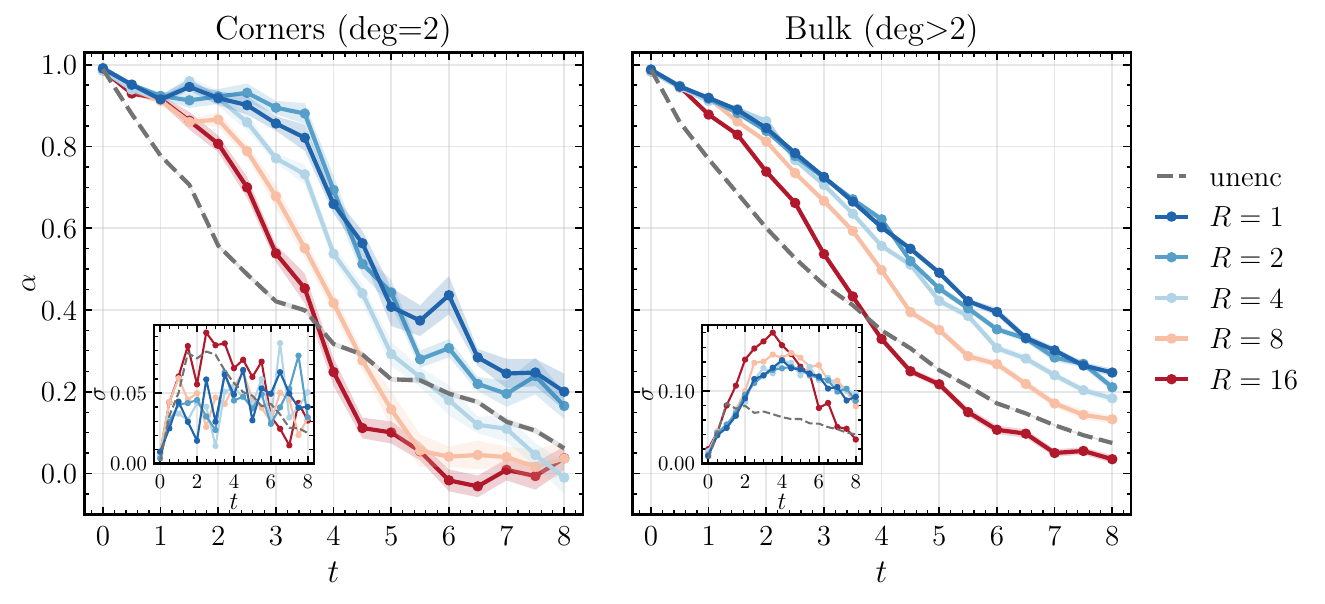}
    \caption{{\it Comparison of $\alpha$ for corner and bulk qubits in 2D simulations.}
    The signal-survival factor $\alpha$ is shown for encoded runs as a function of time $t$ for $R$ syndrome extraction rounds (colors) compared to unencoded runs (grey). 
    The left panel shows $\alpha$ for corners of the grid (logical qubits with two neighbors), and the right panel shows $\alpha$ for bulk qubits (those with more than two neighbors). 
    The insets show the error in the $\alpha$ fits, $\sigma$. 
    Results from the localized regime ($g_z=1.5$) are used.}
    \label{dev_422:fig:corner_vs_bulk}
\end{figure}

Figure~\ref{dev_422:fig:corner_vs_bulk} shows the difference in $\alpha$ between corner and bulk qubits in 2D grid simulations with $g_z=1.5$. 
Corners are defined as logical qubits with two connections, while bulk qubits have $\text{deg}>2$.
Since corners have fewer logical connections, fewer gates act on them and as a result they are less noisy.
In addition, the backwards lightcone on corner qubits is smaller, reducing the number of errors that can affect observables there.
This is reflected in the larger $\alpha$ values in the left panel of Fig.~\ref{dev_422:fig:corner_vs_bulk} compared with the right.
Further, the $\sigma$ values quantifying the spread in the $\alpha$ fit are smaller, indicating the corners more consistently reproduce the MPS expectations.

\subsection{Coupling dependence}
\begin{figure}
    \centering
    \includegraphics[width=0.85\linewidth]{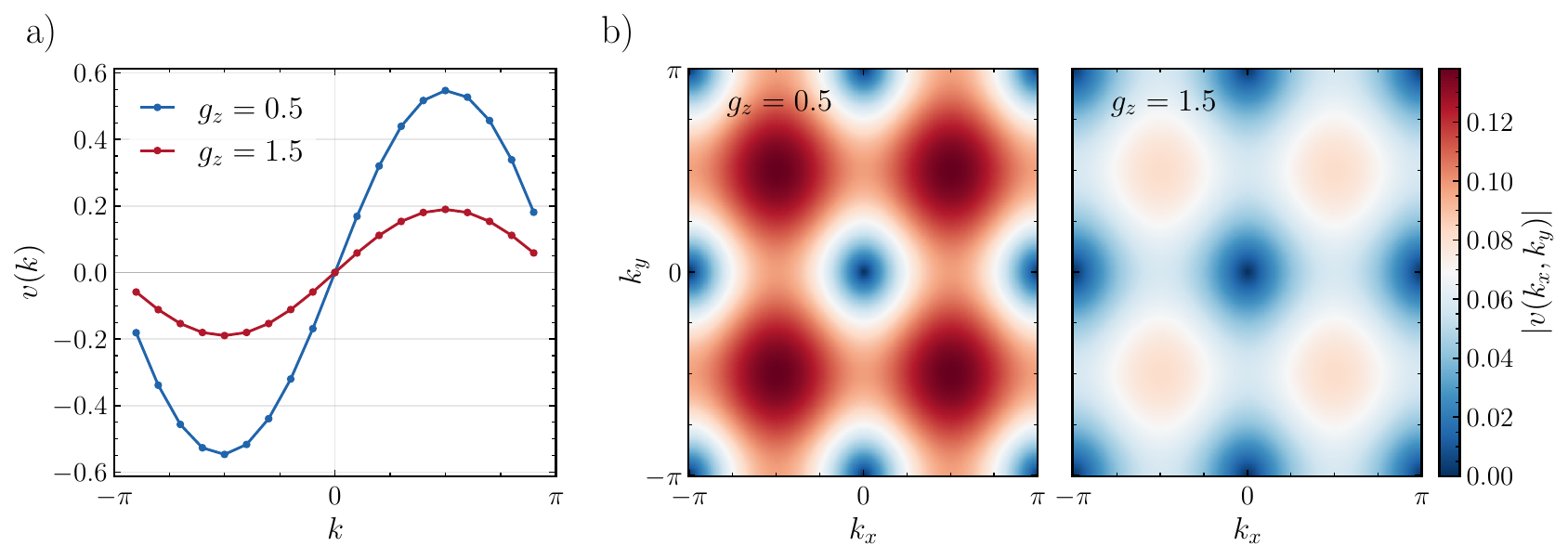}
    \caption{{\it Group velocity in the 1D and 2D Ising model.}
    a)~The group velocity $v(k)$ as a function of momentum $k$ for $g_z=0.5$ (blue) and $g_z=1.5$ (red) calculated in a 20-qubit 1D chain with periodic boundary conditions.
    b)~The group velocity $|v(k_x,k_y)|$ for a $5\times5$ 2D lattice with $g_z=0.5$ (left) and $g_z=1.5$ (right).
    Dispersion relations are calculated using exact diagonalization, and a spectral derivative is used to compute $v(k_x,k_y)$.
    }
    \label{dev_422:fig:v_k_1d_2d}
\end{figure}

\noindent
As shown in Fig.~\ref{dev_422:fig:improvement}, the encoding improvement is roughly independent of $g_z$ for 1D and 2D. 
However, $\alpha$ fits for both encoded and unencoded runs are consistently higher for $g_z=1.5$ than $0.5$, which is seen in Figs.~\ref{dev_422:fig:chain_results}e) and~\ref{dev_422:fig:grid_results}d). 
This difference originates in how far excitations (and errors) can propagate in each system.
As shown in Fig.~\ref{dev_422:fig:v_k_1d_2d}a), the maximum group velocity $v(k) = dE(k)/dk$ is substantially larger at $g_z=0.5$ than at $g_z=1.5$ in 1D, where the stronger longitudinal field slows the propagation front.\footnote{These calculations are done using exact diagonalization with periodic boundary conditions. 
The limits of this method prevent determining the group velocity for our lattice geometry with missing links, so a smaller uniform lattice is used instead.}
A similar, but less dramatic difference exists in 2D, shown in Fig.~\ref{dev_422:fig:v_k_1d_2d}b). 
The consequence of larger $v(k)$ for local observables is a larger backwards lightcone.
This implies more errors have the ability to affect a given $\langle \overline{Z}_n\rangle$ in the melting regime at fixed $t$. 
At a fixed number of syndrome extraction rounds $R$, errors can spread more under faster dynamics before they are caught by error detection.
This is the reason that $\alpha$ is consistently smaller in the melting regime (seen in Figs.~\ref{dev_422:fig:chain_results}e) and~\ref{dev_422:fig:grid_results}d), and degrades faster.
Further, the group velocity in 2D is significantly smaller than in 1D. 
The melting regime in 1D has by far the largest $v(k)$, $\sim2.5$ times larger than $g_z=1.5$ in 1D and $\sim5$ times larger than 2D $v(k_x,k_y)$. 
The fluctuations in $\alpha$ in Fig.~\ref{dev_422:fig:chain_results}e) are attributed to this difference.
Since the difference in $v(k_x,k_y)$ between the regimes is only 0.04 in 2D, errors propagate at more similar speeds, which explains why the $\alpha$ plots in Fig.~\ref{dev_422:fig:grid_results}d) are more similar than in 1D. 
This dependence on the backwards lightcone suggests that more frequent syndrome extraction is necessary for simulations of faster dynamics.
Unfortunately, the coherence time overhead is found to 
suppress most of the advantage of more frequent syndrome measurement in our simulations.

\subsection{Reset dependence}
\begin{figure}
    \centering
    \includegraphics[width=0.85\linewidth]{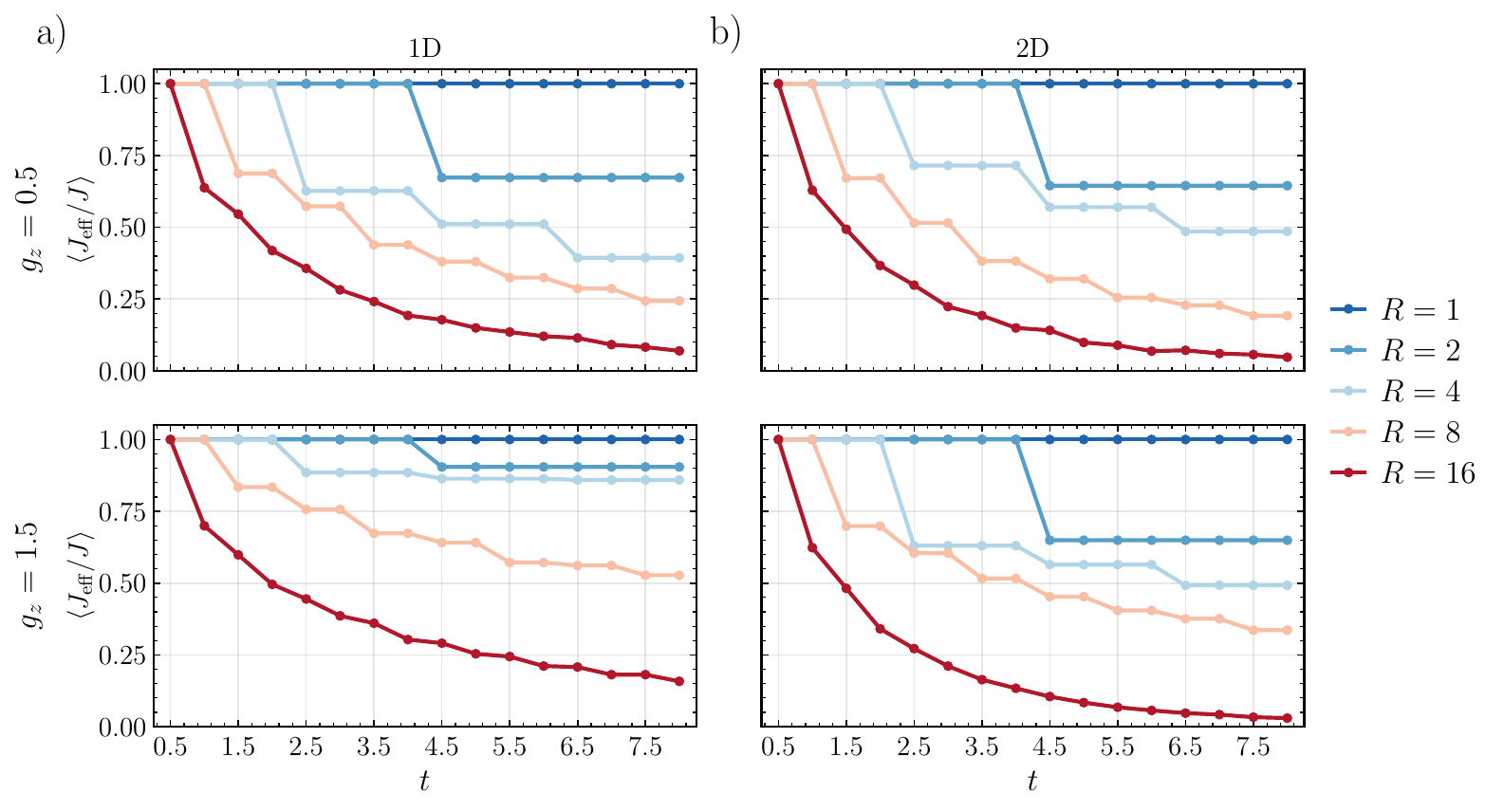}
    \caption{{\it Effective inter-block coupling under residual ancilla sign flips on} {\tt{ibm\_boston}}.
    a)~The effective inter-block coupling $\langle J_\text{eff}/J\rangle$ as a function of time $t$ averaged over all inter-block connections in the 1D simulations shown in Fig.~\ref{dev_422:fig:chain_results} for $g_z=0.5$ (top) and $g_z=1.5$ (bottom).
    Different $R$ values indicate the number of syndrome measurements.
    b)~The same as a) but for 2D results presented in Fig.~\ref{dev_422:fig:grid_results}.}
    \label{dev_422:fig:j_eff_chain_grid}
\end{figure}

\noindent
As a result of omitting resets on ancilla qubits after syndrome measurements, an ancilla that fires remains in the state $|1\rangle$ until a subsequent error measurement flips it back to $|0\rangle$.\footnote{The absence of resets also means that leakage on an ancilla persists across all subsequent rounds, since there is no re-initialization to return a leaked ancilla to the computational subspace.}
Any inter-block $\overline R_{ZZ}(\theta)$ rotation that uses this ancilla in its implementation therefore applies the opposite rotation $\overline R_{ZZ}(-\theta)$.
Interspersing these opposite-angle inter-block rotations among the correct ones partially cancels the intended coupling $J$, producing an effective ``slowed-down'' interaction $J_\text{eff} = J(1-2p_\text{flip})$, where $p_\text{flip}$ is the probability that the mediating ancilla is in $|1\rangle$ at the time of the gate.
The effect is amplified when two ancillas are used to implement the inter-block $\overline R_{ZZ}(\theta)$ rotation.
Figure~\ref{dev_422:fig:j_eff_chain_grid} shows the average $J_\text{eff}$ over all inter-qubit couplings as a function of time for various amounts of syndrome extraction rounds $R$.
In 1D, the reduction in $J$ is smaller in the localized regime, where the stronger longitudinal field $g_z$ suppresses the spread of correlations along the chain.
This results in a lower average $p_\text{flip}$ for $g_z=1.5$.
The differences in $J_\text{eff}$ between regimes and dimensions is again attributed to the group velocity which governs the propagation speed of errors. 
Since errors propagate slower at $g_z=1.5$, the effects of incorrect rotations take longer to spread.
This slowdown could be further studied by running MPS simulations with the couplings $J_\text{eff}$, but this is not done in this work.

The effective reduction in $J$ could be removed entirely with the introduction of ancilla resets after syndrome measurements, and is the primary reason that high-$R$ simulations are found to perform worse than low-$R$ simulations.
Increasing syndrome measurement frequency adds more opportunities for stale ancillas to artificially inject errors into the simulation, and if the corresponding detection events are not selected for postselection (e.g., by ORP), they contaminate the results.
With ancilla resets, this problem is absent. 
However, ancilla resets consume a significant amount of coherence time, and as a result errors are nevertheless introduced.
We find that the implementation of resets does not significantly change the $R$-dependence of the results, with the exception of the highest values of $R$. 
Even for $R=16$, where the improvement from resets is largest, the $\alpha$ values do not exceed those of low $R$, indicating that the added coherence time injects more errors than the syndrome measurement rounds are able to detect. 
The implementation of faster resets on the hardware will likely shift this balance. 
This is another example of the tradeoff between the NISQ considerations of gate depth and coherence time and FT requirements. 

\subsection{Geometric dependence in 1D simulations}
\begin{figure}
    \centering
    \includegraphics[width=0.75\linewidth]{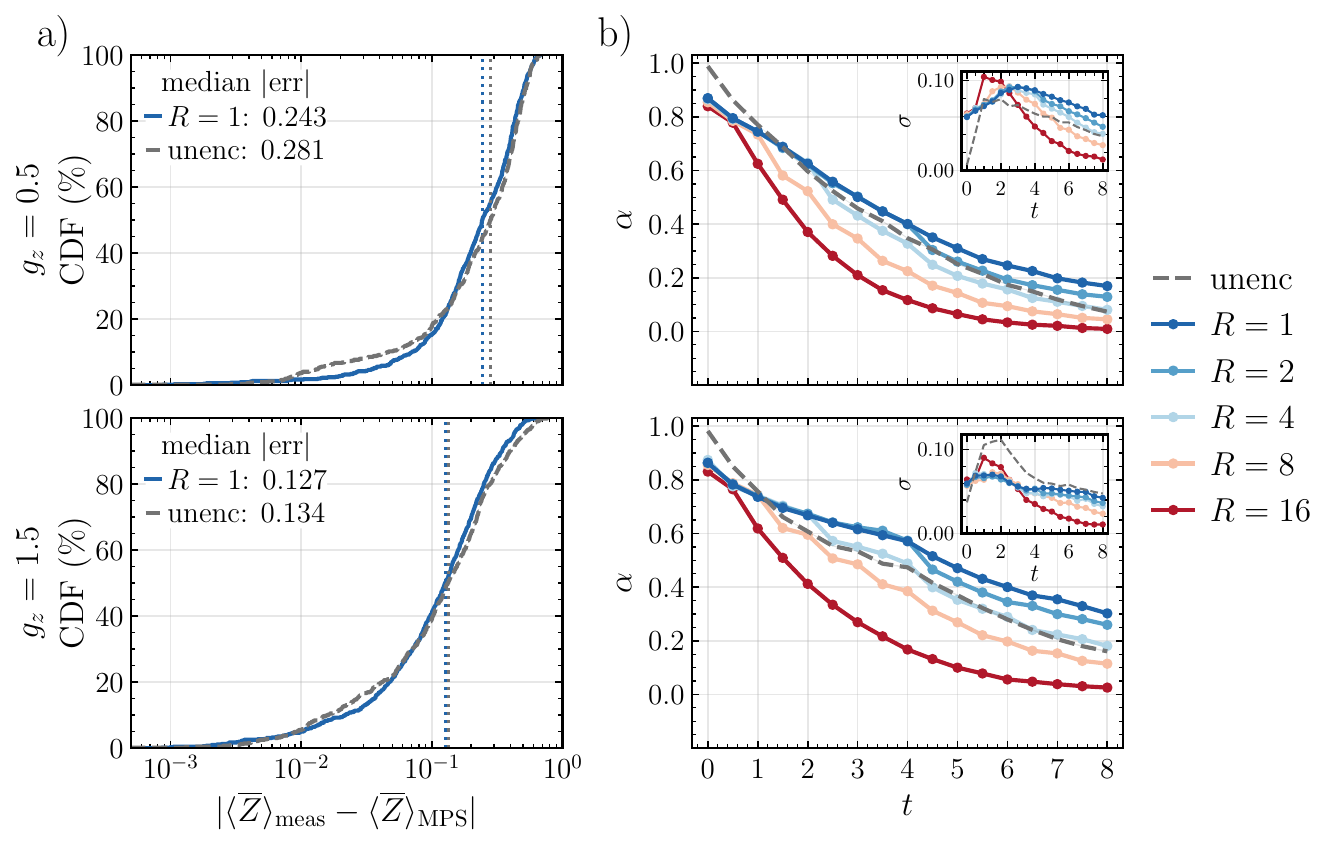}
    \caption{{\it Encoding improvement without postselection in 2D simulations.}
    a)~Cumulative distribution functions (CDFs) of the per-qubit, per-time absolute error $|\langle \overline{Z}\rangle_\text{meas} - \langle \overline{Z}\rangle_\text{MPS}|$ for both values of $g_z$.
    The median error and CDFs corresponding to $R=1$ syndrome measurements are shown compared to unencoded results (solid and dashed lines respectively).
    b)~The signal-survival factor $\alpha$ as a function of $t$ for various $R$ (colored lines) and each value of $g_z$, compared to unencoded results (dashed grey line).
    The insets show the residual error $\sigma$ in the fit. }
    \label{dev_422:fig:grid_results_raw}
\end{figure}

\noindent
The encoded results presented in Fig.~\ref{dev_422:fig:grid_results} show a large improvement over unencoded runs, increasing with time to a greater than $200\%$ improvement at late times.
This is partially due to the geometric gate depth overhead of embedding a square lattice onto heavy-hex connectivity. 
Unencoded circuits are 1.5 times deeper than encoded circuits at the latest time considered (see Table~\ref{dev_422:tab:grid_circ_nums}). 
To isolate the encoding improvement from the geometric overhead, we compare the encoded results without any postselection to unencoded results in Fig.~\ref{dev_422:fig:grid_results_raw}.
The CDF plots show that encoding without postselection has a small improvement at $g_z=0.5$ and a negligible improvement at $g_z=1.5$. 
The decay of the signal-survival factor $\alpha$ shows the performance of the encoded circuits over time without postselection.
At early times, encoded circuits without postselection perform worse than unencoded as a result of added circuit depth to perform syndrome measurements.
At intermediate and late times, low-$R$ encoded runs outperform unencoded due to the geometric overhead.
Comparing to Fig.~\ref{dev_422:fig:grid_results}), the maximum decrease in $\alpha$ from omitting postselection is $\sim 0.4$.
This indicates that much of the improvement seen in Fig.~\ref{dev_422:fig:grid_results} is due to encoding as opposed to the geometric overhead in unencoded circuits.
Thus, removing errors that can potentially spread to $\mathcal O(t^2)$ qubits plays a large role in the improvement observed in 2D simulations.

\section{Classical simulations of false-vacuum decay dynamics}
\label{dev_422:app:physics}
\noindent
Phase transitions play a key role in early-universe dynamics as well as in many condensed matter systems.
For high energy physics, a particularly important question is whether the Higgs field is in a true vacuum or in a false metastable vacuum.
The phase transition from a false vacuum to a true vacuum is known as false-vacuum decay and underlies a range of early-universe phenomena, from cosmological phase transitions and bubble nucleation to the possible fate of the electroweak vacuum~\cite{Coleman:1977py,Callan:1977pt}.
This section describes a simple 1D model for false-vacuum decay in spin systems as intuition for the oscillation and localization behavior observed in the simulations in the main text.
Classical MPS simulations carried out in the same regimes considered in Sec.~\ref{dev_422:sec:results} are shown to support this intuition where analytical calculations are unavailable.

In false-vacuum decay phenomenology, the metastable ``false'' vacuum decays to the stable and energetically favored ``true'' vacuum by quantum tunneling~\cite{Coleman:1977py,Callan:1977pt}.
In the semiclassical regime, thermal or quantum fluctuations typically nucleate localized bubbles of true vacuum.
Bubbles exceeding a critical size then lower the energy through expansion, driving conversion of the system to the true vacuum.
The method in this work provides a lattice realization of false-vacuum decay through quench dynamics of a spin system~\cite{Lagnese:2021grb,Lagnese:2023xjg}.
Rather than directly studying bubble nucleation, we prepare a fixed, finite system in a ferromagnetic initial state with a true-vacuum bubble in the center and follow its unitary time evolution.
The presence of ``decay'' in this setup is indicated by the melting of the true-vacuum bubble, and decay is absent when the bubble remains localized.

Section~\ref{dev_422:sec:results} observes a difference in encoded quantum simulation performance between the $g_z=0.5$ (melting regime) and $g_z=1.5$ (localized regime). 
The melting or localization of true-vacuum bubbles in our simulations is due to domain-wall oscillations known as Bloch oscillations~\cite{Pomponio:2021ltz}.
The following analysis considers two domain walls in a 1D chain.
The Hamiltonian~\eqref{dev_422:eq:h_ising} assigns a fixed energy per flipped spin, so the bubble's potential energy is linear in its size $r$
\begin{align}
    V(r) \ = \ -\chi r \ , \ \chi \ = \ 2m|g_z| \ , \ m = \left(1-\left(\frac{g_x}{|J|}\right)^2 \right)^{1/8} \ ,
\end{align}
where $m$ is the spontaneous magnetization at $g_z=0$~\cite{Pfeuty:1970qrn}.
Unitary time evolution under $\overline H$ conserves total energy, but converts the potential energy into the kinetic energy of expanding bubble walls.
As $V(r)$ changes from bubble expansion, the relative momentum of the bubble walls $k$ changes, as given by Hamilton's equations, 
\begin{align}
    \frac{dk}{dt} \ = \ -\frac{dV}{dr} \  = \ \chi \ .
\end{align}
Neglecting corrections due to nonzero $g_z$, the dispersion relation for a single domain wall is
\begin{align}
    E(k) \ = \ 2 |J| \sqrt{1 + g_x^2 + 2g_x \cos k} \ .
\end{align}
Since $E(k)$ is periodic, the group velocity $v(k) = dE(k)/dk$ changes sign as the momentum is increased in the Brillouin zone, resulting in oscillations of the bubble width
\begin{align}
    r(t) \ = \ \frac{4}{\chi} \left( E\left(k_0 + \frac{\chi}{2}t\right) - E(k_0)\right) \ .
\end{align}
The maximum amplitude of these oscillations is
\begin{align}
    r_\text{max} \ = \ \frac{4}{\chi} (E(\pi) - E(0)) \ = \ \frac{8|J| |g_x|}{m |g_z|}\ ,
    \label{dev_422:eq:r_max}
\end{align}
and the period is 
\begin{align}
    T_B \ = \ \frac{4 \pi}{\chi} \ = \ \frac{2 \pi}{m |g_z|} \ .
\end{align}

\begin{figure}
    \centering
    \includegraphics[width=\linewidth]{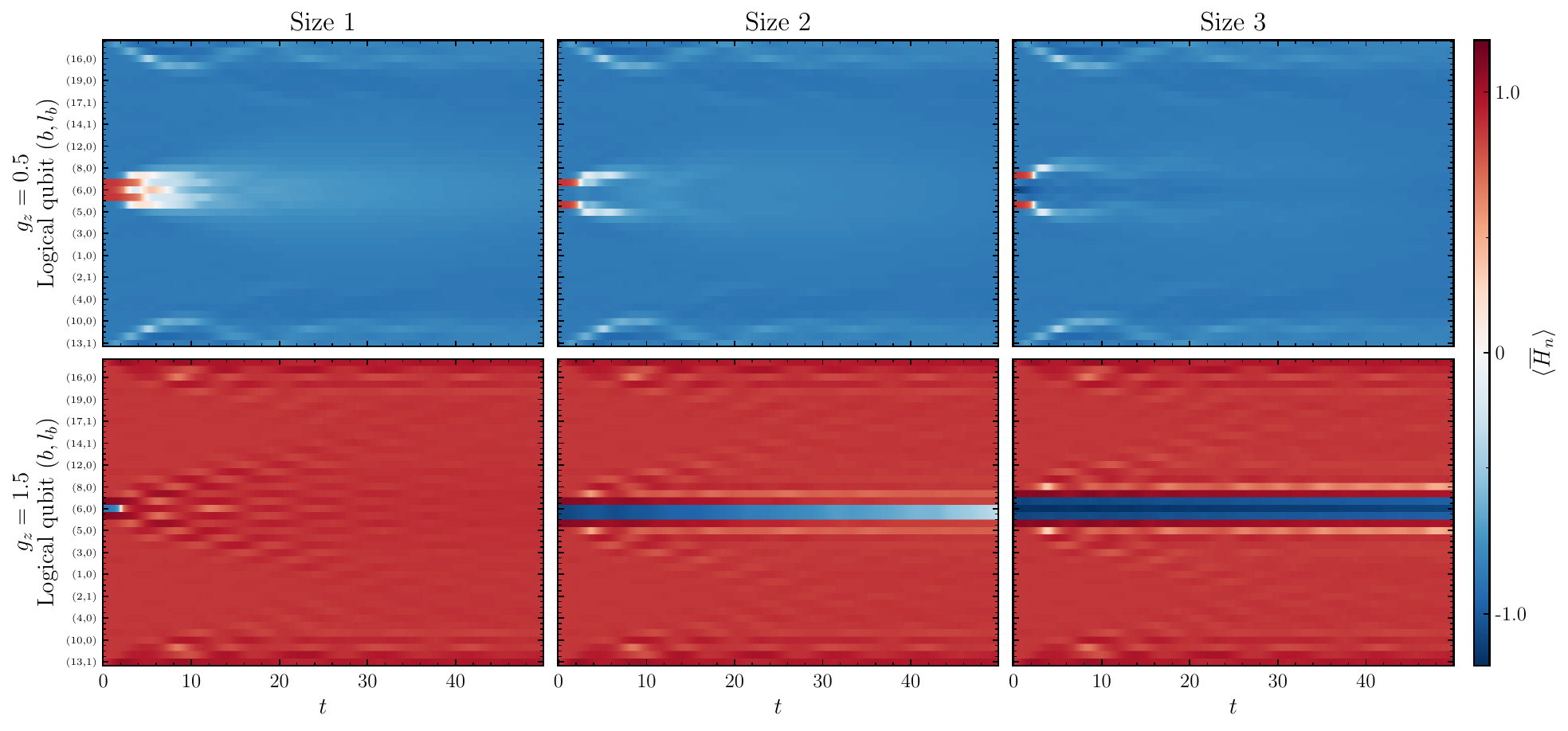}
    \caption{{\it Classical simulations of false-vacuum decay dynamics in 1D.}
    An initial true-vacuum bubble of size one, two, three (columns) at the center of the 42-qubit chain is evolved with $g_z=0.5$ (top row) and $g_z=1.5$ (bottom row).
    The color in the heatmap shows the energy density $\langle \overline{H}_n\rangle$ as a function of logical qubit position $n=(b,l_b)$ and time $t$.
    The time evolution uses $n_T=500$ Trotter steps of size $\delta t=0.1$ and is computed with MPS.}
    \label{dev_422:fig:mps_1d_gz_comparison}
\end{figure}

Although strictly valid only in the analytically tractable $g_z\to 0$ limit, and not at the couplings considered in our simulations, this analysis provides intuition for how oscillations arise in the domain-wall initial state.
An equivalent treatment in the large-$J$ limit maps the system to free fermions in a linear potential~\cite{Balducci:2022zym,Balducci:2022kvd}, where the extent of oscillations is also found to be $r_\text{max}\sim |g_x|/|g_z|$.
Reference~\cite{Balducci:2022zym} finds a transition between the bubble-melting and localized regimes at $|g_z|/|g_x|\sim 1$ for large $J$.
Systems with $|g_z|/|g_x|< 1$ are found to thermalize, while $|g_z|/|g_x|> 1$ shows evidence of slower decay.
Results from our simulations at $J=1$ are consistent with these findings.
Figure~\ref{dev_422:fig:mps_1d_gz_comparison} shows the energy density determined from MPS simulations through the relaxation dynamics following a quench under the Hamiltonian~\eqref{dev_422:eq:h_ising} in 1D.
Initial bubble states ${|\psi(t=0)\rangle = \prod_{i\in B}\overline X_i |\overline 1\rangle^{\otimes N}}$ with bubble sizes $|B|=1,2,3$ are shown in the columns. 
The top row shows the bubble-melting regime ($g_z=0.5$).
Faint oscillations about the center of the bubble are seen for sizes two and three, consistent with Bloch oscillations with a period $T_B\sim12$ and $r_\text{max}\sim10$.
Since our simulations are far outside the small-$g_z$ or large-$J$ regime, we cannot compare these to predictions of $T_B$ or $r_\text{max}$ from the analysis above.
The bottom row ($g_z=1.5$) shows the bubble localized to all accessible times for sizes two and three.
This is consistent with Eq.~\eqref{dev_422:eq:r_max}, where larger $g_z$ makes $r_\text{max}$ comparable to the bubble size and prevents oscillations. 
A size one bubble is seen to decay for both values of the coupling because the Bloch mechanism requires separation between the two bubble walls for the linear potential to act on, which does not exist for a single flipped spin.
This confirms the localization is not due to $Z$-field pinning, in which a strong longitudinal field suppresses interface-moving spin flips when $|g_z|\gg |g_x|$.

\begin{figure}
    \centering
    \includegraphics[width=0.75\linewidth]{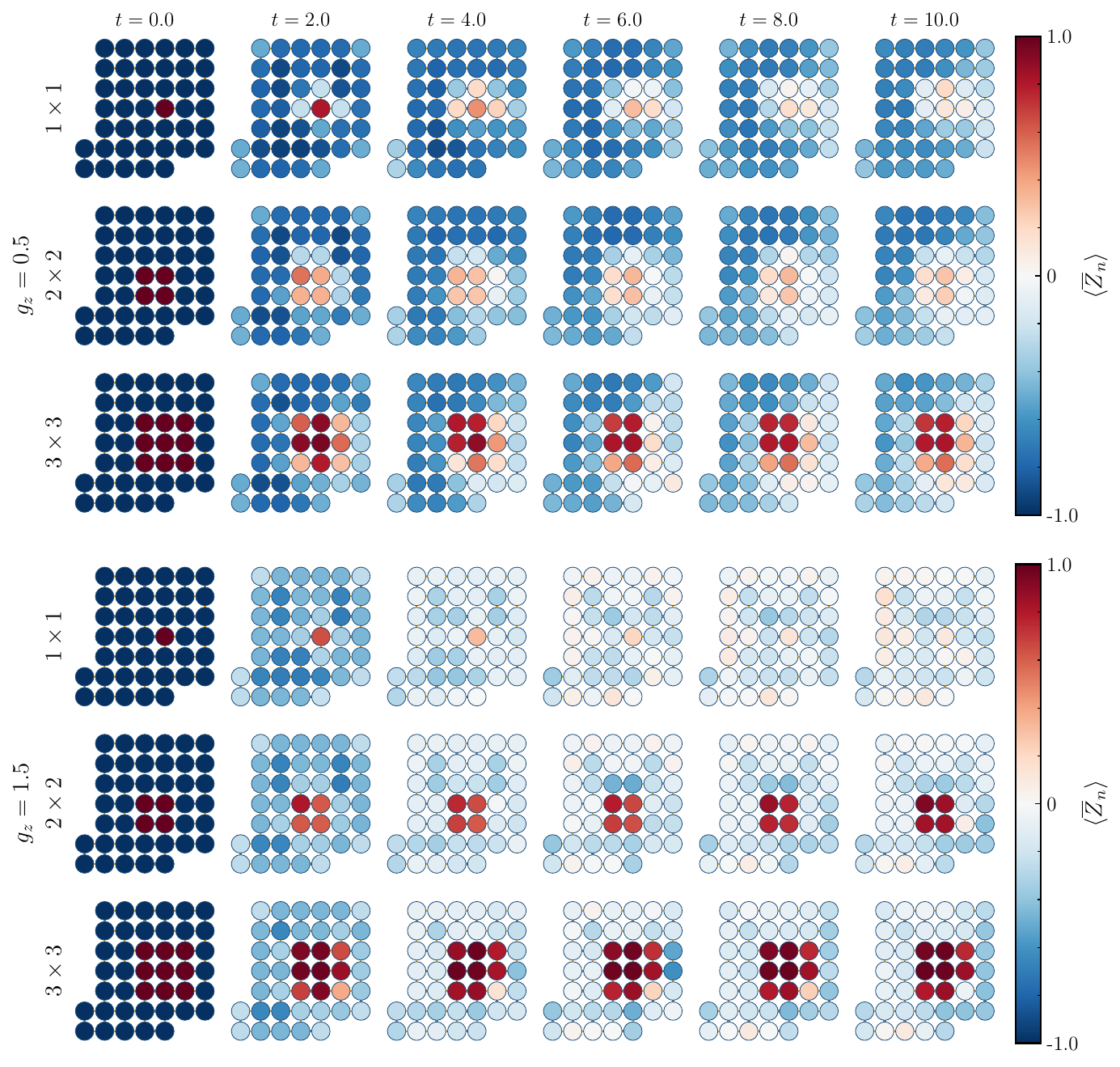}
    \caption{{\it Classical simulations of false-vacuum decay dynamics on a 2D lattice with missing edges.}
    An initial true-vacuum bubble of size $1\times1$, $2\times2$, $3\times3$ (rows) at the center of the 42-qubit lattice is evolved with $g_z=0.5$ (top) and $g_z=1.5$ (bottom).
    The qubit colors show the magnetization $\langle \overline{Z}_n\rangle$ as a function of logical qubit position $n=(b,l_b)$ and time $t$ (columns).
    The time evolution uses $n_T=100$ Trotter steps of size $\delta t=0.1$ calculated with MPS.
    The logical connectivity grid corresponds to  Fig.~\ref{dev_422:fig:chain_results}a) with all available edges used.}
    \label{dev_422:fig:mps_2d_gz_comparison}
\end{figure}

We find similar behavior in 2D simulations. 
Figure~\ref{dev_422:fig:mps_2d_gz_comparison} shows $\langle \overline{Z}_n\rangle$ determined from MPS simulations throughout the lattice for a selection of times.
The dynamics in 2D is found to be slower than in 1D, which is also evidenced by the group velocities shown in Fig.~\ref{dev_422:fig:v_k_1d_2d}.
Further, the melting and localization dynamics is anisotropic due to the irregular geometry realized by the logical qubit connectivity. 
For $g_z=0.5$ sizes $2\times2$ and $3\times3$, the initial bubble is seen to spread more towards the right and bottom as a result of missing edges on the left and top boundaries.
These missing edges also artificially localize the bubble for $g_z=1.5$.
However, physics-driven localization (as opposed to localization driven by missing links) is still seen on the right and bottom, consistent with the dynamics seen in 1D simulations.
Additionally, the bottom right qubit of the $3\times3$ bubble for $g_z=1.5$ melts off as a result of a missing vertical edge that would connect it to the rest of the bubble.\footnote{The global quench of the background contributes a significant amount to the observed dynamics, which is seen in the relaxation dynamics shown in Figs.~\ref{dev_422:fig:mps_1d_gz_comparison} and~\ref{dev_422:fig:mps_2d_gz_comparison}.}

\begin{figure}
    \centering
    \includegraphics[width=0.75\linewidth]{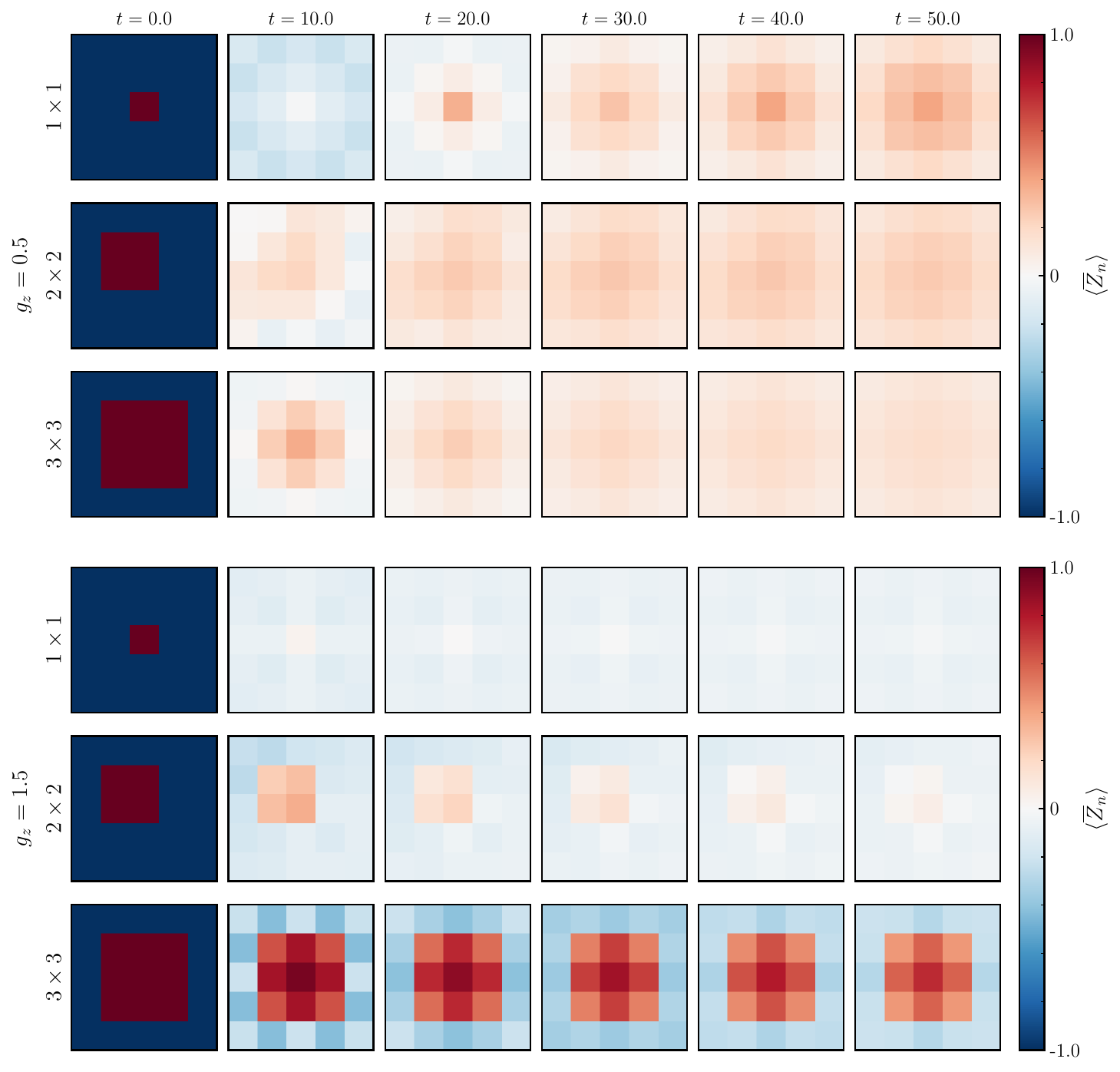}
    \caption{{\it Classical simulations of false-vacuum decay dynamics on a uniform 2D lattice.}
    An initial true-vacuum bubble of size $1\times1$, $2\times2$, $3\times3$ (rows) at the center of the $5\times5$ lattice is evolved with $g_z=0.5$ (top) and $g_z=1.5$ (bottom).
    The qubit colors show the magnetization $\langle \overline{Z}_n\rangle$ as a function of logical qubit position $n=(b,l_b)$ and time $t$ (columns).
    The time evolution uses $n_T=500$ Trotter steps of size $\delta t=0.1$ and is calculated with statevector simulations.}
    \label{dev_422:fig:mps_2d_gz_uniform_comparison}
\end{figure}

To remove the contributions to the melting and localization dynamics from missing links, we compare the results of our realized grid to simulations on a uniform $5\times5$ lattice.
The comparison of $\langle \overline{Z}_n\rangle$ is shown for several selected times in Fig.~\ref{dev_422:fig:mps_2d_gz_uniform_comparison}. 
This figure confirms that melting and localization consistent with Bloch oscillations are present in the absence of grid irregularities. 
The top row of the $g_z=0.5$ panel of Fig.~\ref{dev_422:fig:mps_2d_gz_uniform_comparison} shows a recurrence for the $1\times1$ bubble.
This happens due to boundary effects and is absent in 1D simulations in Fig.~\ref{dev_422:fig:mps_1d_gz_comparison}.
This is additionally verified by MPS simulations of a $7\times7$ lattice.
The $2\times2$ bubble is seen to melt for $g_z=1.5$.
This melting is slower than for $g_z=0.5$ and is also impacted anisotropically by the boundary.
This is consistent with the findings of Ref.~\cite{Balducci:2022kvd} where $|g_z|/|g_x|\gg1$ significantly slows down but does not fully prevent melting.

Together, these classical simulations show that the size-three bubbles (1D) and 
$3\times3$ bubbles (2D) studied in Sec.~\ref{dev_422:sec:results} melt for 
$g_z=0.5$ and remain localized for $g_z=1.5$,
consistent with the Bloch oscillation picture. 
While the dynamics in 2D is significantly slower and the contrast between the melting and localized regimes is less pronounced, the same qualitative behavior holds, supporting the interpretation of the device results in the main text.
Although our couplings lie outside the small-$g_z$ and large-$J$ limits where the oscillation amplitude and period can be predicted analytically, the melting and localization observed here match the Bloch oscillation mechanism qualitatively.

\section{Other layouts}
\label{dev_422:app:other_layouts}
\begin{figure}
    \centering
    \includegraphics[width=\linewidth]{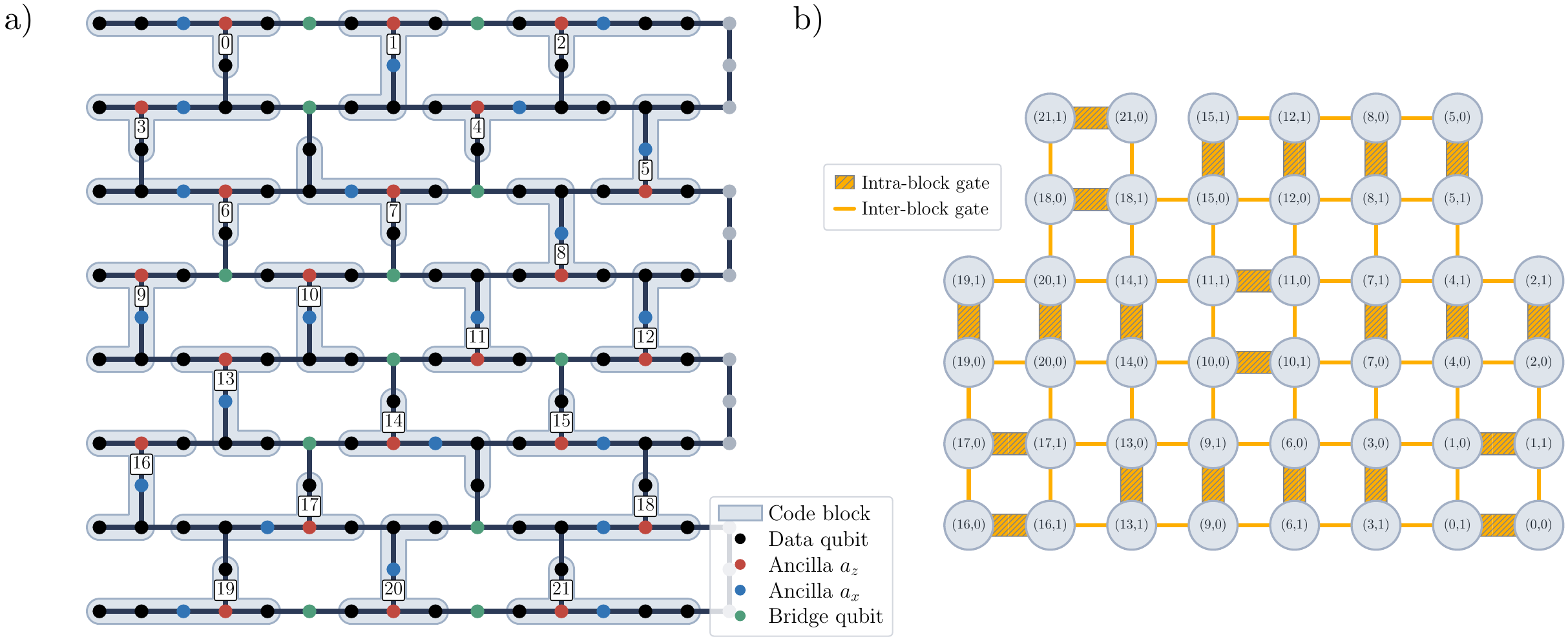}
    \caption{{\it Block placement and logical connectivity graph in the absence of faulty qubits and gates.}
    a)~An example of a block placement (grey) of 21 blocks assuming no faulty qubits and gates on a heavy-hex quantum computer.
    Within each block, the data qubits are shown in black, $Z$-ancillas are shown in red, $X$-ancillas are shown in blue, and bridge qubits are shown in green.
    b)~The resulting connectivity between logical qubits (grey) optimized for similarity to a square lattice.
    Thick orange hatched edges represent intra-block connections and thin orange edges show inter-block connections.}
    \label{dev_422:fig:placement_grid_all_qubits}
\end{figure}

\noindent
As discussed in Sec.~\ref{dev_422:sec:422_device}, 
there is significant freedom in $[[4,2,2]]$ code block placement onto the heavy-hex connectivity. 
In this work, we select a placement that avoids particularly faulty qubits and gates.
The block placement options are further expanded if this constraint is lifted.
While the maximum number of blocks is still 21 (corresponding to 42 logical qubits), the properties of the resulting placement may be more favorable. 
In particular, the resulting logical connectivity graph may be more similar to a regular square lattice.
Figure~\ref{dev_422:fig:placement_grid_all_qubits}a) shows an example of one such placement.
This placement uses 144 active physical qubits, compared to 136 in Fig.~\ref{dev_422:fig:overview}b), a result of more bridge qubits present. 
An example of the resulting logical connectivity graph possible with this placement is shown in Fig.~\ref{dev_422:fig:placement_grid_all_qubits}b). 
This grid has 51 inter-block edges (72 edges total), whereas the grid used in the main text (Fig.~\ref{dev_422:fig:chain_results}a)) has 45 inter-block edges (66 edges total).
Importantly, this grid has fewer missing interior edges, which significantly contribute to the dynamics in our simulations. 
This suggests that more regular square lattice simulations are possible using the current generation of quantum computers with more uniform characteristics.

\begin{figure}
    \centering
    \includegraphics[width=0.475\linewidth]{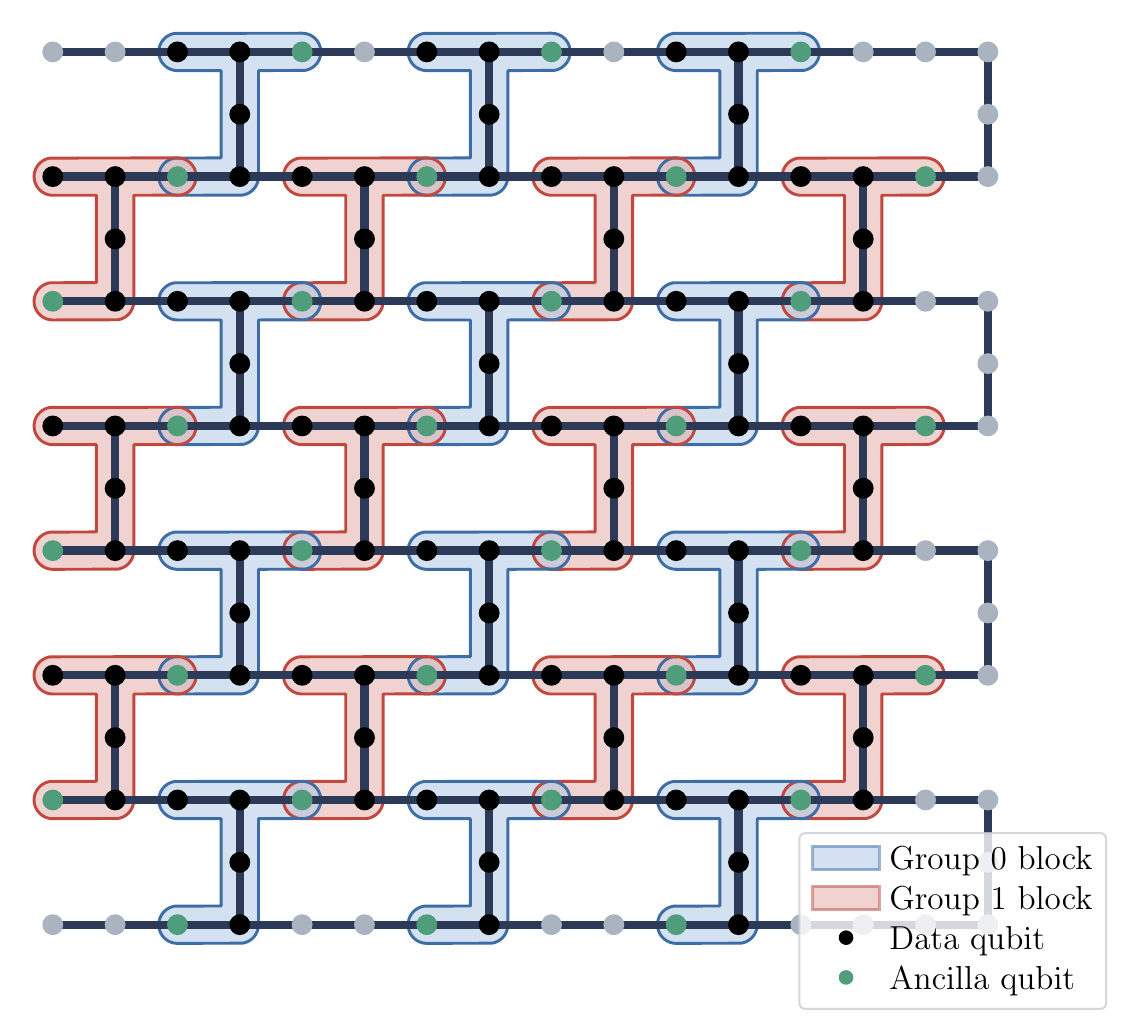}
    \caption{{\it Overlapping block placement.}
    Code blocks are split into two groups (red and blue) for serialized syndrome extraction.
    Data qubits within each block are shown in black and shared ancilla qubits are shown in green.}
    \label{dev_422:fig:block_placement_overlapping}
\end{figure}

It is not possible to regularly tile a heavy-hex connectivity device with blocks that have the topology required for our syndrome extraction circuits (center of Fig.~\ref{dev_422:fig:overview}c)). 
Therefore bridge qubits are necessary, which contribute significantly to the depth of inter-block gates. 
A regular tiling with our block topology is possible if blocks are allowed to share ancillas. 
An example of such a placement is shown in Fig.~\ref{dev_422:fig:block_placement_overlapping}.
This placement supports a higher encoding rate, using 126 physical qubits to encode 48 logical qubits with 24 overlapping blocks.
In addition, the logical connectivity graph possible with this placement is more dense, allowing for three-body gates between triples of neighboring blocks (with ancilla shuttling).
In an overlapping block placement, code blocks share ancilla qubits (shown in green) that are used only for syndrome extraction, and data qubits (black) are permanently tied to a given block.
This removes the overhead of bridge qubits.
Since data qubits of neighboring blocks are located physically closer, inter-block gates are considerably shallower, which is favorable for the total depth budget.
The tradeoff associated with overlapping layouts is that syndrome extraction must now be serialized as a result of ancilla sharing. 
Figure~\ref{dev_422:fig:block_placement_overlapping} splits the blocks into two groups (shown in red and blue), corresponding to the serial order in which syndromes must be measured.
Ancilla shuttling is also required in this scenario, which introduces an overhead larger than that of switching between the compute and syndrome layouts in Fig.~\ref{dev_422:fig:overview}c). 
Further, the conversion to detectors discussed in Sec.~\ref{dev_422:sec:filtering_main_text} and App.~\ref{dev_422:sec:filtering} is modified due to ancilla sharing.
Syndrome measurement is by far the highest overhead in coherence time, which is doubled by serialization.
This work finds that the quality of local observables measured degrades significantly faster with overlapping code blocks compared to non-overlapping placements, and for this reason the non-overlapping placement is used.

\section{Tables of results}
\label{dev_422:app:tables}
\noindent
Tables~\ref{dev_422:tab:chain_results_t_8},~\ref{dev_422:tab:chain_cdf} and~\ref{dev_422:tab:chain_alpha_sigma} show the data from {\tt ibm\_boston} corresponding to Fig.~\ref{dev_422:fig:chain_results}. 
A 1D 42-logical qubit chain is run with $\delta t=0.5$ for $g_z=0.5$ and $g_z=1.5$. 
The initial state in each of these simulations is a size-three true-vacuum bubble product state.
Tables~\ref{dev_422:tab:grid_results_t_4},~\ref{dev_422:tab:grid_cdf} and~\ref{dev_422:tab:grid_alpha_sigma} show the corresponding data for the 2D lattice simulations displayed in Fig.~\ref{dev_422:fig:grid_results}.
The initial state on the 42-logical qubit lattice is a $3\times3$ true-vacuum bubble product state.
Table~\ref{dev_422:tab:acceptance_rates} reports the acceptance rate of $R=1$ runs for 1D and 2D simulations for both values of $g_z$.

\begin{table}[h]
\renewcommand{\arraystretch}{0.7}
\scriptsize
\begin{tabularx}{\linewidth}{|c||Y|Y|Y||Y|Y|Y|}
\hline
\rule{0pt}{10pt} & \multicolumn{6}{c|}{\large$\langle \overline{Z}_n\rangle $}\\\hline\hline
\rule{0pt}{10pt} \multirow{2}{*}{Logical qubit $(b,l_b)$} &  \multicolumn{3}{c||}{$g_z=0.5$}  &  \multicolumn{3}{c|}{$g_z=1.5$} \\\cline{2-7}
\rule{0pt}{10pt} & MPS & Unencoded & $R=1$ & MPS & Unencoded & $R=1$  \\
\hline\hline
$(13,1)$ & 0.217 & 0.100(6) & 0.098(19) & -0.568 & -0.486(5) & -0.329(27) \\
$(13,0)$ & 0.142 & 0.131(5) & 0.113(16) & -0.372 & -0.250(5) & -0.385(26) \\
$(10,1)$ & 0.084 & 0.090(5) & 0.081(17) & -0.138 & -0.066(5) & -0.244(36) \\
$(10,0)$ & -0.079 & -0.006(6) & 0.000(21) & -0.445 & -0.213(5) & -0.256(11) \\
$(9,0)$ & 0.003 & -0.006(6) & -0.019(21) & -0.381 & -0.263(5) & -0.249(11) \\
$(9,1)$ & 0.082 & 0.009(6) & 0.019(34) & -0.330 & -0.260(5) & -0.299(18) \\
$(4,0)$ & 0.080 & -0.043(6) & 0.024(32) & -0.289 & -0.263(6) & -0.248(22) \\
$(4,1)$ & 0.099 & 0.022(6) & 0.071(32) & -0.381 & -0.248(6) & -0.239(33) \\
$(2,0)$ & 0.115 & 0.055(6) & 0.065(18) & -0.407 & -0.283(5) & -0.269(20) \\
$(2,1)$ & 0.130 & 0.109(5) & 0.057(13) & -0.342 & -0.229(5) & -0.229(46) \\
$(0,0)$ & 0.175 & 0.121(6) & 0.067(46) & -0.379 & -0.297(5) & -0.202(31) \\
$(0,1)$ & 0.124 & 0.076(6) & 0.056(15) & -0.351 & -0.286(5) & -0.136(18) \\
$(1,0)$ & 0.071 & 0.044(6) & 0.009(32) & -0.317 & -0.223(5) & -0.132(31) \\
$(1,1)$ & 0.029 & 0.043(6) & 0.030(18) & -0.397 & -0.286(6) & -0.106(42) \\
$(3,1)$ & 0.068 & 0.115(6) & -0.020(17) & -0.347 & -0.166(6) & -0.152(35) \\
$(3,0)$ & 0.071 & 0.109(6) & 0.015(43) & -0.374 & -0.194(5) & -0.268(18) \\
$(7,1)$ & 0.029 & 0.071(5) & -0.010(29) & -0.228 & -0.151(6) & -0.223(11) \\
$(7,0)$ & -0.090 & -0.026(5) & -0.090(50) & -0.371 & -0.271(5) & -0.269(34) \\
$(5,0)$ & -0.011 & -0.028(6) & 0.029(17) & -0.348 & -0.245(5) & -0.237(40) \\
$(5,1)$ & 0.145 & 0.096(6) & 0.082(17) & -0.549 & -0.462(5) & -0.450(26) \\
$(6,1)$ & 0.206 & 0.217(6) & 0.163(41) & 0.623 & 0.549(5) & 0.555(21) \\
$(6,0)$ & 0.170 & 0.219(5) & 0.149(47) & 0.928 & 0.806(3) & 0.813(10) \\
$(11,0)$ & 0.245 & 0.223(5) & 0.211(30) & 0.809 & 0.677(4) & 0.595(12) \\
$(11,1)$ & 0.142 & 0.108(6) & 0.087(17) & -0.705 & -0.499(5) & -0.489(13) \\
$(8,0)$ & -0.071 & -0.058(6) & -0.108(30) & -0.304 & -0.231(6) & -0.179(10) \\
$(8,1)$ & -0.108 & -0.070(6) & -0.089(14) & -0.329 & -0.291(5) & -0.345(32) \\
$(12,1)$ & -0.001 & -0.005(6) & -0.010(28) & -0.252 & -0.206(6) & -0.137(32) \\
$(12,0)$ & 0.084 & 0.038(6) & 0.012(17) & -0.373 & -0.249(5) & -0.309(39) \\
$(15,1)$ & 0.095 & 0.024(6) & 0.083(20) & -0.275 & -0.152(5) & -0.191(13) \\
$(15,0)$ & 0.100 & 0.052(6) & 0.063(28) & -0.403 & -0.318(5) & -0.243(20) \\
$(14,1)$ & 0.101 & 0.096(5) & -0.002(11) & -0.403 & -0.242(6) & -0.237(30) \\
$(14,0)$ & 0.088 & 0.171(5) & -0.010(12) & -0.280 & -0.053(6) & -0.163(39) \\
$(17,0)$ & 0.080 & 0.104(5) & -0.012(11) & -0.333 & -0.199(5) & -0.252(19) \\
$(17,1)$ & 0.066 & 0.076(6) & 0.040(18) & -0.357 & -0.293(5) & -0.246(13) \\
$(20,0)$ & 0.054 & 0.059(5) & 0.032(10) & -0.425 & -0.335(5) & -0.258(26) \\
$(20,1)$ & 0.067 & 0.076(6) & 0.007(15) & -0.274 & -0.200(5) & -0.135(9) \\
$(19,0)$ & 0.088 & 0.067(6) & -0.016(40) & -0.323 & -0.229(5) & -0.288(22) \\
$(19,1)$ & 0.013 & 0.012(5) & -0.048(50) & -0.383 & -0.290(5) & -0.233(9) \\
$(16,1)$ & -0.077 & -0.074(6) & -0.024(20) & -0.443 & -0.278(5) & -0.245(26) \\
$(16,0)$ & 0.083 & 0.012(6) & 0.089(11) & -0.139 & -0.125(6) & -0.247(14) \\
$(18,0)$ & 0.142 & 0.120(6) & 0.099(11) & -0.372 & -0.203(5) & -0.489(35) \\
$(18,1)$ & 0.217 & 0.166(6) & 0.054(23) & -0.568 & -0.301(6) & -0.468(17) \\
 \hline
\end{tabularx}
\caption{{\it Magnetization in 1D simulations at $t=8$.}
For each logical qubit $(b,l_b)$ (first column), $\langle \overline{Z}_n\rangle$ from MPS results (second and fifth columns) is compared to unencoded results (third and sixth columns) and encoded results with $R=1$ syndrome extraction rounds (fourth and seventh columns).
The left columns show $g_z=0.5$ and the right columns show $g_z=1.5$.
The uncertainty is determined through bootstrap resampling.}
\label{dev_422:tab:chain_results_t_8}
\renewcommand{\arraystretch}{1.0}
\end{table}

\begin{table}[h]
\tiny
\setlength{\tabcolsep}{2pt}
\begin{tabularx}{\linewidth}{|c||Y|Y|Y|Y|Y|Y||Y|Y|Y|Y|Y|Y|}
\hline
\rule{0pt}{10pt} & \multicolumn{12}{c|}{\large ${|\langle \overline{Z}\rangle_\text{meas} - \langle \overline{Z}\rangle_\text{MPS}|}$}\\\hline\hline
\rule{0pt}{10pt} \multirow{2}{*}{\makecell{$\%$}} &  \multicolumn{6}{c||}{$g_z=0.5$}  &  \multicolumn{6}{c|}{$g_z=1.5$} \\\cline{2-13}
\rule{0pt}{10pt}  & Unenc. & $R=1$ & $R=2$ & $R=4$  & $R=8$ & $R=16$ & Unenc. & $R=1$ & $R=2$ & $R=4$  & $R=8$ & $R=16$ \\
\hline\hline
0\% & 0.000(0) & 0.000(0) & 0.000(0) & 0.000(0) & 0.000(0) & 0.000(0) & 0.001(0) & 0.000(0) & 0.000(0) & 0.000(0) & 0.000(0) & 0.000(0) \\
10\% & 0.009(0) & 0.005(1) & 0.007(1) & 0.007(1) & 0.008(1) & 0.013(1) & 0.014(1) & 0.009(1) & 0.012(1) & 0.010(1) & 0.011(1) & 0.016(1) \\
20\% & 0.016(1) & 0.014(1) & 0.016(1) & 0.015(1) & 0.017(1) & 0.023(1) & 0.031(1) & 0.018(1) & 0.020(1) & 0.021(1) & 0.023(1) & 0.028(1) \\
30\% & 0.022(1) & 0.021(1) & 0.025(1) & 0.024(1) & 0.027(1) & 0.037(2) & 0.048(1) & 0.033(1) & 0.033(1) & 0.039(2) & 0.041(2) & 0.051(2) \\
40\% & 0.031(1) & 0.031(1) & 0.034(1) & 0.035(1) & 0.040(1) & 0.051(1) & 0.062(1) & 0.047(1) & 0.049(2) & 0.056(2) & 0.062(2) & 0.083(3) \\
50\% & 0.040(1) & 0.039(1) & 0.043(1) & 0.046(1) & 0.055(1) & 0.065(1) & 0.074(1) & 0.067(2) & 0.066(2) & 0.074(2) & 0.084(2) & 0.137(3) \\
60\% & 0.050(1) & 0.049(1) & 0.053(1) & 0.057(2) & 0.068(2) & 0.081(2) & 0.091(1) & 0.084(2) & 0.086(2) & 0.095(2) & 0.124(3) & 0.213(4) \\
70\% & 0.062(1) & 0.061(2) & 0.066(2) & 0.075(2) & 0.082(2) & 0.098(2) & 0.109(1) & 0.108(2) & 0.111(2) & 0.127(3) & 0.167(3) & 0.282(4) \\
80\% & 0.078(1) & 0.076(2) & 0.083(2) & 0.091(2) & 0.105(2) & 0.128(2) & 0.129(1) & 0.137(2) & 0.143(2) & 0.161(3) & 0.215(3) & 0.349(4) \\
90\% & 0.109(1) & 0.104(3) & 0.115(3) & 0.123(3) & 0.136(3) & 0.173(4) & 0.165(1) & 0.170(3) & 0.197(3) & 0.217(4) & 0.283(4) & 0.421(4) \\
100\% & 0.306(5) & 0.398(28) & 0.370(15) & 0.386(22) & 0.429(14) & 0.582(19) & 0.291(5) & 0.330(17) & 0.455(22) & 0.478(19) & 0.548(21) & 0.750(23) \\
 \hline
\end{tabularx}
\caption{{\it Absolute error in 1D simulations.}
The absolute error ${|\langle \overline{Z}\rangle_\text{meas} - \langle \overline{Z}\rangle_\text{MPS}|}$ by decile (first column) is compared between unencoded runs (second and seventh columns) and encoded runs for various numbers of syndrome extraction rounds $R$ (columns 3-6 and 8-11).
The left columns show $g_z=0.5$ and the right columns show $g_z=1.5$.
The uncertainty is determined through bootstrap resampling.}
\label{dev_422:tab:chain_cdf}
\renewcommand{\arraystretch}{1.0}
\end{table}

\begin{table}[h]
\tiny
\setlength{\tabcolsep}{0pt}
\begin{tabularx}{\linewidth}{|c||Y|Y|Y|Y|Y|Y||Y|Y|Y|Y|Y|Y|}
\hline
\rule{0pt}{10pt} & \multicolumn{12}{c|}{ \large Signal-survival factor $\alpha (\sigma)$}\\\hline\hline
\rule{0pt}{10pt} \multirow{2}{*}{\makecell{$t$}} &  \multicolumn{6}{c||}{$g_z=0.5$}  &  \multicolumn{6}{c|}{$g_z=1.5$} \\\cline{2-13}
\rule{0pt}{10pt}  & Unenc. & $R=1$ & $R=2$ & $R=4$  & $R=8$ & $R=16$ & Unenc. & $R=1$ & $R=2$ & $R=4$  & $R=8$ & $R=16$ \\
\hline\hline
0.0 & 0.983(19) & 0.989(12) & 0.987(9) & 0.987(12) & 0.988(9) & 0.986(11) & 0.983(19) & 0.988(10) & 0.987(11) & 0.987(10) & 0.989(10) & 0.984(13) \\
0.5 & 0.931(34) & 0.955(44) & 0.959(44) & 0.947(46) & 0.953(42) & 0.954(46) & 0.929(36) & 0.954(39) & 0.955(44) & 0.955(44) & 0.956(40) & 0.959(35) \\
1.0 & 0.905(45) & 0.948(44) & 0.938(43) & 0.941(44) & 0.937(45) & 0.916(89) & 0.898(47) & 0.944(38) & 0.942(42) & 0.949(43) & 0.943(42) & 0.924(66) \\
1.5 & 0.878(65) & 0.918(79) & 0.893(85) & 0.897(76) & 0.897(73) & 0.870(110) & 0.877(61) & 0.913(71) & 0.908(74) & 0.911(74) & 0.927(62) & 0.922(89) \\
2.0 & 0.870(67) & 0.918(66) & 0.887(69) & 0.883(65) & 0.851(83) & 0.828(136) & 0.888(58) & 0.927(68) & 0.930(66) & 0.915(67) & 0.939(70) & 0.912(106) \\
2.5 & 0.841(78) & 0.875(80) & 0.841(80) & 0.871(80) & 0.823(98) & 0.771(129) & 0.913(55) & 0.960(58) & 0.966(60) & 0.951(71) & 0.983(67) & 0.953(84) \\
3.0 & 0.820(67) & 0.829(90) & 0.802(80) & 0.768(96) & 0.724(99) & 0.694(116) & 0.908(45) & 0.922(76) & 0.945(85) & 0.904(69) & 0.927(72) & 0.839(91) \\
3.5 & 0.777(74) & 0.834(53) & 0.822(62) & 0.802(70) & 0.757(83) & 0.692(95) & 0.893(40) & 0.918(55) & 0.934(55) & 0.914(65) & 0.864(72) & 0.713(77) \\
4.0 & 0.777(62) & 0.923(42) & 0.805(44) & 0.767(44) & 0.721(60) & 0.642(74) & 0.864(37) & 0.886(79) & 0.874(71) & 0.857(74) & 0.798(81) & 0.636(97) \\
4.5 & 0.755(57) & 0.766(47) & 0.876(56) & 0.723(46) & 0.739(55) & 0.639(83) & 0.824(44) & 0.859(72) & 0.858(67) & 0.792(82) & 0.731(116) & 0.539(120) \\
5.0 & 0.842(45) & 0.725(46) & 0.699(56) & 0.682(48) & 0.569(53) & 0.346(58) & 0.801(50) & 0.800(71) & 0.816(88) & 0.784(98) & 0.703(127) & 0.480(140) \\
5.5 & 0.898(51) & 0.592(49) & 0.636(58) & 0.568(49) & 0.448(48) & 0.265(49) & 0.776(61) & 0.810(78) & 0.766(94) & 0.723(107) & 0.645(145) & 0.430(146) \\
6.0 & 0.845(43) & 0.593(52) & 0.591(58) & 0.449(42) & 0.378(46) & 0.127(42) & 0.757(59) & 0.771(77) & 0.729(94) & 0.696(102) & 0.598(129) & 0.371(152) \\
6.5 & 0.822(47) & 0.572(53) & 0.529(51) & 0.327(49) & 0.291(41) & 0.157(38) & 0.750(64) & 0.753(87) & 0.731(86) & 0.670(113) & 0.603(141) & 0.345(146) \\
7.0 & 0.760(44) & 0.542(63) & 0.426(50) & 0.223(56) & 0.152(40) & 0.039(33) & 0.741(60) & 0.757(74) & 0.715(84) & 0.686(100) & 0.583(123) & 0.285(135) \\
7.5 & 0.802(46) & 0.481(45) & 0.465(48) & 0.259(41) & 0.179(34) & -0.019(22) & 0.737(58) & 0.750(74) & 0.701(82) & 0.687(90) & 0.568(126) & 0.255(119) \\
8.0 & 0.779(42) & 0.552(38) & 0.420(39) & 0.253(33) & 0.110(40) & 0.123(30) & 0.735(56) & 0.728(81) & 0.679(79) & 0.667(90) & 0.534(117) & 0.246(115) \\
\hline
\end{tabularx}
\caption{{\it Signal-survival factor for 1D simulations.}
The signal-survival factor $\alpha$ calculated as a function of time $t$ (first column) in unencoded results (second and seventh columns) is compared to encoded results for various numbers of syndrome extraction rounds $R$ (columns 3-6 and 8-11).
The left columns show $g_z=0.5$ and the right columns show $g_z=1.5$.
The uncertainty is the residual error $\sigma$ in the $\alpha$ fits.}
\label{dev_422:tab:chain_alpha_sigma}
\renewcommand{\arraystretch}{1.0}
\end{table}

\begin{table}[h]
\renewcommand{\arraystretch}{0.7}
\scriptsize
\begin{tabularx}{\linewidth}{|c||Y|Y|Y||Y|Y|Y|}
\hline
\rule{0pt}{10pt} & \multicolumn{6}{c|}{\large$\langle \overline{Z}_n\rangle $}\\\hline\hline
\rule{0pt}{10pt} \multirow{2}{*}{Logical qubit $(b,l_b)$} &  \multicolumn{3}{c||}{$g_z=0.5$}  &  \multicolumn{3}{c|}{$g_z=1.5$} \\\cline{2-7}
\rule{0pt}{10pt} & MPS & Unencoded & $R=1$ & MPS & Unencoded & $R=1$  \\
\hline\hline
$(13,1)$ & -0.558 & -0.113(6) & -0.257(16) & -0.093 & 0.094(6) & -0.147(34) \\
$(13,0)$ & -0.725 & -0.159(6) & -0.362(25) & -0.116 & 0.047(6) & -0.094(13) \\
$(10,1)$ & -0.694 & -0.134(6) & -0.357(9) & -0.111 & -0.003(5) & -0.123(13) \\
$(10,0)$ & -0.538 & -0.095(5) & -0.283(14) & -0.138 & -0.017(6) & -0.191(10) \\
$(9,0)$ & -0.524 & -0.139(5) & -0.069(24) & -0.375 & -0.142(5) & -0.127(17) \\
$(9,1)$ & -0.640 & -0.206(6) & -0.284(26) & -0.145 & -0.092(6) & 0.004(18) \\
$(4,0)$ & -0.603 & -0.253(5) & -0.388(17) & 0.005 & -0.078(5) & -0.102(13) \\
$(4,1)$ & -0.617 & -0.262(5) & -0.352(12) & -0.036 & -0.058(5) & -0.115(10) \\
$(2,0)$ & -0.706 & -0.210(5) & -0.418(17) & -0.134 & 0.017(6) & -0.223(28) \\
$(2,1)$ & -0.686 & -0.216(5) & -0.408(17) & -0.279 & -0.102(5) & -0.230(24) \\
$(0,0)$ & -0.627 & -0.271(6) & -0.180(44) & -0.087 & -0.110(6) & -0.119(14) \\
$(0,1)$ & -0.770 & -0.235(5) & -0.278(27) & -0.139 & -0.080(6) & -0.067(26) \\
$(1,0)$ & -0.732 & -0.205(5) & -0.103(24) & -0.137 & -0.025(5) & -0.085(30) \\
$(1,1)$ & -0.554 & -0.166(6) & -0.098(48) & -0.087 & -0.025(5) & -0.148(15) \\
$(3,1)$ & -0.432 & -0.089(6) & -0.228(29) & -0.165 & -0.075(5) & -0.153(32) \\
$(3,0)$ & -0.620 & -0.147(5) & -0.289(23) & -0.186 & -0.056(6) & -0.185(31) \\
$(7,1)$ & -0.457 & -0.037(6) & -0.298(32) & -0.043 & -0.081(6) & -0.115(30) \\
$(7,0)$ & -0.403 & -0.071(6) & -0.296(18) & -0.208 & -0.102(6) & -0.179(29) \\
$(5,0)$ & -0.249 & -0.075(6) & -0.199(19) & -0.283 & -0.097(5) & -0.288(10) \\
$(5,1)$ & 0.182 & 0.166(5) & 0.203(33) & 0.658 & 0.333(5) & 0.420(14) \\
$(6,1)$ & 0.597 & 0.201(5) & 0.579(22) & 0.695 & 0.242(5) & 0.652(29) \\
$(6,0)$ & 0.696 & 0.231(5) & 0.597(16) & 0.804 & 0.371(5) & 0.704(32) \\
$(11,0)$ & 0.762 & 0.343(5) & 0.642(12) & 0.845 & 0.479(5) & 0.656(15) \\
$(11,1)$ & 0.678 & 0.335(5) & 0.579(23) & 0.506 & 0.323(5) & 0.466(9) \\
$(8,0)$ & 0.330 & 0.149(5) & 0.285(23) & 0.486 & 0.267(5) & 0.442(15) \\
$(8,1)$ & -0.263 & -0.077(6) & -0.159(24) & -0.364 & -0.135(5) & -0.304(35) \\
$(12,1)$ & -0.199 & -0.129(6) & -0.182(15) & -0.268 & -0.286(5) & -0.234(18) \\
$(12,0)$ & -0.125 & -0.079(6) & -0.176(11) & -0.120 & -0.099(5) & -0.123(31) \\
$(15,1)$ & -0.219 & -0.083(5) & -0.147(49) & -0.215 & -0.125(6) & -0.216(32) \\
$(15,0)$ & 0.162 & 0.094(6) & 0.164(35) & 0.212 & 0.156(5) & 0.310(17) \\
$(14,1)$ & 0.520 & 0.248(5) & 0.493(12) & 0.751 & 0.350(5) & 0.655(16) \\
$(14,0)$ & 0.228 & 0.168(5) & 0.487(38) & 0.767 & 0.245(5) & 0.649(27) \\
$(17,0)$ & -0.412 & -0.095(6) & -0.288(27) & -0.212 & -0.007(6) & -0.172(13) \\
$(17,1)$ & -0.107 & -0.005(5) & -0.170(15) & -0.297 & -0.068(5) & -0.273(46) \\
$(20,0)$ & -0.289 & -0.065(6) & -0.215(11) & -0.127 & -0.144(6) & -0.013(39) \\
$(20,1)$ & -0.576 & -0.277(5) & -0.313(15) & -0.219 & -0.191(5) & -0.194(11) \\
$(19,0)$ & -0.679 & -0.374(5) & -0.434(10) & -0.102 & -0.081(6) & -0.078(39) \\
$(19,1)$ & -0.709 & -0.287(5) & -0.481(22) & -0.396 & -0.168(5) & -0.246(26) \\
$(16,1)$ & -0.742 & -0.351(5) & -0.618(25) & -0.365 & -0.256(6) & -0.249(20) \\
$(16,0)$ & -0.630 & -0.262(5) & -0.507(21) & -0.076 & -0.023(6) & -0.109(20) \\
$(18,0)$ & -0.282 & -0.092(5) & -0.227(25) & -0.261 & -0.230(6) & -0.298(10) \\
$(18,1)$ & -0.279 & -0.123(6) & -0.200(16) & -0.205 & -0.134(6) & -0.133(31) \\
 \hline
\end{tabularx}
\caption{{\it Magnetization in 2D simulations at $t=4$.}
For each logical qubit $(b,l_b)$ (first column), $\langle \overline{Z}_n\rangle$ from MPS results (second and fifth columns) is compared to unencoded results (third and sixth columns) and encoded results with $R=1$ syndrome extraction rounds (fourth and seventh columns).
The left columns show $g_z=0.5$ and the right columns show $g_z=1.5$.
The uncertainty is determined through bootstrap resampling.}
\label{dev_422:tab:grid_results_t_4}
\renewcommand{\arraystretch}{1.0}
\end{table}

\begin{table}[h]
\tiny
\setlength{\tabcolsep}{2pt}
\begin{tabularx}{\linewidth}{|c||Y|Y|Y|Y|Y|Y||Y|Y|Y|Y|Y|Y|}
\hline
\rule{0pt}{10pt} & \multicolumn{12}{c|}{\large ${|\langle \overline{Z}\rangle_\text{meas} - \langle \overline{Z}\rangle_\text{MPS}|}$}\\\hline\hline
\rule{0pt}{10pt} \multirow{2}{*}{\makecell{$\%$}} &  \multicolumn{6}{c||}{$g_z=0.5$}  &  \multicolumn{6}{c|}{$g_z=1.5$} \\\cline{2-13}
\rule{0pt}{10pt}  & Unenc. & $R=1$ & $R=2$ & $R=4$  & $R=8$ & $R=16$ & Unenc. & $R=1$ & $R=2$ & $R=4$  & $R=8$ & $R=16$ \\
\hline\hline
0\% & 0.000(1) & 0.000(0) & 0.000(0) & 0.000(0) & 0.000(0) & 0.000(0) & 0.000(0) & 0.000(0) & 0.000(0) & 0.000(0) & 0.000(0) & 0.000(0) \\
10\% & 0.047(2) & 0.014(1) & 0.013(1) & 0.016(1) & 0.016(1) & 0.021(2) & 0.016(1) & 0.011(1) & 0.011(1) & 0.010(1) & 0.011(1) & 0.015(1) \\
20\% & 0.109(1) & 0.035(2) & 0.033(2) & 0.037(2) & 0.046(2) & 0.063(3) & 0.042(1) & 0.022(1) & 0.021(1) & 0.022(1) & 0.024(1) & 0.029(2) \\
30\% & 0.163(2) & 0.063(2) & 0.060(2) & 0.067(2) & 0.081(2) & 0.097(3) & 0.072(1) & 0.036(2) & 0.035(1) & 0.036(2) & 0.041(2) & 0.052(2) \\
40\% & 0.215(2) & 0.091(2) & 0.090(2) & 0.102(2) & 0.118(2) & 0.150(3) & 0.102(1) & 0.052(2) & 0.049(2) & 0.051(2) & 0.057(2) & 0.072(2) \\
50\% & 0.281(2) & 0.123(2) & 0.131(3) & 0.137(3) & 0.165(3) & 0.205(4) & 0.134(1) & 0.067(2) & 0.066(2) & 0.065(2) & 0.074(2) & 0.101(2) \\
60\% & 0.347(2) & 0.172(3) & 0.173(3) & 0.193(3) & 0.222(4) & 0.300(4) & 0.177(1) & 0.087(2) & 0.086(2) & 0.084(2) & 0.096(2) & 0.132(3) \\
70\% & 0.403(2) & 0.255(4) & 0.257(4) & 0.277(4) & 0.315(5) & 0.421(5) & 0.222(1) & 0.113(2) & 0.113(2) & 0.110(2) & 0.123(2) & 0.173(3) \\
80\% & 0.456(2) & 0.342(4) & 0.349(5) & 0.373(4) & 0.425(5) & 0.505(4) & 0.279(2) & 0.146(2) & 0.142(2) & 0.147(3) & 0.162(3) & 0.231(3) \\
90\% & 0.527(2) & 0.448(4) & 0.467(4) & 0.495(5) & 0.548(5) & 0.599(4) & 0.398(2) & 0.200(4) & 0.206(4) & 0.222(4) & 0.236(5) & 0.341(6) \\
100\% & 0.669(5) & 0.667(21) & 0.672(20) & 0.747(36) & 0.766(25) & 0.792(34) & 0.763(6) & 0.518(27) & 0.501(15) & 0.535(26) & 0.540(19) & 0.855(28) \\
 \hline
\end{tabularx}
\caption{{\it Absolute error in 2D simulations.}
The absolute error ${|\langle \overline{Z}\rangle_\text{meas} - \langle \overline{Z}\rangle_\text{MPS}|}$ by decile (first column) is compared between unencoded runs (second and seventh columns) and encoded runs for various numbers of syndrome extraction rounds $R$ (columns 3-6 and 8-11).
The left columns show $g_z=0.5$ and the right columns show $g_z=1.5$.
The uncertainty is determined through bootstrap resampling.}

\label{dev_422:tab:grid_cdf}
\renewcommand{\arraystretch}{1.0}
\end{table}

\begin{table}[h]
\tiny
\setlength{\tabcolsep}{0pt}
\begin{tabularx}{\linewidth}{|c||Y|Y|Y|Y|Y|Y||Y|Y|Y|Y|Y|Y|}
\hline
\rule{0pt}{10pt} & \multicolumn{12}{c|}{ \large Signal-survival factor $\alpha (\sigma)$}\\\hline\hline
\rule{0pt}{10pt} \multirow{2}{*}{\makecell{$t$}} &  \multicolumn{6}{c||}{$g_z=0.5$}  &  \multicolumn{6}{c|}{$g_z=1.5$} \\\cline{2-13}
\rule{0pt}{10pt}  & Unenc. & $R=1$ & $R=2$ & $R=4$  & $R=8$ & $R=16$ & Unenc. & $R=1$ & $R=2$ & $R=4$  & $R=8$ & $R=16$ \\
\hline\hline
0.0 & 0.988(6) & 0.988(10) & 0.988(12) & 0.984(16) & 0.986(12) & 0.984(18) & 0.982(37) & 0.987(13) & 0.985(15) & 0.987(11) & 0.985(13) & 0.981(16) \\
0.5 & 0.862(41) & 0.947(37) & 0.948(41) & 0.943(40) & 0.943(44) & 0.944(44) & 0.850(73) & 0.944(44) & 0.942(45) & 0.946(42) & 0.949(33) & 0.940(43) \\
1.0 & 0.770(79) & 0.918(48) & 0.917(52) & 0.912(48) & 0.918(51) & 0.883(78) & 0.757(106) & 0.907(54) & 0.909(48) & 0.918(48) & 0.913(54) & 0.879(75) \\
1.5 & 0.688(76) & 0.896(63) & 0.885(66) & 0.899(70) & 0.861(71) & 0.833(104) & 0.662(109) & 0.892(63) & 0.888(68) & 0.884(65) & 0.881(72) & 0.864(90) \\
2.0 & 0.597(79) & 0.851(84) & 0.845(91) & 0.867(81) & 0.817(95) & 0.744(135) & 0.607(112) & 0.879(69) & 0.866(70) & 0.858(68) & 0.860(78) & 0.810(104) \\
2.5 & 0.524(72) & 0.791(112) & 0.787(106) & 0.773(109) & 0.738(129) & 0.664(151) & 0.553(98) & 0.860(78) & 0.855(64) & 0.854(77) & 0.850(76) & 0.762(102) \\
3.0 & 0.459(73) & 0.734(115) & 0.735(115) & 0.710(122) & 0.667(131) & 0.537(159) & 0.534(85) & 0.854(66) & 0.849(68) & 0.843(78) & 0.836(76) & 0.734(102) \\
3.5 & 0.411(68) & 0.675(126) & 0.684(125) & 0.642(117) & 0.590(140) & 0.434(170) & 0.487(72) & 0.836(72) & 0.845(69) & 0.812(63) & 0.830(68) & 0.725(78) \\
4.0 & 0.348(63) & 0.606(133) & 0.627(123) & 0.555(133) & 0.492(137) & 0.323(153) & 0.474(66) & 0.834(68) & 0.831(74) & 0.814(71) & 0.778(70) & 0.620(89) \\
4.5 & 0.305(60) & 0.551(124) & 0.519(124) & 0.506(130) & 0.387(144) & 0.241(144) & 0.416(61) & 0.788(65) & 0.764(77) & 0.768(73) & 0.730(82) & 0.547(95) \\
5.0 & 0.251(60) & 0.485(121) & 0.452(121) & 0.412(115) & 0.337(139) & 0.209(127) & 0.370(60) & 0.722(64) & 0.703(62) & 0.725(69) & 0.645(86) & 0.417(83) \\
5.5 & 0.215(54) & 0.418(116) & 0.395(115) & 0.374(127) & 0.270(128) & 0.143(119) & 0.322(57) & 0.660(74) & 0.668(66) & 0.621(71) & 0.592(94) & 0.321(90) \\
6.0 & 0.174(53) & 0.399(113) & 0.349(110) & 0.296(109) & 0.248(130) & 0.097(74) & 0.280(58) & 0.618(74) & 0.609(68) & 0.591(67) & 0.558(84) & 0.192(64) \\
6.5 & 0.149(48) & 0.328(99) & 0.321(108) & 0.267(116) & 0.204(110) & 0.087(80) & 0.241(54) & 0.561(70) & 0.575(72) & 0.598(81) & 0.521(87) & 0.123(66) \\
7.0 & 0.119(45) & 0.296(100) & 0.276(95) & 0.229(100) & 0.160(108) & 0.047(47) & 0.206(52) & 0.535(74) & 0.543(69) & 0.531(81) & 0.479(84) & 0.117(43) \\
7.5 & 0.095(41) & 0.263(82) & 0.266(100) & 0.191(96) & 0.133(94) & 0.050(48) & 0.180(50) & 0.505(62) & 0.521(66) & 0.503(79) & 0.444(81) & 0.072(36) \\
8.0 & 0.074(38) & 0.243(87) & 0.207(81) & 0.167(95) & 0.124(75) & 0.035(33) & 0.161(48) & 0.481(64) & 0.482(52) & 0.471(67) & 0.405(78) & 0.055(36) \\
 \hline
\end{tabularx}
\caption{{\it Signal-survival factor for 2D simulations.}
The signal-survival factor $\alpha$ calculated as a function of time $t$ (first column) in unencoded results (second and seventh columns) is compared to encoded results for various numbers of syndrome extraction rounds $R$ (columns 3-6 and 8-11).
The left columns show $g_z=0.5$ and the right columns show $g_z=1.5$.
The uncertainty is the residual error $\sigma$ in the $\alpha$ fits.}
\label{dev_422:tab:grid_alpha_sigma}
\renewcommand{\arraystretch}{1.0}
\end{table}

\begin{table}[h]
\footnotesize
\begin{tabularx}{\linewidth}{|c||Y|Y||Y|Y|}
\hline
\rule{0pt}{10pt} & \multicolumn{4}{c|}{ \large Acceptance rate}\\\hline\hline
\rule{0pt}{10pt} \multirow{2}{*}{\makecell{$t$}} &  \multicolumn{2}{c||}{1D}  &  \multicolumn{2}{c|}{2D} \\\cline{2-5}
\rule{0pt}{10pt}  & $g_z=0.5$ & $g_z=1.5$ & $g_z=0.5$ & $g_z=1.5$ \\
\hline\hline
 $0.0$ & $0.341 \pm 0.339$ & $0.436 \pm 0.347$ & $0.414 \pm 0.355$ & $0.289 \pm 0.314$ \\
 $0.5$ & $0.372 \pm 0.309$ & $0.292 \pm 0.310$ & $0.312 \pm 0.273$ & $0.268 \pm 0.280$ \\
 $1.0$ & $0.307 \pm 0.291$ & $0.343 \pm 0.286$ & $0.339 \pm 0.299$ & $0.298 \pm 0.261$ \\
 $1.5$ & $0.324 \pm 0.300$ & $0.338 \pm 0.264$ & $0.303 \pm 0.228$ & $0.247 \pm 0.251$ \\
 $2.0$ & $0.308 \pm 0.264$ & $0.241 \pm 0.246$ & $0.229 \pm 0.201$ & $0.252 \pm 0.231$ \\
 $2.5$ & $0.297 \pm 0.261$ & $0.294 \pm 0.253$ & $0.226 \pm 0.196$ & $0.274 \pm 0.221$ \\
 $3.0$ & $0.295 \pm 0.281$ & $0.321 \pm 0.258$ & $0.208 \pm 0.188$ & $0.205 \pm 0.183$ \\
 $3.5$ & $0.260 \pm 0.227$ & $0.293 \pm 0.214$ & $0.200 \pm 0.167$ & $0.233 \pm 0.197$ \\
 $4.0$ & $0.235 \pm 0.201$ & $0.249 \pm 0.241$ & $0.176 \pm 0.147$ & $0.208 \pm 0.191$ \\
 $4.5$ & $0.226 \pm 0.209$ & $0.228 \pm 0.177$ & $0.255 \pm 0.179$ & $0.169 \pm 0.160$ \\
 $5.0$ & $0.254 \pm 0.230$ & $0.310 \pm 0.224$ & $0.196 \pm 0.162$ & $0.249 \pm 0.199$ \\
 $5.5$ & $0.210 \pm 0.151$ & $0.183 \pm 0.148$ & $0.255 \pm 0.185$ & $0.211 \pm 0.172$ \\
 $6.0$ & $0.246 \pm 0.196$ & $0.226 \pm 0.203$ & $0.200 \pm 0.177$ & $0.218 \pm 0.141$ \\
 $6.5$ & $0.210 \pm 0.188$ & $0.231 \pm 0.197$ & $0.229 \pm 0.178$ & $0.216 \pm 0.165$ \\
 $7.0$ & $0.209 \pm 0.195$ & $0.214 \pm 0.197$ & $0.139 \pm 0.081$ & $0.227 \pm 0.155$ \\
 $7.5$ & $0.228 \pm 0.190$ & $0.233 \pm 0.207$ & $0.173 \pm 0.136$ & $0.173 \pm 0.124$ \\
 $8.0$ & $0.197 \pm 0.161$ & $0.187 \pm 0.179$ & $0.201 \pm 0.158$ & $0.171 \pm 0.129$ \\
 \hline
\end{tabularx}
\caption{{\it Acceptance rates of $R=1$ encoded simulations}.
The acceptance rates are shown for both 1D simulations (second and third columns) and 2D simulations (fourth and fifth columns) as a function of time $t$ (first column).
These encoded simulations use $R=1$ syndrome extraction rounds, corresponding to the data reported in the main text. 
The uncertainty is the standard deviation over all qubits for each $t$.}
\label{dev_422:tab:acceptance_rates}
\renewcommand{\arraystretch}{1.0}
\end{table}

\end{subappendices}

\chapter{Reflections on quantum simulation}
\label{chap:conclusion}
\noindent
This is a very exciting time to work in this field, and progress is continuously accelerating.
Over the course of the work in this thesis, I have participated in the evolution of quantum computers from the early NISQ era with tens of qubits to the crossover between NISQ and FT, with hundreds of qubits.
Two-qubit gate error rates have dropped to $10^{-4}$, and T1/T2 coherence times have steadily increased. 
Each time I returned to run a new project on hardware, I was struck by how much better the current generation of devices had become since the last.
Practical quantum computation, once a distant goal, is becoming a reality.
Industry roadmaps toward FT are growing more concrete~\cite{quantinuum_roadmap,ionq_roadmap,ibm_quantum_roadmap,google_quantum_roadmap,quera_roadmap,rigetti_roadmap_2025,atomcomputing_whitepaper_2025}, with several companies already demonstrating tens of logical qubits and many planning to reach thousands within the next several years.
Error rates are projected to continue to drop to $10^{-9}-10^{-13}$ with the incorporation of robust FT.
At the same time, the resource requirements for FT itself continue to drop, as better codes, better error rates, and better architectures chip away at the overhead once thought necessary~\cite{Cain:2026rmb,Gidney:2025kie,Babbush:2026nhx,Tripier:2026mfv,Yoder:2025ooz,Bluvstein:2025ped}.
It is exciting to contribute to the research that pushes the current crossover regime closer to full FT.

Hardware progress has been accompanied by progress in algorithms, with state preparation and time evolution methods relevant to fundamental physics improving the efficiency of quantum simulations. 
These efforts collectively attack the problem from opposite ends, quickly narrowing the gap between present-day demonstrations and practical quantum computation.

Our understanding of what makes a given simulation easy or hard, and precisely why, has developed considerably over the course of this thesis. 
None of this progress and understanding would have been possible without working directly with real devices.
The specific behavior of noise on actual hardware significantly shaped the design of the algorithms developed throughout this thesis.
The challenge has shifted from simply confronting the effects of noise to confronting the combined effects of noise and the dynamics being simulated, a shift that itself reflects the improving device landscape. 
This understanding was deepened further by our investigation of quantum information in physical systems. 
Studying entanglement and magic directly sharpened our picture of the physical processes themselves.
In doing so, it also clarified where a simulation's bottlenecks are expected to arise and how noise will interact with them.
The reverse has also proven true. 
Figuring out how to translate a physical system onto a quantum device has in many cases surfaced genuine physical insight that would not have been found otherwise.

Noise is the thread connecting hardware progress, algorithmic progress, and the physical understanding gained along the way, and as error rates and error-management strategies continue to improve, I look forward to the quantum simulation advances the coming years will bring.
As the frontier of quantum computation continues to evolve, I believe quantum simulations are rapidly approaching the point of surpassing the most capable classical methods, and will soon enable scientific discovery in fundamental physics.

\bibliography{bibi}

\end{document}